# A Laser Phase Plate for Transmission Electron Microscopy

by

Jeremy Joseph Axelrod

A dissertation submitted in partial satisfaction of the

requirements for the degree of

Doctor of Philosophy

in

Physics

in the

Graduate Division

of the

University of California, Berkeley

Committee in charge:

Professor Holger Müller, Chair
Professor Alex Zettl
Professor Eva Nogales

Spring 2024

A Laser Phase Plate for Transmission Electron Microscopy

# Abstract

A Laser Phase Plate for Transmission Electron Microscopy

by

Jeremy Joseph Axelrod

Doctor of Philosophy in Physics

University of California, Berkeley

Professor Holger Müller, Chair

Low image contrast is a major limitation in transmission electron microscopy, since samples with low atomic number only weakly phase-modulate the illuminating electron beam, and beam-induced sample damage limits the usable electron dose. The contrast can be increased by converting the electron beam's phase modulation into amplitude modulation using a phase plate, a device that applies a $\pi/2$ radian phase shift to part of the electron beam after it has passed through the sample. Previous phase plate designs rely on material placed in or near the electron beam to provide this phase shift. This results in image aberrations, an inconsistent time-varying phase shift, and resolution loss when the electron beam charges, damages, or is scattered from the material.

In this thesis, I present the theory, design, and implementation of the laser phase plate, which instead uses a focused continuous-wave laser beam to phase shift the electron beam. A near-concentric Fabry-Pérot optical cavity focuses and resonantly enhances the power of the laser beam in order to achieve the high intensity required to provide the phase shift. We demonstrate that the cavity can surpass this requirement and generate a record-high continuous-wave laser intensity of $590\,\mathrm{GWcm^{-2}}$. By integrating the cavity into a transmission electron microscope, we show that the ponderomotive potential of the laser beam applies a spatially selective phase shift to the electron beam. This enables us to make the first experimental observation of the relativistic reversal of the ponderomotive potential.

We then theoretically analyze the properties of the contrast transfer function generated by the laser phase plate. We experimentally determine that resolution loss caused by thermal magnetic field noise emanating from electrically conductive materials in the cavity can be eliminated by designing the cavity with a sufficiently large electron beam aperture. Finally, we show that the laser phase plate provides a stable $\pi/2$ phase shift and concomitant contrast enhancement when imaging frozen hydrated biological macromolecules. We use these images to successfully determine the structure of the molecules. This demonstrates the laser phase plate as the first stable and lossless phase plate for transmission electron microscopy.

To my parents.

*We have friends in other fields—in biology, for instance. We physicists often look at them and say, "You know the reason you fellows are making so little progress?" (Actually I don't know any field where they are making more rapid progress than they are in biology today.) "You should use more mathematics like we do." They could answer us—but they're polite, so I'll answer for them:*

*"What you should do in order for us to make more rapid progress is to make the electron microscope 100 times better."*

Richard P. Feynman
In the lecture "There's Plenty of Room at the Bottom"
Annual American Physical Society Meeting
Caltech, USA
December 29, 1959

# Contents

## Acknowledgments

The work presented in this thesis represents the collective effort of many people, exerted over many years. These acknowledgments should serve as a reminder that progress in science comes from cooperation and collaboration, and that the achievements of an individual are always smaller than those of the greater whole.

The following people and organizations directly contributed to this project. Their contributions were in turn supported by many others, too numerous to individually name. This list is surely incomplete—any omissions are due to my forgetfulness, or ignorance of the scope of the project.

My advisor Professor Holger Müller and our primary collaborator Professor Robert M. Glaeser are both remarkable scientists and, more importantly, wonderfully kind people. Their technical knowledge and collaborative spirit enabled this unusually interdisciplinary project, and made my work as fun and stress-free as I could possibly imagine! Thank you both.

I had the pleasure of working alongside several excellent postdocs. Osip Schwartz's excellent mentorship shaped me as a young scientist. Just the two of us worked together in the first few years of the project, when we did not yet know if the project was technically feasible. We developed a deeply satisfying rapport which allowed us to quickly exchange ideas and make consensus decisions. His ideas laid the foundation for the project.

Sara L. Campbell brought to our project a remarkable energy and depth of practical experience in laser optics. She taught me the right way to do things, inspired me with her work ethic, and put the project on a trajectory to success.

Petar N. Petrov has a deep understanding of microscopy, and is equally good at applying this understanding theoretically and experimentally. Petar has been the source of many new ideas about how to use the laser phase plate, only some of which are discussed in this work. I am excited to see more of his ideas implemented in the future!

Jessie T. Zhang is the most adaptable experimentalist I have met. Though she had no background in transmission electron microscopy, within just a few months of joining our group she had learned what had taken me years. The future of the project is in good hands!

Electron microscopy specialist Jonathan Remis is largely responsible for the success of this project. His expertise in transmission electron microscopy and cryo-electron microscopy has saved the rest of the team uncountable hours of potential frustration and roadblock. Better still, he combines his experience with a contagious and inspirational enthusiasm which makes him very fun to work with.

Visiting scholar Hang Cheng has given the team perspective on the biological applications of the laser phase plate, and will be one of the first researchers to apply it to cryo-electron tomography. He is kind, enthusiastic, and (along with his partner) a great cook!

Graduate student Jeske Y. Dioquino contributed to the construction of our first successful prototype laser phase plate, and laid the groundwork for more advanced mirror inspection techniques which we are just now starting to more fully explore. The entire team was selfishly disappointed to see Jeske leave to pursue an education in science communication, but also excited because we all recognize her astronomical potential in that field.

This project has benefited from the remarkable contributions of several undergraduate researchers. Carter Turnbaugh may as well have been a graduate student for his contributions to the project. He wrote nearly all of the experiment control software, and assisted in virtually every other aspect of the project, including collecting now-published data. I still don't understand how he was able to do all these things while also acing a full undergraduate course load. I am equally impressed by his professionalism and polite demeanor, and am very happy I had the opportunity to work with him.

Shahar Sandhaus has single-handedly and very professionally maintained the computer hardware and software infrastructure of our experiment. As a post-baccalaureate researcher, he is now developing machine-learning image analysis methods for use with the laser phase plate.

Daniel R. Tuthill worked with Osip Schwartz and me very early on in the project, and helped us take initial data showing that the Fabry-Pérot cavity laser phase plate concept was workable.

Other members of the Müller group have contributed either directly or indirectly to the project. Postdoc Philipp Haslinger gave invaluable optics advice early on in the project. I'd also like to thank the truly brilliant graduate students I shared an office with: Matt Jaffe, Victoria Xu, and Ashwin Singh. Our wonderfully open-ended conversations were always extremely useful, wide-ranging, and fun.

This project would not have been successful without the physics department research and development machine shop. Warner T. Carlisle, Gordon W. Long, Abel Gonzalez, Thomas C. Gutierrez, Jesus Lopez, and Jonathan Bradford not only fabricated difficult parts for us, but also helped us immensely with our designs. I would like to thank Gordon W. Long in particular for spending so much time teaching me about machining, when I'm sure he had many other things to do.

The project is similarly indebted to the engineers and technicians at our optics suppliers Laseroptik GmbH, Perkins Precision Developments, Coastline Optics, and FiveNine Optics, who all contributed to making our remarkable cavity mirrors. A special thanks to Emily Kubacki (Perkins Precision Developments), John Tardif (Coastline Optics), and Rob Patterson (FiveNine Optics) for accommodating our odd requests and being willing to try new things.

Our collaboration with Thermo Fisher Scientific has been invaluable in giving us access to their highly experienced research and development team, who know just about everything there is to know about transmission electron microscopy, and who were a pleasure to work with: Bart Buijsse, Pleun Dona, and Afric S. Meijer. Similarly, our local TEM service engineers Sheldon Goobie and Trevor W. Coyle were always knowledgeable, accommodating, and fun to learn from.

A huge thank-you to Professor Eva Nogales, her group, and in particular research specialist Patricia Grob for allowing much of this project to take place in and around their lab space. Without their generous accommodations this project would not have been possible.

Professor Radostin Danev (University of Tokyo), a pioneer of TEM phase plate design, has been a great friend to this project, both as an advisor and advocate. His work has paved a much easier path for us to follow.

Several colleagues from the Lawrence Berkeley National Laboratory have supported this project. Colin Ophus simulated laser phase plate TEM images early on in the project. Howard Padmore performed several profilometry measurements on our cavity mirrors, which were very useful in determining the best mirror manufacturing method. Grant Cutler simulated the thermoelastic deformation of our mirror substrates.

This project has heavily drawn on optics knowledge generated by the Laser Interferometer Gravitational-Wave Observatory collaboration, especially through their publicly-available internal documentation. Several members of the collaboration have also given us extremely helpful advice: Professor Rana X. Adhikari, Dr. Aidan Brooks, Dr. Matthew C. Heintze, Dr. Margot Hennig, and GariLynn Billingsley.

The idea of using a laser to create a phase plate for TEM was originally conceived by Dr. Bryan Reed at Lawrence Livermore National Laboratory. I thank him for his idea which I have had so much fun following up on, and also for providing me with the story of how he came up with the idea.

The confluence of ionizing and non-ionizing radiation hazards in this project gave me the opportunity to work closely with UC Berkeley laser safety officer Eddie Ciprazo and radiation safety specialist/deputy laser safety officer Phil Broughton. Their goal was very clearly to keep us safe, not just fill out forms to cover the university's liability. I am glad I got to learn from their wild stories, and by so doing, not end up in one of them.

Funding for this project was provided by (in chronological order) the National Institutes of Health, the Gordon and Betty Moore Foundation, Thermo Fisher Scientific, and the Chan Zuckerberg Initiative. I was personally supported through the National Science Foundation Graduate Research Fellowship Program from 2017 to 2022. Professor David Agard, Dr. Bridget Carragher, and Dr. Clint Potter at the Chan Zuckerberg Imaging Institute have been particularly supportive of our project. Dr. Shawn Zheng at the Chan Zuckerberg Imaging Institute has provided very helpful advice on TEM image motion correction. Dr. Pavel Olshin at the Chan Zuckerberg Imaging Institute currently helps our team build the next generation of laser phase plate prototypes.

I'd also like to thank my fellow members of UAW Local 2865 for organizing and participating in the 2022 University of California academic workers' strike. Ironically, those 6 weeks away from the lab may have been the most impactful of my PhD.

Finally, thank you to my thesis committee (Professors Holger Müller, Alex Zettl, and Eva Nogales) for making this thesis better through their review and edits.

# Definitions

| | |
|---|---|
| $\simeq$ | Quantities are approximately equal, and exactly equal in some limit |
| $:=$ | Defines variables on the left-hand side of the equation |
| $=:$ | Defines variables on the right-hand side of the equation |
| $\pi$ | Pi, 3.14159 . . . |
| $e$ | Euler's number, 2.71828 . . . |
| $i$ | Imaginary unit, "$\sqrt{-1}$" |
| $\hbar$ | Reduced Planck constant, $6.626\,070\,15 \times 10^{-34}\,\mathrm{J\,s}/\,(2\pi)$ |
| $c$ | Speed of light, $299\,792\,458\,\mathrm{ms}^{-1}$ |
| $\alpha$ | Fine-structure constant, $0.007\,297\,352\,569\,3(11)$ |
| $\mathrm{e}$ | Elementary charge, $1.602\,176\,634 \times 10^{-19}\,\mathrm{C}$ |
| $m_{\mathrm{e}}$ | Electron mass, $9.109\,383\,701\,5(28) \times 10^{-31}\,\mathrm{kg}$ |

# Chapter 1

# Introduction

## 1.1 Phase contrast transmission electron microscopy

Transmission electron microscopy (TEM) extends the capabilities of optical microscopy by using particles (electrons) with much shorter wavelength than optical photons to provide resolutions sufficient to visualize individual atoms. A TEM works by sending a beam of electrons through a thin (typically less than a few hundred nanometers) sample which scatters the beam. The transmitted beam is imaged onto a detector using several electron lenses, which use static electric or magnetic fields to guide the electrons. The detector records the flux of the electron beam as a function of position in the image plane. Regions of the image have higher or lower flux depending on how the electron beam has scattered off the corresponding area in the sample, and how the electron beam has propagated between the sample and the image plane.

For example, blocking some of the scattered electrons before they reach the detector forms an image which has darker regions where there is more scattering in the sample. This is referred to as "amplitude contrast" [1]. However, while blocking more of the scattered electrons increases the contrast, it also decreases the image resolution, because the scattered electrons carry information about the structure of the sample. Blocking few enough scattered electrons to still achieve near-atomic resolution results in weak image contrast, especially for samples containing mostly atoms with a low atomic number, since they scatter more weakly.

If the illuminating electrons are coherent enough to manifest wave-like properties then image contrast is generated by interference of the elastically scattered and unscattered parts of each electron's wave function. This is called "phase contrast", since the contrast is proportional to the phase shift imposed on the illuminating electron wave function by the electromagnetic fields in the sample. Unlike amplitude contrast, phase contrast does not require any of the electrons to be blocked, and can achieve atomic resolution while simultaneously generating the maximum possible contrast for a given amount of scattering.

Realizing phase contrast requires an additional phase shift to be applied to part of the electron's wave function after passing through the sample in order to cause constructive or destructive interference of the scattered and unscattered waves in the image plane. This can be partially accomplished by defocusing the image, which causes the phase of the scattered wave to increase as a function of scattering angle. This means that only some of the scattered wave components will interfere with the unscattered wave, and so only some additional contrast is generated. Regardless, defocus phase contrast is commonly used because defocusing the image is easily done by changing the focal length of the lenses in the TEM or (less commonly) moving the sample along the electron beam.

Optimal phase contrast can be achieved by instead phase shifting the unscattered wave by $\pi/2$ radians with no defocus such that the entirety of the scattered and unscattered waves interfere constructively. This can be done by applying the phase shift to a location in the TEM where the unscattered wave is focused (see figure 1.1).

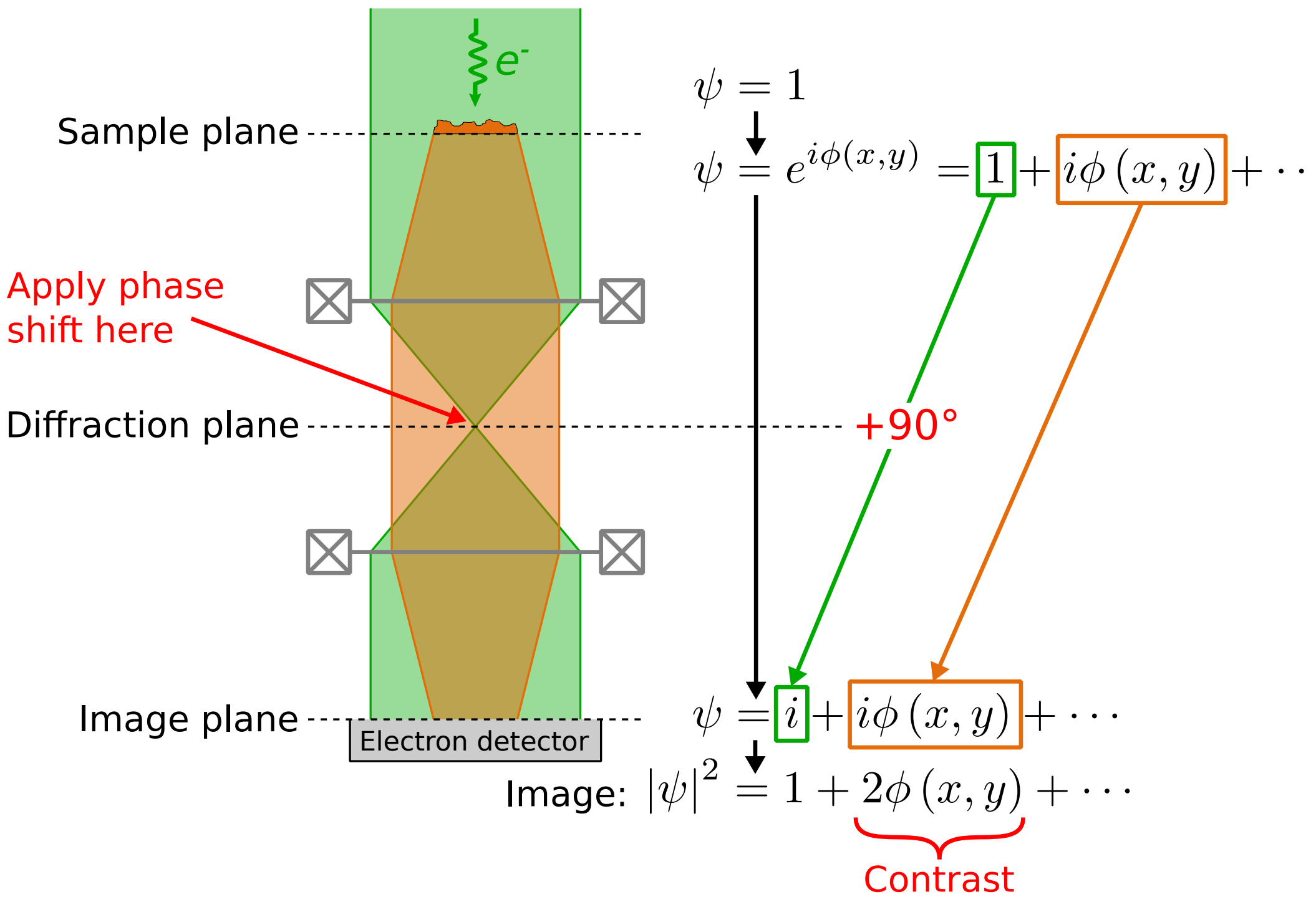

Figure 1.1: The principle of phase contrast TEM imaging. A planar electron wave function (denoted by $\psi$) is incident on a sample. Propagation through the sample spatially phase-modulates the electron wave (denoted by $\phi$). A phase plate applies a 90° phase shift to the unscattered component of the wave (green) relative to the unscattered wave (orange), which causes them to constructively interfere in the image plane, generating an image with phase contrast. Reproduced from [2].

This method of phase contrast imaging was originally developed for optical microscopy in the early 1930s by Frits Zernike, was awarded a Nobel prize in 1953 [3], and is now widely-used, especially for label-free imaging of live cells. The object which applies the phase shift is called a "phase plate", since in optical microscopy the phase shift is applied using a plate of glass. Reliably phase shifting the unscattered wave in TEM is not as trivial. Many phase plate designs have been tested since originally proposed by Boersch in 1947 [4]–[6]. All of these designs rely on putting material in the electron beam to provide an electrostatic or magnetostatic potential to phase shift the unscattered wave (or equivalently, phase shift the entire scattered wave). For example, a thin film with a small hole at the location of the unscattered wave (called a "Zernike phase plate") applies a phase shift to the entire scattered wave due to the attractive electrostatic atomic potentials of the carbon film [6]. The phase shift can be set to $\pi/2$ by choosing the appropriate thickness of film. A microfabricated ring electrode can be used to phase shift the unscattered wave by applying a voltage between the electrode and its surroundings [7]. The phase shift can be tuned by changing the voltage. Similarly, the magnetic vector potential of a magnetized microtoroid will phase shift the wave component which passes through its center via the Aharonov-Bohm effect [8], [9]. However, in all of these cases, the presence of the material in the electron beam causes unwanted side-effects. First, electrons scatter off of the material and no longer contribute to coherent image formation. Second, exposure to the electron beam may damage the device, either changing its phase shifting properties or rendering it completely inoperable. Third, the electron beam will inevitably charge the material, and the electrostatic potential of that charge will induce its own phase shifts in the electron beam.

Electron beam induced charging of the phase plate material is actually exploited in the "Volta phase plate" (VPP), in which a continuous carbon film (like a Zernike phase plate, but with no hole) is placed in the diffraction plane [10], [11]. The focus of the unscattered wave negatively charges a small spot on the film, and the electrostatic potential of this charge phase shifts the unscattered wave. The charge accumulates as more and more electrons pass through the film, which causes the phase shift to increase. In practice, this means that only several images can be collected with a VPP before the phase shift increases far enough beyond $\pi/2$ that the phase contrast is noticeably diminished. The carbon film can then be moved by $\sim 100\,\mu\text{m}$ in the diffraction plane to expose a "fresh" spot on the film, and the process starts over. Regardless, this results in series of images which each have different phase shifts which fluctuate around $\pi/2$ as the film charges and is then moved to a new spot (see figure 1c of [12]). A more serious problem comes from scattering in the film, which substantially reduces the resolution of the TEM when the VPP is used [13]. These issues have limited the applicability of the VPP, and though it is commercially available, it is not widely used.

## 1.2 The laser phase plate

The problems associated with material-based phase plates can be avoided by using a tightly-focused laser beam to phase shift the unscattered wave (see figure 1.2). Such a "laser phase plate" (LPP) consists only of photons, which cannot be charged or damaged by the electron beam. The phase shift arises from the "ponderomotive potential" of the laser beam, which is a repulsive effective potential proportional to the intensity of the laser beam. The high intensity at the focus phase shifts the unscattered wave, while the scattered wave mostly passes around both sides of the beam, or experiences only a small phase shift where it passes through the beam away from the focus.

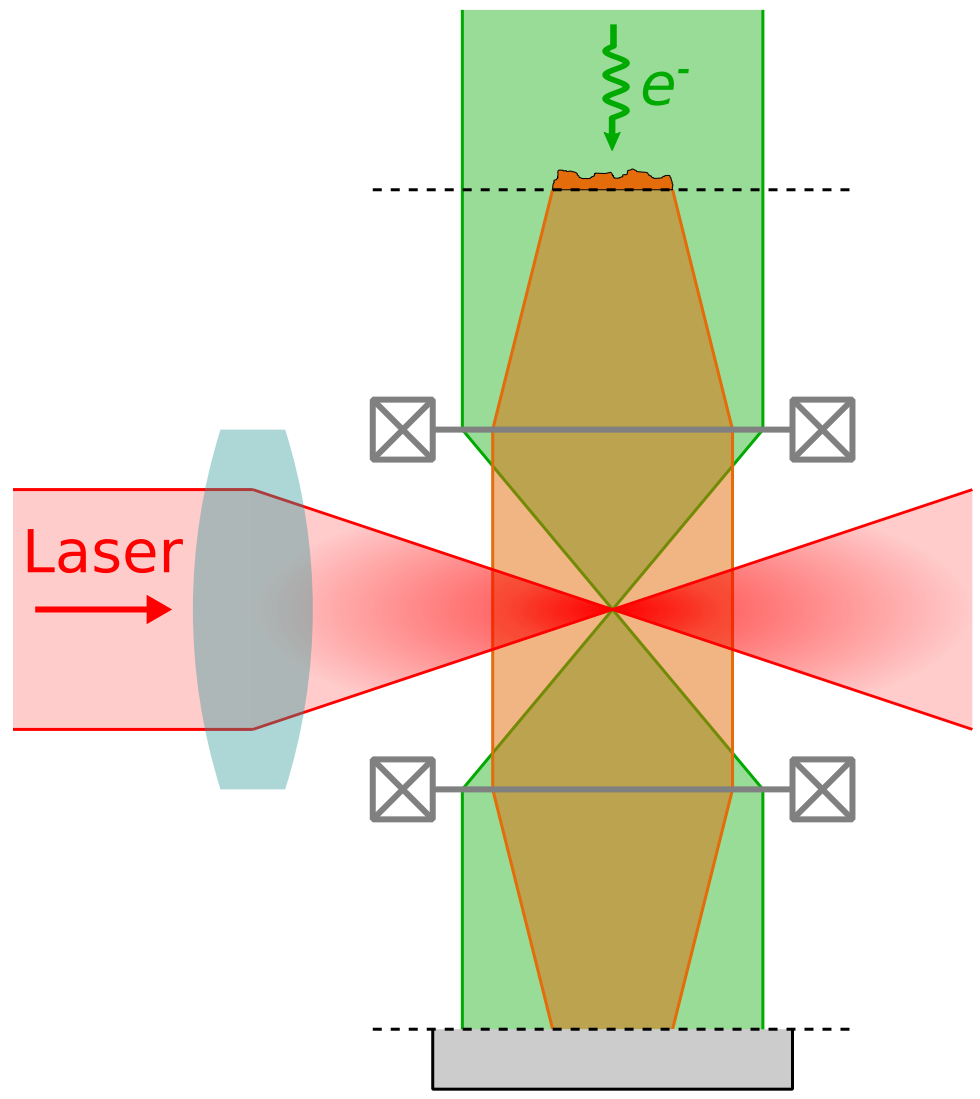

Figure 1.2: The laser phase plate concept. The focus of a laser beam intersects the focus of the unscattered wave to provide the requisite 90° phase shift for phase contrast imaging.

The ponderomotive potential is relatively weak in comparison to the electrostatic potential of a bulk material. Generating a $\pi/2$ phase shift in a small focal spot requires focal intensities of tens or even hundreds of $\mathrm{GW cm^{-2}}$, i.e. kilowatts of laser power focused to a width of several microns. There are two general approaches to achieving these intensities.

### 1.2.1 Pulsed

These intensities are readily achievable over brief ($< 1\,\mathrm{ns}$) periods using commercially available pulsed laser systems. However, if a pulsed laser beam is used, the electron source of the TEM must also be pulsed and synchronized with the LPP so that the phase shift is applied to all electrons. The primary challenge facing a pulsed LPP design is achieving a sufficiently high electron source brightness. Pulsed electron sources, whether using photoemission or RF chopping, can only operate with an average of $\sim 1$ electron per pulse, since Coulomb repulsion between multiple electrons within a pulse causes spatiotemporal decoherence of the beam, resulting in loss of image resolution [14]–[16]. Longer pulses reduce the effect, but increase the pulse energy requirement to achieve a $\pi/2$ phase shift [17]. As such, pulsed source TEMs have so far only demonstrated resolutions of $\sim 1\,\mathrm{nm}$ [18] compared with 0.2 nm or better when using a continuous source. The source brightness is therefore effectively limited by the pulse repetition rate of the pulsed LPP, since each electron pulse must intersect with a phase plate pulse. State-of-the-art pulsed sources have a brightness of $\sim 3.6 \times 10^{20}$ electron sr$^{-1}$m$^{-2}$pulse$^{-1}$ [18]–[20] compared to continuous sources (Thermo Fisher Scientific X-FEG) at $\sim 1.9 \times 10^{32}$ electron sr$^{-1}$m$^{-2}$s$^{-1}$. Therefore, unless the coherence of pulsed sources can be improved, achieving similar coherence and flux as a continuous source requires an extremely (perhaps impossibly) high pulse repetition rate of $5.3 \times 10^{11}$ Hz.

### 1.2.2 Continuous-wave

Generating high enough intensity using a continuous-wave (CW) laser is difficult. High enough power CW lasers do exist (in particular, $CO_2$ lasers with a wavelength of $\sim 10\,\mu$m) but are difficult to focus tightly enough, and also could easily damage the TEM if a malfunction were to misdirect the laser beam. The less dangerous but more complicated approach is to use an optical resonator to passively enhance the power in the laser beam—this is the approach we have taken in the Müller group [21]–[24]. We use one of the simplest types of optical resonator, the Fabry-Pérot cavity, which consists of two curved mirrors facing each other (see figure 1.3). The laser beam is sent through one of the mirrors and then bounces between them many times before it is lost to absorption, scattering, or transmission on the mirrors. The power circulating between the mirrors can be many times greater than that sent in because the beam overlaps on itself many times. It is therefore possible to achieve many kilowatts of circulating power by only sending in several watts. Very little energy is actually stored in the circulating laser beam, and so the laser beam cannot deliver enough optical power to potentially damage the TEM. The challenge of this approach is that it requires highly reflective ($\sim 99.99\%$) mirrors which can withstand the high circulating power, and these mirrors must be aligned relative to one another extremely precisely in order to generate a small enough focal waist at the center of the Fabry-Pérot cavity. Indeed, the required laser focal intensity is higher than has ever been achieved elsewhere for any continuous-wave laser beam. Specifically, the phase shift generated by the ponderomotive potential is proportional to the laser wavelength, electron de Broglie wavelength (1.97 pm for the commonly-used electron kinetic energy of 300 keV), numerical aperture of the laser beam focus (i.e. inversely proportional to the focal width), and circulating power in the laser beam (see equation (9.26)). For reasons discussed in chapter 3, we use a laser wavelength of $\lambda_L = 1064$ nm. As explained in chapter 2, it is difficult to use a Fabry-Pérot optical cavity to achieve a focal numerical aperture larger than $\sim 0.1$. We found that a numerical aperture of 0.05 was achievable. This therefore requires a circulating power of 75 kW to achieve a $\pi/2$ phase shift (see equation (9.26)).

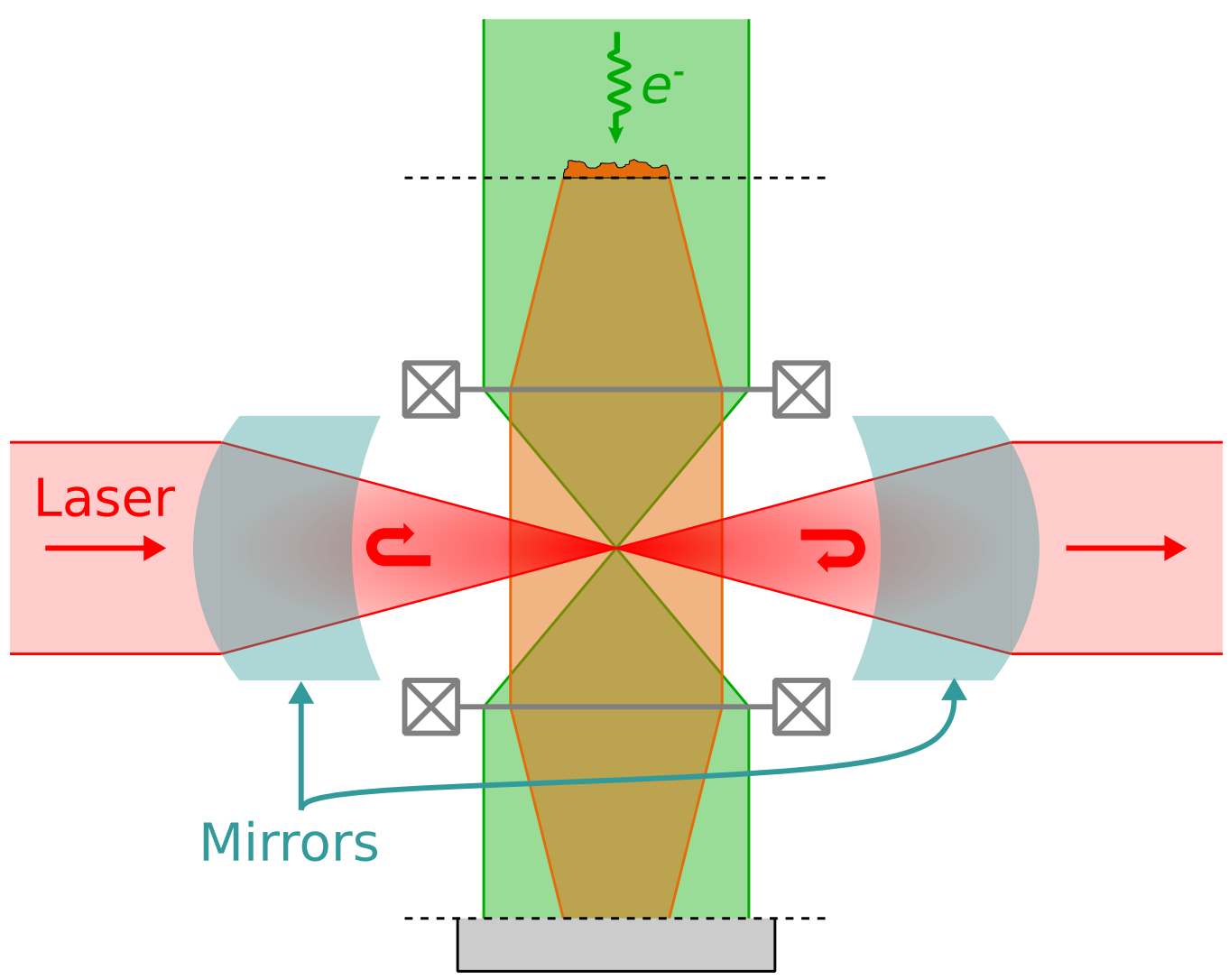

Figure 1.3: A laser phase plate using a Fabry-Pérot cavity. The laser beam bounces between the mirrors of the Fabry-Pérot cavity, enhancing its power.

This thesis describes our design for the Fabry-Pérot cavity LPP, characterizes its performance, and demonstrates that it functions as a nearly ideal phase plate.

### 1.2.3 History

The LPP was conceived in 2007 by Bryan Reed (then at Lawrence Livermore National Laboratory, now at Integrated Dynamic Electron Solutions, Inc.) based off of work done in R. J. Dwayne Miller's group (University of Toronto) which used the ponderomotive potential to characterize the duration of electron pulses used in ultrafast electron diffraction [25]–[28]. Professor Robert M. Glaeser (University of California, Berkeley), who was already working on non-laser phase plate designs, then brought the idea to Professor Holger Müller (University of California, Berkeley), an expert in laser physics and matter wave interferometry. Müller et al. proposed several possible ways to realize a practical LPP [21]. One of these approaches (extremely tight focusing of a $CO_2$ laser beam) was explored for several years, but it was eventually determined that it was too technically challenging to control the laser beam polarization to the degree necessary to provide a small enough focal spot. In 2015, work began on the (successful) Fabry-Pérot cavity approach described in this thesis.

## 1.3 Cryo-electron microscopy

Cryo-electron microscopy (cryo-EM) is the primary application for the LPP. In cryo-EM, biological samples are vitrified and then held at liquid nitrogen temperature while they are imaged via TEM. In "single-particle analysis" (SPA), a solution of a particular macromolecule (typically a protein) is purified and deposited on a TEM sample grid where it is thinned by capillary action and then vitrified. The resulting sample consists of the macromolecules embedded in a thin layer ($< 100$ nm) of amorphous ice. By taking many images of many differently-oriented individual macromolecules in the sample, the three-dimensional atomic structure of the macromolecule can be determined ("reconstructed"). In cryo-electron tomography (cryo-ET), the sample is imaged from multiple angles (by tilting the sample) to reconstruct its three-dimensional structure. This technique can be used to visualize the structure and distribution of biological molecules in their native environment inside of a cell.

The vitrification process fixes the sample and helps protect the relatively delicate biological macromolecules from radiation damage by the TEM's electron beam. However, radiation damage is still a primary limitation for the method, as electron doses of only several electron $\text{Å}^{-2}$ are usually sufficient to alter or destroy the atomic structure of the macromolecules. This means that near-atomic resolution ($< 4\,\text{Å}$) images necessarily suffer from high shot noise. Additionally, motion of the sample induced by exposure to the electron beam blurs the images. These factors severely limit the signal-to-noise ratio in the images—it is therefore crucial to maximize the contrast in the image. Indeed, cryo-EM only became capable of consistently generating near-atomic resolution reconstructions with the advent of direct electron detection cameras, which are capable of recording high-fidelity movies of the sample with a frame rate sufficient to substantially compensate for beam-induced motion [29], [30]. Since then (the early-to-mid 2010s), the already exponentially increasing number of molecular structures determined by SPA cryo-EM has exploded well past 10 000 (see figure 1.4), and the technique has been awarded a Nobel prize (2017). The modern capabilities of cryo-EM are being applied to not only fundamental biological research, but also biomedicine, where it has been used in the development of vaccines for RSV [32] and COVID-19 [33], [34], and in determining the structure of antibody-bound Zika virus [35] and tau filaments from an Alzheimer disease brain [36], to name a few well-known examples.

However, the signal-to-noise ratio in cryo-EM images is still insufficient to determine many types of structures. In SPA, small macromolecules inherently generate less contrast and may not even be visible above the shot noise. Even if they are visible, there may not be enough contrast to determine the position and orientation of the macromolecule to accurately average its structural information with the other macromolecule images in the dataset (such averaging is what enables near-atomic resolution SPA reconstructions). The smallest isolated protein structures that have been determined using cryo-EM have a molecular mass of $\sim 50\,\text{kDa}$ [37] which is roughly the average mass of a eukaryotic protein [38], [39]. There is particular biomedical interest in determining the structures of small membrane proteins, which often serve as drug targets [34], [38], [40]. Proteins embedded in a lipid membrane generate even lower contrast than those isolated in vitreous ice, because of the closer match of the strength of the atomic potentials of the protein and its surroundings. The same challenge of registering low contrast structures applies to structurally heterogeneous ("floppy") macromolecules. Such types of proteins with historically difficult-to-determine structures are not only ideal candidates for use with the LPP, but are also probably less accurately predicted by machine learning protein folding algorithms (e.g. AlphaFold, RoseTTAFold) which rely on the existence previously experimentally determined structures [41], [42]. In cryo-ET, many individual macromolecules or cellular structures are simply not visible at all due to low contrast.

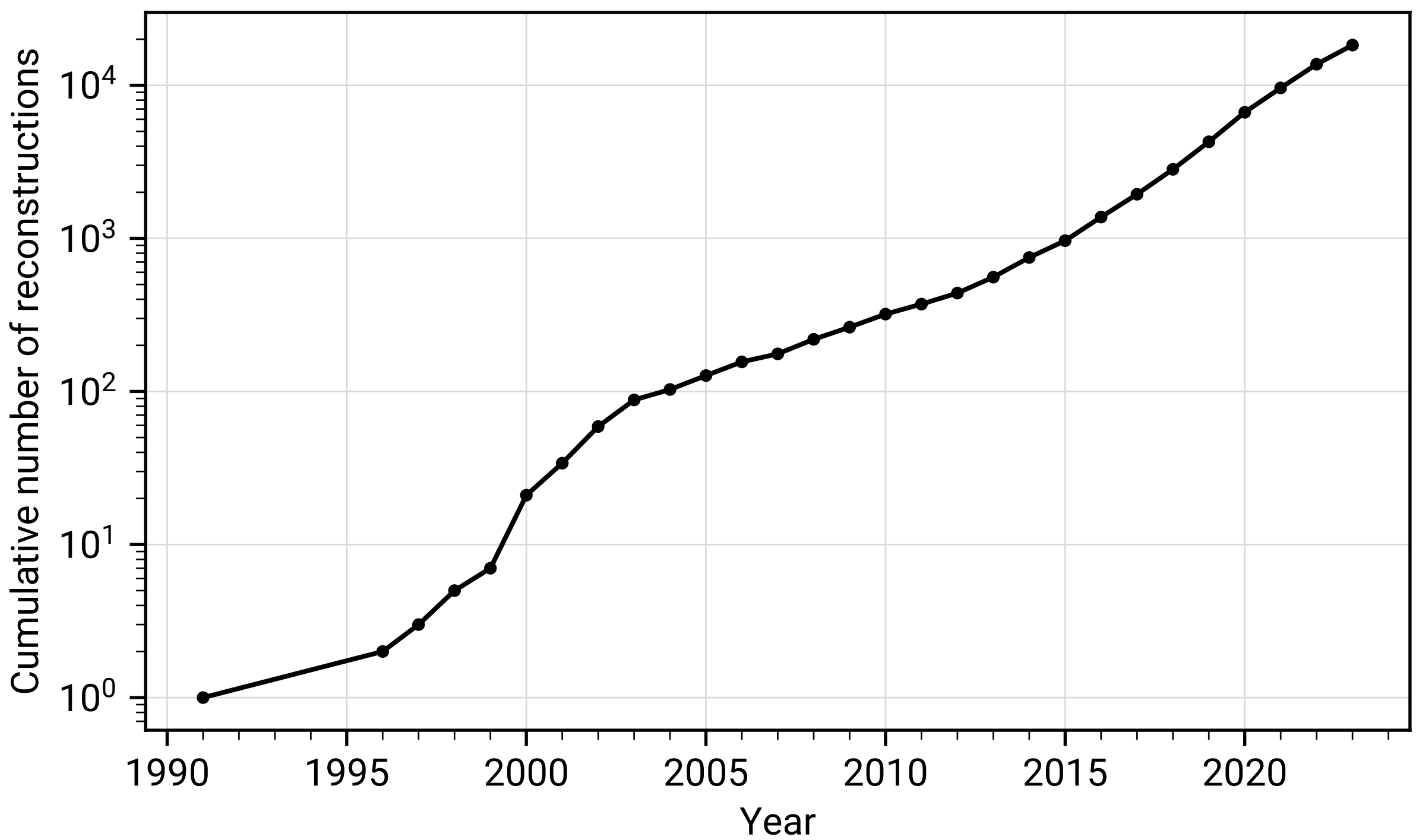

Figure 1.4: Cumulative number of single-particle analysis reconstructions released on the Protein Data Bank as a function of time. Data from [31].

The central goal of this project is to use the additional image contrast provided by the LPP to further extend the capabilities of cryo-EM.

## 1.4 Summary of this thesis

This thesis describes the design, construction, and application of the first LPP. Chapter 2 establishes the properties of Fabry-Pérot optical cavities, with an emphasis on those which determine power enhancement and focusing. In particular, section 2.6 shows how laser-induced thermoelastic deformation of the mirror substrate material limits the maximum achievable focal intensity in a near-concentric Fabry-Pérot cavity. Chapters 3, 4, 5, and 6 describe the design and construction of our LPP Fabry-Pérot cavity and associated systems.

Chapter 7 shows data demonstrating that the LPP Fabry-Pérot cavity can operate stably in the configuration required to generate a $\pi/2$ phase shift. This results in a record-high continuous-wave laser intensity of 590 GWcm$^{-2}$ at the focus of the cavity. Section 7.6 presents measurements of the thermoelastic deformation effect introduced in section 2.6.

Chapter 8 establishes the imaging theory of phase contrast TEM in general, and that of the LPP in particular.

The theory of the interaction between the laser and electron beams which causes the electron beam phase shift is discussed in chapter 9. The interaction is shown to depend not only on the laser intensity, but also its polarization and the electron kinetic energy. We provide the first experimental verification of this effect in section 9.2.

Chapter 10 describes the design of the TEM used in this project, how the TEM is aligned to accommodate LPP imaging, and how phase contrast images can be automatically collected when using the LPP.

Chapter 11 shows theoretically and experimentally that the effects of thermal magnetic field noise must be considered (and can be overcome) when designing a LPP.

Imaging results using the LPP are shown in chapter 12. The LPP is shown to function as the first stable phase plate for TEM, achieving phase shifts of $\frac{\pi}{2}\,(0.989 \pm 0.054)$ rad (i.e. $89.0 \pm 4.9\,\mathrm{deg}$) across all the images taken in a SPA dataset. We find that standard SPA using the LPP does not yield an enhanced reconstruction resolution; potential reasons for this are discussed, and initial data is presented for several imaging methods using the LPP which may yet deliver benefits in SPA.

Finally, a brief outlook on the future directions of the LPP project is presented in chapter 13. In particular, we anticipate that the LPP has huge potential when applied to cryo-ET using a state-of-the art TEM.

# Chapter 2

# Near-concentric Fabry-Pérot cavities

A Fabry-Pérot optical cavity consists of two mirrors that face each other. This arrangement can trap light between the mirrors if the mirrors have the right geometry and relative alignment such that light which reflects off of one mirror is reflected back by the other. Determining the shapes and frequencies of the allowed electromagnetic (EM) fields between and around the mirrors is a question of solving the EM wave equations subject to the boundary conditions imposed by the mirrors. While it is straightforward to write down these governing equations, they are generally impossible to solve analytically, and can even be challenging to solve numerically.

This chapter first introduces several commonly-used simplifying assumptions about the shapes of the EM fields and mirror surfaces. Analytical solutions for the EM fields are then derived using this simplified model. The rest of the chapter is devoted to describing the behavior of these solutions, especially under conditions relevant to the operation of the LPP. Particular attention is paid to the factors which affect the focusing and power enhancement of the laser beam inside of the cavity, since those parameters determine the electron beam phase shift. The physics of the interaction between the electron beam and cavity EM field is described separately in chapter 9. This chapter concludes with a theoretical description of the effects of laser-induced heating of the mirrors on the shape of the EM fields. Those results motivate the design of the cavity mirrors (see chapter 3), and the effect is later explored experimentally in section 7.6.

## 2.1 The model

Consider the fairly general case of two mirrors which face each other (see figure 2.1). We will consider the EM field in several different planes along the optical axis of the cavity, which for now we just define as some line intersecting the two mirrors. Let $z$ denote the coordinate along this axis. We can think of the EM field between the mirrors as a superposition of two free-space EM waves oscillating at frequency $\nu$, which travel back and forth between the mirrors and reflect off of them. For simplicity, we consider the mirror to be an infinitely thin reflecting surface. Denote the electric field amplitude of the right-going ($+z$) wave just to the left of the surface of mirror 1 (the "input mirror") as $\mathbf{E}_\mathrm{i}^+(\mathbf{x}_1)$ where $\mathbf{x}_1$ is the position on the surface of mirror 1. Let the right-going wave just to the right of the surface of mirror 1 be $\mathbf{E}_\mathrm{c}^+(\mathbf{x}_1)$, and let the left-going wave just to the left of the surface of mirror 2 be $\mathbf{E}_\mathrm{c}^-(\mathbf{x}_2)$, where $\mathbf{x}_1$ is the position on the surface of mirror 2.

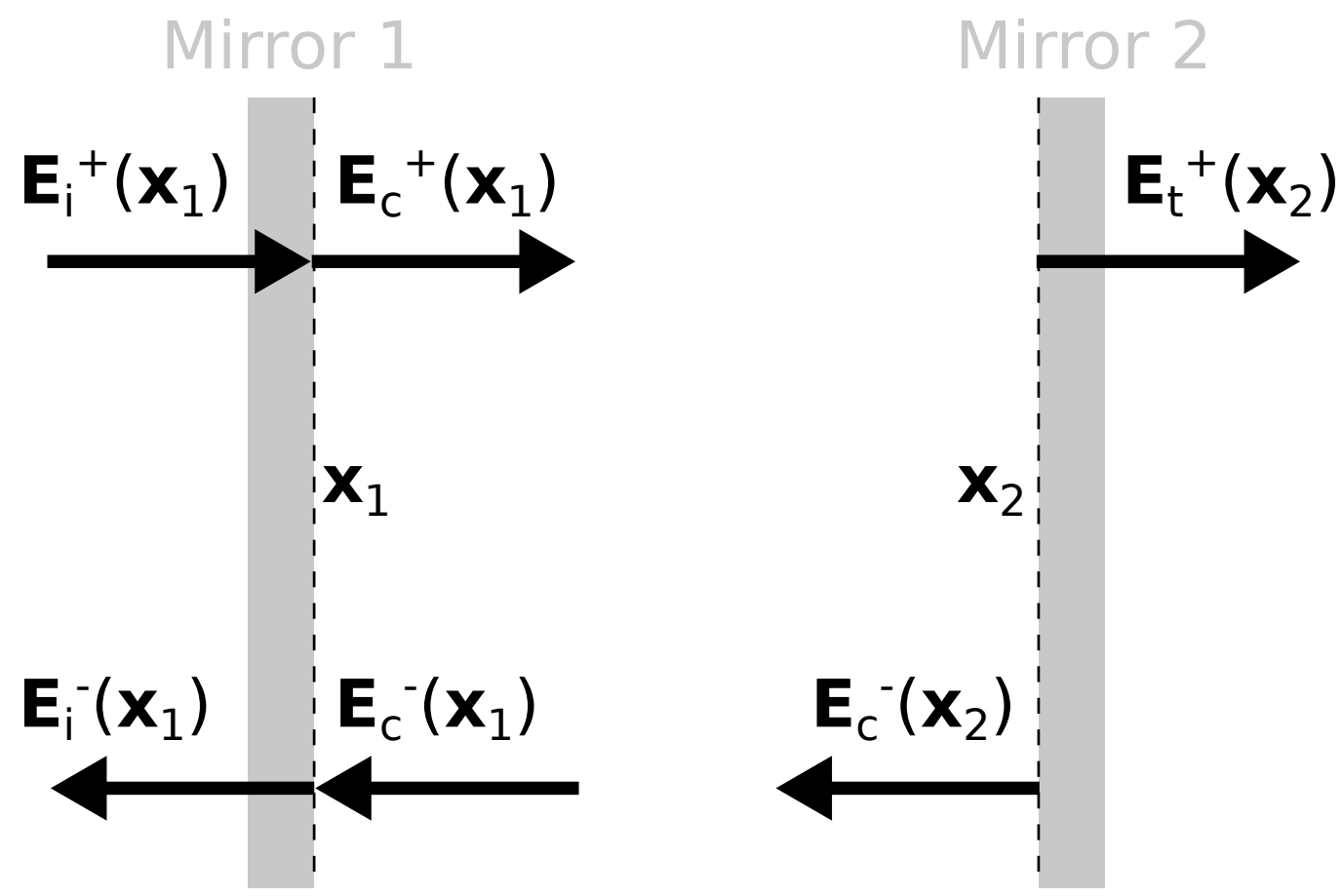

Figure 2.1: A general Fabry-Pérot optical cavity, with electromagnetic wave amplitudes defined on the mirror surfaces.

The right-going wave inside of the cavity at the surface of mirror 1 must be a superposition of the fraction of the wave which enters the cavity from outside and the left-going intra-cavity wave which reflects off of mirror 1. Specifically,

$$\mathbf{E}_\mathrm{c}^+(\mathbf{x}_1) = U_{\mathrm{T},1}\mathbf{E}_\mathrm{i}^+(\mathbf{x}_1) + U_{\mathrm{R},1}U_\mathrm{FS}^-\mathbf{E}_\mathrm{c}^-(\mathbf{x}_2) \tag{2.1}$$

where the operators $U_{\mathrm{T},1}$, $U_{\mathrm{R},1}$, and $U_\mathrm{FS}^-$ represent the effects on the EM field of transmitting through mirror 1, reflecting off of mirror 1, and propagating in free space across the cavity from right to left, respectively. We assume that there is no left-going wave to the right of mirror 2, such that the left-going wave just left of mirror 2 consists only of the right-going wave just right of mirror 1 which has traversed the cavity and reflected off of mirror 2:

$$\mathbf{E}_\mathrm{c}^-(\mathbf{x}_2) = U_{\mathrm{R},2}U_\mathrm{FS}^+\mathbf{E}_\mathrm{c}^+(\mathbf{x}_1) \tag{2.2}$$

Combining these equations yields

$$(1 - U_\mathrm{RT})\,\mathbf{E}_\mathrm{c}^+(\mathbf{x}_1) = U_{\mathrm{T},1}\mathbf{E}_\mathrm{i}^+(\mathbf{x}_1) \tag{2.3}$$

where we have defined the "round-trip operator"

$$U_\mathrm{RT} \coloneqq U_{\mathrm{R},1}U_\mathrm{FS}^-U_{\mathrm{R},2}U_\mathrm{FS}^+ \tag{2.4}$$

which encapsulates how the wave $\mathbf{E}_c^+(\mathbf{x}_1)$ is transformed by propagating across the cavity, reflecting off of mirror 2, propagating back across the cavity, and reflecting off of mirror 1, to return to its starting position and direction.

We now decompose the wave $\mathbf{E}_c^+(\mathbf{x}_1)$ into two components: that which remains unchanged in shape by propagation around the cavity $\overline{\mathbf{E}}_c^+(\mathbf{x}_1)$, and everything else $\tilde{\mathbf{E}}_c^+(\mathbf{x}_1)$. That is,

$$\mathbf{E}_c^+(\mathbf{x}_1) = \overline{\mathbf{E}}_c^+(\mathbf{x}_1) + \tilde{\mathbf{E}}_c^+(\mathbf{x}_1) \tag{2.5}$$

where $\overline{\mathbf{E}}_c^+(\mathbf{x}_1)$ is an eigenvector of the round-trip operator $U_{\mathrm{RT}}$ such that

$$U_{\mathrm{RT}}\overline{\mathbf{E}}_c^+(\mathbf{x}_1) = u_{\mathrm{RT}}\overline{\mathbf{E}}_c^+(\mathbf{x}_1) \tag{2.6}$$

with $u_{\mathrm{RT}}$ a scalar. We call the wave component $\overline{\mathbf{E}}_c^+(\mathbf{x}_1)$ the "cavity mode", and $\tilde{\mathbf{E}}_c^+(\mathbf{x}_1)$ "stray light". We can solve equation (2.3) for the cavity mode wave

$$\overline{\mathbf{E}}_c^+(\mathbf{x}_1) = \frac{U_{\mathrm{T},1}\mathbf{E}_i^+(\mathbf{x}_1) - (1 - U_{\mathrm{RT}})\,\tilde{\mathbf{E}}_c^+(\mathbf{x}_1)}{1 - u_{\mathrm{RT}}} \tag{2.7}$$

though this equation is still too general to be of much use, as on its own it only relates several unknown objects (the wave profiles at different points).

From here, we make several assumptions. First, we assume that the transmission operator $U_{\mathrm{T},1} = \mathcal{T}_1$, where $\mathcal{T}_1$ is simply a scalar representing a spatially uniform transmission amplitude on mirror 1. Most mirrors have a nominally spatially uniform transmission, so this is a reasonable assumption. Implicit in this assumption is that the lateral extent of the mirror is much larger than the input wave profile $\mathbf{E}_i^+(\mathbf{x}_1)$, since the action of the mirror implied by $U_{\mathrm{T},1} = \mathcal{T}_1$ does not itself have a finite lateral extent. Second, we assume that the round-trip operator eigenvalue $u_{\mathrm{RT}} = \mathcal{R}_1\mathcal{R}_2 e^{i\phi_{\mathrm{RT}}}$ where $\mathcal{R}_1$ and $\mathcal{R}_2$ represent reflection amplitudes on mirrors 1 and 2, respectively and $\phi_{\mathrm{RT}} \in \mathbb{R}$. This assumption implies that the only loss in amplitude of the cavity mode in a trip around the cavity is due to the component which is lost upon reflection at each mirror. The round-trip phase shift $\phi_{\mathrm{RT}}$ is an important quantity that we will solve for later. Equation (2.7) now becomes somewhat more interpretable:

$$\overline{\mathbf{E}}_c^+(\mathbf{x}_1) = \frac{\mathcal{T}_1\mathbf{E}_i^+(\mathbf{x}_1) - (1 - U_{\mathrm{RT}})\,\tilde{\mathbf{E}}_c^+(\mathbf{x}_1)}{1 - \mathcal{R}_1\mathcal{R}_2 e^{i\phi_{\mathrm{RT}}}} \tag{2.8}$$

if the reflection amplitudes on both mirrors are near unity magnitude, then the amplitude of the cavity mode may become much larger than that of the input wave or stray light, for values of the round-trip phase which make $\mathcal{R}_1\mathcal{R}_2 e^{i\phi_{\mathrm{RT}}}$ near 1.

## 2.2 Mode shape

To determine the shape of the cavity mode we must make further assumptions. To specialize to the particular case considered in this thesis, we assume both mirrors have concave spherical surfaces with a curvature radius $R > 0$. The mirrors are separated by a distance $L$ measured along along a line which crosses the center of curvature of each mirror (see figure 2.2). We further assume that the transverse profile of the cavity mode only varies slowly along the axis of the cavity, such that we may approximate the free-space propagators $U_{\mathrm{FS}}^{+;-}$ with the Fresnel propagator

$$U_L \coloneqq \exp\left(i\frac{4\pi}{\lambda_\mathrm{L}}L\right)\exp\left(i\frac{\lambda_\mathrm{L}}{4\pi}L\nabla_\perp^2\right) \tag{2.9}$$

where $\nabla_\perp^2 \coloneqq \frac{\partial^2}{\partial x^2} + \frac{\partial^2}{\partial y^2}$, which solves the scalar paraxial wave equation

$$\nabla_\perp^2 + i\frac{4\pi}{\lambda_\mathrm{L}}\frac{\partial}{\partial z} + \frac{8\pi^2}{\lambda_\mathrm{L}^2} = 0 \tag{2.10}$$

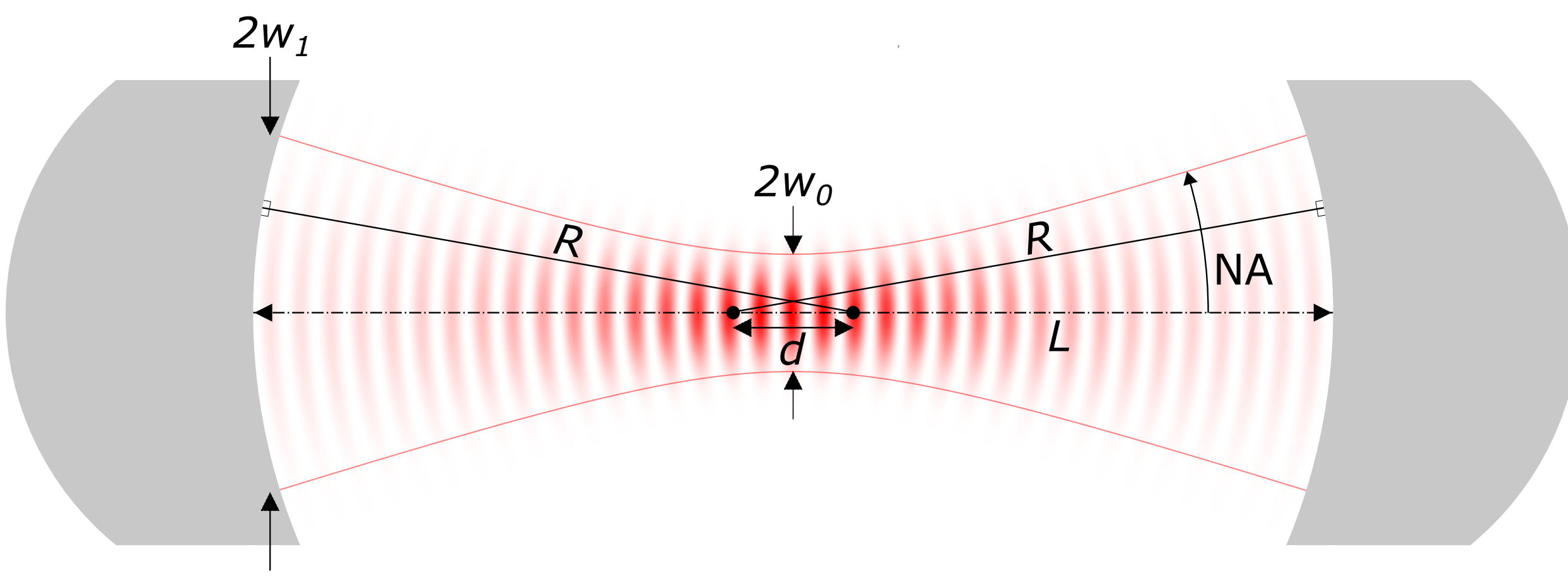

Figure 2.2: A symmetric Fabry-Pérot cavity with a length $L$, mirror curvature radii $R$, and a distance to concentricity $d$ supports modes with a focal waist $w_0$, numerical aperture NA, and mirror waist $w_1$.

Similarly within the paraxial approximation, we approximate the action of reflection off the mirror as simply imposing a spatially quadratic phase shift on the wave (and a spatially uniform reflection coefficient), such that

$$U_{\mathrm{R},j} = \mathcal{R}_j e^{i\frac{2\pi}{\lambda_\mathrm{L}}\frac{1}{R}\left(x^2+y^2\right)} \quad j = 1, 2 \tag{2.11}$$

The easiest way to find the eigenvectors of this round-trip operator is to see that it represents paraxial propagation between two quadratically curved surfaces. The reflection operator effectively enforces that the resulting mode have wavefronts with curvature matching the mirrors' at their surfaces. Such solutions are readily available in the Hermite-Gaussian functions. More rigorously, the exponentials in the round-trip operator can be combined using an algebraic relationship from Mitter and Yamazaki [43] to yield an operator with a single exponential, the exponent of which is equivalent to the quantum Hamiltonian for a massive particle in a harmonic potential. Eigenvectors of this Hamiltonian (Hermite-Gaussian functions) are therefore also eigenvectors of the round-trip operator. I will spare the reader from the full derivation using this approach, but note that it may be useful to cast the round-trip operator into a form equivalent to the "interaction picture" used in quantum mechanics when solving for mode profiles in a cavity with mirror surfaces which deviate slightly from spherical [44].

The Hermite-Gaussian functions are [45]:

$$\begin{aligned}\boldsymbol{\mathcal{E}}_{l,m,p}\left(x,y,z\right) = \hat{\mathbf{x}}_p \mathcal{N} \frac{w_0}{w\left(z\right)} &\exp\left(i\frac{2\pi}{\lambda_{\mathrm{L}}}z - i\psi_{\mathrm{G}}\left(z\right)\right) \\ &H_l\left(\frac{\sqrt{2}x}{w\left(z\right)}\right)\exp\left(-\left(\frac{x}{w\left(z\right)}\right)^2 + i\frac{2\pi}{\lambda_{\mathrm{L}}}\frac{x^2}{2R\left(z\right)}\right) \\ &H_m\left(\frac{\sqrt{2}y}{w\left(z\right)}\right)\exp\left(-\left(\frac{x}{w\left(z\right)}\right)^2 + i\frac{2\pi}{\lambda_{\mathrm{L}}}\frac{y^2}{2R\left(z\right)}\right)\end{aligned} \tag{2.12}$$

where $z$ lies along the optical axis of the cavity (intersecting the mirrors' center of curvature),

$$H_j\left(x\right) \coloneqq \left(-1\right)^j e^{x^2}\frac{d^j}{dx^j}e^{-x^2} \tag{2.13}$$

are physicist's Hermite polynomials and

$$\mathcal{N} \coloneqq \sqrt{\frac{\sqrt{2/\pi}}{l!2^l w_0}}\sqrt{\frac{\sqrt{2/\pi}}{m!2^m w_0}} \tag{2.14}$$

is a normalization factor such that

$$\int_{-\infty}^{\infty} dx \int_{-\infty}^{\infty} dy\, \boldsymbol{\mathcal{E}}^*_{l',m',p'}\left(x,y,z\right)\cdot\boldsymbol{\mathcal{E}}_{l,m,p}\left(x,y,z\right) = \delta_{l',l}\delta_{m',m}\delta_{p',p} \tag{2.15}$$

The function

$$w\left(z\right) \coloneqq w_0\sqrt{1+\left(\frac{z}{z_{\mathrm{R}}}\right)^2} \tag{2.16}$$

is the $1/e^2$ intensity radius of the beam as a function of position along the optical axis,

$$R\left(z\right) = z\left(1+\left(\frac{z_{\mathrm{R}}}{z}\right)^2\right) \tag{2.17}$$

is the radius of curvature of the wavefronts as a function of position along the optical axis,

$$\psi_{\mathrm{G}}(z) \coloneqq (l+m+1)\arctan\left(\frac{z}{z_{\mathrm{R}}}\right) \tag{2.18}$$

is referred to as the "Gouy phase", $l, m = 0, 1, 2, ...$ are the transverse mode indices, $p = 0, 1$ is the polarization index such that $\hat{\mathbf{x}}_0 = \hat{\mathbf{x}}$ and $\hat{\mathbf{x}}_1 = \hat{\mathbf{y}}$, $\lambda_{\mathrm{L}} = c/\nu$ is the light wavelength, $z_{\mathrm{R}} = \pi w_0^2/\lambda_{\mathrm{L}}$ is referred to as the Rayleigh range, and $w_0$ is the $1/e^2$ intensity radius of the beam at its focus.

The boundary conditions require that

$$R\left(\pm\frac{L}{2}\right) = \pm R \tag{2.19}$$

such that the Rayleigh range is determined by the cavity geometry:

$$z_{\mathrm{R}} = \frac{L}{2}\sqrt{\frac{1+g}{1-g}} \tag{2.20}$$

where

$$g \coloneqq 1 - \frac{L}{R} \tag{2.21}$$

is the "cavity stability parameter". It is only possible for the Hermite-Gaussian functions to satisfy the boundary conditions when $-1 \leq g \leq 1$. Under these conditions the cavity is referred to as being "stable".

Defining the "distance to concentricity" $d$ as

$$d \coloneqq 2R - L = L\frac{1+g}{1-g} \tag{2.22}$$

equation (2.20) can be expressed equivalently for the focal waist $w_0 = \sqrt{\frac{z_{\mathrm{R}}\lambda_{\mathrm{L}}}{\pi}}$:

$$w_0 = \sqrt{\frac{\lambda_{\mathrm{L}} L}{2\pi}\sqrt{\frac{1+g}{1-g}}} \tag{2.23}$$

$$= \sqrt{\frac{\lambda_{\mathrm{L}}}{2\pi}\sqrt{Ld}} \tag{2.24}$$

or the numerical aperture $\mathrm{NA} = \frac{\lambda_{\mathrm{L}}}{\pi w_0}$ of the cavity mode,

$$\mathrm{NA} = \sqrt{\frac{2\lambda_{\mathrm{L}}}{\pi L}\sqrt{\frac{1-g}{1+g}}} \tag{2.25}$$

$$= \sqrt{\frac{2\lambda_{\mathrm{L}}}{\pi}\sqrt{\frac{1}{Ld}}} \tag{2.26}$$

which in the paraxial approximation is equal to the divergence half-angle (in radians) of the cavity mode, as illustrated in figure 2.2.

These equations show that as the distance to concentricity approaches zero ($g \to -1$), the focal waist of the mode approaches 0 and the numerical aperture diverges to infinity. Therefore, a near-concentric cavity is required to generate the small focal waist needed for use with the LPP. The LPP uses a $(l, m) = (0, 0)$ mode since it has the smallest focal width of all the modes.

The mode waist at the mirror surface(s) $w_1$ is also of interest:

$$w_1 = \sqrt{\frac{\lambda_\mathrm{L} L}{\pi}\sqrt{\frac{1}{1-g^2}}} \tag{2.27}$$

$$= \sqrt{\frac{\lambda_\mathrm{L} L}{2\pi}\sqrt{\frac{\left(\frac{d}{L}+1\right)^2}{\frac{d}{L}}}} \tag{2.28}$$

which shows that as the cavity approaches concentricity, the waist on the mirror surface increases. Section 7.2 discusses the benefits and challenges that come from having a large area of the mirror surface exposed to the cavity mode. In the near-concentric limit $d \ll L$, the waist on the mirror surface is the numerical aperture times half the cavity length:

$$w_1 \simeq \sqrt{\frac{\lambda_\mathrm{L} L}{2\pi}\sqrt{\frac{L}{d}}} = \frac{L}{2}\mathrm{NA} \tag{2.29}$$

Table 2.1 lists the LPP's mode parameters.

| | |
|---|---|
| $R$ | 10 mm |
| $g$ | −0.99963288 |
| $\lambda_\mathrm{L}$ | 1064 nm |
| NA | 0.05 |
| $d$ | 3.7 µm |
| $w_0$ | 6.8 µm |
| $z_\mathrm{R}$ | 135 µm |
| $w_1$ | 500 µm |

Table 2.1: Mode parameters for the laser phase plate.

### 2.2.1 Astigmatism

A cavity may have mirrors which have ellipsoidal (rather than spherical) surfaces. This astigmatism in the mirrors induces astigmatism in the cavity mode, though it can still be represented by Hermite-Gaussian functions which have different focal waist parameters $w_{0,a}$ and $w_{0,b}$ for each principal axis of the mode, and different focal positions $z_{0,a}$ and $z_{0,b}$ for each of those waists (note that we implicitly set the focal position $z_0 = 0$ in the case of a stigmatic mode in a symmetric cavity). Such functions are still solutions of the paraxial wave equation because both the equation and function are separable, and each component of the separated function is still a one-dimensional Hermite-Gaussian function which solves the corresponding one-dimensional paraxial wave equation. The Gouy phase function is replaced with the more general function

$$\psi_{\mathrm{G}} = \left(l + \frac{1}{2}\right) \arctan\left(\frac{z}{z_{\mathrm{R},a}}\right) + \left(m + \frac{1}{2}\right) \arctan\left(\frac{z}{z_{\mathrm{R},b}}\right) \tag{2.30}$$

where $z_{\mathrm{R},a}$ and $z_{\mathrm{R},b}$ are the Rayleigh ranges corresponding to the two principal axes of the mode.

### 2.2.2 Diffraction loss

The large mode size at the mirror surface in the near-concentric case means that attention must be paid to the approximation that the mirror surface extends far beyond the width of the cavity mode. If this is not the case, then the outer portions of the beam will not be reflected back, which to leading order effectively reduces the reflection coefficient $\mathcal{R}_{1,2}$ of the mirror. This effect is called "diffraction loss" and reduces the amplitude enhancement effect of the cavity as described by equation (2.8). More substantial diffraction loss noticeably changes the shapes of the modes.

In our case, other losses in the mirrors (see chapter 7) result in up to $1 \times 10^{-5}$ of the power in the cavity mode being lost per reflection. Diffraction loss is therefore negligible if it amounts to $< 1 \times 10^{-6}$ power per reflection. The diffraction loss $L_{\mathrm{D}}$ of a Gaussian mode $((l, m) = (0, 0))$ on a cylindrical mirror can be estimated by calculating the fraction of the mode's intensity profile which lands outside of the mirror's outer diameter $D$:

$$L_{\mathrm{D}} = \frac{\int_{D/2}^{\infty} 2\pi r dr \, \left|\boldsymbol{\mathcal{E}}_{0,0,p}\left(x, y, \frac{L}{2}\right)\right|^2}{\int_{0}^{\infty} 2\pi r dr \, \left|\boldsymbol{\mathcal{E}}_{0,0,p}\left(x, y, \frac{L}{2}\right)\right|^2} \tag{2.31}$$

$$= \exp\left(-2\left(\frac{D}{2w_1}\right)^2\right) \tag{2.32}$$

which falls below $1 \times 10^{-6}$ for $\frac{D}{2w_1} > 2.6$. In our design (see chapter 3), $\frac{D}{2w_1} = 7.75$, so diffraction loss is negligible.

### 2.2.3 Alignment sensitivity

The axis of the cavity mode intersects the centers of curvature of the mirrors, such that the mode wavefronts are normally incident on both mirror surfaces (see figure 2.2). If one of the centers of curvature moves relative to the other, then the mode axis changes as well. In the near-concentric limit $d \ll L$, this means that relatively small changes in the position or orientation of one of the mirrors can lead to large changes in the angle of the mode axis.

A transverse shift (in $x$ or $y$) of one mirror by a distance $\delta_{x,y}$ changes the angle of the mode axis by $\Theta = \arctan(\delta_{x,y}/d)$, and also increases the distance to concentricity from $d$ to $d' = \sqrt{d^2 + \delta_{x,y}^2}$ (see figure 2.3). These changes are significant when $\delta_{x,y} \sim d$, so since in the LPP $d = 3.7\,\mu\text{m}$, the transverse alignment of the mirrors must be maintained to $\ll 1\,\mu\text{m}$.

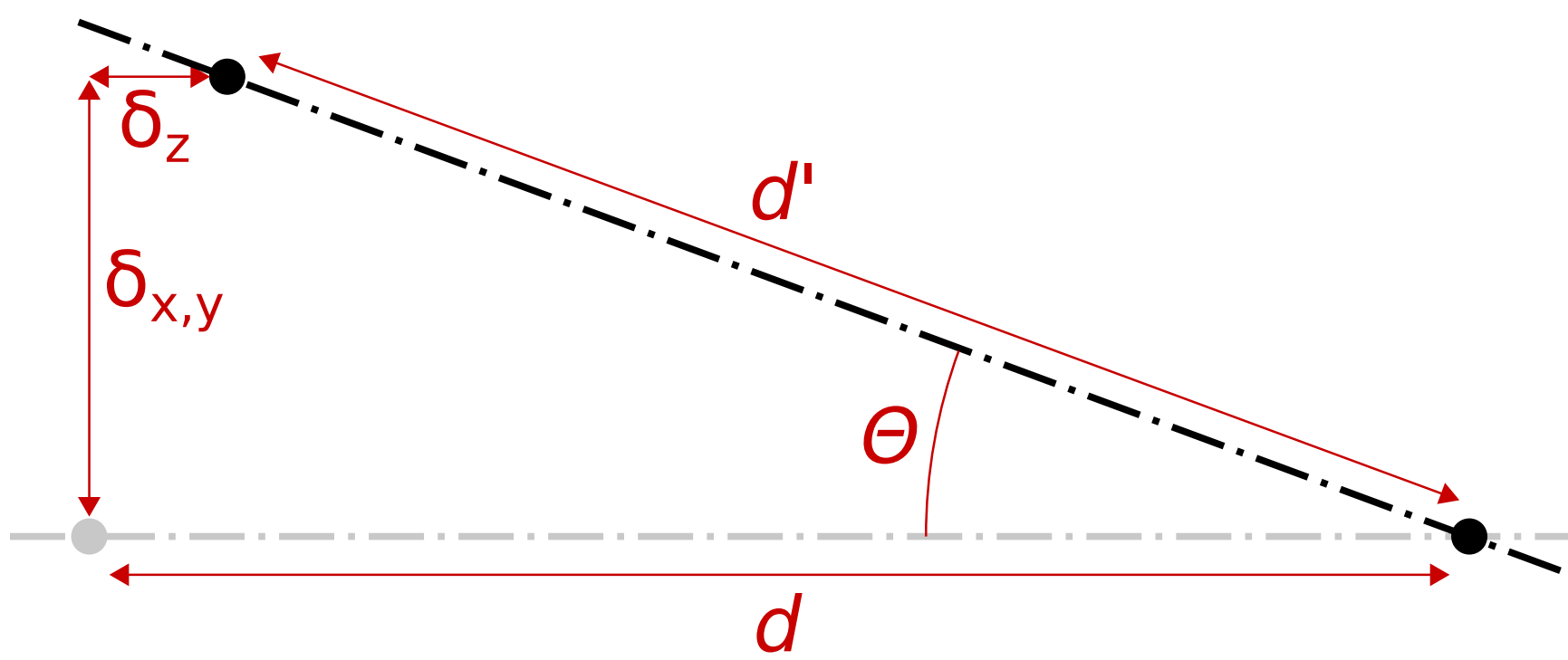

Figure 2.3: A relative lateral ($\delta_{x,y}$) and axial ($\delta_z$) displacement of the mirror curvature centers (black dots) changes the angle of the cavity mode axis by $\Theta$ and the distance to concentricity from $d$ to $d'$.

Similarly, a tilt (about the $x$ or $y$ axes) of one mirror by an angle $\theta_{x,y}$ moves its center of curvature laterally by a distance $\delta_{x,y} = R\sin(\theta_{x,y})$ and longitudinally by a distance $\delta_z = R(1 - \cos(\theta_{x,y}))$, such that the mode axis angle changes by $\Theta = \arctan\left(\frac{\delta_{x,y}}{d-\delta_z}\right)$ and the distance to concentricity increases to $d' = \sqrt{(d-\delta_z)^2 + \delta_{x,y}^2}$. In the limit of a small mirror tilt angle $\theta_{x,y} \ll 1$, this simplifies to $\Theta \simeq \frac{R\theta_{x,y}}{d}$. With $R = 10\,\text{mm}$ in the LPP, this requires that the tilt alignment of the mirrors be maintained to $\ll 100\,\mu\text{rad}$.

## 2.3 Mode spectrum

### 2.3.1 Resonance frequencies

The shape of the cavity mode also determines its amplitude as a function of the EM wave frequency $\nu = c/\lambda_\text{L}$. Applying the paraxial round-trip operator to the Hermite-Gaussian solutions yields the round-trip phase shift (generalizing to the astigmatic case)

$$\phi_\text{RT} = 2L\frac{2\pi}{\lambda_\text{L}} - 2\left(\psi_\text{G}\left(\frac{L}{2}\right) - \psi_\text{G}\left(-\frac{L}{2}\right)\right) \tag{2.33}$$

$$= 2\pi\left(\nu\frac{2L}{c} - \left(l + \frac{1}{2}\right)\frac{\arccos(g_a)}{\pi} - \left(m + \frac{1}{2}\right)\frac{\arccos(g_b)}{\pi}\right) \tag{2.34}$$

where $g_a$ and $g_b$ are the stability parameters corresponding to the mode's principal axes.

Returning to equation (2.8), we then see that the amplitude of the cavity mode is largest (relative to the input wave $\mathbf{E}_i^+(\mathbf{x}_1)$) when the magnitude of the denominator $1 - \mathcal{R}_1\mathcal{R}_2 e^{i\phi_{\mathrm{RT}}}$ is smallest, i.e. when

$$\arg(\mathcal{R}_1) + \arg(\mathcal{R}_2) + \phi_{\mathrm{RT}} = 2\pi(n+1), \quad n \in \mathbb{Z} \tag{2.35}$$

Using the expression for the round-trip phase shift in equation (2.34) we find that the EM wave frequencies which satisfy this resonance condition are

$$\begin{aligned}\nu_{n,l,m} = F&\left(n + 1 - \frac{\arg(\mathcal{R})_1 + \arg(\mathcal{R}_2)}{2\pi}\right)\\ &+ F\frac{\arccos(g_a)}{\pi}\left(l + \frac{1}{2}\right) + F\frac{\arccos(g_b)}{\pi}\left(m + \frac{1}{2}\right)\end{aligned} \tag{2.36}$$

where $F := \frac{c}{2L}$ is the "free spectral range" (FSR) and represents the frequency spacing $\nu_{n+1,l,m} - \nu_{n,l,m}$ between the "longitudinal modes" of the cavity (indexed by $n$). The resonance frequencies between modes of adjacent mode index ($n$, $l$, or $m$) are evenly spaced, and are all offset from zero by an amount which depends on $F$, $g_a$, $g_b$, $\arg(\mathcal{R}_1)$, and $\arg(\mathcal{R}_2)$. The phase shifts $\arg(\mathcal{R}_{1,2})$ due to reflection from the mirrors are usually approximated to be $\pi$ in the limit of highly reflective mirrors where $|\mathcal{R}_{1,2}|$ is close to unity. Under this assumption, $n \geq 0$ in order that $\nu_{n,l,m} \geq 0$, and $n$ represents the number of nodes in the cavity mode field between the mirrors. In the absence of astigmatism, modes with the same "order" $O := l+m$ have the same resonance frequency. Modes of order 0 are called the "fundamental" modes of the cavity.

In the near-concentric limit $g_{a,b} \to -1$, the spacing between modes of different transverse index ($l$ or $m$) approaches the free spectral range, so that the resonance frequencies of modes $(n, l, m)$ and $(n-1, l+1, m)$ (or equivalently $(n-1, l, m+1)$) become adjacent. Figure 2.4 illustrates this effect, though still relatively far from concentricity ($g_a = -0.6$, $g_b = -0.7$) so that the spacings are large enough to be easily visible.

It is useful to define the frequency spacing between these adjacent modes of the cavity as the "transverse mode spacings"

$$\begin{aligned}\Delta_{\nu,a} &:= \nu_{n,l,m} - \nu_{n-1,l+1,m} = F\left(1 - \frac{\arccos(g_a)}{\pi}\right)\\ \Delta_{\nu,b} &:= \nu_{n,l,m} - \nu_{n-1,l,m+1} = F\left(1 - \frac{\arccos(g_b)}{\pi}\right)\end{aligned} \tag{2.37}$$

as it is possible to measure this frequency difference and thereby infer the values of the cavity stability parameters $g_a$ and $g_b$ and by extension the geometric properties of the cavity mode (see section 6.3.2). The definition of $\Delta_{\nu,a}$ and $\Delta_{\nu,b}$ is shown graphically in figure 2.4. Note that for the LPP with $g = -0.99963288$, $\Delta_\nu/F = 8.6 \times 10^{-3}$—much smaller than what is shown in figure 2.4.

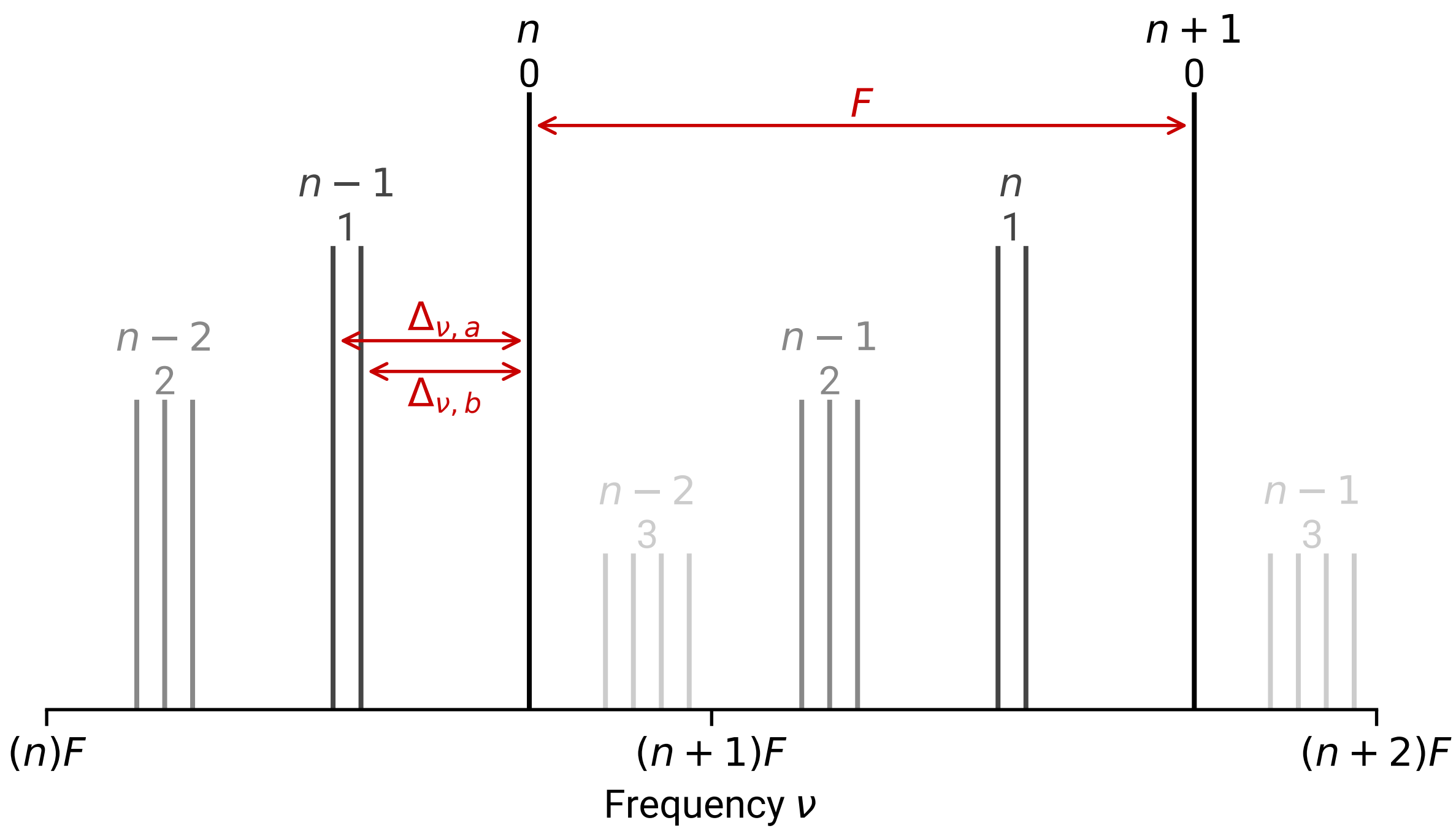

Figure 2.4: Resonance frequencies of an astigmatic cavity with $g_a = -0.6$ and $g_b = -0.7$ shown for modes up to $l + m =$ 3rd order. Annotations above each resonance frequency show the longitudinal mode index (top) and transverse mode order (bottom). The definitions of the free spectral range $F$ and transverse mode spacings $\Delta_{\nu,a}$ and $\Delta_{\nu,b}$ are shown in red.

When two or more modes have the same resonance frequency, they are called "degenerate". We have already seen that modes of the same order are degenerate in the absence of astigmatism, and that all modes of different order become degenerate at concentricity $g = -1$. Modes of different transverse and longitudinal mode indices can also be degenerate if their indices $(n, l, m)$ and $(n', l', m')$ satisfy

$$\frac{\arccos(g)}{\pi} = \frac{-(n - n')}{(l - l') + (m - m')} \tag{2.38}$$

where we ignore astigmatism for the sake of simplicity. This equation only has solutions when the right-hand side is between 0 and 1, and we are primarily interested in the cases where the difference in longitudinal mode index is the smallest, because very high transverse order modes have substantial diffraction loss (see section 2.2.2) which reduces their amplitude. The smallest possible value of $(l - l') + (m - m')$ is ceil $\left((1 - \arccos(g)/\pi)^{-1}\right)$. In the near-concentric limit, $\arccos(g)/\pi$ is only slightly smaller than unity, so the smallest possible value of $(l - l') + (m - m')$ is quite large. For example, in our LPP design $R_0 = 10\,\mathrm{mm}$ and NA $= 0.05$ such that $g = -0.99963288$ so $\arccos(g)/\pi = 0.99137449$ so that we must have $(l - l') + (m - m') \geq 116$ for a degeneracy to occur. By contrast, for $g = -0.9$ the minimum is only 7. This is an advantage of needing to use the near-concentric limit for the LPP, because it prevents the fundamental modes $(l, m) = (0, 0)$ from being degenerate with any low-order modes, which ensures that the shape of the intra-cavity wave is a well-defined Gaussian.

### 2.3.2 Cavity mode transfer function

We can now make concrete statements about both the shape and amplitude of the cavity modes. Define the overall amplitudes of the cavity mode $E_{l,m,p}$ and input beam $E_\mathrm{i}$ such that

$$\overline{\mathbf{E}}_\mathrm{c}^{+}(\mathbf{x}_1) = E_{l,m,p}\boldsymbol{\mathcal{E}}_{l,m,p}(\mathbf{x}_1) \tag{2.39}$$

$$\mathbf{E}_\mathrm{i}^{+}(\mathbf{x}_1) = E_\mathrm{i}\boldsymbol{\mathcal{E}}_\mathrm{i}^{+}(\mathbf{x}_1) \tag{2.40}$$

where $\boldsymbol{\mathcal{E}}_\mathrm{i}^{+}(\mathbf{x}_1)$ is a function (with $L^2$-norm equal to unity) defining the spatial profile of the input beam on the surface of mirror 1. Taking the inner product of equation (2.8) with $\boldsymbol{\mathcal{E}}_{l,m,p}^{*}(\mathbf{x}_1)$ and integrating over the surface of mirror 1 allows us to relate the input beam and cavity mode amplitudes by a transfer function $H_{l,m,p}(\nu)$:

$$E_{l,m,p} = H_{l,m,p}(\nu)\,E_\mathrm{i} \tag{2.41}$$

$$H_{l,m,p}(\nu) \coloneqq \frac{Q_{l,m,p}\mathcal{T}_1}{1 - |\mathcal{R}_1|\,|\mathcal{R}_2|\,e^{i\frac{2\pi}{F}\left(\nu - \nu_{n,l,m}\right)}}, \quad n = 0, 1, 2, ... \tag{2.42}$$

where we define the input overlap integral

$$Q_{l,m,p} \coloneqq \int d\mathbf{x}_1\, \boldsymbol{\mathcal{E}}_{l,m,p}^{*}(\mathbf{x}_1) \cdot \boldsymbol{\mathcal{E}}_\mathrm{i}^{+}(\mathbf{x}_1) \tag{2.43}$$

on the surface of mirror 1, and $\nu_{n,l,m}$ is defined in equation (2.36). The term involving the stray light is eliminated because by definition the stray light's profile must be orthogonal to the cavity mode (otherwise it is considered part of the cavity mode). Within the paraxial approximation, we take $\mathbf{x}_1 = \left(x, y, -\frac{L}{2}\right)$ so the integral is taken over the $xy$-plane at $z = -\frac{L}{2}$. Notice that unless $Q_{l,m,p} = 0$, all modes have a non-zero amplitude.

Figure 2.5 shows an example of the transfer function near a single resonance frequency $\nu_{n,l,m}$, for $\mathcal{T}_1 = \sqrt{100 \times 10^{-6}}$, $\mathcal{R}_1 = \mathcal{R}_2 = -\sqrt{1 - |\mathcal{T}_1|^2}$. The transfer function shows that the amplitude of the cavity mode is larger than that of the input beam near the resonance. The phase of the cavity mode (relative to the input beam) also rapidly changes from $\sim -\pi/2$ to $\sim +\pi/2$ around the resonance.

For high reflectivity mirrors with $|\mathcal{R}_{1,2}| \sim 1$, the transfer function is well-approximated by a square-root-Lorentzian function around each resonance

$$H(\nu) \simeq \frac{Q_{l,m,p}\mathcal{T}_1}{1 - |\mathcal{R}_1|\,|\mathcal{R}_2|}\frac{1}{1 + i\frac{2\pi\left(\nu - \nu_{n,l,m}\right)/F}{1 - |\mathcal{R}_1||\mathcal{R}_2|}} \tag{2.44}$$

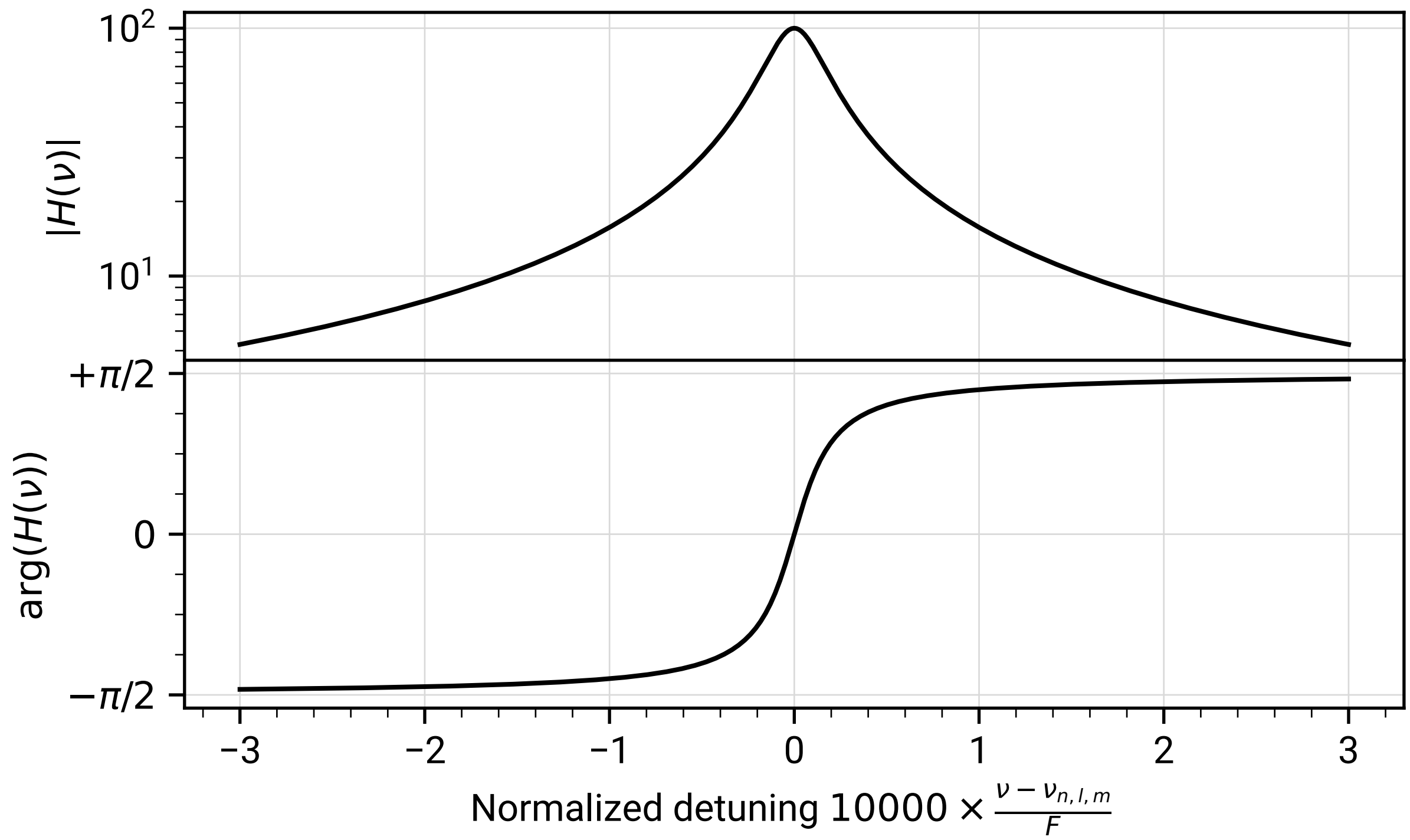

Figure 2.5: Amplitude and phase of the cavity mode transfer function as a function of the detuning from resonance divided by the free spectral range. The cavity mirror transmission and reflection amplitude coefficients are $\mathcal{T}_1 = \sqrt{100 \times 10^{-6}}$, $\mathcal{R}_1 = \mathcal{R}_2 = -\sqrt{1 - |\mathcal{T}_1|^2}$.

such that the "linewidth" of the resonance can be defined as the frequency range for which the squared magnitude of the transfer function is larger than half its maximum value (full-width half-maximum):

$$\Gamma_{\mathrm{FWHM}} \coloneqq \frac{F}{\pi} \left(1 - |\mathcal{R}_1| \, |\mathcal{R}_2|\right) \tag{2.45}$$

The cavity's "finesse" $\mathcal{F}$ is defined as the ratio of the free spectral range to the full-width half-maximum linewidth, so

$$\mathcal{F} \coloneqq \frac{F}{\Gamma_{\mathrm{FWHM}}} = \frac{\pi}{1 - |\mathcal{R}_1| \, |\mathcal{R}_2|} \tag{2.46}$$

The power carried by the EM wave in the cavity mode is proportional to the area integral of the squared magnitude of the wave amplitude, so using equation (2.41) we can write that the power in the cavity mode $P_{\mathrm{c}}$ is

$$P_{\mathrm{c}} = \left|H_{l,m,p}\left(\nu\right)\right|^2 P_{\mathrm{i}} \tag{2.47}$$

where $P_\mathrm{i}$ is the power in the input beam. $P_\mathrm{c}$ is called the "circulating power", because it quantifies the power moving in one direction (left to right) within the cavity. The ratio of the circulating power to the input power is called the "enhancement factor"

$$G_\mathrm{c} \coloneqq \frac{P_\mathrm{c}}{P_\mathrm{i}} = \left| \frac{Q_{l,m,p} \mathcal{T}_1}{1 - |\mathcal{R}_1| \, |\mathcal{R}_2| \, e^{i \frac{2\pi}{F} \left(\nu - \nu_{n,l,m}\right)}} \right|^2 \tag{2.48}$$

A large $\sim 1 \times 10^4$ enhancement factor enables the high circulating power of 75 kW required by the LPP.

### 2.3.3 Cavity transmission transfer function

The wave which exits mirror 2 moving to the right (away from the cavity) $\mathbf{E}_\mathrm{t}^+ (\mathbf{x}_2)$ is related to the intra-cavity wave $\mathbf{E}_\mathrm{c}^+ (\mathbf{x}_1)$ in a straightforward way:

$$\mathbf{E}_\mathrm{t}^+ (\mathbf{x}_2) = \mathcal{T}_2 U_\mathrm{FS}^+ \mathbf{E}_\mathrm{c}^+ (\mathbf{x}_1) \tag{2.49}$$

$$= \mathcal{T}_2 e^{i\phi_\mathrm{RT}/2} E_{l,m,p} \boldsymbol{\mathcal{E}}_{l,m,p} (\mathbf{x}_1) + \mathcal{T}_2 U_\mathrm{FS}^+ \tilde{\mathbf{E}}_\mathrm{c}^+ (\mathbf{x}_1) \tag{2.50}$$

When the wave frequency is near a resonance and the mirrors have near-unity reflectivity, the amplitude of the stray light is negligible compared to the cavity mode, so the transmitted wave can be approximated as

$$\mathbf{E}_\mathrm{t}^+ (\mathbf{x}_2) \simeq \mathcal{T}_2 e^{i\phi_\mathrm{RT}/2} H_{l,m,p} (\nu) \, E_\mathrm{i} \boldsymbol{\mathcal{E}}_{l,m,p} (\mathbf{x}_1) \tag{2.51}$$

and so the power in the transmitted wave is

$$P_\mathrm{t} = |\mathcal{T}_2|^2 \, P_\mathrm{c} \tag{2.52}$$

The cavity transmissivity $T_\mathrm{c}$ is defined as the ratio of the transmitted power and input power:

$$T_\mathrm{c} \coloneqq \frac{P_\mathrm{t}}{P_\mathrm{i}} = |\mathcal{T}_2|^2 \, G_\mathrm{c} \tag{2.53}$$

### 2.3.4 Cavity reflection transfer function

The properties of the light which reflects off of mirror 1 are also of interest. This is because it is difficult to measure the properties of the cavity mode without altering it: any measurement device which absorbs or redirects the intra-cavity EM wave will change the shape, amplitude, and/or phase of the cavity mode. The wave reflecting off of mirror 1 can be written as a superposition of the "promptly-reflected" input wave and left-going intra-cavity field:

$$\mathbf{E}_\mathrm{i}^- (\mathbf{x}_1) = \mathcal{R}_1' \mathbf{E}_\mathrm{i}^+ (\mathbf{x}_1) + \mathcal{T}_1' \mathbf{E}_\mathrm{c}^- (\mathbf{x}_1) \tag{2.54}$$

where $\mathcal{R}_1'$ is the reflection amplitude for a wave reflecting from the left side of mirror 1, and $\mathcal{T}_1'$ is the transmission amplitude for a wave transmitting through mirror 1 from right to left.

Note that phases and amplitudes of $\mathcal{T}_1$, $\mathcal{T}_1'$, $\mathcal{R}_1$, and $\mathcal{R}_1'$ are constrained by the requirement that they represent a physical object (mirror 1) which does not add power to the system. Without loss of generality, we can assume that $\arg(\mathcal{T}_1) = \arg(\mathcal{T}_1') = 0$, in which case the amplitudes must satisfy the inequality [46]

$$\sqrt{|\mathcal{T}_1|^2 |\mathcal{R}_1|^2 + |\mathcal{T}_1'|^2 |\mathcal{R}_1'|^2 + 2 |\mathcal{T}_1'| |\mathcal{R}_1| |\mathcal{R}_1'| |\mathcal{T}_1| \cos(\arg(\mathcal{R}_1) + \arg(\mathcal{R}_1'))} \\ \leq \sqrt{\left(1 - |\mathcal{T}_1|^2 - |\mathcal{R}_1'|^2\right)\left(1 - |\mathcal{T}_1'|^2 - |\mathcal{R}_1|^2\right)} \tag{2.55}$$

In the common case that $|\mathcal{T}_1| = |\mathcal{T}_1'|$ and $|\mathcal{R}_1| = |\mathcal{R}_1'|$ the inequality simplifies to

$$\left|\cos\left(\frac{\arg(\mathcal{R}_1) + \arg(\mathcal{R}_1')}{2}\right)\right| \leq \frac{1 - |\mathcal{T}_1|^2 - |\mathcal{R}_1|^2}{2 |\mathcal{T}_1| |\mathcal{R}_1|} \tag{2.56}$$

For example, if there no loss in mirror 1, $1 - |\mathcal{T}_1|^2 - |\mathcal{R}_1|^2 = 0$ and we must have that $\arg(\mathcal{R}_1) + \arg(\mathcal{R}_1') = \pm\pi$

Neglecting the relatively small amplitude stray light, we can express the left-going intra-cavity wave in terms of the right-going cavity mode defined previously

$$\mathbf{E}_{\mathrm{c}}^{-}(\mathbf{x}_1) = U_{\mathrm{FS}}^{-} U_{\mathrm{R},2} U_{\mathrm{FS}}^{+} \overline{\mathbf{E}}_{\mathrm{c}}^{+}(\mathbf{x}_1) \tag{2.57}$$

$$= \mathcal{R}_2 e^{i\phi_{\mathrm{RT}}} \overline{\mathbf{E}}_{\mathrm{c}}^{+}(\mathbf{x}_1) \tag{2.58}$$

and using the cavity mode transfer function (section 2.3.2) it can then be expressed in terms of just the input field amplitude $E_{\mathrm{i}}$

$$\mathbf{E}_{\mathrm{c}}^{-}(\mathbf{x}_1) = \mathcal{R}_2 e^{i\phi_{\mathrm{RT}}} H_{l,m,p}(\nu) E_{\mathrm{i}} \boldsymbol{\mathcal{E}}_{l,m,p}(\mathbf{x}_1) \tag{2.59}$$

Thus, we see that the reflected wave is a superposition of the input wave and cavity mode profiles:

$$\mathbf{E}_{\mathrm{i}}^{-}(\mathbf{x}_1) = \left(\mathcal{R}_1' \boldsymbol{\mathcal{E}}_{\mathrm{i}}^{+}(\mathbf{x}_1) + \mathcal{T}_1' \mathcal{R}_2 e^{i\phi_{\mathrm{RT}}} H_{l,m,p}(\nu) \boldsymbol{\mathcal{E}}_{l,m,p}(\mathbf{x}_1)\right) E_{\mathrm{i}} \tag{2.60}$$

Integrating the squared magnitude of both sides of equation (2.60) gives an expression for the cavity reflectivity $R_{\mathrm{c}}$:

$$R_{\mathrm{c}} := \frac{P_{\mathrm{r}}}{P_{\mathrm{i}}} = \left(1 - |Q_{l,m,p}|^2\right) |\mathcal{R}_1'|^2 + |Q_{l,m,p}|^2 |\mathcal{R}_1'|^2 \left|\frac{1 - \left(1 - \frac{\mathcal{T}_1' \mathcal{T}_1}{\mathcal{R}_1' \mathcal{R}_1}\right) \mathcal{R}_1 \mathcal{R}_2 e^{i\phi_{\mathrm{RT}}}}{1 - \mathcal{R}_1 \mathcal{R}_2 e^{i\phi_{\mathrm{RT}}}}\right|^2 \tag{2.61}$$

where $P_{\mathrm{r}}$ is the power reflected from the cavity, and recalling that

$$\phi_{\mathrm{RT}} = \frac{2\pi}{F}(\nu - \nu_{n,l,m}) - \arg(\mathcal{R}_1) - \arg(\mathcal{R}_2) \tag{2.62}$$

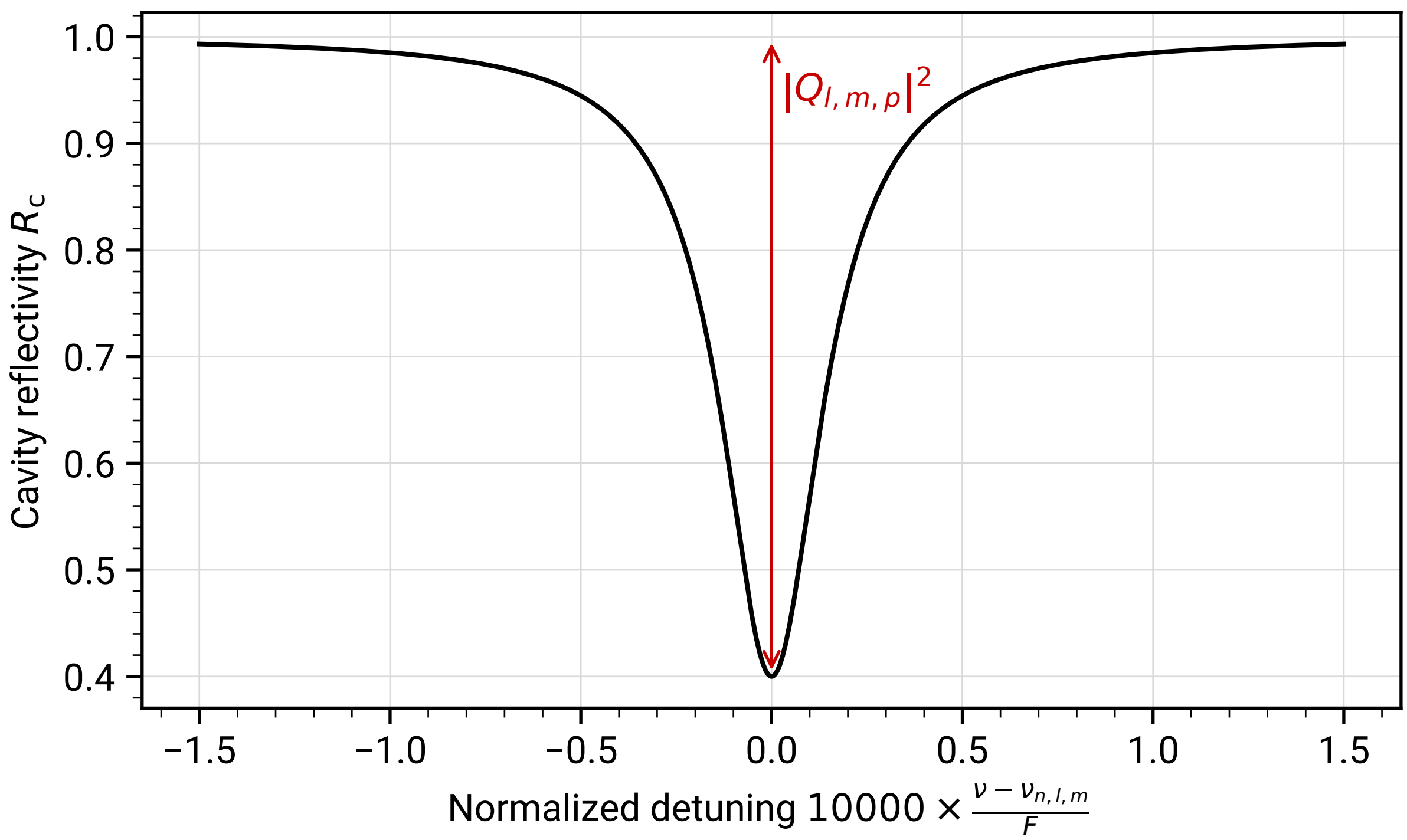

Figure 2.6: Cavity reflectivity as a function of the detuning from resonance divided by the free spectral range. The cavity mirror transmission and reflection amplitude coefficients are $\mathcal{T}_1 = \mathcal{T}_1' = \sqrt{100 \times 10^{-6}}$, $\mathcal{R}_1 = \mathcal{R}_2 = -\mathcal{R}_1' = -\sqrt{1 - |\mathcal{T}_1|^2}$, and $|Q_{l,m,p}|^2 = 0.6$.

An example of the cavity reflectivity as a function of frequency is shown in figure 2.6 for the case of a lossless cavity with $\mathcal{R}_1 = \mathcal{R}_2 = -\sqrt{1 - 100 \times 10^{-6}}$, $\mathcal{R}_1' = -\mathcal{R}_1$, $\mathcal{T}_1 = \mathcal{T}_1' = \sqrt{100 \times 10^{-6}}$, and $|Q_{l,m,p}|^2 = 0.6$. If the mirrors are nearly lossless such that $1 - |\mathcal{R}_j|^2 - |\mathcal{T}_j|^2 \ll 1,\ j = 1, 2$, then on resonance $R_c \simeq \left(1 - |Q_{l,m,p}|^2\right)$. Thus, $|Q_{l,m,p}|^2$ is referred to as the "coupling efficiency" since it represents the fraction of the power in the input beam which can be coupled into the cavity mode. A measurement of the cavity reflectivity can be used to infer the coupling efficiency. Another reason the LPP uses a fundamental mode of the cavity is because it is easiest to achieve a high coupling efficiency to the fundamental mode using a (roughly) Gaussian input beam.

## 2.4 Mode electromagnetic field

The electric and magnetic fields of the on-resonance modes of a symmetric stigmatic cavity with highly reflective mirrors are therefore

$$\mathbf{E}_{l,m,p}(x,y,z,t) = \hat{\mathbf{x}}_p \cos(2\pi\nu_{n,l,m}t)\, E_0 \frac{w_0}{w(z)} H_l\left(\frac{\sqrt{2}x}{w(z)}\right) H_m\left(\frac{\sqrt{2}y}{w(z)}\right)$$
$$\exp\left(-\frac{x^2+y^2}{w(z)^2}\right) \begin{smallmatrix}\sin\\ \cos\end{smallmatrix}\left(\frac{2\pi}{\lambda_\mathrm{L}}z + \frac{2\pi}{\lambda_\mathrm{L}}\frac{x^2+y^2}{2R(z)} - \psi_\mathrm{G}(z)\right) \tag{2.63}$$

$$\mathbf{B}_{l,m,p}(x,y,z,t) = (\hat{\mathbf{z}}\times\hat{\mathbf{x}}_p) \sin(2\pi\nu_{n,l,m}t)\, \frac{E_0}{c}\frac{w_0}{w(z)} H_l\left(\frac{\sqrt{2}x}{w(z)}\right) H_m\left(\frac{\sqrt{2}y}{w(z)}\right)$$
$$\exp\left(-\frac{x^2+y^2}{w(z)^2}\right) \begin{smallmatrix}-\cos\\ \sin\end{smallmatrix}\left(\frac{2\pi}{\lambda_\mathrm{L}}z + \frac{2\pi}{\lambda_\mathrm{L}}\frac{x^2+y^2}{2R(z)} - \psi_\mathrm{G}(z)\right) \tag{2.64}$$

respectively, where $E_0 = 4\sqrt{\frac{1}{\pi c \epsilon_0}\frac{P_\mathrm{c}}{w_0^2}}$ is the amplitude of the standing wave, $\epsilon_0$ is the vacuum permittivity, $\lambda_\mathrm{L} = c/\nu_{n,l,m}$, and for the electric field the sine solution is taken when $n$ is odd, and the cosine solution when $n$ is even (vice versa for magnetic field). Any superposition of these modes is also a solution of the EM wave equations. Note that modes are standing waves, with the magnetic field $\pi/2$ out of phase with the electric field in both space and time.

## 2.5 On-resonance power formulas

The LPP is operated with the EM wave frequency matched to the cavity resonance in order to achieve the highest available enhancement factor. In this condition, the cavity transfer functions derived in the preceding sections can be used to write several useful formulas which relate parameters of the cavity to measurable quantities.

First, define the "mirror loss" as

$$X_j \coloneqq 1 - T_j - R_j, \quad j = 1,2 \tag{2.65}$$

which represents the power not reflected or transmitted (in practice essentially all this power is either scattered or absorbed by the mirror). For brevity we define the mirror power reflection and transmission coefficients as

$$R_j \coloneqq |\mathcal{R}_j|^2 \tag{2.66}$$
$$T_j \coloneqq |\mathcal{T}_j|^2 \tag{2.67}$$

The (on-resonance) enhancement factor can then be written as

$$G_c = |Q|^2 \frac{T_1}{\left(1 - \sqrt{1 - X_1}\sqrt{1 - X_2}\right)^2} \tag{2.68}$$

$$\simeq |Q|^2 \frac{4T_1}{\left(T_1 + X_1 + T_2 + X_2\right)^2} \tag{2.69}$$

where the second line assumes that $X_j \ll 1,\ j = 1, 2$. This approximation will be carried through the rest of this section. $T_1 + X_1 + T_2 + X_2$ is referred to as the "round-trip loss", since it represents the fraction of circulating power lost after the intra-cavity wave completes one round-trip of the cavity.

Clearly, the maximum possible enhancement factor is infinite when $|Q|^2 = 1$ and $X_1 = X_2 = T_2 = 0$, with $T_1 \to 0$. Such a situation is not realistic, since the mirrors will inevitably have some non-zero losses. It is more interesting to consider what values of $T_1$ and $T_2$ maximize the enhancement factor in the presence of some non-zero losses $X_{1,2}$. Clearly, choosing $|Q|^2 = 1$ and $T_2 = 0$ still maximizes the enhancement factor, such that

$$G_c = \frac{4}{X_1 + X_2} \frac{\frac{T_1}{X_1+X_2}}{\left(1 + \frac{T_1}{X_1+X_2}\right)^2} \tag{2.70}$$

which achieves a maximum of $G_c = (X_1 + X_2)^{-1}$ when $T_1 = X_1 + X_2$. This value achieves a balance between low transmission (which reduces loss of the cavity mode power through the input mirror) and high transmission (which increases the rate at which input power can enter the cavity mode). State-of-the-art mirror losses are roughly $1 \times 10^{-6}$, which limits the maximum achievable enhancement factor to $< 5 \times 10^5$. If identical mirrors are used (as in our experiment), then the maximum achievable enhancement factor is $G_c = (X_1 + X_2)^{-1}/2$ when $T = (X_1 + X_2)/2$, where $T \coloneqq T_1 = T_2$. Laser systems which can supply input powers of $P_i \geq 50\,\mathrm{W}$ are available, so the LPP only requires an enhancement factor of 1500 to reach its target circulating power of 75 kW. We design the LPP to achieve this enhancement factor with a generous factor of safety of roughly 8, as there is no apparent advantage to setting the mirror transmissivity to maximize the enhancement factor, and the resulting narrower resonance linewidth may make it more difficult to stabilize the input laser frequency to the center of the resonance (see chapters 3 and 6).

The power lost in the cavity due to mirror loss is the product of the mirror loss and circulating power, summed over both mirrors. Therefore, we can define the cavity loss $L_c$ as the ratio of power lost $P_x$ to input power such that

$$L_c \coloneqq \frac{P_x}{P_i} = (X_1 + X_2) \frac{P_c}{P_i} = (X_1 + X_2) \frac{T_c}{T_2} \tag{2.71}$$

Assuming that the mirrors are the only source of lost power in the cavity, we must also have that $L_c = 1 - R_c - T_c$, so the total mirror loss

$$X_1 + X_2 = T_2 \frac{1 - R_c - T_c}{T_c} \tag{2.72}$$

can be expressed in terms of quantities which can be readily experimentally measured. In principle the mirror loss could be measured directly on and individual mirror by measuring $R_j$ and $T_j$, but these measurements would need to be extremely precise to determine a small mirror loss value. On the other hand, measuring the cavity reflectivity and transmissivity and employing equation (2.72) implicitly leverages the enhancement factor of the cavity to make the measurement more precise. As is discussed in chapter 7, knowing the mirror loss is important for diagnosing potential problems with the cavity mirrors.

Equations (2.69) and (2.71) can also be combined to yield an expression for the coupling efficiency (here abbreviated as $|Q|^2$ for brevity) in terms of other measurable quantities:

$$|Q|^2 = \frac{T_c}{4T_1T_2}\left(T_2\frac{1-R_c-T_c}{T_c}+T_1+T_2\right)^2 \tag{2.73}$$

Assuming that the mirror transmissivities are known, this enables a measurement of the coupling efficiency with the frequency $\nu$ on resonance, as opposed to needing to scan the frequency across the resonance while measuring the cavity reflected power $P_r$ (the resulting signal of which is shown in figure 2.6).

## 2.6 Mirror thermoelastic deformation

Optical power in the cavity mode can be absorbed by the mirror, where it is converted to heat. This generates a temperature gradient in the mirror's substrate material as the heat flows away from the cavity mode, which in turn leads to thermal expansion of the substrate material and a change in the shape of the mirror surface. The change in mirror surface shape changes the shape of the cavity mode, which affects the profile of the heat load on the mirror surface and so the shape of the mirror surface. Thus, the shape of the mode is defined by a set of self-consistent equations describing the mirror surface shape as a function of the cavity mode shape, and the cavity mode shape as a function of the mirror surface shape.

To leading order we can model the effect of the fundamental cavity mode on the mirror surface shape as a change in the mirror's curvature radius [47]–[49]. Define the deformed mirror's curvature radius as

$$R =: \left(R_0^{-1}+R_{th}^{-1}\right)^{-1} \tag{2.74}$$

where $R_0$ is the undeformed curvature radius and $R_{th}^{-1}$ represents the change in curvature induced by heating from the cavity mode. $R_0$ and $R_{th}^{-1}$ are added in inverse because the curvature radius is inversely proportional to the curvature (quadratic coefficient) of the mirror's surface profile. The change in curvature radius is therefore

$$R_\Delta \coloneqq R - R_0 \tag{2.75}$$

$$= -R_0\left(1+\frac{R_{th}}{R_0}\right)^{-1} \tag{2.76}$$

Assuming linear models for the thermal conduction and thermoelastic deformation, dimensional analysis shows that the change in mirror curvature is be proportional to the absorbed power, proportional to the coefficient of thermal expansion (CTE) of the mirror substrate material, inversely proportional to the mirror substrate's thermal conductivity, and inversely proportional to the square of the width of the cavity mode on the mirror's surface $w_1$. That is,

$$R_{\mathrm{th}}^{-1} = -\mathcal{N} \frac{\alpha\,(1+v)}{\kappa} \frac{AP_{\mathrm{c}}}{w_1^2} \tag{2.77}$$

where $\mathcal{N} \geq 0$ is a dimensionless constant that depends on the geometry of the mirror substrate, $\alpha$ is the substrate's CTE, $v$ is the substrate's Poisson's ratio, $\kappa$ is the substrate's thermal conductivity, and $A$ is the fraction of incident power absorbed by the mirror. Though the Poisson's ratio is dimensionless, we separate it from the definition of $\mathcal{N}$ because in the equations for linear thermoelastic deformation it always accompanies the ratio $\alpha/\kappa$ in the so-called "thermal distortivity" parameter [50]

$$\delta \coloneqq \frac{\alpha\,(1+v)}{\kappa} \tag{2.78}$$

which quantifies the susceptibility of a material to thermoelastic deformation. Further defining the "mirror distortivity" as

$$\mathcal{M} \coloneqq \mathcal{N}\delta A \tag{2.79}$$

allows us to encapsulate all the relevant properties of the mirrors in a single parameter, such that

$$R_{\mathrm{th}}^{-1} = -\mathcal{M} \frac{P_{\mathrm{c}}}{w_1^2} \tag{2.80}$$

Note that the sign of $\mathcal{M}$ depends on the sign of the CTE, which for most materials is typically (but not necessarily) positive.

Making the assumption that both cavity mirrors are identical, the cavity stability parameter $g$ can be written in terms of $R_{\mathrm{th}}$:

$$g = g_0 - \frac{L}{R_{\mathrm{th}}} \tag{2.81}$$

$$g_0 \coloneqq 1 - \frac{L}{R_0} \tag{2.82}$$

where $g_0$ is the stability parameter of the undeformed cavity (i.e. with $P_{\mathrm{c}} = 0$). Combining equations (2.80) and (2.27) gives the self-consistent solution for the change in mirror curvature as a function of circulating power:

$$\frac{R_0}{R_{\mathrm{th}}} = -\frac{1+g_0}{g_0 + \mathrm{sgn}\,(\mathcal{M})\sqrt{1+\frac{1-g_0^2}{p^2}}} \tag{2.83}$$

where

$$p \coloneqq \frac{\pi}{\lambda_{\mathrm{L}}} \mathcal{M} P_{\mathrm{c}} \tag{2.84}$$

is a dimensionless parameter characterizing the circulating power. Equivalently, the solution can be written for the stability parameter:

$$g = g_0 + \frac{1 - g_0^2}{g_0 + \operatorname{sgn}(\mathcal{M}) \sqrt{1 + \frac{1-g_0^2}{p^2}}} \tag{2.85}$$

Figure 2.7 shows the dependence of the stability parameter on circulating power. Equation

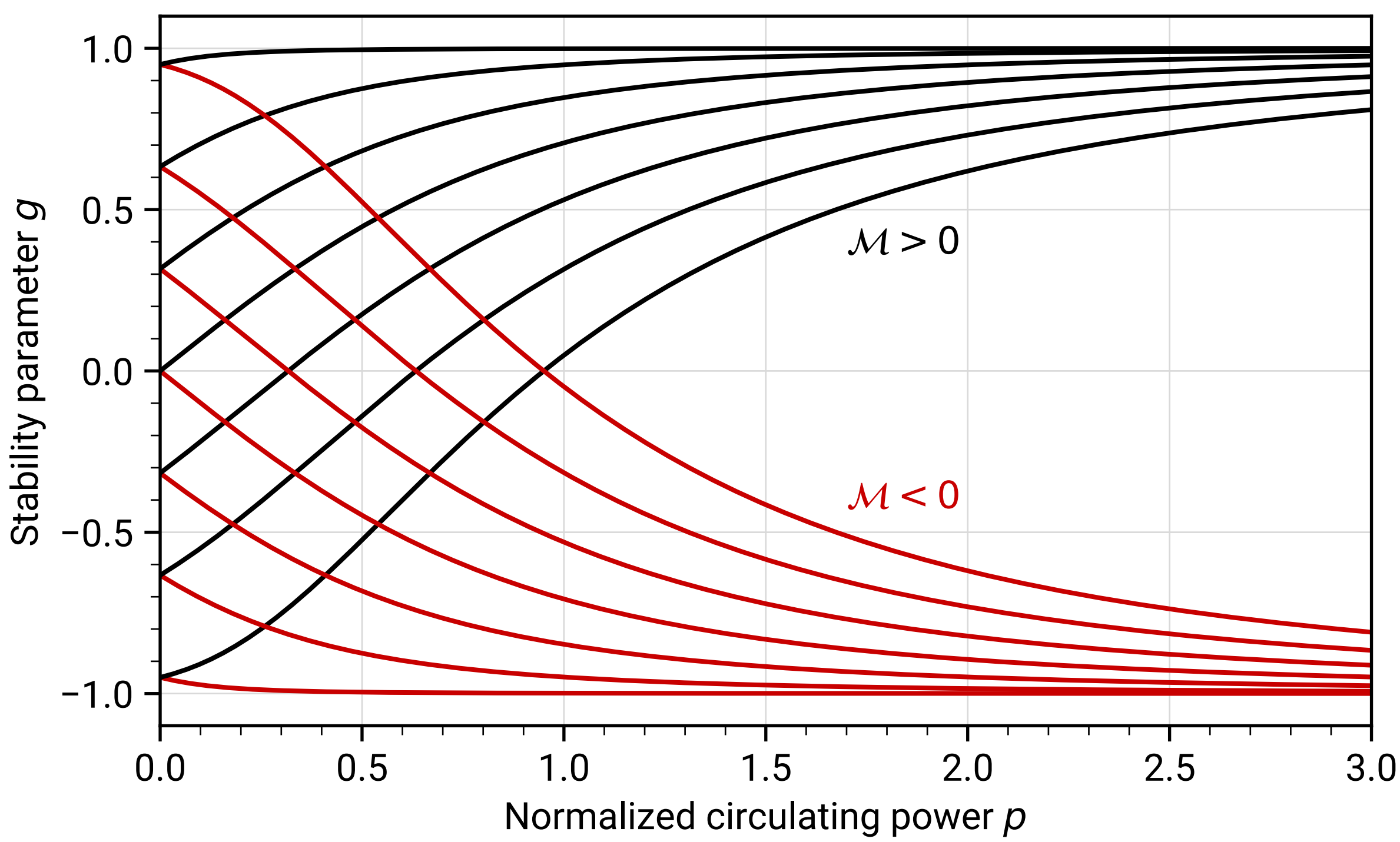

Figure 2.7: Cavity stability parameter $g$ as a function of the normalized circulating power $p$ for several different values of $g_0 = g\,(p = 0)$ and for both a positive (black) and negative (red) mirror distortivity parameter $\mathcal{M}$.

(2.85) can be used to express any other parameter of the cavity mode (e.g. $w_0$) as a function of the circulating power. Increasing circulating power pushes the stability parameter away from its initial value to the edge of its stable range at $-\operatorname{sgn}(\mathcal{M})$. In the common case that $\mathcal{M} > 0$ this means that simultaneously achieving a high circulating power and small focal waist requires that the mirrors have a small distortivity parameter. As is discussed in section 3.3, this forced us to use a non-standard mirror substrate material with a small CTE to achieve our target cavity mode parameters (listed in table 2.1).

When $\mathcal{M} > 0$, in the near-concentric limit $g_0 \to -1$

$$g \simeq -\frac{1-p^2}{1+p^2} \tag{2.86}$$

so that the smallest possible focal waist of the cavity mode is

$$w_{0,\min} \simeq \sqrt{\frac{\lambda_\mathrm{L} R_0}{\pi} p} \tag{2.87}$$

This represents the minimum possible focal waist. Of course, these equations are only valid within the paraxial approximation ($w_0 \gg \lambda_\mathrm{L}\pi$), but $w_{0,\min}$ can easily be substantially larger than this with realistic mirror parameters (see section 7.6).

Interestingly, the fact that $w_{0,\min} \sim \sqrt{p}$ means that the focal intensity has an asymptotic maximum limit. Defining the standing wave focal intensity as

$$I_0 \coloneqq 4\frac{2P_\mathrm{c}}{\pi w_0^2} \tag{2.88}$$

we see that the maximum possible standing wave focal intensity is

$$I_{0,\max} = \frac{8}{\pi}\frac{1}{\mathcal{M}R_0} \tag{2.89}$$

which depends only on the geometry and thermal properties of the mirrors. The existence of this limit implicitly assumes that it is impossible to sustain a stable mode in the deformed cavity if $g_0 < -1$. It is clearly not possible to start the cavity with $g_0 < -1$ and $p = 0$ and then increase the circulating power, because there will be no stable mode to heat the mirror surfaces. However, the cavity could be started with $g_0 > -1$, the power increased, and then the cavity length increased such that $g_0 < -1$ (i.e. the cavity would be beyond concentricity if the mirrors were not deformed). We hypothesize that even in this case the mode will be effectively unstable if $g_0 < -1$. Though not captured by the simple model of the mirror deformation presented in this section, in reality the mirror deformation is localized around the position of the cavity mode on the mirror surface, and the curvature of the mirror away from the cavity mode remains approximately unchanged. This means that the edges of the cavity mode will reflect from mirrors which are in an unstable configuration, and likely be scattered out of the cavity mode, causing increased diffraction loss. The smaller $g_0$ is, the smaller the radius on the mirror surfaces within which the mirror curvature still forms a stable cavity, and the more diffraction loss there will be from outside of this region. We have not yet modeled how large this diffraction loss should be or how rapid its onset is as $g_0$ is reduced below $-1$. In future we may conduct an experiment to verify this behavior, as our previous experiments have only approached, but not passed, the putative focal intensity limit (see section 7.6).

It is also unclear if it is possible for the cavity to remain stable when $g$ passes through 0 as the circulating power increases, since at $g = 0$ any increase in the curvature radius of mirror 1 and decrease in the curvature radius of mirror 2 (or vice versa) results in an unstable (asymmetric) cavity [45]. We have not theoretically explored the stability of thermoelastic deformation effects in asymmetric cavities.

The change of mirror curvature radius with circulating power also results in a change in the mode spectrum of the cavity. In principle this can cause higher order transverse modes (of lower longitudinal mode index) to become degenerate with the fundamental mode, as discussed in section 2.3.1. In this case both modes will be excited by the input beam (assuming non-zero coupling efficiency to each mode) which will change the amount and distribution of the heat load on the mirror surface. This complicates the behavior of the mirror surface shape as a function of circulating power—we have not modeled this behavior. Regardless, operating the LPP under such a condition is undesirable because the shape of the intensity profile at the center of the cavity is more difficult to keep stable or characterize.

Though the model presented in this section is a rudimentary approximation, experimental data presented in section 7.6 shows that it is quite accurate for the conditions relevant to the LPP. Through further work, it should be possible to develop a semi-analytical or numerical model which incorporates a more accurate representation of the mirror surface shape. This would allow us to make predictions about the changes in the mode shape other than just its width, and the behavior of the mode (or lack thereof) when $g_0 < -1$. A fully numerical model would also allow for the simulation of a temperature-dependent CTE, which may be relevant to the substrate material used in the LPP (see section 3.3)

# Chapter 3

# Cavity mirror design

The mirrors are the most critical component in the laser phase plate. The successful design which is described in this chapter was only arrived at after several iterations of prototyping which took several years. In conjunction with the associated optomechanical and laser hardware described in chapters 4 and 6, these mirrors are able to support a circulating power of 99 kW at a numerical aperture of 0.052, corresponding to a record-high continuous-wave laser intensity of 590 $\mathrm{GWcm^{-2}}$ (see section 7.4). They are superlative, and are a testament to the skill and craftsmanship of the people who make them. The manufacturers involved in the mirror fabrication supply chain are listed in section 3.6.

This chapter describes the design criteria for the mirrors, and provides our reasoning for our design decisions. A summary of the mirror specifications is given in section 3.7.

## 3.1 Laser wavelength

The laser wavelength influences the requirements for most of the mirror specifications. Longer wavelengths have the advantage of yielding a stronger ponderomotive potential (see section 9.1), thereby requiring less circulating power to achieve the necessary $\pi/2$ phase shift for phase contrast imaging. Indeed, an earlier incarnation of this project investigated using a tightly-focused $\sim 100$ W $CO_2$ laser with $\lambda_{\mathrm{L}} = 10.6\,\mu\mathrm{m}$ in order to generate a $\pi/2$ phase shift without the need for resonant amplification in a Fabry-Pérot cavity [21]. However, longer wavelengths require a larger numerical aperture to achieve the same focal waist (same second cut-on frequency of the phase plate, see section 8.4), which means that the laser beam illuminates a larger area on the mirror surface. This increases the required clear aperture of the mirror, which is more difficult for manufacturers to achieve. Additionally, the longer wavelength results in a higher initial cut-on spatial frequency of the phase plate (see section 8.4). Ultraviolet and blue wavelengths have been seen to cause photochemical degradation of the mirror surfaces, resulting in lower cavity finesse [51], [52]. Shorter wavelengths also impose stricter requirements on the microroughness and surface figure of the mirrors, as their size as a fraction of the laser wavelength determines the magnitude of their influence on scattering and distortion of the cavity mode, respectively. Cumulatively, these considerations limit the range of reasonable wavelengths to roughly 500 nm to 2 µm.

Within this range the choice of wavelength is further constrained by the availability of lasers with a narrow enough linewidth and high enough output power to pump the cavity. Simultaneously, high-reflectivity mirror coatings must be available at those wavelengths. In 2016, this left us with three options: 532 nm, 1064 nm, and 1550 nm. We chose 1064 nm for several reasons. First, it had the highest power narrow linewidth lasers. We chose a ytterbium-based fiber laser (NKT Koheras ADJUSTIK Y10) for its modularity and convenience, though solid-state neodymium-doped yttrium aluminum garnet (Nd:YAG) lasers are also available at this wavelength. Second, high-reflectivity mirror coating technology at this wavelength was well-developed, in no small part due to the work of the Laser Interferometer Gravitational-Wave Observatory (LIGO) collaboration which uses 1064 nm in their primary interferometer arms [53].

However, laser technology continues to develop quickly, and other wavelengths may become (or already be!) good choices for applications with requirements similar to those of the LPP.

## 3.2 Geometry

### 3.2.1 Mirror radius of curvature

The mirror surface is spherical and concave (CC) to the substrate material. As shown in section 2.2.3, a smaller mirror radius of curvature (ROC) results in lower alignment sensitivity of a near-concentric Fabry-Pérot. This is useful as it reduces the stability requirements of the mirrors' optomechanical mount and the associated mode position stabilization feedback system (see section 6.2.5). It also reduces the cavity mode's sensitivity to thermal deformation of the mirror surfaces (see section 2.6). However, smaller ROC surfaces with smooth enough surfaces to support a high cavity finesse are more difficult to manufacture. Mirror surface polishing requirements are discussed in section 3.4. The smallest "superpolished" (defined in section 3.4) ROC we could find on the market was 10 mm. This required the manufacturer (Coastline Optics) to develop some new tooling and metrology, which cost roughly $10k.

### 3.2.2 Outer diameter

The outer diameter (OD) of the mirror must be large enough that the mirror surface is substantially wider than the cavity mode impinging on it. Portions of the cavity mode which do not hit the mirror surface are lost from the cavity mode, which reduces the finesse of the cavity and heats the surrounding optomechanics. Since superpolished surfaces readily achieve scattering losses of less than $10 \times 10^{-6}$, these "diffraction losses" should be of order $\sim 1 \times 10^{-6}$ to be negligible. Assuming the cavity mode is centered on the optical axis of the mirror, the diffraction loss is

$$L_{\mathrm{D}} = e^{-\frac{D^2}{2w_1^2}} \tag{3.1}$$

where $D$ is the diameter of the mirror surface and $w_1 =\simeq R_0\mathrm{NA}$ is the mode waist on the mirror surface (equation (2.29)). Therefore, for $L_{\mathrm{D}} \leq 1 \times 10^{-6}$, we need $D/w_1 \geq 5.26$, i.e.

$$D \geq 2.63\,\mathrm{mm} \tag{3.2}$$

According to Coastline Optics, it is difficult to fixture such a small optic during superpolishing. It is also difficult to achieve a 100% "clear aperture", i.e. to meet the microroughness, surface figure, and defect specifications across the entire surface. Additionally, a larger usable mirror surface allows some room for the cavity mode to be positioned on the mirror surface to avoid defects. We therefore elected to use a more standard and workable $D = 7.75\,\mathrm{mm}$. Much larger than this diameter and superpolishing would have again become difficult due to the steep sides of the mirror surface at the outer diameter. Tight space constraints on our cavity optomechanics (see chapter 4) also restrict the outer diameter to be not much larger than this.

If a smaller outer diameter is required, it is also possible to have the mirror first superpolished and coated (see section 3.5) and then grind or cut away its outer diameter after applying a temporary protective coating to its optical surfaces [54].

### 3.2.3 Center thickness and lens radius of curvature

The mirror substrate is designed such that it focuses throughgoing light. This allows us to send a nominally collimated laser beam through the back side of the mirror while still achieving a high coupling efficiency to the focused cavity mode. The coupling efficiency can then be optimized by finely adjusting the collimation of the input laser beam using external lenses. We design the substrate such that it focuses light at the center of curvature of the CC surface, since this is nearly exactly at the focus of the mode in a near-concentric Fabry-Pérot cavity. This requires the back surface of the mirror substrate to be convex (CX), where the relationship between the required CX ROC $R_1$, center thickness $T_{\mathrm{c}}$, and refractive index of the substrate material $n$ are illustrated in figure 3.1a.

To keep the design of the mirror from becoming too complicated, we used a spherical CX surface. However, this causes spherical aberration of the focal point, as shown in figure 3.1a. In our design (see section 3.7) this only slightly impacts the maximum coupling efficiency that can be achieved using a Gaussian input beam. But if required, this limitation can be overcome by selecting the center thickness and CX ROC such that the lens gives no spherical aberration. The trivial case is illustrated in figure 3.1b, where the CC and CX surfaces are concentric. All rays incident normal to the CX surface are focused at the center of the CC ROC. The one nontrivial case, known as the aplanatic condition [55], is illustrated in figure 3.1c, where incident rays convergent towards a virtual focal point are focused to the center of curvature of the CC surface. Both cases require a converging (rather than collimated) beam be input to the cavity. A separate optic can be used to generate this converging beam from a collimated beam. The numerical aperture of this beam is lower with the aplanatic design, which reduces spherical aberration from the separate optic. The optic can also have an aspherical surface to further reduce spherical aberration.

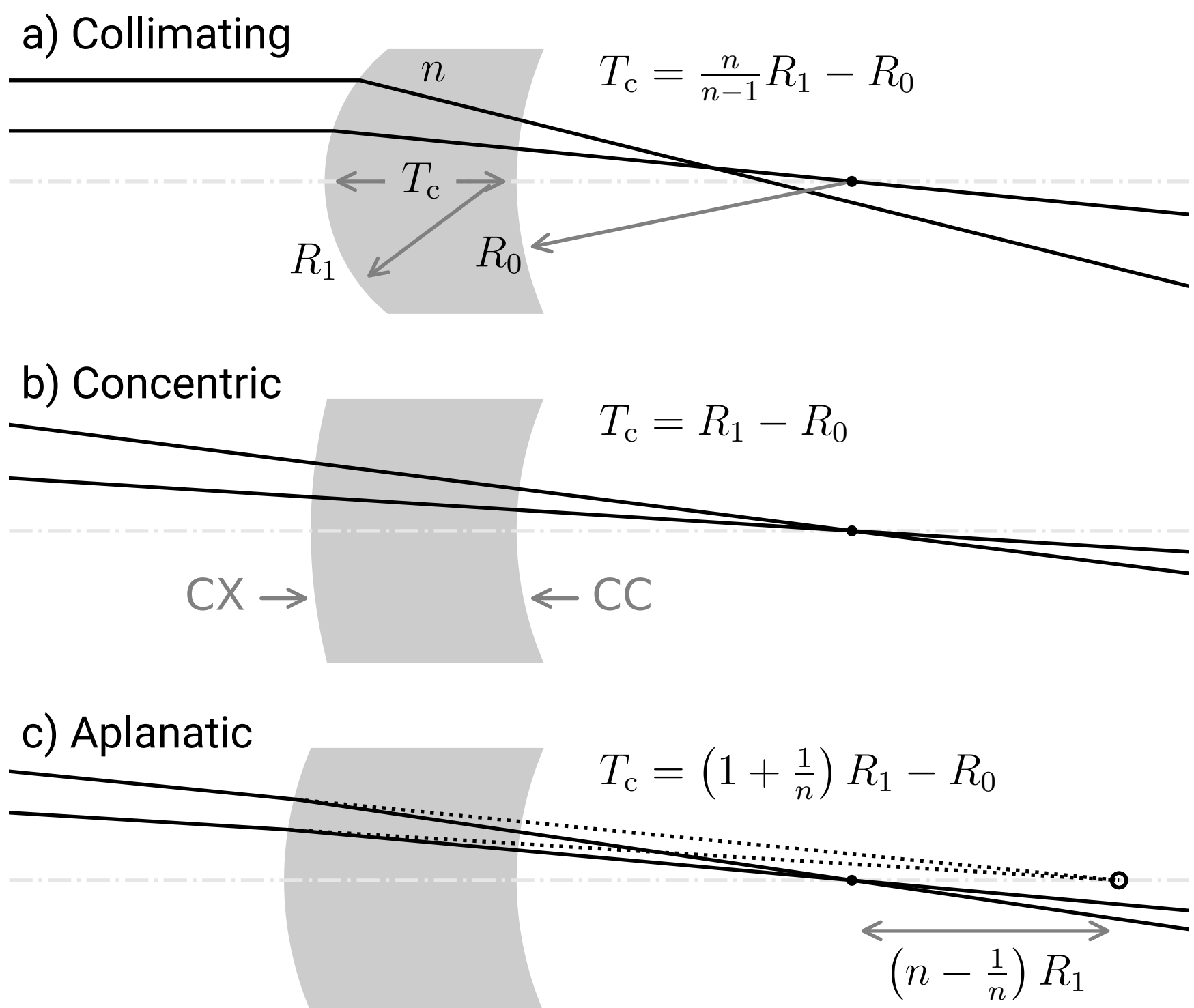

Figure 3.1: Three designs for a mirror substrate. Relationships between the convex (CX) surface radius of curvature (ROC) $R_1$, concave (CC) surface ROC $R_0$, substrate center thickness $T_c$, and the substrate refractive index $n$ are also shown. **a)** The CX surface focuses a collimated beam at the center of curvature of the CC surface. There is spherical aberration. **b)** The CX surface is concentric with the CC surface. There is no focusing of a wave convergent at the CC surface center of curvature, but also no spherical aberration. **c)** The CX ROC, center thickness, and refractive index of the substrate are such that the CX surface focuses a convergent beam at the CC surface center of curvature with no spherical aberration. The virtual focal point of the incident beam is beyond the CC surface center of curvature.

We found that the extra complication of a separate coupling lens was not worth the small increase in maximum coupling efficiency, and so we use the collimating design of figure 3.1a. The wedge angle between the CX and CC surfaces is specified to be $< 1.5\,\mathrm{mrad}$, which is equivalent to a centration error of the CX surface of $< 15\,\mu\mathrm{m}$. Such a small error negligibly increases the aberrations of the lens formed by the mirror substrate.

### 3.2.4 Chamfers

The intersection of the mirror surface and outer diameter of the substrate is doubly-chamfered to reduce the chance of chipping. Figure 3.13 shows the design and dimensions of the chamfer. The chamfer at 90° to the optical axis is added when the mirror surface is superpolished by Coastline Optics. This helps reduce the chance of chipping at the otherwise acute angle between the CC surface and 45° chamfer. An example of chipping on the edge of a mirror substrate with too small of a chamfer is shown in figure 3.2. Since the dimensions of the chamfers are not well-toleranced, I would recommend increasing the size of these chamfers by about a factor of two in future designs.

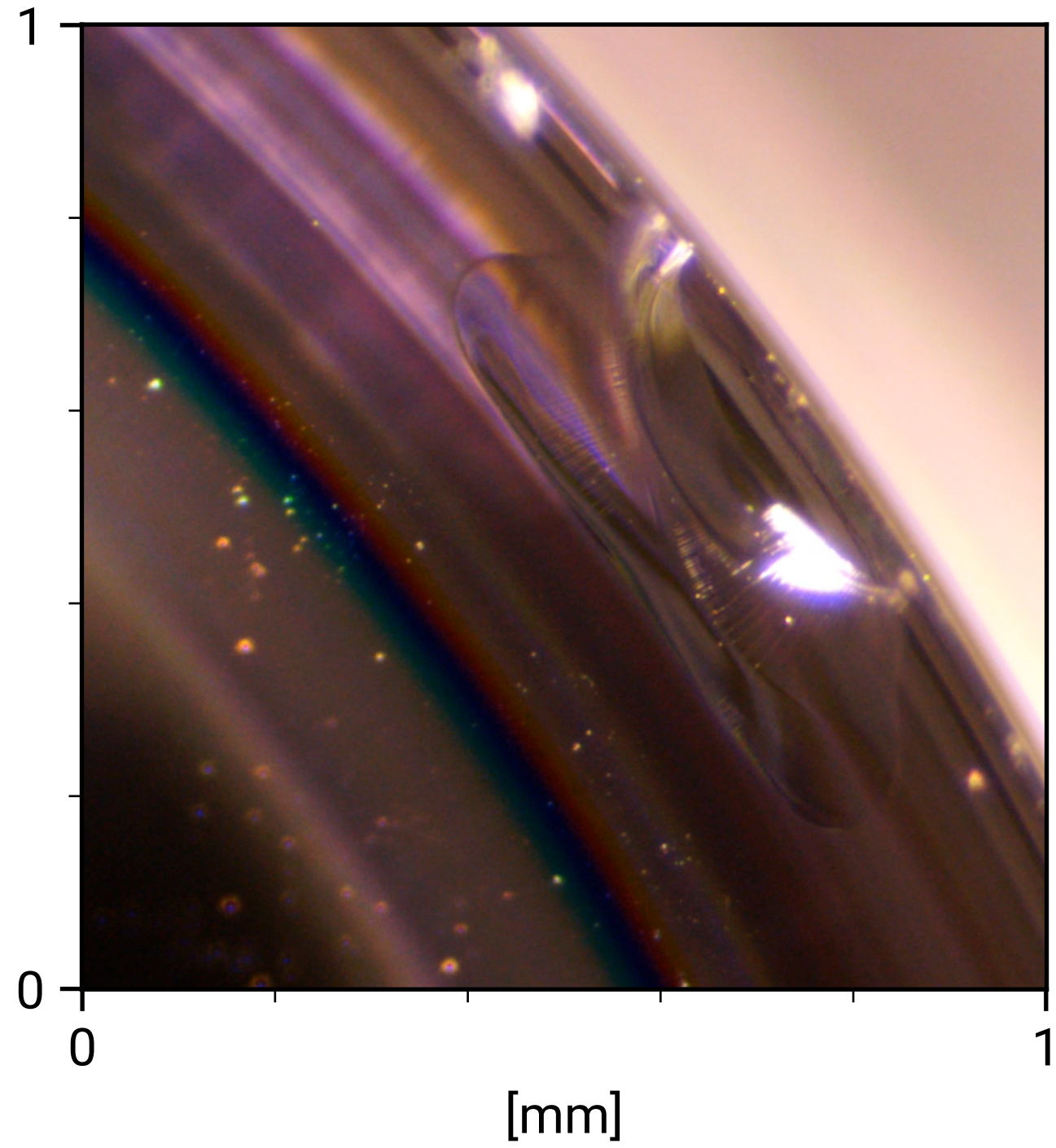

Figure 3.2: An optical microscope image of the chipped edge of a mirror. From lower-left to upper-right: coated mirror surface with point defects, likely from the chipped edge; rainbow pattern of the edge of the coating; stress fractures in the mirror substrate; chip in the edge; edge of the mirror.

## 3.3 Substrate material

The mirror substrate material must meet several requirements. First, it must be transparent to the laser wavelength being used, otherwise light cannot be coupled into the cavity. Second, it must be amenable to superpolishing. Third, it should have the smallest possible ratio of coefficient of thermal expansion (CTE) to thermal conductivity, so as to reduce the amount of thermal deformation due to heating from the cavity mode (see section 2.6). This ratio can strongly depend on temperature, so the possibility of heating or cooling the entire cavity to optimize the mirror performance should be considered. Indeed, many materials have a negative CTE at low temperature and positive CTE at room temperature, and thus a CTE zero-crossing below room temperature [56]. The amount of thermal deformation is also proportional to the material's Poisson's ratio, but in common optical materials its value does not depend as strongly on temperature as the CTE and thermal conductivity [50], [57]. The ratio of CTE to thermal conductivity is shown in figure 3.3 for several different possible substrate materials. References for the data used are given in table 3.1.

| Material | CTE | TC |
|---|---|---|
| Fused silica | [58] | [59] |
| Sapphire | [60] for $\leq 300\,\mathrm{K}$,<br>[58] for $> 300\,\mathrm{K}$ | [59] |
| ULE7972 | [61] | [61] (assumed to not depend on temperature) |
| Silicon | [62] | [63] |
| Diamond | [58] | [59] |

Table 3.1: References for the data shown in figure 3.3.

Diamond has a remarkably high thermal conductivity, even at room temperature. It is also transparent for optical wavelengths $> 220\,\mathrm{nm}$. However, it is unclear if it can be superpolished, especially on a small concave radius of curvature.

Sapphire has a high thermal conductivity, but only at cryogenic ($< 100\,\mathrm{K}$) temperatures. Additionally, its CTE crosses zero at $32\,\mathrm{K}$. Like diamond, it is transparent for optical wavelengths between the ultraviolet and mid-infrared. Initial tests done on two sapphire substrates by Coastline Optics indicate that it may be possible to superpolish at $10\,\mathrm{mm}$ ROC, though there is residual threefold astigmatism in the surface profile from the differing hardnesses of the different crystal planes (see figure 3.4).

Silicon, like sapphire, has a high thermal conductivity at cryogenic temperatures, with two CTE zero-crossings. Its thermal conductivity and CTE are somewhat higher and lower, respectively, than sapphire near room temperature, giving it a roughly order of magnitude smaller CTE-to-conductivity ratio there. It can possibly be superpolished with a $10\,\mathrm{mm}$ ROC, given that it can be superpolished at longer ($\sim 1\,\mathrm{m}$) ROC [64]. But it is not transparent for wavelengths $< 1.2\,\mu\mathrm{m}$, so the only wavelength option it is compatible with is $1550\,\mathrm{nm}$.

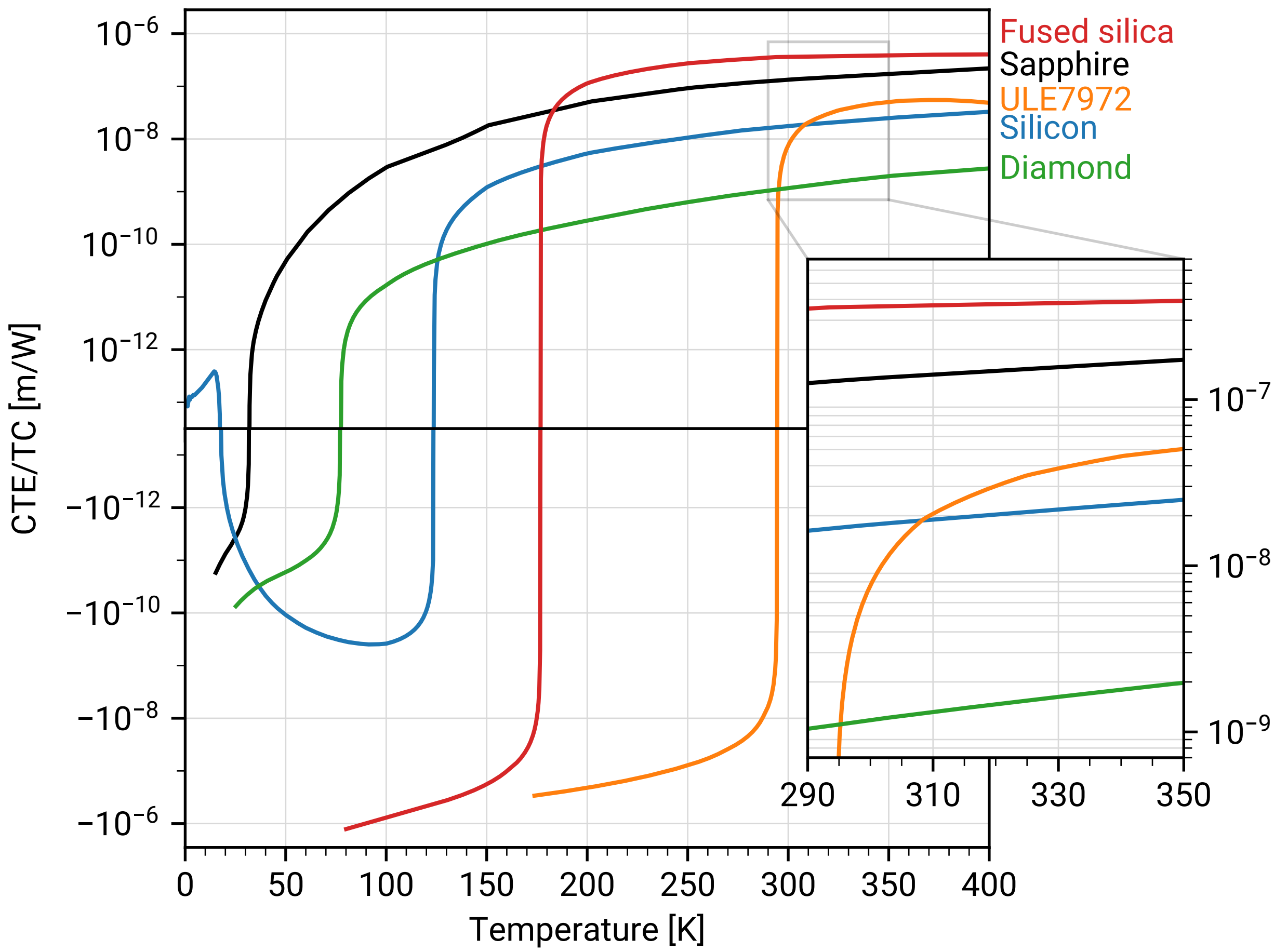

Figure 3.3: The ratio of coefficient of thermal expansion (CTE) to thermal conductivity (TC) for several different materials as a function of temperature. References for the CTE and TC data used are given in table 3.1.

Fused silica also has good transparency from ultraviolet to near-infrared, and, as one of the most common optical substrate materials, can be superpolished well. However, it has a poor ratio of CTE to thermal conductivity compared to the alternatives, except in a narrow range of cryogenic temperatures around the zero-crossing of its CTE.

ULE Corning Code 7972 ("ULE7972"), and ULE Corning Code 7973 ("ULE7973"), manufactured by Corning Incorporated, exploit a zero-crossing in the CTE as a function of temperature to achieve a low CTE-to-thermal-conductivity ratio for a relatively small range of temperatures. They are mostly transparent between the visible and near-infrared. However, we found that ULE7972 (and presumably ULE7973, as they share very similar chemistry) does attenuate 1064 nm appreciably, with an absorption coefficient of roughly $0.01\,\mathrm{cm}^{-1}$. Fortunately, since our mirror substrates are only 5.59 mm thick, this does not cause an appreciable loss in power of the input laser beam, and at most causes some additional thermal lensing in the mirror substrate. Of greater concern is that the manufacturing process for these titania-doped silica glasses results in a striated composition. This causes both the material and optical properties of the substrate to change along one axis with a periodicity of $\sim 160\,\mu m$ [65]. Fortunately, this does not appear to noticeably affect the results of superpolishing, even on a 10 mm ROC surface which may cut across several striae (see section 3.4). However, it does substantially degrade the optical clarity of the substrate. Figure 3.5 shows noticeable distortion due to the striae in a ULE7972 mirror substrate which was manufactured such that the normal axis to the striation planes was perpendicular to the optical axis. This requires that the substrates be manufactured with the optical axis aligned with the normal axis to the striation planes, as illustrated in figure 3.6. If this were not the case, the refractive index variations across the width of the laser beam inside the substrate would distort the wavefront of the beam, likely severely diminishing the achievable coupling efficiency to the cavity mode. As such, ULE is rarely used as a transmissive optic. ULE7973 is marketed as having less dramatic striae [66], and also can have a higher CTE zero-crossing temperature (it can be adjusted somewhat on a batch-by-batch basis) than ULE7972. The higher zero-crossing temperature makes it easier to explore the behavior of the cavity mode in the presence of mirror substrate thermal deformation when there is a negative CTE (see section 2.6 for a longer discussion of this effect).

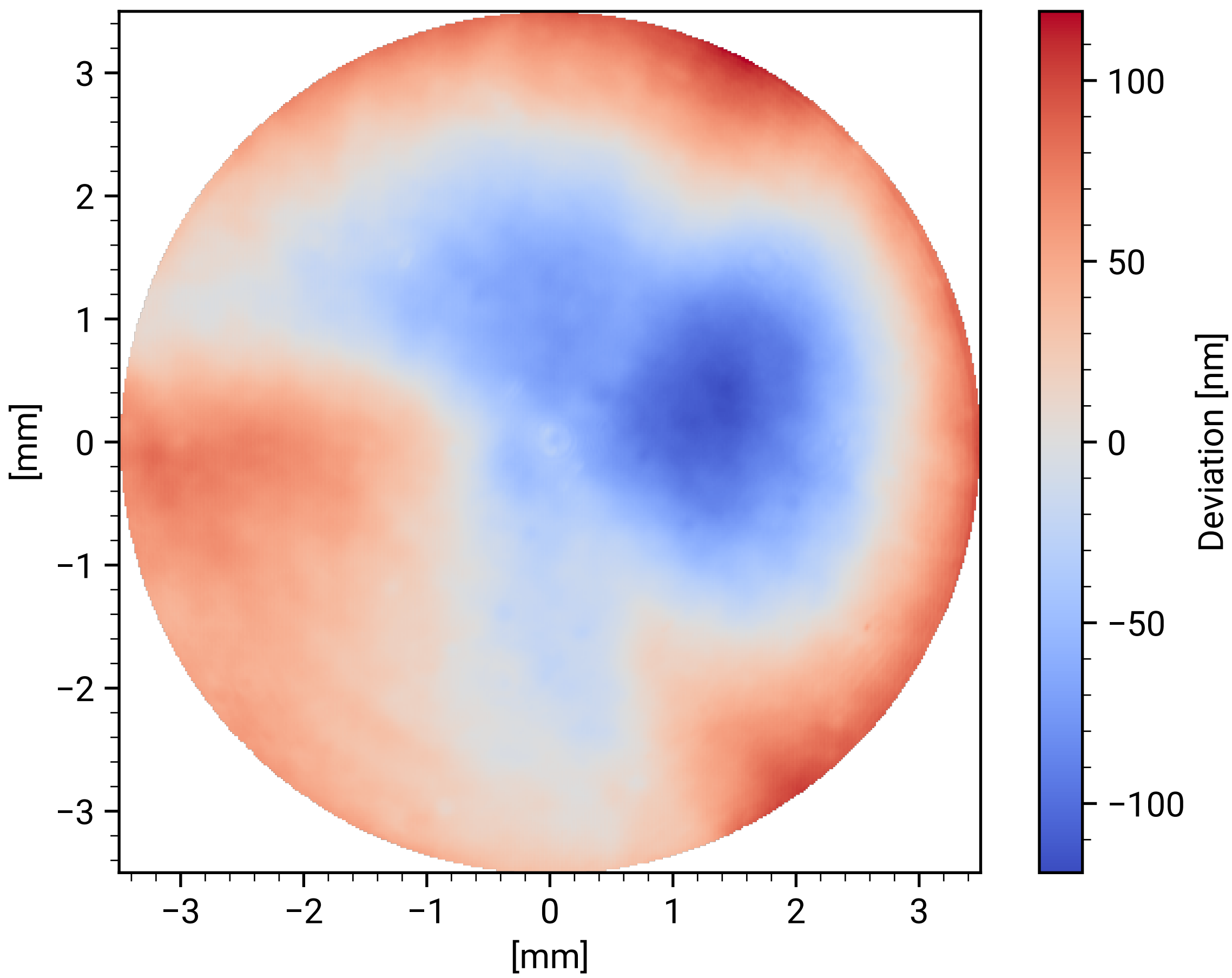

Figure 3.4: Deviation of the mirror surface from spherical for a 10 mm curvature radius superpolished sapphire substrate. The optical axis of the substrate is aligned with the $c$-axis of the sapphire crystal.

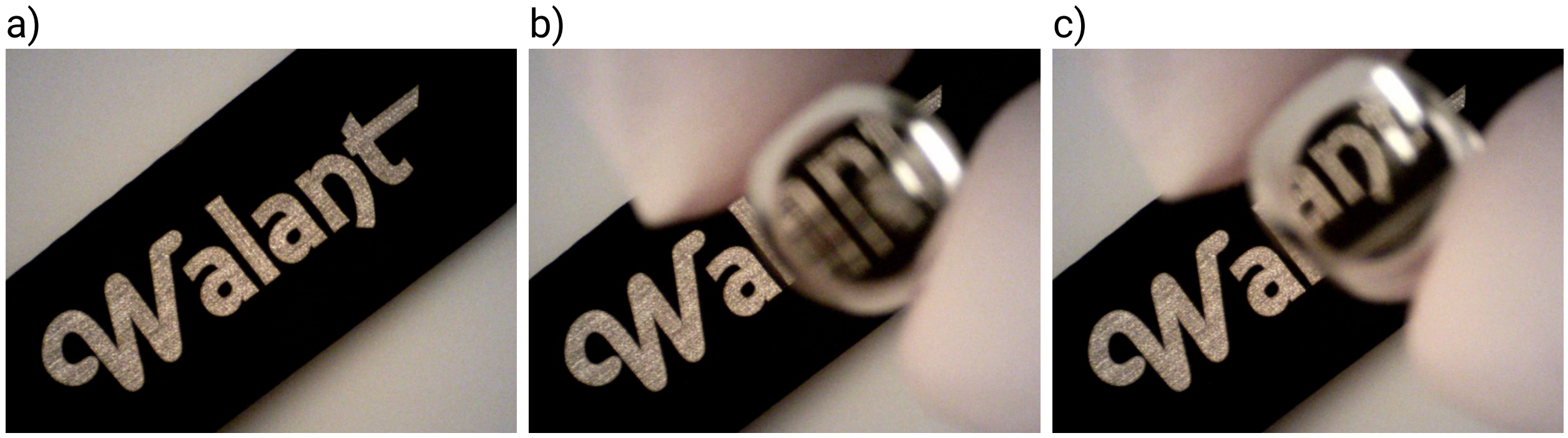

Figure 3.5: Striation planes in the refractive index of ULE7972 are visible to the eye. **a)** Reference image showing lettering on a black background. **b)** Gloved fingers (white) hold a ULE7972 optic such that the line of sight through it to the lettering below is nearly parallel to the striation planes. Striations in the image of the lettering are visible, running along the length of the lettering. **c)** The optic is tilted slightly such that the striation planes are no longer aligned with the line of sight, and the striations are no longer visible. The diameter of the optic is 7.75 mm.

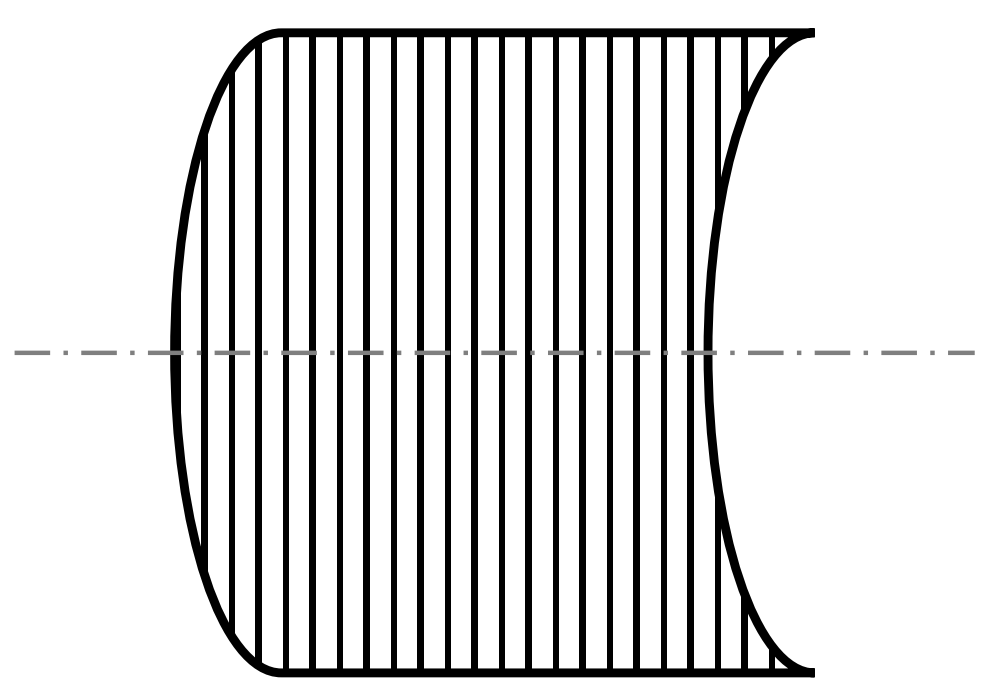

Figure 3.6: ULE striation planes (vertical stripes) must be oriented perpendicular to the optical axis (horizontal dash-dotted line) to avoid transmitted wavefront aberrations.

Initial prototypes of our optical cavity used fused silica mirror substrates [22], [23], but the focal intensity (electron beam phase shift) was limited by thermal deformation of the substrate. Our successful prototypes [24], [67] now use ULE7972. We are in the process of testing ULE7973.

## 3.4 Polishing

Polishing of the CC mirror surface must address three criteria: surface figure, microroughness, and defects.

### 3.4.1 Surface figure

The surface figure describes the deviation of the surface from spherical over lateral scales much larger than the laser wavelength, and determines the shape of the cavity mode. When the surface profile is approximated with a power series, the lowest order surface figure aberration possible is (twofold) astigmatism, where the surface is not azimuthally symmetric about the optical axis and there are different radii of curvature on two orthogonal principal axes. In this case, the cavity mode (in the paraxial approximation) will still be Gaussian, but no longer azimuthally symmetric. In the near-concentric regime, a difference in principal radii of curvature on the order of the distance to concentricity results in a noticeably elliptical mode shape. In principle, this effect can be used to create a focal spot which is still narrow perpendicular to the electron beam axis, but extended parallel to it. This generates a larger phase shift per circulating power while retaining the same CTF. We have not tried to do this—we instead rely on the mirrors to be astigmatic enough that we don't have to align the "long" axis of the focus to the electron beam axis.

The near-concentric configuration of the cavity sets stringent limits on the astigmatism. Figure 3.7 shows the aspect ratio of the elliptical mode generated by mirrors with identical astigmatic surfaces as a function of the larger of the two principal numerical apertures $\mathrm{NA}_+$ and the difference in radii of curvature $\Delta_R$ of the two principal axes of the mirror curvature. The cavity length and laser wavelength are taken to be 20 mm and 1064 nm, respectively. This is an upper limit on the effect of astigmatism because in general the principal axes of the two mirrors will not be aligned. Regardless, it shows that achieving a reasonably round cavity mode aspect ratio of $< 1.2$ with a maximum numerical aperture of $\mathrm{NA}_+ = 0.05$ requires the difference in radii of curvature along the principal axes to be $< 2\,\mu\mathrm{m}$.

Higher order aberrations make the cavity mode non-Gaussian [44]. Some amount of these aberrations are present on our mirrors (which we infer by the presence of anharmonicity in the transverse mode frequency spectrum of the cavity, see section 6.3.2), but they do not noticeably affect the shape of the focal spot.

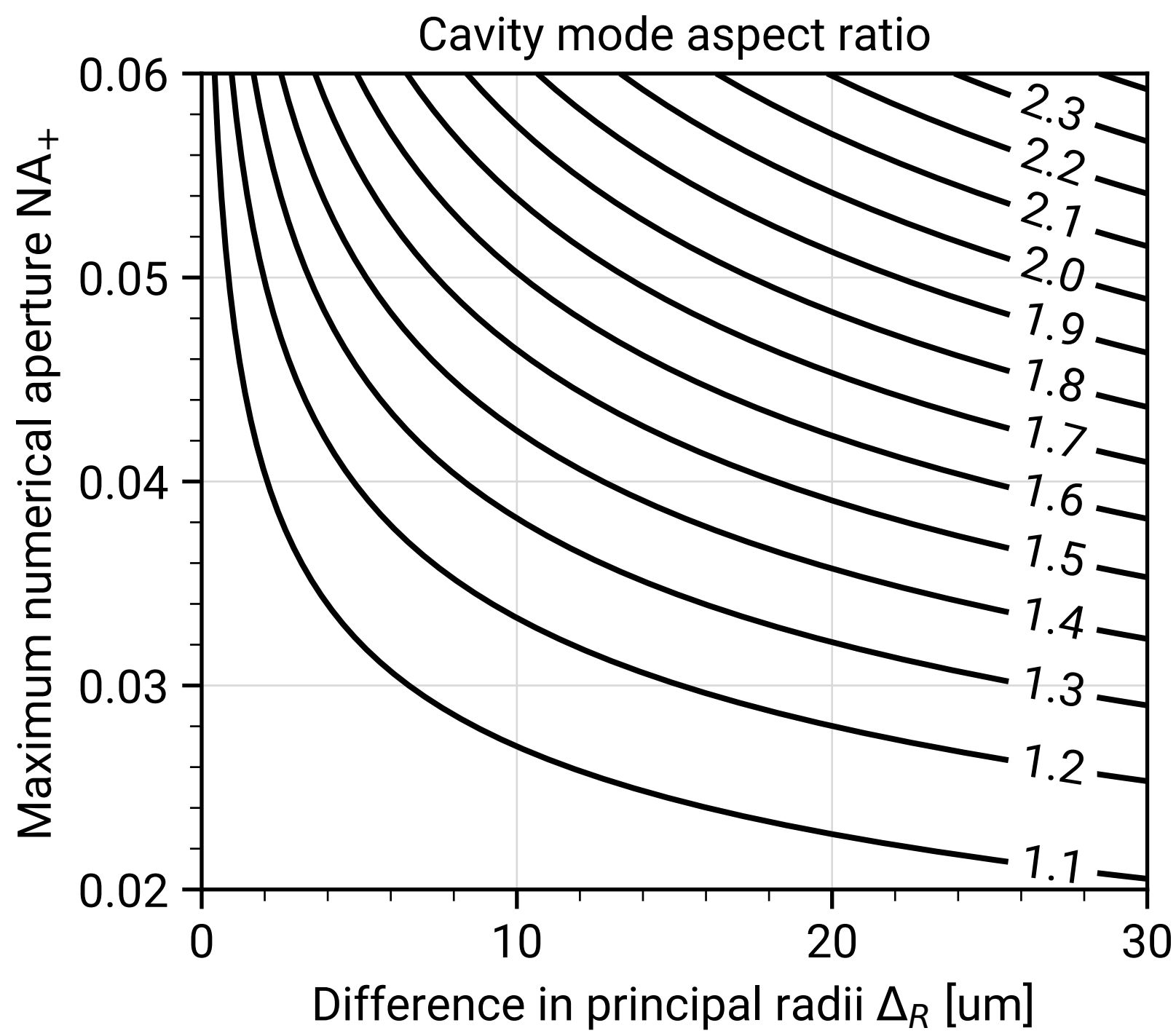

Figure 3.7: Aspect ratio of the cavity mode as a function of the larger of the two principal numerical apertures $\mathbb{NA}_+$ and the difference in radii of curvature $\Delta_R$ of the two principal axes of the mirror curvature. This assumes a cavity length of 20 mm and a laser wavelength of 1064 nm.

Optical profilometry of our mirror surfaces (using a Zygo NewView 9000) show that the surfaces of our mirrors deviate from spherical by no more than $\sim 1$ nm over an area of $0.85\,\mathrm{mm} \times 0.85\,\mathrm{mm}$ about the optical axis.[1] Since the optics of the profilometer are not this close to spherical, their contribution to the surface profile measurement must be removed. This can be done by rotating the surface under test by 90° between two measurements, and subtracting the resulting surface profiles. The resulting surface profile only represents components of the mirror surface profile which are not $C_{4n}$-symmetric, with $n \in \mathbb{N}$. Therefore, astigmatism ($C_2$-symmetric) is still shown by the subtracted profile. Dividing the subtracted profile by two normalizes the astigmatism component so that the profile represents the deviation from spherical. The results of this surface measurement are shown in figure 3.8. The concentric rings in the surface profile data are likely a measurement artifact. Otherwise, the main aberration present is twofold astigmatism. The difference in the two principal radii of curvature is roughly 2 µm, right at the requirement for achieving a $< 1.2$ mode aspect ratio at a maximum numerical aperture of 0.05. This is likely an overestimate of the actual astigmatism due to systematic errors in the measurement. Indeed, we have measured the astigmatism in our cavity (though using different mirrors) to be roughly 1/4 of this value (see section 7.3.3).

[1]A special thanks to Howard A. Padmore of Lawrence Berkeley National Lab for performing the Zygo NewView 9000 measurements shown in this section!

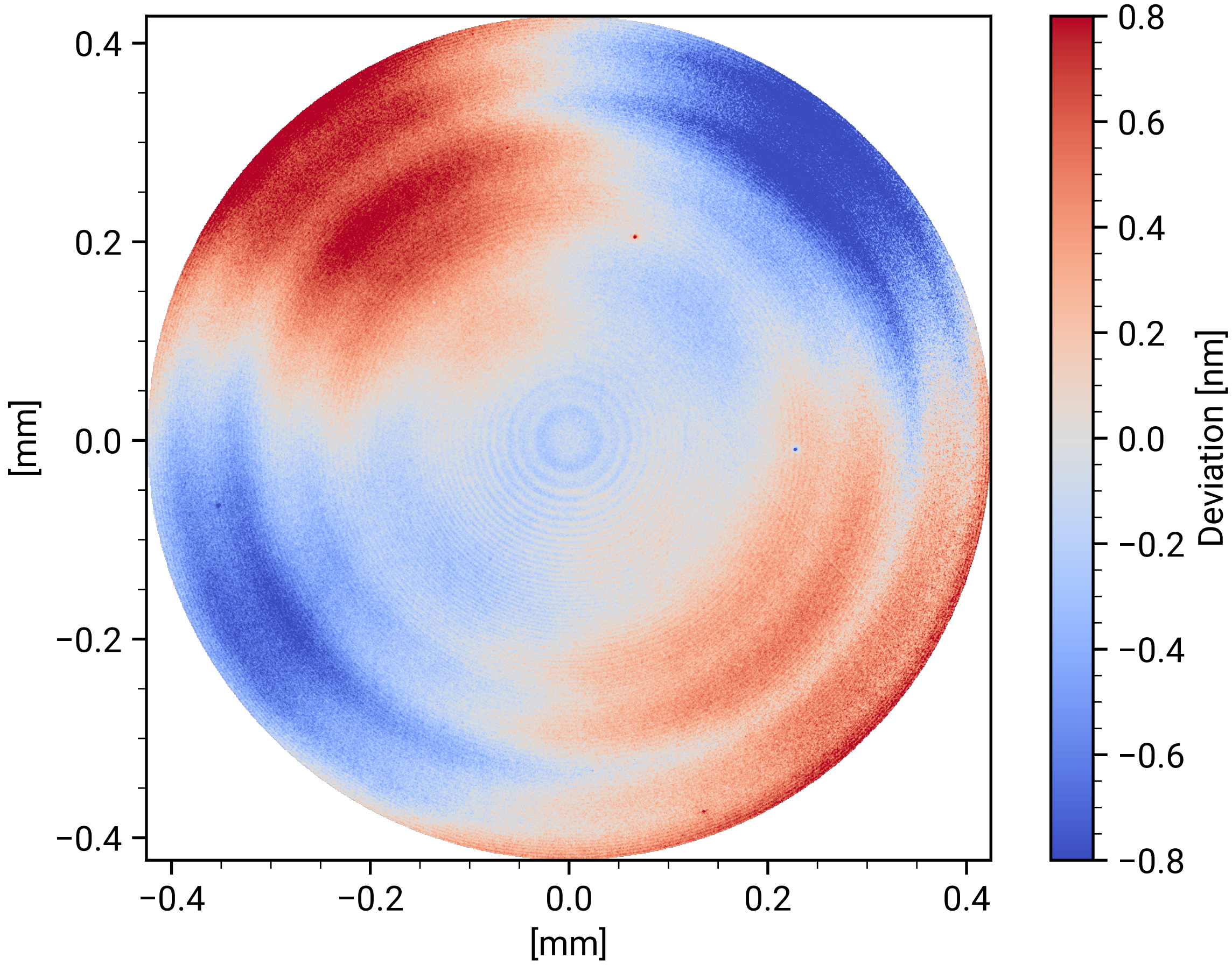

Figure 3.8: Deviation of the mirror surface from spherical, derived from two measurements taken with the mirror rotated around its optical axis by 90°.

Note that the deviation of 1 nm over a 0.85 mm × 0.85 mm can be quadratically extrapolated to infer that the mirror surface deviation is no larger than 13 nm over the 3 mm clear aperture.

### 3.4.2 Microroughness

The microroughness describes the deviation of the surface from planar over lateral scales of a few wavelengths and smaller, and determines the amount of light that is scattered out of the cavity mode. The relationship between the standard deviation in surface height $\sigma_z$ and the fraction of incident light scattered $S$ is

$$S = 1 - e^{-\left(\frac{4\pi\sigma_z}{\lambda_L}\right)^2} \tag{3.3}$$

assuming that the surface has a reflectivity near unity, the light is normally incident, and the distribution of heights is Gaussian [68]. This relationship is plotted in figure 3.9 for $\lambda_L = 1064\,\text{nm}$. The maximum possible power amplification factor of a Fabry-Pérot cavity with mirrors which each lose a fraction $L$ of the incident light to scattering and/or absorption is $\frac{1}{4L}$ (see section 2.5). Given the input power available from our laser system ($\sim 50\,\text{W}$, see section 6.1), our cavity only requires a mirror loss fraction of $L < 1.7 \times 10^{-4}$ in order to achieve a circulating power of 75 kW, which in conjunction with a NA $= 0.05$ cavity mode provides a $\pi/2$ phase shift of 300 keV electrons. Since the absorption of the mirrors can be made $< 1 \times 10^{-6}$, this means that the scattering needs to be $S < 1.7 \times 10^{-4}$, corresponding to a surface roughness of 10 Å, which is readily achievable by standard optical polishing processes.

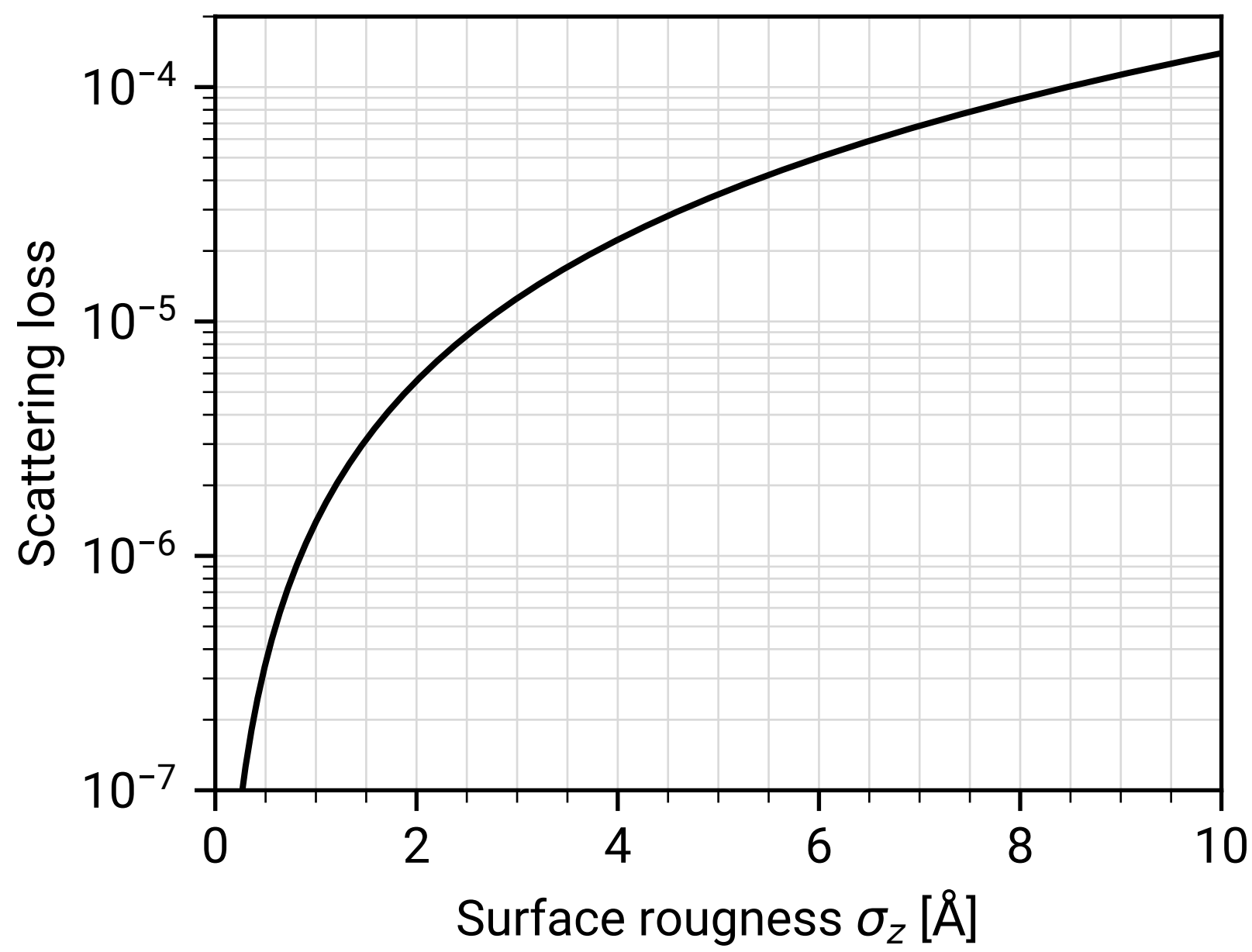

Figure 3.9: Predicted scattering losses as a function of the surface roughness $\sigma_z$ for a wavelength of $\lambda_L = 1064\,\text{nm}$.

However, scattering losses lead to heating of the cavity mount, which causes thermal misalignment of the cavity. This can be (and is, see chapter 6) compensated for using various feedback mechanisms, but we still strive to have the lowest possible scattering losses. The microroughness of our coated mirrors is measured (Zygo NewView 9000) to be 0.5 Å, which should result in scattering losses of only $0.4 \times 10^{-6}$. Such a surface is referred to as "superpolished", though there is no precise definition of the term. Measurements of the microroughness over a 87.5 µm × 87.5 µm area near the center of one of our mirrors is shown in figure 3.10. The microroughness of the profilometer's optics are non-negligible, so the measurement is taken twice in nearby areas of the mirror separated by $\sim 0.2$ mm, and the resulting surface profiles (after fitting out the overall spherical height profile) are subtracted and then divided by $\sqrt{2}$. Assuming that the surface profiles in the two measurements are uncorrelated, the surface roughness can be inferred to be the standard deviation of the surface heights in this normalized difference. The distribution of heights is Gaussian, as shown in figure 3.10b.

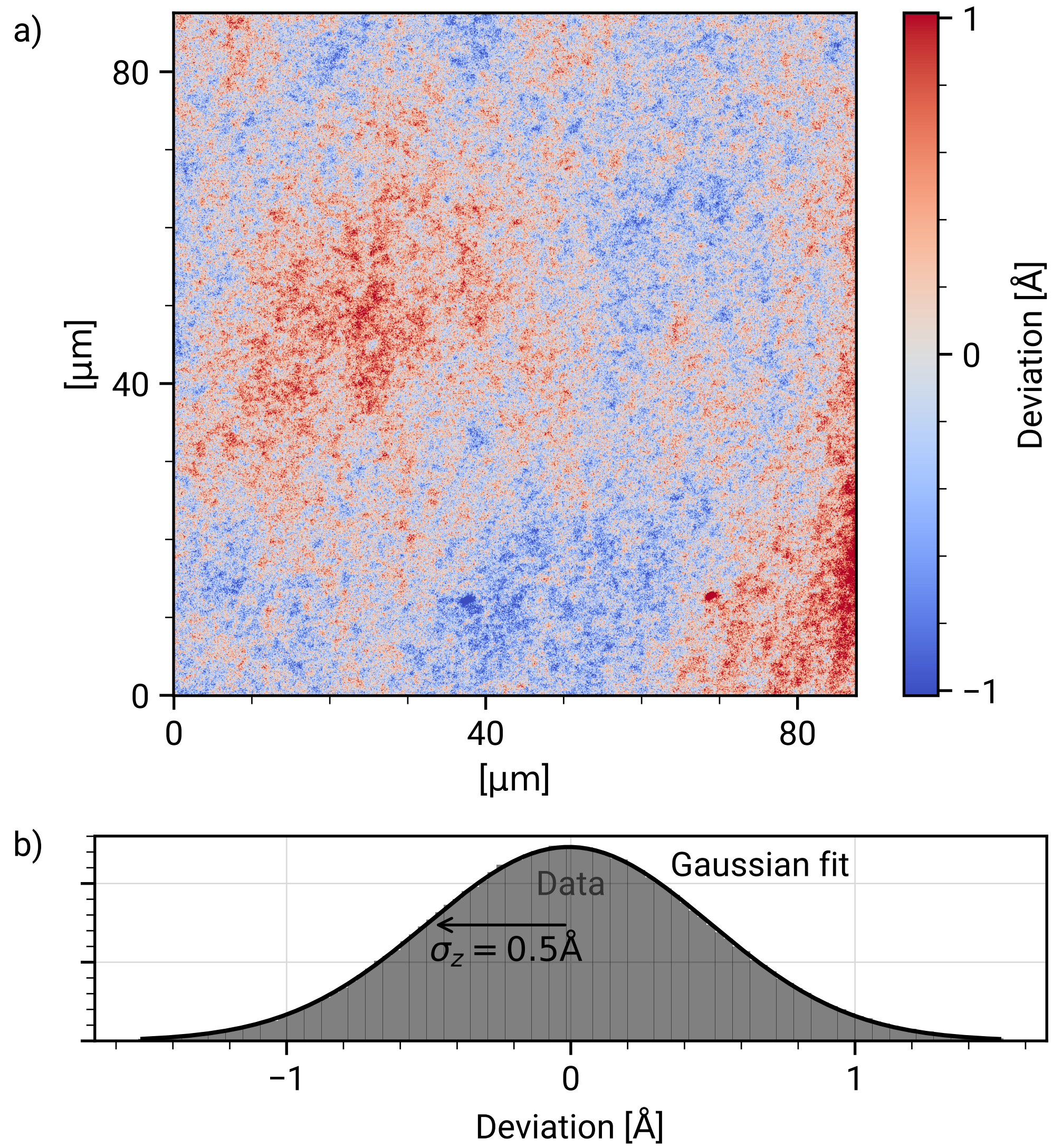

Figure 3.10: **a)** Microroughness of a coated mirror surface, derived from the difference in profiles of two nearby areas of the mirror surface. **b)** Histogram of the height values shown in a), along with a Gaussian fit.

### 3.4.3 Defects

In practice, we find the scattering losses to be higher than predicted from the measured surface roughness, which we attribute to surface defects. Surface defects are for us perhaps the most important aspect of the mirror polishing process. A single defect near the optical axis can cause $\sim 10 \times 10^{-6}$ if not $\sim 100 \times 10^{-6}$ of scattering, ruining the performance of the mirror. Since the mode diameter on the mirror surface is $\sim 1$ mm, a large are of the surface must be defect-free. We therefore require a 0-0 scratch-dig surface quality specification (as defined in [69]) over the clear aperture (3 mm) of our mirrors, though in principle the mirrors may be able to function with a smaller clear aperture (see section 7.2)

### 3.4.4 Convex surface and outer diameter

The CX surface of the mirror substrate does not need to be polished to nearly as high of a quality as the CC surface, since it only needs to focus a $\sim 50$ W laser beam to a diffraction limited spot at a numerical aperture of $\sim 0.05$. We specify a surface quality of 20-10 scratch-dig with a surface figure of $< 55$ nm peak-valley deviation from spherical. This roughly corresponds to a "$\lambda/10$" specification.

The outer diameter of the substrate is also polished to a low (non-optical) quality. This aids in cleaning the substrate and reducing the surface area available for harboring or generating particulate contamination which may migrate to the critical CC surface, as compared to a standard ground glass finish.

## 3.5 Coating

The high-reflectivity (HR) mirror coatings on the CC surface of the mirror substrate are ion beam sputtered (IBS) dielectric multilayers consisting of $SiO_2$ and $Ta_2O_5$, manufactured by FiveNine Optics. The coatings have a nominal transmissivity of $80 \times 10^{-6}$, which supports a high enough power amplification factor to achieve a $\pi/2$ phase shift in the LPP using $> 6$ W of input laser power, which is readily commercially available (see section 6.1). A lower transmissivity would reduce the input power requirement, but may make frequency stabilizing the laser to the (narrower) cavity resonance more difficult. We have all mirrors coated with the same transmissivity coating so that any mirror can be used as the input or output mirror. This reduces the number of mirrors we need to have manufactured in order to find a pair of mirrors which are sufficiently defect-free to successfully use in the cavity. The coating is designed such that reflected light generates a near-zero electric field amplitude at the coating surface, which reduces the scattering and absorption of surface contaminants and defects. The surface layer of the coating is $Ta_2O_5$ which reduces the electric field amplitude inside of the coating and so reduces absorption. The HR coating does not substantially add to the scattering loss due to the microroughness in the substrate.

We had the absorption of the HR coating characterized by photothermal common-path interferometry [70] at Island Interferometry/Stanford Photo-Thermal Solutions, using a plano-plano fused silica substrate witness sample from the same coating run as the mirror substrates. The absorption in the coatings was typically measured to be state-of-the-art at $< 0.5 \times 10^{-6}$, though occasionally a coating run would have considerably higher absorption, in which case we would not use the mirrors from that run. An example of a two-dimensional map of absorbance as a function of position on a low-absorption coating surface is shown in figure 3.11. The few isolated spikes in absorption are of unknown origin. They could be inherent defects in the coating, or contamination which has accumulated on top of the coating [71], [72]. The latter option is certainly possible, as the mirrors were tested outside of a cleanroom environment.

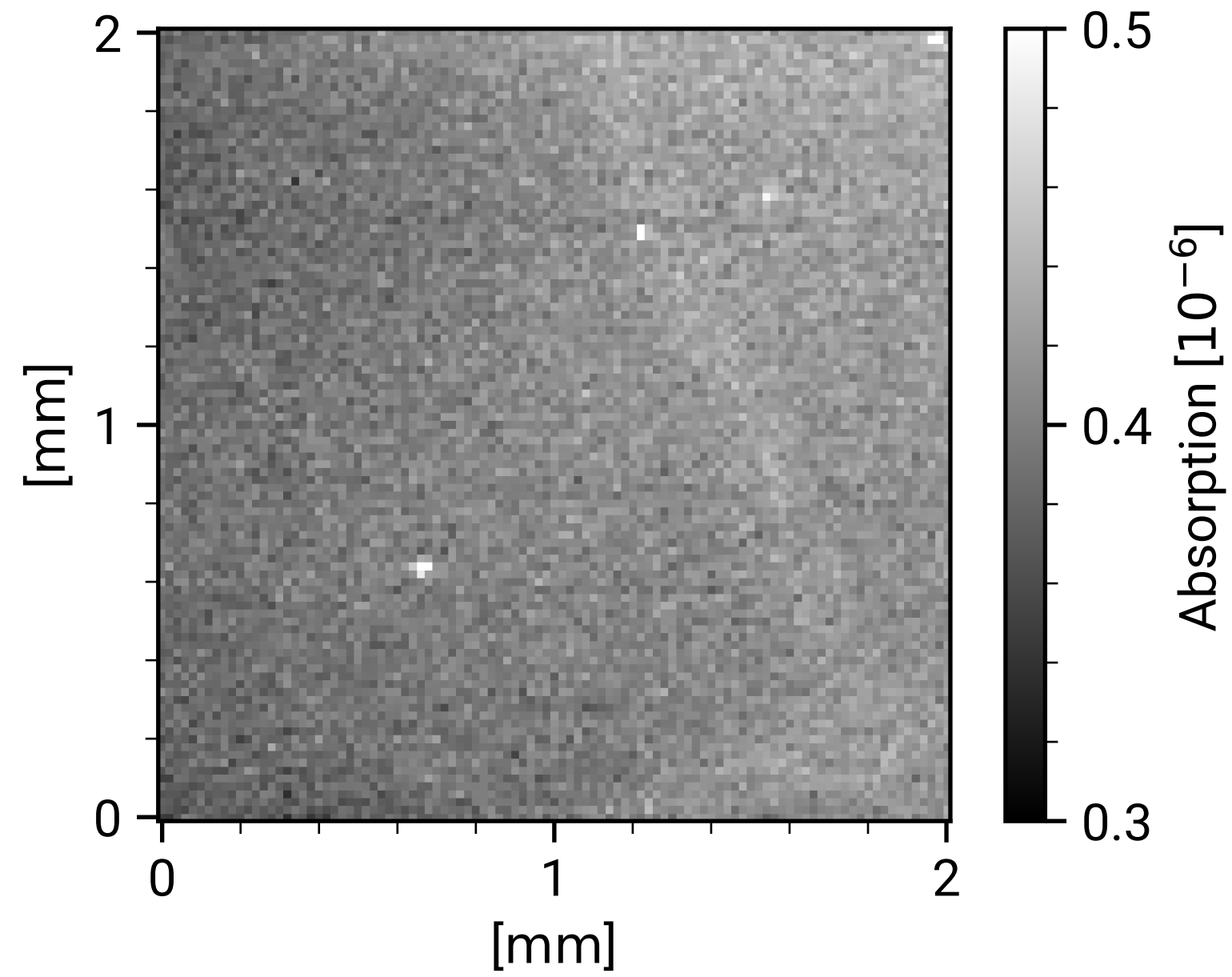

Figure 3.11: Absorption as a function of position of a high-reflectivity ion beam sputtered coating on a planar fused silica witness sample.

An IBS anti-reflective (AR) coating with a nominal reflectance of $< 1 \times 10^{-3}$ was applied to the CX surface of the mirror substrate to reduce unwanted reflections and increase the input coupling efficiency.

## 3.6 Supply chain

We have been unable to find a optics manufacturer that can perform all the steps in our mirror's manufacturing process. This requires us to distribute the process between three separate companies.

### 3.6.1 Substrate

The first step involves sourcing the substrate material (typically from a Corning Incorporated distributor) and shaping the substrates, with 20-10 scratch-dig surface quality on both the CC and CX surfaces. At this point, only a single 45° chamfer is cut at the edge of the CC surface. Initially, this work was done by Perkins Precision Developments, and then later by Laseroptik GmbH. We switched to Laseroptik GmbH because they are able to supply larger numbers of substrates more quickly and cheaply than Perkins Precision Developments.

### 3.6.2 Superpolishing

The second step is superpolishing of the CC surface, which is performed by Coastline Optics using a pitch polishing process. This step removes between 5 µm and 50 µm of material from the CC surface. Coastline Optics also add the flat annular chamfer surface at the edge of the CC surface.

### 3.6.3 Coating

The third step of coating the CC and CX surfaces is done by FiveNine Optics.

### 3.6.4 Inspection

We inspect the mirrors using a darkfield optical microscope between every step of the manufacturing process. This gives us the opportunity to reject mirrors with obvious defects. Relevant defects to look for after the initial shaping of the substrates are chipped or small chamfers (see figure 3.2), and incorrectly oriented ULE striation planes (see figure 3.5). After superpolishing and after coating, the most relevant defects are chipped chamfers and defects in the clear aperture of the CC surface (see section 7.2). Finally, Island Interferometry/Stanford Photo-Thermal Solutions measures the absorption of the mirror coating using a plano-plano fused silica witness sample from the coating run.

### 3.6.5 Throughput

This supply chain is illustrated in figure 3.12. Rough costs and (more importantly) lead times are also shown. The main bottleneck in the supply chain is the superpolishing: due to the small CC ROC, substrates can only be superpolished one at a time, and Coastline

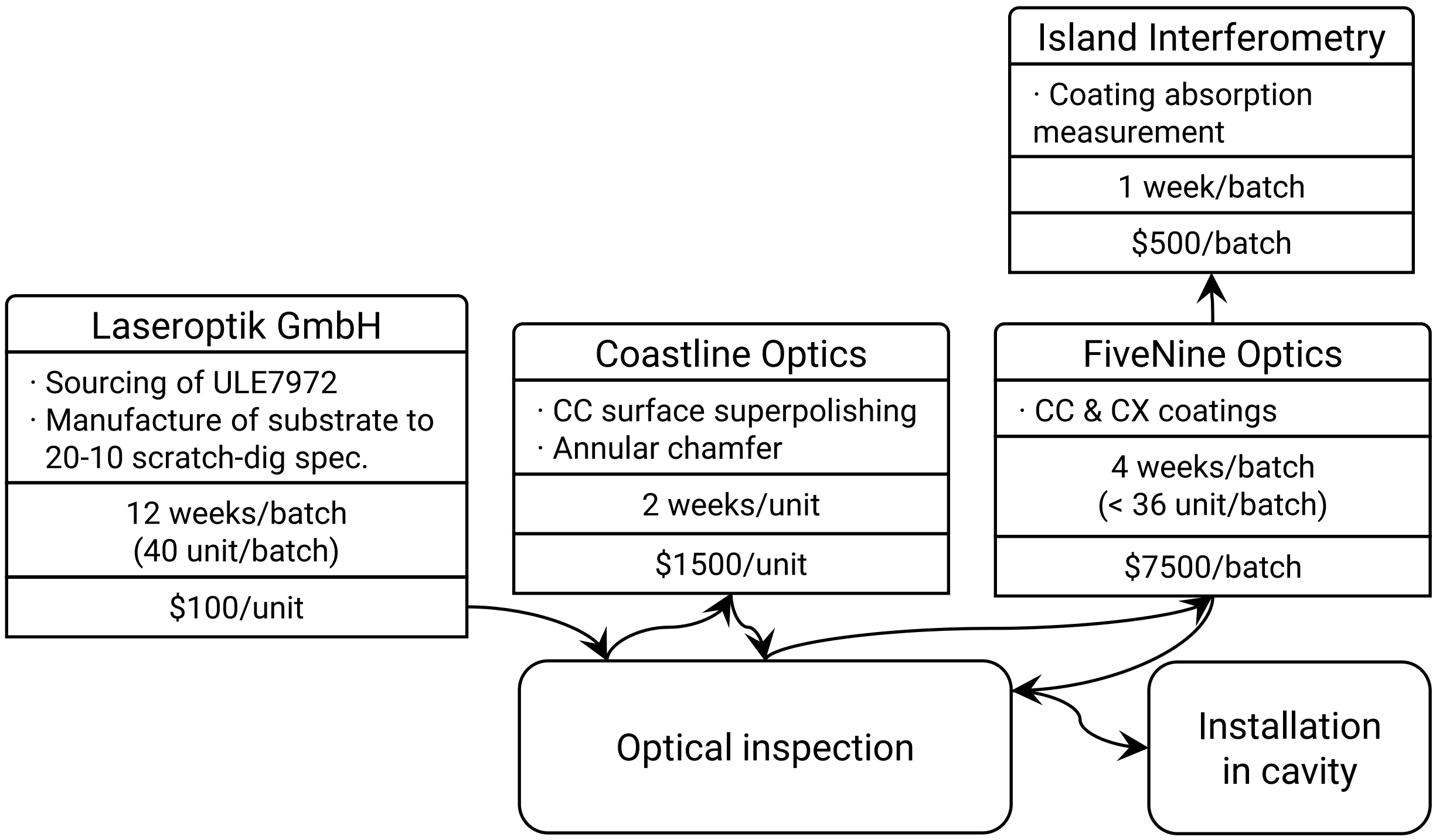

Figure 3.12: Supply chain for our mirrors.

Optics currently only one set of tooling for our small CC ROC substrates. This makes the throughput of the superpolishing step only about one substrate per two weeks. This would not be much of a problem if we only needed two mirrors, but we have found that not every mirror which is superpolished ends up having a satisfactorily defect-free CC surface by the time it has been coated. Some fraction of the mirrors still have defects on the CC surface after superpolishing. The coating process is also liable to introduce defects, with some coating runs being much worse than others. We believe a combination of handling technique by the coating chamber technicians and the cleanliness of the coating chamber and tooling are responsible for this inconsistency. Based on our experience, we roughly estimate that we get one usable cavity mirror for every five superpolished substrates we receive from Coastline Optics.

## 3.7 Specifications

A summary of specifications for our cavity mirrors are shown in figure 3.13.

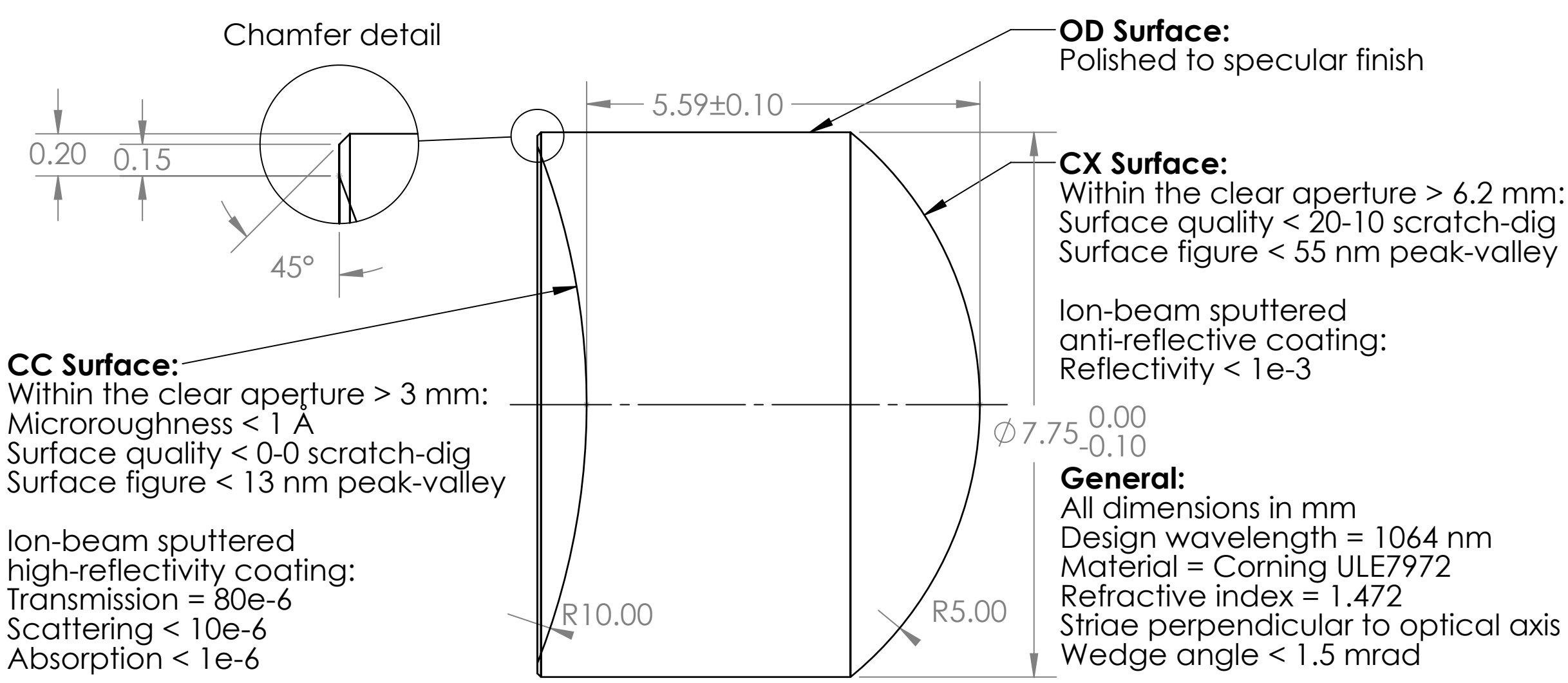

Figure 3.13: Specifications for our cavity mirrors.

# Chapter 4

# Cavity mount design

## 4.1 General requirements

The cavity mirrors must be mounted such that they are aligned in the near-concentric condition. This "cavity mount" must in turn be mounted to our custom TEM such that the electron beam and cavity mode focus overlap. For this purpose, our TEM has a pair of 25 mm diameter access ports which lie in the magnified diffraction plane. These ports are separated by 170° about the electron beam axis, as shown in figure 4.1.

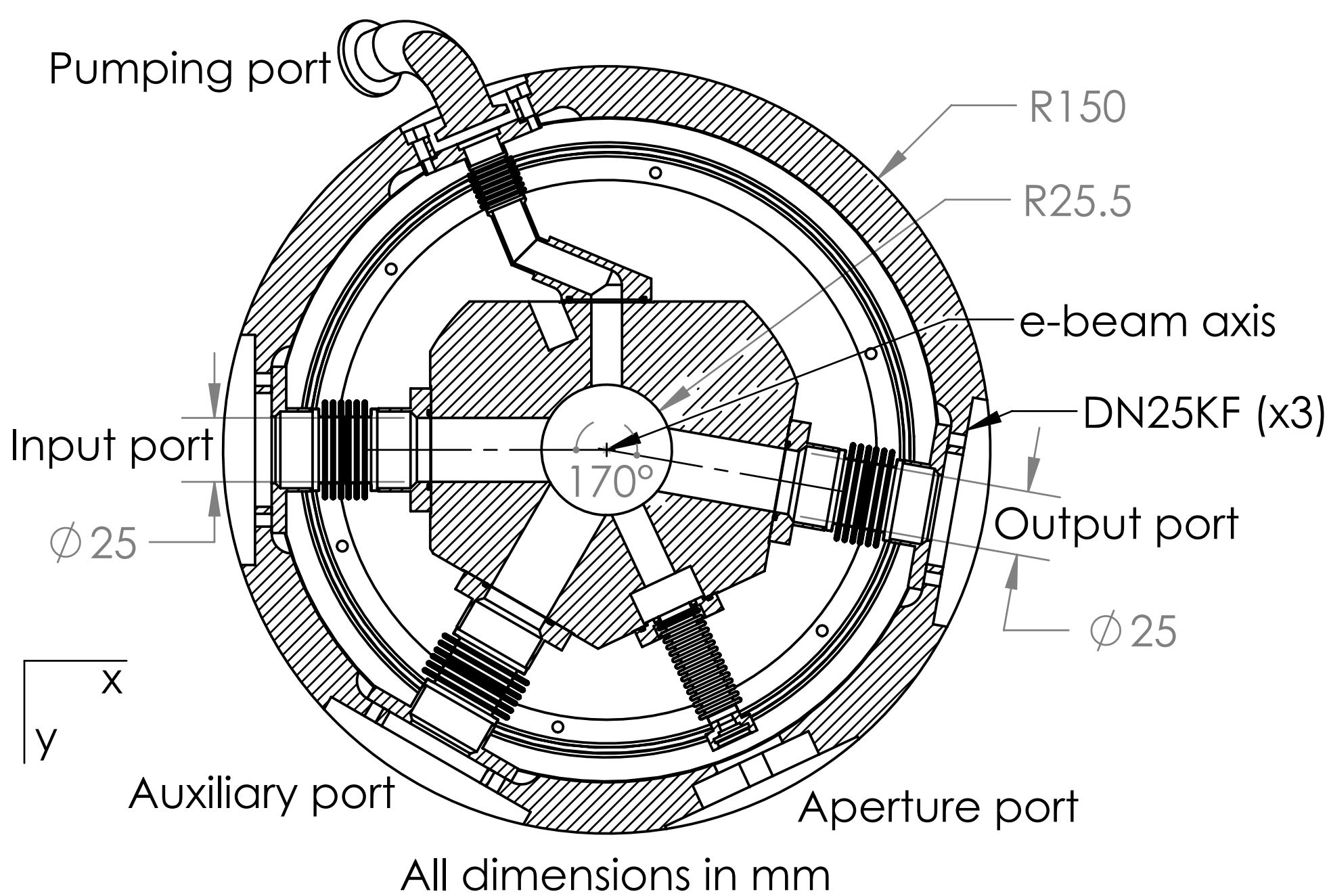

Figure 4.1: A cross section of the transmission electron microscope column in the laser phase plate plane.

We use one port ("input") to insert the cavity mount on a support arm, and the other port ("output") to send the light transmitted through the cavity to various feedback and diagnostic electro-optical systems (see chapter 6). Working within the space constraints of these access ports was the single biggest challenge we faced when designing the cavity mount.

There are restrictions on the types of materials that can be used in the mount. Its materials must be vacuum-compatible, and though the pressure in the phase plate section of the TEM is only $\sim 1 \times 10^{-6}$ mbar, we choose ultra-high vacuum (UHV) compatible materials wherever possible. This helps reduce outgassing and may help keep the mirror surfaces clean. The materials must also be non-ferromagnetic with a low magnetic susceptibility. This prevents the mount from magnetizing in the magnetic field of the electron lenses and thereby introducing aberrations in the TEM. For this reason, we avoid using even "non-magnetic" stainless steels like alloys 304 or 316 inside of the magnetic field shielding of the TEM column. Parts should also be made from materials that are easy to machine.

We did not use any silicone-based adhesives or lubricants, as they are known to migrate throughout vacuum systems and contaminate surfaces [73]. An earlier prototype of our cavity did use silicone-based adhesive Kapton tape (Accu-Glass Products, Inc. 110791), and we did not see any problems obviously attributable to the silicone.

Our design for the cavity mount is shown in figure 4.2. Specific features of its design are discussed in the following sections.

## 4.2 Flexure mount

To avoid requiring difficult-to-achieve sub-micron tolerances, and to enable adjustment of the cavity mode position and numerical aperture, our cavity mount incorporates three axes of optomechanical adjustment between the two cavity mirrors. This allows us to initially align the cavity mirrors relatively far from concentricity ($\sim 500\,\mu$m) and then incrementally reduce the distance to concentricity until the desired near-concentric condition ($\sim 3.7\,\mu$m). The cavity mount consists of a 23 mm diameter monolithic piece of 6061 aluminum, with a leaf-spring flexure cut into it to enable the relative alignment of the two mirrors (see figure 4.2). The leaf spring flexure allows three degrees of freedom between the parts of the cavity mount it connects: two degrees of angular adjustment (rotation about the $y$- and $z$-axes in figure 4.2), and one degree of overall axial displacement (along the $x$-axis in figure 4.2). These three degrees of freedom control the position of the center of curvature of the output mirror surface, which (relative to the fixed center of curvature of the input mirror surface) defines the position and numerical aperture of the cavity mode (see chapter 2). As such, the angular degrees of freedom can be used to compensate for any relative lateral offset in the mirror positions.

We refer to the part of the mount on the output side of the flexure as the "head" of the cavity mount. We use 6061 aluminum as the mount material because it is the most easily machinable material which satisfies our requirements. Bronze, titanium, or even tungsten may be possible alternatives. The relatively large coefficient of thermal expansion of alu-minum sets more stringent, but still manageable, requirements on the temperature stability of the cavity mount (see section 6.2.3). We choose a monolithic design for several reasons. First, it saves space because no additional fasteners are required to join together multiple parts. Fasteners could be replaced with epoxy, but this would still require some form of alignment guide to be designed into the parts, and as mentioned earlier, we avoid using epoxies as much as possible due to the potential for it to outgas and/or contaminate the mirrors. Second, the monolithic design has good thermal stability. A single part means there are no interfaces to increase thermal resistivity, and results in a more homogeneous temperature profile across the part. Perhaps more importantly, there should be little to no hysteresis in the thermal expansion of a monolithic mount. This is important in our use case, because turning the phase plate on and off cycles a thermal load of $\sim 1\,\mathrm{W}$ on the mount. The primary disadvantage of the monolithic design is that it is more difficult to manufacture, as there are dozens of delicate machining operations on a single part. I believe that a non-monolithic design should also be possible so long as consideration is paid to thermal stability.

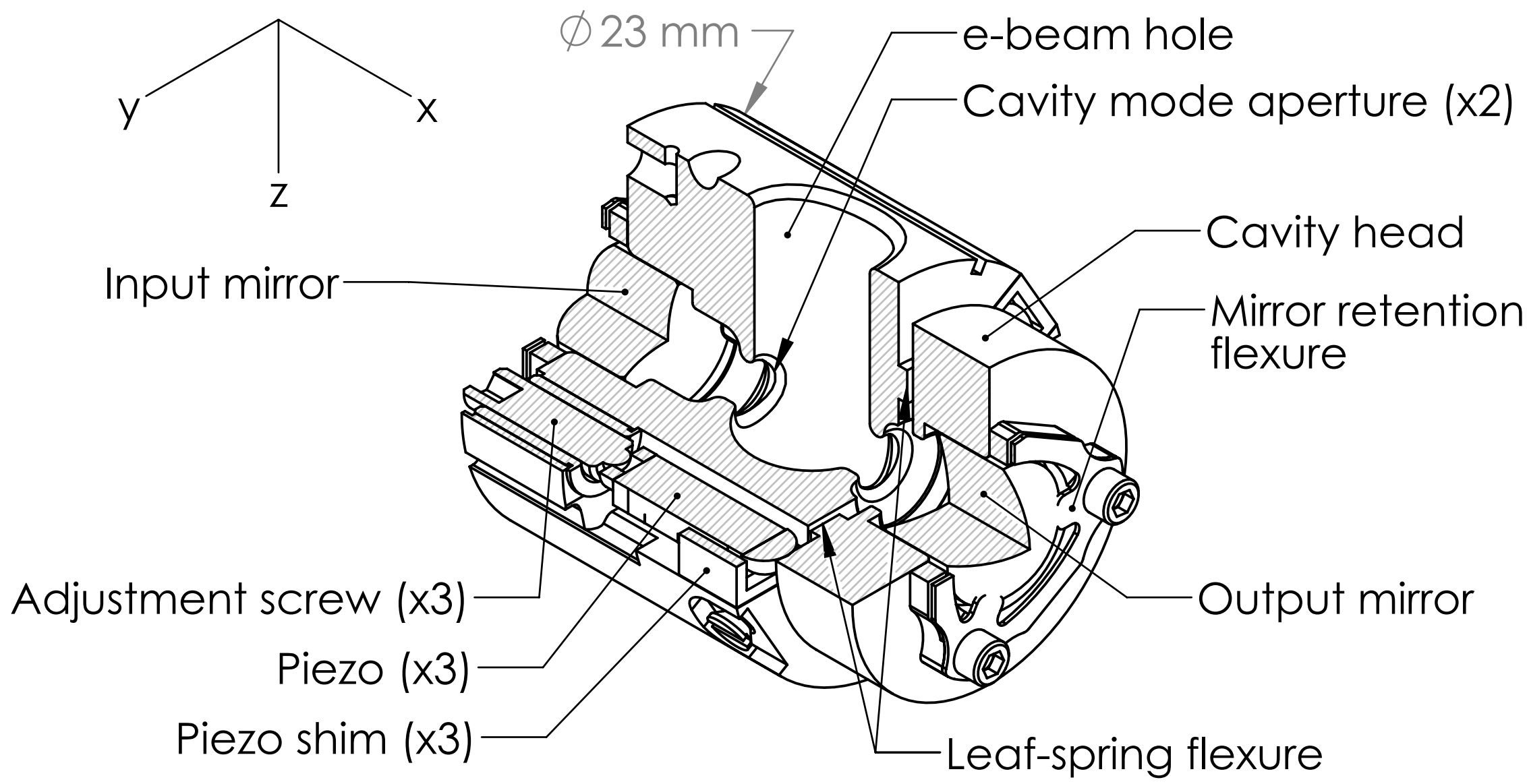

Figure 4.2: A cutaway view of the cavity mount, including cavity mirrors and associated optomechanics.

The leaf-spring flexure is manufactured by cutting two pairs of slits into the cylindrical mount at right angles and a small axial distance apart. This is done with a 1/64 inch slitting saw on a milling machine. During the operation, the two ends of the mount are held together with a sacrificial external fixture that is also partially cut through by the slitting saw. Dimensions of the flexure are shown in figure 4.3.

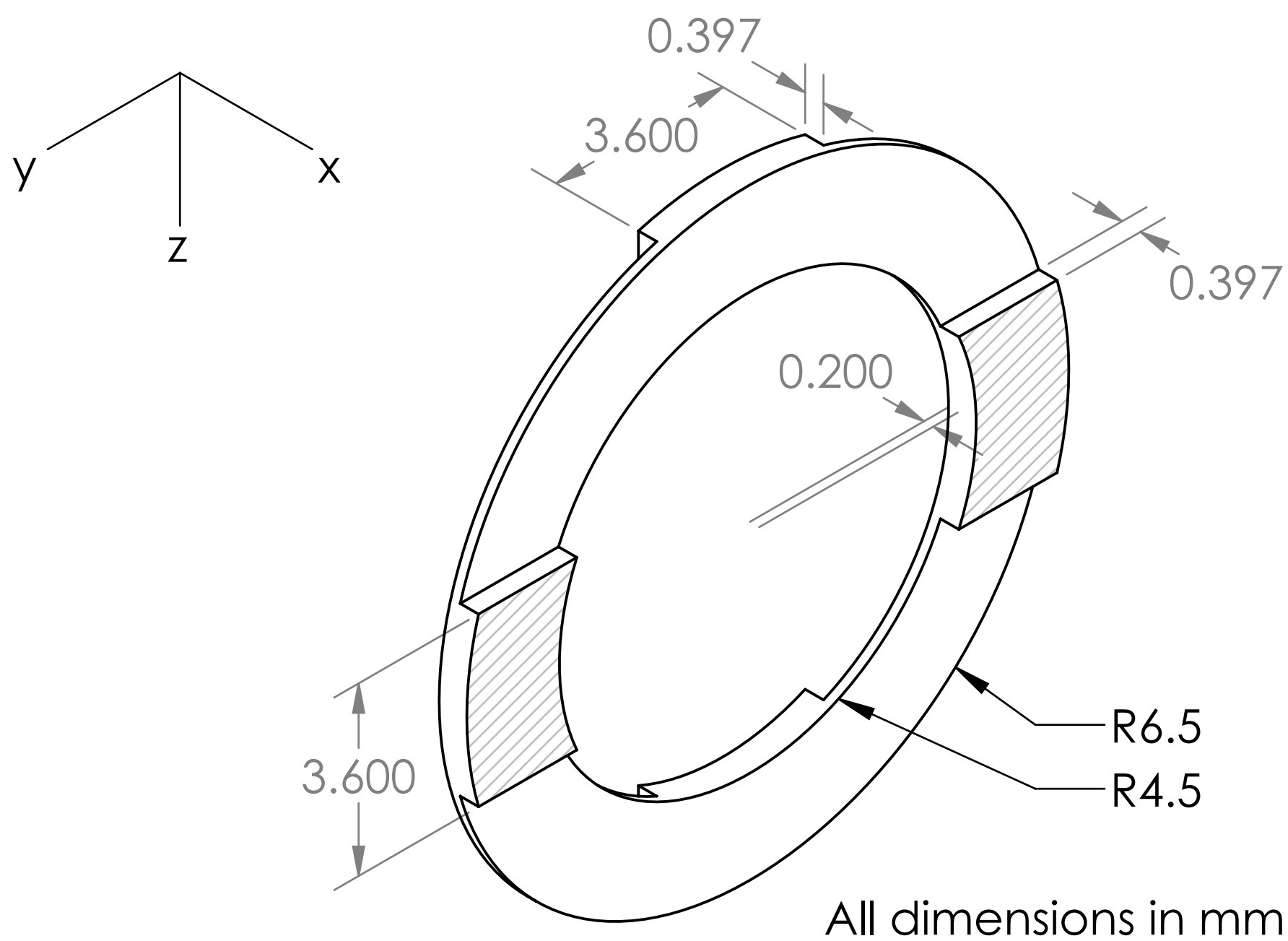

Figure 4.3: An isolated view of the leaf-spring flexure.

The flexure is designed to move axially by $\sim 500\,\mu\text{m}$ from its machined position to its nominal operating state near concentricity. In doing so, it is designed to deform plastically, which ensures a relatively constant counterforce throughout the adjustment range. However, it means that if the cavity must be reset to its far-from-concentricity alignment and then realigned to near concentricity (for example, if replacing a cavity mirror), the flexure must first be deformed back into place. We have successfully done this once on a prototype cavity, but it is likely that multiple resets will fatigue the flexure and eventually lead to it cracking.

We tested a few versions of the flexure which used different inner and outer diameters in order to obtain a reasonably high stiffness while still maintaining enough range of motion before failure. Though we have not characterized the stiffness of the flexure, I estimate (based on tactile experience building and aligning the cavity mount) that it provides $\sim 10\,\text{N}$ of counterforce when aligned near concentricity.

## 4.3 Mirror mounting

The cavity mirrors are mounted in cylindrical sockets at the ends of the cavity mount. Figure 4.4 shows the socket on the cavity head; the other socket has the same design.

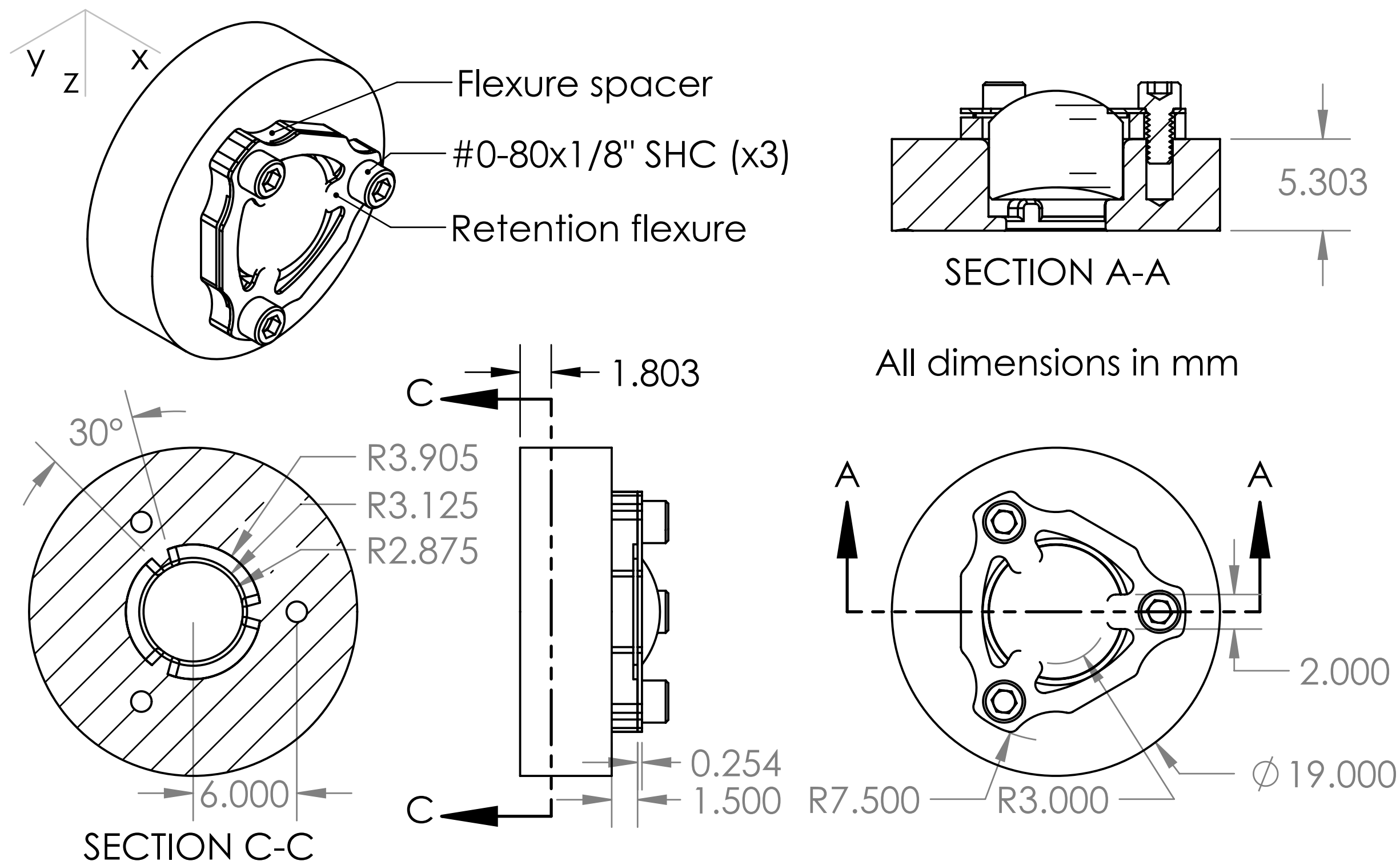

Figure 4.4: The mirror mounting socket and retention flexure on the cavity head.

The sockets have a diameter of 0.3075 inch, giving a radial clearance of 30 µm to the nominal OD surface of the mirror. Three raised areas machined into the bottom of the socket provide contact points for the "front" (intersection of CC and OD surfaces) edge chamfer of the mirror (see chapter 3 for the design of the mirrors). The mirror is held in place by a plastically deformed flexure spring which presses axially on the CX surface at three contact points, aligned with the front edge contact pads. This flexure also provides some amount of force which keeps the mirror centered on the optical axis, since the edges of the CX surface where the flexure contacts are fairly steep. We tested several different thicknesses and materials of the flexure to get the largest possible retention force without chipping the CX mirror surface, settling on 0.01 inch thick 1000-series aluminum. The flexure does still scratch the mirror surface at the contact points, but this is not optically relevant as the contact points are outside of the clear aperture of the CX surface of the mirror. Such scratching may still generate unwanted particulate contamination, but we have not yet seen that this contaminates the CC surface of the mirror.

We initially tested an alternative mounting scheme which used a radial canted coil spring to provide both a centering restoring force and additional contact points to the mount for thermal conduction. However, our design required the front edge chamfer of the mirror to be pushed past the canted coil spring. This resulted in chipping of the mirror chamfer and scratching of the canted coil, which contaminated the CC surface of the mirror with particulates, so we abandoned this approach. We then found that simply omitting the canted coil spring resulted in a design with adequate centering and thermal conductivity.

## 4.4 Electron beam hole

A hole for the electron beam to pass through is placed at right angles to, and intersecting with, the optical axis of the cavity mount (see figure 4.2). Initial prototypes used a relatively small hole diameter of 2 mm in an effort to reduce the chance of particulate contamination making its way inside of the cavity mount and to the mirror surfaces. However, we found that thermal magnetic field noise emanating from the electrically conductive cavity mount can limit the resolution of the TEM (see chapter 11), and that the noise can be reduced by increasing the diameter of the electron beam hole. Therefore, we redesigned for the largest practical electron beam hole (8 mm diameter). Any larger than this and the electron beam hole would start interfering with the piezoelectric actuators used for alignment of the cavity mirrors (see section 4.6).

There are 3 mm diameter holes where the cavity axis intersects the electron beam hole. These apertures were made as small as possible without noticeably clipping the cavity mode to reduce the chance that outside contaminants enter the electron beam hole and contaminate the CC surface of the mirrors. The edges of these apertures are filleted using a single-hemisphere cutter (J.W. Done 14000-074) to eliminate burrs which may lead to particulate generation and mirror surface contamination. We considered using electropolishing for this purpose, but could not find any local companies which could provide a specular finish on 6061 aluminum.

## 4.5 Adjustment screws

The cavity mount flexure is driven by three sets of adjustment screws which push on piezoelectric actuators, which in turn push on the flexure head. This is shown in detail in figure 4.5. The adjustment screws provide the coarse, large range adjustment, and are used to align the cavity to within several micrometers of its near-concentric condition. As such, they need to have a range of $> 500\,\mu\text{m}$, but also be adjustable in sub-micron increments. The limited space inside the TEM access port makes it difficult to fit differential micrometers or motorized screws, so we use manually-driven matched screw/bushing adjusters with a 3 mm screw diameter and a remarkably fine 50 µm thread pitch (Kozak Micro Adjusters TSBM3-05-10/7). These commercially-available off-the-shelf components come standard with a 360 brass bushing, 303 stainless steel screw, and 52100 steel ball tip. However, brass is not typically considered to be vacuum compatible (especially above room temperature) [74], and we avoid the steels due to their magnetizability. Because of this, we order custom versions of these screws which use a phosphor bronze bushing and titanium screw and silicon carbide ball tip with a venting slot cut in the screw tip for the small volume of air otherwise trapped below the ball tip. The custom screws perform just as well as the commercially-available off-the-shelf version as provided by the manufacturer. However, we found that they cannot be used after UHV cleaning and degreasing, as the titanium screw tends to gall and bind inside of the bushing after just a few millimeters of travel. We then tested several different screw materials for this issue and found that screws made of 304 stainless steel do not have this problem, and work well even after UHV cleaning. Screws made of aluminum bronze and phosphor bronze performed even more poorly than titanium. To avoid magnetization issues, we decided to use the titanium screws after regreasing them with Apiezon AP100, a vacuum-compatible silicone-free grease. The regreasing was performed by rubbing a small volume of grease between clean nitrile-gloved fingers until there was just a sheen of grease left on the gloves. The outer diameter of the screw was then spun in between the gloved fingers. Only a barely-visible amount of grease was transferred to the screw in this process, but the amount was sufficient to prevent galling.

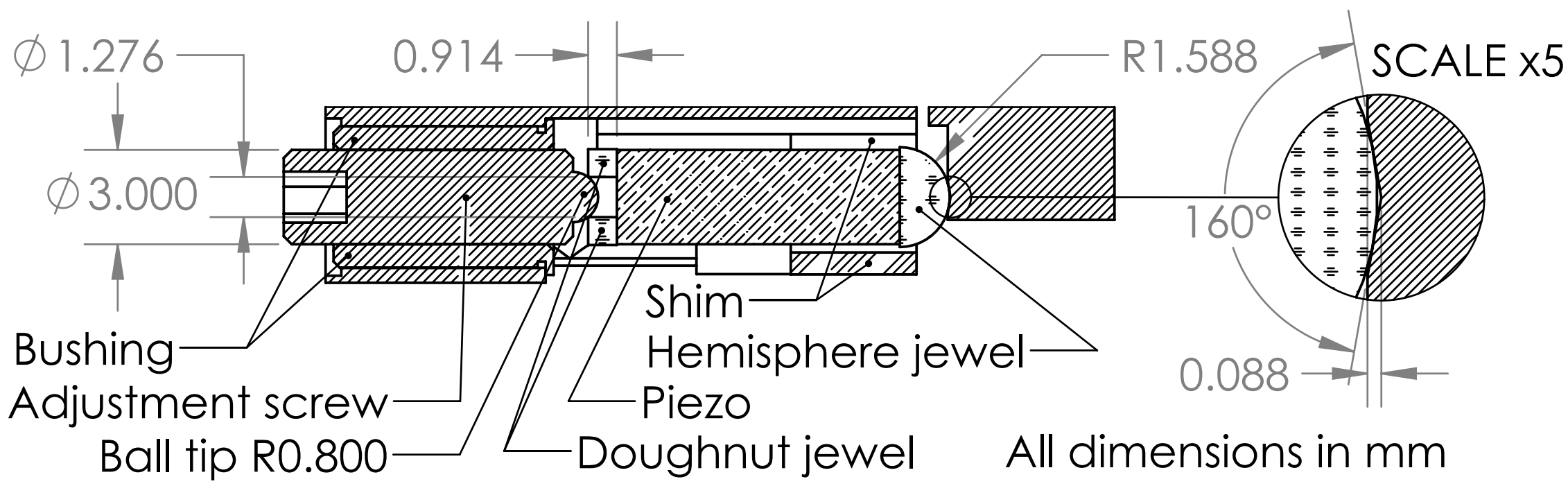

Figure 4.5: Detailed cross-section view of one of three identical cavity mount adjustment mechanisms.

The bushings are mounted in the cavity mount such that the bushing lip prevents the bushing from being pushed out of place by the flexure counterforce. A set screw pressing on the outer diameter of the bushing prevents the bushing from rotating along with the screw.

The screws are driven by 1.5 mm ball-end hex keys which have had the diameters of their head slightly ground down to allow them to engage and disengage from the screws' sockets more easily. This is critical for not disturbing the alignment of the cavity when removing the hex keys after the near-concentric alignment has been achieved. The keys are mounted at the end of a longer rod which allows the screws to be turned with the cavity mount attached at the end of its support arm.

## 4.6 Piezoelectric actuators

The piezoelectric actuators (PI P-883.11) provide fine control over the cavity alignment. They also enable active control of the alignment when the cavity is operating, since we have not designed a mechanism by which the adjustment screws can be turned while the cavity is under vacuum (they are only used for initial alignment). They are $3\,\mathrm{mm} \times 3\,\mathrm{mm} \times 9\,\mathrm{mm}$ in size and have a nominal travel range of $6.5\,\mu\mathrm{m}$ along their long axis. Each piezo is mounted between the ball tip of its adjustment screw and a small conical socket drilled in the head of the cavity mount (see figure 4.5). As such, each piezo needs a ball head socket and ball head on its respective end faces. We use a doughnut shaped sapphire jewel as the socket (Swiss Jewel Company R128.0) and a hemispherical sapphire jewel as the ball head (Edmund Optics 48-430). Both jewels are epoxied to the end face using a thin layer of Loctite Hysol 1C (marketed as "Torr Seal") between the jewel surface and the piezo end face. The jewels ensure a kinematic coupling between the head and body of the cavity mount, through the piezos and adjustment screws. That is, extending each screw and/or piezo only provides an axial force on the cavity head, which is constrained to move with only three degrees of freedom, as discussed in section 4.2.

We initially designed the cavity mount such that the piezos would not contact the cavity mount if they were to rotate around their long axis, so that there would be no chance that the electrodes printed on their side would short-circuit. However, this also allowed the piezos to rotate with the adjustment screw during initial alignment of the cavity, which would eventually pull the piezo's electrode wires out of place, electrically disconnecting the piezo. To prevent this from happening, we installed a small U-shaped shim made of polyether ether ketone (PEEK) plastic around the piezo which prevents the piezo from rotating more than $\sim 10^\circ$ but does not cause its electrodes to short-circuit. This piece is held in place with acrylic adhesive Kapton tape (Accu-Glass Products, Inc. 110794). This solution works, but I would recommend a re-design in future versions to avoid needing it.

The fringing electric field of the piezo slightly deflects the electron beam. We considered electrically shielding the piezos from the electron beam by covering the sockets in the cavity mount they reside in with a layer of aluminum foil. However, during normal operation of the cavity the piezo voltages only change slowly, and by small amounts, so their effect on the electron beam is negligible. Still, in future versions of the design we plan to incorporate electrostatic shielding of the piezos.

## 4.7 Output deflection prism

The cavity's output beam provides information about the state of the cavity and is useful for both diagnostics and active feedback (see chapter 6). To get the output beam out of the TEM, we need to deflect it by roughly $10^\circ$ so it aligns with the TEM's second access port. We do this using a simple wedge prism (Thorlabs, Inc. PS814-C) which was cut to size by Perkins Precision Developments (see figure 4.6). The prism's substrate material is N-BK7 glass. In future, fused silica would be a better material choice to reduce thermal lensing. The prism is mounted at a slight angle to the optical axis to optimize the deflection angle so that the beam exits the TEM near the middle of the access port, as well as to prevent reflections from coupling back into the cavity ("etaloning").

The prism is held (using a set screw) in a "cap" which fits over the head of the cavity mount and is fixed to it using several fasteners. This cap serves both to prevent the relatively delicate cavity head from being damaged during installation of the cavity into the TEM, and to further isolate the output mirror from external contaminants. Figure 4.6 shows the cap with deflection prism mounted on the cavity mount.

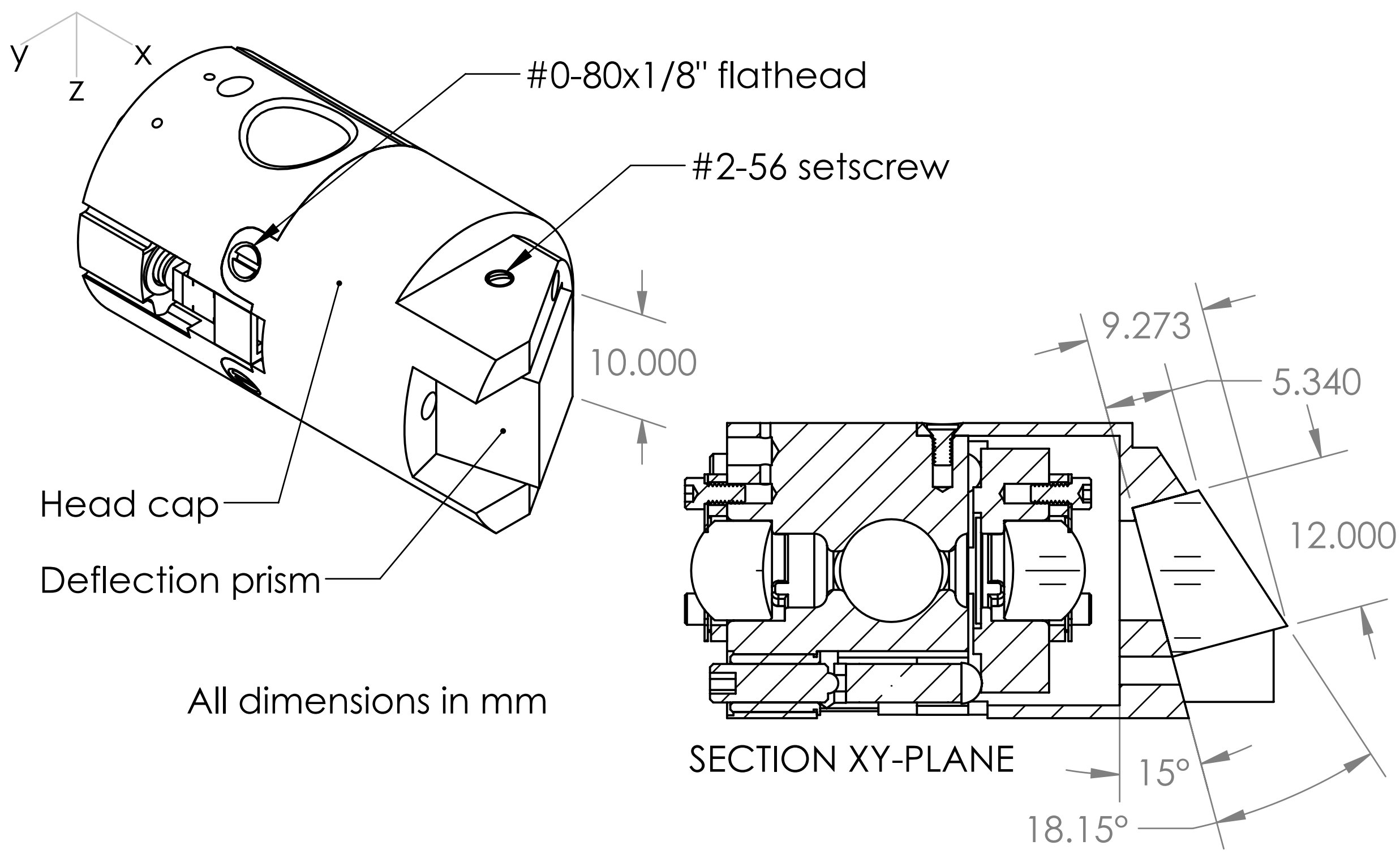

Figure 4.6: The deflection prism and cavity head cap mounted on the cavity mount.

## 4.8 Support arm

The cavity mount and associated components are mounted at the end of a support arm made of 6061 aluminum. This holds the cavity mount in place at the center of the TEM column in the phase plate module section (see figure 4.7). The cavity mount is screwed to the support arm with three #2-56×0.5" titanium socket head cap screws (McMaster-Carr 95435A225), using two custom phosphor bronze alignment pins for precise alignment of the two parts. For simplicity, we do not use any thermal contact gasket material between the parts.

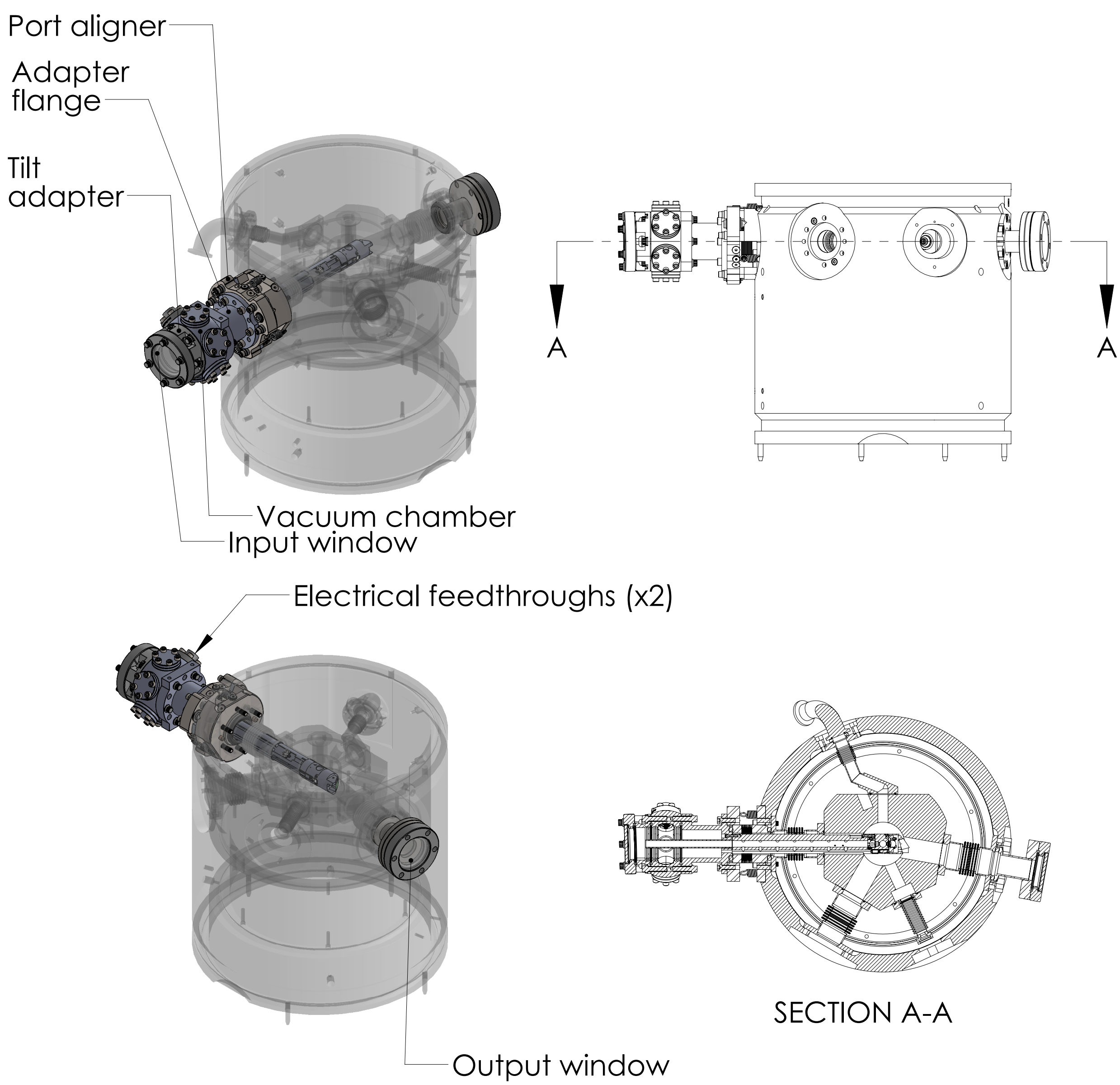

Figure 4.7: The entire cavity assembly installed in the phase plate module section of the transmission electron microscope column.

At its other end, the support arm is connected to a set of vacuum flanges which mount to the TEM access port's vacuum flange (see section 4.9). The support arm is roughly cylindrical, and has a 7.75 mm diameter hole bored along its axis to let the cavity input beam through. Holes between the outer and inner diameter allow for vacuum pumping out of the central bore. There are six slots running along the length of the support arm. Three of these slots accommodate the cavity mount adjuster screw hex keys, so that they can be inserted through the support arm during initial alignment of the cavity. The three other slots carry UHV-compatible electrically shielded twisted pair wire bundles (Accu-Glass Products, Inc. 112140) which connect to custom-made electrical screw terminal blocks which in turn connect to the six piezo wires and four thermistor wires (see section 4.12). The terminal blocks are made from PEEK, inlaid with six tubular phosphor bronze conductors. Wires inserted inside of the conductors are held in contact with the conductor using titanium #000-120 slotted binding head screws (Antrin Miniature Specialties, Inc.), which are machined to size.

## 4.9 Vacuum flanges

The support arm is screwed to a custom DN25KF-to-DN40CF vacuum flange adapter made of 316 stainless steel, using three #2-56×0.5” titanium socket head cap screws (McMaster-Carr 95435A225). We used stainless steel instead of aluminum so that metal gaskets could be used on both the KF and CF flanges, which reduces the vacuum leak/outgassing rate compared to fluorocarbon gaskets. However, we currently only use a metal gasket on the CF flange, so that we do not need to replace the KF gasket when unmounting the assembly from the TEM. We have also tested a 6061 aluminum adapter, which has the advantage that its higher thermal conductivity results in the operating temperature of cavity mount being lower by several degrees. Holes are drilled in the flange to allow the adjustment screw hex keys and wire bundles to pass through. A simple aluminum tube slip-mounts into the flange around the optical axis to prevent any wires from impinging on the laser beam. The adapter also has two external planar mounting surfaces with #8-32 tapped holes which allows it to be mounted onto jigs for assembly and insertion into the TEM.

The adapter mounts to a custom vacuum port aligner which allows the entire DN25KF flange (and therefore, the support arm and cavity mount) to be moved by up to 1 mm in any direction relative to the microscope column (see figure 4.8). The port aligner consists of a modified commercially-available off-the-shelf edge-welded vacuum bellows (Metal-Flex Welded Bellows, Inc. 17510AA-1) attached to an optomechanical-style kinematic coupling which uses three 3/8”-100 adjustment screws (Thorlabs, Inc. F38SS100). In principle, screws with piezo-actuated ball tips (Thorlabs, Inc. POLARIS-P20) could be fitted to provide an extra degree of fine control over the alignment. Counterforce is provided by three extension springs. The port aligner can also be locked into position with six #8-32×1.375” socket head cap screws fitted with leveling washers (McMaster-Carr 91944A026). We initially aligned the port aligner such that it would place the cavity at the nominal center of the TEM column. Since then, we have not needed to realign it.

The DN40CF side of the adapter is mounted to a commercially-available off-the-shelf DN40CF-flanged vacuum chamber (Kimball Physics MCF275-SphHex-C2A6) which incor-porates six DN16CF flanges, two of which we use to mount electrical feedthroughs (Accu-Glass Products, Inc. 100010) for the piezos, thermistors, and heater tape. Vacuum-side electrical connections to the feedthrough pins are made with socket connectors (Accu-Glass Products, Inc. 100181) crimped to the ends of the UHV-compatible wires.

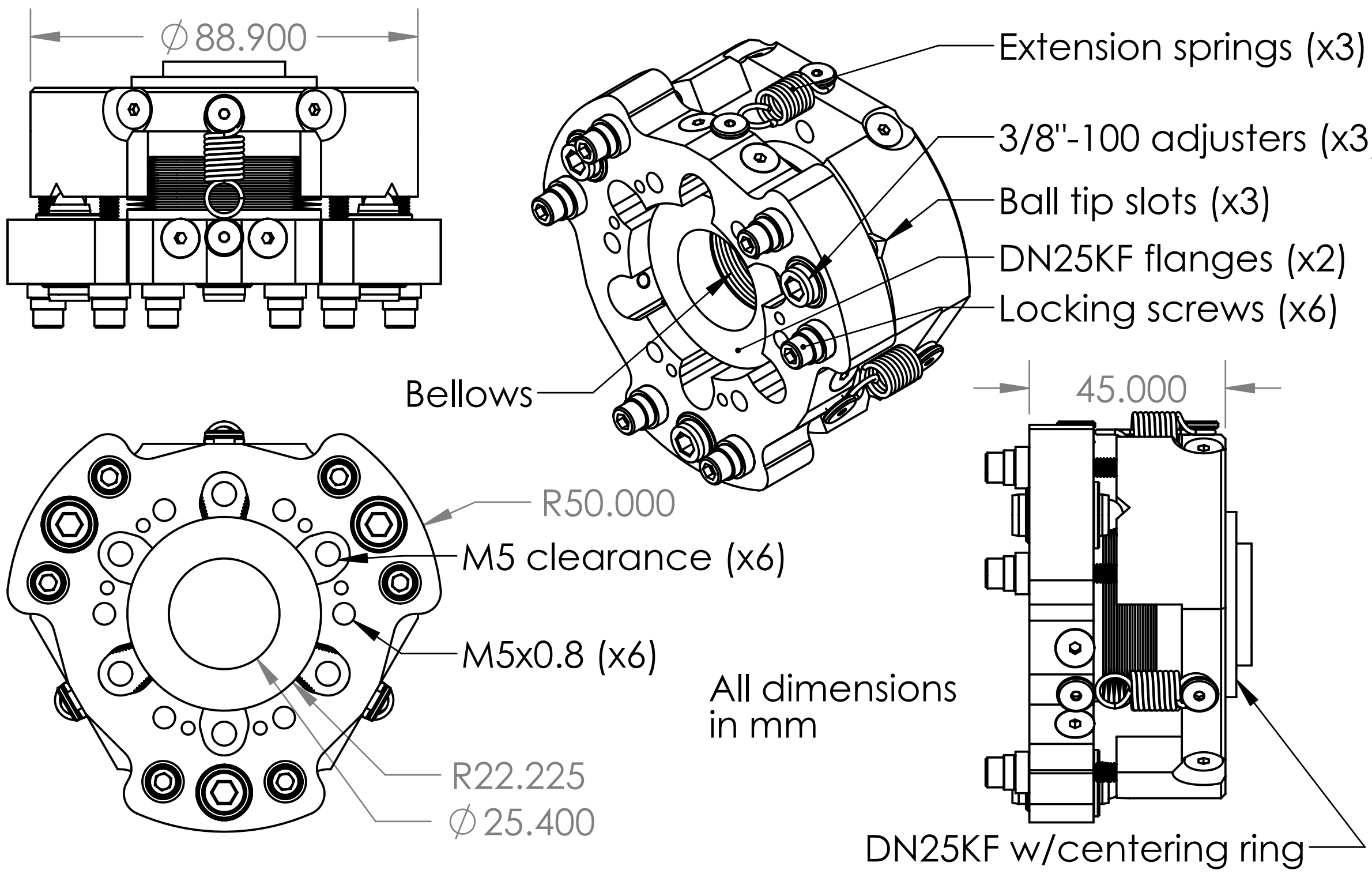

Figure 4.8: Custom DN25KF-to-DN25KF vacuum port aligner for positioning the cavity mount inside of the transmission electron microscope column. Extension springs are shown retracted.

The other four DN16CF flanges are sealed with blank flanges. The opposing DN40CF flange of this chamber is mounted to a DN40CF flange 2° tilt adapter (Kimball Physics MCF275-TiltAdptr-C2-2Deg), on which is mounted the laser beam input window flange (Manufacturing Precision Feedthrough Products A4545-1-CF). The tilt adapter prevents reflections off of the window from coupling back into the cavity. The input window is made of fused silica and anti-reflection coated. The output access port of the TEM was mounted with the same model of window, on a commercially-available off-the-shelf DN25KF-to-DN40CF adapter (Kurt J. Lesker Company F0275XQF25).

## 4.10 X-ray safety

Though leaded glass is often used in viewports on TEMs to keep X-rays from leaving the column, we found that even with a fused silica window there were no detectable X-rays (above background) outside of the column. This is likely because the electron beam does not hit any material objects in the phase plate, so no bremsstrahlung is generated there. The 8 mm diameter electron beam hole used in our design is large enough that it is probably impossible for the electron beam to directly hit the cavity mount. All commercially-available off-the-shelf flanges are made from 304 stainless steel, with the exception of the vacuum chamber which is made from 316 stainless steel. Though we did not find it necessary, additional lead shielding could be easily added to the outside of the TEM column around the cavity assembly, with just a small hole to admit the laser beam.

## 4.11 A note on fasteners

Fasteners are used wherever possible to reduce the need for epoxy. We avoid using epoxy to avoid having to consider its outgassing properties, and the potential for its volatiles to contaminate the mirror surfaces. Also, epoxying parts in place makes it more difficult to disassemble the parts if changes need to be made in the prototyping process.

To reduce the potential for galling, we used fasteners made from a material different from that of their tapped hole. Since our in-column parts are 6061 aluminum, and we avoid using stainless steel inside the TEM column, we mostly use titanium fasteners. A few fasteners are beryllium copper instead. We found that, when degreased, these materials still have a tendency to gall in a 6061 aluminum hole. We were able to overcome this by tapping holes with a relatively loose "H3" tolerance, which adds between 0.001" and 0.0015" to the nominal thread dimensions.

## 4.12 Temperature control

Controlling the temperature of the cavity mount helps control its alignment stability. The cavity mount temperature is roughly 10 K above ambient with a circulating laser power of 75 kW in the cavity. We use a Kapton-encapsulated electrical heater tape with a resistance of 25 Ω and acrylic adhesive backing (Accu-Glass Products, Inc. 112237) wrapped around the support arm to maintain the cavity mount at a roughly (±2 K) constant temperature, regardless of the laser circulating power. When the laser is off, the heater is set to output roughly 1 W of heat. When the laser is turned on, the heater power is reduced to zero or near zero keep the cavity mount temperature roughly stable.

The cavity mount temperature is measured using a thermistor (Amphenol Advanced Sensors A96N4-GC11KA143L/37C) which is taped to its outer diameter roughly halfway between the two mirrors. The thermistor has exceptionally small wires (#44AWG), which need to be crimped in a small piece of aluminum foil to get a good electrical connection when inserting the wires into the support arm terminal blocks. Another thermistor of the same model is placed on the outer diameter of the support arm near its connection with the flange adapter.

## 4.13 Campus machine shops are important!

The cavity mount and other auxiliary parts were manufactured by the UC Berkeley Physics Research and Development Machine Shop. I cannot say enough good things about this shop and the people who work there! I strongly believe that without their easily accessible expertise and advice, this project would not have succeeded. While there is nothing “special” about the shop in terms of its equipment, the fact that it brings together highly skilled machinists and researchers under the same roof means that we were able to rapidly iterate on design ideas, test design variations during the manufacturing process, and learn more about machining techniques which informed our future designs. Because intangibles like these do not appear on an administrator’s balance sheet, I worry that without continued advocacy the shop’s physical and human resources will be slowly stripped away, leading to reduced research output or a migration away from hardware-intensive projects.

# Chapter 5

# Cavity assembly and installation

This chapter describes the procedures we use to assemble and install the cavity mount and associated hardware. Since all the parts are connected by fasteners, this usually just requires simple tools like screwdrivers, tweezers, and a pair (sometimes more) of steady hands. The procedure is only made complicated by the need to keep the mirrors free of contamination. Even well-designed mirrors are rendered useless by contamination on their reflecting surface. Therefore, much of this chapter focuses on our methods for making and keeping the mirrors clean.

Contaminants, whether solid particulates, liquids, or volatiles, increase scattering and absorption on the mirror surface. Scattering reduces the enhancement factor of the cavity, which means that more input power is required to achieve the same circulating power. The required input power may exceed that which is available from commercial laser systems. For example, a particulate with a relatively large but realistic scattering cross-section of $500\,\mu\text{m}^2$ positioned at the center of the cavity mode will cause $\sim 5 \times 10^{-3}$ of scattering. This limits the amplification factor to a maximum of 100 (see section 2.5), which would require 750 W of input power to impart a $\pi/2$ phase shift to 300 keV electrons with a cavity mode numerical aperture of 0.05 and a laser wavelength of 1064 nm. At the time of writing, commercially available narrow-linewidth single-mode lasers at 1064 nm have a maximum output power of $\sim$ 100 W. Scattering also increases the heat load on the cavity mount, which causes more thermal misalignment of the cavity. Though this can be counteracted by introducing means to stabilize the cavity temperature and/or the cavity alignment (see section 6.2.5), such systems typically work better when they do not need to compensate for large inputs.

Absorption causes the mirror surface to thermoelastically deform, which can put an upper limit on the numerical aperture of the cavity mode (see section 2.6). But perhaps most importantly, absorption in small particulate contamination can melt the underlying mirror coating and damage an area of the coating surrounding the contaminant [72]. This can cause otherwise small contaminants with negligible scattering to become larger defects when they are exposed to high laser intensity. An example of this effect is shown in section 7.2.

An important note: do not assume that all of our contamination control procedures are necessary. Our goal was not to determine which procedures result in a functioning laser phase plate—our goal was simply to make a functioning laser phase plate. The procedures described in this chapter achieved that goal. It is therefore likely that some (perhaps even many) of our procedures are excessive. On the other hand, many of our procedures are based off of those used by the LIGO collaboration, which has similar mirror cleanliness requirements to us. They have put much more research into mirror contamination control than our small team ever could, and we have heavily leaned on their knowledge through their publicly available documentation [75].

## 5.1 Assembly laboratory

The laboratory we used for assembling and optically testing the cavity was remarkable, but only due to its small size. In order to have a space for assembling and testing the cavity near the TEM, we borrowed a roughly 5 m length of workbench in a storage room just down the hall from the TEM. Many thanks to Eva Nogales' group for generously offering us this space! This was much more convenient than the alternative of having to move our experiment between buildings when transitioning from assembling the cavity to installing it on the TEM. Roughly half of this space was dedicated to equipment used for contamination control while assembling the cavity, and the other half was for testing the cavity. A photo of the lab is shown in figure 5.1.

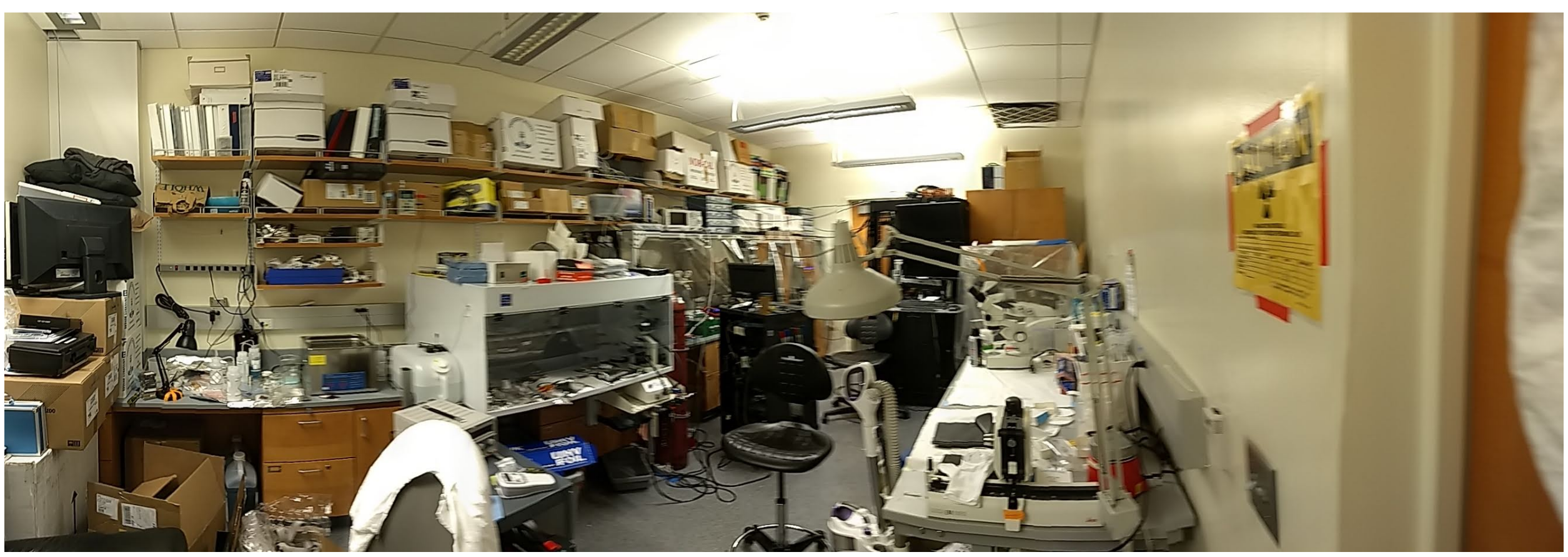

Figure 5.1: A stitched panorama of the laboratory where we assembled and tested the cavity. We were allocated the workbench space along the far wall. From left to right: the sonication area where parts were cleaned, the horizontal laminar flow hood where the cavity was assembled, and the clean enclosure in which the assembled cavity was optically tested.

The optical test setup consisted of a T-slotted aluminum extrusion ("80/20") frame loosely enclosed with clear vinyl sheeting. An ultra-low particulate air (ULPA) fan filter unit (Terra Universal, Inc. 6601-24-U) blowing into this enclosure from on top kept the air particulate density inside the enclosure at roughly ISO 14644-1 Class 5 (equivalent to FED STD 209E Class 100) standard. This was done to keep dust from building up on uncovered non-cavity optics; the cavity itself was never directly exposed to the air in this enclosure. A vibration isolation platform (Minus K Technology, Inc. 100BM-8) and 24" ×18" optical breadboard (Thorlabs, Inc. B1824F) were placed inside of the enclosure. The optical breadboard supported a small vacuum chamber into which the cavity could be mounted, as well as two other breadboards for the input- and output-side optics for the cavity. The vacuum chamber was pumped by an ion getter pump (IGP) with a nominal pumping speed of $20\,\mathrm{L s^{-1}}$. Initial pumping was done using a portable turbomolecular pumping station which used a dry diaphragm roughing pump and a $\sim 1\,\mathrm{m}$ long bellows. The vacuum pressure with cavity installed was comparable to that of the phase plate module of the TEM at $\sim 1 \times 10^{-6}\,\mathrm{mbar}$. The vacuum chamber was vented through a needle valve flange (Kurt J. Lesker Company F0275XVALVE) using the building's nitrogen supply, filtered through a $0.01\,\mathrm{\mu m}$ 99.99% efficiency filter (Parker 9922-05-BQ) and delivered by 1/4" outer diameter tubing (made of Saint-Gobain Versilon, or Saint-Gobain Tygon).

## 5.2 Horizontal laminar flow hood

Assembly of the cavity was done in a 4′ wide horizontal laminar flow hood (Terra Universal, Inc. 1688-43) with a $0.12\,\mathrm{\mu m}$ 99.999% efficient ULPA fan filter unit (Terra Universal, Inc. 6601-24-U). As an aside, I am not a fan (pun intended) of Terra Universal, Inc.'s products. The build quality is mediocre, and I would recommend buying from another company if you can afford the additional cost and lead time that might entail. We measured the particulate density in the flow hood using a laser particle counter (Met One Instruments, Inc. GT-526S). It consistently detected no particles, and since it does not register non-zero densities lower than $1\,\mathrm{ft}^3$ for particles larger than $0.3\,\mathrm{\mu m}$ in diameter, we believe that the inside of the flow hood achieves ISO 14644-1 Class 3 (equivalent to FED STD 209E Class 1) or better. We used an ionizer bar (Terra Universal, Inc. 2005-07A) downstream of the filter to reduce static electric buildup in the workspace, since particulates can be attracted to charged surfaces via an induced dipole force. Figure 5.2 shows a photo of the flow hood ready for cavity assembly work.

## 5.3 Cleanroom garments

We wore cleanroom garments when working in the flow hood to avoid introducing contamination from human hair, skin, oils, and clothing fibers.

We used ISO 14644-1 Class 4 rated nitrile gloves (Ansell Nitrilite 93-401). We used nitrile over latex both because it is hypoallergenic and because we found that our latex cleanroom gloves reacted mildly with the pure isopropyl alcohol (IPA) we used to clean our parts (the cleaning procedure is described in section 5.4): after letting the gloves soak in IPA overnight, the solution became a light yellow color, similar to that of the latex. To absorb sweat and make glove changes somewhat easier, we typically used nylon glove liners (Valutek VTGNLR-1/2). We did not use two layers of gloves, as we could change gloves just outside of the flow hood without risking contamination of the parts in the hood. We stored gloves inside the flow hood by taping their bag to the interior wall of the hood, such that the gloves could be grabbed at the wrist from outside of the hood, but they were still surrounded by filtered airflow. This does mean that the outside of the cuffs of our gloves were touched with bare fingers.

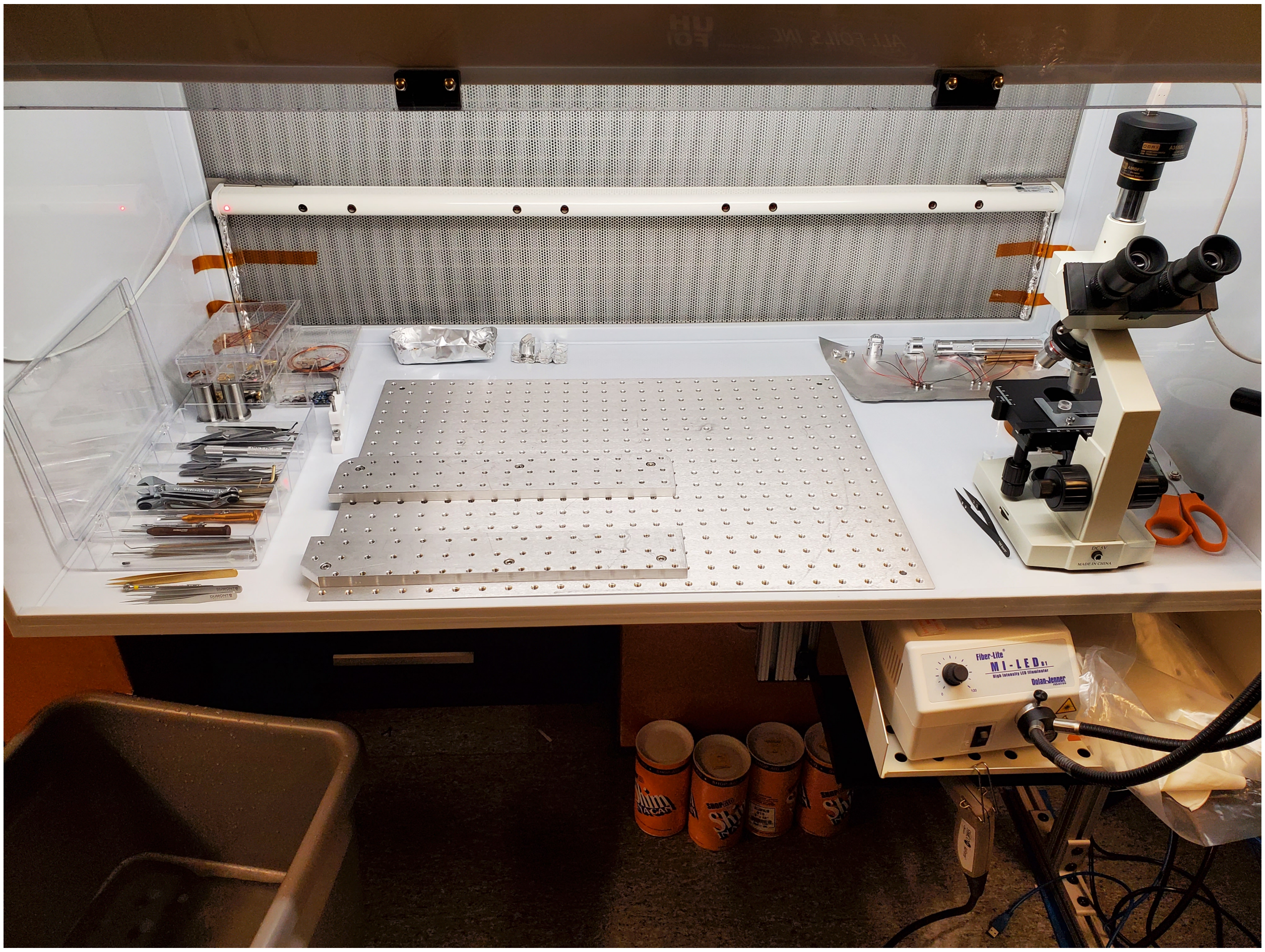

Figure 5.2: The horizontal laminar flow hood. Left: tools; center: aluminum breadboard for mounting assembly jigs; right: optical darkfield microscope for inspecting mirrors. Some cleaned parts ready for assembly sit on ultra-high vacuum aluminum foil at the right rear.

In addition to the gloves, we wore Tyvek cleanroom frocks (DuPont C270BWHLG00300C)

along with Tyvek cleanroom hoods (DuPont IC668BWH0001000C). Both were rated to ISO 14644-1 Class 5. The frock mainly served to isolate the inside of the hood from our arms. The hood was typically unnecessary, as we usually kept our heads outside of the hood, and in general, we tried to avoid placing any body parts (even protected) above or upstream of the parts being worked on. We also always wore simple surgical or particulate masks (no cleanroom rating). Breathing, and especially talking, generates a spray of droplets, and we were concerned that some of these droplets may be large enough to travel upstream into the flow hood far enough to land on parts. Even before the recent COVID-19 pandemic put such ideas into the zeitgeist, I remember seeing a particularly disgusting and ingenious demonstration at the SPIE Photonics West Exhibition in which an optics cleaning company needing dirty demonstration pieces simply placed optics face up on a table between conversing sales representatives and prospective customers. A photo of the author wearing our standard cleanroom garments is shown in figure 5.3.

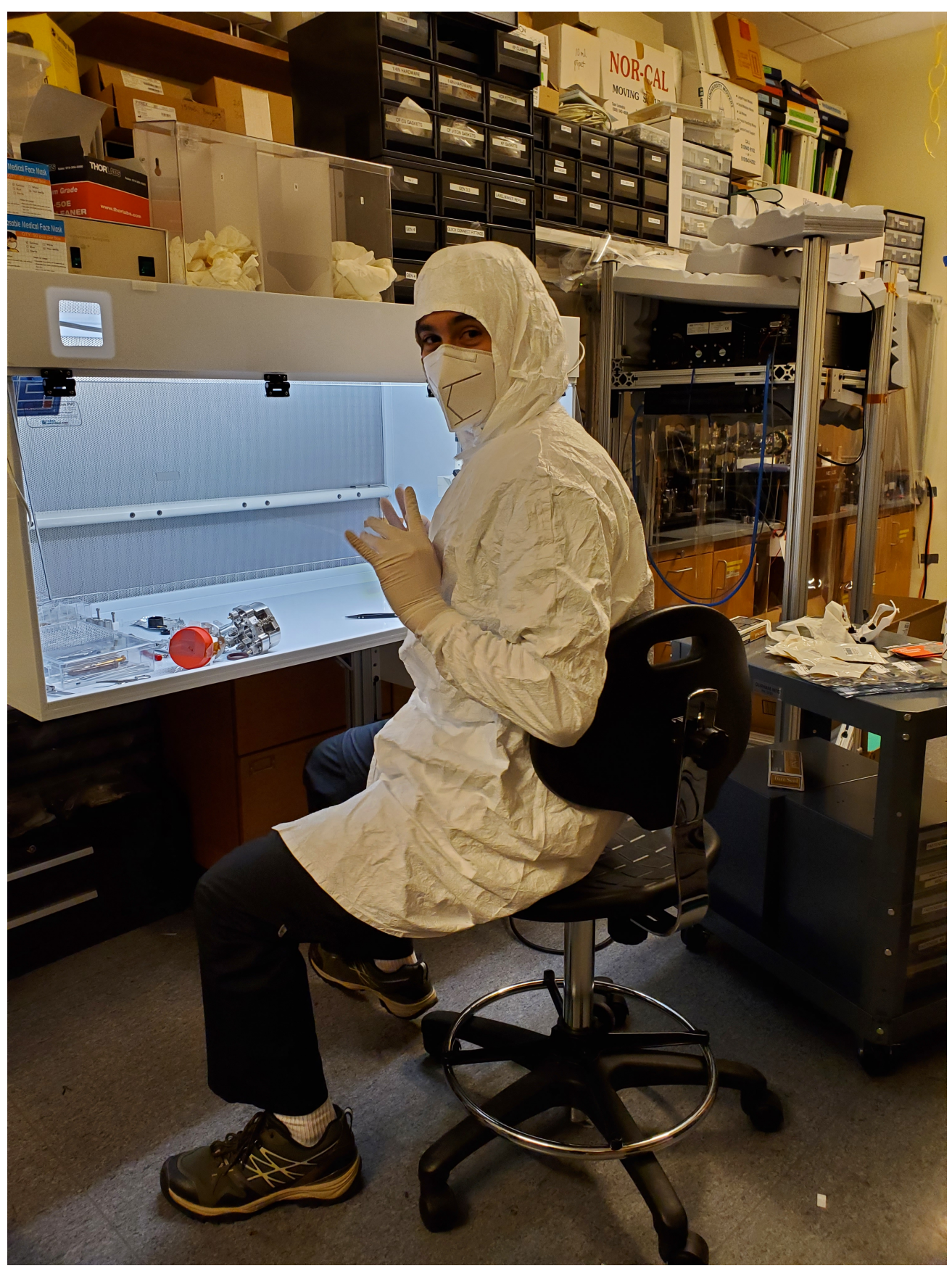

Figure 5.3: The author wearing cleanroom apparel, ready to work in the horizontal flow hood.

Remarkably, we found that even cleanroom gloves still can leave behind particulate con-tamination on the surfaces they touch. We tested this by touching a clean optical surface with the gloves and then inspecting it in a simple darkfield microscope mounted inside the flow hood (see section 5.5). A clean, untouched surface is show in figure 5.4a. A similar surface after being touched by a brand-new, uncross-contaminated glove is shown in figure 5.4b. We found that simply rinsing the gloves after donning them with IPA and then drying them with an ionizing nitrogen blow-off gun (Simco-Ion, Technology Group AirForce Model 6115) substantially reduced the contamination (figure 5.4c). On the advice of colleagues in the LIGO collaboration (thank you Aidan Brooks, Matthew Heintze, and Margot Hennig!) we achieved similar results by cleaning our gloves after donning but before use by wiping them down with cleanroom wipes pre-soaked in IPA (Texwipe AlphaSat with Vectra Alpha 10 TX8410)[1]. Regardless, we never directly touched the mirrors, instead handling them by gripping their outer diameter with precleaned tweezers. Similarly, we did not touch the mirror sockets on the cavity mount, though we did extensively touch the outer diameter of the cavity mount during the assembly process.

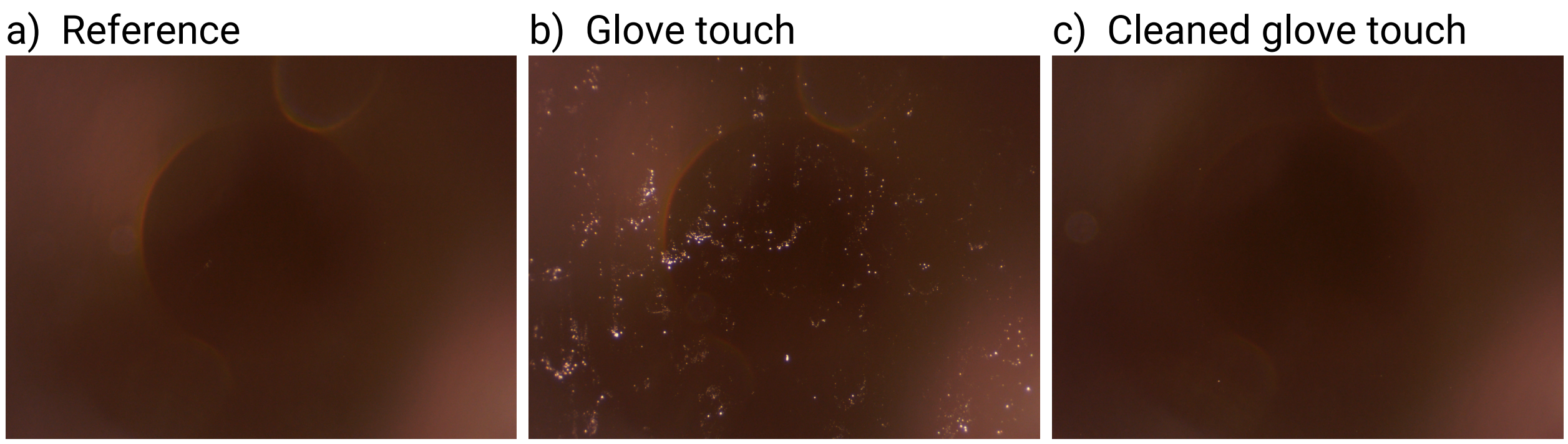

Figure 5.4: Darkfield images of **a)** a clean optical surface, as a reference, **b)** a surface after being touched with a cleanroom glove, immediately after donning, **c)** a surface after being touched with a cleanroom glove that was donned and then rinsed with isopropyl alcohol and dried with ionized nitrogen. The field of view of all images is 2.8 mm × 2.1 mm.

## 5.4 Parts cleaning

Before being assembled, the non-optical parts (e.g. cavity mount, support arm, fasteners, etc.), as well as tools, were all cleaned to remove surface contaminants. This reduces the chance of the mirrors becoming contaminated by removing potential contaminants from their environment. It also improves the vacuum pressure by reducing outgassing. The cavity alignment piezos and shielded twisted pair wire bundles were not cleaned before use, as they came ultra high vacuum (UHV)-cleaned from the manufacturer.

[1] Yes. That is what the product is called. I have spared you the registered trademark symbols.

Our cleaning procedure is similar to most standard UHV cleaning procedures. We first sonicated parts in a dilute solution of detergent (Brulin AquaVantage 815 GD) in deionized water (DI) at 60 °C for at least 15 min. The detergent was roughly diluted to the manufacturer's specifications. We chose this particular detergent because it is compatible with a broad range of metals, ceramics, and plastics, and is non-toxic. The sonicator had a bath volume of 22 L and operated at 40 kHz with a power of 480 W (VEVOR 22L). Parts were placed in precleaned glass beakers inside of the sonicator bath to prevent cross-contamination.

After sonicating in detergent, the parts were thoroughly rinsed with DI. They were then sonicated in DI for at least 10 min at 60 °C. The parts were then removed from their beaker and rinsed with pure high-performance liquid chromatography (HPLC)-grade IPA to remove any remaining DI. They were then placed into another precleaned beaker containing pure HPLC-grade IPA and sonicated for at least 15 min at 60 °C.

The sonicator and its contents were exposed to unfiltered room air (see the location of the sonicator to the left of the flow hood in figure 5.1). It is possible that because of this, particles fall into the beakers and then adhere to the part when it is removed. To counteract this, in the final step of cleaning the parts were removed from their IPA beakers just outside of the flow hood and given a final rinse with IPA from an LDPE wash bottle. In future, I would recommend eliminating any uncertainty associated with this procedure by instead using two 4′ wide flow hoods side by side: one for cleaning parts and one for assembling them. Also, using an LDPE wash bottle to deliver the IPA was not ideal. After the bottle has been squeezed, it backfills with air through its spout. This bubbles air back through the solution. Pure IPA is hygroscopic, and we found that bottles of IPA that had not been recently refilled would tend to leave small droplets, presumably water, on the surface of parts after the IPA had evaporated. These increase drying time, and might deposit a residue on the surface if they are allowed to evaporate while on the part. To avoid this, I would suggest using a glass gas wash bottle pressurized with clean nitrogen. A valve on the nitrogen line could be used to dispense the liquid from the bottle.

The part was then blown dry using ionized filtered nitrogen from a blowoff gun operating at roughly 2 bar (Simco-Ion, Technology Group AirForce Model 6115). The nitrogen supply for the blowoff gun was delivered via 1/4" tubing (made of Saint-Gobain Versilon, or Saint-Gobain Tygon) which was not precleaned, and was prefiltered using a 0.01 µm 99.99% efficiency filter (Parker 9922-05-BQ). We would have preferred a higher nitrogen pressure for faster and perhaps more effective blowoff, but 2 bar was the maximum supplied by the building and we didn't want to use a compressed gas cylinder which we would have had to replace frequently.

Dried parts were placed on sheets of UHV aluminum foil (Kurt J. Lesker Company FOILA24.0015) inside of the flow hood, upstream or to the side of the assembly area.

## 5.5 Mirror inspection and cleaning

Getting the mirrors clean is just as important as keeping them clean. To characterize the cleanliness of each mirror, the effectiveness of our cleaning techniques, and any manufacturing defects we inspected each mirror using a simple darkfield microscope, consisting of a trinocular upright optical microscope (OMAX M83EZ-C50U) with color camera and white LED gooseneck illuminators (Dolan-Jenner Industries MILEDUSDG MI-LED-US-DG) which output roughly 2 W ($\sim$ 2000 lm) of optical power. The mirror was placed in a custom acrylic holder which held it with its concave (CC) surface (the primary surface of interest) facing up. During inspection, cleaning, and assembly the mirrors were handled by gripping their outer diameter with precleaned ceramic tip tweezers. The illuminators were positioned above and to the side of the mirror (see figure 5.5). This achieved an approximately darkfield illumination, though the highly curved surface of the mirror and relatively small outer diameter generated bright reflections. We usually used the microscope's 4$\times$ magnification, 0.1 numerical aperture objective, which resulted in a camera field of view of 2.8 mm $\times$ 2.1 mm. The depth of field was roughly 50 µm, which meant that only part of the field of view would be in focus, especially near the edge of the CC surface of the mirror. Since the superpolished CC surface itself is ideally featureless, it was only possible to focus when there were scatterers present. Inspecting a mirror therefore first consisted of inspecting the CC surface edge chamfers, then focusing roughly 50 µm below the chamfer and moving the microscope stage around to inspect the edge of the CC surface near the chamfer. This edge is outside of the specified clear aperture of the mirror (see section 3.7), so there would typically be at least several scatterers present. By then focusing on these scatterers, it was possible to work the defocus closer to that required at the center of the mirror. Once at the center of the mirror, we would slowly scan the focus looking for scatterers within the clear aperture. An example image of the center of a mirror surface with few scatterers is shown in figure 5.6a. Such a mirror would be a candidate for installation into the cavity. Figure 5.6b shows a mirror with many scatterers. It is not usually obvious from the darkfield images if scatterers are contaminants (which in principle can be removed by cleaning) or defects (which cannot).

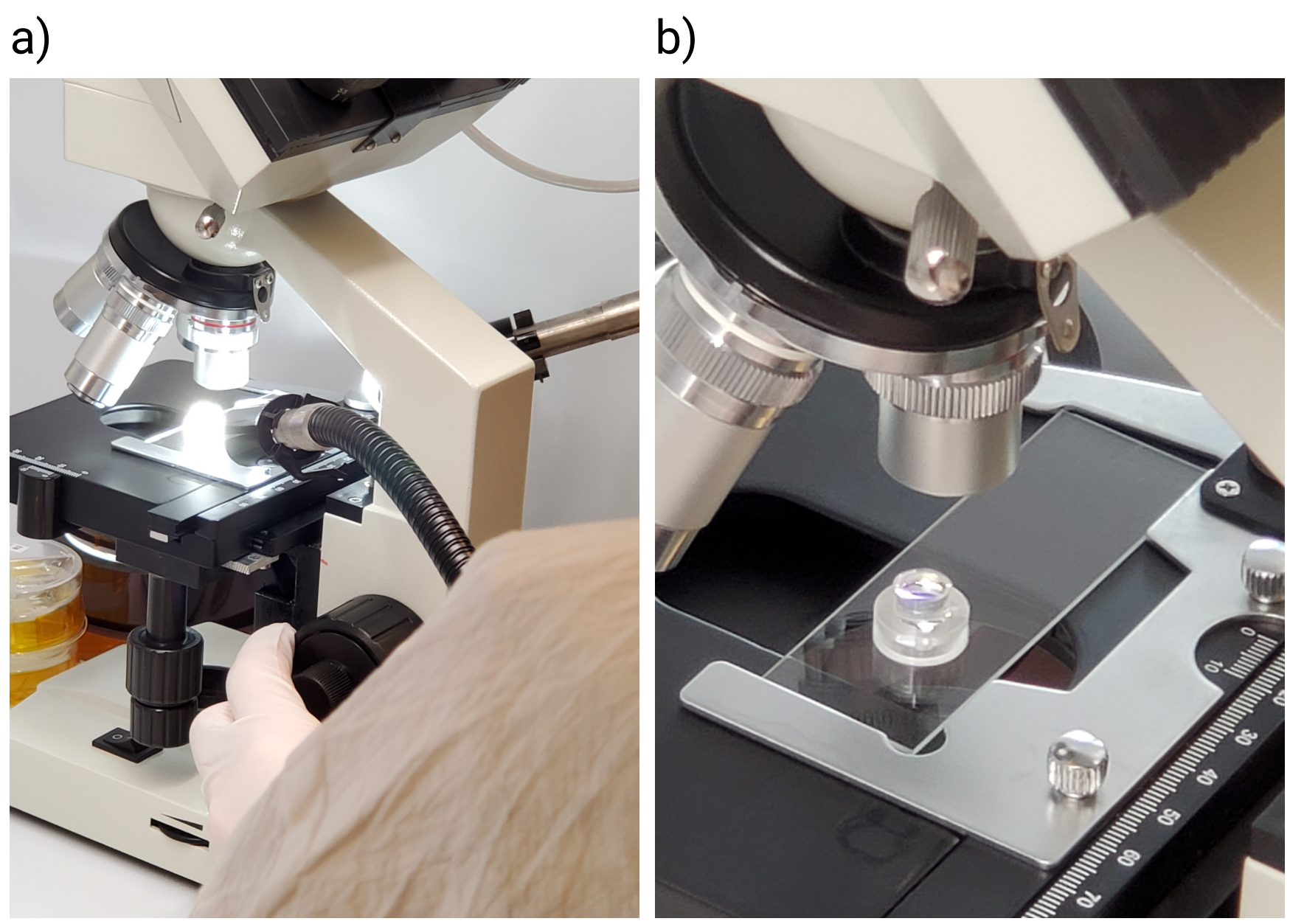

Figure 5.5: **a)** Darkfield microscope with a mirror in place and gooseneck illuminators turned on. **b)** Detail view of the mirror sitting in its acrylic holder on top of a microscope slide, with the illuminators turned off.

a) b)

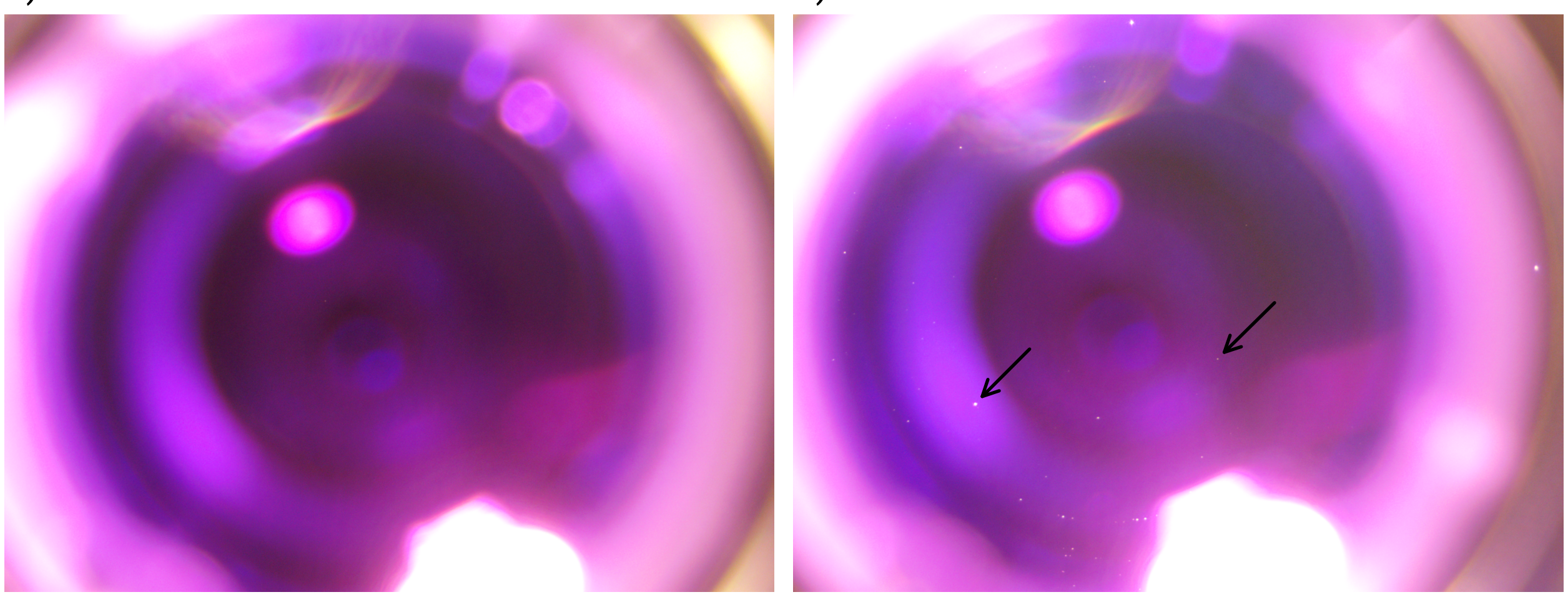

Figure 5.6: Darkfield images of **a)** a mirror with few scatterers; a good candidate for installation in a cavity, and **b)** a mirror with many scatterers; two are annotated with arrows for reference. The field of view of both images is 2.8 mm × 2.1 mm.

Though imperfect, this darkfield system was cheap and easy to set up in the early days of the project. Once we were accustomed to the system we could easily make subjective comparisons between mirrors, and this familiarity disincentivized us from upgrading the inspection system to something with a more uniform darkfield illumination. However, I would strongly recommend making this upgrade. As is shown in section 7.2, the current system is only just capable of seeing some potentially problematic (albeit weak) scatterers.

We had our mirror coating manufacturer (FiveNine Optics) package the mirrors in PET-G optics packaging containers (Safe-Guard Optics Packaging Containers EW0.31-01500350) after they had been coated and cleaned. These containers prevent both optical surfaces of the mirror from contacting the container, and do not contain any particulating material like foam or lens tissue. We and/or FiveNine Optics precleaned the inside of the containers by blowing it off with ionized nitrogen. I would also recommend first degreasing the containers with the method described in section 5.4, as we were told by a sales representative from Safe-Guard Optics Packaging Containers that they do not degrease the containers after manufacture. The containers were sealed with polyimide tape and shipped inside of a resealable plastic cleanroom bag. We would unpackage the containers just in front of the flow hood and blow off their outside with ionized nitrogen before bringing them into the flow hood. With these containers we found that it was sometimes unnecessary to clean the mirrors ourselves after unpackaging them.

We tested several methods of cleaning the mirrors, with varying degrees of success. Holding the mirror in place with tweezers and blowing it off with ionized nitrogen at a pressure of 2 bar was only successful in removing the largest contaminants. Spraying HPLC-grade IPA from an LDPE wash bottle directly onto the CC surface, followed by immediate drying with ionized nitrogen was typically effective in removing most, but not all particulate contaminants. It came with the usual added complication of solvent cleaning: it was difficult to avoid developing "drying streaks" on the convex (CX) surface of the mirror since that surface could not be easily blown off in our setup where the mirror was held with the CC surface facing up. Our most successful cleaning technique used Red First Contact Polymer (made by Photonic Cleaning Technologies). First, a small O-ring made of FKM ("Viton") or FFKM ("Kalrez") fluoroelastomer with an outer diameter roughly that of the mirror was cleaned according to the procedure described in section 5.4. The O-ring was then dipped into the First Contact Polymer (FCP) one half at a time, while holding the other half with tweezers. In between dips, the deposited polymer was allowed to dry. This built up a thin $< 1$ mm layer of FCP on the O-ring. The O-ring was then placed on the CC surface of the mirror, and more FCP was dropped onto the center of the CC surface with a plastic pipette until the bowl formed by the CC surface and O-ring was filled. Figure 5.7 illustrates a cross-section of the mirror, FCP, and O-ring in this state. The FCP was then allowed to fully dry overnight. Once dry, the mirror was gripped on its outer diameter with tweezers. A second pair of tweezers gripped one side of the O-ring and peeled the O-ring and dried FCP layer off of the CC surface. In this way, the O-ring served both to contain the FCP on the CC surface of the mirror, and act as a handle by which the FCP could be gripped to pull it off of the surface. During the peeling process we blew ionized nitrogen across the mirror to discharge any static electric buildup. Many thanks to former Müller group graduate student Matt Jaffe for telling us about this specific approach to using FCP! Coating the O-ring prevented any contaminants from being deposited on the CC surface by the O-ring itself, and also maintained a thick layer of FCP even directly under the O-ring. This prevented the FCP from ripping during the peeling process and leaving behind thin patches of FCP near the contact point of the O-ring on the CC surface, which we had observed when using uncoated O-rings. This cleaning method never eliminated all scatterers on the CC surface, but we believe that the remaining scatterers were defects, not contamination. However, the method is time intensive and requires some practice.

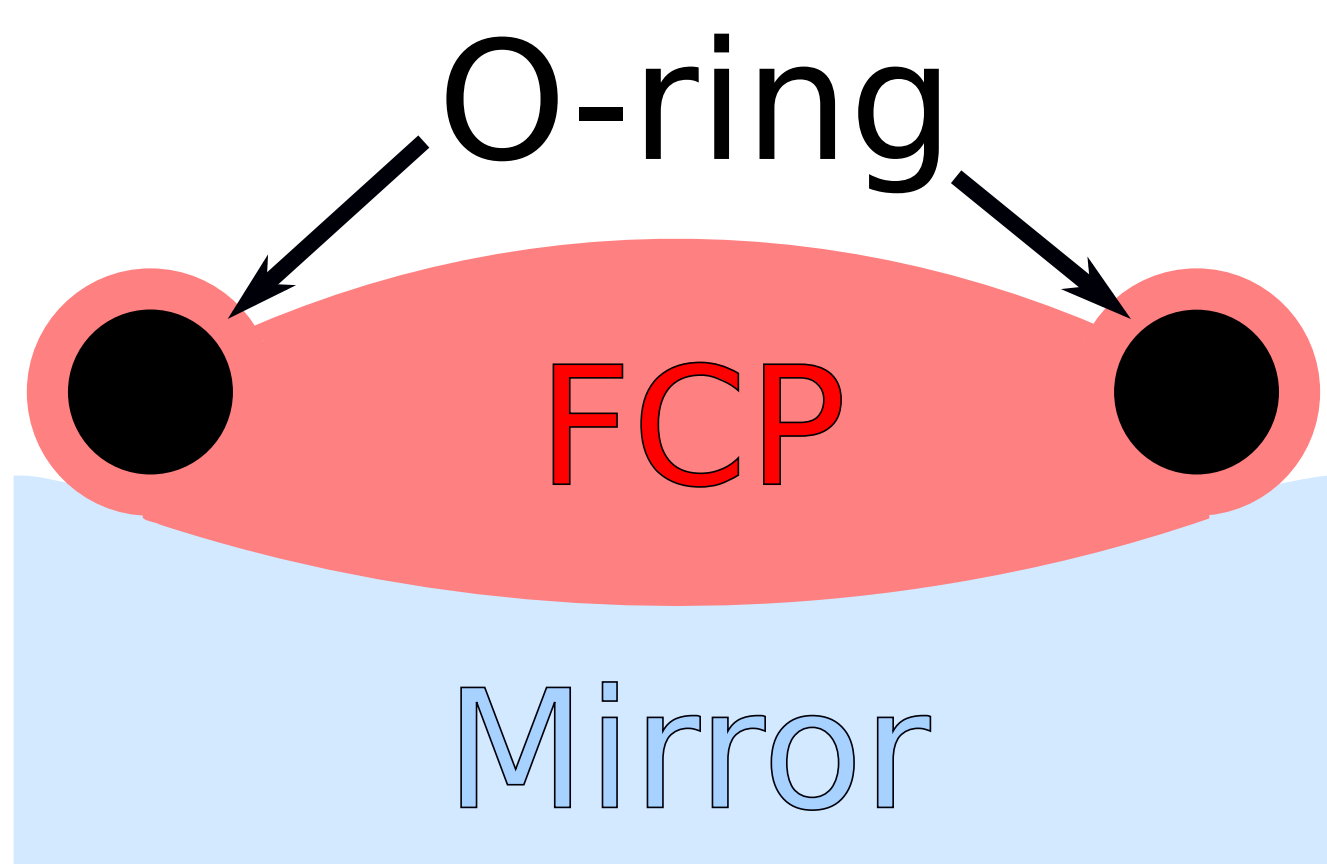

Figure 5.7: Cross-section of an O-ring coated with First Contact polymer (FCP) containing a reservoir of FCP on the concave surface of the mirror, before drying.

FiveNine Optics cleans the mirrors after coating by spinning the mirror around its optical axis at $> 1000\,\text{rpm}$ in a modified semiconductor wafer spin coater and spraying the CC surface with a sequence of solvents: first DI, then acetone, then IPA. A cotton swab wipes the wet surface as it spins. The spinning helps move the solvent across the surface and prevent it from evaporating in place and thereby leaving a residue. The cleaning is performed in a horizontal laminar flow hood. I would recommend setting up a cleaning system like this to increase the cleaning throughput beyond what is practical with the FCP method.

Figure 5.8 shows the center of the CC surface of the two mirrors installed in the most recent (and most successful) iteration of our cavity. All discernible scatterers are annotated with arrows. The mirrors are not entirely free of scatterers, but the few scatterers that are present are at least several hundred microns from the center of the mirror, which reduces the laser intensity they are exposed to and therefore the fraction of the cavity mode they scatter. As is discussed in section 7.2, we found that it is better to have several scatters farther from the mirror center than to have even a single very weak scatterer close to the mirror center. The input mirror (serial number 3.4.4, figure 5.8b) also had a large chip in its chamfer, due to a manufacturing defect which resulted in there being no protective flat annular chamfer (see section 3.2.4). An image of that defect is shown in figure 5.9. Both mirrors were cleaned using FCP.

## 5.6 Assembly

After the parts were cleaned the cavity was assembled in the flow hood. We used an unanodized aluminum optical breadboard as a mounting platform for several assembly jigs. The optical breadboard was too big to fit in our sonicator, so we washed it by hand with detergent, DI, and IPA. Once inside the flow hood it was wiped down with TX8410 wipes. One jig mounted the vacuum flange adapter (see section 4.9) to the breadboard so that the entire assembly could be built around this part (see figure 5.10).

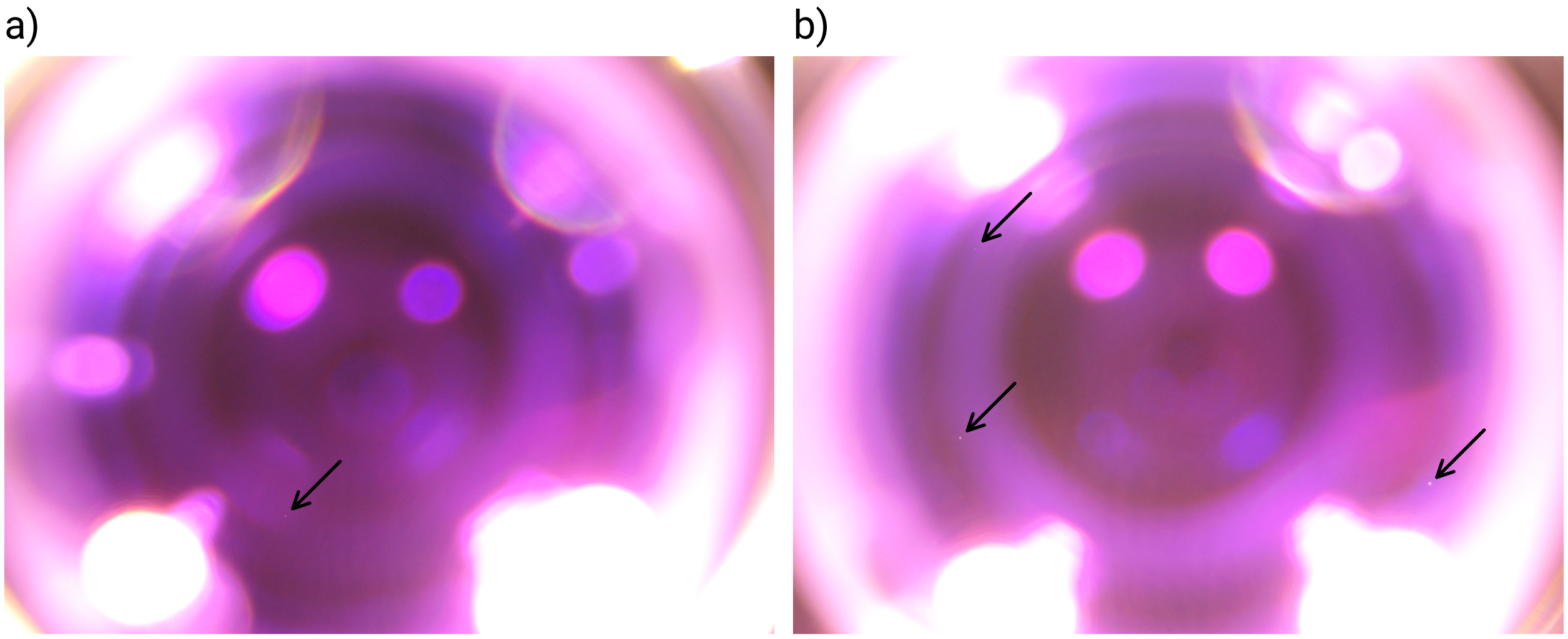

Figure 5.8: Darkfield images of center of the CC surface of the two mirrors installed in our cavity. All discernible scatterers are annotated with arrows. The field of view of both images is 2.8 mm × 2.1 mm. **a)** The output mirror, serial number 3.3.7. **b)** The input mirror, serial number 3.4.4.

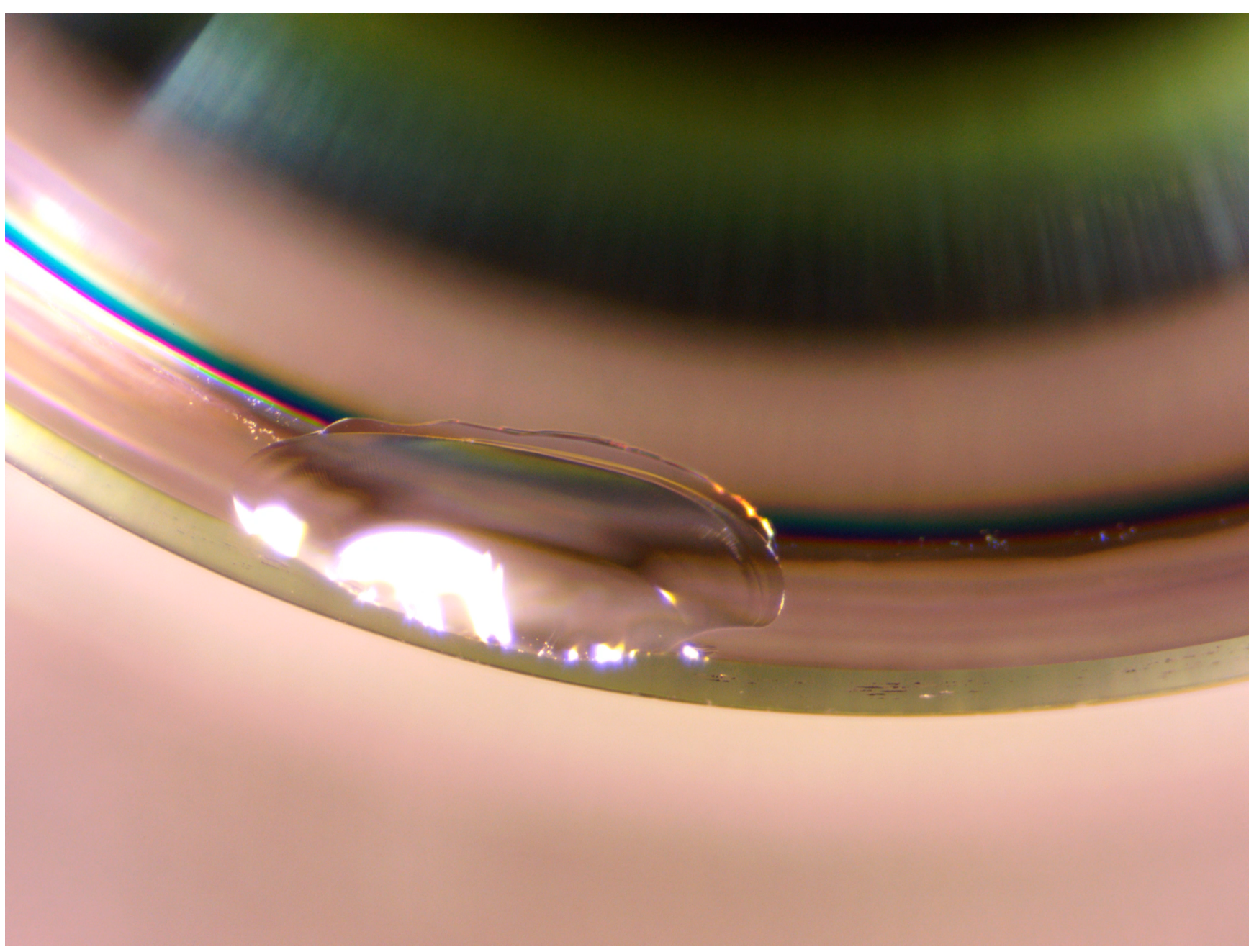

Figure 5.9: Darkfield image of a chip in the chamfer of the cavity input mirror, serial number 3.4.4. The field of view is 2.8 mm × 2.1 mm.

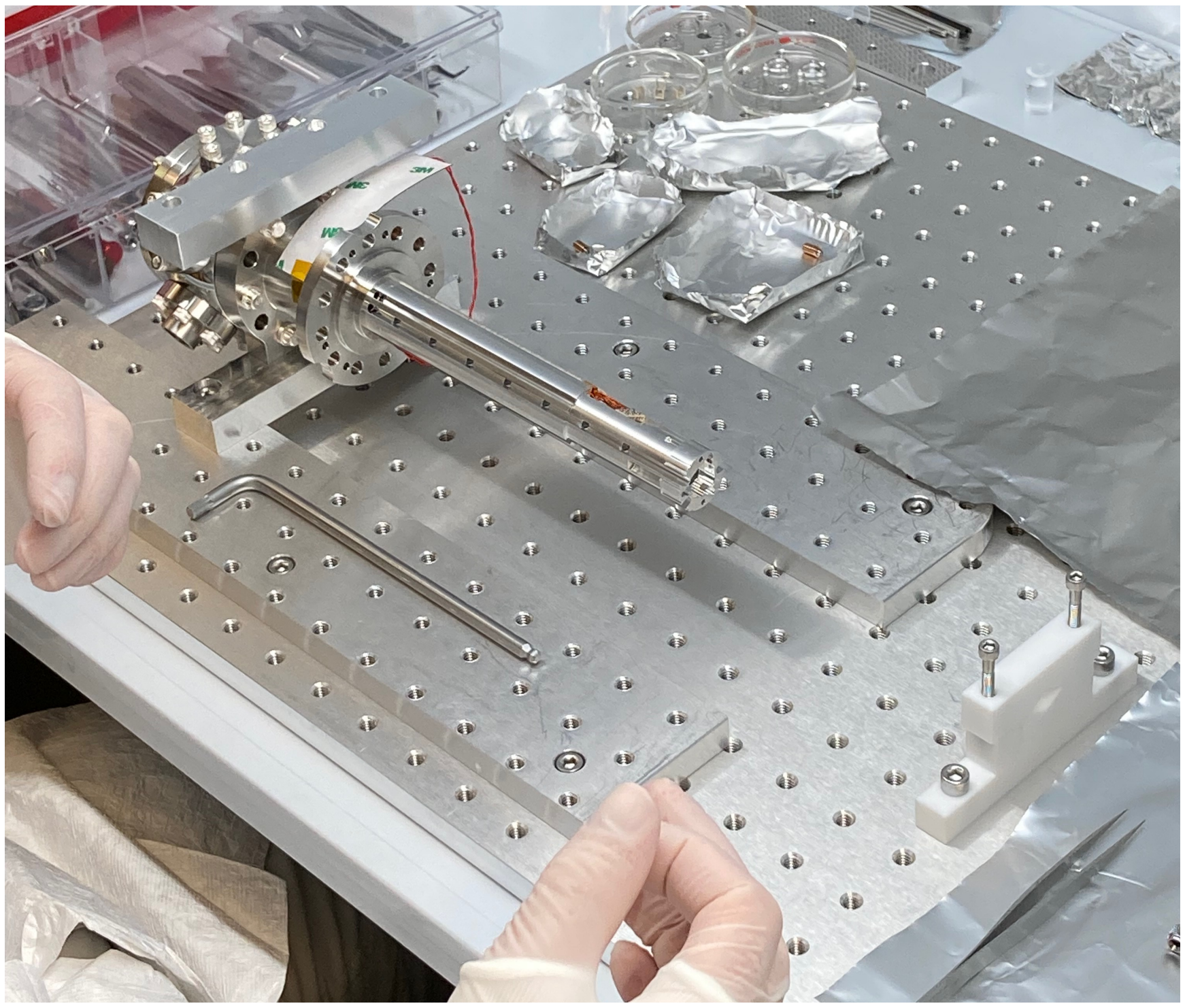

Figure 5.10: The support arm, vacuum flange adapter, and vacuum flanges being assembled while mounted to the optical breadboard by an assembly jig.

We started by assembling the support arm and vacuum flanges around the flange adapter, and connecting electrical feedthroughs to the wire bundles running to the support arm terminal blocks. Each end of the shielding on the wire bundles was wrapped with polyimide tape to prevent fraying. Feedthrough connector pins were crimped to the wires using a crimping tool (Daniels Manufacturing Corporation K155, paired with Daniels Manufacturing Corporation AFM8, sold as Accu-Glass Products, Inc. 100190). We could not degrease this tool so instead we tried to clean the crimping contact points with IPA as best as possible to avoid cross-contamination.

Assembly of the cavity mount then began. Sapphire jewels were epoxied to the piezo ends by depositing a small spot of epoxy (Loctite Hysol 1C) on the end of the piezo and positioning the jewels by eye. This is the only part of the cavity mount where epoxy is used. A photo of a completed piezo assembly with jewels is shown in figure 5.11. The piezos and micrometer screws were then installed into the cavity mount.

Another jig, made from polytetrafluoroethylene (PTFE) was used to gently hold the cavity mount in place while the mirrors were being inserted into their sockets (see figure 5.12). Inserting the mirrors was the most complicated part of the assembly process. The cavity mount was held with its optical axis horizontal so that when the second mirror was inserted there would be no chance of debris falling down onto the CC surface of the first mirror. We experienced this issue with a previous cavity, though our procedure then also likely led to more environmental contamination to begin with. Each mirror was inserted by first gripping its outer diameter with tweezers and slotting the mirror into its socket. The mirror was then pushed the rest of the way into the socket by pushing on its CX surface with a small metal rod with an annular PTFE tip, which had a 3 mm inner diameter and 5 mm outer diameter. This kept contact point of the rod tip away from the optical axis thereby preventing damage to the relevant part of the CX surface, while still allowing the mirror retention spacer and flexure (see section 4.3) to be slid down along the length of the rod so that the retention flexure could be installed while the mirror was being held in place by the rod. This process was finicky, and required more than two hands. I would recommend designing an additional jig that is inserted along the axis of the electron beam hole and blocks any contaminants that fall from one mirror to another. This would allow the mirrors to be installed with the cavity mounted vertically and the mirror CC surfaces facing down.

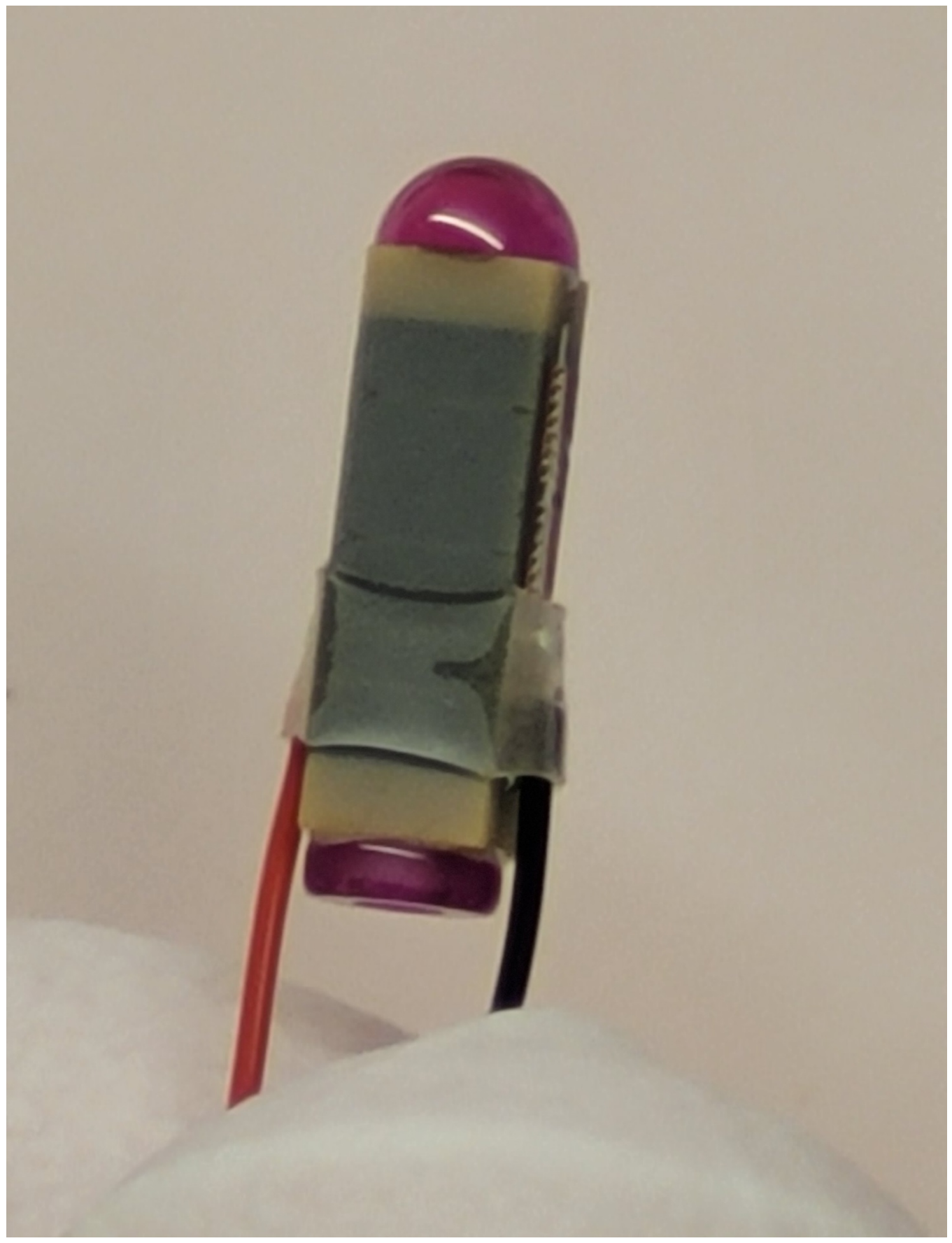

Figure 5.11: A cavity alignment piezo with sapphire jewels epoxied in position.

The cavity mount head cap (see section 4.7) was then installed on the cavity mount, and the cavity mount attached to the end of the support arm. The piezo wires were then attached.

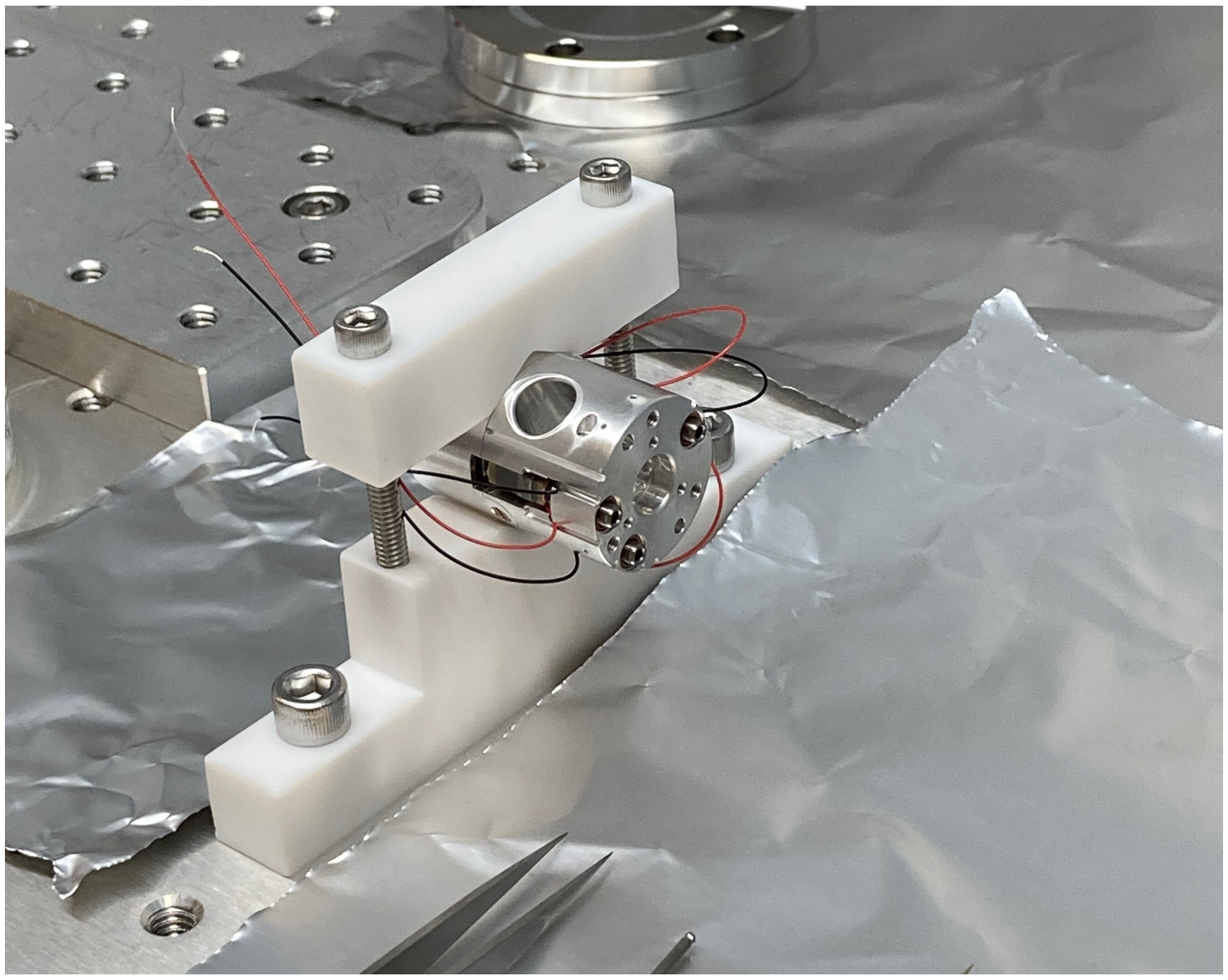

Figure 5.12: The cavity mount is held in a polytetrafluoroethylene jig for the mirror insertion procedure. Here, the cavity mount positioned for the insertion of the input mirror. The input mirror socket is visible on the right face of the cavity mount.

A thermistor was taped (using polyimide tape) to the outer diameter of the cavity mount. Another was mounted similarly on the outer diameter of the support arm near where it connected to the adapter flange. The 0.3 mm diameter of the thermistors made them easy to install anywhere, but also meant that they came attached to two extremely small #44AWG wires bundled together in a polyimide insulated cable. To strip the insulation off this cable we briefly exposed the cable to a butane flame until the insulation caught fire and burnt off. The charred insulation was then cleaned off through a combination of mechanical agitation and sonication using IPA as a solvent. The two stripped ends of the wires could then be separated. The wires were too small in diameter to be pinched by the #000-120 screws in the terminal block, so we crimped each wire end inside of a small piece of UHV vacuum foil. This effectively increased the diameter of the end of the wire and allowed it to be retained by the terminal block screw.

We avoided using coated tools wherever possible, since they usually generate particulate contamination. Wera Werkzeuge GmbH makes an excellent set of stainless steel hex L-keys in both imperial (05022721001) and metric (05022720001) sizes.

A photo of the assembled cavity, support arm, and vacuum flanges is shown in figure 5.13.

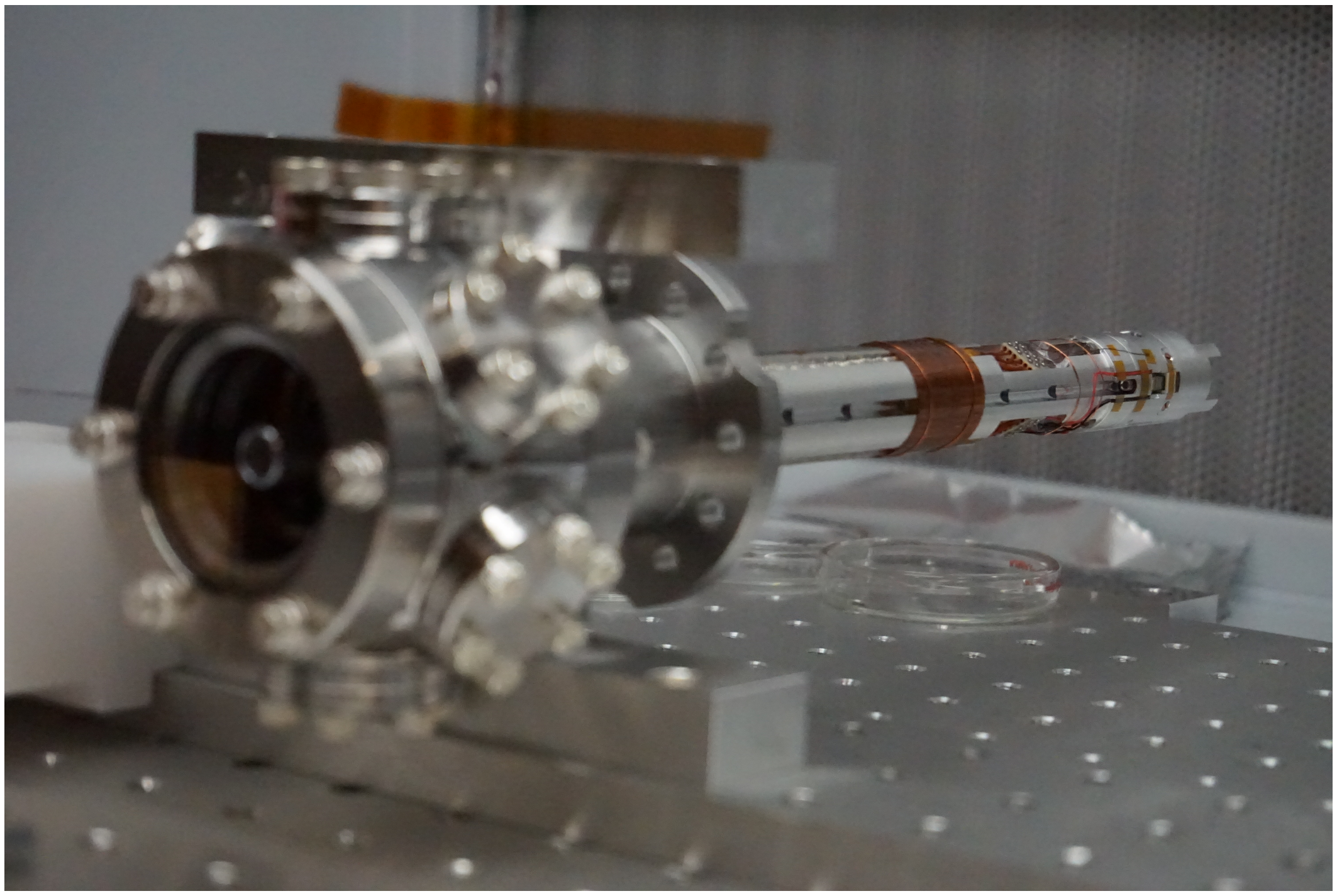

Figure 5.13: Fully assembled cavity, support arm, and vacuum flanges.

## 5.7 Testing

A small vacuum chamber designed for testing and transporting the cavity was then brought into the flow hood. The completed cavity assembly was slid into this chamber and the mating KF flange connected.

At this point, the cavity assembly was enclosed in a vacuum-tight container with ULPA-filtered air from the flow hood. Under this condition, the entire vacuum chamber was carried from the flow hood to the optical testing area. There, the test chamber was connected to the rest of the vacuum system, and the gate valve on the test chamber the between the two was opened. The chamber was pumped down, and the input and output optical breadboards were installed and aligned to the cavity assembly. The cavity was then aligned and tested. That process is discussed in chapter 7.

## 5.8 Installation

After initial alignment and testing, the cavity was installed in the TEM. A jig for inserting the cavity assembly was mounted to the TEM column. It consists of a platform with two slots which run parallel to the optical axis of the cavity assembly. A PTFE adapter connects to the cavity assembly via the vacuum flange adapter and slides in these slots. This allows the cavity to be slid smoothly into the TEM without bumping into the sides of the vacuum port. A photo of the cavity assembly and jig taken just before inserting the cavity into the TEM is shown in figure 5.14. The jig is rudimentary, and still requires some manual positioning of the cavity assembly to properly mate it with the port aligner flange. A photo of the author installing the cavity assembly using the jig is shown in figure 5.15.

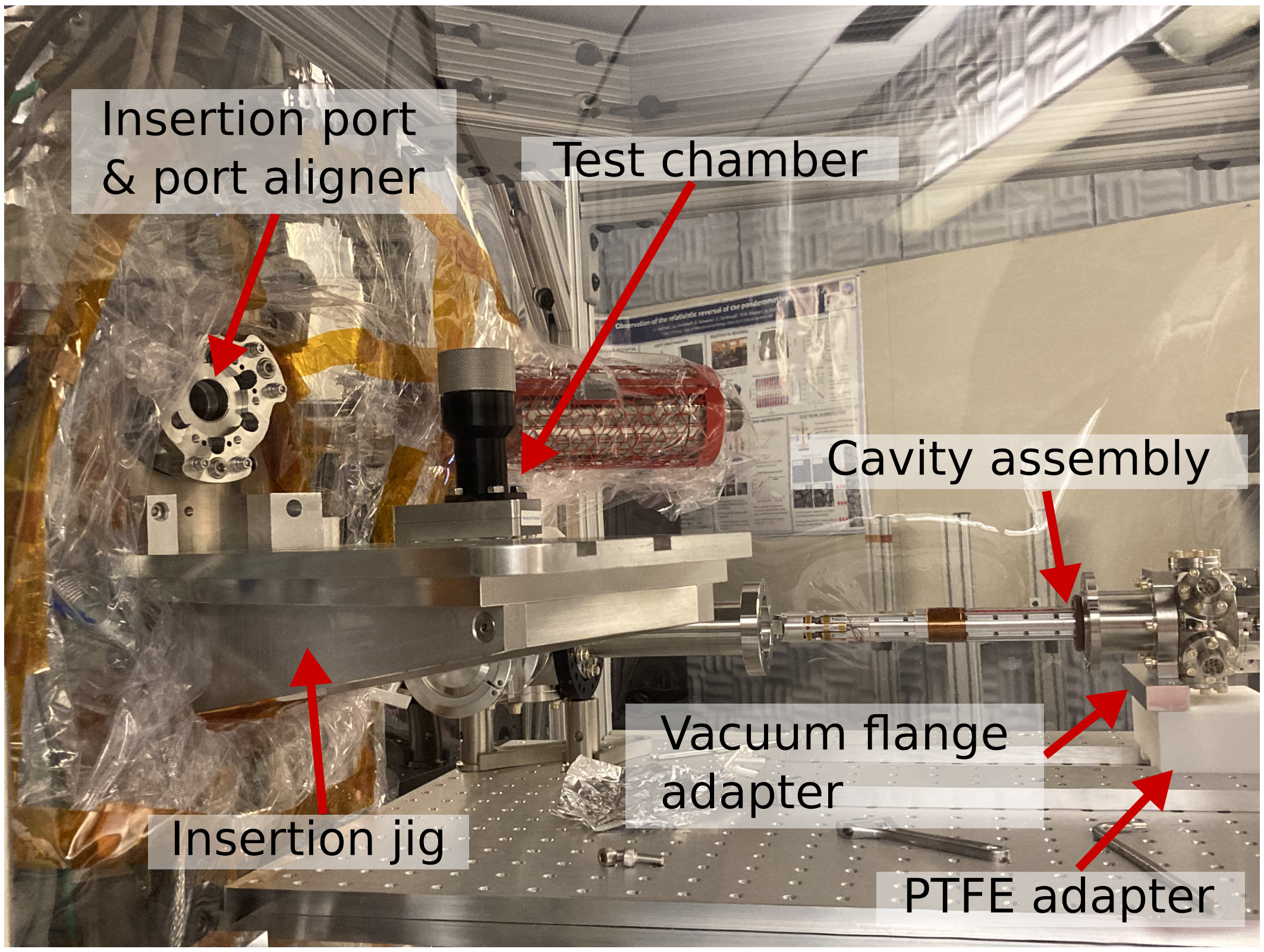

Figure 5.14: Cavity assembly insertion jig. A PTFE adapter block attaches to the cavity assembly and is slid along slots in the insertion jig platform.

To preserve a clean environment for the cavity mirrors during installation, we built a ULPA-filtered clean enclosure which was then mounted just next to the TEM column. This provided a volume of clean air enclosing the vacuum flange on the TEM column where the cavity was to be installed. The enclosure frame was made of T-slotted aluminum extrusion, with clear vinyl panels. A hole was cut in one of the vinyl panels to provide access to the insertion port on the TEM, and the edges of the hole were sealed to the microscope with antistatic stretch wrap (McMaster-Carr Supply Company 1899T75) and polyimide tape. A $2' \times 2'$ ULPA fan filter unit (Terra Universal, Inc. 6601-22-U) mounted on the top provided the clean airflow. A photo of this system is show in figure 5.16 (it is also visible in figures 5.14 and 5.15).

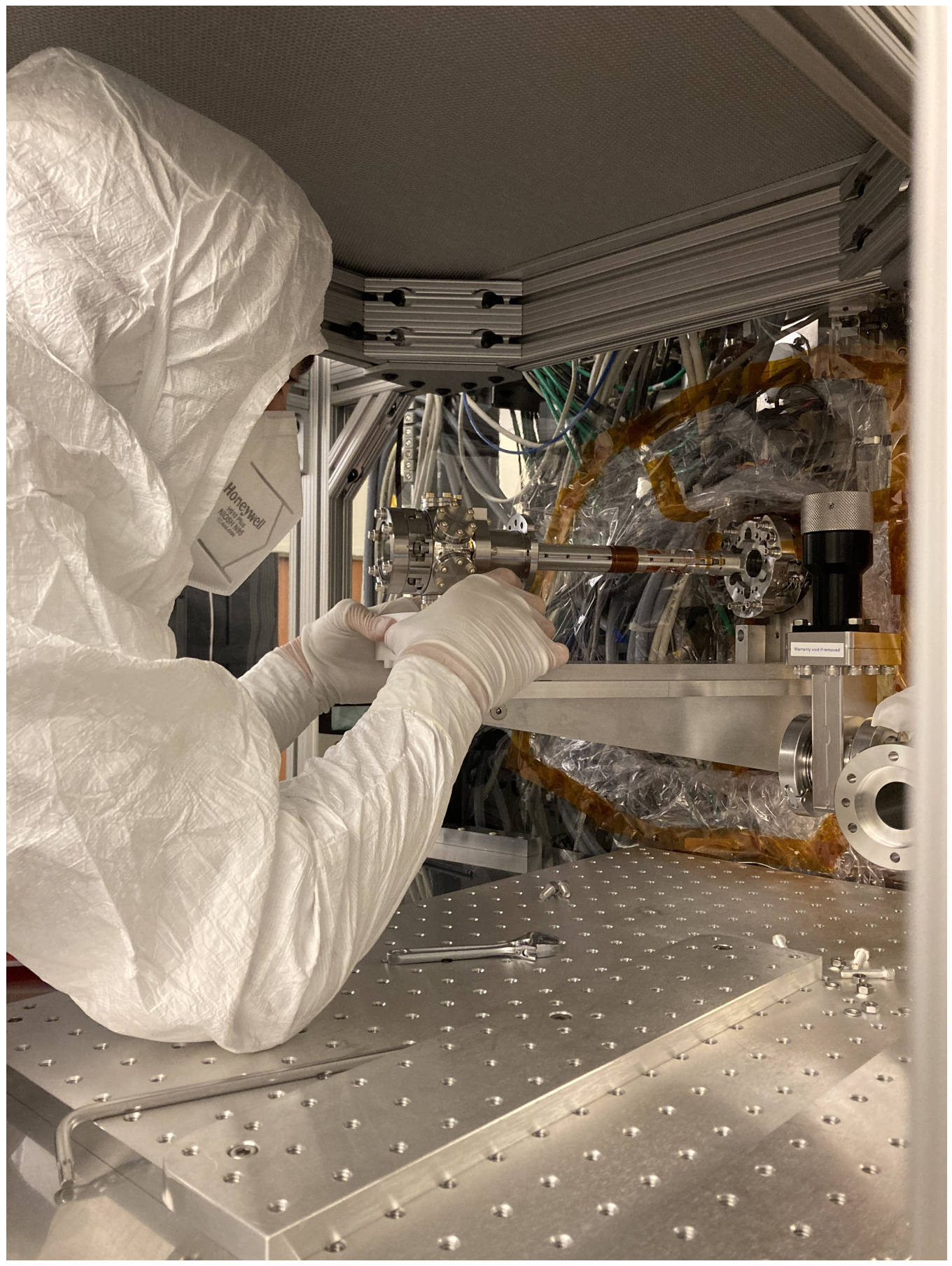

Figure 5.15: Sliding the cavity assembly and PTFE adapter along the insertion jig.

With the clean enclosure in place, the cavity assembly could be installed. First, the TEM column was vented (using the TEM software's standard venting procedure) with filtered nitrogen. While maintaining a continual flow of nitrogen through the TEM and out of the open vacuum port, the port aligner (see section 4.9) was installed. Similarly, the cavity test chamber was vented with filtered nitrogen. Its gate valve was then closed and the test chamber disconnected from the rest of the test vacuum system. The test chamber with cavity assembly was then carried to the clean enclosure in the TEM lab, just down the hall from the assembly and testing laboratory. Once there, the cavity assembly was removed from the test chamber, mounted to the insertion jig, and slid into place. The flange was screwed together, and the TEM vacuum was pumped down using the TEM software's standard sequence. The clean enclosure and insertion jig were then removed, and the external optical breadboards moved from the test setup onto the TEM. The external optical breadboard hardware is discussed in section 6.5. We found that it was usually not necessary to realign the cavity alignment micrometer screws after the installation procedure. Uninstalling the cavity followed the same steps in reverse.

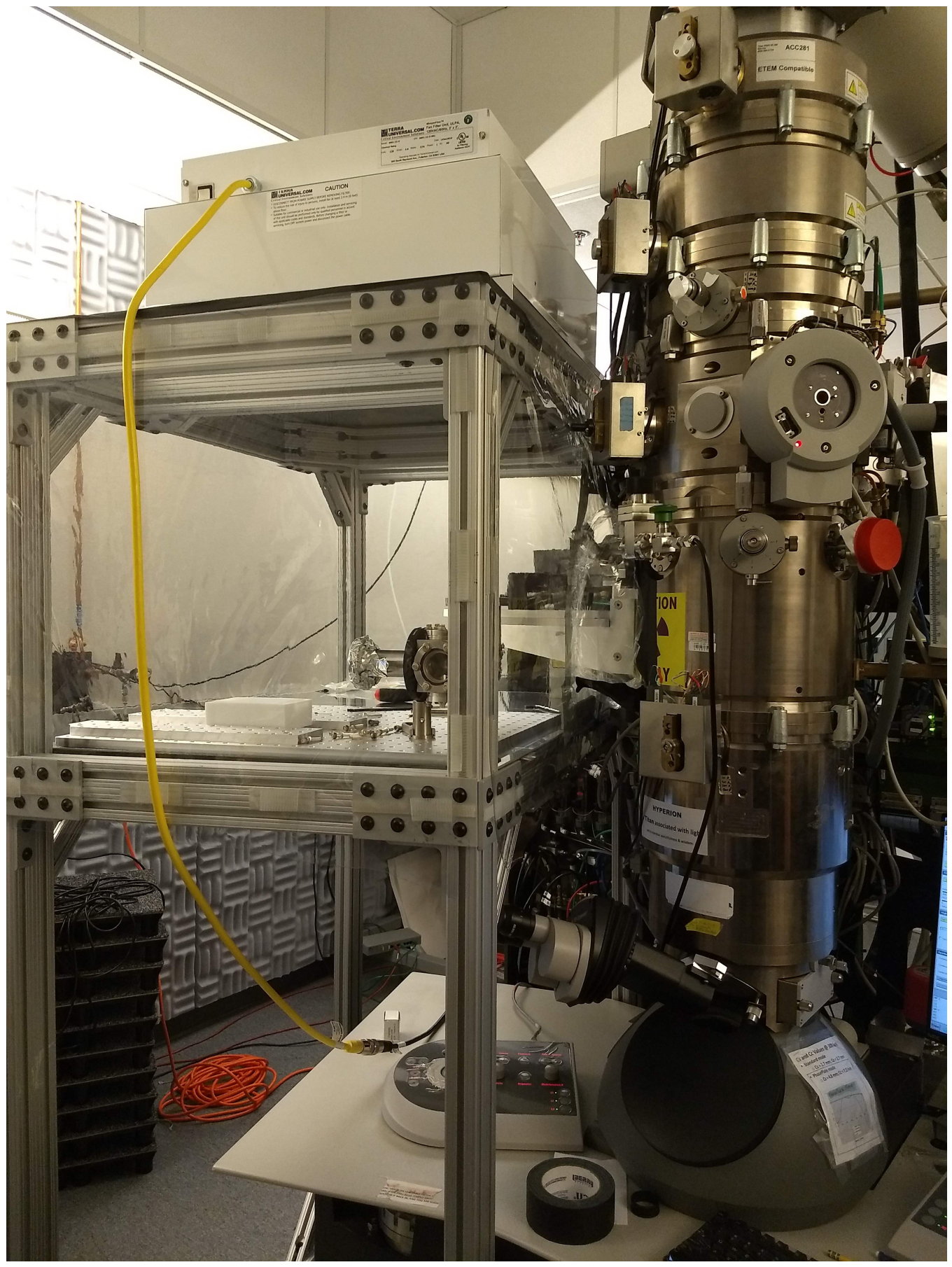

Figure 5.16: The ultra-low particulate air-filtered clean enclosure (left) mounted next to the transmission electron microscope column (right), ready for installation of the cavity assembly.

## 5.9 Operation in the transmission electron microscope

At start of the project we were concerned that the vacuum environment of the TEM might not be clean enough to maintain the cleanliness of our mirrors. This is partly because previous (non-laser) phase plate experiments had been done using this TEM, where devices were inserted in the same port we used for our cavity. These devices were not as dust-sensitive as our cavity mirrors and so their installation procedures may have introduced contaminants.

Also, the TEM is designed to have samples inserted into its vacuum system on a regular basis. These samples often intentionally or unintentionally carry particulate matter.

However, after the roughly 900 days that the current generation of LPPs have been installed in our TEM we have not seen any degradation of the mirrors (cavity operating time statistics are shown in section 7.5). A previous design which used epoxy to hold the mirrors in place, vacuum oil to lubricate sliding parts, and less careful contamination control procedures suffered an unexplained loss of cavity finesse while operating in the TEM. We do not know the cause, but suspect it was due to contamination from the oil or epoxy.

We do not change how we use the TEM to accommodate the LPP, other than the special TEM alignment procedures described in section 10.3. On one occasion, the microscope column was accidentally vented (through its filtered nitrogen venting line) while the cavity was operating at a circulating power of 120 kW; the cavity was undamaged. Several times, small volumes of unfiltered room air have been introduced the the TEM vacuum system due to faulty sample holder seals or user error. These events show that the cavity is at least not infinitely delicate. Determining how robust it really is will take more time, and more mistakes.

# Chapter 6

# Apparatus

To function as a laser phase plate (LPP), light must be sent in to the cavity such that its amplitude can be resonantly enhanced. This requires a laser system, as well as associated electro-optical feedback systems which keep the laser frequency on-resonance with the cavity. Additional feedback systems are required to maintain the alignment of the cavity. Other electro-optical systems are used as diagnostic tools to evaluate the performance of the cavity and troubleshoot potential problems. This chapter describes these systems as we have implemented them on our prototype LPP. The performance of the LPP using these systems (with the cavity hardware described in chapters 3 and 4) is described in chapter 7.

Figure 6.1 shows a schematic of the entire LPP system, including both optical and electrical components. The position of some optical components have been adjusted for aesthetic purposes, but the topology of the layout is the same as in the actual system, and all optics have been included, other than alignment irises and absorptive neutral densities filters which are used on some cameras and photodetectors to reduce the beam power. **Figure 6.1 will be heavily referenced throughout this chapter. For brevity, I will refer to component abbreviations shown in this figure without explicitly referencing the figure.**

Tables 6.1 and 6.2 list the manufacturers and product numbers for the commercial off-the-shelf optical and electrical components (respectively) shown in figure 6.1. For brevity I will not list this information in the text when each component is referenced.

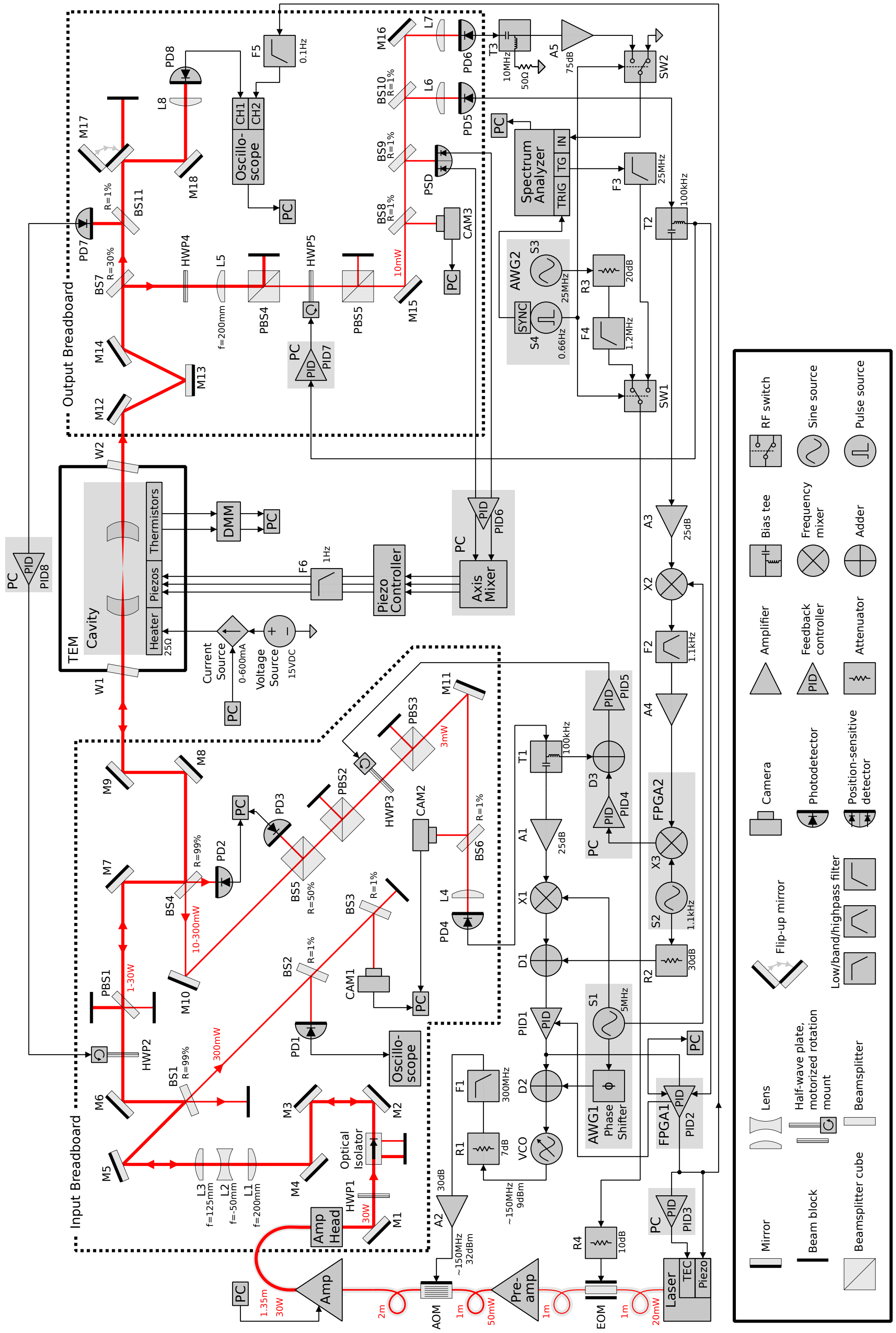

Figure 6.1: Schematic of the laser phase plate apparatus, including both optical and electrical components.

| | |
|---|---|
| Laser | NKT Photonics Koheras ADJUSTIK Y10 |
| Preamp | Azurlight Systems ALS-IR-1064-0.05-A-SF |
| Amp, Amp Head | Azurlight Systems ALS-IR-1064-50-A-SF |
| EOM | EOSPACE Inc. PM-0S5-10-PFA-PFA-106-UL |
| AOM | G&H SFO3452-T-M150-0.4C2G-3-F2P-01 |
| M1-M18 | Thorlabs, Inc. BB1-E03. M17 is mounted on a manual flip-up mount. |
| HWP1-HWP5 | Thorlabs, Inc. WPH05M-1064. PC-controlled rotation mounts on HWP2, HWP3, and HWP5 are Thorlabs, Inc. K10CR1. |
| Optical Isolator | Thorlabs, Inc. IO-5-1064-VHP |
| L1 | Newport Corporation SPX028AR.1 |
| L2 | Newport Corporation SBC034AR.33 |
| L3 | Newport Corporation SPX023AR.1 |
| L5 | Newport Corporation SPX028AR.1 |
| BS1 | Edmund Optics 87-381 |
| BS2, BS3, BS6, BS8, BS9, BS10 | Thorlabs, Inc. BSF10-C |
| BS4 | Thorlabs, Inc. BB1-E03P |
| BS5 | Thorlabs, Inc. BS014 |
| BS7 | Thorlabs, Inc. BSS11 |
| BS11 | Thorlabs, Inc. PS810-C |
| PBS1 | Thorlabs, Inc. PBSW-1064 |
| PBS2-PBS5 | Thorlabs, Inc. PBS25-1064-HP |
| W1, W2 | Manufacturing Precision Feedthrough Products A4545-1-CF |
| CAM1-CAM3 | Thorlabs, Inc. DCC3240M |
| PD2, PD3, PD7 | Thorlabs, Inc. S121C |
| PD4, PD5 | Hamamatsu G12180-series photodiode with transimpedance amplifier based on Texas Instruments OPA843 op-amp |
| PD6 | Thorlabs, Inc. FGA01, reverse-biased with $50\,\Omega$ load. |
| PD8 | Thorlabs, Inc. FGA10, reverse-biased with $50\,\Omega$ load. |
| PSD | Thorlabs, Inc. PDP90A |

Table 6.1: Manufacturer and model numbers for the optical components shown in figure 6.1.

| | |
|---|---|
| PC | Personal computer running Windows operating system, with a Texas Instruments analog-to-digital and digital-to-analog interface with bandwidth $< 100\,\text{Hz}$. |
| AWG1<br>AWG2 | Rigol DG832<br>Rigol DG1022Z |
| FPGA1, FPGA2 | Red Pitaya STEMlab 125-14 |
| Spectrum Analyzer | Rigol DSA815-TG |
| Piezo Controller | Thorlabs, Inc. MDT693B |
| DMM | Keithley Instruments, Inc. 2701, with input multiplexer |
| Oscilloscope | Rigol DS1054Z |
| T1, T2<br>T3 | Mini-Circuits ZFBT-4R2GW+<br>Mini-Circuits ZFBT-4R2G+ |
| A1, A3<br>A2<br>A5 | Mini-Circuits ZFL-500LN+<br>Mini-Circuits ZHL-1-2W+<br>Mini-Circuits ZX60-P103LN+, and two Mini-Circuits ZFL-500LN+ in series |
| X1, X2 | Mini-Circuits ZRPD-1+ |
| R1<br>R2<br>R3<br>R4 | Mini-Circuits HAT-4+ and Mini-Circuits HAT-3+ in series<br>Mini-Circuits VAT-30W2+<br>Mini-Circuits VAT-20W2+<br>Mini-Circuits VAT-10W2+ |
| F1<br>F3<br>F4 | Mini-Circuits BLP-300+<br>Mini-Circuits BHP-25+<br>Mini-Circuits ZFHP-1R2-S+ |
| SW1, SW2 | Mini-Circuits ZASWA-2-50DRA+ |
| VCO | Mini-Circuits ZX95-200A+ |

Table 6.2: Manufacturer and model numbers for the electrical components shown in figure 6.1.

## 6.1 Optics

The optical system is conceptually split between those on the input side of the cavity and those on the output side. The input optics are primarily responsible for shaping the input laser beam to best overlap with the cavity mode, and for detecting the optical signals necessary for keeping the laser frequency on-resonance with the cavity. The output optics are used for feedback and diagnostic systems.

### 6.1.1 Input

The input optics laser beam path starts at the laser ("Laser", lower left corner of figure 6.1). The laser is a fiber laser operating at a wavelength of 1064 nm. It outputs approximately 20 mW of optical power. We chose this laser for several reasons. First, its wavelength can be continuously tuned ("mode-hop-free") through a range of $\sim \pm 300$ pm. When we were first designing the system, we were unsure how much the cavity would thermally expand due to heating from light scattered and absorbed by the cavity mirrors, and for simplicity we did not want to commit to stabilizing the cavity temperature. The large wavelength tuning range meant that we could keep the laser on-resonance with the cavity even as it underwent relatively large ($\sim 1\,\mu$m) length changes due to thermal expansion. Second, the laser has relatively good frequency stability, with a free-running linewidth of $\sim 3$ kHz. In conjunction with the laser frequency stabilization feedback system, this helps keep the laser frequency on-resonance with the cavity. Third, the system is compact and natively fiberized. This reduced the complexity of our physical installation of the system in the TEM room (see section 6.5), and made the system more robust than it would have been if the laser were mounted on an optical breadboard and the light then coupled into an optical fiber.

The light from the laser is sent to a fiber electro-optic modulator (EOM), through a distance of $\sim 1$ m of polarization-maintaining optical fiber. The EOM is used to phase modulate the laser light as part of a diagnostic system described in section 6.3.2.

The output light from the EOM is then amplified in a fiber amplifier (Preamp) which increases the power to $\sim 50$ mW. An acousto-optical modulator (AOM) up-shifts the laser frequency by between 145 MHz and 155 MHz. This control is used in part of the laser frequency feedback system. Frequency-modulated 5 MHz sidebands of this carrier frequency serve to phase modulate the light at 5 MHz. These sidebands are used in the laser frequency feedback system described in section 6.2.3. Polarization-maintaining single-mode optical fiber is used to conduct light between components.

A second larger fiber amplifier (Amp) then increase the laser power to up to 50 W. We do not need this much power, so to prolong the life of the amplifier we typically operate it at 30 W of output power. The output power of the amplifier is controllable through a voltage set by a personal computer and digital-to-analog converter (PC). The amplifier output is carried through a 1.35 m polarization-maintaining fiber until it is coupled into free space inside of the amplifier head unit (Amp Head). The unit contains at least one coupling lens, an optical isolator (to prevent back-reflections from entering the amplification stage), and at least one photodetector used to set the power or possibly actuate an interlock (I'm not sure of the exact internal design). While the amplifier body (Amp) is mounted on a 19" rack next to the TEM, the amplifier head is mounted on an optical breadboard which is mounted directly to the TEM column (see section 6.5 for details). We refer to this breadboard as the "input breadboard". The optical components placed on this breadboard are shown in figure 6.1 enclosed by the dashed line labeled "Input breadboard".

The output of the amplifier head is a free-space laser beam with a diameter of roughly 1 mm and vertical polarization (in/out of the page in figure 6.1). A half-wave plate (HWP1) rotates the polarization by 45° so that it emerges from the subsequent optical isolator (Optical Isolator) with horizontal polarization. The optical isolator is required because the amplifier head is only specified to handle a retroreflected beam with $\leq$ 10% of its output power, and the cavity reflects up to 100% of the input beam. We set the optical isolator to output horizontally polarized light so that we do not need an additional half-wave plate later in the beam path, since the LPP is conventionally used with horizontally polarized light (polarization perpendicular to electron beam axis).

A three lens Galilean telescope (L1, L2, L3) is used to change the size and divergence of the input beam to best match that of the cavity mode. The three lens design allows for magnifications around unity (as is required in our system) without requiring a beam focus (Keplerian telescope) which can cause fluctuating wavefront distortions due to laser heating of air. L1 and L3 are mounted in longitudinal translation mounts (Thorlabs, Inc. SM2NR1) such that the alignment of the telescope can be easily adjusted. The lens substrate material is chosen to be UV fused silica to minimize thermal lensing.

A beamsplitter (BS1) with 99% reflectivity allows a small fraction of the beam to be sent to two diagnostic systems. A photodetector (PD1) is used to measure fluctuations in the input laser power (see section 6.3.6). A camera (CAM1) images the spatial profile of the input beam to check that it is nominally the appropriate size for coupling to the cavity (1 mm diameter beam at the CC surface of the cavity input mirror).

After reflecting off of BS1, the beam continues through a half-wave plate (HWP2) and a polarizing beamsplitter (PBS1). HWP2 is mounted in a motorized rotation mount which is controlled by the PC. PBS1 reflects vertically polarized light. These two optics enable computer control of the laser power sent to the cavity input. We originally controlled the input laser power by adjusting the output power of the laser amplifier. However, we found that the pointing and divergence of the input beam changed with input power, presumably due to thermal lensing or thermal misalignment in the amplifier head or subsequent optics. These changes were large enough to substantially reduce the input coupling efficiency to the cavity. The polarization-based attenuator using HWP2 and PBS1 is much more stable. The rotation mount for HWP2 is mounted to the optical breadboard using a rubber post (McMaster-Carr 94955K31, or similar), since we found that when using a standard steel post the vibrations of the mount's stepper motors would cause the cavity's length to vibrate and disrupt the laser frequency feedback system.

Beamsplitter BS4 splits off 1% of the input light and sends it to a photodetector (PD2) which is calibrated to measure the input power sent to the cavity.

Mirrors M8 and M9 are mounted in 2-axis optomechanical mounts (Newport Corporation SU100-F2K) which are fitted with high-precision adjusters (Newport Corporation DS-4F) that are used to (manually) adjust the pointing of the input beam to optimize its overlap with the cavity mode. Without the high-precision actuators this would be a tedious and difficult task, as the coupling efficiency is sensitive to sub-milliradian changes in the input beam angle. Optimizing the input coupling efficiency requires these mirrors to be adjusted while the laser frequency is locked to the cavity resonance and the input power is nominal for operating the LPP ($\sim$ 10 W). On the current prototype system we need to perform this alignment once every few months. We do not know what causes the misalignments that require these realignments. To reduce the risk of a mistake causing damage to the laser optics or laser operator, we plan to use computer-controlled motorized mounts in a future version of the LPP.

The input light enters the vacuum of the TEM column through the vacuum window described in section 4.9 (W1). Some fraction of the light (essentially all when off-resonance) then reflects off of the CC surface of the input mirror of the cavity. This light travels backwards along the input beam path. When it encounters BS4 again, 1% is transmitted where it is analyzed. First, 50% of this light is split off by beamsplitter BS5 and its power measured by photodetector PD3. This photodetector provides a calibrated measurement of the power reflected off of the cavity. When calibrating this photodetector, we set the laser frequency off-resonant with the cavity, measure the power of the beam sent in the forward direction to the cavity input (after M9), and then assume that all of that power is reflected from the cavity. This overestimates the reflected power by likely not more than a few percent, due to losses in the beam path from M9 to the cavity input mirror and back to BS4. The ratio of the input power reading to that given by PD3 establishes the calibration.

A polarizing beamsplitter (PSB2) cleans up the polarization of the beam so that the polarization-based attenuator consisting of motorized half-wave plate HWP3 and polarizing beamsplitter PBS3 can achieve high enough attenuation to leave just 3 mW at its output. Similarly to HWP2, the rotation mount for HWP3 is mounted to the optical breadboard using a rubber post. The attenuator is used to stabilize the amount of power sent to photodetector PD4, which is used in the laser frequency feedback system to detect the phase of the 5 MHz modulation of the reflected light. Lens L4 has a focal length of roughly 50 mm and serves to focus the beam onto the active area of PD4. Camera CAM2 is used for diagnostic purposes to examine the spatial profile of the reflected beam.

The cavity reflected light which reflects off of BS4 travels all the way back along the input beam path to the optical isolator where it is diverted into a beam block.

Figure 6.2 shows a top-down view of the input breadboard, with components (other than standard mirrors) labeled with the abbreviations used in figure 6.1.

### 6.1.2 Output

The laser beam exits the output cavity mirror and moves from vacuum to air in the vacuum window W2, which is mounted to a flange on the outer diameter of the TEM column. The remaining output optics are all mounted on the "output breadboard" which is mounted to the TEM column. Mirrors M12-M14 are used to align the beam into the remaining optics, and are useful when moving the breadboard between our optical test setup and the TEM.

70% of the beam is transmitted through the beamsplitter BS7. Roughly 1% of that light is then split off to photodetector PD7 using beamsplitter BS11. PD7 provides a calibrated measurement of the cavity output power.

A manual flip-up mirror (M17) can be positioned to send the beam to photodetector PD8, which is used in a diagnostic system to measure the reflectivity of the cavity mirrors (see section 6.3.1). Lens L8 has a focal length of $\sim 50\,\text{mm}$ and is used to focus the beam onto the active area of the PD8 photodiode. During normal operation of the LPP M17 is positioned such that it does not reflect the beam to PD8, and instead the beam encounters a beam block.

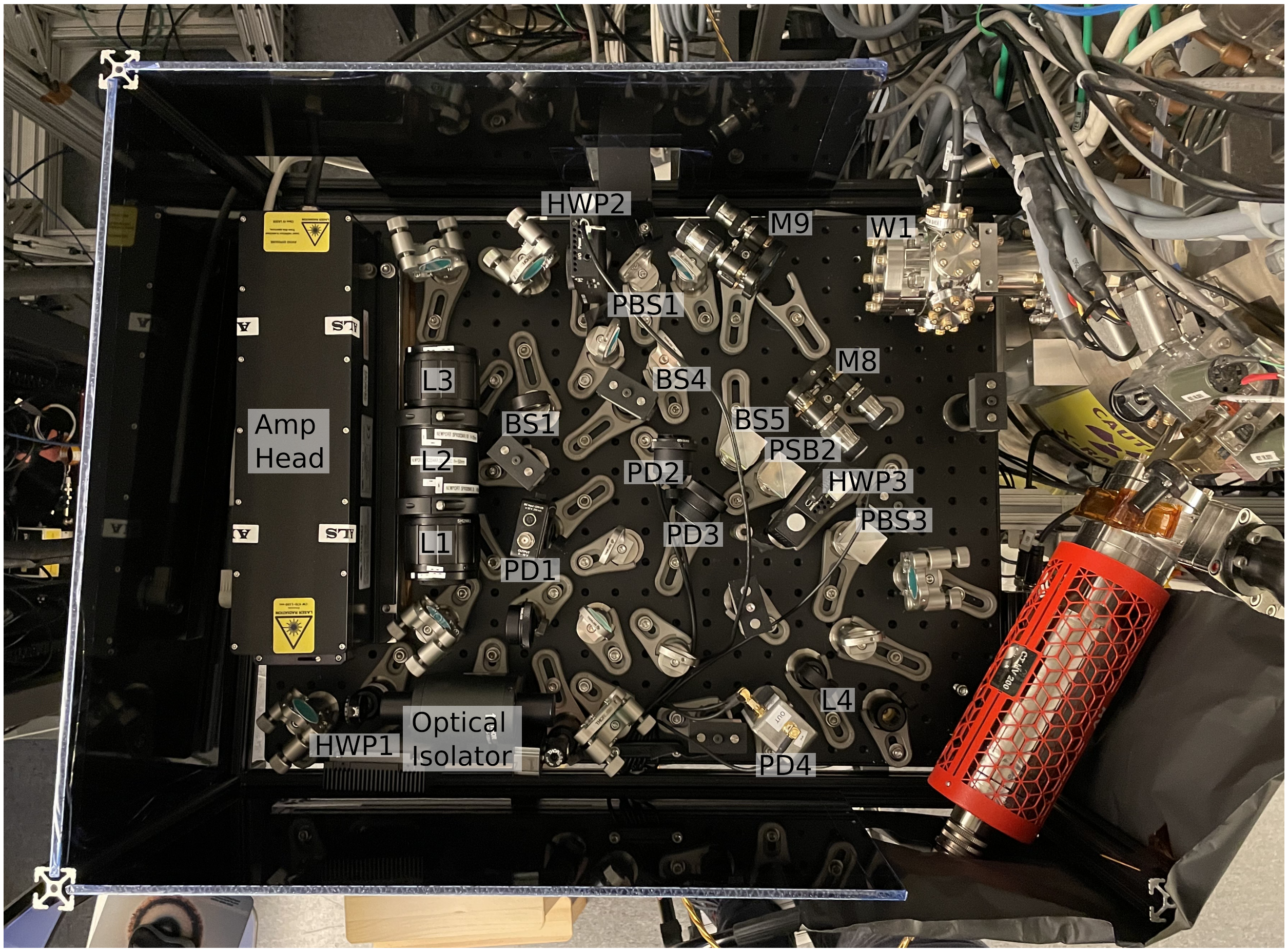

Figure 6.2: Top-down view of the input breadboard. Components are labeled with abbreviations corresponding to the schematic diagram shown in figure 6.1. The transmission electron microscope column is seen in the upper right corner of the image.

The light reflected from BS7 encounters a half-wave plate (HWP4) and polarizing beam-splitter (PBS5) which are used to both attenuate the beam power and clean up its polarization. HWP4 is mounted in a non-motorized rotation mount. The lens L5 mounted between them approximately images the CC surface of the output cavity mirror onto the position-sensitive detector (PSD). It is positioned between HWP4 and PBS5 because there were no other available locations which would satisfy the imaging condition with the available focal lengths of lenses.

The subsequent motorized half-wave plate (HWP5) and polarizing beamsplitter (PBS5) constitute another computer-controlled polarization-based attenuator, which serves to provide $\sim 10\,\mathrm{mW}$ of light to the remaining sensors. Like the other motorized waveplate mounts, the rotation mount for HWP5 is mounted to the optical breadboard using a rubber post.

A camera (CAM3) monitors the spatial profile of the output beam for diagnostic purposes.

The position-sensitive detector (PSD) measures the position of the beam and is used in a feedback system to maintain the alignment of the cavity (see section 6.2.5).

Photodetector PD5, like PD4, detects the 5 MHz modulation of the beam and is used to stabilize the gain of the laser frequency feedback system (see section 6.2.4). Lens L6 has a focal length of roughly 50 mm and serves to focus the beam onto the active area of PD5.

Photodetector PD6 is used to detect modulation of the beam induced by the EOM, which is used as part of a diagnostic system described in section 6.3.2.

Figure 6.3 shows a top-down view of the output breadboard, with components (other than standard mirrors) labeled with the abbreviations used in figure 6.1.

## 6.2 Feedback systems

The LPP relies on several feedback systems to stabilize the parameters of the intracavity laser beam such that the resulting electron beam phase shift profile is stable. Some of these systems operate independently of one another, while some are interdependent.

A feedback system controls a parameter of some process by taking the difference between the value of that parameter (called the “process variable”) and its desired value (called the “setpoint”) to calculate the “error signal”. From the error signal it then calculates a corresponding “control variable” which affects the process, thereby changing the process variable.

Often, the control variable is made to be a weighted sum of the error signal, time derivative of the error signal, and integral of the error signal up to the current time. Such a feedback system is called a “proportional-integral-derivative controller”, or “PID controller” for short. This calculation can be performed in analog or digital circuitry. All of the feedback systems we use are PID controllers, although some incorporate additional terms in the weighted sum, and some have additional functionality. The details of each feedback system used in the LPP are discussed in the following subsections.

### 6.2.1 Photodetector optical power

When operating the LPP it is necessary to first stabilize the laser frequency to the cavity resonance when the cavity input power is relatively low ($< 5\,\mathrm{W}$) and then increase the input power to the nominal operating condition ($\sim 10\,\mathrm{W}$). Above this threshold power, the laser frequency feedback system will not successfully engage. The reason for this is still under investigation, but may be due to the cavity length (resonance frequency) changing rapidly (on the order of the cavity ringdown time of $\sim 1\,\mu\mathrm{s}$) due to radiation force on the mirrors from the high circulating power. However, several photodetectors used in the system (PD4-PD6, PSD) need to be illuminated with a constant optical power to provide consistent amplitude electrical signals. Given a changing input power this requires that those detectors be placed behind variable optical attenuators which can maintain a constant output optical power. We make these variable optical attenuators using a half-wave plate followed by a polarizing beamsplitter. Control of the attenuation is provided by mounting the half-wave plate in a motorized computer-controlled rotation mount.

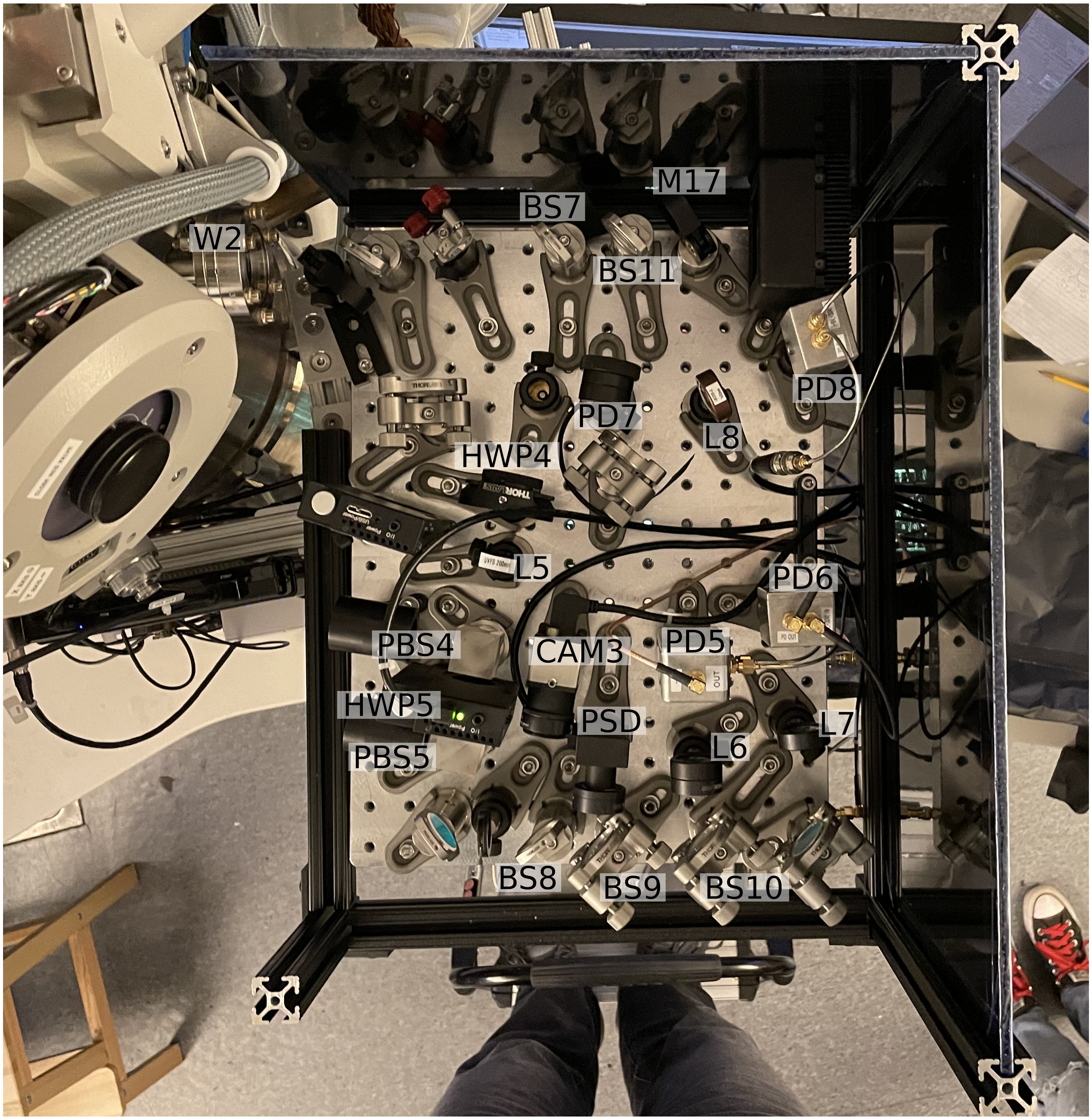

Figure 6.3: Top-down view of the output breadboard. Components are labeled with abbreviations corresponding to the schematic diagram shown in figure 6.1. The transmission electron microscope column is seen in the upper left corner of the image.

The associated (digital) feedback system reads the DC voltage from the photodetector as the process variable and sends this signal through a PID controller. The control variable output of the PID controller commands a rotation of the half-wave plate rotation mount such that the photodetector DC voltage is stabilized to its setpoint value.

On the input breadboard, the DC voltage is read from photodetector PD4 (through bias tee T1) and passed through PID controller PID5 which controls the rotation of half-wave plate HWP3. This stabilizes the optical power on photodetector PD4 and camera CAM2. The setpoint of PID controller PID5 is the control variable of another PID controller PID4, which is discussed in section 6.2.4.

On the output breadboard, the DC voltage is read from the photodetector PD5 (through bias tee T2) and passed through PID controller PID7 which controls the rotation of half-wave plate HWP5. This stabilizes the optical power on camera CAM3, position-sensitive detector PSD, and photodetectors PD5 and PD6. The setpoint of PID controller PID7 is provided by the user.

### 6.2.2 Circulating power

A similar system is used to stabilize the circulating power in the cavity. This is necessary because there are often slow drifts in the input laser power and cavity input coupling efficiency, and a stable circulating power is required to achieve a stable electron beam phase shift.

The circulating power measurement provided by photodetector PD7 is digitized and passed through PID controller PID8, which controls the rotation of the half-wave plate HWP2. This adjusts the cavity input power to stabilize the circulating power. The setpoint of PID controller PID8 is provided by the user.

### 6.2.3 Laser frequency

We stabilize the laser frequency to the cavity resonance using the Pound-Drever-Hall (PDH) method [76], [77]. A sine wave voltage source (S1) modulates the output frequency of a voltage-controlled oscillator (VCO) at 5 MHz (the signal is first passed through a phase shifter and adder D2). This generates sidebands on the VCO's output carrier frequency, which is adjustable from 145 MHz to 155 MHz. The VCO output is then attenuated from 9 dBm to 2 dBm (using attenuator R1) and low-pass filtered (using filter F1) to remove harmonics of the carrier frequency. The signal is then amplified to 32 dBm (using amplifier A2) and sent to the fiber acousto-optic modulator (AOM). The AOM phase modulates the laser beam at a frequency of 5 MHz due to the sidebands on the AOM drive signal. When this phase modulated light's carrier frequency is within the full-width half-maximum of the cavity resonance, the light reflected from the cavity becomes amplitude-modulated at 5 MHz such that the amplitude of the modulation quadrature in phase with the VCO modulation signal depends approximately linearly on the difference between the laser carrier frequency and cavity resonance ("detuning"). Photodetector PD4 measures this amplitude-modulated signal, which is then high-pass filtered by bias tee T1 and amplified by amplifier A1. The signal is then demodulated by mixing the signal with the VCO modulation signal from source S1 using mixer X1. A constant phase shift is applied to the VCO modulation signal so that the mixer demodulates the appropriate quadrature of the photodetector signal.

The output of mixer X1 constitutes the error signal of the laser frequency feedback system, as its voltage is proportional to the detuning. Typically, a low-pass filter is placed after the mixer to eliminate the summed frequency component of the mixer output (10 MHz in our case), and any residual input frequency components (5 MHz in our case). We omit this component to reduce group delay in the feedback path since "brickwall" low-pass filters which can efficiently filter out these higher frequencies necessarily incur relatively large group delays within their passband. For example, the 2.5 MHz filter Mini-Circuits BLP-2.5+ has over 300 ns of delay within the passband. This is important because delay in the feedback path limits the bandwidth of the feedback system, and we found that a higher feedback bandwidth resulted in fewer "unlocking" events in which the detuning would increase beyond the cavity linewidth and not return without human intervention. Some of these unlocking events are caused by mechanical agitation of the cavity (e.g. opening or closing the pneumatically actuated vacuum gate valves on the TEM, or a loud noise in the room where the cavity is installed), but others are not obviously attributable to this cause.

The error signal output of mixer X1 is added (in adder D1) to a small amplitude sinusoidal wave signal which is used as part of the PDH gain feedback system (see section 6.2.4). The resulting (modified) error signal is then processed by an analog feedback controller (PID1) which calculates a control variable which is the sum of proportional, integral, and double-integral terms. The double integral terms give a higher feedback gain at low frequencies, which we found to be helpful in reducing unlocking events for unknown reasons.

The output of the VCO ultimately drives the AOM and controls the laser frequency. While the AOM can rapidly change the laser frequency (changes in the laser frequency occur about 200 ns after changing the RF signal), it can only change it by roughly ±5 MHz before the optical output power of the AOM becomes too low to seed the subsequent fiber amplifier. This range is too small to track changes in the cavity length larger than $\sim \pm 400$ pm. To avoid having to stabilize the cavity length to this degree, we cascade the output of the feedback controller PID1 with two more feedback controllers which each control actuators with larger ranges (though lower bandwidths) than the AOM. In total, this results in a feedback system with approximately the bandwidth of the fastest controller and the range of the largest range controller.

Specifically, the control variable of controller PID1 is used as the error signal for the digital feedback controller PID2. This feedback controller acts to keep the control variable of PID1 within its operating range. The control variable of PID2 drives a piezoelectric actuator in the seed laser which adjusts the length of the seed laser fiber cavity, thereby controlling its frequency. The range of this actuator is considerably larger than that of the AOM at $\sim \pm 50\,\mathrm{MHz}$, though its bandwidth is substantially smaller at $\sim 35\,\mathrm{kHz}$. Its range can be made as large at 10 GHz at the expense of bandwidth, though we did not find this to be helpful. The controller PID2 consists of a Red Pitaya STEMlab 125-14 (FPGA1), which incorporates a microcontroller, field-programmable gate array (FPGA), and analog-to-digital/digital-to-analog converters. The FPGA is programmed with custom software which implements a PID controller. It also implements a scheme for automatically sweeping the laser frequency to find the cavity resonance, using a triangle wave output signal with a progressively increasing amplitude. When a cavity resonance is encountered, light transmits through the cavity, which generates a signal at photodetector PD5. The low-frequency component ($< 100\,\mathrm{kHz}$) of this signal is detected by FPGA1 through bias tee T2. If the signal voltage rises above a user-specified threshold, FPGA1 switches its output from the triangle wave signal to the PID controller and feedback commences. Conversely, if an unlocking event occurs and the signal from PD5 falls below the threshold, FPGA1 switches its output from the PID controller to the triangle wave in hopes of recapturing the resonance frequency.

This "autolocking" architecture works robustly when the circulating power is $< 20\,\mathrm{kW}$, typically relocking in $< 1\,\mathrm{sec}$. However, at higher circulating powers $> 60\,\mathrm{kW}$ it is usually unable to recapture the resonance frequency. The reason for this is still under investigation. We believe that it is due to the cavity resonance frequency changing due to radiation force on the mirrors and/or thermal expansion of the cavity in the first few microseconds after the laser frequency scan encounters the cavity resonance. It may be possible to adapt the output of FPGA1 to anticipate such shifts.

The microcontroller of FPGA1 is connected to the PC. This allows for user control of the FPGA1 controller parameters. The connection is also used to automatically disable the outputs of the feedback controllers running on the PC when FPGA1 detects that the laser is not on resonance with the cavity (via photodetector PD5). This prevents their control variables from running away and needing to be reset before relocking the laser frequency.

The circuit diagram of feedback controller PID1 and how it integrates with adjacent components is shown in figure 6.4. The adder D1 is a simple op-amp (U2A) adder circuit which adds the error signal output of mixer X1 to the signal from source S2 after attenuation in attenuator R2. An additional user-controlled potentiometer (RV1) controls the setpoint. The summed signal is passed to an output port (OUT1) for diagnostic purposes by a buffer op-amp (U3A). It continues into the proportional-integral-integral (PII) gain stage (U5A and U6A) where the gain coefficients are user controllable via potentiometers RV2, RV3, and RV4. The integrating capacitors (C3 and C5) are short-circuited by two voltage-controlled switches (U4A, U4.2A). When FPGA1 detects (via photodetector PD5) that the laser frequency is not on resonance with the cavity, it turns on its digital output D_IO1 which closes the switches U4A and U4.2A. This prevents the integrating capacitors from charging and thus the op-amp (U5A, U6A) outputs from railing. In this uncharged state, the integrators are ready to start computing the control variable from a neutral starting configuration when the cavity resonance is encountered and the switches opened.

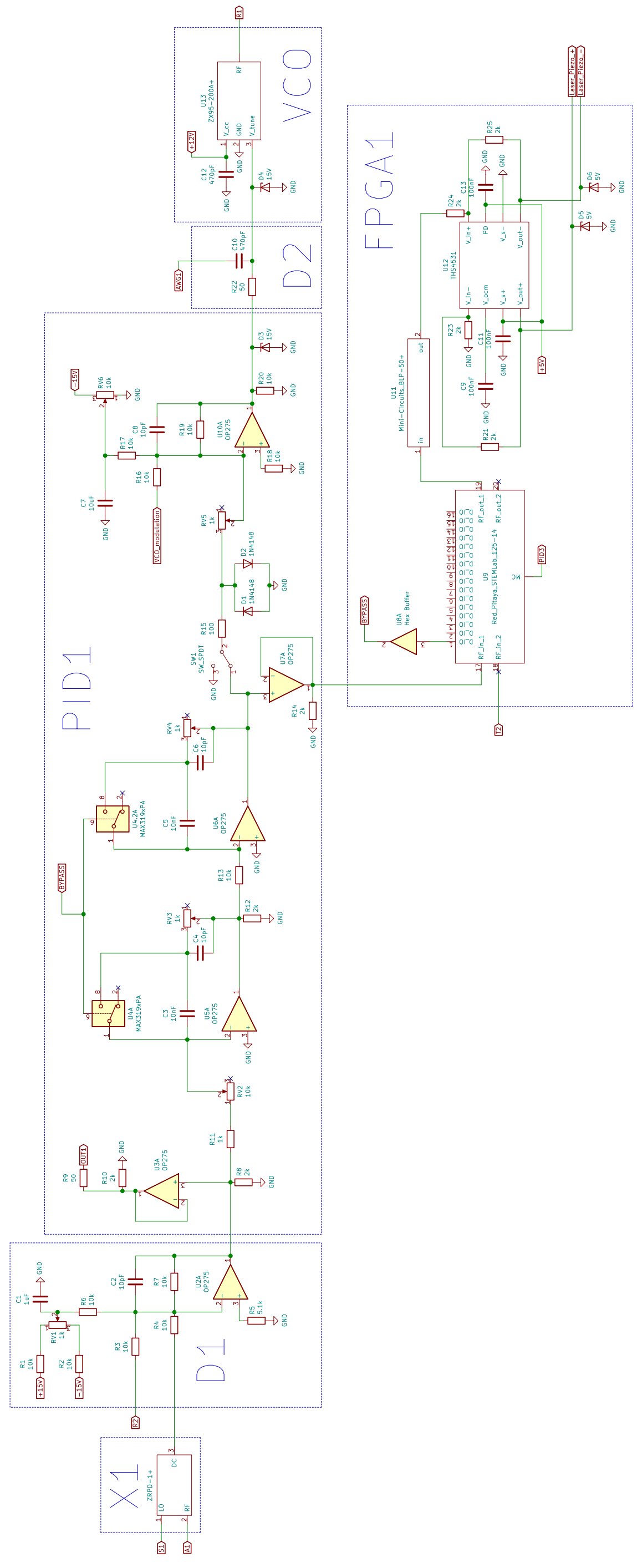

Figure 6.4: Circuit diagram of the analog feedback controller PID1 (see annotations in figure 6.1) and adjacent components.

After the PII gain stage the signal passes through a user-controlled switch (SW1) which is used to turn the output signal path to the AOM on and off for diagnostic purposes. Resistor R15 and diodes D1 and D2 act as a clipper which prevents the voltage at the input of the next stage of the circuit from exceeding roughly 0.6 V. The subsequent potentiometer (RV5) then sets the gain this signal is amplified by in the following op-amp (U10A) adder stage, which results in an adjustable clipping voltage at the output of the adder. This adder also serves to add a user-controllable DC voltage to the signal via potentiometer RV6. This is used to set the central frequency of the VCO output in order to maximize the optical power transmitted through the AOM and reduce the residual amplitude modulation on the transmitted light. The adder also includes an input port (VCO_modulation) for diagnostic purposes.

The output voltage of the adder is restricted to be positive and not greater than 15 V by Zener diodes D3 and D4. This protects the input to the VCO. The 5 MHz PDH modulation signal is added using a simple diplexer (resistor R22 and capacitor C10). The resulting signal is sent to the input of the VCO (V_tune).

Op-amp U7A passes the PII output to one of the RF inputs of FPGA1. The other RF input reads the photodetector PD5 signal through bias tee T2. The RF output (RF_out_1) of FPGA1 is passed through a 50 MHz low-pass filter (Mini-Circuits BLP-50+) to reduce high-frequency noise from the FPGA's digital-to-analog converter. This signal is then converted to the 0 V-to-5 V unipolar differential format required by the laser piezo driver input using a Texas Instruments THS4531 op-amp. Each of the differential outputs (Laser_Piezo_+ and Laser_Piezo_-) is protected by a 5 V Zener diode (D5 and D6).

The cascaded combination of controllers PID1 and PID2 provides a laser frequency range of $\sim \pm 50$ MHz, which on its own would still require that the cavity length be stabilized within $\sim \pm 4$ nm. To further relax this requirement, the output of controller PID2 is cascaded with yet another digital feedback controller PID3. This is done by passing the digital value of the RF output of FPGA1 through its microcontroller (MC) to the PC, where it is used as the error signal for controller PID3. Its control variable sets the temperature of the seed laser fiber cavity via a thermoelectric cooler (TEC), which changes the laser frequency through a range of $\sim \pm 80$ GHz with a low bandwidth of $\sim 1$ Hz. This allows the cavity length to be tracked through a range of $\sim \pm 5.7$ µm. For reference, a 20 mm long piece of aluminum expands by 5.7 µm when its temperature changes by 12.3 K (near room temperature). We found that our cavity has an effective thermal expansion coefficient of 0.3 µmK$^{-1}$, so that its temperature only need to be stabilized within about $\pm 19$ K in order that its length remain within the $\sim \pm 5.7$ µm range required by the laser frequency feedback system. When using an electrical in vacuum heater tape (see section 4.12) we are easily able to manually stabilize the cavity temperature within $\pm 1$ K, so this large of a laser tuning range is really no longer necessary. It may be worth revisiting the selection of seed laser to get a higher feedback bandwidth in pursuit of an even more robust laser frequency feedback system.

In total, the cascaded feedback architecture gives the laser frequency feedback system a bandwidth of ∼ 100 kHz with a range of ∼ ±80 GHz. A Bode plot of the (small signal) closed-loop transfer function of the feedback system is shown in figure 6.5, which demonstrates the bandwidth of the system. The sharp peak in the transfer function amplitude near 35 kHz is due to the bandwidth-limiting resonance in the laser piezo actuator. The broad peak around 200 kHz is due to positive feedback generated by a ∼ 180° phase delay at the ultimate bandwidth of the system. The bandwidth is partially limited by delay in the AOM feedback system, as it takes ∼ 70 ns for the VCO output to respond to a change in its input voltage, ∼ 200 ns for the AOM output laser frequency to respond to a change in its input drive frequency, and ∼ 80 ns for the laser to travel from the AOM output through the fiber amplifier to the cavity input and back to photodetector PD4. The transfer function was measured by injecting a swept-frequency sinusoidal signal to the error signal at adder D1, and then observing the resulting change in error signal just before D1 (between mixer X1 and adder D1).

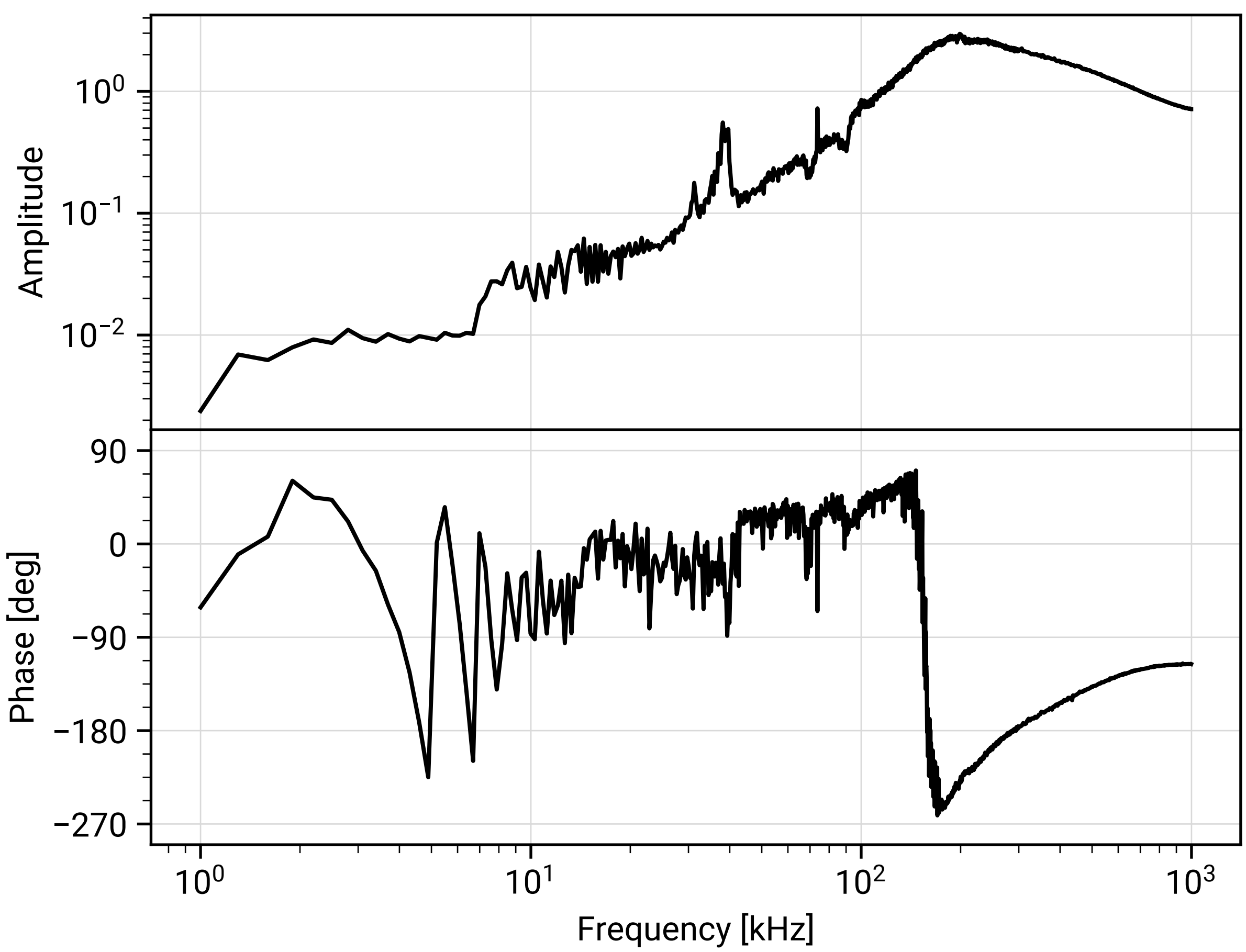

Figure 6.5: Small signal closed-loop transfer function of the laser frequency feedback system. The amplitude of the transfer function is shown in the top panel, and the phase of the transfer function is shown in the bottom panel. Reproduced from [24].

The power spectral density of the laser frequency detuning (with the cavity operating at a circulating power of 77 kW) is shown in figure 6.6. The detuning is inferred from the error signal voltage and a calibration of the peak-to-peak error signal amplitude when the laser frequency is scanned across the cavity resonance. The corresponding laser linewidth (full-width half-maximum) is 22.7 kHz over a measurement time of 10 s.

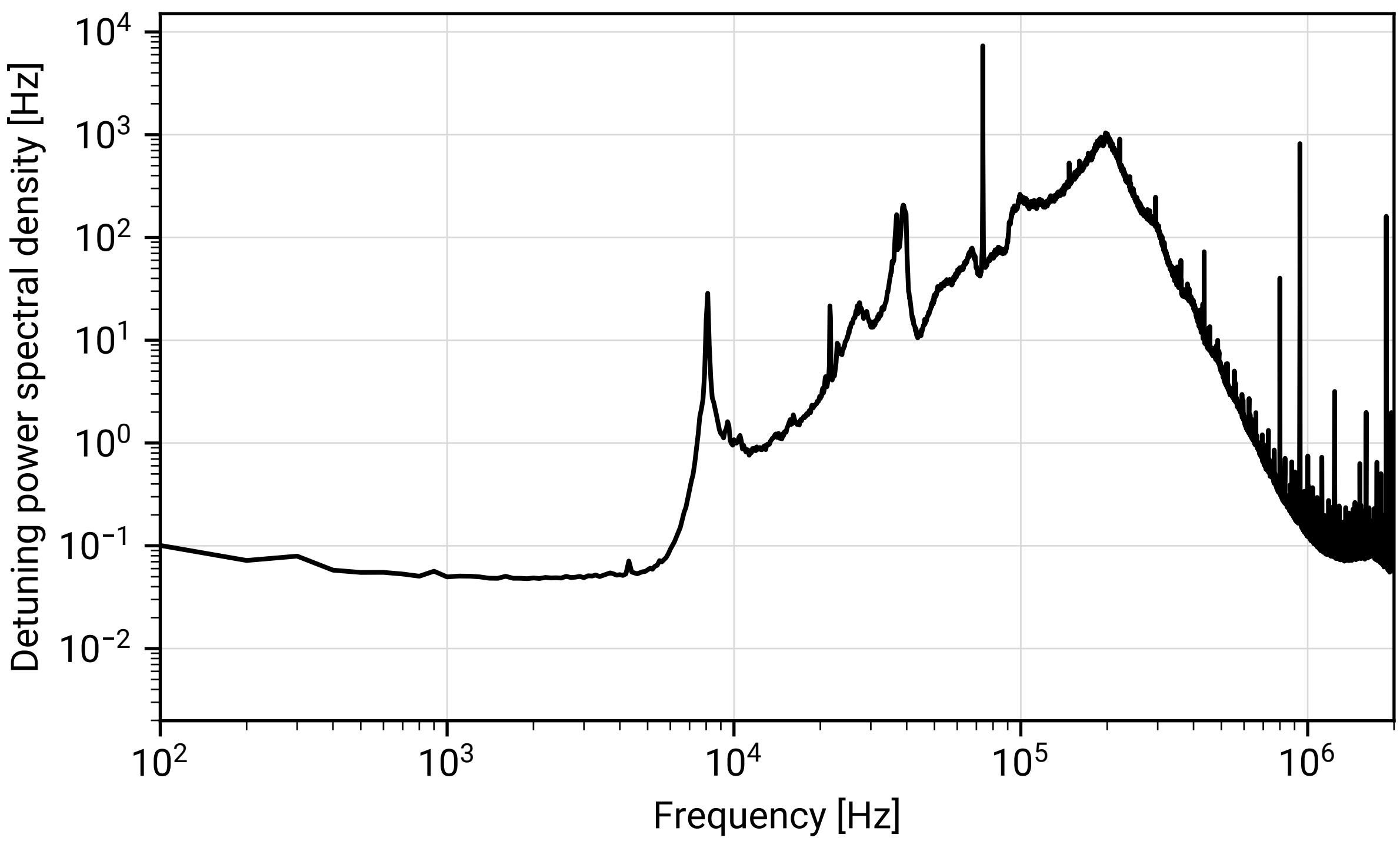

Figure 6.6: Power spectral density of the laser frequency detuning as measured using the Pound-Drever-Hall method, with a circulating power of 77 kW.

### 6.2.4 Pound-Drever-Hall gain

The light incident on photodetector PD4 can be conceptually decomposed into several components. The first component is the light at the carrier frequency which has reflected off of the input cavity mirror. This "promptly-reflected" light has the same spatial profile as the input beam. A second component is the light at the carrier frequency which emanates from inside the cavity and exits through the input mirror. This light has the spatial profile of the cavity mode. Note that the spatial profile of the input beam and cavity mode are not necessarily (and typically not) the same—their similarity is quantified by the overlap integral $Q_{l,m,p}$ (see section 2.3.2). The third component is the light at the PDH modulation sideband frequency. Assuming the modulation frequency is much larger than the cavity linewidth, to good approximation the spatial mode of this light will be the same as the input beam, since little of this light enters the cavity.

These three components interfere on the photodetector PD4. However, interference between the promptly-reflected light and the sidebands does not generate amplitude modulation, as these two components represent the prompt reflection of the purely phase-modulated input light. The only interference term which generates amplitude modulation at photodetector PD4 is that between the cavity mode leakage light and the (promptly-reflected) sidebands. Since the spatial modes of these two components are in general different, the depth of the amplitude modulation depends on the spatial overlap between the cavity mode and input beam. The PDH error signal voltage therefore depends on not only the detuning, but also the squared magnitude of the input beam's overlap integral with the cavity mode (i.e. input coupling efficiency). Increasing the input coupling efficiency increases the PDH error signal voltage at a fixed detuning, which increases the gain of the feedback system. Conversely, a smaller input coupling efficiency leads to a lower feedback gain [24].

In our system, we found that the cavity coupling efficiency $|Q|^2$ (defined in section 2.3.4) would change by up to $\sim 0.3$ as the circulating power in the cavity was increased from its initial setting to its ultimate setpoint ($\sim 75\,\mathrm{kW}$). We suspect this is due to 1) thermal lensing in optics on the input breadboard which are subject to changing optical power, 2) thermal lensing in the relatively absorptive ULE7972 mirror substrate, and 3) thermal deformation of the cavity mount and support arm. This amount of change could cause nearly a factor of 2 change in the feedback gain. In the case of increasing gain this sometimes resulted in the loop becoming unstable due to positive feedback, and in the case of decreasing gain typically caused the feedback system to unlock more often. We endeavored to reduce thermal lensing in the input optics by using reflective optics when possible, and using fused silica substrates (which is less absorptive than other optical glasses) when not. To improve on our current design, I would recommend replacing the optical isolator with a model which uses a lower-absorption material in the Faraday rotator.

In principle, it should be possible to stabilize the input coupling efficiency by measuring the input coupling efficiency using photodetectors PD2, PD3, and PD7 (see section 2.5) and then using an optimization algorithm to achieve a setpoint value by adjusting the input beam pointing by motorizing the optomechanical mounts of mirrors M8 and/or M9. Such a system can only compensate for changes in the relative position and angle of the input beam and cavity mode, and cannot compensate for changes in relative beam divergence due to thermal lensing. Control of the beam divergence would require motorizing at least one optomechanical mount of lenses L1-L3. In both cases, care must be taken such that the position, angle, and/or divergence of the input beam does not "run away" to an undesirable configuration (e.g. hitting non-optical surfaces), and considerations must be made for vibration isolation if oscillatory motors are used to drive the optomechanics, since these vibrations could cause the laser frequency feedback system to unlock, or reduce the resolution of the TEM.

To avoid these issues we decided to accept the change in coupling efficiency and instead directly stabilize the gain of the laser frequency feedback system by adjusting the optical power incident on photodetector PD4 using the variable attenuator (HWP3 and PBS3). We measure the gain of the PDH error signal by dithering the error signal setpoint by a small amount at a frequency of 1.1 kHz (using sinusoidal source S2 attenuated by attenuator R2, then added to the error signal in adder D1). The relatively low frequency ensures that the laser frequency will mostly follow the commanded change in setpoint. A PDH detection system (photodetector PD5, bias tee T2, amplifier A3, mixer X2) installed on the output side of the cavity measures the detuning independently of the input-side PDH system. The principle of operation of the PDH method when used at the cavity output is fundamentally the same as when used at the input, however, it is not preferred for laser frequency feedback because the signal is weaker (and therefore noisier) due to attenuation of the sidebands when transmitted through the cavity, and the feedback bandwidth is lower because both the carrier and sidebands must traverse the cavity (which takes time on the order of an inverse cavity linewidth) in order to generate a change in the error signal. Input-side PDH systems immediately detect changes in the cavity light since the sidebands are promptly reflected from the input mirror and do not traverse the cavity.

The depth of detuning modulation caused by the dither signal is measured from the output-side PDH error signal by filtering it through a bandpass filter centered around the dither frequency (filter F2), amplifying the signal (amplifier A4), and then performing a digital lock-in measurement using FPGA2 (this operation is represented by mixer X3). Since all light leaving the cavity has the same spatial profile (the cavity mode), the output PDH system's gain does not depend on the input coupling efficiency. Therefore, since the amplitude of the dither signal does not change, an increase in the detuning modulation depth measured by the output PDH system implies a decrease in the input PDH system gain, and vice versa.

The feedback controller PID4 takes as inputs the measured detuning modulation depth as its error signal, and a user-controlled setpoint (which sets the desired input PDH gain). The control variable of PID4 is added (in adder D3) to the error signal of the photodetector optical power feedback system for the variable optical attenuator which controls the power incident on photodetector PD4. This effectively changes the setpoint of that feedback system to keep the input PDH gain constant.

This system also stabilizes the detuning resulting from any non-zero setpoint of the feedback controller PID1, which may result from intentional user input or unintentional residual amplitude modulation on the input laser beam.

We initially implemented this feedback system on an earlier prototype of the cavity which suffered from roughly 10× higher scattering losses than the cavity described in the majority of this work. This resulted in a cavity temperature change (between low and high circulating power states) on the order of tens of degrees Kelvin, and concomitantly higher changes in the input coupling efficiency. With this cavity, the feedback system was necessary to maintain the stability of the laser frequency feedback system. However, on the newer cavity, the better inherent stability of the input coupling efficiency means that this feedback system is likely no longer strictly necessary.

### 6.2.5 Cavity mode position

Laser-induced heating and the concomitant thermoelastic deformation of the cavity mount changes the relative alignment of the cavity mirrors. In the near-concentric configuration, even small changes in the relative positions and/or angles of the cavity mirrors cause large changes in the position of the cavity mode within the cavity (see section 2.2.3). Stabilizing the cavity temperature helps reduce the change in alignment as laser power is increased, but is an imperfect solution since it may not exactly replicate thermal gradients that exist across the cavity mount arising from the specific distribution of heat delivered by the light scattered from and absorbed by the mirrors.

We do (manually) keep the cavity temperature roughly constant using the heat delivered by an in-vacuum heater tape installed on the cavity support arm (Heater, Current Source, Voltage Source in figure 6.1). But we also directly stabilize the position of the cavity mode.

A change in the position of the cavity mode results in the cavity output beam exiting the cavity output mirror at a different position and angle. Our cavity mode position feedback system detects the position of the output beam some distance from the cavity output mirror using a position-sensitive detector (PSD). A lens (L5) is used to approximately image the CC surface of the cavity output mirror onto the PSD so that the system is largely insensitive to spurious deflections of the beam near the output mirror (perhaps due to thermal lensing in the mirror substrate or output vacuum window W2). The two ($x$ and $y$) position output signals from the PSD are each separately processed through a feedback controller (PID6) with user-controlled setpoints. The resulting control variables are used to adjust the cavity alignment piezoelectric actuators (Piezos) to maintain the position of the output beam on the PSD.

The two control variable axes are mixed (Axis Mixer) such that they drive all three piezos in a coordinated motion which leaves the cavity length nominally unchanged. This mixing is accomplished via a linear transformation which models the position of the cavity head as a function of the cavity alignment piezo lengths. Specifically, the positions of the tips of the three alignments piezos $\mathbf{r}_j = (r_{j,x}, r_{j,y}, r_{j,z})\,,\ j = 1, 2, 3$ define a plane on which the cavity head (and therefore output mirror) lies. Taking the $z$-axis to lie along the extension axis of the piezos, this plane satisfies the system of equations

$$\begin{pmatrix} r_{1,x} & r_{1,y} & -1 \\ r_{2,x} & r_{2,y} & -1 \\ r_{3,x} & r_{3,y} & -1 \end{pmatrix} \begin{pmatrix} n_x \\ n_y \\ z_0 \end{pmatrix} = - \begin{pmatrix} r_{1,z} \\ r_{2,z} \\ r_{3,z} \end{pmatrix} \tag{6.1}$$

where the normal vector of the plane is $\mathbf{n} \coloneqq (n_x, n_y, 1)$ and the plane intersects the $z$-axis at $z_0$. In our case, $n_{x,y} \ll 1$ so that their values are approximately equal to the tip/tilt angles (in radians) of the plane. $z_0$ represents the length of the cavity due to the extension lengths of the piezos $r_{j,z}$.

We assume that the piezo lengths are linearly related to the respective piezo voltages $V_j$ by scaling factors $S_j$ such that $r_{j,z} = S_j V_j$, which in principle are all equal but in practice may be slightly different. Thus, the (three) piezo voltages required to achieve the (two) tip/tilt angles are

$$\begin{pmatrix} V_1 \\ V_2 \\ V_3 \end{pmatrix} = -\begin{pmatrix} 1/S_1 & 0 & 0 \\ 0 & 1/S_2 & 0 \\ 0 & 0 & 1/S_3 \end{pmatrix} \begin{pmatrix} r_{1,x} & r_{1,y} & -1 \\ r_{2,x} & r_{2,y} & -1 \\ r_{3,x} & r_{3,y} & -1 \end{pmatrix} \begin{pmatrix} n_x \\ n_y \\ z_0 \end{pmatrix} \tag{6.2}$$

Setting $z_0 = 0$ enforces the condition that the output voltages doe not change the cavity length.

The transformation is completely general in the sense that it has 9 independent input parameters ($S_j$, $r_{j,x}$, $r_{j,y}$). In principle these parameters could be experimentally determined by changing each piezo voltage and measuring the resulting change in $n_x$, $n_y$, and $z_0$ (through their effect on the cavity mode position and cavity length, respectively). However, we simply assumed that $S_1 = S_2 = S_3$, and used the nominal positions of the piezo tips from the cavity design (see figure 6.7) for $r_{j,x}$ and $r_{j,y}$. This approximation was accurate enough to not cause any noticeable problems with the feedback system.

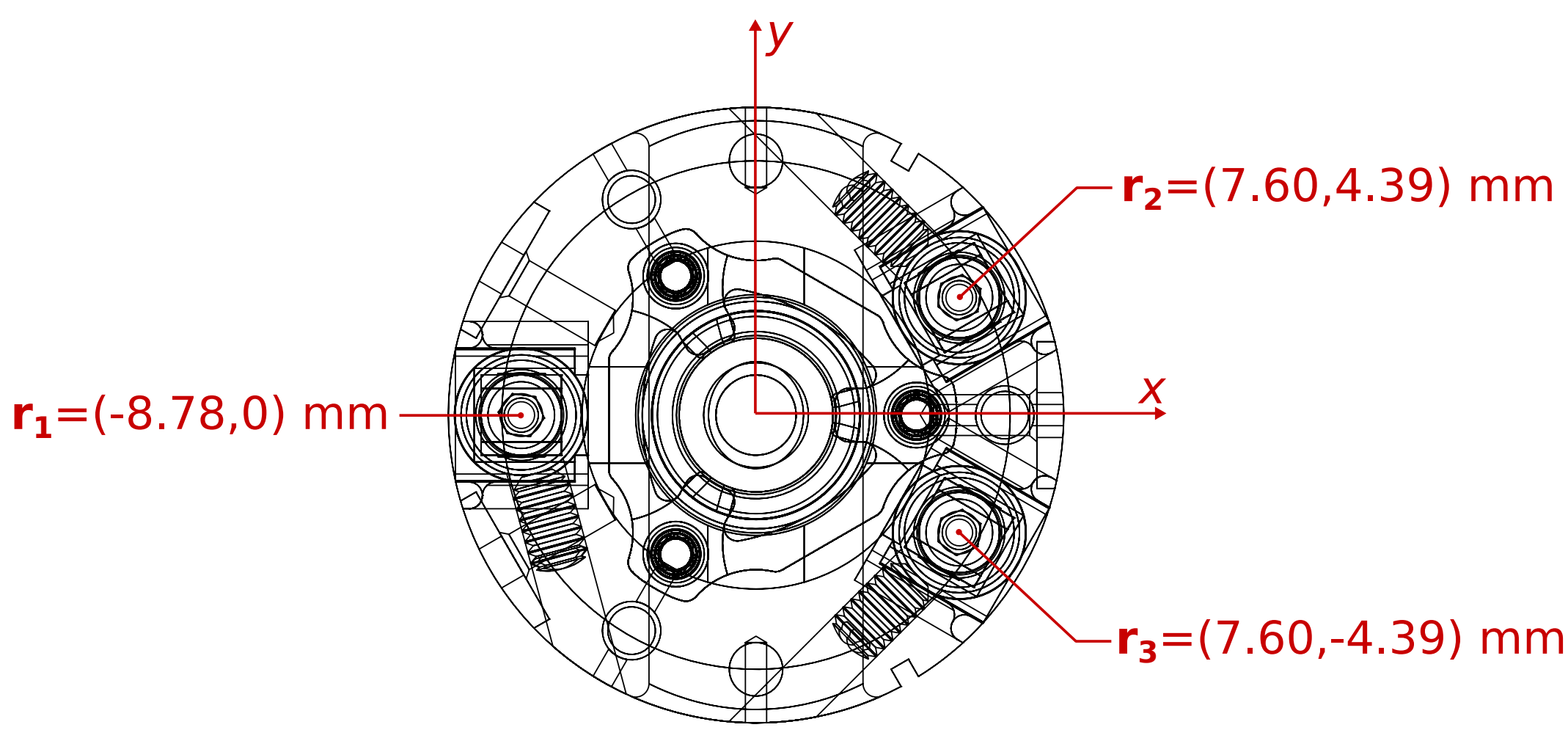

Figure 6.7: An end view of the cavity mount assembly looking from the output side to the input side. The electron beam travels in the $-y$-direction through the center of the mount. The optical axis lies on the $z$-axis (not annotated). The nominal positions of the piezo tips in the $x, y$-plane are annotated.

The output of the axis mixer commands the output voltage of the piezo controller, and the output voltages are low-pass filtered at $\sim 1\,\mathrm{Hz}$ to reduce cavity length fluctuations due to (primarily high-frequency) voltage noise from the piezo controller. This filtering is likely unnecessary if a lower noise piezo controller is used. We elected to substantially filter the controller output since the feedback system did not require a high bandwidth. The filtered voltage is sent to the piezos through a vacuum feedthrough (see section 4.9).

The feedback controller PID6 is programmed such that it is only active when the laser frequency feedback system has determined that the laser frequency is on resonance. This prevents the cavity mode position from "running away" when the laser frequency unlocks.

## 6.3 Diagnostic systems

Several diagnostic systems are installed on the LPP. While these systems are not strictly necessary to operate the LPP, they are helpful in initially configuring the system, characterizing its performance, and diagnosing problems.

### 6.3.1 Cavity ringdown

We measure the round-trip loss (equivalently, finesse) of the cavity by coupling light into the cavity for a short period of time and then observing the subsequent decay of optical power exiting the cavity output mirror. This is referred to as the "cavity ringdown" (CRD) method [78]. The easiest way for us to briefly couple light into the cavity is to rapidly sweep the laser frequency across the cavity resonance, which is referred to as "rapidly-swept continuous-wave cavity ringdown" [79]. This method results in a slightly more complicated ringdown signal than a method in which the input light is simply switched off, since light of slightly different frequencies within the cavity linewidth interfere, generating oscillations in the output power.

Assuming the input light wave $E_{\mathrm{i}}^{+}(t)$ is linearly chirped in time $t$,

$$E_{\mathrm{i}}^{+}(t) = E_{\mathrm{i}} \exp\left(i2\pi\left(\nu_0 + \frac{\nu'}{2}t\right)t\right) \tag{6.3}$$

where $\nu_0$ is its frequency at $t = 0$ and $\nu'$ is the rate at which its frequency changes, its Fourier transform is

$$\begin{aligned} \mathcal{F}\left[E_{\mathrm{i}}^{+}\right](\nu) &\coloneqq \sqrt{2\pi}\int_{-\infty}^{\infty} d\nu\, e^{-i2\pi\nu t} E_{\mathrm{i}}^{+}(t) && (6.4) \\ &= \frac{E_{\mathrm{i}}}{F}\sqrt{\frac{F^2}{\nu'}} e^{i\pi/4} \exp\left(-i2\pi^2\left(\frac{\nu-\nu_0}{F}\right)^2 \frac{F^2}{\nu'}\right) && (6.5) \end{aligned}$$

where $F$ is the free spectral range and $\nu$ is the frequency variable. The cavity transmission light wave $E_{\mathrm{t}}^{+}(t)$ is therefore the inverse Fourier transform of the product of the input wave Fourier transform and the cavity transmission transfer function (see section 2.3.3)

$$H_{\mathrm{t}}(\nu) = e^{i\phi_{\mathrm{RT}}/2} \frac{Q_{l,m,p}\mathcal{T}_1\mathcal{T}_2}{1 - |\mathcal{R}_1||\mathcal{R}_2|} \frac{1}{1 + i\frac{2\pi\left(\nu-\nu_{n,l,m}\right)/F}{1-|\mathcal{R}_1||\mathcal{R}_2|}} \tag{6.6}$$

such that the transmitted power as a function of time is proportional to

$$\left|E_{\mathrm{t}}^{+}(t)\right|^2 = \left|\sqrt{2\pi}\int_{-\infty}^{\infty} d\nu\, e^{i2\pi\nu t} H_{\mathrm{t}}(\nu)\,\mathcal{F}\left[E_{\mathrm{i}}^{+}\right](\nu)\right|^2 \tag{6.7}$$

$$\propto \left|\int_{-\infty}^{\infty} d\xi\, e^{i2\pi\xi F t}\frac{1}{1+i\frac{4\pi\xi}{X_{\mathrm{RT}}}}\exp\left(-i2\pi^2\xi^2\frac{F^2}{\nu'}\right)\right|^2 \tag{6.8}$$

after setting $\nu_0 = \nu_{n,l,m}$ such that the resonance frequency is crossed at $t = 0$, ignoring the time delay caused by $\phi_{\mathrm{RT}}$, and recalling that

$$1 - |\mathcal{R}_1||\mathcal{R}_2| \simeq \frac{X_{\mathrm{RT}}}{2} \tag{6.9}$$

in the limit that the round-trip loss $X_{\mathrm{RT}} \coloneqq T_1 + X_1 + T_2 + X_2 \ll 1$. The expression in equation (6.8) can either be numerically integrated, or it can be expressed in terms of the complex Dawson function

$$D(z) \coloneqq e^{-x^2}\int_0^x dt\, e^{t^2} \tag{6.10}$$

such that

$$\left|E_{\mathrm{t}}^{+}(t)\right|^2 \propto \left|\sqrt{\frac{\pi}{2}}\exp\left(-\frac{1}{2}X_{\mathrm{RT}}\bar{t} + \frac{i}{2}\bar{\nu}'\bar{t}^2 - \frac{i}{8}\frac{X_{\mathrm{RT}}^2}{\bar{\nu}'}\right) + i\sqrt{2}D\left(\frac{i + 2\bar{\nu}'\bar{t}/X_{\mathrm{RT}}}{\sqrt{8i\bar{\nu}'/X_{\mathrm{RT}}^2}}\right)\right|^2 \tag{6.11}$$

where $\bar{\nu}' \coloneqq \nu'/F^2$ and $\bar{t} \coloneqq Ft$ [80]. Either equation (6.8) or (6.11) can be used to fit the measured ringdown signal, where the non-trivial fit parameters are the round-trip loss $X_{\mathrm{RT}}$ and the (normalized) chirp rate $\bar{\nu}'$. Trivial fit parameters are the overall amplitude of the signal and the time at which $t = 0$ is defined. The free spectral range $F$ must be known. We approximate $F$ by assuming that the cavity length $L = 20\,\mathrm{mm}$, though in principle the free spectral range can also be independently measured via the cavity mode spectrum.

We perform the fitting by first fitting the Fourier transform of the signal as a function of the non-trivial parameters and signal amplitude, since the Fourier transform eliminates the need to fit the $t = 0$ point. This fit result is then used as an initial guess for a nonlinear least squares fitter which fits the time-domain signal as a function of all fit parameters simultaneously. An example of a measured CRD signal and its corresponding fit in both real and Fourier space are shown in figure 6.8.

By adjusting the voltages on the cavity alignment piezos it is possible to generate a two-dimensional map of the round-trip loss of the cavity as a function of the cavity mode position. The position of the mode on the mirror surface is estimated from the piezo voltages and cavity's distance to concentricity, as measured by the transverse mode spectrum frequency spacing. It could also be directly measured using a camera or position-sensitive detector, though we have not gone to the trouble to calibrate such a method on our system. We use this "CRD map" to determine if the cavity mirrors have any major defects within $\sim 1$ mm of the optical axis of the cavity (examples shown in section 7.2). Even though we visually inspect the mirror surfaces just before installing them in the cavity, we have seen that the subsequent handling of the mirrors and installation process can contaminate the mirrors, so the CRD map is a good double-check that the mirrors are still clean in their final configuration installed in the cavity. A contaminated mirror may be damaged by the high circulating power in the cavity, so if contamination can be detected in the CRD map before subjecting the mirror to high power, the mirror(s) may be removed, cleaned, and saved from permanent damage. On the other hand, given a sufficient supply of replacement mirrors it may be more efficient to simply attempt to achieve high circulating power with suspect mirrors and replace the mirror(s) only if they become damaged.

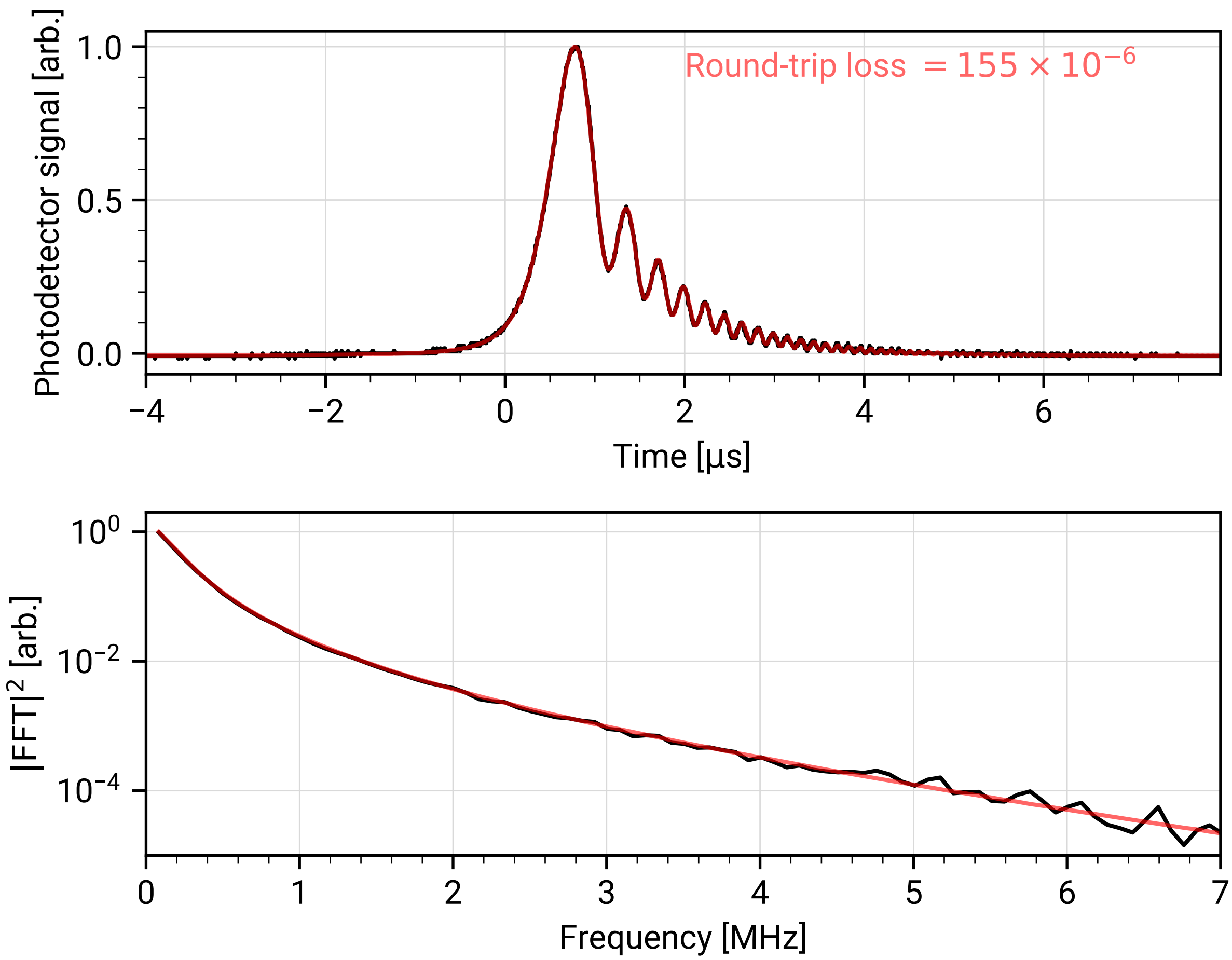

Figure 6.8: A measured cavity ringdown signal (black) as a function of time (top panel) and frequency (bottom panel). The best fit to the model given in equation (6.11) is shown in both panels in red. The best fit value of the round-trip loss is annotated, assuming a free spectral range of 7.49 GHz.

The ringdown signal is captured by photodetector PD8 with flip-up mirror M17 in place to divert the cavity output beam. The signal from this photodetector is read by an oscilloscope (Oscilloscope). The oscilloscope is triggered such that it only records ringdown signals which occur on the half of the laser frequency scan when the laser frequency is decreasing. This is accomplished by differentiating the laser frequency scan triangle wave signal with a low-pass filter (F5) and setting the oscilloscope to only trigger when that signal is either above (or below) zero (depending on whether an increasing scan voltage corresponds to an increasing or decreasing laser frequency).

On the decreasing frequency portion of the laser scan, the fundamental mode resonance will be encountered before any other modes. Triggering on the first rising edge signal ensures that the recorded ringdown signal is from the fundamental mode, at least until the coupling efficiency to the fundamental mode becomes so low that the fundamental mode ringdown signal barely rises above the photodetector dark noise. At this point the signal-to-noise ratio of the resulting signal would be too low to use anyways. The oscilloscope trigger is set to then ignore any rising edge events for a subsequent time which is slightly more than half the period of the laser frequency scan, so that all other (higher order mode) ringdown signals are ignored. This scheme allows us to always capture the ringdown signal from the fundamental mode, up to the point where the coupling efficiency to the fundamental mode becomes quite small ($< 0.05$). Still, the requirement to maintain at least a small coupling efficiency is what limits the mode position range of the CRD map.

The accuracy and precision of our round-trip loss measurement made via CRD is approximately $\pm 1 \times 10^{-6}$. One factor limiting both the accuracy and precision is that retroreflections from other optics in the laser beam path slightly change the shape of the ringdown signal ("etaloning") [81]. These retroreflections effectively form extremely low finesse parasitic cavities which slightly modify the linewidth of the combined cavities. In our system we found that retroreflections from window W1 caused etaloning. To eliminate the effect we mounted the window at a $2°$ angle to the optical axis. We also found that the effect of the etaloning could be quantified by making CRD measurements as a function of the (primary) cavity length. This changes the primary cavity resonance frequency, but does not change the resonance frequency of the parasitic cavities. This results in a CRD round-trip loss measurement which sinusoidally varies with periods equal to the free spectral range of the parasitic cavities. This variation can be fit to a (multiple) sinusoidal model and the mean value used as the "actual" primary cavity round-trip loss. Indeed, using this method is what allowed us to identify that window W1 was responsible for the etaloning because it was the only optic an appropriate distance from the cavity to cause etaloning with the observed free spectral range.

The second factor limiting the measurement precision is noise in the ringdown profile data. In our case the primary noise source was likely from the low (8) bit-depth digitization of the ringdown signal, but at some point technical noise from the photodetector and eventually shot noise will become limiting. To combat this noise source, we record several ringdown traces at each position in the CRD map and average the fitted round-trip loss values.

### 6.3.2 Transverse mode spectrum

To measure the numerical aperture (equivalently, focal waist) of the cavity mode during operation of the LPP we measure the transverse mode spectrum of the cavity. Assuming that the the cavity modes are Hermite-Gaussian the frequency spacing between higher order transverse modes of the cavity is a simple function of the numerical aperture of the fundamental mode (see section 2.3.1). We use this measurement of the numerical aperture to manually set the cavity's distance to concentricity during initial alignment of the system. It then occasionally requires small changes every few weeks or months to maintain approximately the same numerical aperture.

We measure the transverse mode spectrum by phase modulating the input laser beam using an electro-optic modulator (EOM) driven by a swept frequency source (Spectrum Analyzer) and detecting the resulting amplitude modulation of the light at the cavity output using photodetector PD6. In the near-concentric configuration, the transverse modes of the cavity have resonance frequencies just below the fundamental mode resonance frequency, so only one sideband of the phase modulated input beam passes through the cavity when it encounters a transverse mode resonance. This frequency component interferes with the fundamental mode carrier frequency at photodetector PD6, generating a fluctuating signal at the sideband frequency. The two modes are spatially orthogonal, so to generate a non-zero signal the laser beam must not be entirely captured by photodetector PD6. In practice, we achieve this by intentionally misaligning the beam so that only a portion of it hits PD6. The depth of amplitude modulation is dependent on the sideband's detuning from the transverse mode resonance, so that the depth of amplitude modulation measured as a function of the sideband frequency approximately traces out the cavity transfer function of the transverse modes. An example of a measured transverse mode spectrum taken at a circulating power of 72 kW is shown in figure 6.9. The spectrum shows several notable features. First, the cavity mode is clearly astigmatic, as the two first order modes ($(1, 0)$ and $(0, 1)$) are not degenerate. Second, the mode spectrum is anharmonic, as the 2nd (and higher) order modes do not have resonance frequencies which are integer multiple sums of the first order mode frequencies. The expected (harmonic) frequencies are indicated by vertical dashed lines. This anharmonicity is a general indicator of spherical aberration of the cavity mode. It is also present at low circulating powers, though to a lesser extent, indicating that at least some of the aberration is caused by thermoelastic deformation of the mirror due to heating from the cavity mode (see section 2.6). Modes beyond 2nd order are not readily identifiable with particular mode indices due to the combination of astigmatism and anharmonicity.

The swept frequency source driving the EOM is provided by the tracking generator (TG) output of the spectrum analyzer using $-20$ dBm of output power. It is high-pass filtered at 25 MHz to reduce spurious low-frequency fluctuations near the start and end of each frequency sweep. Still, we found that these fluctuations, and the brief cessation of signal to the EOM between sweeps, would destabilize the laser frequency feedback system. To maintain a constant power RF signal to the EOM, we installed an RF switch (SW1) on the tracking generator output, which switches the EOM drive signal to a constant 25 MHz standby signal (S3) between frequency sweeps. This signal is provided by one of the outputs of an arbitrary wave generator (AWG2), after passing it through a 20 dB attenuator (R3) and 1.2 MHz high-pass filter. The other channel of AWG2 provides a 0.66 Hz, $\sim 80\,\%$ duty cycle pulse wave (S4) to switch SW1 such that the tracking generator output is only sent to the EOM during each frequency sweep (which has a duration of $\sim 1.5\,$s), with the standby signal sent otherwise. The output of switch SW1 is further attenuated by 10 dB (R4) before it is sent to the EOM. The sync output (SYNC) of the pulse channel on AWG2 triggers (TRIG) the start of each spectrum analyzer sweep.

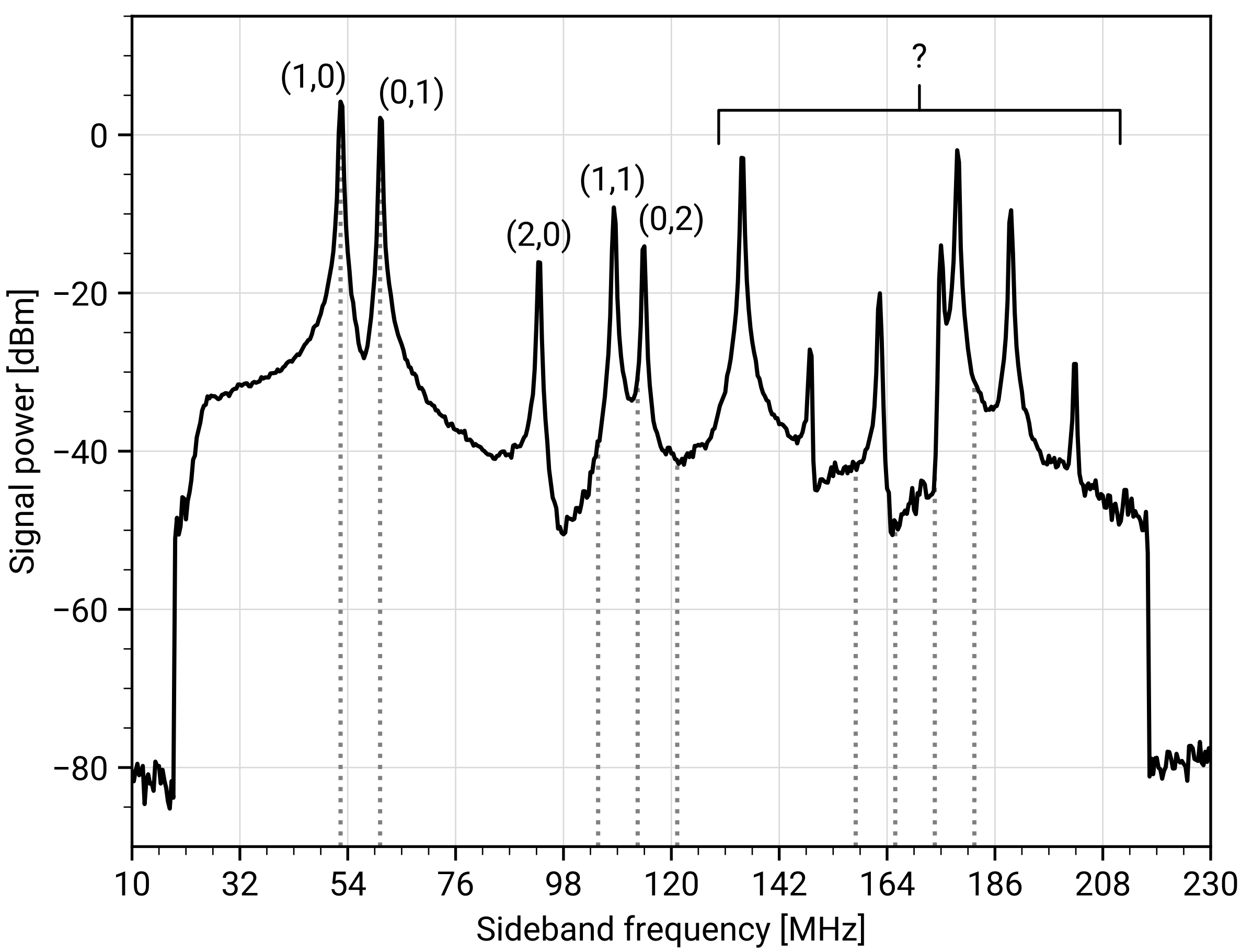

Figure 6.9: Transverse mode spectrum of the cavity with a circulating power of 72 kW. The putative transverse mode indices are annotated up to order 2, beyond which unambiguous identification of the modes becomes difficult. The vertical dashed lines show the expected positions of the 2nd and 3rd order modes if the spectrum were harmonic, based on the 1st order mode frequencies.

A (DC-terminated) bias tee (T3) and sequence of RF amplifiers (A5) relay the amplified

AC component of the photodetector (PD6) signal to a second RF switch (SW2). This switch (also driven by the pulse source S4) prevents the input signal from reaching the spectrum analyzer input unless its tracking generator output is being sent to the EOM. The spectrum shown in figure 6.9 shows this cutoff on the low and high frequency ends of the sweep. The output of the switch is connected to the input of the spectrum analyzer. The (digital) spectrum analyzer is connected to the PC so that its data can be read and recorded.

The system records 0.66 spectra per second, and higher throughput can be achieved at the expense of signal-to-noise ratio in the spectrum amplitudes. It should therefore be possible to use this system in feedback to the cavity alignment piezos to adjust the cavity length to stabilize the numerical aperture of the cavity mode. We have not yet found this to be necessary for the LPP, but it would be straightforward to implement. A phase-sensitive measurement of a single transverse mode frequency could likely provide an even higher measurement speed, if necessary. A feedback system is likely necessary if a higher numerical aperture ($> 0.05$) and/or circulating power is used, or if it is important to stabilize the numerical aperture.

This system also gives a coarse measurement of the linewidth of the transverse modes, though its precision is not high enough to be particularly useful.

### 6.3.3 Cavity reflectance and transmittance

The cavity reflectance and transmittance is measured using the calibrated power readings from photodetectors PD2, PD3, and PD7. These measurements are used with equations (2.72) and (2.73) to provide real-time estimates of the round-trip mirror loss and input coupling efficiency, respectively.

### 6.3.4 Beam profiler cameras

Cameras CAM1, CAM2, and CAM3 are used to examine the input, reflected, and transmitted beam profiles, respectively. We only use them for troubleshooting, to see if the beam is unexpectedly clipped or distorted.

### 6.3.5 Cavity thermistors

The thermistors installed on the cavity mount and support arm (Thermistors) provide a measurement of the cavity and support arm temperature, respectively. We use these measurements to manually set the current sent to the vacuum heater tape (Heater) to roughly stabilize the cavity temperature. This system could be readily reconfigured with a feedback controller. The temperature measurements also let us see when the cavity temperature has equilibrated—we typically wait for this to happen (a few minutes) before starting to use the LPP so that the cavity is not moving relative to the TEM column due to thermal expansion in the support arm.

### 6.3.6 Cavity input beam photodetector

Photodetector PD1 is used when initially setting up the system to minimize the amount of residual amplitude modulation on the cavity input beam caused by the PDH modulation in the AOM. The frequency modulation of the AOM drive signal caused by the 5 MHz PDH modulation nominally phase modulates the light at the AOM output. However, since different optical frequencies leave the fiber AOM at slightly different angles, they couple with different efficiencies into the AOM's output fiber. This in general leads to unequal amplitudes of the optical sidebands, resulting in a mix of amplitude and phase modulation. We minimize this source of residual amplitude modulation by adjusting the central frequency of the VCO (using potentiometer RV6 in figure 6.4) while monitoring the depth of amplitude modulation at 5 MHz as measured by photodetector PD1.

## 6.4 Software

The PC runs custom software which controls the feedback and diagnostic systems, provides a user interface for manually adjusting experimental parameters, and logs data. The software uses a client-server architecture, where separate server programs (written in C++) are responsible for interfacing with hardware, running feedback controller calculations, and logging data to disk. The client program is a graphical user interface (written in Python using PyQt) which communicates with the servers by sending a receiving text strings via TCP/IP. A screenshot showing the graphical user interface is shown in figure 6.10. The client can therefore also seamlessly communicate with servers running on other machines, as is the case for the Red Pitayas. This architecture also allows for flexible Python scripting control of the servers, which is implemented to execute particular experimental protocols like generating a CRD map (see section 6.3.1). Each server can write its internal variable values to a comma-separated value file.

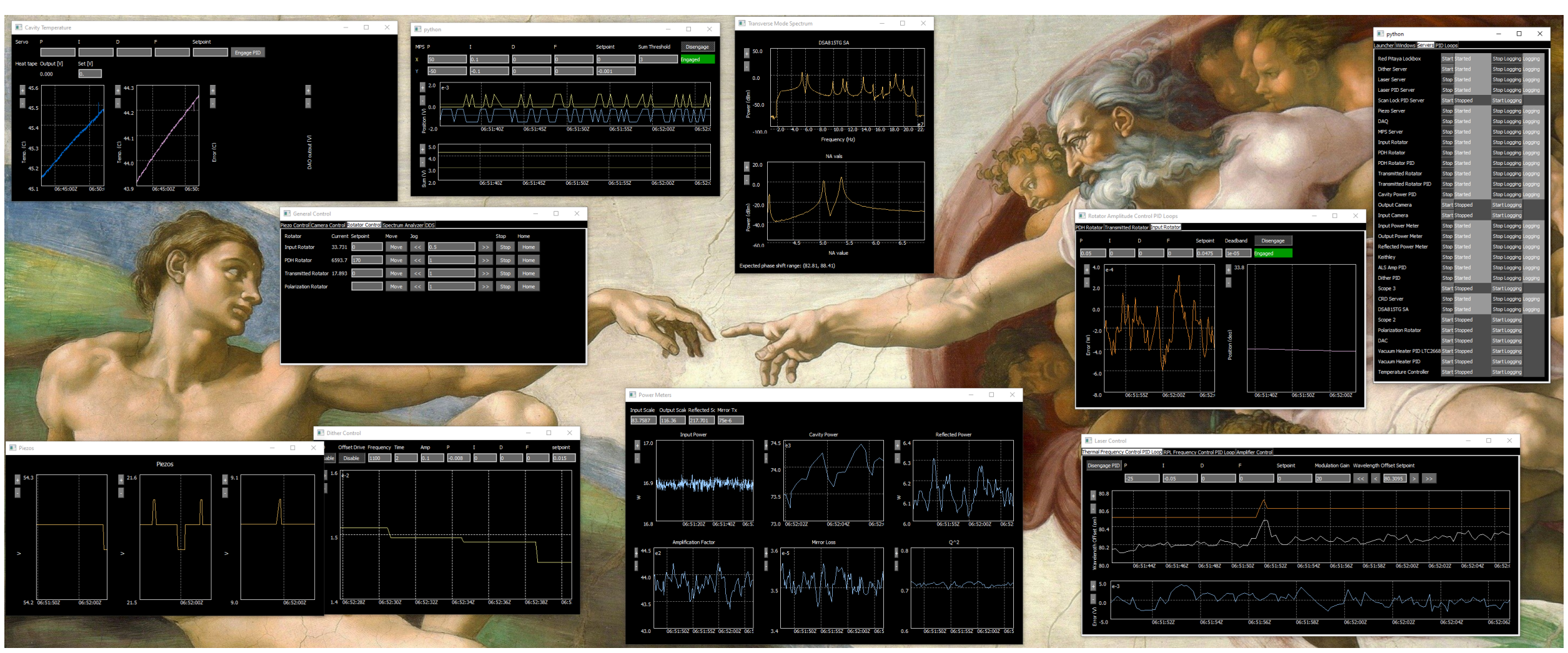

Figure 6.10: Screenshot of the graphical user interface of the client program used to interface with the laser phase plate.

## 6.5 Physical installation

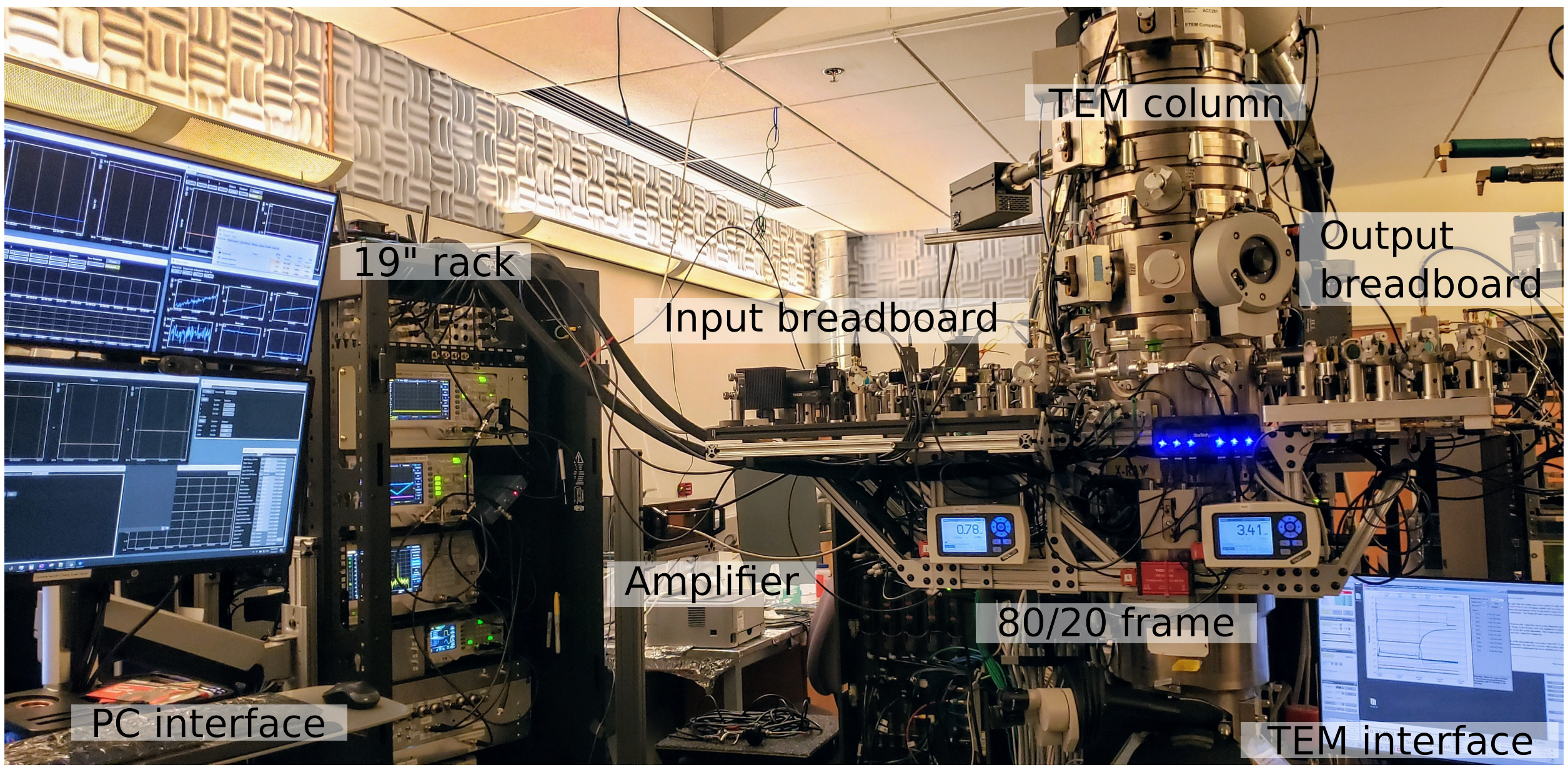

Figure 6.11: Physical installation of the laser phase plate on our transmission electron microscope.

The input and output breadboards are mounted on a frame made of 1” T-slotted aluminum extrusion (“80/20”). The input breadboard is commercial off-the-shelf (Thorlabs, Inc. PBG1824F), while the output breadboard is a custom machined 16” × 12” × 1” solid piece of 6061 aluminum. The 80/20 frame is screwed to the outer diameter of the TEM column using several custom machined adapter blocks. These blocks use pre-existing tapped holes on the outer diameter of the TEM column, some of which were provided by the manufacturer for this purpose, and some of which were originally designed to connect the microscope to its vibration isolation platform. These mounting points became available because the additional TEM column section for the LPP necessitates a lower mounting point.

A laser safety enclosure made of 1/4” thick Alumalite paneling surrounds both optical breadboards. Gaps between the are patched with black anodized aluminum foil. Foil is also used to cover the output breadboard, since there are tight clearances between the top of the optics and the volume needed to install the TEM’s side-entry sample holder.

The laser amplifier sits on a separate rack made of 80/20 near the TEM column. There is no system for vibration isolation along the amplifier output optical fiber between the amplifier and amplifier head—the fiber dangles freely, respecting the minimum bend radius (15 cm) specified by the manufacturer. We did not find that this degraded the resolution of the TEM, though it is worth noting that this could be a limiting effect at higher resolutions ($< 3$ Å) than our TEM is capable of.

The laser and fiber preamplifier are installed in a 19" rack, along with most of the rest of the electronics. A few miscellaneous pieces of electronics equipment (e.g. USB hubs, bias tee T3, amplifier A5) are mounted to the 80/20 frame on the TEM column. The entire rack and laser amplifier are powered through a consumer-grade 120 V uninterruptible power supply (UPS). This does form a large ground loop between the mains power ground and TEM ground, but we have yet to see any negative impacts from this. In principle, the entire LPP system can be made to share the TEM ground by galvanically isolating its UPS from the mains ground. The entire system, including fiber amplifier running at its nominal optical output power of 30 W, draws approximately 600 W of electrical power from the mains.

A photograph of the system (with laser safety enclosures removed to show the optical breadboards) is shown in figure 6.11.

# Chapter 7

# Cavity performance data

## 7.1 Initial alignment

After the cavity has been assembled, it must be aligned to near-concentricity. The cavity mount is designed such that the cavity is roughly 500 µm from concentricity before alignment. Achieving the target cavity mode numerical aperture of 0.05 requires a distance to concentricity of just 3.7 µm. To align the cavity to this configuration, the cavity is placed in a small vacuum chamber on an optical test bench along with the input and output breadboards (see chapter 6). The input laser beam is aligned (at low power) such that it retroreflects off of the center of the input cavity mirror. The chamber is vented with filtered nitrogen, and the input vacuum window removed. Long custom-made M1.5 hex adjusters are inserted along the length of the cavity support arm into the cavity alignment micrometer screws from the input side of the cavity assembly (see figure 7.1).

A camera positioned immediately at the output vacuum window of the test chamber images the spatial profile of the beam transmitted through the cavity. We initially align the cavity by turning the cavity alignment screws in order to make the profile of the output beam (mostly just transmitted non-resonant light) as circularly symmetric as possible. At this point, the cavity modes become visible when scanning the laser frequency over more than one cavity free spectral range ($\sim 7.5$ GHz). We then slowly advance each micrometer screw one at a time in rotational increments of roughly 90°, decreasing to just a few degrees as the cavity approaches concentricity. This keeps the cavity roughly in alignment, though the mode position moves as each micrometer screw is advanced. Even with the 50 µm thread pitch micrometer screws, this process becomes quite sensitive as the cavity approaches concentricity. Once the target distance to concentricity of 3.7 µm is well within the 6.5 µm extension range of the cavity alignment piezos, we extend the piezos to achieve the target cavity mode numerical aperture of 0.05. We infer the numerical aperture by measuring the resonance frequency spacing between transverse modes of the cavity (see section 6.3.2). In this case we measure the frequency spacing by scanning the laser frequency and measuring the optical power reflected from the cavity input as a function of time using a photodetector (PD4 in figure 6.1). Dips in the cavity reflectivity correspond to cavity modes. We phase modulate the input laser beam (using by the spectrum analyzer shown in figure 6.1) such that reflectivity dips also appear at the cavity modes' sideband frequencies. We then adjust the sideband frequency such that the dips from the sidebands of the fundamental mode overlap with the dips (from the carrier frequency) of the first order modes. The resulting sideband frequency represents a measurement of the frequency spacing between the fundamental and first order modes, accurate to within the cavity linewidth.

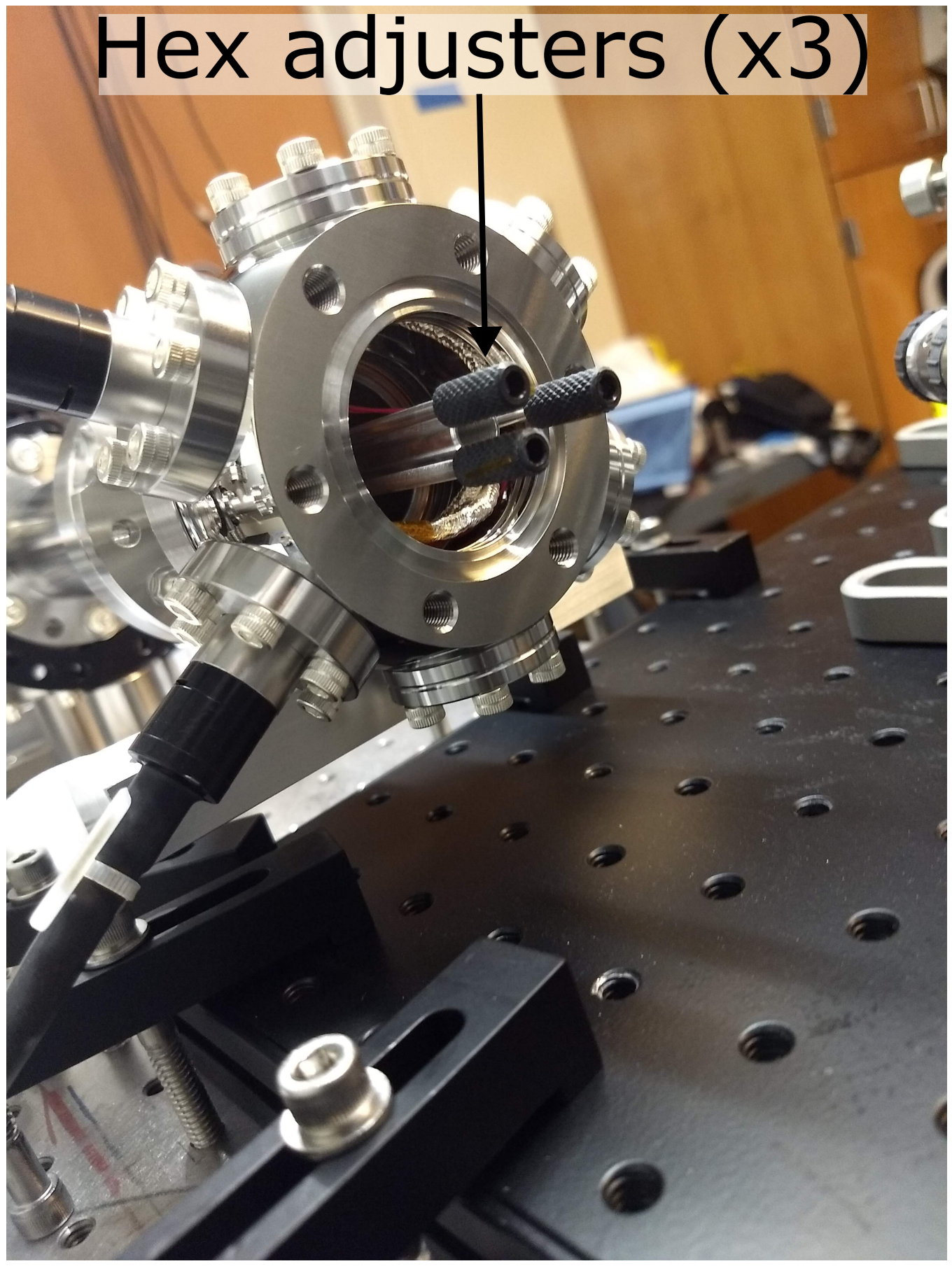

Figure 7.1: Hex adjusters in position for turning the cavity alignment micrometer screws. The input laser beam passes between the handles of the three adjusters.

Once the target numerical aperture of 0.05 has been achieved, the hex adjusters are removed, the input vacuum window replaced, and the vacuum system pumped down. We then measure the cavity finesse as a function of mode position using the cavity ringdown map method (see section 6.3.1). This lets us find the cavity mode position with the lowest round-trip losses. Higher scattering from the mirrors leads to more heating of the cavity mount, which (in absence of our feedback systems, see chapter 6) destabilize the cavity alignment and lead to longer thermal equilibration times. Higher absorption on the mirrors leads to more thermoelastic deformation of the mirror which reduces the numerical aperture of the cavity mode (see section 2.6).

We have successfully used the CRD map measurements to identify defects on the cavity mirrors (see section 7.2). When we observe a defect via the CRD map, we either move the cavity mode as far as practicable from the defect (to reduce the laser intensity it will be exposed to), or replace the cavity mirror(s).

A CRD map of our cavity is shown in figure 7.2. Darkfield images of the mirror surfaces taken before assembly are shown in figure 5.8. The CRD map shows that the round-trip loss is near its minimum possible value of $\sim 160 \times 10^{-6}$ (due to the nominal mirror transmission of $80 \times 10^{-6}$) over much of the measured area. However, areas towards the top of the map have noticeably higher losses. In this case, we moved the mode away from this area by as much as possible, while still keeping the mode close enough to the optical axis to achieve a reasonable input coupling efficiency. This resulted in us using a mode position of approximately ($-0.2$ mm, $-0.5$ mm), which places the center of the mode roughly 1.2 mm from the putative defect in the upper right corner of the map. This would make the laser intensity at the defect only $1 \times 10^{-5}$ that at the center of the cavity mode, assuming a mode waist of $w_1 = 0.5$ mm (NA = 0.05) on the mirror surface.

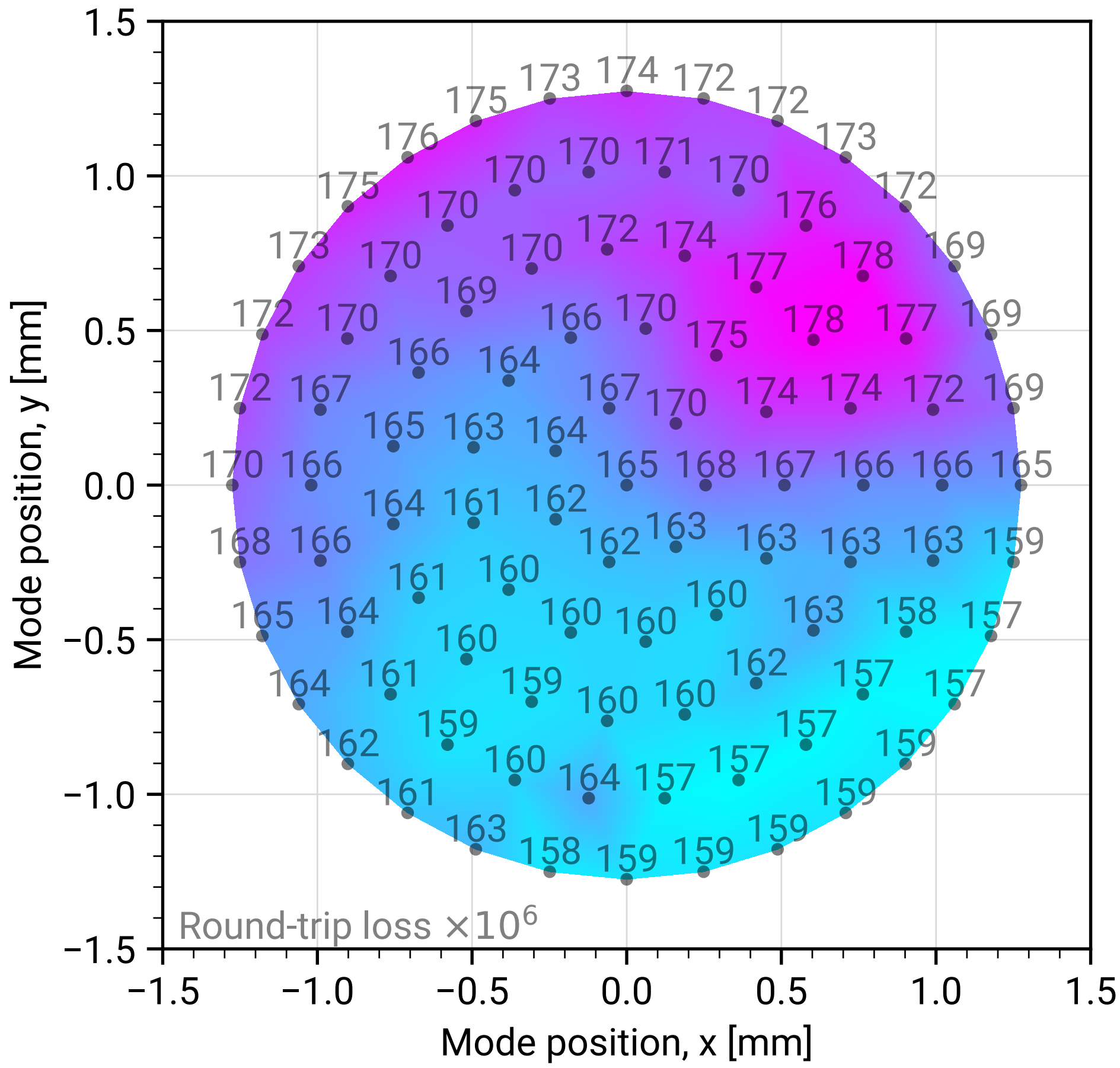

Figure 7.2: Round-trip losses in the cavity as a function of mode position on the mirror surfaces. The center of the map does not necessarily correspond with the center of the mirror surfaces.

The input coupling efficiency is then optimized for the chosen mode position by adjusting the input beam position, angle, divergence, and diameter. This 6-parameter optimization is done by optimizing the input beam position and angle via a standard "beam walking" procedure using mirrors M8 and M9 (figure 6.1) for 9 different configurations of the input beam telescope (3 different positions of each of the lenses L1 and L3 in figure 6.1, two at the opposite ends of their adjustment ranges and one at the middle). The telescope configuration which yields the maximum input coupling efficiency is then used. This sometimes requires changing to different focal length telescope lenses, if the optimum appears to be outside of the adjustment range of the lens positions. We measure the input coupling efficiency by measuring the cavity reflectivity on photodetector PD4 while scanning the laser frequency over the fundamental mode resonance frequency (see section 2.3.4). We typically achieve an input coupling efficiency of $\sim 70\%$ while operating with a circulating power of 75 kW.

The photodetectors used to measure the input, reflected, and output optical power (PD2, PD3, PD7 in figure 6.1) are calibrated by using a separate photodetector power meter to measure the ratio between the power in the beam at the desired location and at the location of the measurement photodetector. For the input power, this measurement is made just outside of window W1. For the reflected power, we measure the power of the input beam and assume that 100% of it is reflected when the laser frequency is off the cavity resonance (strictly speaking, we actually lose a few percent due to reflections off of the convex surface of the input cavity mirror and window W1). For the output power, we measure just outside of window W2. We infer the circulating power in the cavity by dividing the calibrated output power by the nominal transmission of the output mirror ($80 \times 10^{-6}$), as measured by FiveNine Optics on a plano-plano witness sample from the same coating run.

## 7.2 Laser-induced mirror damage

The surface of the mirrors can be damaged by exposure to the high circulating power in the cavity. We have seen this occur on at least three separate occasions, each time when the mirrors were first exposed to high ($> 10$ kW) circulating powers. This section will review one of these incidents in detail.

During assembly of the LPP prototype discussed in this thesis, inspections of the mirror surfaces before installation showed several small defects within the clear aperture. We have found this to be fairly typical. We believe that these defects are introduced during the coating process since we do not see them in visual inspections of the mirror surfaces before coating, and coating-related defects have been reported elsewhere [72], [82]. They are not removed by cleaning of the mirror surface with First Contact polymer (see section 5.5 for details on that cleaning procedure). As such, we simply try to choose mirrors for the cavity which have the fewest, smallest defects. As will be discussed later in this section, it is also very important that these defects be as far from the presumptive cavity mode position as possible.

An initial CRD map of the cavity formed by these mirrors is shown in figure 7.3a. It is evident that a relatively small defect (or defects) incurring additional round-trip losses of $> 8 \times 10^{-6}$ is present off the lower-right edge of the map. The first time we increased the circulating power in the cavity, at least one of the mirrors was damaged by the laser beam at a circulating power of 30 kW and numerical aperture of 0.048. The damage occurred essentially instantaneously on the timescales of our measurements ($< 1$ s), and resulted in the laser frequency feedback system unlocking. When we re-locked the laser frequency to the cavity resonance, we saw that the cavity's power enhancement factor was considerably lower than nominal, and the transverse mode spectrum showed resonances with noticeably larger linewidth (see figure 7.4).

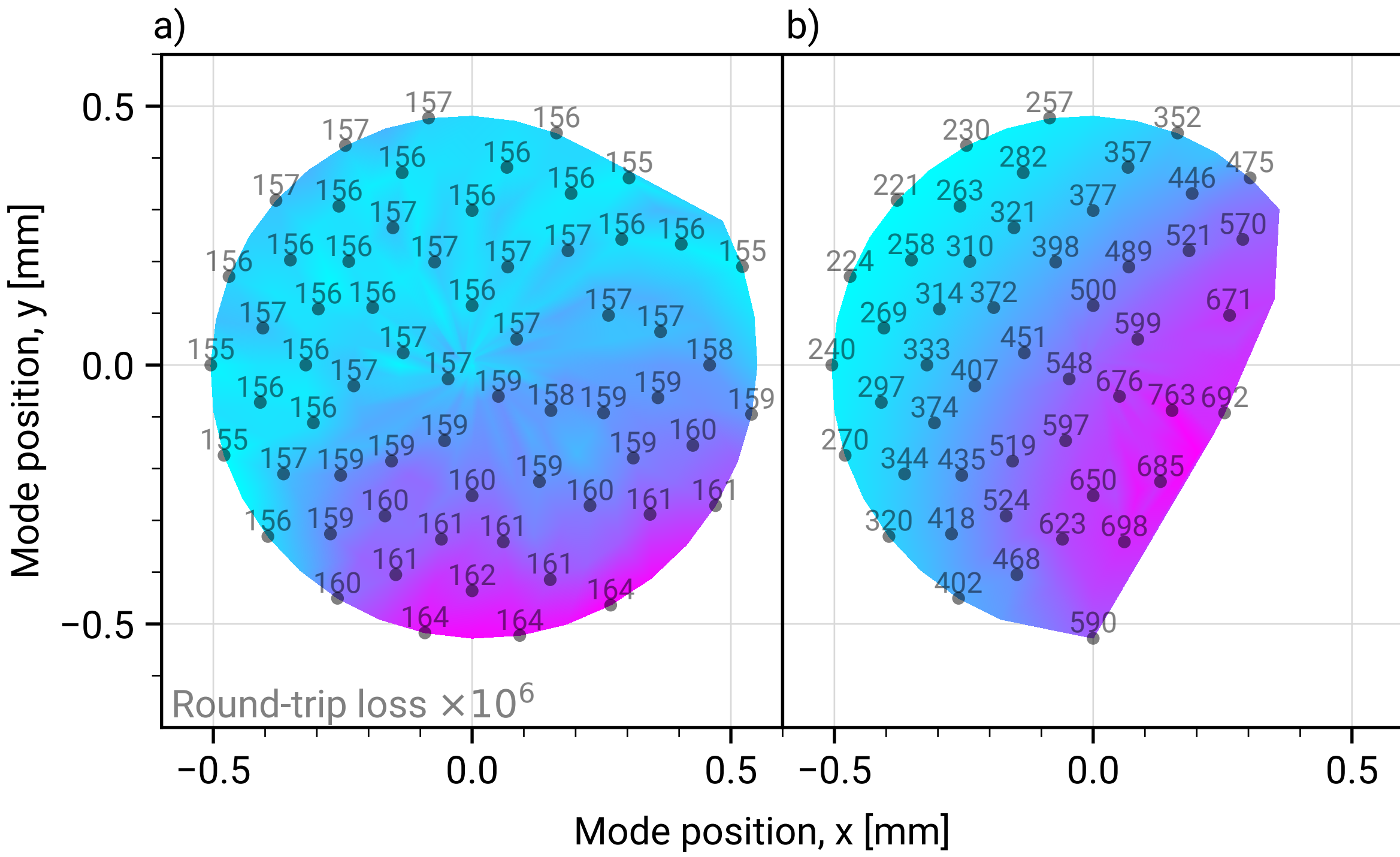

Figure 7.3: Cavity ringdown maps of the cavity **a)** before, and **b)** after laser-induced damage. The origins of each map represent the same nominal position on the mirror surfaces. Note that the color scales are different for panels a) and b). Annotations show the round-trip losses for some of the data points making up the color map.

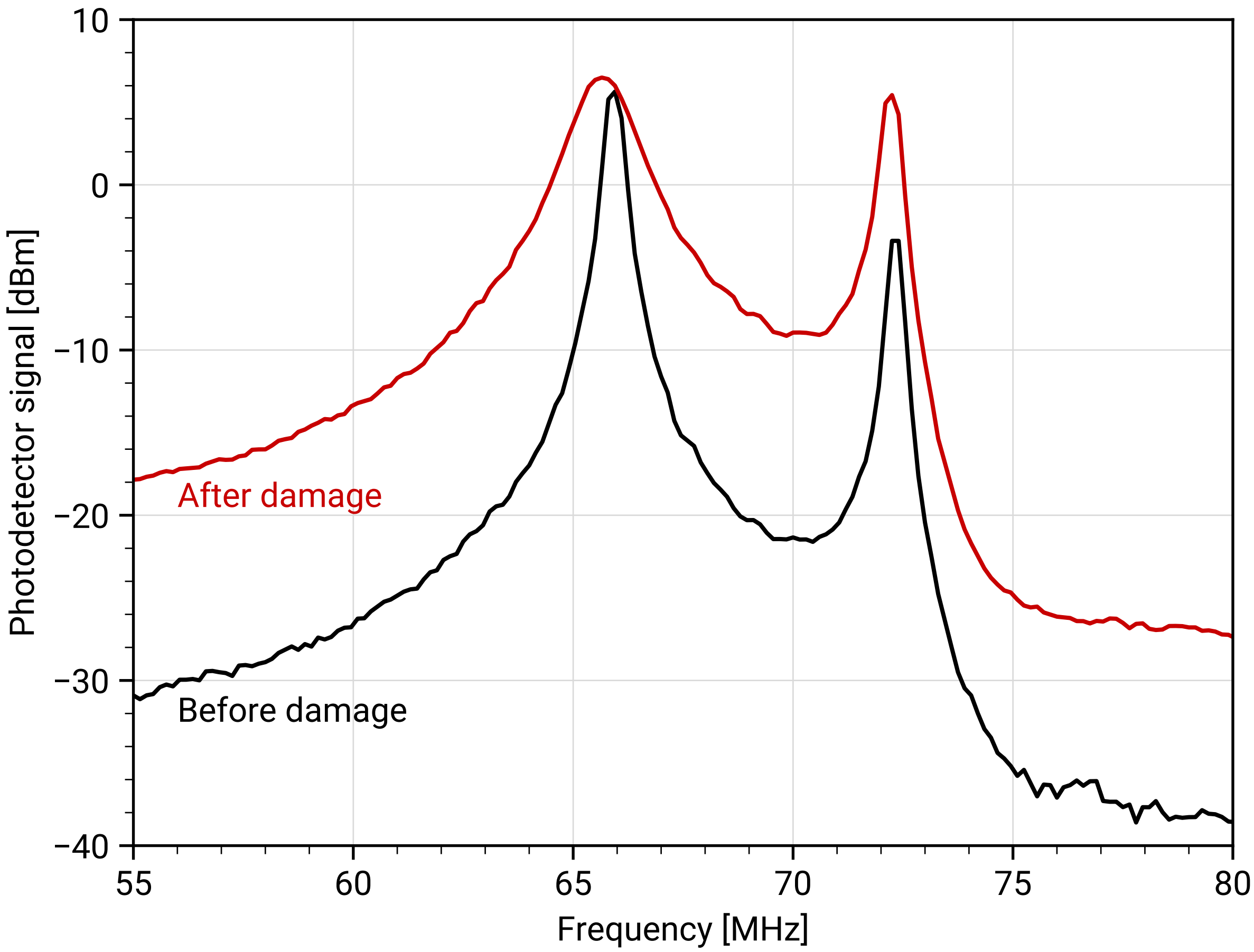

Figure 7.4: Spectrum of the first order cavity modes immediately before (black) and then after (red) the laser-induced damage event.

A subsequent CRD map showed dramatically higher round-trip losses in excess of $700 \times 10^{-6}$, with a spatial profile roughly corresponding to the same location as the small defect which was present before the high power run (see figure 7.3b). A map over a larger range of mode positions showed the full extent of the defect, which caused up to $\sim 1000 \times 10^{-6}$ of round-trip loss (see figure 7.5). Note that the CRD map represents a convolution of the cavity mode profile (Gaussian) with the defects on the mirror surfaces, so that even a point defect appears in the map with roughly the size and shape of the cavity mode.

The cavity mirrors were then removed from the cavity mount and inspected. We found that one pre-existing defect on the input mirror (serial number 3.4.2) had become damaged by the laser beam, causing the additional round-trip losses in the cavity. Figure 7.6 shows images of this defect (annotated “1”) before and after exposure to the high power beam. The output mirror (serial number 3.3.7) was undamaged. The images have been registered based on the locations of other defects on the mirror surface and chamfer edge. Figure 7.7 shows a higher-magnification image of the damaged defect, which exhibits a concentric structure and surrounding small defects (which is perhaps material ejected from the primary defect during the damage process) which seems to be characteristic of laser-induced damage on optical surfaces [72], [83].

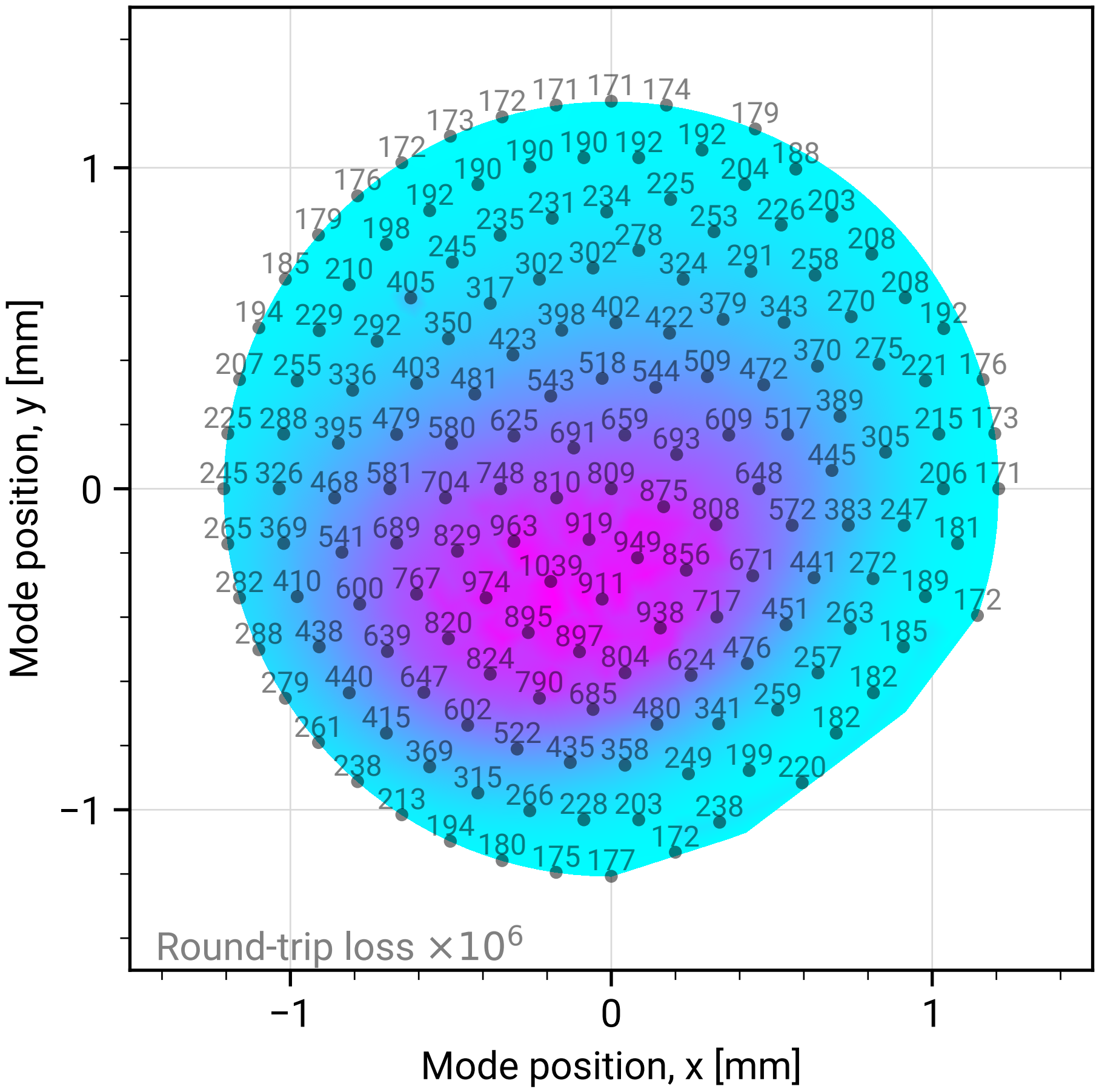

Figure 7.5: Cavity ringdown map of the cavity after laser-induced damage. This map covers a larger mirror area than that in figure 7.3, but is not centered on the cavity mode position that was used when the damage event occurred.

The mirror was then replaced (by serial number 3.4.4) and the cavity reassembled. Both mirrors were first cleaned using First Contact polymer (see section 5.5). The surfaces of both of the mirrors used in the reassembled cavity are shown in figure 5.8, and the corresponding

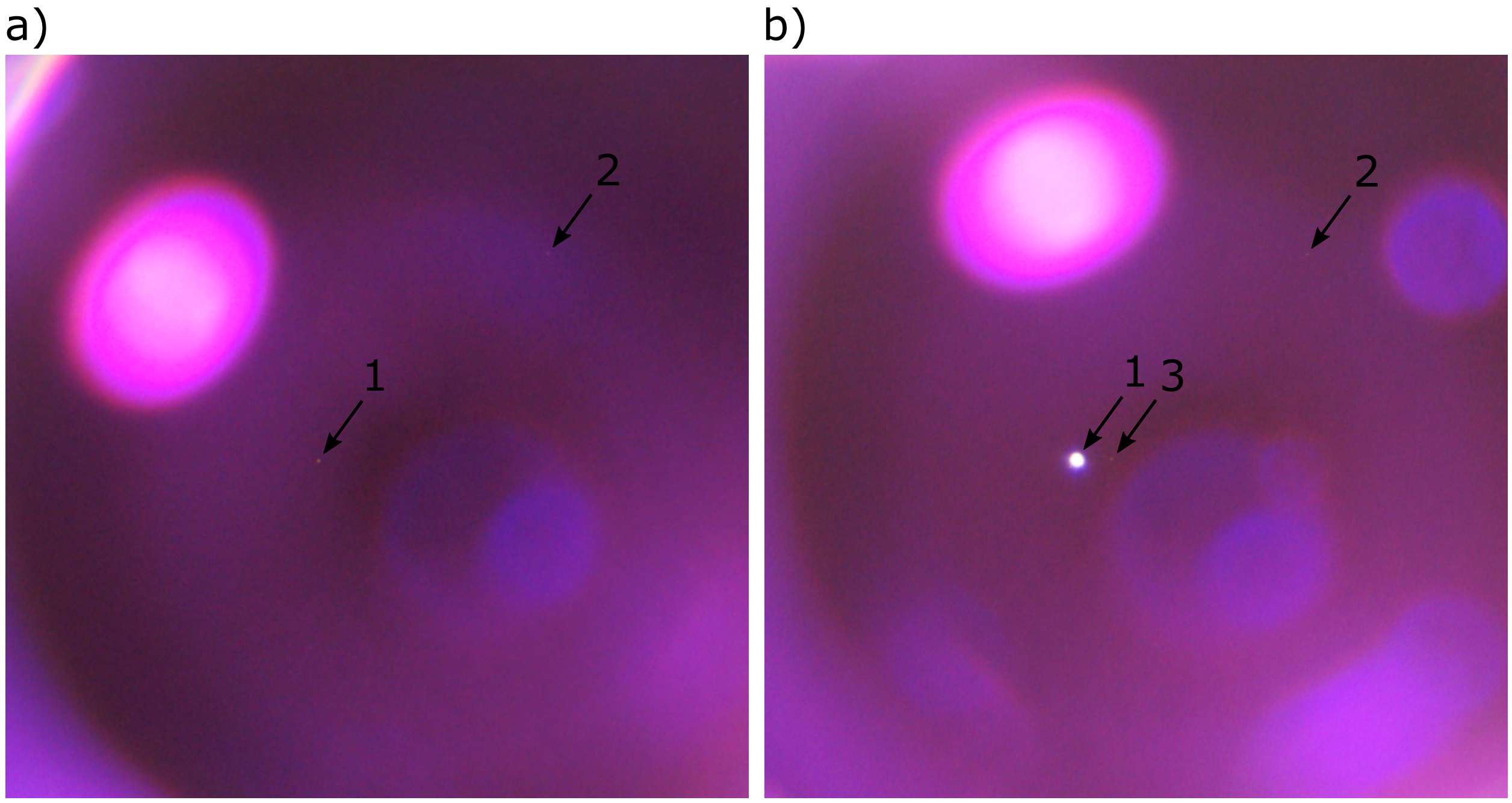

Figure 7.6: Microscope images of the center of the cavity input mirror's (serial number 3.4.2) surface **a)** before, and **b)** after being damaged by the cavity mode. The field of view of both images is 1 mm × 1 mm. Arrows indicate the location of defects. Defect 1 was damaged by the cavity mode, while defect 2 was not. Defect 3 appears after the damage event.

CRD map in figure 7.2.

Clearly, achieving a functioning LPP requires that the area around the center of the cavity mode be free of defects which may be damaged by the laser beam. By assuming that such a defect is damaged at some threshold laser intensity $I_{\mathrm{LID}}$, we can estimate how large this area must be. If the LPP is to provide a $\pi/2$ phase shift, its circulating power $P_{\mathrm{c}}$ and numerical aperture NA must satisfy (see equation (9.26))

$$P_{\mathrm{c}} = \sqrt{\frac{\pi^5}{8}} \frac{\hbar c^2}{\alpha} \frac{1}{\lambda_{\mathrm{e}} \lambda_{\mathrm{L}} \mathrm{NA}} \tag{7.1}$$

In the near-concentric limit, the mode waist on the mirror surfaces $w_1$ is

$$w_1 \simeq \mathrm{NA} R_0 \tag{7.2}$$

where $R_0$ is the radius of curvature of the mirror surface. The laser intensity on the mirror surface $I_1$ is a function of the distance from the center of the cavity mode $r_1$ such that

$$I_1 (r_1) = \frac{2P_{\mathrm{c}}}{\pi w_1^2} e^{-2\left(\frac{r_1}{w_1}\right)^2} \tag{7.3}$$

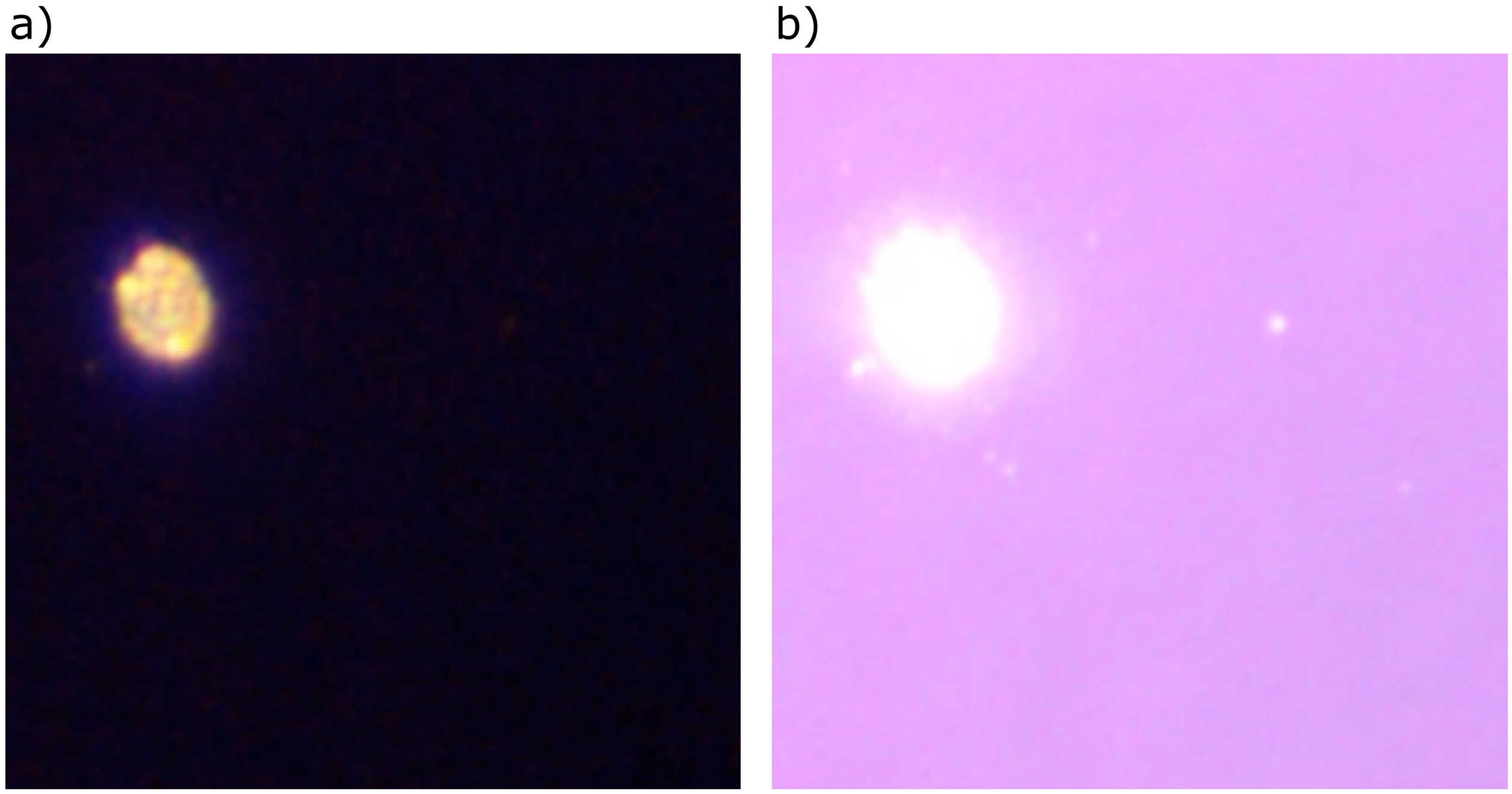

Figure 7.7: Microscope images of the damaged defect on the cavity input mirror (serial number 3.4.2) with a **a)** short, and **b)** long camera exposure. The field of view of both images is 91 µm × 91 µm.

The radial distance $r_{\mathrm{LID}}$ at which this intensity falls below the threshold intensity $I_{\mathrm{LID}}$ can therefore be expressed in terms of the numerical aperture of the cavity mode:

$$r_{\mathrm{LID}} = \sqrt{-\frac{R_0^2 \mathrm{NA}^2}{2} \ln\left(\sqrt{\frac{2}{\pi^3}} \frac{\alpha}{\hbar c^2} R_0^2 \mathrm{NA}^3 \lambda_{\mathrm{e}} \lambda_{\mathrm{L}} I_{\mathrm{LID}}\right)} \tag{7.4}$$

Thus, $r_{\mathrm{LID}}$ represents the radius of the required clear aperture for an LPP which operates at a numerical aperture NA. This relationship is plotted in figure 7.8 for several different values of $I_{\mathrm{LID}}$, using parameters from our cavity ($R_0 = 10\,\mathrm{mm}$, $\lambda_{\mathrm{e}} = 1.97\,\mathrm{pm}$, $\lambda_{\mathrm{L}} = 1064\,\mathrm{nm}$). It shows that there is a numerical aperture for which the clear aperture requirement is maximal. This is because a low numerical aperture results in a higher intensity, but over a smaller area (and vice-versa). In our case, the CRD maps in figures 7.3 and 7.5 show that the defect lay roughly 0.5 mm from the center of the cavity mode at the time of the damage event. Since the damage occurred with a circulating power of 30 kW at a cavity mode numerical aperture of 0.048, we estimate that the defect was exposed to a traveling-wave laser intensity of approximately $I_{\mathrm{LID}} \sim 1\,\mathrm{MWcm^{-2}}$ at the time the damage occurred. At this threshold intensity and a numerical aperture of 0.05, a clear aperture radius of $\geq 0.6\,\mathrm{mm}$ is required to achieve a $\pi/2$ electron beam phase shift. Defects with threshold damage intensities of $\geq 20\,\mathrm{MWcm^{-2}}$ (if such defects exist) should not be damaged by the laser beam, regardless

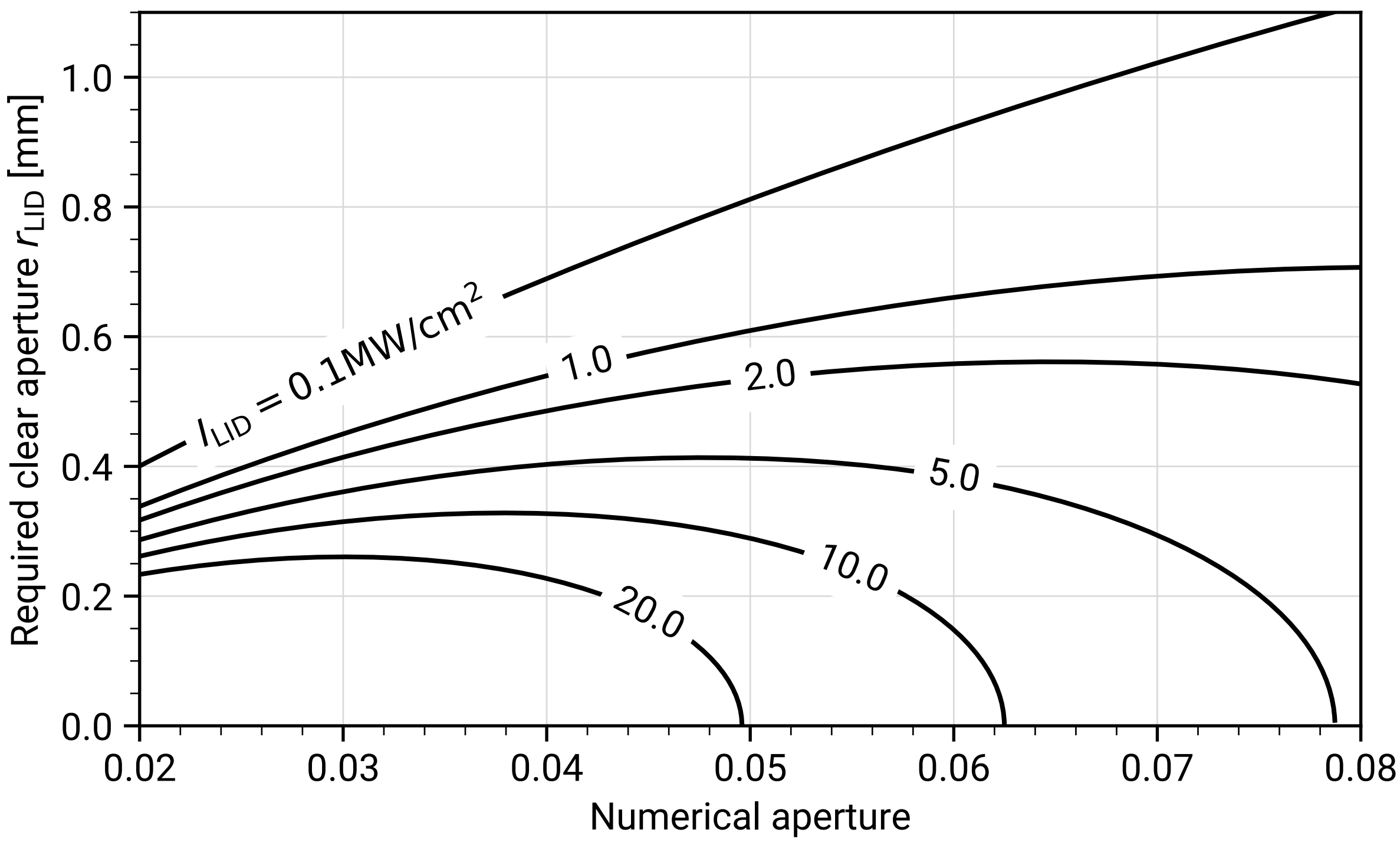

Figure 7.8: The required clear aperture radius $r_{\mathrm{LID}}$ to avoid damaging a defect susceptible to a threshold intensity $I_{\mathrm{LID}}$ as a function of the cavity mode numerical aperture. The corresponding circulating power is set such that the resulting cavity mode focus provides a $\pi/2$ phase shift for electrons with a wavelength of $\lambda_{\mathrm{e}} = 1.97\,\mathrm{pm}$. The laser wavelength $\lambda_{\mathrm{L}}$ and mirror radius of curvature $R_0$ are taken to be 1064 nm and 10 mm, respectively.

of their location.

In summary, we have found that choosing mirrors with the largest possible clear aperture is more important than choosing mirrors which have fewer or smaller defects in total. That is, a mirror with a large defect $> 0.6\,\mathrm{mm}$ or so from the optical axis is likely preferable to one with a small defect near the optical axis. Indeed, on prior cavity designs we have seen that even mirrors with laser-induced damage spots can still perform adequately so long as those damaged spots are sufficiently far from the cavity mode as to not severely increase the round-trip losses. Another important point is that we have so far only seen this kind of damage occur the first time the circulating power is increased. Once a cavity has been tested at high power, its mirrors do not appear to degrade in quality (see section 7.5 for details).

## 7.3 Representative data

We only run the LPP's laser system when we are using the LPP. Given that we have not seen any long-term degradation of the cavity performance (see section 7.5), it should also be possible to leave the entire LPP system running 24/7 at a circulating power of 75 kW and numerical aperture of 0.05. In practice, this would require making the automatic laser frequency re-locking system (see section 6.2.3) more robust so that the system could recover from an unlocked state without human intervention. For now, we simply turn on the system at the beginning of each working day, and turn it off at the end. We refer to the time the system is operating as a "run". Starting a run involves first turning on the laser system and letting the fiber amplifier and input breadboard optics thermally equilibrate for $\sim 15$ min. Engaging the laser frequency (and other) feedback systems takes another minute or so. Increasing the input laser power to reach the target 75 kW circulating power takes another minute or so, at which point we typically wait a few minutes for the cavity temperature to stabilize (after manually turning off the current to the vacuum heater tape). After this, the system rarely requires any human intervention, unless there is a laser frequency unlocking event, which occurs perhaps every few days or so. Recovering the system from an unlocking event takes a few minutes, unless it goes unnoticed for long enough for the cavity to substantially cool, in which case it can take tens of minutes for the cavity temperature to re-equilibrate. Again, we do not see this as a fundamental limitation of the technology—rather, it is simply more time efficient for us at the current stage in the project to manually deal with an unlocking even than work on preventing them or automatically recovering from them.

Turning off the system at the end of the day takes only a minute or so, since it is not necessary to wait for anything to thermally equilibrate. The input laser power is first decreased so that the circulating power drops to $\sim 5$ kW, then the feedback loops are disengaged, and the laser system turned off.

### 7.3.1 Circulating power & numerical aperture long-term stability

Data from a representative run is shown in figure 7.9. The circulating power was set to 75 kW, and is kept at that level using the circulating power feedback system (see section 6.2.2). The cavity mode numerical aperture is not directly stabilized by any feedback loop, and is seen to slowly decrease slightly as the cavity temperature (only roughly stabilized by human feedback) decreases. However, such a small change is inconsequential to the performance of the LPP. The numerical aperture of each principal axis of the cavity mode is shown in figure 7.9 and annotated with the corresponding first-order mode index from which the measurement was made.

The measured circulating power and numerical aperture can be used to calculate the expected electron beam phase shift using equation (9.26). The expected phase shift is shown in figure 7.10, using the data from figure 7.9. The black line represents the expected phase shift based on the geometric mean of the numerical aperture principal axes, while the shaded region indicates the range of possible phase shift values depending on the orientation of the cavity mode astigmatism relative to the electron beam axis (which is extremely stable over time). The standard deviation of the expected phase shift value during the entire run is 1.2°, which has a negligible effect on the LPP imaging properties. This level of stability far exceeds the best that has been achieved using a Volta phase plate [84].

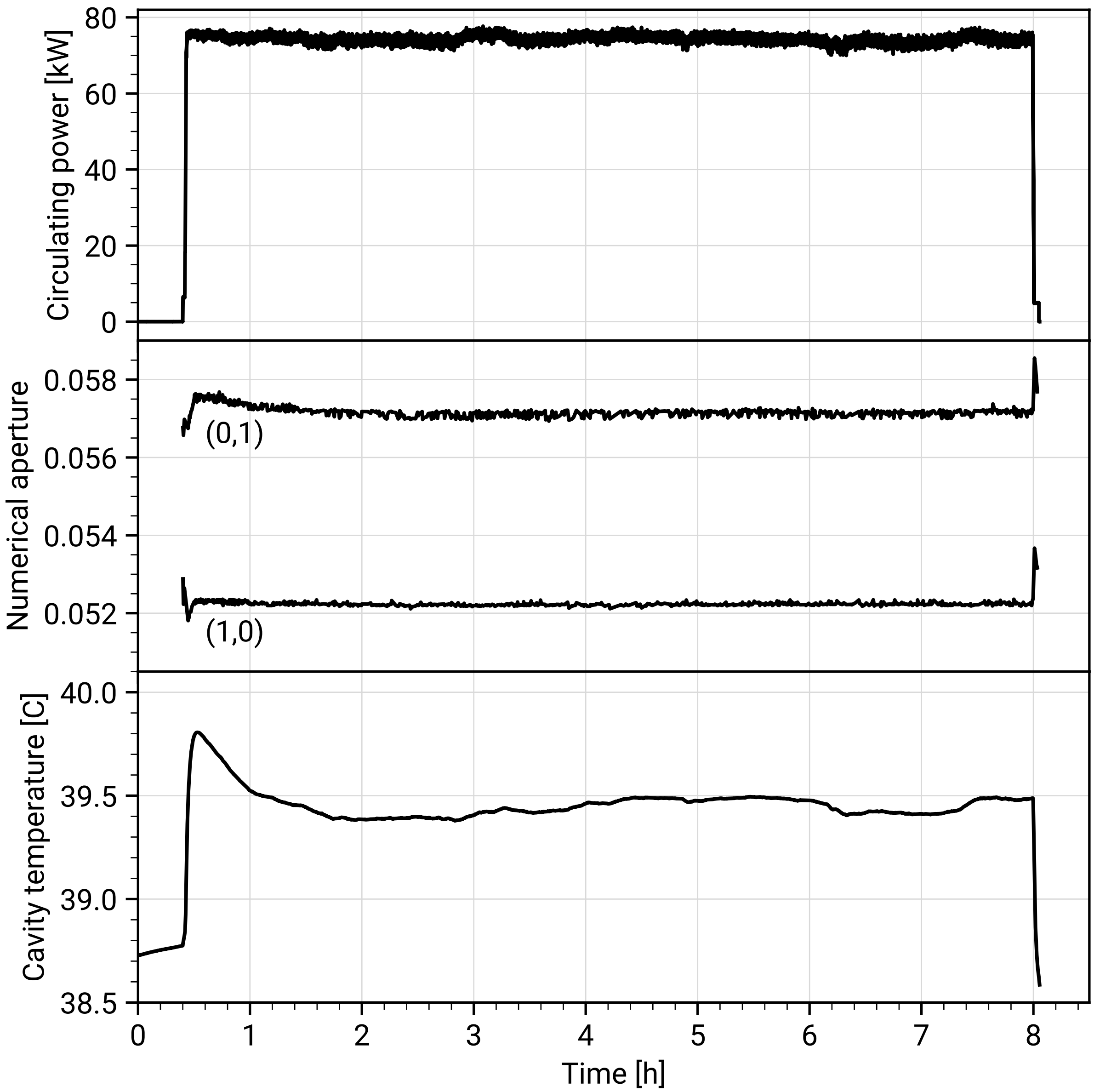

Figure 7.9: Cavity parameters as a function of time over the course of a typical high power run. Top panel: circulating power. Middle panel: cavity mode numerical aperture along each principal axis (measured via transverse mode spectrum of the annotated mode indices). Bottom panel: cavity temperature.

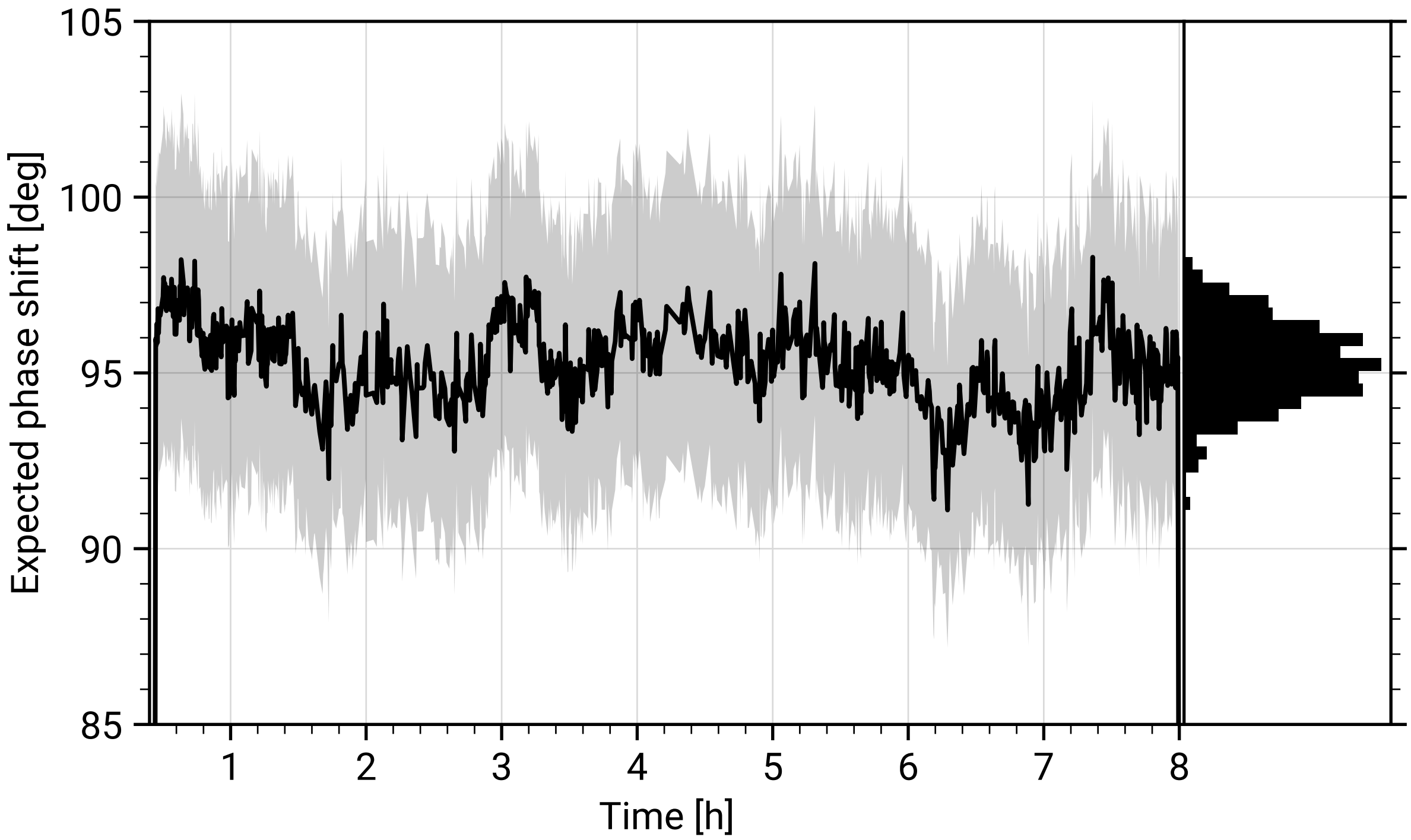

Figure 7.10: Expected electron beam phase shift calculated from the measured cavity circulating power and numerical aperture shown in figure 7.9, using equation (9.26). The black line shows the expected phase shift calculated from the geometric mean of the numerical aperture principal axes, while the shaded region shows the range of possible phase shifts depending on the orientation of the astigmatic cavity mode relative to the electron beam axis. The right panel shows a histogram of the expected phase shift values.

We have found that the electron beam phase shift stability is not limited by the stability of the cavity mode parameters, but rather by the relative alignment between the electron and laser beams. By developing data collection procedures to maintain this alignment (see chapter 10) we have demonstrated excellent (±4.9°) phase shift stability on single-particle analysis cryo-EM samples (see section 12.1).

### 7.3.2 Circulating power short-term stability

It is important that the electron beam phase shift generated by the LPP is stable over the typical exposure time (1 s to 10 s) for a TEM image. This ensures both that the phase shift is as close as possible to the desired value, and that there is no blurring of the image (resolution loss) from a changing image transfer function (see section 8.7.5.2). The LPP parameters which affect the electron beam phase shift are the laser wavelength, cavity mode numerical aperture, and circulating power (see equation 9.26). We know that the fractional change in laser wavelength over the exposure timescale is negligibly small ($\ll 1 \times 10^{-6}$). We also know that the cavity mode numerical aperture must be relatively stable on these timescales, as our transverse mode spectrum measurement operates on a $< 1\,\mathrm{s}$ timescale and shows realistically small linewidths which do not appear to have been blurred by a fluctuating transverse mode frequency spacing.

We quantified the stability of the circulating power by measuring the cavity output power. Since the cavity output power is proportional to the circulating power (via the output mirror transmissivity), the relative noise on the output power (relative to the mean output power) should be the same as that on the circulating power, with the exception of shot noise, which has a smaller relative magnitude on the circulating power because the circulating power is higher than the output power. We used photodetector PD6 (see figure 6.1) directly connected to a $50\,\Omega$ impedance and oscilloscope to measure the fluctuations in the output power. The measurement bandwidth of 5 MHz was limited by the oscilloscope. The output breadboard photodetector optical power feedback system (see section 6.2.1) and circulating power feedback system (see section 6.2.2) were disabled for this test so that they would not affect the measurement. That is, the measurement represents the stability of the circulating power without any feedback, other than that keeping the laser frequency on resonance with the cavity.

Figure 7.11 shows the relative root-mean-square (RMS) of the noise on the output power as a function of the measurement time, with a circulating power of 75 kW. The output power photodetector signal is normalized to have a mean of unity over the measurement time, so that the RMS is relative to the mean output power. Though the RMS grows appreciably as the measurement (exposure) time increases through 1 s, even its largest value is negligibly small: a relative RMS of 0.00285 on the electron beam phase shift (proportional to the cavity output power) around a mean value of $\pi/2$ is an absolute phase shift RMS of just 0.0045, which induces a contrast transfer function envelope with a value of $1 - 1 \times 10^{-5}$, barely different from unity (see section 8.7.5.2). The stability of the alignment between the electron and laser beams is a much greater source of noise in the electron beam phase shift on the timescale of a TEM image exposure time (see chapter 10 and section 12.1). This measurement also shows that the considerably larger relative noise on the circulating power measurements (e.g. figure 7.9) made by photodetector PD7 must be measurement noise not present on the output power. We have not investigated the cause.

Still, the majority of the noise comes from timescales $< 1 \times 10^{-2}\,\mathrm{s}$. To visualize these noise sources, figure 7.12 shows the "relative intensity noise" spectrum of the cavity input and output light with a circulating power of 75 kW. Photodetector PD1 was used to simultaneously measure the noise on the cavity input light. The relative intensity noise spectrum shown is the power spectral density of the photodetector signal after it has been normalized to a mean value of unity. It shows that the output power (circulating power) is noisier than the input power, which is to be expected because the transfer function of the cavity converts frequency (phase) noise on the input light into amplitude noise on the output light. We did not investigate which discrete noise peaks were from technical noise in the photodetectors. We suspect the broad noise peak in both signals around 200 kHz is due to residual amplitude modulation generated by the laser frequency feedback system, since it has a similar peak in its noise spectrum at that frequency (see section 6.2.3). The noise floor at high frequencies is due to shot noise and is different for the two signals because the PD6 was exposed to a higher optical power than PD1.

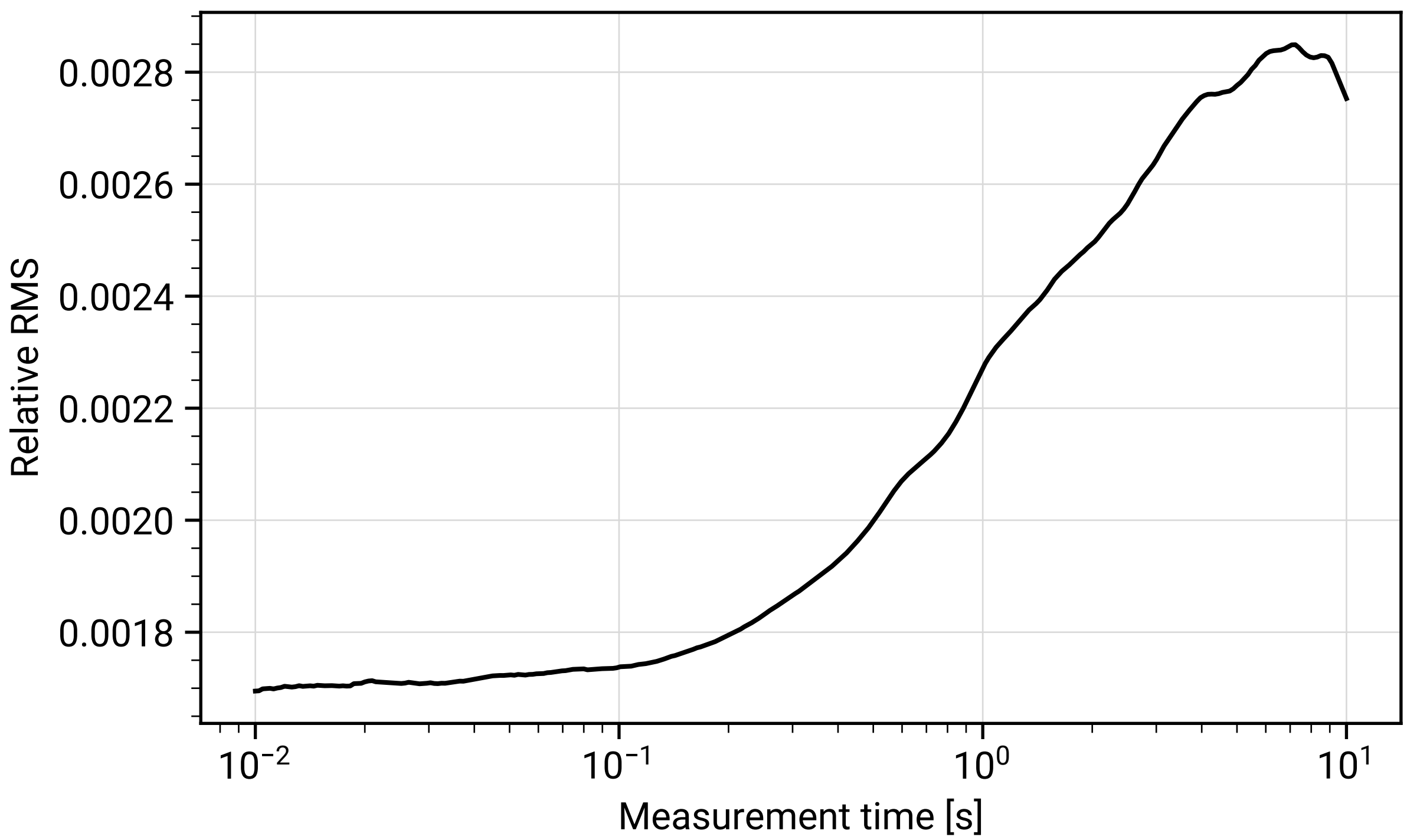

Figure 7.11: Root-mean-square of the noise on the output power relative to the mean output power as a function of the measurement time, with a circulating power of 75 kW.

### 7.3.3 Cavity mode astigmatism

The mode in our cavity is slightly astigmatic. Based on the measurements shown in figure 7.9, we estimate that the astigmatism causes a difference in distance to concentricity between the two principal axes of the mode of 0.93 µm. This value depends somewhat on the cavity mode position. This amount of astigmatism only causes our mode to have an aspect ratio of 1.094 in the data shown in figure 7.9.

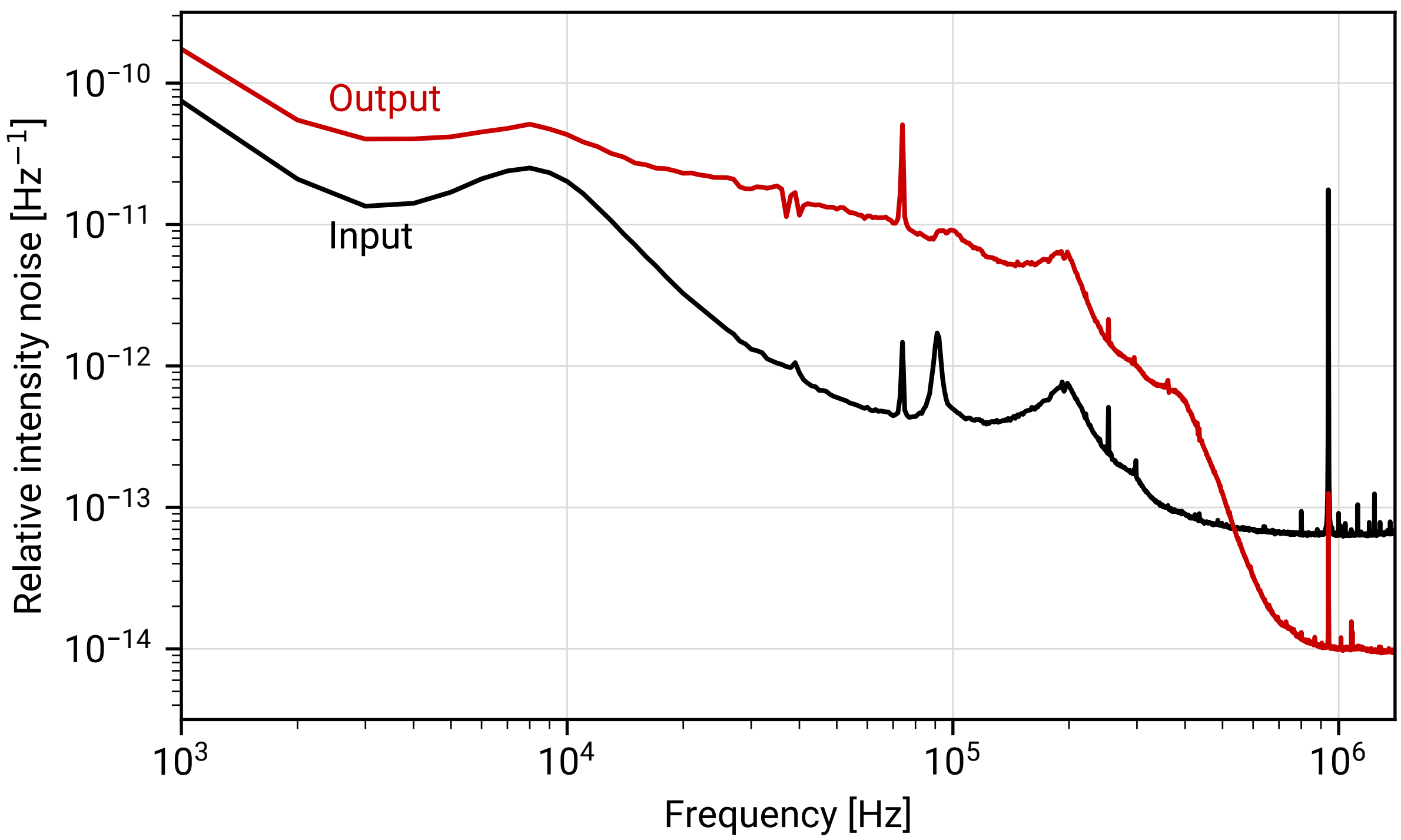

Figure 7.12: Relative intensity noise spectrum of the cavity input and output light, with a circulating power of 75 kW.

## 7.4 Superlative data

For use as an LPP the cavity only needs to achieve a circulating power of 75 kW at a numerical aperture of 0.05. However, we have occasionally operated the cavity at higher powers and/or different numerical apertures. This section presents some of this "superlative" data. These occasions were often unintentional, and we have not systematically explored the maximum possible circulating power and/or numerical aperture, since it is possible that such exploration could damage the cavity mirrors and replacing the mirrors would take time away from our primary goal of developing laser phase contrast TEM. Therefore, the data presented in this section should be regarded as a lower limit on the ultimate values that may be achieved in our system.

### 7.4.1 Highest focal intensity

The highest focal intensity we have achieved in the cavity is 590 GWcm$^{-2}$, when the circulating power was increased to 99 kW, with a cavity mode numerical aperture of 0.052. The higher circulating power was used in an effort to achieve a $\pi/2$ electron beam phase shift in spite of an imperfect alignment between the electron and laser beams (see chapter 10).

We believe that this is the highest continuous-wave or time-averaged laser intensity ever achieved. Other projects have achieved higher circulating powers [48], [85] but never with as small focal waists. For context, this light intensity would exert a photon pressure of 97 atm on a reflecting object. It is equal to the surface intensity of blackbody radiation emitted by an object with a temperature of 570 000 K. As such, the focus of our optical cavity is the continuously brightest place in the solar system outside of the core of the sun.

### 7.4.2 Highest circulating power

The highest circulating power of 175 kW was achieved on the cavity prototype previous to the one described in this thesis. Its design is nearly identical, as only small changes were made when upgrading it to the current design. Importantly, it uses the same mirror design (see chapter 3).

However, we found that these mirrors underwent substantially more thermoelastic deformation due to surface heating from the cavity mode, likely due to contamination on the mirror surface leading to higher absorption. This caused the numerical aperture of the cavity mode to decrease with increasing circulating power much more quickly than in the current prototype cavity (see section 7.6). Rather than attempt to maintain the numerical aperture by lengthening the cavity using the cavity alignment piezos, we simply increased the circulating power further to achieve a $\pi/2$ electron beam phase shift (with numerical aperture $< 0.05$). This resulted in typical circulating powers of $\sim 125$ kW.

The highest circulating power was achieved when an incorrect value for the cavity output mirror transmissivity was entered in the experiment control software graphical user interface. This made the displayed value of circulating power an underestimate, and so when the circulating power was set to its nominal value, it was actually higher than intended.

### 7.4.3 Highest mirror intensity

The highest circulating power also resulted in the highest laser intensity at the mirror surfaces of 92 $\mathrm{MWcm^{-2}}$, as thermoelastic deformation of the mirror surface at that high power had substantially reduced the cavity mode numerical aperture to 0.0345, thereby decreasing the size of the cavity mode on the mirror surfaces.

## 7.5 Operational statistics

We have operated the LPP in the TEM at or near its nominal circulating power and numerical aperture for hundreds of hours. In that time, we have not seen any degradation of the cavity performance. This is in spite of several incidents in which the TEM vacuum was accidentally vented (either totally to a filtered nitrogen source, or partially to room air) while the LPP was operating. This demonstrates that the LPP is not so delicate as to be useless as an imaging tool. Time will tell if there is an ultimate operational lifetime for the LPP. There is also no obvious limitation on how long an individual run can be—we have demonstrated continuous runs as long as 10 h.

Figure 7.13 shows the number of hours each LPP prototype has been operated with a power of at least the value shown on the abscissa. Because the current prototype has mirrors with lower absorption, it can be operated at higher numerical aperture (0.05) and therefore lower circulating power (75 kW) as compared to the previous prototype. Between the two prototypes, the system has been operated at a circulating power $> 50$ kW for more than 1000 h. Strictly speaking, the data shown in figure 7.13 is a lower limit on the hours in operation because occasionally data from a run is accidentally not logged or is corrupted.

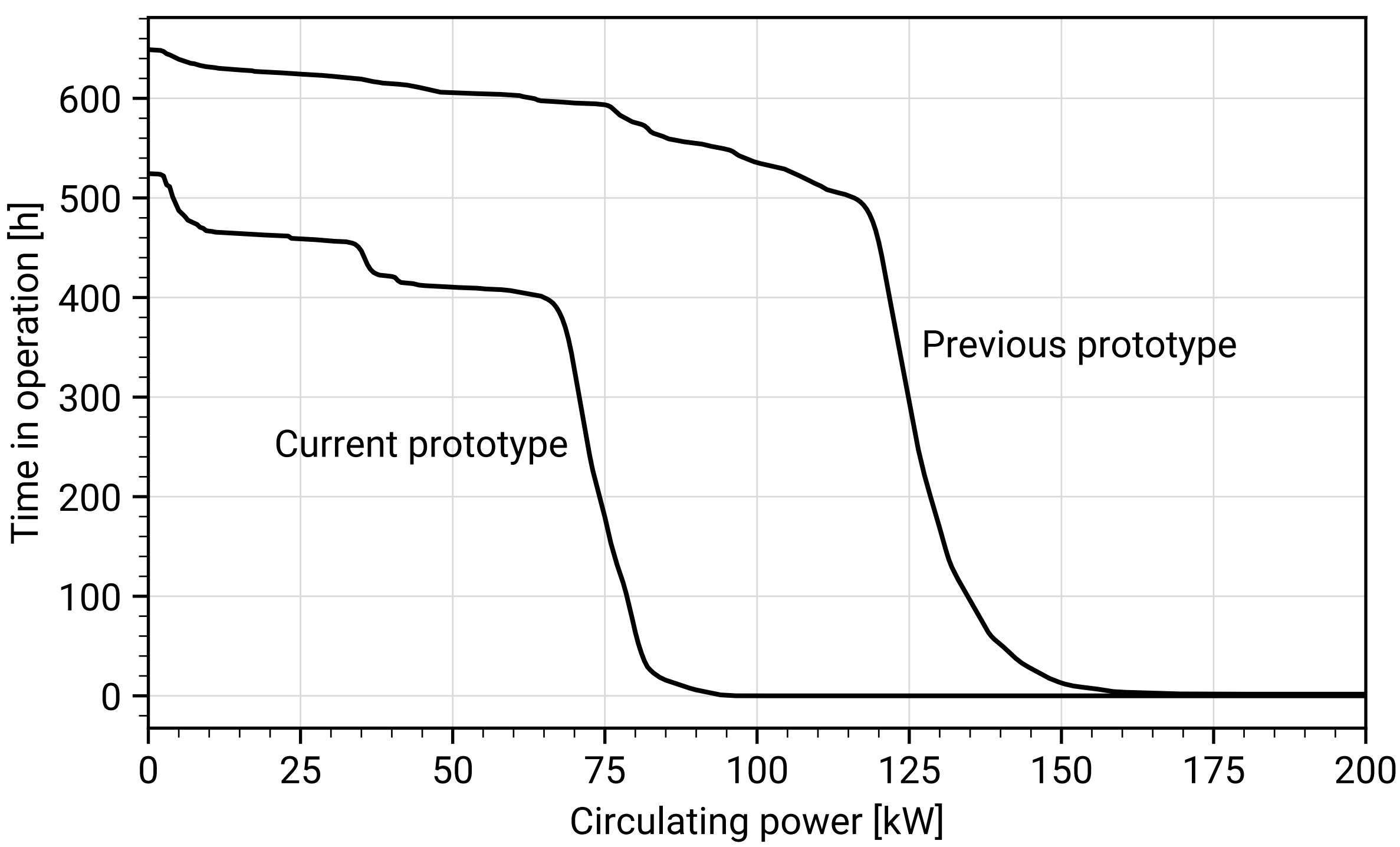

Figure 7.13: The number of hours that the laser phase plate has been operated with a circulating power of at least that shown on the abscissa. Data is shown for both the current and previous prototypes of the laser phase plate.

## 7.6 Mirror thermoelastic deformation

The cavity mode heats the concave surface of the mirrors due to absorption in the reflective coating. As is shown in section 2.6, this leads to thermoelastic deformation of the mirror substrate which changes the curvature of the mirror surface in the vicinity of the cavity mode. This in turn changes the spatial profile of the mode. To leading order, this effect results in a cavity mode numerical aperture which depends on the circulating power. We have quantified the magnitude of this effect in both our previous and current cavity prototypes.

We measure the change in mirror radius of curvature (ROC) by measuring the cavity's transverse mode spectrum and laser wavelength. The frequencies of the (first-order) transverse modes are translated to the corresponding cavity stability parameter via equation 2.37. Changes in the cavity length during the experiment (due to thermal expansion of the cavity mount when the circulating power is changed) are inferred from the measured changes in the laser wavelength. A nominal cavity length of 20 mm is assumed since the results of the measurement only depend on changes in the cavity length around this nominal value. The mirror radius of curvature can then be calculated from the cavity stability factor and cavity length measurement via equation (2.21). This procedure is applied to both the $(1, 0)$ and $(0, 1)$ cavity modes, and the resulting radii of curvature averaged.

This data is then fit with the model described by equation (2.83), where the laser wavelength $\lambda_{\mathrm{L}}$, cavity length $L$, and circulating power $P_{\mathrm{c}}$ are measured for each data point, and the zero-circulating power mirror radius of curvature $R_0$ (equivalent to the zero-circulating power stability parameter $g_0$) and the mirror distortivity parameter $\mathcal{M}$ are used as fit parameters.

The results of the measurement for the previous prototype cavity are shown in figure 7.14. The fit shows reasonable agreement with the data, though there is a small systematic error in the residuals which causes the fit to underestimate the radius of curvature change at low and high circulating powers. We have not yet determined the source of this error. One possibility is that it is due to the positive temperature dependence of the ULE7972 mirror substrates' coefficient of thermal expansion (CTE), which should further increase the radius of curvature change at high circulating power. It could also be from the model approximating the change in mirror surface due to thermoelastic deformation as being simply a change in radius of curvature.

As expected, the data shows an increase in mirror radius of curvature with increasing circulating power. The nonlinearity of the increase is due to the lower numerical aperture mode (at higher radius of curvature) causing yet more thermal deformation of the mirror substrate. The fit model has a mirror distortivity parameter of $\mathcal{M} = 4.21 \times 10^{-14}\,\mathrm{mW^{-1}}$. According to the argument presented in section 2.6, this should restrict the cavity focal intensity to a maximum of $I_{0,\mathrm{max}} = 605\,\mathrm{GWcm^{-2}}$. Figure 7.15 shows the cavity mode focal waist $w_0$ as a function of circulating power from the data used in figure 7.14. The corresponding standing-wave focal intensities are shown up to the predicted maximum $I_{0,\mathrm{max}}$ of equation 2.89, which illustrates how the cavity achieved an intensity of roughly 75% of the predicted limit.

The mirrors in the current prototype experience much less thermoelastic deformation. We believe that this is because the mirrors in the previous prototype have a higher absorption. During the first attempt to assemble the previous prototype cavity, the mirrors became contaminated with small gold flakes from a radial canted coil spring we intended to use to hold the mirrors in the cavity mount sockets. We cleaned off the contamination by washing the mirror surfaces with pure HPLC-grade isopropanol, but suspect that either the isopropanol left behind a residue, or small gold particulates remain. We believe that the magnitude of thermal deformation ($\mathcal{M} = 9.32 \times 10^{-15}\,\mathrm{mW^{-1}}$) exhibited by the current prototype, shown in figure 7.16 represents the ultimate performance of a ULE7972 mirror substrate with a low-absorption 1064 nm ion-beam sputtered reflective coating. As shown in figure 3.11, the witness sample for the coating run for the current prototype mirrors has an absorption of $0.4 \times 10^{-6}$. The much lower mirror distortivity measured in our current prototype mean that the change in numerical aperture with circulating power is now negligible for LPP applications. It also increases the predicted maximum intensity limit to 2.7 TWcm$^{-2}$. Still, if an application requires higher circulating power, higher numerical aperture, and/or a longer cavity (larger mirror curvature radius), this limit may again become an issue. There are currently no options for high-reflective coatings with absorption $\ll 0.4 \times 10^{-6}$, so in this case the only option is to further reduce the thermal distortivity of the mirror substrate material. Figure 3.3 shows the ratio of CTE to thermal conductivity of several other possible substrate materials as a function of temperature. However, the simplest option may be to simply cool the ULE7972 (or equivalently, ULE7973) substrates to a temperature below their CTE zero-crossing, such that the CTE is negative. As is discussed in section 2.6, a negative CTE should not result in an obvious upper limit to the focal intensity. However, further experimental or simulation work will need to be done to quantify the effects of such thermoelastic deformation on the shape of the cavity mode.

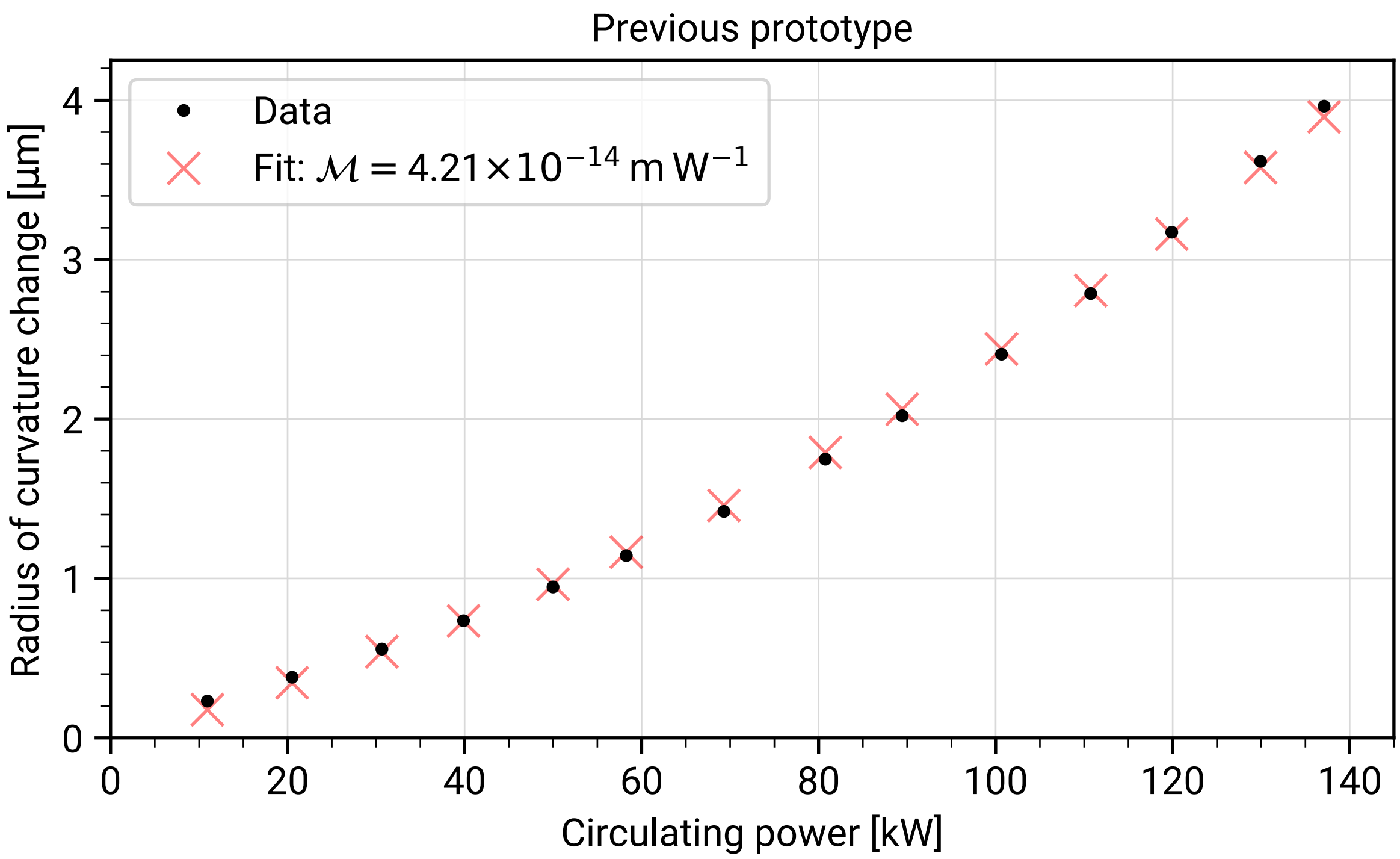

Figure 7.14: Change in the radius of curvature of the mirrors as a function of circulating power using our previous prototype cavity.

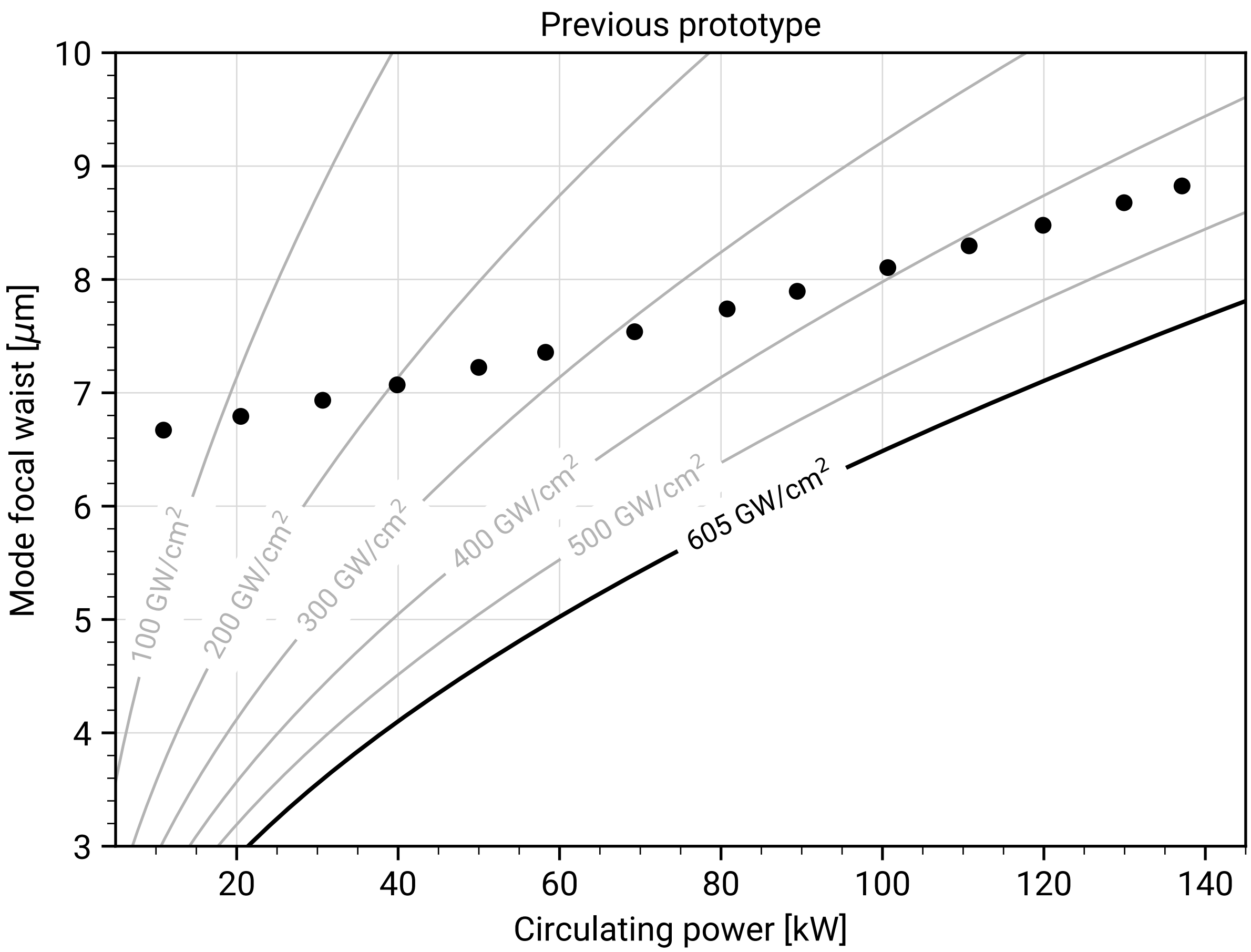

Figure 7.15: Measurements of the mode focal waist as a function of circulating power (black dots) using our previous prototype cavity. The corresponding standing-wave focal intensities are shown with gray lines. The predicted maximum focal intensity is shown with the black line.

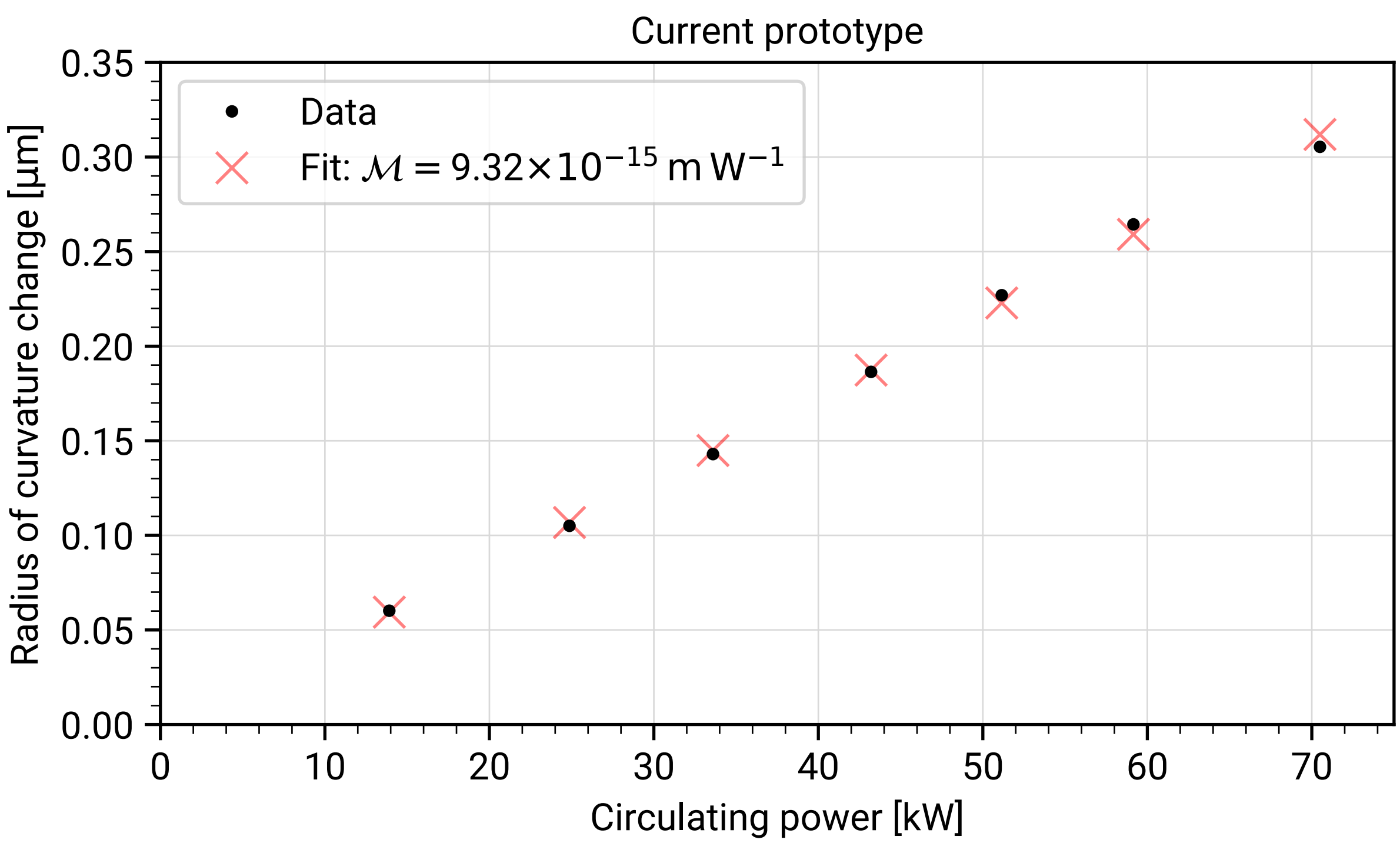

Figure 7.16: Change in the radius of curvature of the mirrors as a function of circulating power using our current prototype cavity.

# Chapter 8

# The contrast transfer function

The contrast transfer function is the core mathematical concept in phase contrast TEM. It describes the extent to which the microscope converts the electron wave's phase modulation induced by the sample in the object plane into amplitude modulation in the image plane. As such, it is of central importance in the understanding and design of phase plates.

This chapter starts by reviewing the derivation and definition of the contrast transfer function in sections 8.1, 8.2, and 8.3. The specific form of the LPP's contrast transfer function is considered in section 8.4, requirements for in-focus imaging using the LPP are given in section 8.5, and image aberrations generated by the LPP are discussed in section 8.6. Section 8.7 reviews well-known mechanisms of resolution loss in phase contrast imaging, and also discusses several new ones specific to the LPP (sections 8.7.5, 8.7.6, and 8.7.8). Finally, section 8.8 models the images which are formed when the LPP does not lie in the diffraction plane.

## 8.1 Electron beam wave mechanics

A description of the contrast transfer function first requires a physical model for the propagation of the wave function of the TEM's electron beam. The most complete description of an electron wave function's propagation uses quantum electrodynamics, which describes the electron as a quantized excitation of a fundamental field. Describing the evolution of a system of charges and electromagnetic (EM) fields between some initial and final state requires calculating the action of the fields in all intermediate configurations and summing the complex exponentials of the actions. This formalism is exceedingly unwieldy.

Approximate solutions can be found by making several assumptions about the system. First, we assume that there is always only one electron. Of course, a TEM's electron beam consists of more than one electron, but they are typically separated by $\sim 10 - 100\,\mathrm{mm}$ along the beam axis (Coulomb interaction energy $\sim 140 - 14\,\mathrm{neV}$) for commonly used beam currents, and so we assume that they are non-interacting, not entangled, and so can be treated separately. The assumption also means that our model cannot take into account any processes which create or destroy electrons, but such processes are negligible for the low-strength static EM field shapes of a TEM's electron-optical components. Second, we assume that the EM field can be treated classically. Again, because the EM fields in the TEM are static and do not vary rapidly in space, this should be an accurate assumption. Third, we assume that the interaction of the electron's spin with the EM field does not change the spatial component of the electron's wave function (its trajectory). This assumption largely holds true in a TEM, which has its EM fields designed such that the Lorentz force is the dominant interaction with the electron [86].

Under these assumptions, the electron's wave function $\bar{\Psi}(\mathbf{x}, t)$ is a scalar function which must obey the Klein-Gordon equation

$$\left(c^2 \left(-i\hbar\boldsymbol{\nabla} + e\mathbf{A}(\mathbf{x})\right)^2 + m_{\mathrm{e}}^2 c^4\right) \bar{\Psi}(\mathbf{x}, t) = \left(i\hbar\frac{\partial}{\partial t} + e\phi(\mathbf{x})\right)^2 \bar{\Psi}(\mathbf{x}, t) \tag{8.1}$$

where $\mathbf{A}(\mathbf{x})$ and $\phi(\mathbf{x})$ are the magnetic vector and electrostatic potentials of the EM field, respectively [87]. Since the potentials are assumed to not vary in time, the time-dependence of the wave function can be separated out by defining

$$\bar{\Psi}(\mathbf{x}, t) =: e^{-i\left(m_{\mathrm{e}}c^2+E\right)t/\hbar}\Psi(\mathbf{x}) \tag{8.2}$$

where $E$ is the energy of the wave function (not including the electron's rest energy), and $\Psi(\mathbf{x})$ represents the spatial dependence of the wave function, which describes the trajectory of the electron. The Klein-Gordon equation can then be rewritten as a generalized version of the Helmholtz equation for $\Psi(\mathbf{x})$:

$$\left(\tilde{\boldsymbol{\nabla}}^2 + k(\mathbf{x})^2\right)\Psi(\mathbf{x}) = 0 \tag{8.3}$$

$$\tilde{\boldsymbol{\nabla}} \coloneqq \boldsymbol{\nabla} + \frac{i}{\hbar}e\mathbf{A}(\mathbf{x}) \tag{8.4}$$

$$k(\mathbf{x}) \coloneqq \frac{1}{c\hbar}\sqrt{\left(m_{\mathrm{e}}c^2 + K(\mathbf{x})\right)^2 - m_{\mathrm{e}}^2 c^4} \tag{8.5}$$

where $K(\mathbf{x}) \coloneqq E + e\phi(\mathbf{x})$ is the kinetic energy of the electron, so that $\lambda_{\mathrm{e}}(\mathbf{x}) \coloneqq \frac{2\pi}{k(\mathbf{x})}$ should be interpreted as the de Broglie wavelength of the wave function.

Notice that since the magnetic vector potential has gauge freedom and appears along with the gradient operator, the wave function $\Psi(\mathbf{x})$ depend on the choice of gauge. This means that, strictly speaking, the wave function does not have an unambiguously defined shape or "wavefronts" as a classical wave does. But since this gauge freedom can have no physical consequences, we always (implicitly or explicitly) fix the gauge of the vector potential, thereby defining a single unambiguous solution to the generalized Helmholtz equation.

The generalized Helmholtz equation is simple enough to be useful, but in this work we can afford to further simplify it to apply to the case of a beam-like wave function. Such a wave function travels primarily along one direction (defined to be the $+z$-axis) such that it is well-approximated as

$$\Psi(\mathbf{x}) \simeq \psi(\mathbf{x})\, e^{ikz} \tag{8.6}$$

where $\psi(\mathbf{x})$ is assumed to vary slowly along $z$ relative to the wavelength $\frac{2\pi}{k}$. We also only need consider the propagation of the beam in the absence of the static EM fields, because we can model the action of the fields of the TEM as that of ideal thin lenses, and then later add aberrations to the wave function. In this model, the continually-varying magnetic field inside of the TEM is reduced to a series of thin lenses, deflectors, etc. which are separated by field-free regions.

In this case, in the field-free regions the generalized Helmholtz equation reduces to the paraxial wave equation for the slowly-varying component of the wave function $\psi(\mathbf{x})$:

$$\frac{\partial}{\partial z}\psi(\mathbf{x}) = i\frac{\lambda_{\mathrm{e}}}{4\pi}\boldsymbol{\nabla}_{\perp}^{2}\psi(\mathbf{x}) \tag{8.7}$$

where $\boldsymbol{\nabla}_{\perp} := \hat{x}\frac{\partial}{\partial x} + \hat{y}\frac{\partial}{\partial y}$, and $\lambda_{\mathrm{e}} = 2\pi/k(\mathbf{x})$ where the electron kinetic energy $K$ is taken to be constant over the field-free region.

Solutions to the paraxial wave equation are formally given by the propagator

$$U(Z) := \exp\left(iZ\frac{\lambda_{\mathrm{e}}}{4\pi}\boldsymbol{\nabla}_{\perp}^{2}\right) \tag{8.8}$$

such that

$$\psi(x, y, z + Z) = U(Z)\,\psi(x, y, z) \tag{8.9}$$

though this form is only occasionally useful. Solutions can also be formulated in terms of a convolution with the Fresnel kernel

$$h_{Z}(x, y) := \frac{1}{i\lambda_{\mathrm{e}}Z}e^{i\frac{\pi}{\lambda_{\mathrm{e}}Z}\left(x^{2}+y^{2}\right)} \tag{8.10}$$

such that

$$\psi(x, y, z + Z) = h_{Z}(x, y) * \psi(x, y, z) \tag{8.11}$$

### 8.1.1 Fourier transform by a lens

This simple model for describing the electron beam propagation can be made even simpler by combining field-free propagation of the beam with the action of an ideal thin lens. Such a lens imparts a quadratic phase profile to the electron wave function, such that for a lens with focal length $f$ at position $z$, the transmitted wave function is

$$\psi_{\mathrm{t}}(x, y, z) = e^{-i\frac{\pi}{\lambda_{\mathrm{e}}f}\left(x^{2}+y^{2}\right)}\psi_{\mathrm{i}}(x, y, z) \tag{8.12}$$

where $\psi_{\mathrm{i}}$ is the incident wave function. Therefore, given a known wave function at $z = -d_1$ and the lens (assumed $f > 0$) at $z = 0$, the wave function at $z = d_2 > 0$ will be

$$\psi(x, y, d_2) = h_{d_2} * \left(e^{-i\frac{\pi}{\lambda_{\mathrm{e}} f}\left(x^2+y^2\right)}\left(h_{d_1} * \psi(x, y, -d_1)\right)\right) \tag{8.13}$$

$$= \int dx'dy'\, h_{d_2}(x - x', y - y')\, e^{-i\frac{\pi}{\lambda_{\mathrm{e}} f}\left(x'^2+y'^2\right)} \cdot \int dx''dy''\, h_{d_1}(x' - x'', y' - y'')\, \psi(x'', y'', -d_1) \tag{8.14}$$

$$= iD\frac{e^{i2\pi(d_1+d_2)/\lambda_{\mathrm{e}}}}{\lambda_{\mathrm{e}} d_1 d_2} e^{i\frac{\pi}{\lambda_{\mathrm{e}}}\left(\frac{1}{d_2}+\frac{D}{d_2^2}\right)\left(x^2+y^2\right)} \cdot \int dx''dy''\, e^{i\frac{\pi}{\lambda_{\mathrm{e}}}\left(\frac{1}{d_1}+\frac{D}{d_1^2}\right)\left(x''^2+y''^2\right)} e^{i\frac{2\pi}{\lambda_{\mathrm{e}}}\frac{D}{d_1 d_2}\left(xx''+yy''\right)}\psi(x'', y'', -d_1) \tag{8.15}$$

after some algebra including a Fourier transform of a complex Gaussian function, and defining

$$D := \left(\frac{1}{f} - \frac{1}{d_2} - \frac{1}{d_1}\right)^{-1} \tag{8.16}$$

In imaging applications, the input wave $\psi(x'', y'', -d_1)$ is often primarily a spherical wave with wavefront curvature radius $R$ ($R < 0$ is a diverging wave), so that the input wave can be written in separate terms of the spherical wave component and the remaining wave component $\psi_1(x'', y'')$

$$\psi(x'', y'', -d_1) = e^{-i\frac{\pi}{\lambda_{\mathrm{e}} R}\left(x''^2+y''^2\right)}\psi_1(x'', y'') \tag{8.17}$$

There is no loss in generality in rewriting $\psi(x'', y'', -d_1)$ in this form. This yields

$$\psi(x, y, d_2) = iD\frac{e^{i2\pi(d_1+d_2)/\lambda_{\mathrm{e}}}}{\lambda_{\mathrm{e}} d_1 d_2} e^{i\frac{\pi}{\lambda_{\mathrm{e}}}\left(\frac{1}{d_2}+\frac{D}{d_2^2}\right)\left(x^2+y^2\right)} \cdot \int dx''dy''\, e^{i\frac{\pi}{\lambda_{\mathrm{e}}}\left(\frac{1}{d_1}+\frac{D}{d_1^2}-\frac{1}{R}\right)\left(x''^2+y''^2\right)} e^{i\frac{2\pi}{\lambda_{\mathrm{e}}}\frac{D}{d_1 d_2}\left(xx''+yy''\right)}\psi_1(x'', y'') \tag{8.18}$$

Importantly, we now see that if

$$\frac{1}{d_1} + \frac{D}{d_1^2} - \frac{1}{R} = \frac{1}{f} - \frac{1}{d_2} - \frac{1}{d_1 - R} = 0 \tag{8.19}$$

the output wave is proportional to the Fourier transform of the non-spherical component of the input wave. That is, the Fourier transform is formed in the plane in which the incident spherical wave is focused by the lens. Such a plane is referred to as a "diffraction plane" in imaging contexts. The output wave is in general also superimposed with a spherical wave unless

$$\frac{1}{d_2} + \frac{D}{d_2^2} = 0 \tag{8.20}$$

i.e.

$$d_2 = d_1 \left(1 - \frac{d_1}{R}\right) \tag{8.21}$$

## 8.2 Definition of the contrast transfer function

Consider an electron beam propagating in the $+z$-direction and incident on some sample lying in the $x$-$y$ plane. Denote the electron wave function exiting the sample in the object plane as $\psi_{\mathrm{E}}(\mathbf{x})$, where $\mathbf{x} = (x, y)$. The electron wave then propagates through the remainder of the microscope to the image plane. At any level of approximation, the equations describing the electron wave's propagation through the microscope are linear. Combined with the approximation that the imaging properties of the microscope do not vary as a function of position in the object plane (isoplanatism), the wave function in the image plane $\psi_{\mathrm{im}}$ may be written as a convolution of some point spread function $H(\mathbf{x})$ with the sample exit wave function, such that

$$\psi_{\mathrm{im}}(\mathbf{x}) = \psi_{\mathrm{E}}(\mathbf{x}) * H(\mathbf{x}) \tag{8.22}$$

For simplicity, any magnification and/or rotation of the wave function is ignored and so $\mathbf{x}$ can be thought of as the object plane coordinates for both the exit and image wave functions. Without loss of generality, we may write

$$\psi_{\mathrm{E}}(\mathbf{x}) = \psi_0(\mathbf{x}) + \psi_{\mathrm{S}}(\mathbf{x}) \tag{8.23}$$

where $\psi_0(\mathbf{x})$ is the wave function that would exist in the object plane if the sample were not present, and $\psi_{\mathrm{S}}(\mathbf{x})$ represents the wave components which been generated via interactions with the sample. We refer to $\psi_0(\mathbf{x})$ as the "unscattered wave" and $\psi_{\mathrm{S}}(\mathbf{x})$ as the "scattered wave". The interaction with the sample in general modifies the phase and amplitude of the wave function such that

$$\psi_{\mathrm{E}}(\mathbf{x}) = e^{i\phi(\mathbf{x})} e^{-\mu(\mathbf{x})} \psi_0(\mathbf{x}) \tag{8.24}$$

where $\phi(\mathbf{x}), \mu(\mathbf{x}) \in \mathbb{R}$, and so

$$\psi_{\mathrm{S}}(\mathbf{x}) = S(\mathbf{x}) \psi_0(\mathbf{x}) \tag{8.25}$$

$$S(\mathbf{x}) \coloneqq e^{i\phi(\mathbf{x})} e^{-\mu(\mathbf{x})} - 1 \tag{8.26}$$

Note that the integrated amplitude of the electron's entire wave function is not changed by the interaction with the sample since the electrons are not created or destroyed. However, the amplitude of the wave function component with the same energy as the illuminating wave function may be changed by inelastic scattering in the sample, which is what is represented here. We only consider the component of the exit wave with the same energy as the unscattered wave since only it is able to coherently interfere with the unscattered wave. We will see that this is a key requirement of phase contrast.

Here I will follow the convention where the propagation phase of the electron beam is defined to be positive, i.e. $\psi(\mathbf{x};z) \propto e^{+i2\pi z/\lambda_{\mathrm{e}}}$, where $\lambda_{\mathrm{e}}$ is the electron wavelength. Therefore, $\phi$ is more positive in the more attractive areas of the sample, because the momentum of the electron (proportional to $+\frac{d\phi}{dz}$) increases in those regions. Note however that only the variations in $\phi$ as a function of $\mathbf{x}$ are relevant to the image, because the global phase of the exit wave is fundamentally undetectable.

The image formed by the electron density in the image plane can then be written in terms of $\psi_0(\mathbf{x})$, $S(\mathbf{x})$, and $H(\mathbf{x})$:

$$\begin{aligned}\left|\psi_{\mathrm{im}}(\mathbf{x})\right|^2 = {} & \left|\psi_0(\mathbf{x}) * H(\mathbf{x})\right|^2 \\ & + \left|\left(S(\mathbf{x})\,\psi_0(\mathbf{x})\right) * H(\mathbf{x})\right|^2 \\ & + \left(\psi_0(\mathbf{x}) * H(\mathbf{x})\right)\left(\left(S(\mathbf{x})\,\psi_0(\mathbf{x})\right) * H(\mathbf{x})\right)^* \\ & + \left(\psi_0(\mathbf{x}) * H(\mathbf{x})\right)^*\left(\left(S(\mathbf{x})\,\psi_0(\mathbf{x})\right) * H(\mathbf{x})\right)\end{aligned} \tag{8.27}$$

where the first two lines represent the images formed by the unscattered and scattered waves, respectively, and the third and fourth lines represent the images formed by the interference between them. This expression can be simplified by assuming that (as is typically the case in TEM) the unscattered wave is planar such that $\psi_0(\mathbf{x}) = e^{i2\pi\mathbf{s}_0\cdot\mathbf{x}}$ where $\mathbf{s}_0 = \boldsymbol{\theta}_0/\lambda_{\mathrm{e}}$ is the transverse spatial frequency of the wave, incident at an angle of $\boldsymbol{\theta}_0$ relative to the optical axis of the imaging system. In this case

$$\psi_0(\mathbf{x}) * H(\mathbf{x}) = \tilde{H}(\mathbf{s}_0)\, e^{i2\pi\mathbf{s}_0\cdot\mathbf{x}} \tag{8.28}$$

$$\left(S(\mathbf{x})\,\psi_0(\mathbf{x})\right) * H(\mathbf{x}) = e^{i2\pi\mathbf{s}_0\cdot\mathbf{x}}\mathcal{F}^{-1}\left[\mathcal{F}\left[S(\mathbf{x})\right](\mathbf{s})\cdot\tilde{H}(\mathbf{s}_0+\mathbf{s})\right](\mathbf{x}) \tag{8.29}$$

where the wave transfer function $\tilde{H}(\mathbf{s}) \coloneqq \mathcal{F}\left[H(\mathbf{x})\right](\mathbf{s})$ is the Fourier transform of the point spread function, and $\mathbf{s}$ is the non-angular spatial frequency Fourier conjugate to the real-space position in the object plane. As such,

$$\begin{aligned}\left|\psi_{\mathrm{im}}(\mathbf{x})\right|^2 = {} & \left|\tilde{H}(\mathbf{s}_0)\right|^2 \\ & + \left|\mathcal{F}^{-1}\left[\mathcal{F}\left[S(\mathbf{x})\right](\mathbf{s})\cdot\tilde{H}(\mathbf{s}+\mathbf{s}_0)\right](\mathbf{x})\right|^2 \\ & + \tilde{H}(\mathbf{s}_0)\,\mathcal{F}^{-1}\left[\mathcal{F}\left[S^*(\mathbf{x})\right](\mathbf{s})\cdot\tilde{H}^*(\mathbf{s}_0-\mathbf{s})\right](\mathbf{x}) \\ & + \tilde{H}^*(\mathbf{s}_0)\,\mathcal{F}^{-1}\left[\mathcal{F}\left[S(\mathbf{x})\right](\mathbf{s})\cdot\tilde{H}(\mathbf{s}_0+\mathbf{s})\right](\mathbf{x})\end{aligned} \tag{8.30}$$

and so the Fourier transform of the image is

$$\begin{aligned}\mathcal{F}\left[\left|\psi_{\mathrm{im}}(\mathbf{x})\right|^2\right](\mathbf{s}) = {} & \left|\tilde{H}(\mathbf{s}_0)\right|^2 \delta(\mathbf{s}) \\ & + \mathcal{F}\left[\left|\mathcal{F}^{-1}\left[\mathcal{F}\left[S(\mathbf{x})\right](\mathbf{s}) \cdot \tilde{H}(\mathbf{s}+\mathbf{s}_0)\right](\mathbf{x})\right|^2\right](\mathbf{s}) \\ & - 2\mathcal{F}\left[\mathrm{Re}\left[S\right](\mathbf{x})\right](\mathbf{s}) \cdot \mathrm{ATF}(\mathbf{s}) \\ & + 2\mathcal{F}\left[\mathrm{Im}\left[S\right](\mathbf{x})\right](\mathbf{s}) \cdot \mathrm{CTF}(\mathbf{s}) \end{aligned} \tag{8.31}$$

where $\delta(\mathbf{s})$ is the Dirac delta function, and the amplitude transfer function (ATF) and contrast transfer function (CTF) are defined as

$$\mathrm{ATF}(\mathbf{s}) = -\frac{1}{2}\left(\tilde{H}(\mathbf{s}_0)\tilde{H}^*(\mathbf{s}_0-\mathbf{s}) + \tilde{H}^*(\mathbf{s}_0)\tilde{H}(\mathbf{s}_0+\mathbf{s})\right) \tag{8.32}$$

$$\mathrm{CTF}(\mathbf{s}) = -\frac{i}{2}\left(\tilde{H}(\mathbf{s}_0)\tilde{H}^*(\mathbf{s}_0-\mathbf{s}) - \tilde{H}^*(\mathbf{s}_0)\tilde{H}(\mathbf{s}_0+\mathbf{s})\right) \tag{8.33}$$

respectively. The wave transfer function may be further decomposed into magnitude and phase components such that $\tilde{H}(\mathbf{s}) = A(\mathbf{s})\, e^{i\chi(\mathbf{s})}$ where $A(\mathbf{s}) \in \mathbb{R}$ is the aperture function and $\chi(\mathbf{s}) \in \mathbb{R}$ is the aberration function. The aberration function describes the phase accumulated by the different spatial frequency components of the wave function after propagating through the microscope, while the aperture function describes the degree to which different spatial frequency components in the image are attenuated. In practice, this attenuation is usually due to the electron beam being blocked by an opaque aperture. This may be done intentionally to enable dark field imaging, but in the case of bright field phase contrast imaging no portion of the beam is intentionally blocked and it is safe to assume that $A(\mathbf{s}) = 1$, at least up to the highest relevant spatial frequencies in the image. This allows the amplitude and contrast transfer functions to be written in a simpler form:

$$\mathrm{ATF}(\mathbf{s}) = -\frac{1}{2}\left(e^{-i(\chi(\mathbf{s}_0-\mathbf{s})-\chi(\mathbf{s}_0))} + e^{i(\chi(\mathbf{s}_0+\mathbf{s})-\chi(\mathbf{s}_0))}\right) \tag{8.34}$$

$$\mathrm{CTF}(\mathbf{s}) = -\frac{i}{2}\left(e^{-i(\chi(\mathbf{s}_0-\mathbf{s})-\chi(\mathbf{s}_0))} - e^{i(\chi(\mathbf{s}_0+\mathbf{s})-\chi(\mathbf{s}_0))}\right) \tag{8.35}$$

or equivalently, with terms which are symmetric or antisymmetric about $\mathbf{s}_0$:

$$\mathrm{ATF}(\mathbf{s}) = -e^{i(\chi(\mathbf{s}_0+\mathbf{s})-\chi(\mathbf{s}_0-\mathbf{s}))/2}\cos\left(\frac{1}{2}\left(\chi(\mathbf{s}_0+\mathbf{s})+\chi(\mathbf{s}_0-\mathbf{s})\right)-\chi(\mathbf{s}_0)\right) \tag{8.36}$$

$$\mathrm{CTF}(\mathbf{s}) = -e^{i(\chi(\mathbf{s}_0+\mathbf{s})-\chi(\mathbf{s}_0-\mathbf{s}))/2}\sin\left(\frac{1}{2}\left(\chi(\mathbf{s}_0+\mathbf{s})+\chi(\mathbf{s}_0-\mathbf{s})\right)-\chi(\mathbf{s}_0)\right) \tag{8.37}$$

In this form, it is easier to see that both the ATF and CTF consist of a phase component (the exponential) and an amplitude component (the sinusoid). Both components generally cause delocalization of sample structure information across the image (colloquially, “blur”), though the amplitude component also may attenuate some of this information. Image processing algorithms may be able to recover the sample structure from delocalized information (so long as it is not delocalized outside of the field of view of the image, see section 8.7.2), but attenuated information is irretrievably lost. As such, the characteristics of the amplitude component are usually considered to be more important than those of the phase component. Still, the electron optical systems of TEMs are usually aligned to minimize any dependence of the phase component on spatial frequency, which is commonly referred to as "coma-free alignment" (see section 10.3.7).

Equation (8.31) is still relatively unwieldy. Though there are simple transfer functions for the real and imaginary components of the sample structure, the second line (representing the image formed by the scattered wave) cannot be readily simplified. However, if the sample consists of "weak phase objects" such that $\phi(\mathbf{x}), \mu(\mathbf{x}) \ll 1$, then

$$S(\mathbf{x}) \simeq i\phi(\mathbf{x}) - \mu(\mathbf{x}) \tag{8.38}$$

and so we can reasonably approximate equation (8.31) to first order in $\phi$ and $\mu$,

$$\mathcal{F}\left[|\psi_{\text{im}}(\mathbf{x})|^2\right](\mathbf{s}) \simeq \delta(\mathbf{s}) + 2\mathcal{F}[\mu(\mathbf{x})](\mathbf{s}) \cdot \text{ATF}(\mathbf{s}) + 2\mathcal{F}[\phi(\mathbf{x})](\mathbf{s}) \cdot \text{CTF}(\mathbf{s}) \tag{8.39}$$

which eliminates the scattered wave image term. This is referred to as the "weak phase approximation". A further simplification is possible if we assume that $\mathcal{F}[\mu(\mathbf{x})](\mathbf{s})$ is proportional to $\mathcal{F}[\phi(\mathbf{x})](\mathbf{s})$. This is only a rough approximation, but is usually accurate enough for the weak phase object samples in cryo-EM [40, p. 1-12]. If we define the amplitude contrast ratio $\mathcal{A} \geq 0$ such that

$$\mathcal{F}[\mu(\mathbf{x})](\mathbf{s}) = \mathcal{A}\mathcal{F}[S(\mathbf{x})](\mathbf{s}) \tag{8.40}$$

$$\mathcal{F}[\phi(\mathbf{x})](\mathbf{s}) = \sqrt{1-\mathcal{A}^2}\mathcal{F}[S(\mathbf{x})](\mathbf{s}) \tag{8.41}$$

then

$$\mathcal{F}\left[|\psi_{\text{im}}(\mathbf{x})|^2\right](\mathbf{s}) \simeq \delta(\mathbf{s}) + 2\mathcal{F}[\phi(\mathbf{x})](\mathbf{s}) \cdot \text{CTF}(\mathbf{s}) \tag{8.42}$$

or equivalently

$$|\psi_{\text{im}}(\mathbf{x})|^2 \simeq 1 + 2\phi(\mathbf{x}) * \mathcal{F}^{-1}[\text{CTF}(\mathbf{s})](\mathbf{x}) \tag{8.43}$$

where the weak phase approximation contrast transfer function is defined to be

$$\begin{aligned} \text{CTF}(\mathbf{s}) = &-e^{i(\chi(\mathbf{s}_0+\mathbf{s})-\chi(\mathbf{s}_0-\mathbf{s}))/2} \\ &\cdot \sin\left(\frac{1}{2}(\chi(\mathbf{s}_0+\mathbf{s})+\chi(\mathbf{s}_0-\mathbf{s})) - \chi(\mathbf{s}_0) + \arctan\left(\frac{\mathcal{A}}{\sqrt{1-\mathcal{A}^2}}\right)\right) \end{aligned} \tag{8.44}$$

The amplitude contrast ratio phase shifts the sinusoid, providing non-zero contrast at zero spatial frequency. Since $\mathcal{A} \approx 0.07$ for biological macromolecules in vitreous ice [40, p. 1-12], we often approximate the amplitude contrast term to leading order in $\mathcal{A}$, such that

$$\text{CTF}(\mathbf{s}) = -e^{i(\chi(\mathbf{s}_0+\mathbf{s})-\chi(\mathbf{s}_0-\mathbf{s}))/2} \sin\left(\frac{1}{2}(\chi(\mathbf{s}_0+\mathbf{s})+\chi(\mathbf{s}_0-\mathbf{s})) - \chi(\mathbf{s}_0) + \mathcal{A}\right) \tag{8.45}$$

Equation (8.43) shows that in the weak phase approximation the image (the electron wave probability density in the image plane) is simply a convolution of the inverse Fourier transformed CTF with the phase profile of the sample. The CTF describes the transfer of information from the sample phase profile (the Fourier transform of which is usually called the "structure factor") as a function of spatial frequency. I will refer to equations (8.42), (8.43), and (8.45) extensively throughout the rest of this work, as they provide a simple yet accurate model for phase contrast image formation.

## 8.3 The aberration function

As of yet, the form of the aberration function $\chi(\mathbf{s})$ has remained unspecified. In practice, the aberration function depends on the design and alignment of the TEM's imaging system. As such, the aberration function is often written as a power series in the spatial frequency's polar coordinates ($s = |\mathbf{s}|, \theta_s = \arg(s_x + is_y)$), where the series coefficients quantify the aberrations [88, p. 61]. The first few terms of this power series readily capture the effects of the most common kinds of aberrations. For a discussion on why this is the case, see [89, ch. 23]. However, such a power series does not easily capture changes in the aberration function near the origin like those of a phase plate, so we represent these terms in a separate function $\eta(\mathbf{s})$. In total then, the aberration function can be written as

$$\chi(\mathbf{s}) = -\gamma(\mathbf{s}) - \eta(\mathbf{s}) \tag{8.46}$$

$$\gamma(\mathbf{s}) = \frac{2\pi}{\lambda_e} \sum_{l=1}^{\infty} \frac{(\lambda_e s)^l}{l} \sum_m C_{l-1,m} e^{im\theta_s}, \quad m = -l, -l+2, \cdots, l-2, l \tag{8.47}$$

where $C_{l,m}$ are the aberration coefficients and $C_{l,-m} = C_{l,m}^*$ such that $\gamma(\mathbf{s}) \in \mathbb{R}$. The signs in the first equation are chosen by convention. Note that terms with even $m$ are symmetric about the origin and so only appear in the argument of the sine in the CTF, while the odd $m$ terms are antisymmetric and so only appear in the complex exponential. Common names for several of the lowest order aberration coefficients are shown in table 8.1.

| | |
|---|---|
| $C_{0,\pm 1}$ | Image shift |
| $C_{1,0}$ | Defocus |
| $C_{1,\pm 2}$ | (Twofold) Astigmatism |
| $C_{2,\pm 1}$ | Coma |
| $C_{2,\pm 3}$ | Threefold astigmatism |
| $C_{3,0}$ | Spherical aberration, "$C_s$" |
| $C_{3,\pm 2}$ | Third order twofold astigmatism |
| $C_{3,\pm 4}$ | Third order fourfold astigmatism |

Table 8.1: Commonly-used names for the lowest-order aberration coefficients [88, p. 63].

Image shift just results in a shift in the position of the image in the image plane, and as such, is not typically considered to be an "aberration" and is usually ignored. As we will see shortly, a non-zero defocus is often intentionally used to increase image contrast. Defocus is so often referred to that it is often abbreviated with the symbol $Z := -C_{1,0}$ where $Z > 0$ is referred to as "underfocus" while $Z < 0$ is referred to as "overfocus". Astigmatism, sometimes referred to as twofold astigmatism, is usually an unwanted aberration and can be removed by aligning the microscope. The same is true for coma, though the microscope alignment procedures are somewhat more involved (see section 10.3.7). Threefold astigmatism is often small enough to be ignored. Spherical aberration, like defocus, plays an important role in determining the shape of the CTF, and is given its own symbol $C_\mathrm{s} := C_{3,0}$. In non-aberration-corrected TEMs its value is essentially fixed by the design of the electron optics. Cylindrically symmetric magnetic lenses (like those used in non-aberration-corrected TEMs) always have $C_\mathrm{s} > 0$ [89, p. 319], [90]. Given all this, it is usually sufficient to model only the effects of defocus and spherical aberration, such that

$$\gamma(\mathbf{s}) = \frac{2\pi}{\lambda_\mathrm{e}}\left(-\frac{1}{2}Z\lambda_\mathrm{e}^2 s^2 + \frac{1}{4}C_\mathrm{s}\lambda_\mathrm{e}^4 s^4\right) \tag{8.48}$$

Assuming that the sample is illuminated with an on-axis plane wave such that $\mathbf{s}_0 = 0$, and taking $\eta(\mathbf{x}) = 0$, the CTF is then

$$\mathrm{CTF}(\mathbf{s}) = \sin\left(\frac{2\pi}{\lambda_\mathrm{e}}\left(-\frac{1}{2}Z\lambda_\mathrm{e}^2 s^2 + \frac{1}{4}C_\mathrm{s}\lambda_\mathrm{e}^4 s^4\right) - \mathcal{A}\right) \tag{8.49}$$

This formula illustrates the fundamental limitations of conventional phase contrast TEM. First, the CTF remains near zero for low spatial frequencies. Since the structure factor of nearly all real-world samples is largest at low spatial frequencies, this substantially limits the total contrast in the image, and images appear to be high-pass filtered versions of the sample structure factor. Second, the CTF oscillates increasingly quickly with increasing spatial frequency. These oscillations generate concentric rings in the power spectral density of the image called "Thon rings". This means that, averaged over high frequencies, only half of the information in the structure factor is present in the image (since the average of a squared sinusoid is 1/2). These oscillations also blur the image. This blur can be partially corrected for by image processing, but information about the sample which is blurred beyond the edge of the field of view is irretrievably lost. This effect is a major limitation in conventional TEM, and is explored in section 8.7.2.

Figure 8.1 shows examples of the conventional CTF for several different values of defocus, assuming $\lambda_\mathrm{e} = 1.97\,\mathrm{pm}$ (corresponding to an electron energy of 300 keV), $C_\mathrm{s} = 2.7\,\mathrm{mm}$ (typical of a modern TEM), and $\mathcal{A} = 0$ (for simplicity). This illustrates why underfocus ($Z > 0$) is more commonly used than overfocus: the opposing signs of the $C_{1,0}$ and $C_{3,0}$ aberration coefficients reduces the frequency of CTF oscillations, which results in less image blur. If no defocus is used, the CTF remains close to zero until relatively high frequency, resulting in little total contrast. At the "Scherzer defocus"

$$Z_\mathrm{S} := \sqrt{C_\mathrm{s}\lambda_\mathrm{e}} \tag{8.50}$$

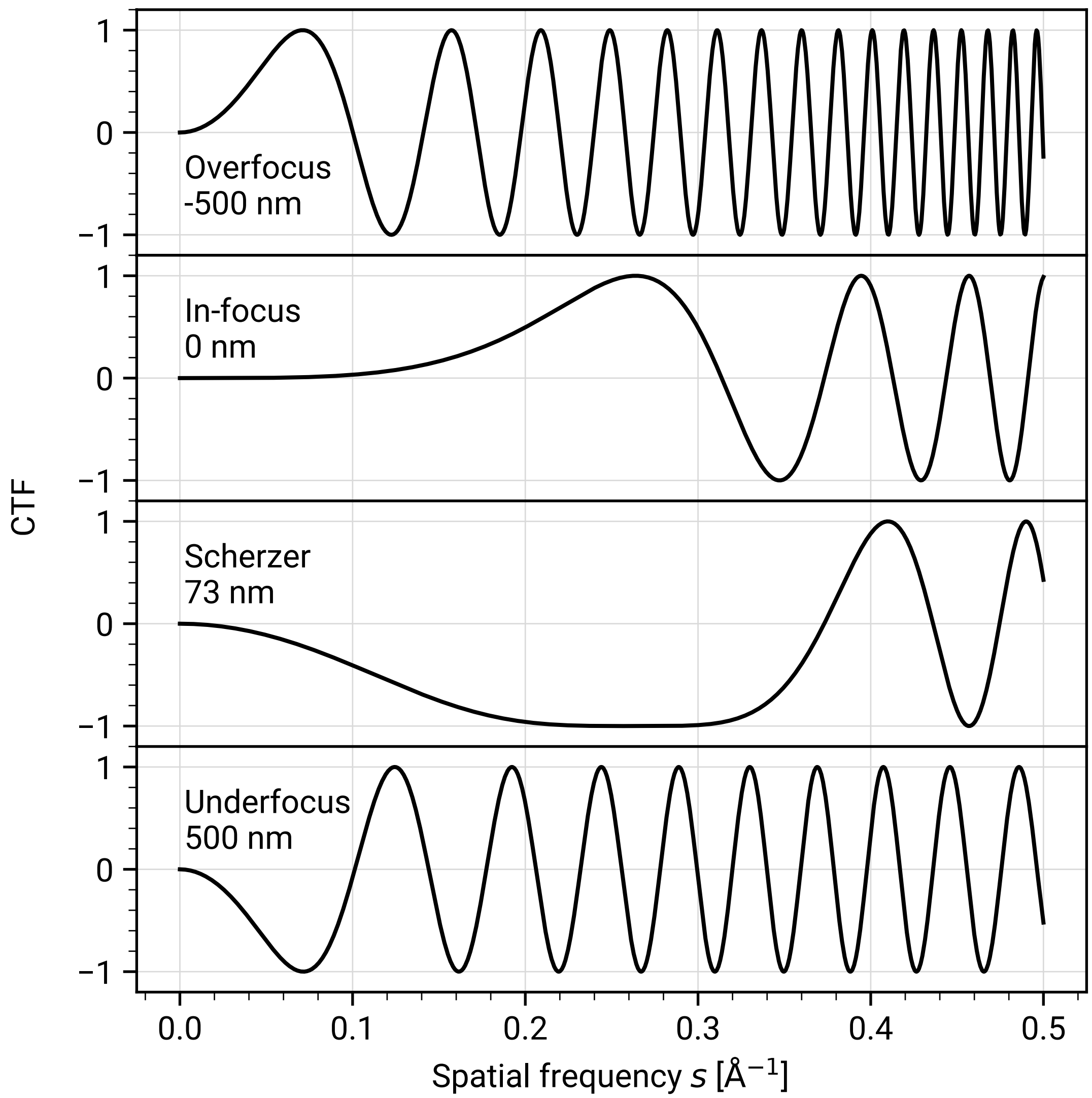

Figure 8.1: The conventional contrast transfer function (CTF) as a function of spatial frequency for four characteristic defocus values.

the passband in which the CTF does not oscillate is maximized [86, p. 1735], [91]. However, this defocus setting is rarely used in practice, especially in cryo-EM. Instead, a relatively large ($\sim 1\,\mathrm{um}$) underfocus is used in an effort to further increase the low-frequency contrast in the image to the point where individual macromolecules are identifiable over the shot noise.

## 8.4 The laser phase plate

The limitations of the conventional CTF can be removed by using a phase plate. As seen in the argument of the sine function in equation (8.45), non-zero contrast at frequency $\mathbf{s}$ fundamentally arises from the difference between the aberration function at the location of the unscattered wave $\mathbf{s}_0$, and the average of the aberration function at frequencies $\mathbf{s}_0 \pm \mathbf{s}$. A phase plate introduces this difference by adding an additional phase shift to the aberration function to points near the location of the unscattered wave. Indeed, an ideal aberration function, generating a CTF of unity magnitude for all frequencies, would consist of a phase shift of $-\chi(\mathbf{s}_0) = \pi/2 - \mathcal{A} + \pi n,\ n \in \mathbb{Z}$ applied only to the location of the unscattered wave (or, equivalently, to all points except $\mathbf{s}_0$). Such an ideal phase plate allows for detection of the maximum possible amount of phase information from the (weak phase) sample [92]. Therefore, the smaller the size of the phase shifted region around the unscattered wave, the closer to ideal the CTF becomes.

Letting the function $\eta(\mathbf{s})$ represent this localized phase shift, and taking $\gamma(\mathbf{s})$ as in equation (8.48) with $\mathbf{s}_0 = 0$ for simplicity, we see that in the region away from $\mathbf{s}_0$ where $\eta(\mathbf{s}) = 0$ the CTF can be expressed as

$$\mathrm{CTF}(\mathbf{s}) = \sin\left(\frac{2\pi}{\lambda_\mathrm{e}}\left(-\frac{1}{2}Z\lambda_\mathrm{e}^2 s^2 + \frac{1}{4}C_\mathrm{s}\lambda_\mathrm{e}^4 s^4\right) - (\eta(0) + \mathcal{A})\right) \tag{8.51}$$

so that if, for example, $\eta(0) + \mathcal{A} = \pi/2$, the CTF has a magnitude near unity at low frequencies until the phase due to defocus and spherical aberration starts to become appreciable at higher frequencies, again causing the CTF to oscillate. I will colloquially refer to $\eta(0) + \mathcal{A}$ as the "phase shift" of the phase plate.

Realizing a phase plate requires selectively phase shifting only the low spatial frequency components of the electron wave. This is done by applying the phase shift in a diffraction plane of the imaging system. A diffraction plane is defined as a plane in which the unscattered wave $\psi_0(\mathbf{x})$ is brought to a focus. As shown in section 8.1.1, the wave function in a diffraction plane is proportional to the Fourier transform of the exit wave in the object plane. Thus, any phase shift applied to the wave there simply adds to the aberration function, which by definition is applied to the Fourier transform of the exit wave (convolution theorem applied to equation (8.22)). The relationship between position in the diffraction plane and the corresponding spatial frequency in the image depends on the effective focal length of the imaging system $f_\mathrm{eff}$ and the electron wavelength $\lambda_\mathrm{e}$. Information at frequency $\mathbf{s}$ is carried by a wave component which makes an angle $\vartheta = \lambda_\mathrm{e}\mathbf{s}$ with the optical axis in the object plane, and such a wave is focused at position $\mathbf{x} = f_\mathrm{eff}\vartheta$ in the diffraction plane. Therefore, the correspondence between position in the diffraction plane $\mathbf{x}$ and (object-referred) spatial frequency $\mathbf{s}$ is

$$\mathbf{x} = f_\mathrm{eff}\lambda_\mathrm{e}\mathbf{s} \tag{8.52}$$

in the diffraction plane.

As shown in section 9.1 (equation (9.25)), the ponderomotive potential of a standing wave Gaussian laser beam imparts a position-dependent phase shift to a throughgoing electron wave. As a function of equivalent spatial frequency, this means that the phase plate component of the aberration function is

$$\eta(\mathbf{s}) = \frac{\eta_0}{2} \frac{e^{-2\frac{S_y^2}{1+S_x^2}}}{\sqrt{1+S_x^2}} \left[ 1 + \rho(\theta,\beta) e^{-\Theta^2 \frac{2}{\mathrm{NA}^2}\left(1+S_x^2\right)} \left(\frac{1}{1+S_x^2}\right)^{1/4} \cdot \cos\left(2\frac{S_x}{1+S_x^2} S_y^2 + \frac{4}{\mathrm{NA}^2} S_x - \frac{3}{2}\arctan(S_x) - \varphi\right)\right] \tag{8.53}$$

where laser beam is centered at $\mathbf{s} = \mathbf{s}_l$, the $x$-axis has been chosen to lie along the axis of the laser beam, and

| | |
|---|---|
| $S_x := \frac{\pi \mathrm{NA}^2 f_{\mathrm{eff}} \lambda_{\mathrm{e}}}{\lambda_{\mathrm{L}}} (s_x - s_{\mathrm{l},x})$ | is the normalized spatial frequency $x$-component |
| $S_y := \frac{\pi \mathrm{NA} f_{\mathrm{eff}} \lambda_{\mathrm{e}}}{\lambda_{\mathrm{L}}} (s_y - s_{\mathrm{l},y})$ | is the normalized spatial frequency $y$-component |
| $\rho(\theta,\beta) := 1 - 2\beta^2 \cos^2(\theta)$ | describes the laser polarization and electron velocity dependence of the phase shift |
| $\eta_0$ | is the maximum phase shift (when $S_x = S_y = 0$, $\theta = \pi/2$, $\Theta = 0$, $\varphi = 0$) |
| $\beta$ | is the electron beam velocity divided by the speed of light |
| $\theta$ | is the angle between the electric field polarization axis of the laser beam and the electron beam axis |
| $\Theta$ | is the angle between the laser beam axis and the $x$-$y$ plane |
| NA | is the numerical aperture of the laser beam |
| $\varphi$ | $= 0$ $(= \pi)$ if the laser beam has antinode (node) at $s_x = 0$ |

Equations (8.45), (8.46), (8.47), and (8.53) describe the laser phase plate's CTF, including conventional aberrations. We will mostly consider the case that the LPP is centered in the diffraction plane, i.e. $\mathbf{s}_l = 0$. It is also possible for the laser beam to lie above or below the diffraction plane, the consequences of which will be further explored in section 8.8.

An example of the LPP CTF is shown in figure 8.2 for the parameters listed in table 8.2. Figure 8.2a shows the CTF up to a resolution of 2 Å. Thon rings due to spherical aberration are visible for frequencies above $\left(5\,\text{Å}\right)^{-1}$. The phase shift generates a relatively flat CTF down to low frequency. Side-effects of the LPP are seen near the $x$-axis, where the standing wave structure makes a series of phase shifted fringes. These fringes are seen in detail in figure 8.2b, which shows the area delineated by the black box in figure 8.2a. The periodicity of the fringes is $\frac{\lambda_{\mathrm{L}}}{2 f_{\mathrm{eff}} \lambda_{\mathrm{e}}} = (74\,\mathrm{nm})^{-1}$. The transverse width of the fringes ($y$-direction) is determined by the NA of the laser beam.

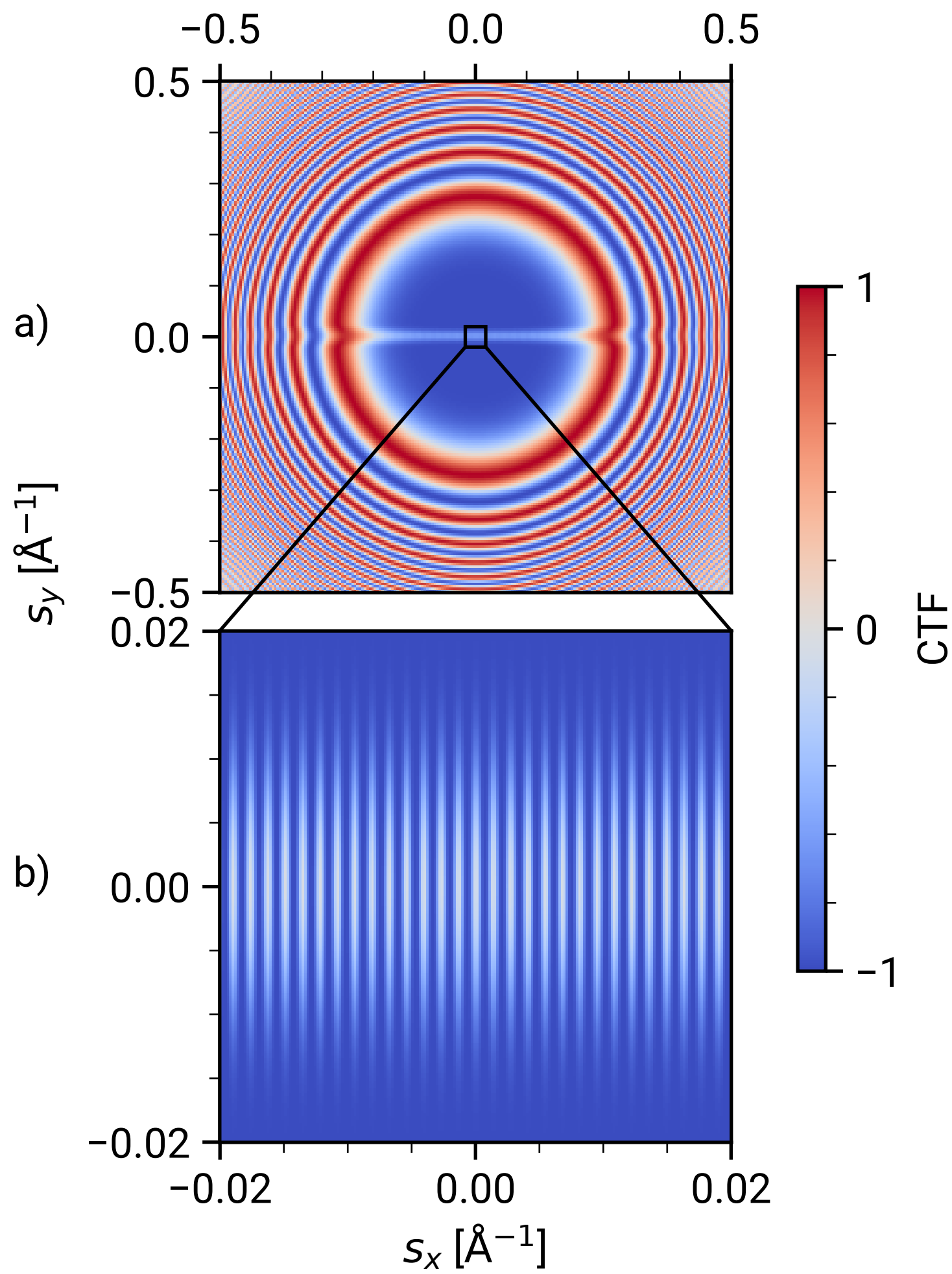

Figure 8.2: **a)** The contrast transfer function (CTF) of a laser phase plate (equation (8.53)) as a function of $s_x$ and $s_y$, using the parameters as listed in table 8.2. **b)** A magnified view of the CTF near the origin, as delineated by the black box in a). The color scale is shown on the right.

In single particle analysis cryo-EM, data from many objects in different orientations is averaged together. An azimuthal root-mean-squared average of the LPP CTF

$$\mathrm{ctf}\,(s) := \sqrt{\frac{1}{2\pi}\int_0^{2\pi} d\theta_s\,\left(\mathrm{CTF}\,(\mathbf{s})\right)^2} \tag{8.54}$$

represents the averaged effects of the anisotropy in the CTF. The azimuthally averaged CTF of figure 8.2 is shown in black in figure 8.3. Note the logarithmic scaling of the frequency

| | |
|---|---|
| $f_{\mathrm{eff}}$ | 20 mm |
| $\lambda_{\mathrm{e}}$ | 1.97 pm |
| $Z$ | 0 |
| $C_{\mathrm{s}}$ | 4.8 mm |
| $\eta_0$ | $\pi/2$ |
| NA | 0.05 |
| $\varphi$ | 0 |

Table 8.2: Parameters for the CTF shown in figure 8.2.

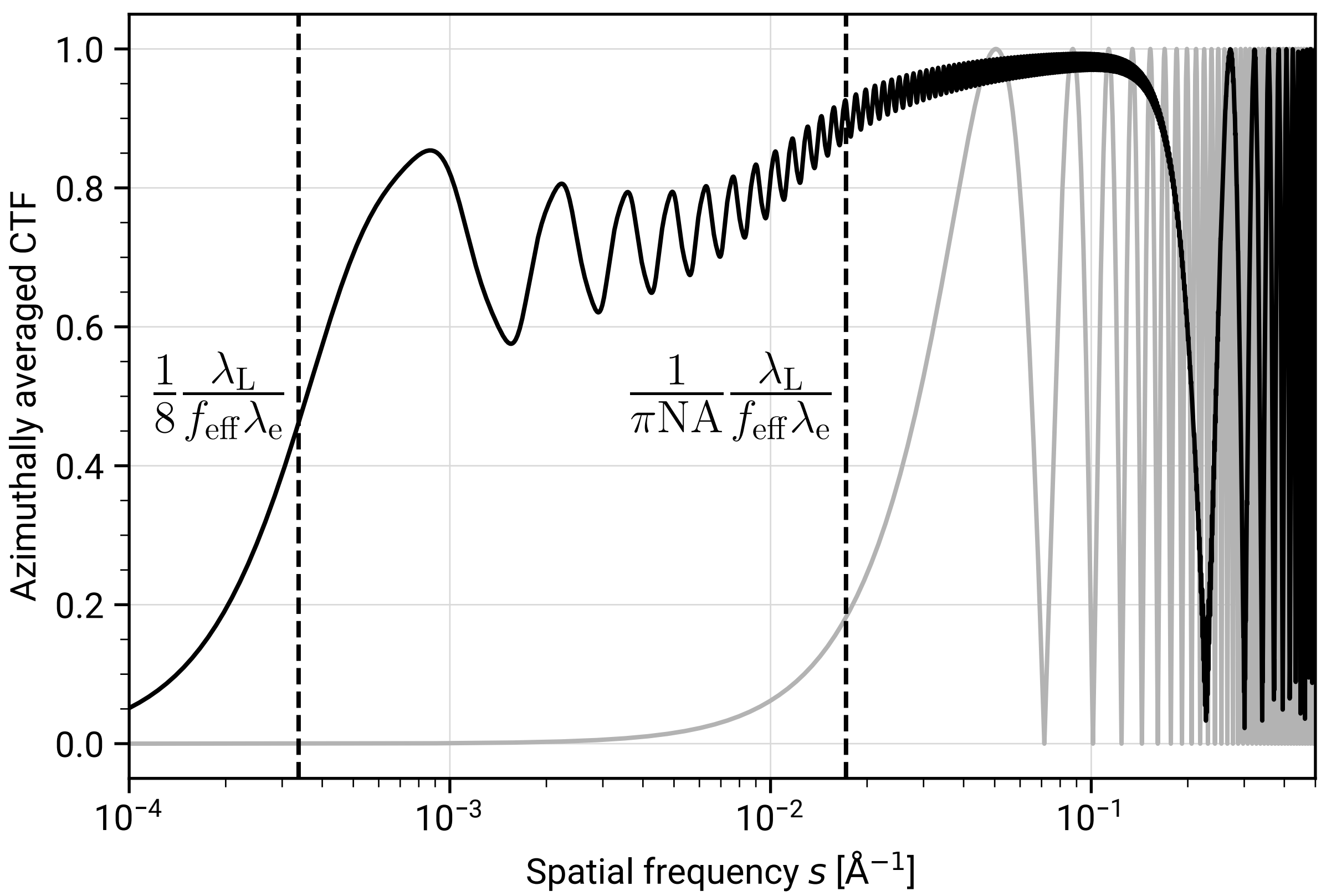

Figure 8.3: The azimuthal average of the laser phase plate's contrast transfer function (CTF) from figure 8.2 is shown in black as a function of spatial frequency. The two characteristic cut-on frequencies of the laser phase plate are denoted by the dashed vertical lines and annotated with formulas for their values. The azimuthally averaged conventional CTF with a defocus of 1 µm is shown in gray as a comparison.

axis. This helps show the low cut-on frequency

$$s_{-,\lambda_\mathrm{L}} \coloneqq \frac{1}{8}\frac{\lambda_\mathrm{L}}{f_\mathrm{eff}\lambda_\mathrm{e}} \approx (300\,\mathrm{nm})^{-1} \tag{8.55}$$

of the LPP, which is due to the standing wave structure in the phase plate. This first cut-on plateaus below unity because only a portion of the wave components with spatial frequencies of equal magnitude have a substantially different phase shift from the unscattered wave. The oscillations in the CTF in this region come the azimuthal average over the laser fringes. A second cut-on occurs at

$$s_{-,\mathrm{NA}} \coloneqq \frac{1}{\pi\mathrm{NA}}\frac{\lambda_\mathrm{L}}{f_\mathrm{eff}\lambda_\mathrm{e}} \approx (5.8\,\mathrm{nm})^{-1} \tag{8.56}$$

which brings the azimuthally averaged CTF near unity. This second cut-on comes from the transverse width of the laser beam. The azimuthally averaged conventional CTF is shown for comparison in gray, with a defocus of 1 µm (and $C_\mathrm{s} = 4.8\,\mathrm{mm}$, same as the LPP case). The first cut-on frequency of the LPP is lower than that of conventional defocus by roughly two orders of magnitude. Specifically, the conventional azimuthally averaged CTF first reaches 0.5 at a cut-on frequency of

$$s_{-,Z} = \sqrt{\frac{1}{6\lambda_\mathrm{e}\,|Z|}} \tag{8.57}$$

The defocus which generates cut-on frequencies of $s_{-,\lambda_\mathrm{L}}$ and $s_{-,\mathrm{NA}}$ are 7.4 mm and 2.9 µm, respectively. Thus, for NA = 0.05 as used in figures 8.2 and 8.3, the transverse width of the LPP only generates somewhat more low frequency contrast than a reasonable defocus value ($\sim$ 1 µm), while the defocus required to achieve the low frequency contrast due to the standing wave structure is extremely large.

## 8.5 In-focus imaging

Imaging with the defocus at or near zero ("in-focus") reduces the number of oscillations in the CTF. When using a phase plate this increases the total information about the sample contained in the image by up to a factor of two compared to a defocused image. It also reduces the delocalization of sample structure information outside of the field of view, which is a potential limitation on the ultimate resolution of the image (see section 8.7.2).

Consider an ideal phase plate where $\eta_0 \coloneqq \eta(\mathbf{s}_0) + \mathcal{A} \geq 0$ is the phase shift at the location of the unscattered wave including the effect of amplitude contrast. For a given $\eta_0$ and $C_\mathrm{s}$, the highest possible spatial frequency $s_\mathrm{S}$ (the "Scherzer frequency") of the first zero-crossing of the CTF is

$$s_\mathrm{S} = \left(\frac{4}{C_\mathrm{s}\lambda_\mathrm{e}^3}\right)^{1/4}\left(\sqrt{\frac{\pi - \arcsin(\mathrm{CTF}_0) - \eta_0}{2\pi}} + \sqrt{\frac{\pi - \arcsin(\mathrm{CTF}_0)}{2\pi}}\right)^{1/2} \tag{8.58}$$

where $0 \leq \mathrm{CTF}_0 \leq 1$ is the smallest magnitude of CTF allowed at the first frequency where the derivative of the aberration function is zero. As shown in figure 8.4, this effectively sets a limit on the depth of the dip in the CTF caused by underfocus overcompensating spherical aberration at medium frequencies. Figure 8.4 shows examples of ideal phase plate CTFs as a function of spatial frequency for different values of $\eta_0$ and $\mathrm{CTF}_0$. Notice that the phase plate reduces the Scherzer frequency $s_\mathrm{S}$, and a smaller $\mathrm{CTF}_0$ increases $s_\mathrm{S}$ at the expense of smaller CTF values for frequencies smaller than $s_\mathrm{S}$. The corresponding defocus generalizes the Scherzer defocus formula (equation (8.50)) to

$$Z_\mathrm{S} = \sqrt{C_\mathrm{s}\lambda_\mathrm{e}}\left(\frac{\pi - \arcsin\left(\mathrm{CTF}_0\right) - \eta_0}{\pi/2}\right)^{1/2} \tag{8.59}$$

where the conventional Scherzer defocus is recovered when $\eta_0 = 0$, $\mathrm{CTF}_0 = 1$. The formulas assume that $0 \leq \eta_0 < \pi$ and $C_\mathrm{s} \geq 0$. Defocus values associated with each set of $\eta_0$ and $\mathrm{CTF}_0$ are annotated in figure 8.4.

Figure 8.5 shows $s_\mathrm{S}$ divided by the conventional Scherzer frequency of $\left(\frac{4}{C_\mathrm{s}\lambda_\mathrm{e}^3}\right)^{1/4}$ as a function of $\mathrm{CTF}_0$ and the phase shift $\eta_0$. The Scherzer frequency is maximized at $(2)^{1/4}\left(\frac{4}{C_\mathrm{s}\lambda_\mathrm{e}^3}\right)^{1/4}$ when both $\eta_0$ and $\mathrm{CTF}_0$ are zero. For $\eta_0 = \pi/2$, the Scherzer frequency increases from $\sqrt{\frac{1}{2}}\left(\frac{4}{C_\mathrm{s}\lambda_\mathrm{e}^3}\right)^{1/4}$ to $\sqrt{\frac{1}{2}+\sqrt{\frac{1}{2}}}\left(\frac{4}{C_\mathrm{s}\lambda_\mathrm{e}^3}\right)^{1/4}$ as $\mathrm{CTF}_0$ decreases from 1 to 0. Solutions do not exist in the gray region of the plot.

In principle, spherical aberration is the main limitation on the passband of the CTF. But for dose-sensitive samples which cannot be directly focused on before imaging, in practice the main limitation comes from the accuracy of the focusing [10]–[12]. To see this, define the cutoff frequency of the CTF $s_+$ such that $|\mathrm{CTF}(s)| \geq \mathrm{CTF}_0$ for $s \leq s_+$. Figure 8.6a shows the range of defocus values (in nm) which satisfy this requirement, as a function of $s_+$ and $\mathrm{CTF}_0$, for $\eta_0 = \pi/2$, $C_\mathrm{s} = 4.8\,\mathrm{mm}$ and $\lambda_\mathrm{e} = 1.97\,\mathrm{pm}$. Notice that achieving a cutoff frequency of $\left(4\,\text{Å}\right)^{-1}$ while keeping the CTF above $\mathrm{CTF}_0 = 0.8$ requires the focus to be set within a $\sim 20\,\mathrm{nm}$ range. Even in an aberration-corrected TEM with $C_\mathrm{s} = 0$, the required defocus precision for these conditions is still $< 50\,\mathrm{nm}$ (figure 8.6b). The sensitive dependence of the cutoff on defocus also sets an upper limit on the thickness of the sample. If the sample is substantially thicker than the required defocus range for a given cutoff frequency, then parts of the sample will not be sufficiently in focus. This is well within the realm of possibility for cryo-EM samples (cell sections in particular), which may be as thick as several hundred nanometers [93]. Similarly, if the sample is substantially tilted (as in cryo-electron tomography), then the defocus will vary across the field of view. Our progress towards achieving consistent in-focus cryo-EM imaging is described in section 12.3

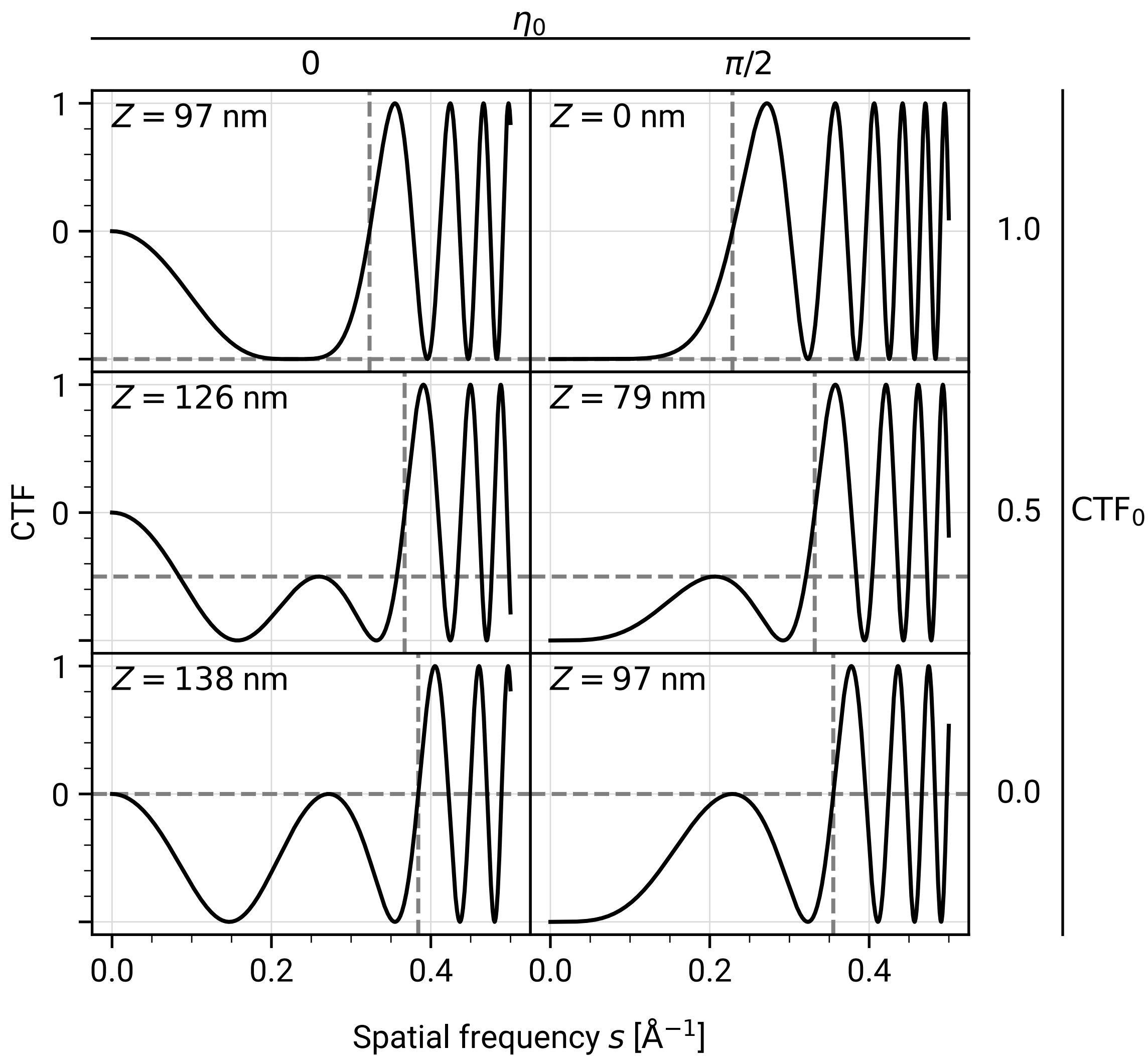

Figure 8.4: The contrast transfer function (CTF) of an ideal phase plate as a function of spatial frequency $s$, for several different values of phase shift $\eta_0$ and minimum CTF value in the passband $CTF_0$.

## 8.6 Ghost images

### 8.6.1 A simple model

The LPP differs noticeably from an ideal phase plate in that it consists of a series of separate phase shifting regions (at the anti-nodes of the standing wave). Assuming the LPP is centered in the diffraction plane ($\mathbf{s}_l = 0$) and NA $\ll 1$, the phase shift near the center of the diffraction plane ($|\mathbf{s}| \ll \frac{\lambda_L}{\mathrm{NA} f_{\mathrm{eff}} \lambda_e}$) is approximately sinusoidal:

$$\eta(\mathbf{s}) \simeq \frac{\eta_0}{2} \left(1 + \rho(\theta, \beta) \cos\left(\frac{4\pi}{\lambda_L} f_{\mathrm{eff}} \lambda_e s_y\right)\right) \tag{8.60}$$

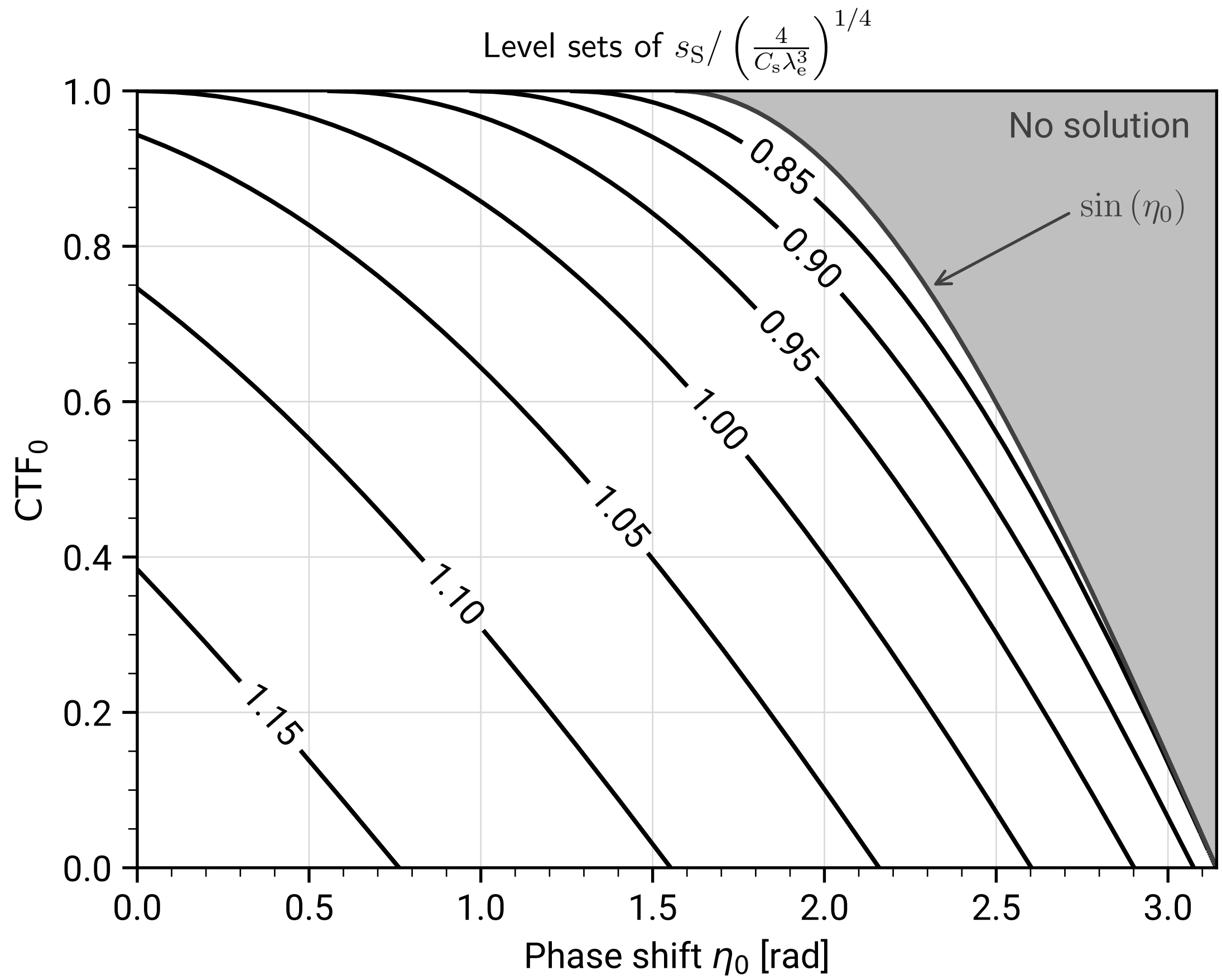

Figure 8.5: Level sets of the Scherzer frequency $s_S$ (equation (8.58)) divided by the conventional Scherzer frequency $\left(\frac{4}{C_s \lambda_e^3}\right)^{1/4}$, as a function of the minimum contrast transfer function in the passband $\mathrm{CTF}_0$ and the phase shift $\eta_0$. The value of each level set is annotated on the corresponding contour line.

As shown in section 8.4, this sinusoidal profile appears in the CTF, and is responsible for two striking features of images taken using the LPP. First, the LPP generates contrast for a small set of very low spatial frequencies (see section 8.4). Second, the image contains several copies of all features in the image. These copies are displaced in both directions along the laser axis, separated by integer multiples of a fixed distance. We refer to these copies as “ghost

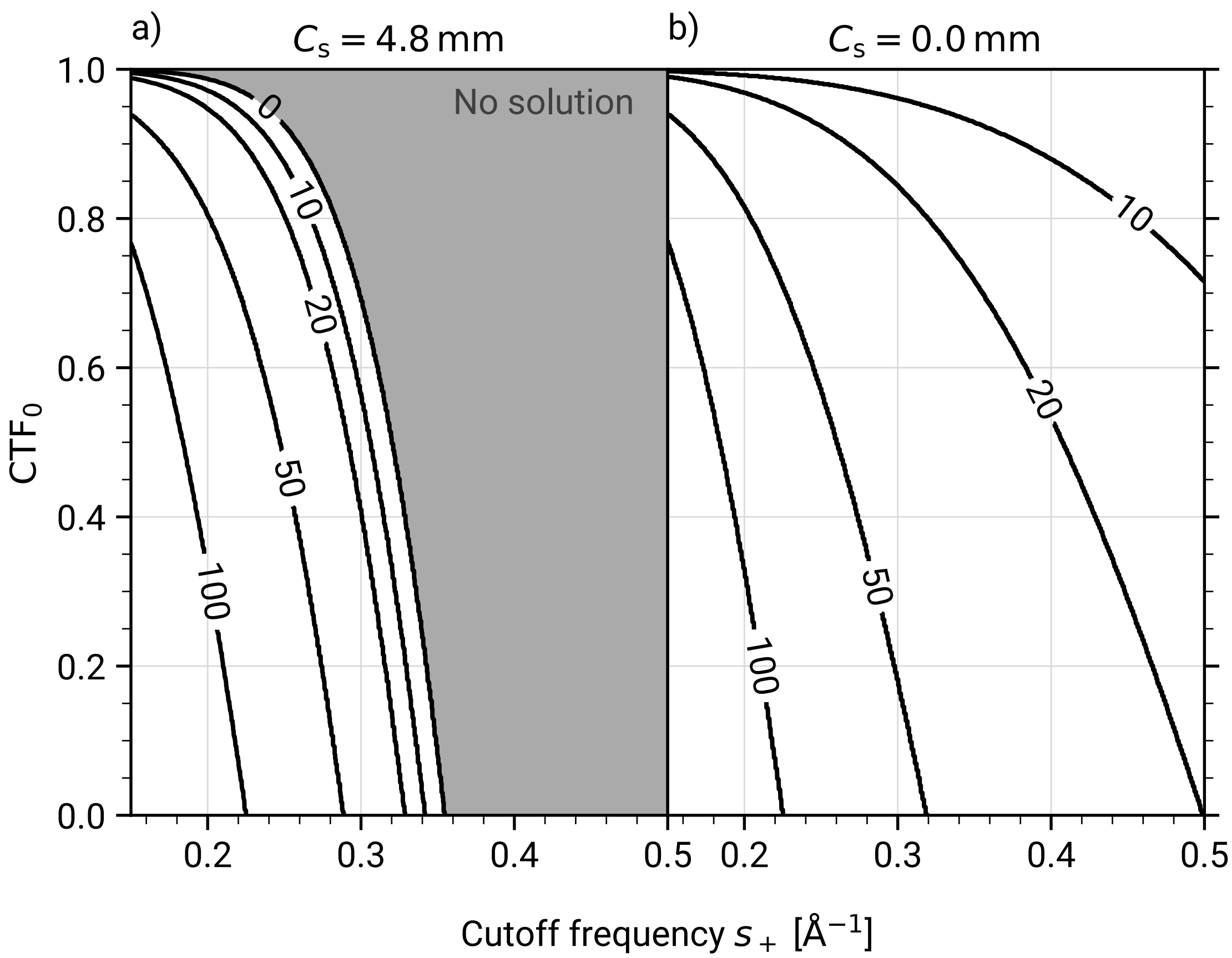

Figure 8.6: Level sets of the defocus range (in nanometers) required to keep the magnitude of the contrast transfer function above $\mathrm{CTF}_0$ until the cutoff frequency $s_+$. The value of each level set is annotated on the corresponding contour line. **a)** For $C_s = 4.8$ mm as in our custom transmission electron microscope. The conditions cannot be satisfied for any defocus value in the gray region. **b)** For $C_s = 0$ mm as in an aberration-corrected transmission electron microscope.

images", as they are low-contrast, low-pass filtered versions of the primary image objects. They come from diffraction of the electron wave on the laser's standing wave, which serves as a sinusoidal phase grating. Therefore, the wave components which form the ghost images are still phase coherent with the unscattered wave, and so their relative contrast varies as a function of the diffraction phase. To quantify these effects, we ignore the conventional aberration function $\gamma(\mathbf{s})$, and write the wave transfer function as

$$\tilde{H}(\mathbf{s}) = e^{-i\eta(\mathbf{s})} \tag{8.61}$$

$$= e^{-i\frac{\eta_0}{2}} \sum_{n=-\infty}^{\infty} (-i)^n J_n\left(\frac{\eta_0\rho}{2}\right) e^{-in\frac{4\pi}{\lambda_\mathrm{L}} f_\mathrm{eff}\lambda_\mathrm{e} s_y} \tag{8.62}$$

by using the Jacobi-Anger expansion

$$e^{iz\cos(\phi)} = \sum_{n=-\infty}^{\infty} i^n J_n(z)\, e^{in\phi} \tag{8.63}$$

where $J_n(z)$ is the $n$-th Bessel function of the first kind, and $\rho \coloneqq \rho(\theta,\beta)$ for brevity. Therefore, the electron wave in the image plane is

$$\psi_\mathrm{im}(\mathbf{x}) = \mathcal{F}^{-1}\left[\tilde{H}(\mathbf{s})\,\mathcal{F}[\psi_\mathrm{E}](\mathbf{s})\right](\mathbf{x}) \tag{8.64}$$

$$= e^{-i\frac{\eta_0}{2}} \sum_{n=-\infty}^{\infty} (-i)^n J_n\left(\frac{\eta_0\rho}{2}\right) e^{-i2\pi n\mathbf{d}_\mathrm{g}\cdot\mathbf{s}_0}\psi_\mathrm{E}(\mathbf{x} - n\mathbf{d}_\mathrm{g}) \tag{8.65}$$

which is an infinite sum of shifted and attenuated copies of the exit wave $\psi_\mathrm{E}(\mathbf{x})$. Note that this formula is valid for arbitrary exit waves, not just ones satisfying the weak phase approximation. The difference in positions between the ghost images in the object plane is

$$\mathbf{d}_\mathrm{g} \coloneqq 2f_\mathrm{eff}\frac{\lambda_\mathrm{e}}{\lambda_\mathrm{L}}\hat{\mathbf{y}} \tag{8.66}$$

where $d_\mathrm{g} \coloneqq |\mathbf{d}_\mathrm{g}| = 74\,\mathrm{nm}$ for $f_\mathrm{eff} = 20\,\mathrm{mm}$, $\lambda_\mathrm{L} = 1064\,\mathrm{nm}$ and $\lambda_\mathrm{e} = 1.97\,\mathrm{pm}$ corresponding to an electron kinetic energy of $K = 300\,\mathrm{keV}$.

We are mostly interested in the typical LPP configuration where $\eta_0 = \pi/2$ and $\rho = 1$. In this case, $|J_n(\eta_0\rho/2)| \sim (\eta_0\rho/2)^{|n|} < 0.074$ for $|n| > 1$, which lets us approximate the image in using only terms which are at most linear in $J_1$:

$$\begin{aligned}|\psi_\mathrm{im}(\mathbf{x})|^2 \simeq J_0^2\left(\frac{\eta_0\rho}{2}\right)\Bigg(&|\psi_\mathrm{E}(\mathbf{x})|^2\\ &+ 2\mathcal{J}_1\left(\frac{\eta_0\rho}{2}\right)\mathrm{Re}\left[ie^{-i\xi}\psi_\mathrm{E}^*(\mathbf{x}+\mathbf{d}_\mathrm{g})\,\psi_\mathrm{E}(\mathbf{x})\right]\\ &+ 2\mathcal{J}_1\left(\frac{\eta_0\rho}{2}\right)\mathrm{Re}\left[ie^{i\xi}\psi_\mathrm{E}^*(\mathbf{x}-\mathbf{d}_\mathrm{g})\,\psi_\mathrm{E}(\mathbf{x})\right]\Bigg)\end{aligned} \tag{8.67}$$

where $\mathcal{J}_n(x) \coloneqq J_n(x)/J_0(x)$ and $\xi \coloneqq 2\pi\mathbf{d}_\mathrm{g}\cdot\mathbf{s}_0$ is the "diffraction phase". This retains just the two "first-order" ghost images on each side of the primary image. We will examine the behavior of the ghost images within this approximation in the particular case where the ghost images do not overlap with the primary image. The case where the ghost images do overlap with the primary image is more complicated an obscures the key features of the ghost images, though it is still well-modeled by equation (8.67).

### 8.6.2 Non-overlapping images

Consider the case where $\psi_{\mathrm{E}}(\mathbf{x})$ is generated by an object centered about $\mathbf{x} = 0$ with a width smaller than $d_{\mathrm{g}}$, such that

$$\psi_{\mathrm{E}}(\mathbf{x}) = \begin{cases} e^{i2\pi\mathbf{s}_0\cdot\mathbf{x}} e^{i\phi(\mathbf{x})} e^{-\mu(\mathbf{x})} & |\mathbf{x}| < d_{\mathrm{g}}/2 \\ e^{i2\pi\mathbf{s}_0\cdot\mathbf{x}} & |\mathbf{x}| \geq d_{\mathrm{g}}/2 \end{cases} \tag{8.68}$$

The resulting image, as modeled by equation (8.67), will then be

$$|\psi_{\mathrm{im}}(\mathbf{x})|^2 = J_0^2\left(\frac{\eta_0\rho}{2}\right) \cdot \begin{cases} 1 & -\infty < \mathbf{d}_{\mathrm{g}}\cdot\mathbf{x} < -\frac{3}{2}d_{\mathrm{g}} \\ 1 + 2\mathcal{J}_1\left(\frac{\eta_0\rho}{2}\right)\left(e^{-\mu(\mathbf{x}+\mathbf{d}_{\mathrm{g}})}\sin\left(\phi\left(\mathbf{x}+\mathbf{d}_{\mathrm{g}}\right)+\xi\right)-\sin\left(\xi\right)\right) & -\frac{3}{2}d_{\mathrm{g}} < \mathbf{d}_{\mathrm{g}}\cdot\mathbf{x} < -\frac{1}{2}d_{\mathrm{g}} \\ e^{-2\mu(\mathbf{x})} - 4\mathcal{J}_1\left(\frac{\eta_0\rho}{2}\right)e^{-\mu(\mathbf{x})}\cos\left(\xi\right)\sin\left(\phi\left(\mathbf{x}\right)\right) & -\frac{1}{2}d_{\mathrm{g}} < \mathbf{d}_{\mathrm{g}}\cdot\mathbf{x} < \frac{1}{2}d_{\mathrm{g}} \\ 1 + 2\mathcal{J}_1\left(\frac{\eta_0\rho}{2}\right)\left(e^{-\mu(\mathbf{x}-\mathbf{d}_{\mathrm{g}})}\sin\left(\phi\left(\mathbf{x}-\mathbf{d}_{\mathrm{g}}\right)-\xi\right)+\sin\left(\xi\right)\right) & \frac{1}{2}d_{\mathrm{g}} < \mathbf{d}_{\mathrm{g}}\cdot\mathbf{x} < \frac{3}{2}d_{\mathrm{g}} \\ 1 & \frac{3}{2}d_{\mathrm{g}} < \mathbf{d}_{\mathrm{g}}\cdot\mathbf{x} < \infty \end{cases} \tag{8.69}$$

The first and last lines describe the image outside of the regions to which the ghost images are projected, that is, the uniform background. The second and third lines describe the ghost images, one on each side of the object. They may be brighter or darker than the background, depending on the object and the diffraction phase. The object itself is described by the third line. Notably, the wave structure of the phase plate generates phase contrast, and the phase contrast is modulated by the cosine term depending on the phase shift encountered by the unscattered wave.

In the case of an object with no amplitude contrast,

$$|\psi_{\mathrm{im}}(\mathbf{x})|^2 = J_0^2\left(\frac{\eta_0\rho}{2}\right) \cdot \begin{cases} 1 & -\infty < \mathbf{d}_{\mathrm{g}}\cdot\mathbf{x} < -\frac{3}{2}d_{\mathrm{g}} \\ 1 + 4\mathcal{J}_1\left(\frac{\eta_0\rho}{2}\right)\cos\left(\xi + \frac{\phi(\mathbf{x}+\mathbf{d}_{\mathrm{g}})}{2}\right)\sin\left(\frac{\phi(\mathbf{x}+\mathbf{d}_{\mathrm{g}})}{2}\right) & -\frac{3}{2}d_{\mathrm{g}} < \mathbf{d}_{\mathrm{g}}\cdot\mathbf{x} < -\frac{1}{2}d_{\mathrm{g}} \\ 1 - 4\mathcal{J}_1\left(\frac{\eta_0\rho}{2}\right)\cos\left(\xi\right)\sin\left(\phi\left(\mathbf{x}\right)\right) & -\frac{1}{2}d_{\mathrm{g}} < \mathbf{d}_{\mathrm{g}}\cdot\mathbf{x} < \frac{1}{2}d_{\mathrm{g}} \\ 1 + 4\mathcal{J}_1\left(\frac{\eta_0\rho}{2}\right)\cos\left(\xi - \frac{\phi(\mathbf{x}-\mathbf{d}_{\mathrm{g}})}{2}\right)\sin\left(\frac{\phi(\mathbf{x}-\mathbf{d}_{\mathrm{g}})}{2}\right) & \frac{1}{2}d_{\mathrm{g}} < \mathbf{d}_{\mathrm{g}}\cdot\mathbf{x} < \frac{3}{2}d_{\mathrm{g}} \\ 1 & \frac{3}{2}d_{\mathrm{g}} < \mathbf{d}_{\mathrm{g}}\cdot\mathbf{x} < \infty \end{cases} \tag{8.70}$$

which illustrates several of the key behaviors of the ghost images. First, when the LPP is aligned ($\xi = 0$), the ghost images have the opposite contrast as the primary image (assuming a weak phase object). As defined in section 8.2, $\phi(\mathbf{x})$ is more positive in more attractive regions of the sample (higher atomic number and/or density). The primary image of these regions will be darker than the background, while the ghost images will be brighter. Second, the contrast of the ghost images oscillates asymmetrically as a function of the diffraction phase about $\xi = 0$. This means that the ghosts may have the same contrast, opposite contrast, or zero contrast depending on $\xi$ and $\phi(\mathbf{x})$. This effect can be used as a visual aid when aligning the LPP—the position of the unscattered wave is adjusted until the image shows increased contrast and any visible ghost images on each side of an object have equal brightness.

### 8.6.3 Discussion

In an actual LPP, the ghost images are not perfect copies of the primary image because the wave transfer function is only sinusoidal in a narrow strip (the laser beam) of spatial frequencies about the origin. The ghosts are composed only of spatial frequencies which lie within that strip, so they are low-pass filtered relative to the primary image, and primarily in the direction orthogonal to the laser axis. In the case of a weak phase object and an ideal phase contrast phase shift of $\eta_0\rho = \pi/2$, the contrast of the ghost images is very weak, and typically they are not visible in noisy low-dose images. For stronger phase and/or amplitude objects they become quite visible, and substantially impact the interpretability of cluttered images (see figure 9.5). It may be possible to address this issue to some extent in image processing since the forward model for the ghost image formation is well-described, but there will surely be limitations to this approach from limited signal-to-noise ratio in the image and ghost images being projected into the field of view by unseen objects outside of the field of view. It is also possible to completely eliminate the ghost images by removing the standing wave structure from the LPP phase profile by changing the polarization of the laser light. This effect is discussed further in chapter 9. However, this has the side effect of increasing the cut-on frequency of the LPP, since there is no longer contrast being generated at frequencies within the width of the laser beam. A third solution is to use multiple crossed laser beams to form the LPP. In this case, each beam can have a lower maximum phase shift $\eta_0$, since their phase shifts add where they cross. This reduces the contrast of each ghost image, though the number of total ghost images is increased (one set of ghost images per laser beam, see section 13.3).

## 8.7 Resolution & the contrast transfer function envelope

Several effects act to reduce the magnitude of the CTF as spatial frequency increases, thereby limiting the resolution of the imaging system. The achievable resolution is often of fundamental importance to the study of the sample. Therefore, it is useful to quantitatively model these resolution-loss effects to predict the expected resolution of the TEM. This also enables the experimental identification of any unexpected resolution limitations. Some of the effects discussed here are specific to phase plates.

### 8.7.1 Partially coherent illumination

The most common resolution limitation affecting state-of-the-art conventional TEMs is incoherence of the electron source. This causes each electron which illuminates the sample to have slightly different properties (e.g. kinetic energy or incidence angle). This in turn causes each electron wave to be imaged with a slightly different CTF. When integrated over the millions of electrons which may constitute a single image, the high resolution information can average out, resulting in resolution loss. I will start by presenting a general framework for modeling this effect, and then specialize to several specific cases.

The distribution of any ensemble of electron properties (e.g. momentum, energy) can be described by a normalized probability density function $p\left(\mathbf{X}\right)$, where the components of vector $\mathbf{X}$ represent the different properties. The image $I\left(\mathbf{x}\right)$ formed by many electrons with properties taken from this probability distribution is just a weighted sum over the individual images formed by each electron:

$$I\left(\mathbf{x}\right) = \int d\mathbf{X}\, p\left(\mathbf{X}\right) \left|\psi_{\text{im}}\left(\mathbf{x}\right)\right|^2 \tag{8.71}$$

noting that $\left|\psi_{\text{im}}\left(\mathbf{x}\right)\right|^2$ implicitly depends on $\mathbf{X}$. Making the assumption that the phase shift imparted by the sample $\phi\left(\mathbf{x}\right)$ does not depend on the incident electron beam properties,

$$\mathcal{F}\left[I\left(\mathbf{x}\right)\right]\left(\mathbf{s}\right) = \delta\left(\mathbf{s}\right) + 2\mathcal{F}\left[\phi\left(\mathbf{x}\right)\right]\left(\mathbf{s}\right) \cdot \overline{\text{CTF}}\left(\mathbf{s}\right) \tag{8.72}$$

$$\overline{\text{CTF}}\left(\mathbf{s}\right) \coloneqq \int d\mathbf{X}\, p\left(\mathbf{X}\right) \text{CTF}\left(\mathbf{s};\mathbf{X}\right) \tag{8.73}$$

where the dependence of the CTF on $\mathbf{X}$ has been made explicit as a second argument. This assumption is not generally true, but is usually a good approximation as long as the spread in the parameters $\mathbf{X}$ is not too large. In order to keep the equations analytically tractable, we approximate the probability density function as a normal distribution, such that

$$p\left(\mathbf{X}\right) = \frac{1}{\sqrt{\left(2\pi\right)^N \det\left(\mathbf{\Sigma}\right)}} \exp\left(-\frac{1}{2}\left(\mathbf{X}-\boldsymbol{\mu}\right)\cdot\mathbf{\Sigma}^{-1}\left(\mathbf{X}-\boldsymbol{\mu}\right)\right) \tag{8.74}$$

where $\mathbf{\Sigma}$ is the distribution's covariance matrix and $\boldsymbol{\mu}$ is its mean. $N$ is the dimension (number of components) of $\mathbf{X}$. By writing the CTF of equation (8.45) as

$$\text{CTF}\left(\mathbf{s};\mathbf{X}\right) = -\frac{i}{2}\left(e^{i\xi_-\left(\mathbf{s};\mathbf{X}\right)} - e^{i\xi_+\left(\mathbf{s};\mathbf{X}\right)}\right) \tag{8.75}$$

$$\xi_\pm\left(\mathbf{s};\mathbf{X}\right) \coloneqq \pm\left(\chi\left(\mathbf{s}_0 \pm \mathbf{s}\right) - \chi\left(\mathbf{s}_0\right) + \mathcal{A}\right) \tag{8.76}$$

(recalling that $\chi\left(\mathbf{s}\right)$ is the wave aberration function as defined in equation (8.46)) we can Taylor expand the function $\xi_\pm\left(\mathbf{s};\mathbf{X}\right)$ in $\mathbf{X}$ about the mean of the probability distribution $\boldsymbol{\mu}$ such that

$$\xi_\pm\left(\mathbf{s};\mathbf{X}\right) \simeq \xi_\pm\left(\mathbf{s};\boldsymbol{\mu}\right) + \left(\mathbf{X}-\boldsymbol{\mu}\right)\cdot\frac{\partial\xi_\pm\left(\mathbf{s};\boldsymbol{\mu}\right)}{\partial\mathbf{X}} + \frac{1}{2}\left(\mathbf{X}-\boldsymbol{\mu}\right)\cdot\frac{\partial^2\xi_\pm\left(\mathbf{s};\boldsymbol{\mu}\right)}{\partial\mathbf{X}^2}\left(\mathbf{X}-\boldsymbol{\mu}\right) + \cdots \tag{8.77}$$

and so

$$\begin{aligned}\overline{\mathrm{CTF}}(\mathbf{s}) &\simeq -\frac{i}{2}\int d\mathbf{X}\,\frac{1}{\sqrt{(2\pi)^N\det(\boldsymbol{\Sigma})}}\exp\left(-\frac{1}{2}(\mathbf{X}-\boldsymbol{\mu})\cdot\boldsymbol{\Sigma}^{-1}(\mathbf{X}-\boldsymbol{\mu})\right)\\ &\cdot\left[\exp\left(\frac{i}{2}(\mathbf{X}-\boldsymbol{\mu})\cdot\frac{\partial^2\xi_-(\mathbf{s};\boldsymbol{\mu})}{\partial\mathbf{X}^2}(\mathbf{X}-\boldsymbol{\mu})\right)\exp\left(i(\mathbf{X}-\boldsymbol{\mu})\cdot\frac{\partial\xi_-(\mathbf{s};\boldsymbol{\mu})}{\partial\mathbf{X}}\right)e^{i\xi_-(\mathbf{s};\boldsymbol{\mu})}\right.\\ &\quad\left.-\exp\left(\frac{i}{2}(\mathbf{X}-\boldsymbol{\mu})\cdot\frac{\partial^2\xi_+(\mathbf{s};\boldsymbol{\mu})}{\partial\mathbf{X}^2}(\mathbf{X}-\boldsymbol{\mu})\right)\exp\left(i(\mathbf{X}-\boldsymbol{\mu})\cdot\frac{\partial\xi_+(\mathbf{s};\boldsymbol{\mu})}{\partial\mathbf{X}}\right)e^{i\xi_+(\mathbf{s};\boldsymbol{\mu})}\right]\end{aligned}\tag{8.78}$$

We can perform the integral over $\mathbf{X}$ to yield

$$\overline{\mathrm{CTF}}(\mathbf{s}) \simeq -\frac{i}{2}\left(M_-(\mathbf{s})\,e^{i\xi_-(\mathbf{s};\boldsymbol{\mu})} - M_+(\mathbf{s})\,e^{i\xi_+(\mathbf{s};\boldsymbol{\mu})}\right)\tag{8.79}$$

$$M_\pm(\mathbf{s}) \coloneqq \left(\frac{\det\left(\left(\boldsymbol{\Sigma}^{-1}-i\frac{\partial^2\xi_\pm(\mathbf{s};\boldsymbol{\mu})}{\partial\mathbf{X}^2}\right)^{-1}\right)}{\det(\boldsymbol{\Sigma})}\right)^{1/2} e^{-\frac{1}{2}\frac{\partial\xi_\pm(\mathbf{s};\boldsymbol{\mu})}{\partial\mathbf{X}}\cdot\left(\boldsymbol{\Sigma}^{-1}-i\frac{\partial^2\xi_\pm(\mathbf{s};\boldsymbol{\mu})}{\partial\mathbf{X}^2}\right)^{-1}\frac{\partial\xi_\pm(\mathbf{s};\boldsymbol{\mu})}{\partial\mathbf{X}}}\tag{8.80}$$

where $M_\pm(\mathbf{s})$ are the envelope functions for the positive and negative frequency components of the CTF. If $M(\mathbf{s}) \coloneqq M_+(\mathbf{s}) = M_-(\mathbf{s})$ then there is a single envelope function for the CTF:

$$\overline{\mathrm{CTF}}(\mathbf{s}) = M(\mathbf{s})\,\mathrm{CTF}(\mathbf{s};\boldsymbol{\mu})\tag{8.81}$$

The following sections calculate these envelope functions in the particular cases of limited temporal and spatial coherence of the electron beam.

#### 8.7.1.1 Temporal coherence

In general, the focal length of a magnetic lens depends on the electron kinetic energy $K$. In the context of an imaging system, this means that the defocus depends on $K$. Therefore, a spread in illuminating electron kinetic energies will result in a spread in defocus values for the electron waves constituting an image. The resulting "temporal coherence" CTF envelope is the primary resolution limitation in state-of-the-art TEMs. The exact form of the envelope can be derived by setting $\mathbf{X} = K$ and writing the defocus as a function of kinetic energy to first order about the point $\boldsymbol{\mu} = K_0$ such that

$$Z(K) = Z_0 + C_\mathrm{c}\epsilon\frac{K-K_0}{K_0}\tag{8.82}$$

where $Z_0 \coloneqq Z(K_0)$, $C_\mathrm{c}$ is defined as the coefficient of chromatic aberration, and

$$\epsilon = \frac{1+\frac{K_0}{m_\mathrm{e}c^2}}{1+\frac{K_0}{2m_\mathrm{e}c^2}}\tag{8.83}$$

is a relativistic correction factor which is not included in the definition of $C_\mathrm{c}$ [86, p. 1723]. $m_\mathrm{e}$ is the electron mass, and $c$ is the speed of light. Thus,

$$\frac{\partial \xi_\pm(\mathbf{s}; K_0)}{\partial K} = \pm \frac{\pi \epsilon C_\mathrm{c} \lambda_\mathrm{e}}{K_0} \left( |\mathbf{s}_0 \pm \mathbf{s}|^2 - |\mathbf{s}_0|^2 \right) \tag{8.84}$$

noting that we ignore the relatively small dependence of $\lambda_\mathrm{e}$, $C_\mathrm{s}$, and the LPP's $\eta_0$ (see equation (9.26)) on $K$, such that only the chromatism of the defocus aberration is modeled [86, p. 1723], [89, ch. 24]. Using the result of equation (8.80) only up to first order then gives

$$M_{\pm,\mathrm{TC}}(\mathbf{s}) = \exp\left( -\frac{1}{2} \pi^2 \lambda_\mathrm{e}^2 \sigma_Z^2 \left( |\mathbf{s}_0 \pm \mathbf{s}|^2 - |\mathbf{s}_0|^2 \right)^2 \right) \tag{8.85}$$

where $\sigma_Z \coloneqq \epsilon C_\mathrm{c} \frac{\sigma_K}{K_0}$ is the standard deviation in defocus values resulting from the chromatic aberration, and $\sigma_K$ is the standard deviation of the electron kinetic energy distribution. Since the entire derivation of the CTF has implicitly assumed that the incidence angle of the electron wave in the object plane is small such that $|\mathbf{s}_0| \ll 1/\lambda_\mathrm{e}$, the envelope function is approximately the same for both CTF components:

$$M_\mathrm{TC}(\mathbf{s}) = \exp\left( -\frac{1}{2} \pi^2 \lambda_\mathrm{e}^2 \sigma_Z^2 |\mathbf{s}|^4 \right) \tag{8.86}$$

This is the canonical formula for the CTF temporal coherence envelope. The envelope starts at unity at zero spatial frequency and falls through a value of 0.5 at a frequency of

$$s_{0.5,\mathrm{TC}} = \left( \frac{2\ln(2)}{\pi^2} \frac{1}{\lambda_\mathrm{e}^2 \sigma_Z^2} \right)^{1/4} \tag{8.87}$$

The typical full width at half maximum (FWHM) of the kinetic energy distribution for a modern field emission electron gun is roughly $\mathrm{FWHM}_K = 0.8\,\mathrm{eV}$. Noting that $\sigma_K = \mathrm{FWHM}_K / \left(2\sqrt{2\ln(2)}\right)$, this means that for a commonly used kinetic energy of $K_0 = 300\,\mathrm{keV}$ and typical $C_\mathrm{c} = 2.7\,\mathrm{mm}$, $s_{0.5,\mathrm{TC}} = \left(1.4\,\text{Å}\right)^{-1}$. This is high enough of a frequency to not severely impact most cryo-EM work, which is concerned with frequencies $< \left(2\,\text{Å}\right)^{-1}$. However, improving the temporal coherence envelope further by using a more monochromatic source has still been shown to improve the resolution of single particle reconstructions [94]. In our custom TEM, the higher $C_\mathrm{c} = 7.6\,\mathrm{mm}$ reduces this cutoff to $s_{0.5,\mathrm{TC}} = \left(2.4\,\text{Å}\right)^{-1}$, which notably impacts the image resolution.

Terms are sometimes added ad hoc to the defocus standard deviation $\sigma_Z$ to represent defocus spread originating from sources other than the native electron beam energy spread [95, p. 72]. For example,

$$\sigma_Z = C_\mathrm{c} \sqrt{\left( \epsilon \frac{\sigma_V}{V_0} \right)^2 + \left( \epsilon \frac{\sigma_K}{K_0} \right)^2 + \left( 2 \frac{\sigma_\mathcal{I}}{\mathcal{I}_0} \right)^2} \tag{8.88}$$

where $\sigma_V$ represents a spread in accelerating voltage of the microscope, and $\sigma_{\mathcal{I}}$ a spread in electron lens coil currents (the factor of 2 coming from the fact that the focal length of the magnetic lens is proportional to the square of the current). However, in a modern TEM these additional terms are usually negligible as $\frac{\sigma_{\mathcal{I}}}{\mathcal{I}_0} \approx 0.1 \times 10^{-6}$, and $\frac{\sigma_V}{V_0} \approx 0.07 \times 10^{-6}$, while for $\mathrm{FWHM}_K = 0.8\,\mathrm{eV}$, $\frac{\sigma_K}{K_0} \approx 1.13 \times 10^{-6}$. However, it is worth noting that monochromated or cold field emission sources have a FWHM energy spread of $< 0.3\,\mathrm{eV}$, in which case the contribution of the other terms becomes more substantial since then $\frac{\sigma_K}{K_0} < 0.42 \times 10^{-6}$. It is also worth noting that the energy spread distribution of field emission sources is only roughly Gaussian [96, ch. 4.1], [97, ch. 44], but this is rarely accounted for in models of the CTF envelope.

#### 8.7.1.2 Spatial coherence

Each electron wave emanates from a slightly different location in the electron source. This causes there to be a spread in illumination angles $\boldsymbol{\theta}_0$ in the object plane. Noting that illuminating waves with different illumination angle are represented in our formulas by waves with different diffraction plane positions $\mathbf{s}_0 = \boldsymbol{\theta}_0/\lambda_\mathrm{e}$ of the unscattered wave, we can again use the general formula developed in this section with $\mathbf{X} = \mathbf{s}_0$. We assume that the unscattered electron beam is centered in the diffraction plane so that $\boldsymbol{\mu} = 0$. This assumption is made both to keep the resulting formulas simple, and because this is typically how the microscope is aligned in order to minimize coma (see section 10.3.7). We include defocus, spherical aberration, and the laser phase plate in the aberration function. Since we are interested in the high frequency behavior of the envelope function, we can restrict $\mathbf{s}$ to the region outside of the phase plate such that

$$\frac{\partial \eta\left(\pm \mathbf{s}\right)}{\partial \mathbf{s}_0} \simeq 0 \tag{8.89}$$

and so

$$\frac{\partial \xi_\pm\left(\mathbf{s};0\right)}{\partial \mathbf{s}_0} = 2\pi Z \lambda_\mathrm{e} \mathbf{s} - 2\pi C_\mathrm{s} \lambda_\mathrm{e}^3 \mathbf{s} \left|\mathbf{s}\right|^2 \pm \frac{\partial \eta\left(0\right)}{\partial \mathbf{s}_0} \tag{8.90}$$

Assuming an azimuthally symmetric distribution of incidence angles, we can take $\boldsymbol{\Sigma} = \sigma_s^2 \mathbf{1}$ where $\sigma_s$ is the standard deviation of the unscattered beam distribution over spatial frequency. This leaves

$$M_{\pm,\mathrm{SC}}\left(\mathbf{s}\right) = \exp\left(-\frac{1}{2}\sigma_s^2 \left|\left(2\pi Z \lambda_\mathrm{e} - 2\pi C_\mathrm{s} \lambda_\mathrm{e}^3 \left|\mathbf{s}\right|^2\right)\mathbf{s} \pm \frac{\partial \eta\left(0\right)}{\partial \mathbf{s}_0}\right|^2\right) \tag{8.91}$$

The frequency-dependent term represents the effect of defocus and spherical aberration, while the gradient of the phase plate profile at the center of the unscattered beam shifts and asymmetrizes the envelope function. It also causes a frequency-independent reduction of the envelope function, which is fundamentally due to the concomitant spread in phase shifts imparted to the unscattered beam. We will see in section 8.7.5 that such a frequency-independent envelope is a general consequence of incoherence when a phase plate is used.

The spatial coherence of field emission electron sources is good enough that for realistic parameters, the spatial coherence envelope is insignificant compared to the temporal coherence envelope. For example, the Thermo Fisher Scientific X-FEG source has a brightness at $K = 300\,\mathrm{keV}$ of $3 \times 10^{13}\,\mathrm{Asr^{-1}m^{-2}}$, meaning that for a typical electron flux in the object plane of $\sim 1\,\mathrm{nA\mu m^{-2}}$, the angular spread is $\sigma_{\theta_0} \sim 3\,\mathrm{\mu rad}$, and the corresponding spatial frequency spread is $\sigma_s \sim (700\,\mathrm{nm})^{-1}$ so that even with a large defocus $Z = 10\,\mathrm{\mu m}$ and $C_\mathrm{s} = 4.8\,\mathrm{mm}$, the cutoff frequency is still relatively high at $s_{0.5,\mathrm{SC}} = \left(1.3\,\text{Å}\right)^{-1}$. However, defocus and spherical aberration do strongly limit the resolution through the mechanism explained in section 8.7.2.

### 8.7.2 Delocalization beyond the field of view

Only a finite-sized area of the wave function in the image plane is actually detected (e.g. using a camera sensor). This means that the image is actually the product of a boxcar function $B(\mathbf{x})$ (representing a rectangular detector) and the squared magnitude of the wave function:

$$I(\mathbf{x}) = B(\mathbf{x})\left|\psi_\mathrm{im}(\mathbf{x})\right|^2 \tag{8.92}$$

Therefore, using the convolution theorem and equation (8.42),

$$\mathcal{F}[I(\mathbf{x})](\mathbf{s}) = \mathcal{F}[B(\mathbf{x})](\mathbf{s}) + 2\mathcal{F}[B(\mathbf{x})](\mathbf{s}) * (\mathcal{F}[\phi(\mathbf{x})](\mathbf{s})\,\mathrm{CTF}(\mathbf{s})) \tag{8.93}$$

If the object being imaged is substantially smaller than the field of view, then variations in its Fourier transform on the scale of the inverse of the field of view will be primarily due to its position within the field of view, rather than its inherent structure. Specifically, we can write the Fourier transform of an object centered at any position $\mathbf{x}_0$ as a function of its Fourier transform if it was centered at the origin $\mathcal{F}[\phi_0(\mathbf{x})](\mathbf{s})$:

$$\mathcal{F}[\phi(\mathbf{x})](\mathbf{s}) = e^{-i2\pi\mathbf{x}_0\cdot\mathbf{s}}\mathcal{F}[\phi_0(\mathbf{x})](\mathbf{s}) \tag{8.94}$$

Since $\mathcal{F}[\phi_0(\mathbf{x})](\mathbf{s})$ varies negligibly across the width of $\mathcal{F}[B(\mathbf{x})](\mathbf{s})$ we can make the approximation that

$$\mathcal{F}[I(\mathbf{x})](\mathbf{s}) = \mathcal{F}[B(\mathbf{x})](\mathbf{s}) + 2\mathcal{F}[\phi_0(\mathbf{x})](\mathbf{s})\left(\mathcal{F}[B(\mathbf{x})](\mathbf{s}) * \left(e^{-i2\pi\mathbf{x}_0\cdot\mathbf{s}}\mathrm{CTF}(\mathbf{s})\right)\right) \tag{8.95}$$

such that the convolution acts only on the combination of the object's position-dependent phase and the CTF. Note that

$$B(\mathbf{x}) = \mathrm{rect}\left(\frac{x}{L_x}\right)\mathrm{rect}\left(\frac{y}{L_y}\right) \tag{8.96}$$

$$\mathcal{F}[B(\mathbf{x})](\mathbf{s}) = \frac{\sin(\pi L_x s_x)}{\pi s_x}\frac{\sin(\pi L_y s_y)}{\pi s_y} \tag{8.97}$$

where $L_x$ and $L_y$ are the length and width of the rectangular detector (as referred to the object plane), and the CTF of equation (8.45) can be written in terms of its positive and negative frequency components

$$\mathrm{CTF}\left(\mathbf{s}\right) = \frac{i}{2}\left(e^{i(\chi(\mathbf{s}_0+\mathbf{s})-\chi(\mathbf{s}_0)+\mathcal{A})} - e^{-i(\chi(\mathbf{s}_0-\mathbf{s})-\chi(\mathbf{s}_0)+\mathcal{A})}\right) \tag{8.98}$$

so that the convolution can be expressed as

$$\begin{aligned}\mathcal{F}\left[B\left(\mathbf{x}\right)\right]\left(\mathbf{s}\right) * \left(e^{-i2\pi\mathbf{x}_0\cdot\mathbf{s}}\mathrm{CTF}\left(\mathbf{s}\right)\right) = \frac{1}{\pi^2}\int d\sigma_x d\sigma_y \frac{\sin\left(\pi L_x\sigma_x\right)}{\sigma_x}\frac{\sin\left(\pi L_y\sigma_y\right)}{\sigma_y}e^{-i2\pi\mathbf{x}_0\cdot(\mathbf{s}-\boldsymbol{\sigma})}\\ \cdot\frac{i}{2}\left(e^{i(\chi(\mathbf{s}_0+\mathbf{s}-\boldsymbol{\sigma})-\chi(\mathbf{s}_0)+\mathcal{A})} - e^{-i(\chi(\mathbf{s}_0-\mathbf{s}+\boldsymbol{\sigma})-\chi(\mathbf{s}_0)+\mathcal{A})}\right)\end{aligned} \tag{8.99}$$

If the aberration function does not change too quickly on the scale of the inverse field of view $1/L_{x,y}$, then the aberration function can be Taylor expanded to first order about $\boldsymbol{\sigma} = 0$ such that

$$\chi\left(\mathbf{s}_0 \pm \mathbf{s} \mp \boldsymbol{\sigma}\right) \simeq \chi\left(\mathbf{s}_0 \pm \mathbf{s}\right) \mp \boldsymbol{\sigma}\cdot\boldsymbol{\nabla}\chi\left(\mathbf{s}_0 \pm \mathbf{s}\right) \tag{8.100}$$

and the convolution can be directly evaluated, resulting in

$$\mathcal{F}\left[I\left(\mathbf{x}\right)\right]\left(\mathbf{s}\right) = \mathcal{F}\left[B\left(\mathbf{x}\right)\right]\left(\mathbf{s}\right) + 2\mathcal{F}\left[\phi\left(\mathbf{x}\right)\right]\left(\mathbf{s}\right)\overline{\mathrm{CTF}}\left(\mathbf{s}\right) \tag{8.101}$$

$$\overline{\mathrm{CTF}}\left(\mathbf{s}\right) \coloneqq \frac{i}{2}\left(M_{+,\mathrm{D}}\left(\mathbf{s}\right)e^{i(\chi(\mathbf{s}_0+\mathbf{s})-\chi(\mathbf{s}_0)+\mathcal{A})} - M_{-,\mathrm{D}}\left(\mathbf{s}\right)e^{-i(\chi(\mathbf{s}_0-\mathbf{s})-\chi(\mathbf{s}_0)+\mathcal{A})}\right) \tag{8.102}$$

where the delocalization envelope function is defined as

$$M_{\pm,\mathrm{D}}\left(\mathbf{s}\right) \coloneqq \mathrm{rect}\left(\frac{\nabla_x\chi\left(\mathbf{s}_0 \pm \mathbf{s}\right) - 2\pi x_{0,x}}{2\pi L_x}\right)\mathrm{rect}\left(\frac{\nabla_y\chi\left(\mathbf{s}_0 \pm \mathbf{s}\right) - 2\pi x_{0,y}}{2\pi L_y}\right) \tag{8.103}$$

The envelope function has hard cutoffs which depend on the derivative of the aberration function, and position of the object within the field of view. This is because under the first order approximation made in equation (8.100), information at a particular spatial frequency is displaced from the object by an amount proportional to the gradient of the aberration function. If that information is displaced outside of the field of view, it is lost. Information from the positive and negative frequency components of the CTF is displaced in opposite directions, and so if the object is closer to one side of the field of view then one frequency component will be lost before the other [98], [99]. Including terms beyond first order in the Taylor expansion somewhat softens the cutoff.

To get a sense of scale, consider a defocus-only aberration function with $\mathbf{s}_0 = 0$, a square field of view $L \coloneqq L_x = L_y$, and a centered object $\mathbf{x}_0 = 0$, such that

$$M_{\pm,\mathrm{D}}\left(\mathbf{s}\right) = \mathrm{rect}\left(\frac{|Z|\,\lambda_{\mathrm{e}}s_x}{L}\right)\mathrm{rect}\left(\frac{|Z|\,\lambda_{\mathrm{e}}s_y}{L}\right) \tag{8.104}$$

The highest spatial frequency present in the image is

$$s_{\max} = \frac{L}{\sqrt{2}\,|Z|\,\lambda_{\mathrm{e}}} \tag{8.105}$$

so if for example $L = 200\,\mathrm{nm}$, $\lambda_{\mathrm{e}} = 1.97\,\mathrm{pm}$ (typical values for high-resolution work) then $s_{\max} = \left(2.8\,\text{Å}\right)^{-1}$ when $Z = 20\,\mu\mathrm{m}$. This places the cutoff right in the near-atomic resolution frequency range, and illustrates why defocus values of $< 20\,\mu\mathrm{m}$ are essentially always used in practice.

Since pixelated detectors are now commonly used, it is useful to require the delocalization envelope cutoff frequency to be above the Nyquist frequency of the detector. Assuming a square $N \times N$ detector with a sample-referred pixel size of $d$, the delocalization cutoff frequency will be larger than the Nyquist frequency so long as

$$|Z| < \frac{d^2 N}{\lambda_{\mathrm{e}}} \tag{8.106}$$

The pixel size must be small enough that the Nyquist frequency is higher than the highest desired resolution of the image. For example, if a near-atomic resolution of 2 Å is desired, the pixel size must no larger than 1 Å. Typical detectors have $N \approx 4000$, so with $\lambda_{\mathrm{e}} = 1.97\,\mathrm{pm}$, this again means that

$$|Z| < 20\,\mu\mathrm{m} \tag{8.107}$$

Note that this represents the upper limit on the maximum usable defocus, because objects are not necessarily centered in the field of view, and the pixel size may be chosen to be even smaller than the Nyquist requirement to compensate for the effects discussed in sections 8.7.3 and 8.7.4.

### 8.7.3 Pixel size

When an image is captured using a pixelated detector, the electron flux in the image plane is integrated over each pixel's area. Assuming rectangular pixels of size $d_x \times d_y$ (referred to the object plane), the resulting image pixel values (normalized per pixel area) are

$$I_{ij} = \frac{1}{d_x d_y} \int dx dy \,\mathrm{rect}\left(\frac{x - x_i}{d_x}\right) \mathrm{rect}\left(\frac{y - y_j}{d_y}\right) I\left(\mathbf{x}\right) \tag{8.108}$$

where $x_i$ and $y_j$ are the coordinates of the center of the $(i, j)$-th pixel. This integral can be expressed as a convolution

$$I_{ij} = \overline{I}\left(x_i, x_j\right) = \left[I * B\right]\left(x_i, y_j\right) \tag{8.109}$$

where

$$B\left(\mathbf{x}\right) = \frac{\mathrm{rect}\left(x/d_x\right)}{d_x} \frac{\mathrm{rect}\left(y/d_y\right)}{d_y} \tag{8.110}$$

Therefore,

$$\mathcal{F}\left[\bar{I}\right](\mathbf{s}) = M_{\mathrm{PS}}(\mathbf{s}) \cdot \mathcal{F}[I](\mathbf{s}) \tag{8.111}$$
$$= \delta(\mathbf{s}) + 2\mathcal{F}[\phi(\mathbf{x})](\mathbf{s}) \cdot M_{\mathrm{PS}}(\mathbf{s})\,\mathrm{CTF}(\mathbf{s}) \tag{8.112}$$

where

$$M_{\mathrm{PS}}(\mathbf{s}) := \mathcal{F}[B](\mathbf{s}) \tag{8.113}$$
$$= \frac{\sin(\pi d_x s_x)}{\pi d_x s_x}\frac{\sin(\pi d_y s_y)}{\pi d_y s_y} \tag{8.114}$$

is the pixel size envelope function. Though the effect does not depend on the CTF, it causes the CTF to be multiplied by an envelope function, similarly to the other CTF envelopes we have seen. Noting that the Nyquist frequency along the $j$-th axis is $s_{\mathrm{N},j} := \frac{1}{2d_j}$, we see that the pixel size envelope drops to 0.63 at the Nyquist frequency along the $j$-th axis.

### 8.7.4 Detective quantum efficiency

An ideal pixelated detector registers the number of electrons that enter each pixel during the exposure time. Real-world detectors are not able to precisely determine the location of each electron, or the number of electrons that have entered a pixel. Roughly speaking, the former results in an additional CTF envelope $M_{\mathrm{DQE}}(\mathbf{s})$, while the latter introduces noise into the image, which can be described by its power spectral density $\mathrm{NPS}(\mathbf{s})$. Both of these effects reduce the signal-to-noise ratio in the images, which can be quantified by defining the detective quantum efficiency (DQE) as the ratio of the signal-to-noise ratio of the detector output to that of the detector input (i.e. the electron beam incident on the detector). The DQE can then be written as

$$\mathrm{DQE}(\mathbf{s}) := \frac{N_{\mathrm{d}}^2}{N_{\mathrm{i}}}\frac{\left(M_{\mathrm{DQE}}(\mathbf{s})\,M_{\mathrm{PS}}(\mathbf{s})\right)^2}{\mathrm{NPS}(\mathbf{s})} \tag{8.115}$$

where $N_{\mathrm{d}}$ is the number of electrons detected and $N_{\mathrm{i}}$ is the number of electrons incident on the detector [30], [100].

In direct electron detectors, the DQE typically falls monotonically as the spatial frequency increases towards the Nyquist frequency. Therefore, using larger image magnifications (i.e. smaller object plane-referred pixel sizes) results in less resolution loss in the image due to the effects of the detector's DQE.

### 8.7.5 Phase shift spread

The phase shift $\eta(\mathbf{s}_0)$ may depend on variables $\mathbf{X}$ external to the imaging system (e.g. time), in addition to its dependence on the location of the unscattered wave $\mathbf{s}_0$. In either case, there will be a probability density function $p(\mathbf{X})$ describing the distribution of $\mathbf{X}$ over the many electrons constituting the image, and so just as in section 8.7.1 the image can be described as an incoherent sum over the electron waves, such that there exists an averaged CTF:

$$\overline{\mathrm{CTF}}(\mathbf{s}) := \int d\mathbf{X}\, p(\mathbf{X})\, \mathrm{CTF}(\mathbf{s};\mathbf{X}) \tag{8.116}$$

$$\mathrm{CTF}(\mathbf{s};\mathbf{X}) = \frac{i}{2}\left(e^{i(\chi(\mathbf{s}_0+\mathbf{s})-\chi(\mathbf{s}_0)+\mathcal{A})} - e^{-i(\chi(\mathbf{s}_0-\mathbf{s})-\chi(\mathbf{s}_0)+\mathcal{A})}\right) \tag{8.117}$$

If $\mathbf{X} \neq \mathbf{s}_0$, then $\chi(\mathbf{s}_0 \pm \mathbf{s})$ does not depend on $\mathbf{X}$ outside of the phase plate, since the phase plate provides no phase shift there. If $\mathbf{X} = \mathbf{s}_0$ then $\chi(\mathbf{s}_0 \pm \mathbf{s})$ does depend on $\mathbf{X}$, but less so at intermediate frequencies which are outside of the phase plate but still low enough that the defocus and spherical aberration envelope (section 8.7.1.2) is negligible. In both cases we have that

$$\overline{\mathrm{CTF}}(\mathbf{s}) = M_{\mathrm{PP}} \cdot \frac{i}{2}\left(e^{i(\chi(\mathbf{s}_0+\mathbf{s})+\eta_{\mathrm{eff}}+\mathcal{A})} - e^{-i(\chi(\mathbf{s}_0-\mathbf{s})+\eta_{\mathrm{eff}}+\mathcal{A})}\right) \tag{8.118}$$

$$= M_{\mathrm{PP}} \cdot -e^{i(\chi(\mathbf{s}_0+\mathbf{s})-\chi(\mathbf{s}_0-\mathbf{s}))/2} \sin\left(\frac{1}{2}\left(\chi(\mathbf{s}_0+\mathbf{s})+\chi(\mathbf{s}_0-\mathbf{s})\right)+\eta_{\mathrm{eff}}+\mathcal{A}\right) \tag{8.119}$$

where

$$M_{\mathrm{PP}} := \left|\int d\mathbf{X}\, p(\mathbf{X})\, e^{-i\chi(\mathbf{s}_0)}\right| \tag{8.120}$$

is the (notably frequency-independent) phase shift spread envelope, and

$$\eta_{\mathrm{eff}} := \arg\left(\int d\mathbf{X}\, p(\mathbf{X})\, e^{-i\chi(\mathbf{s}_0)}\right) \tag{8.121}$$

is the effective phase shift provided by the phase shift distribution. Thus, a spread in phase shifts attenuates the CTF equally at all frequencies (outside of the phase plate), and modifies the amount by which the CTF argument is phase shifted.

In the following subsections I will calculate $M_{\mathrm{PP}}$ and $\eta_{\mathrm{eff}}$ in the cases of limited spatial coherence of the electron beam and a temporally fluctuating phase shift.

#### 8.7.5.1 Unscattered beam spread

If $\mathbf{X} = \mathbf{s}_0$ then we can calculate $M_{\mathrm{PP}}$ and $\eta_{\mathrm{eff}}$ for the laser phase plate profile of equation (8.53) by numerical integration. We take $p(\mathbf{s}_0)$ to be a normal distribution with a standard deviation of $\sigma_s$, the numerical aperture of the laser beam to be an experimentally achievable (see chapter 7) NA = 0.05, the phase shift to be $\eta_0 = \pi/2$, and ignore the relatively small contributions of the conventional aberration function $\gamma(\mathbf{s}_0)$ such that $\chi(\mathbf{s}_0) \simeq -\eta(\mathbf{s}_0)$. The results are shown in figure 8.7 as a function of the ratio of the unscattered beam size to the laser wavelength. The envelope $M_{\mathrm{PP}}$ drops from unity to about 0.85 as the unscattered beam size increases. It then plateaus at that value. Similarly, the effective phase shift $\eta_{\mathrm{eff}}$ drops from $\pi/2$ to roughly $\pi/4$, where it plateaus. Recall from section 8.7.1.2 that a typical unscattered beam size for low-dose cryo-EM imaging is $\sigma_s \sim (700\,\mathrm{nm})^{-1}$. With $f_{\mathrm{eff}} = 20\,\mathrm{mm}$, $\lambda_{\mathrm{e}} = 1.97\,\mathrm{pm}$, and $\lambda_{\mathrm{L}} = 1.064\,\mathrm{\mu m}$, this means that $\sigma_s f_{\mathrm{eff}}\lambda_{\mathrm{e}}/\lambda_{\mathrm{L}} \approx 0.05$, so that $M_{\mathrm{PP}} \approx 0.98$ and $\eta_{\mathrm{eff}} \approx 0.91 \cdot \pi/2$. The effect on $M_{\mathrm{PP}}$ is negligible. The relative effect on $\eta_{\mathrm{eff}}$ is somewhat larger, though in practice the effective phase shift can be made to be $\pi/2$ again simply by increasing $\eta_0$ by a similar fraction.

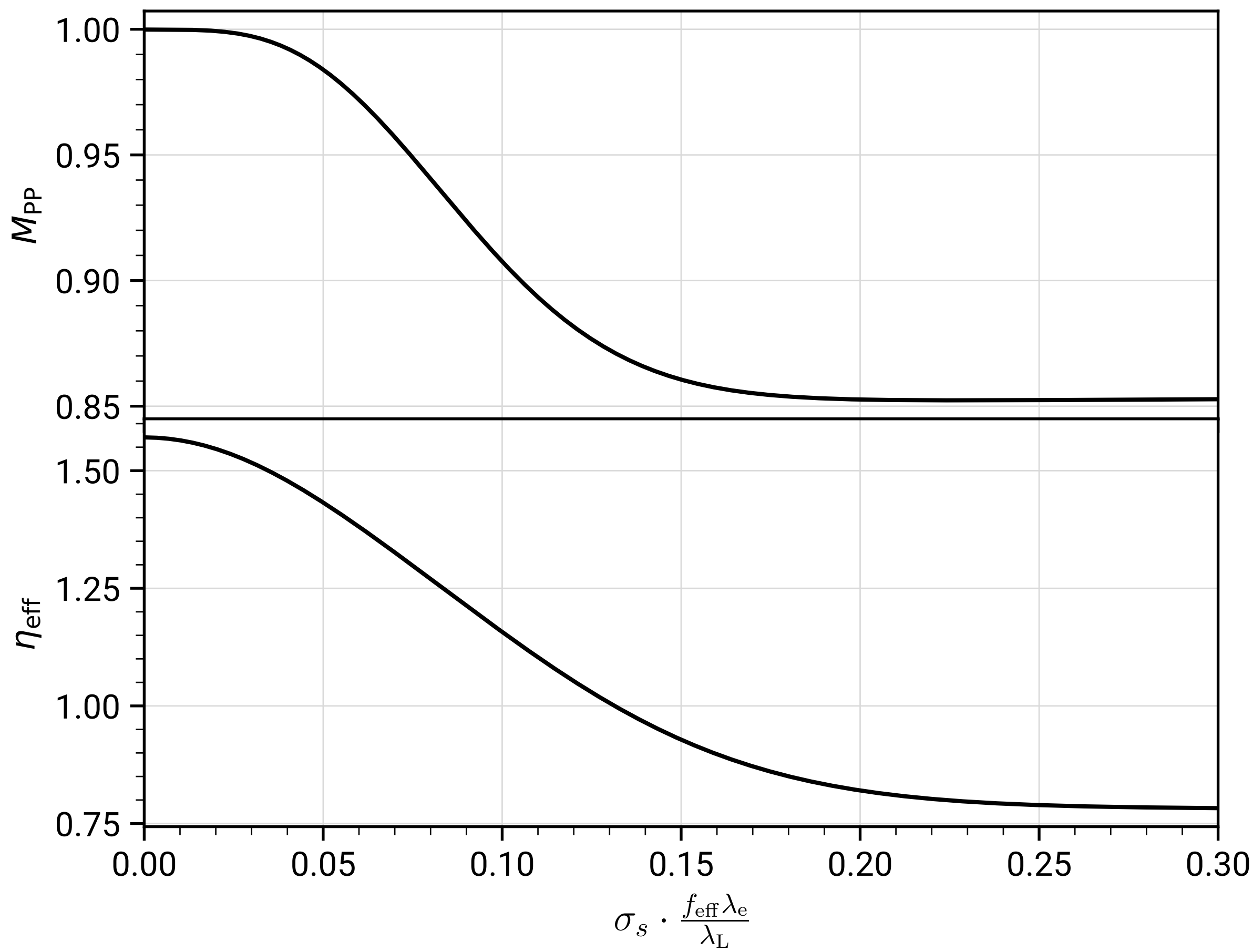

Figure 8.7: The phase plate envelope function (top) and effective phase shift (bottom) of the laser phase plate as a function of the size of the unscattered electron beam in the diffraction plane $f_{\mathrm{eff}}\lambda_{\mathrm{e}}\sigma_s$ divided by the laser wavelength $\lambda_{\mathrm{L}}$, assuming NA $= 0.05$ and $\eta_0 = \pi/2$.

The unscattered beam size may also contain contributions from stochastic deflection of the beam by the sample, in which case the preceding derivation is still valid so long as the unscattered beam size $\sigma_s$ is adjusted appropriately. As of yet, we have not observed any significant contributions from this effect in untilted single-particle analysis cryo-EM samples, though they may become more significant in thicker, tilted cryo-ET samples.

#### 8.7.5.2 Temporally fluctuating phase shift

Fluctuations of the phase shift as a function of time also cause a spread in phase shifts. In the case of the LPP, such fluctuations are primarily due to fluctuations in the laser power (see equation (9.26) for the dependence of the laser phase shift on the parameters of the laser and electron beams). By defining $\eta := \eta(\mathbf{s}_0)$ for brevity, taking $\mathbf{X} = \eta(\mathbf{s}_0)$, $\chi(\mathbf{s}_0) \simeq -\eta(\mathbf{s}_0)$, and assuming that the phase shift values are normally distributed about some mean $\overline{\eta}$ with a standard deviation of $\sigma_\eta$, we see that

$$\int d\mathbf{X}\, p(\mathbf{X})\, e^{-i\chi(\mathbf{s}_0)} = \int d\eta\, p(\eta)\, e^{i\eta} \tag{8.122}$$

$$= e^{-\frac{1}{2}\sigma_\eta^2} e^{i\overline{\eta}} \tag{8.123}$$

Thus,

$$M_{\mathrm{PP}} = e^{-\frac{1}{2}\sigma_\eta^2} \tag{8.124}$$

$$\eta_{\mathrm{eff}} = \overline{\eta} \tag{8.125}$$

so that the envelope factor depends only on the standard deviation of the phase shift, and the effective phase shift is equal to the mean phase shift. Note that the standard deviation of the phase shift must be relatively large compared to the typical $\overline{\eta} = \pi/2$ to have an appreciable impact on the envelope. For example, $M_{\mathrm{PP}}$ is only reduced to 0.9 when $\sigma_\eta \approx 0.3 \cdot \frac{\pi}{2}$. The typical laser power (and thus, phase shift) fluctuations in our LPP can be measured by recording the power exiting the Fabry-Pérot cavity as a function of time using a photodetector. Such measurements (see section 7.3.2) show that the relative standard deviation of the phase shift distribution is approximately 0.0045 rad, which only causes the envelope to decrease to $M_{\mathrm{PP}} = 1 - 1 \times 10^{-5}$. As such, the effect is negligible.

### 8.7.6 Phase plate induced electron loss

Material-based phase plates such as the Volta phase plate (VPP) generally lose some fraction of the throughgoing electrons to scattering. This reduces the signal-to-noise ratio in the image. In the case of the VPP the effect leads to a CTF envelope which falls from near unity at zero frequency to roughly 0.6 at 0.5 $\text{Å}^{-1}$ [13].

The LPP suffers from no such effect. In principle, it is possible for electrons to be elastically scattered from the laser beam in random directions due to spontaneous Compton scattering, but the scattering probability is extremely low. It can be roughly estimated by integrating the Thomson scattering cross section $A_{\mathrm{T}} \approx 6.65 \times 10^{-29}\,\mathrm{m}^2$ times the laser photon flux over time as the electron traverses the laser beam. For a (standing) laser beam with a circulating power of $P_{\mathrm{c}}$, wavelength of $\lambda_{\mathrm{L}}$, and focal waist of $w_0$, the photon flux as a function of position across the beam $z$ is $\frac{2P_{\mathrm{c}}\lambda_{\mathrm{L}}}{\pi^2\hbar c w_0^2} e^{-2z^2/w_0^2}$. The velocity of the electron with kinetic energy $K$ is $c\sqrt{1-\left(\frac{K}{m_\mathrm{e}c^2}+1\right)^{-2}}$ where $m_\mathrm{e}$ is the electron mass and $c$ is the speed of light, so that the scattering probability is

$$p_\mathrm{T} = \int_{-\infty}^{\infty} \frac{dz}{c\sqrt{1-\left(\frac{K}{mc^2}+1\right)^{-2}}} \cdot A_\mathrm{T} \cdot \frac{2P_\mathrm{c}\lambda_\mathrm{L}}{\pi^2\hbar c w_0^2} e^{-2z^2/w_0^2} \tag{8.126}$$

$$= \frac{1}{c\sqrt{1-\left(\frac{K}{mc^2}+1\right)^{-2}}} \cdot A_\mathrm{T} \cdot \frac{2P_\mathrm{c}\lambda_\mathrm{L}}{\pi^2\hbar c w_0^2} \cdot \sqrt{\frac{\pi}{2}} w_0 \tag{8.127}$$

For typical LPP values of $P_\mathrm{c} = 75\,\mathrm{kW}$, $\lambda_\mathrm{L} = 1064\,\mathrm{nm}$, $w_0 = \frac{\lambda_\mathrm{L}}{\pi \mathrm{NA}}$ where $\mathrm{NA} = 0.05$, and $K = 300\,\mathrm{keV}$, we have

$$p_\mathrm{T} \approx 2.7 \times 10^{-8} \tag{8.128}$$

i.e. only about one in every 30 million electrons is randomly scattered, so the effect is negligible.

### 8.7.7 Motion blur

Apparent motion of the sample during an image exposure blurs the image. Apparent motion of the sample can be due to actual motion of the sample relative to the electron optics, or motion of the electron beam due to external influences like electromagnetic interference or dynamical charging of the sample. If the motion is uniform across the field of view and has a constant velocity $\mathbf{v} = v\hat{\mathbf{x}}$ during the exposure time $\tau$, the blurred image can be written as a spatial convolution:

$$I(\mathbf{x}) = \frac{1}{\tau}\int_{-\tau/2}^{\tau/2} dt\ \left|\psi_\mathrm{im}(\mathbf{x}-\mathbf{v}t)\right|^2 \tag{8.129}$$

$$= \left[B * \left|\psi_\mathrm{im}\right|^2\right](\mathbf{x}) \tag{8.130}$$

where the convolution kernel is

$$B(\mathbf{x}) = \frac{\mathrm{rect}\left(\frac{x}{v\tau}\right)}{v\tau}\delta(y) \tag{8.131}$$

Thus, the Fourier transform of the image is

$$\mathcal{F}[I](\mathbf{s}) = \mathcal{F}[B](\mathbf{s}) \cdot \mathcal{F}\left[\left|\psi_\mathrm{im}\right|^2\right](\mathbf{s}) \tag{8.132}$$

$$= \delta(\mathbf{s}) + 2\mathcal{F}[\phi](\mathbf{s}) \cdot M_\mathrm{M}(\mathbf{s})\,\mathrm{CTF}(\mathbf{s}) \tag{8.133}$$

where the motion blur CTF envelope is

$$M_\mathrm{M}(\mathbf{s}) \coloneqq \mathcal{F}[B](\mathbf{s}) \tag{8.134}$$

$$= \frac{\sin(\pi v\tau s_x)}{\pi v\tau s_x} \tag{8.135}$$

Modern direct detection electron cameras capture movies, the frames of which can be registered ("motion corrected") and summed to form an image. Such "fractionation" of the image means that motion is blurred only over the timescale $\tau$ of a single frame's exposure time, rather than the total exposure time of the movie, which greatly reduces motion blurring effects. However, at high resolutions motion blurring during the frame time may still be relevant. Beam-induced motion is a function of electron dose [101], so reducing the intra-frame motion blur can be achieved either by reducing the frame time or decreasing the electron flux. Both cases require more frames to be recorded in order to collect the same total number of electrons on the camera. If the per-frame electron dose is too low, there will not be sufficient signal-to-noise ratio in the single-frame image to perform accurate motion correction, negating the benefit of the nominally reduced intra-frame motion blur. The additional contrast provided by the LPP may be used to boost this signal-to-noise ratio and enable motion correction at lower per-frame electron doses in order to further reduce motion blur. Such "superfractionation" of the image is discussed further in section 12.5.

### 8.7.8 Thermal magnetic field fluctuation blur

Magnetic field fluctuations arising from thermal currents in the conductive materials surrounding the electron beam blurs the image. This effect is normally only limiting for aberration-corrected TEMs which achieve a resolution $< 1\,\text{Å}$. However, in our custom TEM the additional diffraction plane magnification increases the instrument's sensitivity to these fluctuations, making them relevant at resolutions $< 3\,\text{Å}$. The effects of thermal magnetic field fluctuations are discussed in detail in chapter 11, so I will only reproduce the resulting CTF envelope from equation (11.18) here:

$$M_{\mathrm{J}}\left(\mathbf{s}\right) \coloneqq e^{-2\pi^2 \left\langle \delta x^2 \right\rangle |\mathbf{s}|^2} \tag{8.136}$$

where $\langle \delta x^2 \rangle$ is the variance in apparent object positions generated by the blurring effect. Formulas for calculating $\langle \delta x^2 \rangle$ are also given in chapter 11.

### 8.7.9 Summary table

Below is a quick-reference table listing formulas for the different CTF envelopes discussed in this section. As a leading-order approximation, envelopes should be multiplied when multiple effects are simultaneously present.

| $\overline{\mathrm{CTF}}(\mathbf{s}) = \frac{i}{2}\left(M_{+}(\mathbf{s})\, e^{i(\chi(\mathbf{s}_0+\mathbf{s})-\chi(\mathbf{s}_0)+\mathcal{A})} - M_{-}(\mathbf{s})\, e^{-i(\chi(\mathbf{s}_0-\mathbf{s})-\chi(\mathbf{s}_0)+\mathcal{A})}\right)$ | | |
|---|---|---|
| Name: | Formula: | Section: |
| Temporal coherence | $M_{\pm,\mathrm{TC}}(\mathbf{s}) = \exp\left(-\frac{1}{2}\pi^2\lambda_\mathrm{e}^2\sigma_Z^2\left(\lvert\mathbf{s}_0 \pm \mathbf{s}\rvert^2 - \lvert\mathbf{s}_0\rvert^2\right)^2\right)$ | 8.7.1.1 |
| Spatial coherence | $M_{\pm,\mathrm{SC}}(\mathbf{s}) = \exp\left(-\frac{1}{2}\sigma_s^2\left\lvert\left(2\pi Z\lambda_\mathrm{e} - 2\pi C_\mathrm{s}\lambda_\mathrm{e}^3\lvert\mathbf{s}\rvert^2\right)\mathbf{s} \pm \frac{\partial\eta(0)}{\partial\mathbf{s}_0}\right\rvert^2\right)$ | 8.7.1.2 |
| Delocalization | $M_{\pm,\mathrm{D}}(\mathbf{s}) = \mathrm{rect}\left(\frac{\nabla_x\chi(\mathbf{s}_0\pm\mathbf{s})-2\pi x_{0,x}}{2\pi L_x}\right)\mathrm{rect}\left(\frac{\nabla_y\chi(\mathbf{s}_0\pm\mathbf{s})-2\pi x_{0,y}}{2\pi L_y}\right)$ | 8.7.2 |
| Pixel size | $M_{\mathrm{PS}}(\mathbf{s}) = \frac{\sin(\pi d_x s_x)}{\pi d_x s_x}\frac{\sin(\pi d_y s_y)}{\pi d_y s_y}$ | 8.7.3 |
| Phase shift | $M_{\mathrm{PP}} = \left\lvert\int d\mathbf{X}\, p(\mathbf{X})\, e^{-i\chi(\mathbf{s}_0)}\right\rvert$ | 8.7.5 |
| Motion blur | $M_{\mathrm{M}}(\mathbf{s}) = \frac{\sin(\pi v\tau s_x)}{\pi v\tau s_x}$ | 8.7.7 |
| Thermal magnetic field noise | $M_{\mathrm{J}}(\mathbf{s}) = e^{-2\pi^2\left\langle\delta x^2\right\rangle\lvert\mathbf{s}\rvert^2}$ | 8.7.8 |

## 8.8 Laser Ronchigrams

If the phase plate lies above or below the diffraction plane, an image of its phase shift profile will be projected into the image plane. Such an image is called a "Ronchigram", and is well-known in scanning TEM where the sample plays the role of the phase plate [102]. The Ronchigram forms due to interference between components of the convergent (or divergent) electron wave which have scattered from the phase plate. This is illustrated for the case of the LPP in figure 8.8.

It is straightforward to formulate a model of the Ronchigram image formation. Let the laser phase plate lie in a plane a distance $\Delta$ downstream of the electron diffraction plane (figure 8.8 depicts the case of $\Delta < 0$). $\Delta$ is called the "plane offset". Denote the electron wave function just before the phase plate plane as $\psi_1$, and the wave function just after the phase plate plane as $\psi_2$. The phase plate imprints its phase profile $\eta(x, y)$ on the wave function such that

$$\psi_2 = e^{-i\eta} \cdot \psi_1 \tag{8.137}$$

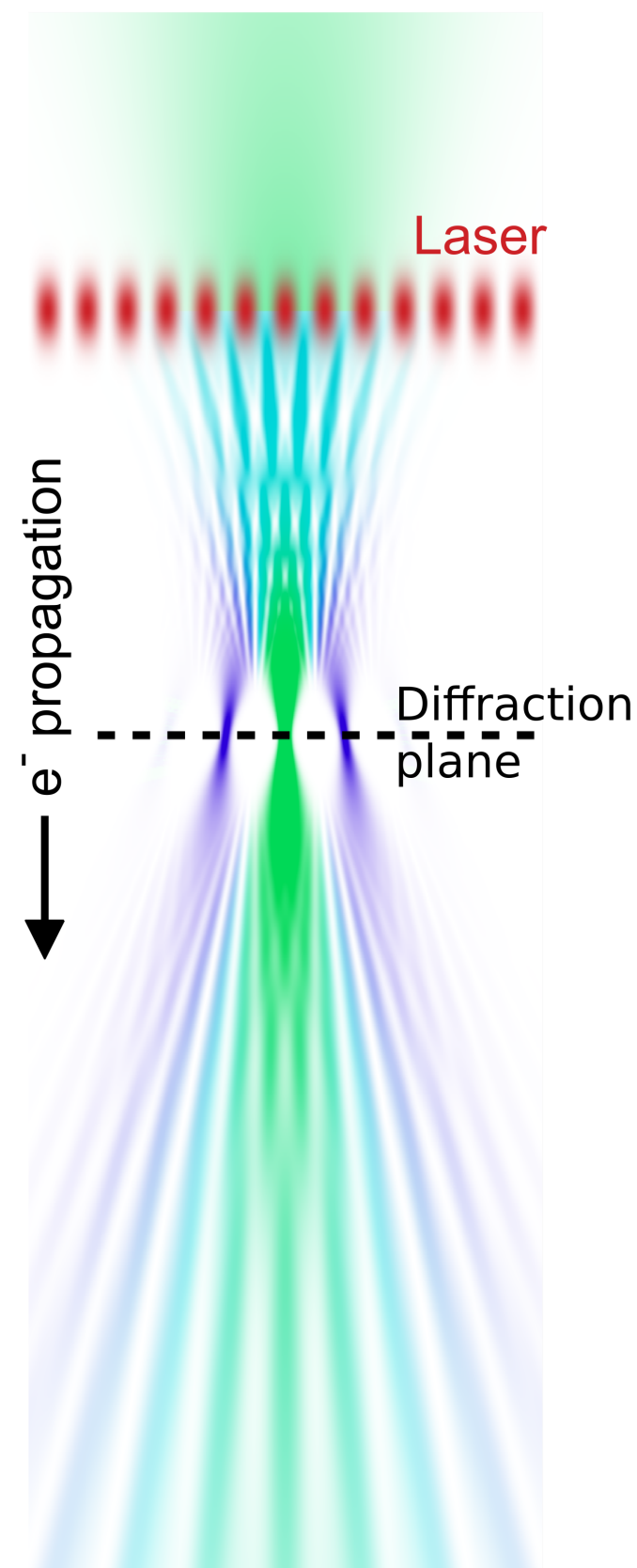

Figure 8.8: Simulation of the electron beam propagation through the laser phase plate. The horizontal scale is exaggerated relative to the vertical scale. Top to bottom, a converging Gaussian electron beam (green) is diffracted by the laser phase plate (red), which generates density modulation in the electron beam (teal) as it propagates downstream from the laser wave. At the diffraction plane (dashed line) the diffraction orders spatially separate, each forming an isolated focal point. As the diffraction orders expand beyond their focal points, they overlap again and their interference pattern forms a Ronchigram image of the phase plate. The hue of the electron beam represents the fraction of diffracted (blue) or undiffracted (green) wave function amplitudes. Reproduced from [23].

noting that here $\eta$ is taken to be a function of real-space coordinates and not the corresponding Fourier-space spatial frequencies in the imaging system. Propagation of $\psi_2$ to the image plane can be broken down into two steps: propagation from the phase plate plane to the diffraction plane, and then propagation from the diffraction plane to the image plane. The second step can be expressed as a Fourier transform, so that the wave function at the image plane, $\psi_{\mathrm{im}}$, can be written as

$$\psi_{\mathrm{im}}\left(\mathbf{x}\right) \propto \mathcal{F}\left[h_{-\Delta} * \psi_2\right]\left(\frac{1}{f_{\mathrm{eff}}\lambda_{\mathrm{e}}}\mathbf{x}\right) \tag{8.138}$$

where $\mathbf{x}$ is the sample-referred image coordinate, $f_{\mathrm{eff}}$ is the effective focal length of the microscope's combined objective/Lorentz/transfer lens system (see chapter 10), the operator $*$ represents a two-dimensional convolution, and

$$h_z(\mathbf{x}) \coloneqq \frac{1}{i\lambda_{\mathrm{e}} z} e^{i\frac{\pi}{\lambda_{\mathrm{e}} z}|\mathbf{x}|^2}, \quad \mathbf{x} \coloneqq (x, y) \tag{8.139}$$

is the Fresnel kernel representing paraxial propagation of the electron wave function through a distance $z$.

Combining equations (8.137) and (8.138), we can express the wave function in the image plane as

$$\psi_{\mathrm{im}}\left(\mathbf{x}\right) \propto \mathcal{F}\left[h_{-\Delta} * \left(e^{-i\eta} \cdot \psi_1\right)\right]\left(\frac{1}{f_{\mathrm{eff}}\lambda_{\mathrm{e}}}\mathbf{x}\right) \tag{8.140}$$

Some algebraic manipulation (including application of the convolution theorem) allows this equation to be rewritten in terms of only Fourier transforms:

$$\psi_{\mathrm{im}}\left(\mathbf{x}\right) \propto \int d^2\mathbf{x}_2\, e^{-i2\pi\frac{1}{f_{\mathrm{eff}}\lambda_{\mathrm{e}}}\mathbf{x}\cdot\mathbf{x}_2} h_{-\Delta}(\mathbf{x}_2) \int d^2\mathbf{x}_1\, e^{i2\pi\frac{1}{\Delta\lambda_{\mathrm{e}}}\mathbf{x}_2\cdot\mathbf{x}_1} e^{-i\eta(\mathbf{x}_1)}\psi_1\left(\mathbf{x}_1\right) h_{-\Delta}\left(\mathbf{x}_1\right) \tag{8.141}$$

We assume that $\psi_1$ is a converging or diverging wave with a uniform amplitude: $\psi_1(\mathbf{x}_1) \propto e^{i\frac{\pi}{\Delta\lambda_{\mathrm{e}}}|\mathbf{x}_1|^2}$. This implicitly omits the presence of a sample, though in principle a sample can be accounted for by including a convergent or divergent version of its scattered wave. This leaves

$$\psi_{\mathrm{im}}\left(\mathbf{x}\right) \propto \int d^2\mathbf{x}_2\, e^{-i2\pi\frac{1}{f_{\mathrm{eff}}\lambda_{\mathrm{e}}}\mathbf{x}\cdot\mathbf{x}_2} h_{-\Delta}(\mathbf{x}_2) \int d^2\mathbf{x}_1\, e^{i2\pi\frac{1}{\Delta\lambda_{\mathrm{e}}}\mathbf{x}_2\cdot\mathbf{x}_1} e^{-i\eta(\mathbf{x}_1)} \tag{8.142}$$

Integrating over $\mathbf{x}_2$ allows $\psi_{\mathrm{im}}$ to be expressed in terms of a single convolution with a Fresnel kernel:

$$\psi_{\mathrm{im}}\left(\mathbf{x}\right) \propto \int d^2\mathbf{x}_1\, h_{\Delta}\left(\frac{\Delta}{f_{\mathrm{eff}}}\mathbf{x} - \mathbf{x}_1\right) \cdot e^{-i\eta(\mathbf{x}_1)} \tag{8.143}$$

This equation has the same form as the expression for image formation via conventional defocus-based phase contrast ($h_\Delta$ is the point spread function with a quadratic aberration function), where $\Delta$ is interpreted as the defocus and $f_{\mathrm{eff}}/\Delta$ is interpreted as the magnification. Therefore, the Ronchigrams are simply defocus phase contrast images of the laser's phase shift profile (equation (8.53)).

The contrast of the Ronchigram oscillates as a function of $\Delta$. To see this, consider equation (8.53) in the limit of NA $\ll 1$ and $\Theta = 0$ so that $\eta(\mathbf{x}) \simeq \frac{\eta_0}{2}\left(1+\cos\left(\frac{4\pi}{\lambda_{\mathrm{L}}}z\right)\right)$. Inserting this expression into equation (8.142) and rewriting $e^{-i\eta}$ using the leading two terms of the Jacobi-Anger expansion gives the image intensity

$$|\psi_{im}(\mathbf{x})|^2 \propto 1 - 4\frac{J_1\left(\frac{\eta_0}{2}\right)}{J_0\left(\frac{\eta_0}{2}\right)}\sin\left(4\pi\Delta\frac{\lambda_{\mathrm{e}}}{\lambda_{\mathrm{L}}^2}\right)\cos\left(\frac{4\pi}{\lambda_{\mathrm{L}}}\frac{\Delta}{f_{\mathrm{eff}}}z\right) \tag{8.144}$$

where $J_n(x)$ is the $n$th Bessel function of the first kind. The amplitude of the cosine wave in the image oscillates sinusoidally as a function of $\Delta$, with a period of $\frac{\lambda_{\mathrm{L}}^2}{2\lambda_{\mathrm{e}}} = 287.3\,\mathrm{mm}$ for $\lambda_{\mathrm{L}} = 1064\,\mathrm{nm}$ and $\lambda_{\mathrm{e}} = 1.97\,\mathrm{pm}$. In order to do phase contrast imaging without forming Ronchigram images, we require that $|\Delta| \ll \frac{\lambda_{\mathrm{L}}^2}{2\lambda_{\mathrm{e}}}$. Essentially any value of $|\Delta|$ can be generated in a TEM because the electron beam can either be focused in the diffraction plane ($\Delta = 0$), collimated in the diffraction plane ($|\Delta| = \infty$), or something in between. The dependence of the Ronchigram amplitude on $\eta_0$ can be used to estimate $\eta_0$ if $\Delta$ is known (usually done by maximizing the amplitude as a function of $\Delta$) or as an aid in aligning the electron beam to near the center of the laser beam (see chapter 10).

Ronchigrams taken at three different plane offsets are shown in figure 8.9, which illustrates the change in effective magnification of the phase plate profile.

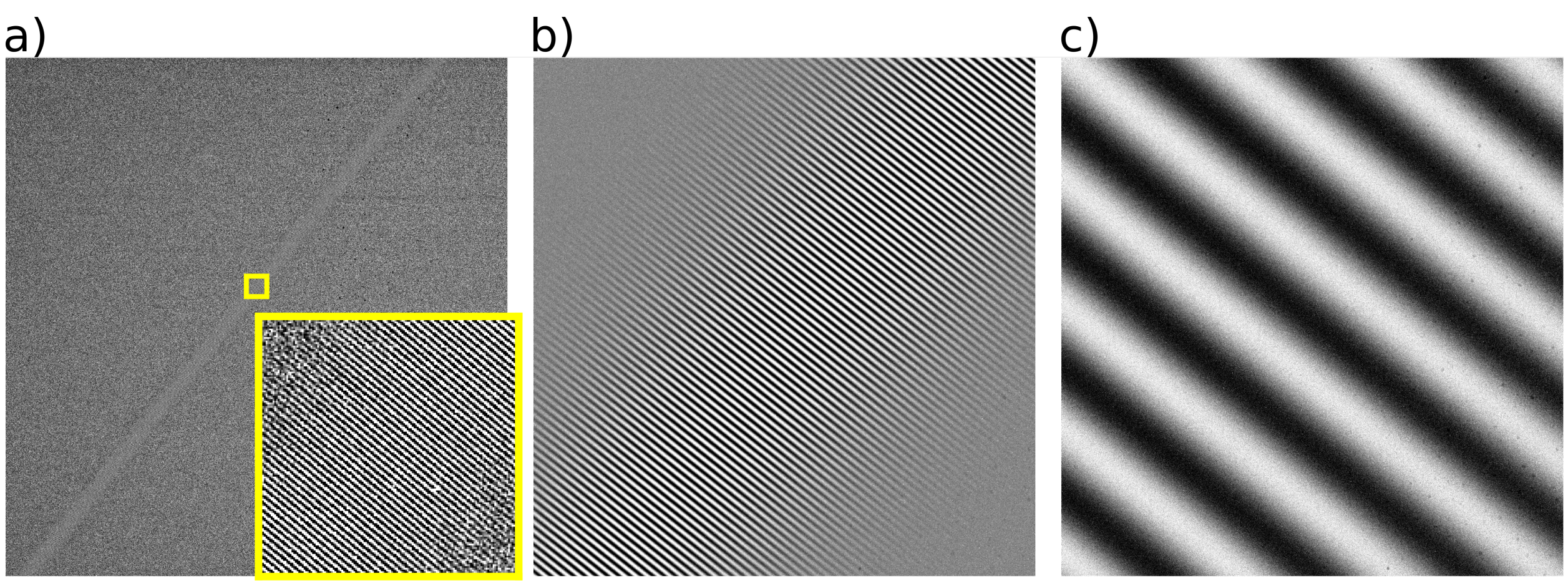

Figure 8.9: Ronchigrams of the laser phase plate taken with decreasing plane offset magnitude, left to right. The phase plate profile is visualized on the scale of the **a)** Rayleigh range **b)** focal waist, and **c)** laser wavelength. The inset in panel a) shows a magnified view of the focal waist. The numerical aperture of the cavity mode in this case was roughly 0.026, and the electron wavelength used was $\lambda_{\mathrm{e}} = 4.18\,\mathrm{pm}$. Reproduced from [23].

A partially coherent electron beam serves to generate Ronchigrams over a spread of positions in the image, dependent on the position of the electron wave focus for each electron.

Similar to the effect of partial coherence on phase contrast imaging, this results in the Ronchigram image being convolved with a kernel with the shape of the electron beam in the diffraction plane. This reduces the amplitude of the Ronchigram, and must be taken into account if precise estimates of $\eta_0$ are to be made from the Ronchigram. Similarly, the effects of camera DQE and pile-up nonlinearity must be accounted for [23]. Astigmatism in the electron wave primarily results in anisotropic magnification of the laser phase plate profile.

We primarily use the laser Ronchigrams as an aid in aligning the electron unscattered wave focus to the laser focus (details in chapter 10).

# Chapter 9

# The relativistic ponderomotive potential

The phase shift experienced by the electron beam as it passes through the laser beam can be described by the "ponderomotive potential", which is an effective potential that arises from the interaction of the laser's oscillating electromagnetic (EM) field and the electron's concomitant oscillatory motion driven by the field. When the electron passes through an EM wave, its position oscillates around some mean trajectory at the EM wave frequency due to the Lorentz force. Since the EM wave in general has an amplitude which varies with position, the oscillatory motion in general moves the electron between regions with lower and higher field amplitude, in phase with the oscillation of the EM wave. When in the lower field portion of the oscillation cycle, the electron experiences a smaller Lorentz force returning it to the high field portion of the cycle. This results in an overall force which pushes the electron away from the high field regions of the wave. This force can be written as the gradient of the ponderomotive potential.

The ponderomotive potential is usually formulated assuming that the electron's velocity is much smaller than the speed of light. This means that the EM wave must have a small enough amplitude to not accelerate the electron's velocity to a substantial fraction of the speed of light, and also that the electron's initial velocity as it enters the EM wave must be much smaller than the speed of light. The former requirement is satisfied even for the high light intensities used in the LPP, as electron motion for optical frequencies does not become appreciably relativistic until intensities of $\sim 1 \times 10^{18}\,\mathrm{Wcm^{-2}}$, at least 6 orders of magnitude higher than that of the LPP [103]. However, the latter requirement is not, since the electrons in a TEM move with a speed of $0.78c$ (assuming a kinetic energy of 300 keV i.e. a wavelength of $\lambda_\mathrm{e} = 1.97\,\mathrm{pm}$). When such relativistic effects are included it is in general not possible to write the effective force on the electron as the gradient of a potential [104]. However, recent theoretical work [105]–[107] indicated that if an initially relativistic electron traverses a standing laser wave (with non-relativistic intensity) at nearly right angles to the wave axis, the interaction may still be written as a pseudo-potential which depends on the initial speed of the electron and the polarization of the EM wave. A derivation of this relativistic ponderomotive potential is presented in section 9.1. We provided the first experimental verification of this theory by imaging Ronchigrams of the LPP's focal waist in our TEM (see section 8.8), and measuring their amplitude as a function of the electron speed and laser polarization [108]. These results are presented in section 9.2.

## 9.1 Theoretical model

The phase factor $e^{-i\eta}$ accumulated by a wave function moving between two points can be approximated with $\eta = -\frac{\mathcal{S}}{\hbar}$ where $\mathcal{S}$ is the classical action evaluated along the classical trajectory $\mathbf{r}(t)$ connecting those two points. This is known as the "stationary-phase approximation". The sign of the phase shift is chosen so that we will have $\eta \geq 0$ for the interaction. The action is a Lorentz scalar, and so must be the same when evaluated in any reference frame. Since in the case of the LPP only the initial electron speed is relativistic, all motion in the reference frame which co-moves with the initial electron velocity $\mathbf{v} = c\beta\hat{\mathbf{z}}$ is non-relativistic. $\beta$ is the electron's speed in units of the speed of light $c$. In this reference frame, the action is equal to the time integral of the electron's non-relativistic Lagrangian along its classical trajectory, such that

$$\eta = -\frac{1}{\hbar}\int dt' \left(\frac{1}{2}m_{\mathrm{e}}\mathbf{v}'^2(t') - \mathrm{e}\mathbf{A}'(\mathbf{r}'(t'), t')\cdot\mathbf{v}'(t')\right) \tag{9.1}$$

where $m_{\mathrm{e}}$ is the electron mass, $e$ is the elementary charge, $\mathbf{v}'(t') = \frac{d\mathbf{r}'}{dt'}$ is the electron velocity in the co-moving frame, $\mathbf{A}'(\mathbf{r}'(t'), t')$ is the magnetic vector potential (in the Coulomb gauge) of the EM wave evaluated at the electron's position $\mathbf{r}'(t')$. Variables in the co-moving frame are denoted by an apostrophe.

We evaluate this expression perturbatively to second order in the field strength parameter $\mathrm{e}|\mathbf{A}'|/m_{\mathrm{e}}c$, which from a quantum perspective corresponds to simulated Compton scattering in which the electron absorbs and then re-emits (or vice-versa) a photon from and into the same EM wave. In the co-moving frame, the electron is initially at rest at position $\mathbf{r}'_0$. The electric field of the EM wave accelerates it to a velocity $\mathbf{v}'_1(t')$ which, to first order in $\mathrm{e}|\mathbf{A}'|/m_{\mathrm{e}}c$ is given by $\mathbf{v}'_1(t') = \mathrm{e}\mathbf{A}'(\mathbf{r}'_0, t')/m_{\mathrm{e}}$. Using this expression in equation (9.1), the phase can then be expressed (to second order in $\mathrm{e}|\mathbf{A}'|/m_{\mathrm{e}}c$) in terms of only $\mathbf{A}'$:

$$\eta = \frac{1}{\hbar}\int dt' \frac{\mathrm{e}^2}{2m_{\mathrm{e}}}\mathbf{A}'^2(\mathbf{r}'_0, t') \tag{9.2}$$

The first order contribution to the phase shift vanishes in the limit of the integral extending over many cycles of the EM wave, since its integrand is sinusoidal as a function of time, which integrates to zero. This is a effectively a consequence of energy-momentum conservation, which precludes a single photon from being absorbed or emitted by a free electron.

To express the phase shift in terms of the laboratory frame Coulomb gauge vector potential, we perform a Lorentz transformation and then restore the Coulomb gauge by a gauge transformation. The Lorentz transformation of the field can be written as

$$0 = \frac{\gamma}{c}\tilde{\Phi}\left(\mathbf{x}, t\right) - \gamma\beta\tilde{A}_z\left(\mathbf{x}, t\right) \tag{9.3}$$

$$\mathbf{A}'_\perp\left(\mathbf{x}', t'\right) = \tilde{\mathbf{A}}_\perp\left(\mathbf{x}, t\right) \tag{9.4}$$

$$A'_z\left(\mathbf{x}', t'\right) = -\frac{\gamma}{c}\beta\tilde{\Phi}\left(\mathbf{x}, t\right) + \gamma\tilde{A}_z\left(\mathbf{x}, t\right) = \frac{1}{\gamma}\tilde{A}_z\left(\mathbf{x}, t\right) \tag{9.5}$$

where $\gamma := \left(1 - \beta^2\right)^{-1}$, $\tilde{A}_z$ and $\tilde{\mathbf{A}}_\perp$ are the longitudinal and transverse components of the laboratory frame vector potential, and $\tilde{\Phi}$ is the laboratory frame scalar potential. Fields not in Coulomb gauge are denoted by a tilde. We then restore the Coulomb gauge in the laboratory frame by a gauge transformation:

$$\tilde{\mathbf{A}} \to \mathbf{A} = \tilde{\mathbf{A}} + \nabla G \tag{9.6}$$

$$\tilde{\Phi} \to \Phi = \tilde{\Phi} - \partial_t G = 0 \tag{9.7}$$

where, using equations (9.3) and (9.6), we see that equation (9.7) is satisfied when

$$\left[\partial_t + c\beta\partial_z\right] G\left(\mathbf{x}, t\right) = c\beta A_z\left(\mathbf{x}, t\right) \tag{9.8}$$

This partial differential equation can be directly integrated to give a solution for the gauge function, which is defined for all $\mathbf{x}$ and $t$:

$$G\left(\mathbf{x}, t\right) = c\beta \int_{-\infty}^{t} dT\, A_z\left(\mathbf{x} - c\beta\left(t - T\right)\hat{z}, T\right) \tag{9.9}$$

We can then rewrite equation (9.2) in the laboratory frame in Coulomb gauge:

$$\eta = \frac{1}{\hbar}\int dt'\, \frac{\mathrm{e}^2}{2m_\mathrm{e}}\mathbf{A}'^2\left(\mathbf{r}'_0, t'\right) \tag{9.10}$$

$$= \frac{1}{\hbar}\int dt\, \frac{\mathrm{e}^2}{2m_\mathrm{e}\gamma}\left[\tilde{\mathbf{A}}_\perp^2\left(\mathbf{r}_0\left(t\right), t\right) + \frac{1}{\gamma^2}\tilde{A}_z^2\left(\mathbf{r}_0\left(t\right), t\right)\right] \tag{9.11}$$

$$= \frac{1}{\hbar}\int dt\, \frac{\mathrm{e}^2}{2m_\mathrm{e}\gamma}\left[\tilde{\mathbf{A}}^2\left(\mathbf{r}_0\left(t\right), t\right) - \beta^2\tilde{A}_z^2\left(\mathbf{r}_0\left(t\right), t\right)\right] \tag{9.12}$$

$$= \frac{1}{\hbar}\int dt\, \frac{\mathrm{e}^2}{2m_\mathrm{e}\gamma}\left[\left(\mathbf{A}\left(\mathbf{r}_0\left(t\right), t\right) - \nabla G\left(\mathbf{r}_0\left(t\right), t\right)\right)^2 - \beta^2\left(A_z\left(\mathbf{r}_0\left(t\right), t\right) - \nabla_z G\left(\mathbf{r}_0\left(t\right), t\right)\right)^2\right] \tag{9.13}$$

where $\mathbf{r}_0\left(t\right)$ is the unperturbed electron trajectory in the laboratory frame.

Using the slowly-varying envelope approximation, which assumes that the amplitude of the EM wave varies slowly along the electron trajectory relative to its oscillation period, we can time-average the integrand of equation (9.13) over one cycle of the field, leaving an effective potential

$$U\left(\mathbf{x}\right) = \frac{\mathrm{e}^2}{2m_\mathrm{e}\gamma}\left\langle\left(\mathbf{A}\left(\mathbf{x}, t\right) - \nabla G\left(\mathbf{x}, t\right)\right)^2 - \beta^2\left(A_z\left(\mathbf{x}, t\right) - \nabla_z G\left(\mathbf{x}, t\right)\right)^2\right\rangle \tag{9.14}$$

such that

$$\eta(x,y) = \frac{1}{\hbar}\int dt\, U(x,y,c\beta t) \tag{9.15}$$

where the angle brackets indicate a time-average over the EM wave cycle and $\mathbf{x} =: (x,y,z)$. To zeroth order in $\beta$, this potential is simply the well-known non-relativistic ponderomotive potential $\frac{\mathrm{e}^2}{2m}\left\langle \mathbf{A}^2(\mathbf{x},t)\right\rangle$. However, equation (9.13) remains valid even if the electromagnetic field does not have a slowly-varying envelope in time and space.

When the electron is relativistic, the $\beta$-dependent terms in equation (9.13) cannot be neglected. In particular, the $\nabla G$ terms become relevant if the amplitude of the EM wave varies substantially over distances comparable to its wavelength, such as in a standing wave. In the case of a monochromatic standing wave with its wave axis parallel to the $x$-axis and polarization specified by angle $\theta$ and ellipticity parameter $\epsilon$, the Coulomb gauge vector potential can be written as

$$\begin{aligned}\mathbf{A}(\mathbf{x},t) &= A_0(y,z)\cos(2\pi x/\lambda_{\mathrm{L}}) \\ &\quad\times\left[\cos(\theta)\cos(2\pi\nu t)\,\hat{\mathbf{z}} + \sin(\theta)\cos(2\pi\nu t - \epsilon)\,\hat{\mathbf{y}}\right]\end{aligned} \tag{9.16}$$

where $A_0(y,z)$ is the wave's amplitude envelope and $\nu = c/\lambda_{\mathrm{L}}$ is its frequency. Note that since the electric field $\mathbf{E} = -\frac{\partial \mathbf{A}}{\partial t}$, the electric field amplitude envelope is $E_0(y,z) = -2\pi\nu A_0(y,z)$. If the slowly-varying envelope approximation is satisfied, we may approximate

$$G(\mathbf{x},t) \simeq \frac{c\beta}{2\pi\nu}\cos(\theta)\,A_0(y,z)\cos(2\pi x/\lambda_{\mathrm{L}})\sin(2\pi\nu t) \tag{9.17}$$

where

$$\nabla G(\mathbf{x},t) \simeq -\hat{x}\beta\cos(\theta)\,A_0(y,z)\sin(2\pi x/\lambda_{\mathrm{L}})\sin(2\pi\nu t) \tag{9.18}$$

so that time-averaging the integrand of equation (9.13) results in the relativistic effective potential

$$U(\mathbf{x}) = \frac{\mathrm{e}^2 A_0^2(y,z)}{4m_{\mathrm{e}}\gamma}\frac{1}{2}\left[1 + \rho(\theta,\beta)\cos(4\pi x/\lambda_{\mathrm{L}})\right] \tag{9.19}$$

where

$$\rho(\theta,\beta) = 1 - 2\beta^2\cos^2(\theta) \tag{9.20}$$

describes the relative depth of the standing wave structure of the potential. An electron beam passing through such an EM wave will acquire a spatial phase modulation

$$\eta(x,y) = \eta_y(y)\frac{1}{2}\left[1 + \rho(\theta,\beta)\cos(4\pi x/\lambda_{\mathrm{L}})\right] \tag{9.21}$$

$$\eta_y(y) := \frac{1}{\hbar}\int dz\,\frac{\mathrm{e}^2 A_0^2(y,z)}{4m_{\mathrm{e}}c\beta\gamma} \tag{9.22}$$

where $\eta_y(0)\,\rho(\theta,\beta)$ is the depth of the phase modulation along the wave axis.

Equations (9.19) and (9.20) show that the relativistic interaction is strongly dependent on both the electron speed $\beta$ and EM wave polarization angle $\theta$, though not on the ellipticity parameter $\epsilon$. Interestingly, if $\beta \geq 1/\sqrt{2}$, there exists a polarization angle $\theta_\mathrm{r}$, which we call the "relativistic reversal angle", such that $\rho(\theta_\mathrm{r},\beta) = 0$. At this angle, the standing wave structure of the phase shift disappears entirely, and therefore no Kapitza-Dirac diffraction occurs.

The relativistic interaction also modifies the laser-induced group delay of the electron wave function. The group delay, equivalent to the retardation of a classical particle and defined as $\tau = -\hbar\frac{d\eta}{dK}$, can be calculated from the energy dependence of the electron phase shift:

$$\tau(x,y) = \frac{\hbar}{m_\mathrm{e}c^2}\frac{\eta_y(y)}{\beta^2\gamma}\frac{1}{2}\left[1 + \varrho(\theta,\beta)\cos(4\pi x/\lambda_\mathrm{L})\right], \tag{9.23}$$

$$\varrho(\theta,\beta) \coloneqq 1 + 2\beta^2\left(1 - 2\beta^2\right)\cos^2(\theta). \tag{9.24}$$

In particular, at $\theta = \theta_\mathrm{r}$, when the standing-wave structure in the potential of equation (9.19) vanishes, the standing wave structure is still present in the spatial profile of the group delay. Furthermore, when $0 < \beta < 1/\sqrt{2}$, the group delay is negative for portions of the standing wave around the electric field nodes. This negative group delay corresponds to an attractive potential, in contrast to the non-relativistic formulation of the ponderomotive potential which is always repulsive.

Applying the full expression for the fundamental mode of the cavity (equation (2.63)) to equation (9.13) is tedious. Instead, we can approximate the relativistic effects by simply matching the behavior of the solution near the focus of the mode, where the wave looks most planar. This results in a solution which matches that of the non-relativistic ponderomotive potential, except that we include the factor $\rho(\theta,\beta)$ in the modulation depth of the phase shift along the laser axis, such that

$$\eta(x,y) = \frac{\eta_0}{2}\frac{e^{-2\frac{Y^2}{1+X^2}}}{\sqrt{1+X^2}}\Bigg[1 + \rho(\theta,\beta)\,e^{-\Theta^2\frac{2}{\mathrm{NA}^2}\left(1+X^2\right)}\left(\frac{1}{1+X^2}\right)^{1/4} \cdot\cos\left(2\frac{X}{1+X^2}Y^2 + \frac{4}{\mathrm{NA}^2}X - \frac{3}{2}\arctan(X) - \varphi\right)\Bigg] \tag{9.25}$$

where laser beam is centered at $\mathbf{x} = \mathbf{x}_\mathrm{l}$, the $x$-axis has been chosen to lie along the axis of the laser beam,

| | |
|---|---|
| $X \coloneqq \frac{\pi\mathrm{NA}^2}{\lambda_\mathrm{L}}(x - x_{\mathrm{l},x})$ | is the normalized $x$-coordinate |
| $Y \coloneqq \frac{\pi\mathrm{NA}}{\lambda_\mathrm{L}}(y - x_{\mathrm{l},y})$ | is the normalized $y$-coordinate |
| $\Theta$ | is the angle between the laser beam axis and the $x$-$y$ plane, assumed to be small |
| NA | is the numerical aperture of the laser beam |
| $\varphi$ | $= 0$ $(= \pi)$ if the laser beam has antinode (node) at $s_x = 0$ |

and

$$\eta_0 = \sqrt{\frac{2}{\pi^3}}\frac{\alpha}{\hbar c^2}\lambda_{\mathrm{L}}\lambda_{\mathrm{e}}\mathrm{NA}P_{\mathrm{c}} \tag{9.26}$$

is the maximum phase shift (when $X = Y = 0$, $\theta = \pi/2$, $\Theta = 0$, $\varphi = 0$), where $\alpha \approx 1/137.035999084$ is the fine-structure constant, $\lambda_{\mathrm{e}}$ is the electron wavelength, and $P_{\mathrm{c}}$ is the circulating power in the cavity mode. Note that $\eta_0$ only increases linearly with NA since while the potential is proportional to the optical intensity, the phase shift is proportional to the intensity times the width of the beam along the electron beam axis.

The expression for the phase shift profile in equation 9.25 likely becomes more inaccurate further from the focus, since in those areas the electron beam crosses multiple wavefronts of the standing wave due to their curvature, and so the slowly-varying envelope approximation is no longer satisfied. The accuracy of this expression in those limits could be evaluated by numerically integrating equation (9.13). We have not yet done this because the profile of the phase shift far from the focus is not particularly relevant to the formation of the phase contrast images (see chapter 8).

This derivation neglects the interaction between the electron's spin and the EM wave. Spin effects should be considered if the electron's Larmor frequency in the magnetic field of the EM wave is comparable to the EM wave frequency [109], [110]. For the LPP, the magnetic field amplitude at the $450\,\mathrm{GWcm^{-2}}$ focus is still only about $6.1\,\mathrm{T}$, so the Larmor frequency is $170\,\mathrm{GHz} \ll \nu = 280\,\mathrm{THz}$, and so spin effects should negligible.

This derivation assumes that the electron only experiences the second-order interaction between it and the laser beam. This neglects not only higher-order terms, but also the second order interaction in which the electron couples the laser beam and other modes of the EM field. The effect of this spontaneous Compton scattering is considered in section 8.7.6 and shown to be negligible for the LPP.

## 9.2 Experimental data

We provided the first experimental verification of the model presented in section 9.1 [108]. A schematic of the experiment is shown in Fig. 9.1a. The electron beam of our TEM passes through the standing laser wave of our optical cavity, where the axis of the standing wave $\hat{\mathbf{x}}$ is perpendicular to the propagation direction of the electron beam $\hat{\mathbf{z}}$. The interaction imprints a spatial phase modulation on the electron wave function, as described by equation (9.21). The electron beam then propagates away from the interaction region before it is imaged using a direct electron detection camera (Gatan K2) [111]. The electron beam is brought to a focus before it crosses the standing laser wave at the cavity focus such that a point-projection image, known as a "Ronchigram," is formed on the camera (see section 8.8) [95], [102]. As illustrated in Fig. 9.1b, paraxial propagation of the electron beam from the interaction region to the camera partially converts the phase modulation of the electron wave function to amplitude modulation, allowing the phase modulation to be imaged. The electron's kinetic energy $K = (\gamma - 1)\, m_e c^2$ can be adjusted between 80 keV ($\lambda_e = 4.18$ pm) and 300 keV ($\lambda_e = 1.97$ pm) by changing the TEM's accelerating voltage.

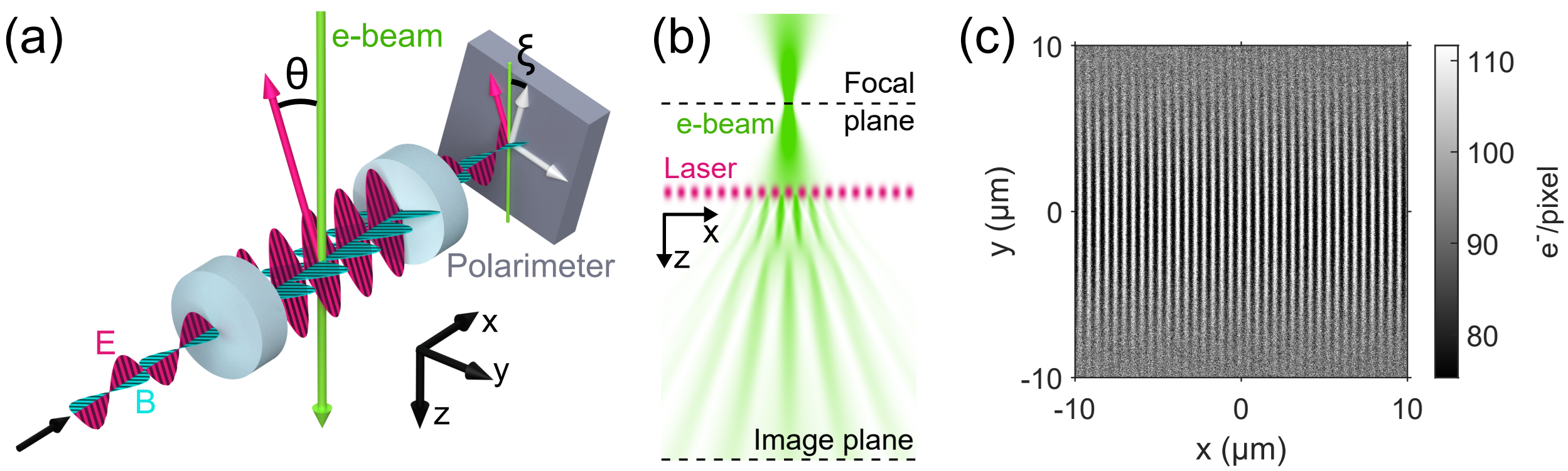

Figure 9.1: **a)** Schematic. An electron beam intersects a standing laser wave (electric field shown in magenta, magnetic field shown in cyan) formed between the two mirrors (blue cylinders) of a Fabry-Pérot optical cavity. The dimensions of the cavity are not shown to scale. The polarization axis of the standing wave (magenta arrows) makes an angle $\theta$ with the electron beam axis. The polarimeter is tilted by an angle $\xi$ relative to the electron beam axis. **b)** Phase modulation detection scheme. The electron beam crosses the standing laser wave after passing through a focus. The interaction with the standing wave imprints a spatial phase modulation on the electron beam which is converted to intensity modulation as the electron beam propagates to the image plane. **c)** Ronchigram of the standing laser wave. The direct electron detection camera records the number of electrons landing on each of its pixels. In the image plane, the standing wave structure of the phase modulation manifests as a series of bright and dark fringes. Reproduced from [108].

The phase profile imprinted on the electron beam by the laser beam is given by equation (9.25). Since a relativistic electron spends little time interacting with the EM wave, the laser intensity must be high in order for the electron wave function to accumulate appreciable phase. We achieve phase shifts on the order of 1 rad with a circulating power of 44 kW focused to a focus waist of $w_0 = 8\,\mu$m (NA = 0.042), corresponding to a peak standing wave intensity of 175 GWcm$^{-2}$.

The apparatus for this experiment deviates slightly from that described for the LPP used for imaging applications in chapter 6. A half-wave plate placed at the input of the optical cavity is used to control the linear polarization angle of the light entering the cavity. A portion of the light transmitted through the cavity is sent to a polarimeter which measures the optical power in the two orthogonal polarization components relative to the polarimeter axis. The polarimeter was placed at the location of photodetector PD7 as shown in figure 6.1. The polarimeter employs polarizing beamsplitter cubes to separate the orthogonal polarization components, and calibrated photodetectors to measure the optical power of each component [108]. The absolute value and sign of the polarization angle $\theta$ are determined from the polarimeter reading and orientation of the half-wave plate, respectively. The polarization at the polarimeter is assumed to be the same as that inside the cavity, as the cavity was measured to not appreciably change the polarization between its input and output [108].

Ronchigrams were collected at electron beam energies of $K = 80$, 150, 215, 230, 245, 260, 275, 290, and 300 keV. At each electron energy the rotation angle of the half-wave plate was incremented between 10 s electron camera exposures. The half-wave plate was rotated through 90° in one direction and then rotated back to the original position, thereby rotating the polarization angle from $\theta \approx -90°$ to $\theta \approx 90°$ and back again. Rotation of the polarization angle through a full 180° allowed for the determination of any misalignment between the polarimeter axis and the electron beam axis (angle $\xi$ shown in figure 9.1a).

Figure 9.1c shows a typical unprocessed Ronchigram. Both the standing wave structure and the transverse Gaussian profile of the cavity mode are clearly evident. Each Ronchigram was fit in Fourier space using the phase modulation depth of the standing wave as a fit parameter. As shown in section 8.8, the electron wave function in the image plane due to a spatial phase modulation $\eta(\mathbf{x})$ applied a distance $\Delta$ beyond the diffraction (focal) plane is

$$\psi_{\mathrm{im}}(\mathbf{x}) \propto \int d^2\mathbf{x}_1\, h_\Delta\left(\frac{\Delta}{f_{\mathrm{eff}}}\mathbf{x} - \mathbf{x}_1\right) e^{-i\eta(\mathbf{x}_1)} \tag{9.27}$$

$$h_z(\mathbf{x}) \coloneqq \frac{e^{i2\pi z/\lambda_{\mathrm{e}}}}{i\lambda_{\mathrm{e}} z} e^{i\frac{\pi}{\lambda_{\mathrm{e}} z}\mathbf{x}^2}, \quad \mathbf{x} \coloneqq (x, y) \tag{9.28}$$

where $f$ is the effective focal length of the microscope's objective lens, and $\lambda_{\mathrm{e}}$ is the electron's wavelength. Using equation (9.19) and the fact that the numerical aperture $\mathrm{NA} = \frac{\lambda_{\mathrm{L}}}{\pi w_0}$ of the cavity mode is small, we can write that

$$\eta(x, y) \simeq \bar{\eta} e^{-\frac{1}{2}(2\pi \mathrm{NA}\cdot y/\lambda_{\mathrm{L}})^2} \cdot \frac{1}{2}\left[1 + e^{-2\left(\frac{\Theta}{\mathrm{NA}}\right)^2} \rho(\theta, \beta) \cos\left(\frac{4\pi}{\lambda_{\mathrm{L}}} x\right)\right] \tag{9.29}$$

where $\bar{\eta}$ is the mean phase shift along the $x$-direction at $y = 0$, $w_0$ is the cavity mode waist, and $\Theta$ is the angle between the cavity mode axis and the electron beam axis ($\hat{\mathbf{z}}$) minus $\pi/2$ rad. Inserting this expression into equation (9.27), using the Jacobi-Anger identity, and integrating under the assumption that $\mathrm{NA} \ll 1$ gives

$$\psi_{\mathrm{im}}(x, y) \propto \sum_{n=-\infty}^{\infty} e^{i\left(-\frac{\pi n}{2} - 4\pi \frac{\lambda_{\mathrm{e}}}{\lambda_{\mathrm{L}}^2} n^2\right)} e^{i4\pi n \frac{\Delta}{f_{\mathrm{eff}}}\frac{x}{\lambda_{\mathrm{L}}}} J_n\left(\frac{\bar{\eta}}{2} e^{-\frac{1}{2}\left(2\pi \mathrm{NA}\frac{\Delta}{f_{\mathrm{eff}}}\frac{y}{\lambda_{\mathrm{L}}}\right)^2} e^{-2\left(\frac{\Theta}{\mathrm{NA}}\right)^2} \rho(\theta, \beta)\right) \tag{9.30}$$

where $J_n(x)$ is the $n$th Bessel function of the first kind. The resulting intensity distribution

is therefore

$$\begin{aligned}\left|\psi_{\mathrm{im}}(x,y)\right|^2 = &\sum_{n=-\infty}^{\infty} e^{i4\pi n\frac{\Delta}{f_{\mathrm{eff}}}\frac{x}{\lambda_{\mathrm{L}}}} \sum_{m=-\infty}^{\infty} e^{-i\pi n/2} e^{-i4\pi\frac{\lambda_{\mathrm{e}}}{\lambda_{\mathrm{L}}^2}\left(n^2+2mn\right)} \\ &\cdot J_{n+m}\left(\frac{\bar{\eta}}{2} e^{-\frac{1}{2}\left(2\pi \mathrm{NA}\frac{\Delta}{f_{\mathrm{eff}}}\frac{y}{\lambda_{\mathrm{L}}}\right)^2} e^{-2\left(\frac{\Theta}{\mathrm{NA}}\right)^2}\rho(\theta,\beta)\right) \\ &\cdot J_{m}\left(\frac{\bar{\eta}}{2} e^{-\frac{1}{2}\left(2\pi \mathrm{NA}\frac{\Delta}{f_{\mathrm{eff}}}\frac{y}{\lambda_{\mathrm{L}}}\right)^2} e^{-2\left(\frac{\Theta}{\mathrm{NA}}\right)^2}\rho(\theta,\beta)\right)\end{aligned} \tag{9.31}$$

This expression is normalized such that $\left|\psi_{\mathrm{im}}(x,y)\right|^2 = 1$ when $\bar{\eta} = 0$. The image is expressed in terms of a Fourier series along the $x$-axis, so that the power spectral density in the $n$-th order term (as determined by an $N$-point discrete Fourier transform) is

$$\begin{aligned}N^2 \Bigg| &\sum_{m=-\infty}^{\infty} e^{-i8\pi\frac{\Delta\lambda_{\mathrm{e}}}{\lambda_{\mathrm{L}}^2}nm} J_{n+m}\left(\frac{\bar{\eta}}{2} e^{-\frac{1}{2}\left(2\pi \mathrm{NA}\frac{\Delta}{f_{\mathrm{eff}}}\frac{y}{\lambda_{\mathrm{L}}}\right)^2} e^{-2\left(\frac{\Theta}{\mathrm{NA}}\right)^2}\rho(\theta,\beta)\right) \\ &\cdot J_{m}\left(\frac{\bar{\eta}}{2} e^{-\frac{1}{2}\left(2\pi \mathrm{NA}\frac{\Delta}{f_{\mathrm{eff}}}\frac{y}{\lambda_{\mathrm{L}}}\right)^2} e^{-2\left(\frac{\Theta}{\mathrm{NA}}\right)^2}\rho(\theta,\beta)\right)\Bigg|^2\end{aligned} \tag{9.32}$$

The following procedure was used to fit each Ronchigram:

1. Remove dead pixels from the image and normalize the background level to unity.
2. Determine the value of $\frac{\Delta}{f_{\mathrm{eff}}\lambda_{\mathrm{L}}}$ and direction of the $x$-axis. This was done taking a 2D discrete Fourier transform of the image and locating the non-zero spatial frequency at which its magnitude squared was largest.
3. Interpolate the image onto a grid aligned with the axis of the standing wave in the image.
4. Perform a 1D discrete Fourier transform of the interpolated image along the standing wave axis and extract the power spectral density of the $n = 1, 2$ orders as a function of the transverse coordinate, $y$.
5. Fit these power spectral density profiles to the model described by equation (9.32) as a function of the variables $\frac{\Delta\lambda_{\mathrm{e}}}{\lambda_{\mathrm{L}}^2}$, $\bar{\eta}\rho(\theta,\beta)$, and NA.

For the purposes of fitting, it was assumed that $\Theta = 0$, since during the experiment $\Theta$ was held constant and so the factor $e^{-2\left(\frac{\Theta}{\mathrm{NA}}\right)^2}$ was effectively absorbed into the fit's estimation of $\bar{\eta}$, which was then normalized to extract a measurement of $\rho(\theta,\beta)$. By comparing the measured value of $\bar{\eta}$ to that predicted by equation (9.22), the tilt angle $\Theta$ was estimated to be between 1.2° and 1.6°. Contamination or defects on the cavity mirror surfaces prevented the mode from being aligned such that $\Theta = 0$ while maintaining sufficient mirror reflectivity to support the required circulating cavity power.

To correct for small variations in the laser wave parameters during the experiment, the phase modulation depth was normalized by the optical power at the polarimeter (proportional to the circulating power $P_c$) and the mode waist $w_0$, which were both measured at the time the Ronchigram was taken. The mode waist was determined from a measurement of the cavity's transverse mode frequency spacing (see section 6.3.2). The fractional change in $\lambda_L$ during a typical experiment was measured to be small enough ($\sim 10^{-6}$) that it was assumed to be constant for the purpose of normalization.

Each set of normalized modulation depth versus polarization angle data was fit to equation (9.20), with an angle-independent normalization constant, a polarization angle axis offset, and the electron speed $\beta$ as fit parameters. The electron speed was used as a fit parameter because the nominal electron energies $K$ are only accurate to approximately $\pm 1\%$ (per the TEM manufacturer's specifications). The polarization angle axis offset accounts for the polarimeter misalignment angle $\xi$, as well as any linear polarization rotation induced by optics between the cavity output and polarimeter [108].

This data is presented in figure 9.2a; for clarity, only the $K = 80$, 150, 215, and 300 keV data sets are shown. The remaining data sets are shown in the supplementary materials of [108]. The relative phase modulation depth exhibits a dependence on the polarization angle $\theta$ that is well-modeled by equation (9.20); the root-mean-squared difference between the fit and data across all data sets is $7.1 \times 10^{-3}$. To show the relative phase modulation depth's $\beta$-dependence, the fit values at $\theta = 0$ are plotted as a function of the nominal values of $\beta^2$ for all data sets in Fig. 9.2b, where they are compared with the linear dependence expected from equation (9.20). Again, there is a good correspondence between the measured values and theoretical model.

Imaging of the spatial phase modulation profile allows the relativistic reversal effect to be directly observed in the Ronchigrams. As the polarization angle is rotated through the relativistic reversal angle, the standing wave structure in the Ronchigram diminishes in amplitude until it disappears entirely and then re-emerges with the opposite sign. This is demonstrated in Fig. 9.2c, using Ronchigrams from the $K = 300$ keV data set shown in figure 9.2a. A temporally linear drift in the fringe position of $0.4\,\mathrm{nms^{-1}}$ due to thermal expansion of the cavity support structure has been removed from the displayed data and is shown in figure 9.3. The change in fringe position around $\theta_r$ was used to infer the sign of the relative phase modulation depth for the $K \geq 215$ keV data sets in figures 9.2a and 9.2b.

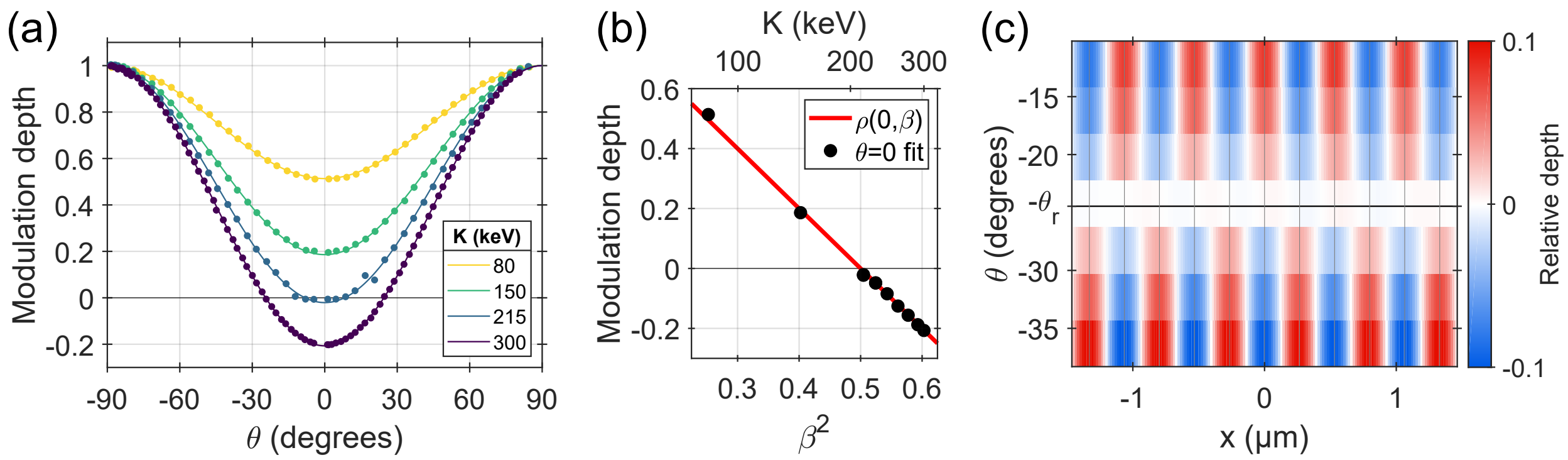

Figure 9.2: **a)** Relative phase modulation depth as a function of polarization angle. Measurements of the relative phase modulation depth (dots) are plotted along with fitted theory curves (lines) as a function of the laser wave polarization angle $\theta$ for several values of the nominal electron energy $K$ (equivalently, electron speed $\beta$). The fitted theory (solid lines) is described by equation (9.20). **b)** Relative phase modulation depth as a function of electron speed. Fit values for the relative phase modulation depth at $\theta = 0$ are shown as a function of $\beta^2$ for each of the nine electron energies examined (black dots). The theoretical dependence on $\beta^2$ predicted by equation (9.20) is shown in red. **c)** Relativistic reversal. Ronchigram standing wave fringes are shown as a function of polarization angle $\theta$ and position along the laser beam axis $x$ for the $K = 300\,\text{keV}$ data set used in panel (a). The bright (red) and dark (blue) fringes reverse positions around the relativistic reversal angle $\theta_r$ (horizontal black line). The vertical lines are a guide-to-eye. Reproduced from [108].

Equation (9.21) predicts that the spatial phase modulation profile should be independent of the EM standing wave's ellipticity angle, $\epsilon$, as defined in equation (9.16). To verify this prediction, the input half-wave plate was replaced with a quarter-wave plate. Rotating the quarter-wave plate then changed the polarization from either horizontal linear or vertical linear (depending on the linear polarization set before the quarter-wave plate) to left- or right-hand circular polarization. Intermediate angles of the quarter-wave plate generated elliptical polarizations. Therefore, $\epsilon$ is a function of the polarization angle $\theta$ in this configuration. Since the polarimeter used in this experiment was only capable of measuring $\theta$ (e.g. $\theta = 0°$ for vertical linear polarization, $\theta = 45°$ for circular polarization), the independence of the phase modulation depth on $\epsilon$ can only be inferred by comparing the measured values of the normalized phase modulation depth to those predicted by equation (9.20). This data is shown in figure 9.4 for two data sets—one taken with an elliptical polarization varying between horizontal linear and circular (red dots), and one taken with an elliptical polarization varying between vertical linear and circular (blue dots). Both data sets were taken at $K = 300\,\text{keV}$. The measurements agree well with the $\epsilon$-independent theory.

These results show that the relativistic ponderomotive potential exhibits a strong depen-dence on the electron velocity and the EM wave polarization, and that this dependence is well-described by the semiclassical theory of the interaction presented in section 9.1. The most striking feature of the polarization dependence is that the standing wave structure of the phase shift reverses its sign at a particular polarization angle when $\beta \geq 1/\sqrt{2}$. Therefore, this experiment can also be understood as an observation of the relativistic reversal of the amplitude of Kapitza-Dirac diffraction [112], [113].

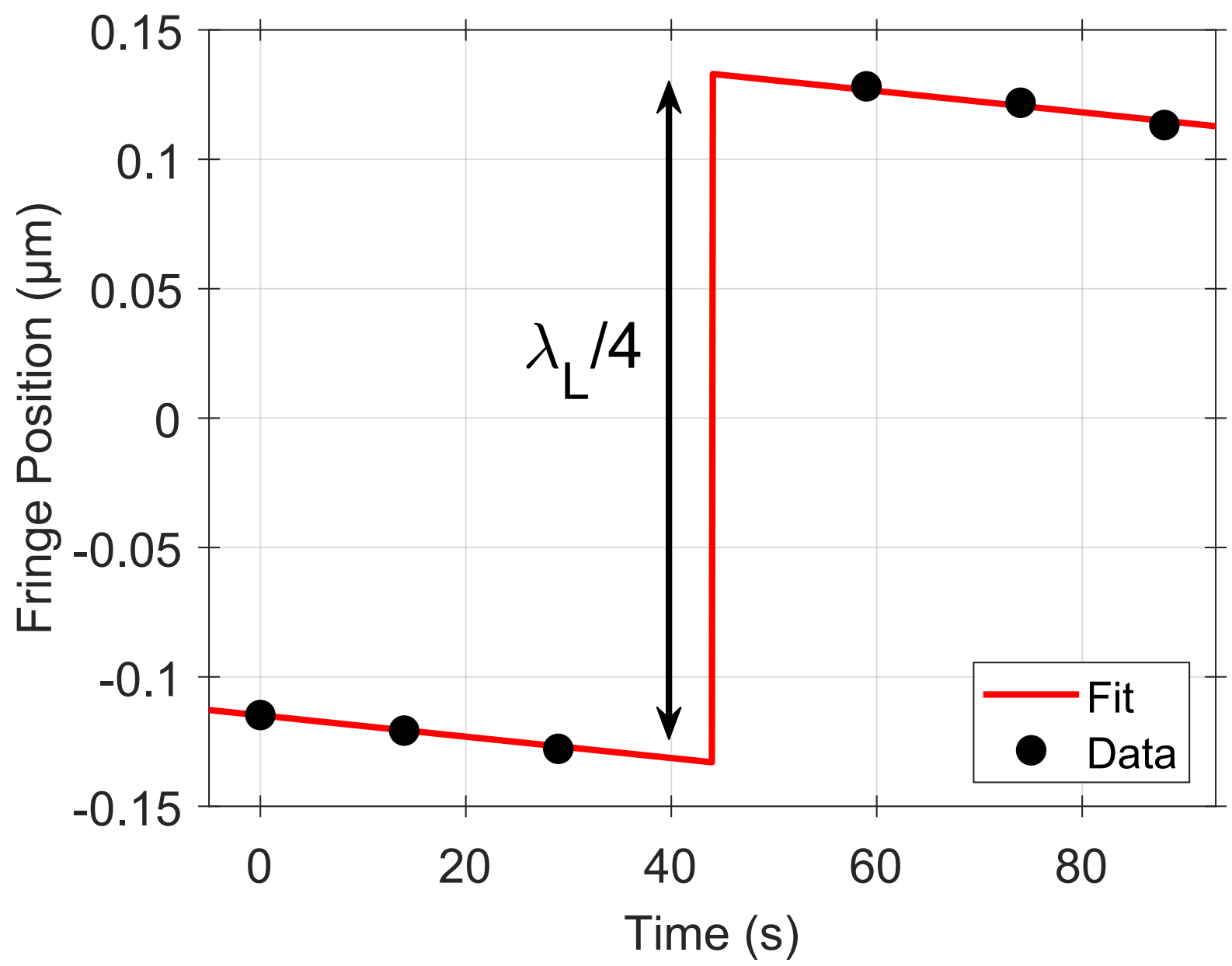

Figure 9.3: Ronchigram fringe position versus time of Ronchigram capture for the data shown in figure 9.2c (black dots). The linear fit (including $\lambda_L/4$ jump across the relativistic reversal angle) is shown in red. The fitted drift rate is $0.4\,\mathrm{nms}^{-1}$. Reproduced from [108].

The dependence of the relativistic ponderomotive potential on the polarization of the EM wave provides an avenue for dynamical optical control of relativistic electron beams. When the standing wave structure of the phase modulation is eliminated, the electron wave function does not diffract from the EM wave. Therefore, varying the polarization of the standing wave could be used to make a rapidly-switchable electron beamsplitter, or implement electron pulse slicing [114]. The same effect could be used to temporally phase modulate an electron beam focused through a single antinode of the standing wave.

In the context of the LPP, operation at the relativistic reversal angle eliminates the presence of "ghost" images due to diffraction (see section 8.6). Figure 9.5 shows an example of how ghost images from a strong-phase object (thick polymer of a "lacey carbon" TEM test standard) can be entirely eliminated by operating at the relativistic reversal angle, while still maintaining some phase contrast. The data shown was taken with $K = 80\,\mathrm{keV}$ and a phase shift of $\eta_0 < \pi/2$. Ghost artifacts appear to the lower-left and upper-right of the primary object image, aligned with the standing wave axis (figure 9.5a). However, as the polarization is rotated to $|\theta| < 90°$, the phase shift modulation depth decreases, resulting in less electron diffraction and weaker ghost artifacts (figure 9.5b). At the relativistic reversal angle $\theta_{\mathrm{r}}$, the electron diffraction is eliminated entirely, and the ghost artifacts disappear (figure 9.5c). However, the phase shift profile $\eta(x, y) = \eta_y(y)/2$ still remains to provide phase contrast enhancement in the image (compare figure 9.5c and figure 9.5d—the former image has higher contrast). The remaining contrast enhancement is evident as increased low spatial frequency power spectral density in the Fourier transforms of the images (figure 9.5e-h, detail in figure 9.5i-l). Note that achieving a phase shift of $\pi/2$ when using the reversal angle requires $\eta_0 = \pi$. This may be possible on our current cavity prototype, though we have not yet tried. If $K < 212\,\mathrm{keV}$ it should be possible to approximate this effect by tilting the cavity mode relative to the electron beam axis (angle $\Theta \gg \mathrm{NA}$ in equation (9.25)).

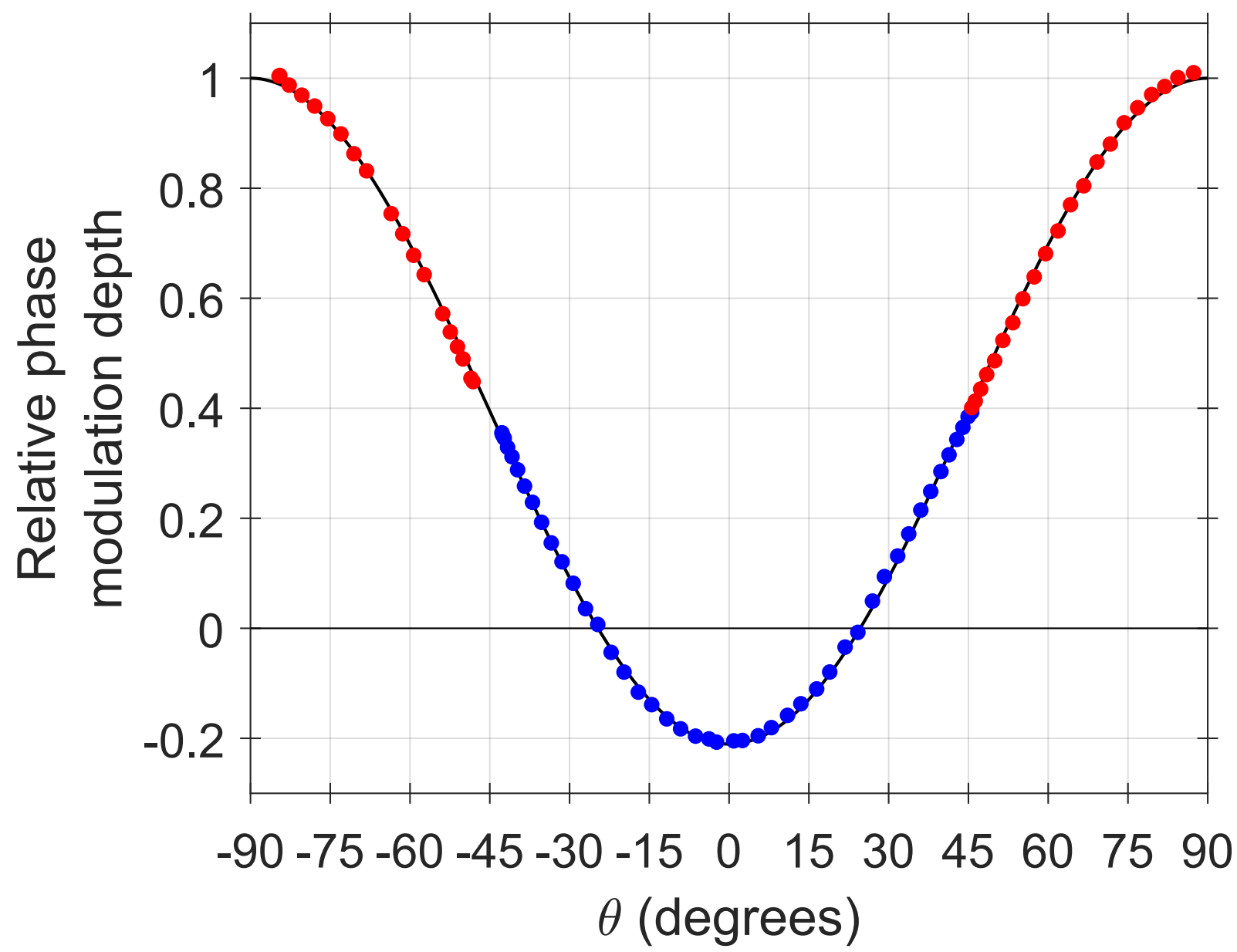

Figure 9.4: Measurements of the relative phase modulation depth (dots) are plotted along with the fitted theory curve (black line) for elliptical polarizations with $0° \leq \theta \leq 45°$ (blue dots) and $45° \leq \theta \leq 90°$ (red dots). $K = 300\,\mathrm{keV}$ in both cases. Reproduced from [108].

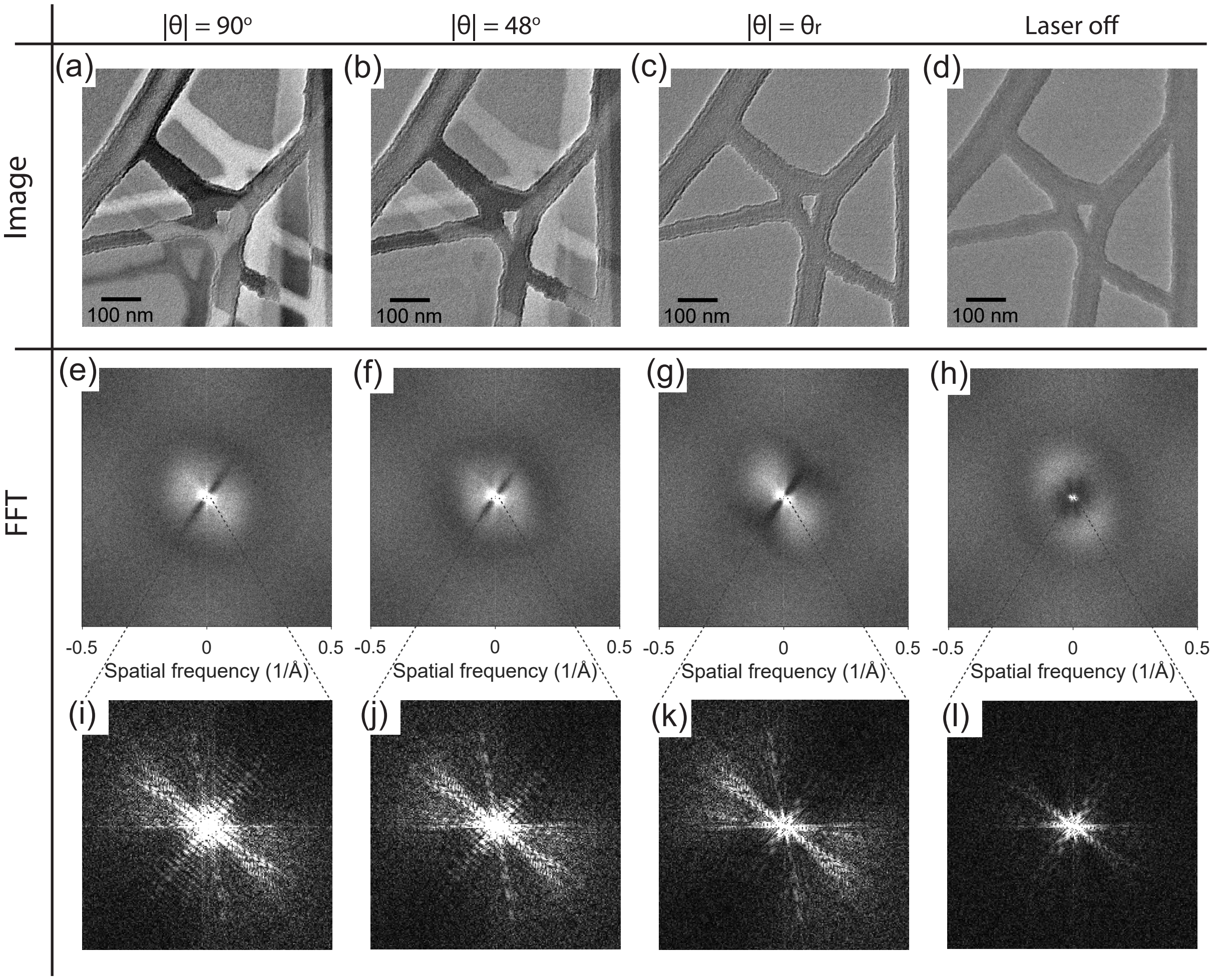

Figure 9.5: Laser phase contrast electron microscopy: ghost image artifacts and phase contrast as a function of polarization angle. **a-c)** Images of a lacey carbon sample using different polarization angles (annotated, top). **d)** Comparison with the laser off. Ghost artifacts are present when the laser is on and the polarization angle is not equal to the relativistic reversal angle $\theta_{\mathrm{r}}$. **e-h)** Fourier transforms of panels (a-d), respectively. Phase contrast enhancement for all polarization angles is seen as increased power spectral density at low spatial frequencies, compared to the "laser off" case (panel (h)). **i-l)** Detail of the lowest spatial frequencies of the Fourier transforms in panels (e-h), respectively. In panels (e) and (f), the standing wave of the laser beam is evident. At the relativistic reversal angle (panel (k)), the standing wave structure disappears from the Fourier transform. Reproduced from [108].

# Chapter 10

# Laser phase plate transmission electron microscopy

## 10.1 Custom transmission electron microscope

Our TEM is a custom modified version of a Thermo Fisher Scientific Titan which includes an additional section in the microscope column. This "phase plate module" is installed below the standard "octagon" section of the column which houses the sample stage and objective lens. A schematic of our TEM is shown in figure 10.1. The phase plate module provides the space and vacuum access ports needed to install the LPP in the column, and also houses electron optics which relay the magnified diffraction plane formed in the plane of the phase plate to the subsequent (standard) electron optics in the column. When using these relay optics, the standard Lorentz lens (focal length 22 mm) is used to create a 5.7-fold magnified diffraction pattern 180 mm below the back focal plane of the objective lens. In that plane, two 25 mm diameter ports in the microscope column, at nearly opposite locations with respect to each other, allow installation of the phase plate (see section 4.8). An additional "transfer lens" located just below the laser phase plate forms an image of the sample in the selected area aperture plane, which is then further magnified onto the electron camera or fluorescent screen by the standard projection lens optics.

This relay optics system maintains a relatively low coefficient of spherical aberration $C_\mathrm{s}$: the Lorentz lens contributes negligibly to $C_\mathrm{s}$, being located near a plane where the marginal ray crosses the optical axis, while the transfer lens operates under low numerical aperture and so only contributes modestly. Unfortunately, however, the chromatic aberration caused by the relay optics cannot be reduced in a similar manner. All in all, the combination provides an effective focal length $f_\mathrm{eff} = 20\,\mathrm{mm}$ (compared to 3.5 mm for the objective lens alone), $C_\mathrm{s} = 4.7\,\mathrm{mm}$ (compared to 2.7 mm), and an increased chromatic aberration coefficient of $C_\mathrm{c} = 7.6\,\mathrm{mm}$ (compared to 2.7 mm). The microscope can also be operated with the relay optics deactivated, where it has the parameters of a standard Titan TEM. In this case, both the Lorentz and transfer lenses are turned off, and the objective lens is set to form an image

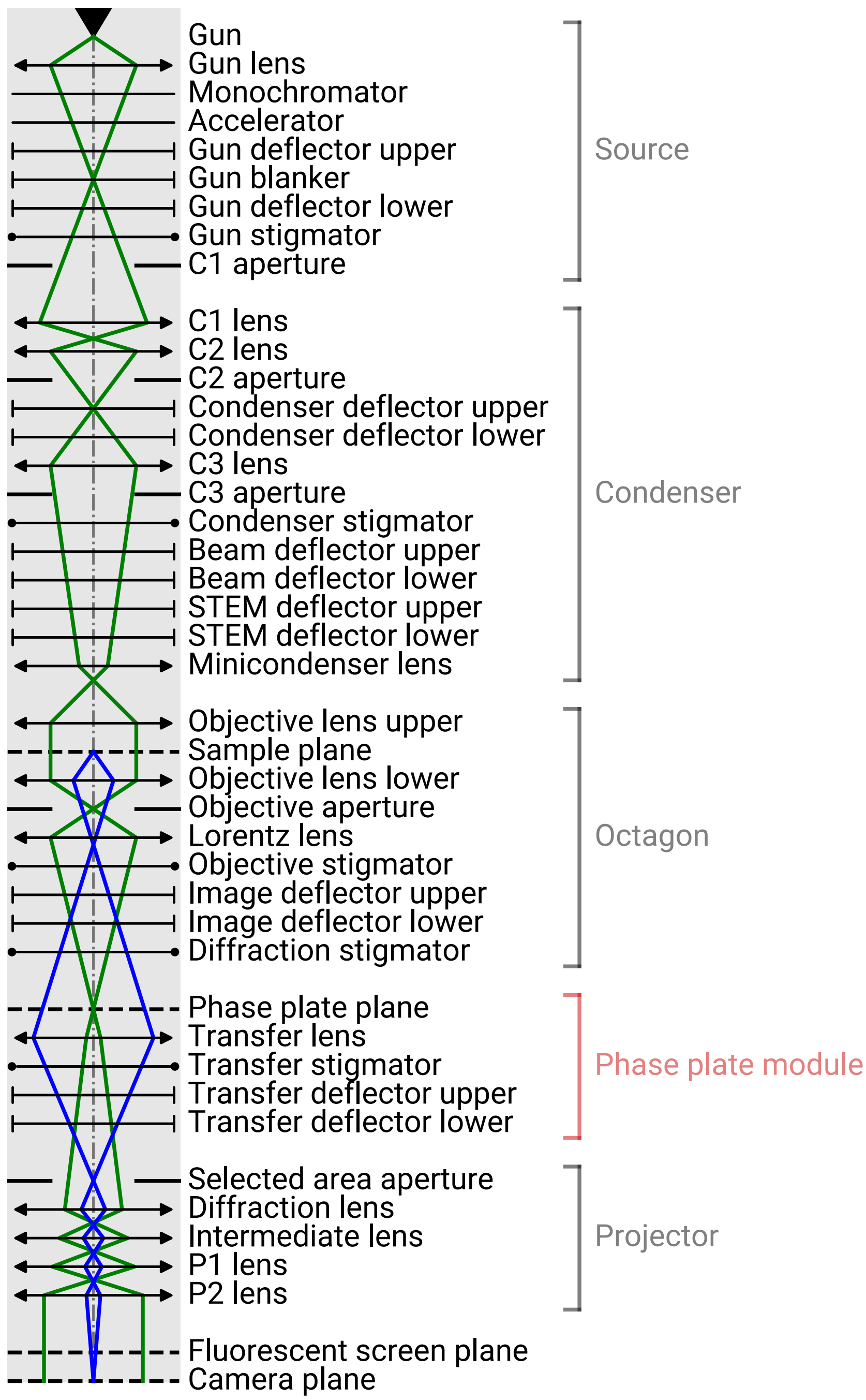

Figure 10.1: Schematic of our transmission electron microscope's electron optical components. Components in the phase plate module are custom. The illuminating beam ray is shown in green; the marginal image ray is shown in blue. Not to scale.

in the selected area aperture plane.

The extra space provided by the phase plate module is necessary because a standard Titan (or similar) TEM has a focal length of 3.5 mm, which leaves little space for the LPP's hardware to extend above (upstream) of the diffraction plane. There is similarly little space below the standard diffraction plane, as it is usually very near or even below the upper surface of the lower pole piece of the objective lens. As is discussed in chapter 8, the 5.7-fold magnified diffraction plane in our TEM beneficially decreases the cut-on frequency of the phase plate. However, it also makes the TEM more susceptible to resolution loss due to thermal magnetic field fluctuations inside of the column (see chapter 11).

The TEM is otherwise standard. A field emission gun (FEG) emits electrons which are focused by the electrostatic gun lens. The beam then passes through a Wien filter monochromator [115] which (when active) deflects the beam by an amount which depends on its energy. A small C1 aperture can then block all but a portion of the dispersed beam, resulting in a reduced energy spread which can improve the resolution of the TEM (see section 8.7.1.1). The monochromator also includes an electrostatic deflector and stigmator. Using the monochromator results in a less spatially uniform beam intensity in the sample plane, which can be a challenge when interpreting and processing images recorded by the camera. It also reduces the available adjustment range of beam diameters and convergence angles in the sample plane, which can provide constraints on how the images are collected. Additionally, we have found that the alignment of the monochromator is too unstable to use without continuous human supervision to keep the beam aligned with the C1 aperture. This reduces the image collection rate. For these reasons, we have not yet used the monochromator with the LPP.

After the monochromator, the beam passes through the electrostatic accelerator, which accelerates the electrons to a user-specified energy from 80 keV to 300 keV. The magnetic gun deflectors align the beam into the subsequent condenser lenses. The gun blanker is a magnetic deflector which is used to deflect the beam far from the optical axis such that it is blocked by either the C1, C2, or C3 aperture before it encounters the sample. This keeps the beam from damaging the sample until a camera exposure is started, at which time the beam is "unblanked" by de-energizing the beam blanker. The gun stigmator stigmates the beam before it enters the condenser. When the monochromator is not being used, the C1 aperture is used to reduce the electron beam flux at the sample. We use the largest available C1 aperture (2 mm diameter) which does not clip the beam.

The C1, C2, C3, and minicondenser lenses form the condenser system, which is used in conjunction with the C2 and C3 apertures to adjust the diameter, collimation, and flux of the beam in the sample plane. The relative strengths of the C1 and C2 lenses determines the size of the beam in the C2 aperture plane and (when the C2 aperture is chosen to be smaller than the beam diameter) is used to control the beam flux. The C2 and C3 lens strengths are adjusted together to change the diameter and/or collimation of the beam in the sample plane. The C3 aperture can be used to limit the illuminated area in the sample plane. The minicondenser lens is used in one of two states (effectively, "on" and "off") to change the magnification of the condenser system, which changes the range of sample plane beam diameters that can be achieved while keeping the beam collimated. The condenser deflectors are used to align the beam to the optical axis of the compound C2/C3 lens system so that the beam does not tilt or shift when the C2 and C3 lenses are adjusted to change the sample plane beam diameter. The condenser stigmator is used to stigmate the beam in the sample plane. The beam deflectors tilt and/or shift the beam in the sample plane. The STEM deflectors do the same, but are designed to be used for fast scanning of a probe beam for scanning TEM. In our system they are de-energized.

The upper objective lens nominally collimates the divergent beam from the condenser system onto the sample. The lower objective lens then forms the first image of the sample. The objective aperture is used to block electrons which have scattered from the sample at a relatively high angle to generate scattering contrast—however, we primarily use it to reduce the rate of ice accumulation on the sample (see section 10.2). The Lorentz lens images the back focal plane of the lower objective lens onto the phase plate plane with a magnification of 5.7. The objective stigmator stigmates the image, and also includes a threefold stigmator ($C_{2,\pm 3}$, see table 8.1). The image deflectors tilt and/or shift the beam, and are used to move the electron beam in the phase plate plane. The diffraction stigmator is used to stigmate the image of the diffraction plane, and is typically only when the projection system is set to image the diffraction plane rather than the sample plane.

The transfer lens images the image plane of the lower objective lens onto the selected area aperture plane. The selected area aperture can be used to project only a selected area onto the camera. We do not use a selected area aperture. The transfer stigmator can also be used to stigmate the image—we use the transfer stigmators rather than the objective stigmators for this purpose because doing so does not affect the position of the electron beam in the phase plate plane, which is important for maintaining proper alignment to the phase plate (see section 10.3). The transfer deflectors tilt and/or shift the beam, and are usually used in tandem with the image deflectors to move the beam in the phase plate plane while keeping it in the same place in the camera plane.

The diffraction, intermediate, P1, and P2 lenses comprise the projector, which magnifies the selected area aperture plane onto the fluorescent screen and camera plane with a user-controlled range of magnifications.

Note that each deflector and stigmator has two degrees of freedom, so that the beam can be deflected/stigmated in both directions perpendicular to the beam axis.

## 10.2 Ice contamination control

Our custom Thermo Fisher Scientific Titan TEM is fitted with a cryo-pump (referred to as the “cryo-box”) that surrounds the specimen, similar to that provided in the Thermo Fisher Scientific Krios product line. Nevertheless, the ice contamination rate in our Titan is at least 30 times as high as it is in Krios microscopes (which typically achieve contamination rates of less than $0.5\,\text{Å}\text{h}^{-1}$). As a result, cryo-EM specimens can be used in our Titan for only a relatively small number of hours, after which the ice contamination becomes too thick for high-resolution data collection.

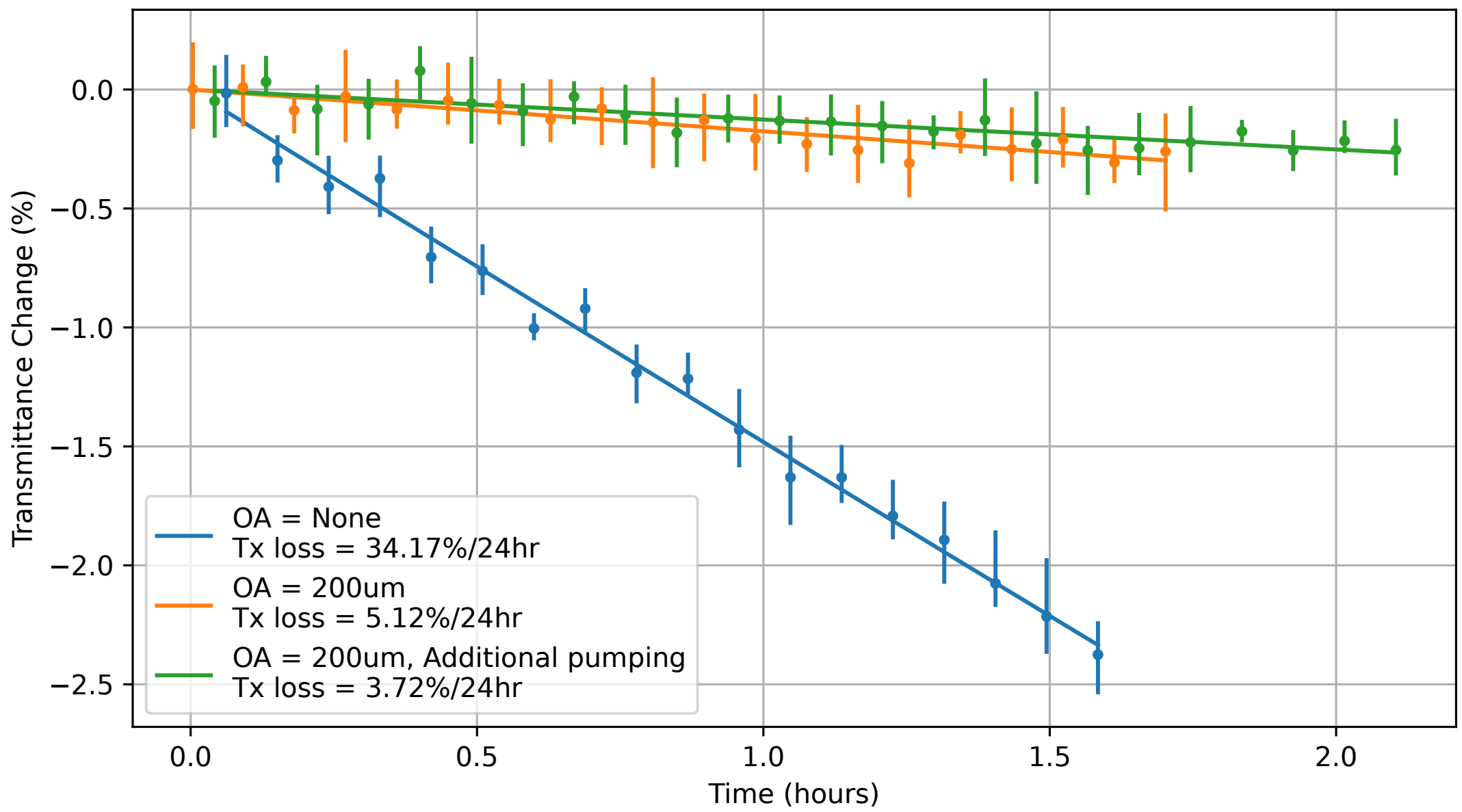

Figure 10.2: Change in percent transmittance as a function of time when i) no objective aperture (OA) is inserted and no additional vacuum pumping is used (blue), ii) a 200 µm diameter objective aperture is inserted (orange), and iii) 200 µm diameter objective aperture is inserted and additional vacuum pumping is used (green). The dots with error bars represent measurements taken at discrete intervals, while the lines represent exponential best fits to the data. The corresponding percent transmittance (Tx) change rates are shown in the legend. Reproduced from [67].

Fortunately, we found that the contamination rate is decreased by roughly an order of magnitude by combining the use of an objective aperture together with adding more vacuum pumping capacity (using a non-evaporable getter pump) to the lower portion of the column of the microscope near the X-lens. The contamination rate was estimated by measuring the decrease in transmittance of as-supplied holey-carbon Quantifoil grids (mounted in a Gatan 626 cryo-holder) as a function of time. Transmittance values were measured by recording low-magnification images, in which the field of view was $\sim 30$ µm at the specimen. Under the imaging conditions used, the initial transmittance of the carbon film was typically $\sim 88\%$, measured as the number of electrons per unit area passing through the carbon film relative to the number per unit area passing through a hole. Figure 10.2 shows the changes in percent transmittance values when measured as a function of time, showing the resulting reduction in contamination rate that was achieved by the use of an objective aperture, together with the additional improvement achieved by increasing the pumping capacity.

The biggest benefit is gained by inserting a 200 µm diameter objective aperture, which is inserted into the space between the cryo-box and the lower pole piece of the objective lens, and thus remains at room temperature. Such an aperture is large enough that inevitable charging of the aperture edge, which occurs with use, does not have a detectable effect on the image quality. We infer that this aperture reduces the contamination rate by reducing the solid angle over which water molecules can approach the sample from the room-temperature surfaces of the lower part of the microscope column. Though the specimen is still exposed to the surface of the room-temperature objective aperture, we expect that water molecules on this surface are quickly pumped away by the surrounding cryo-box, after which the upper, specimen-facing surface of the objective aperture, although at room temperature, is no longer a source of water molecules.

All of the cryo-EM data presented in chapter 12 of this work was collected with the 200 µm objective aperture inserted. We opted to not use the NEG pump for these experiments because it only provides a small additional benefit, and is slightly inconvenient to use since it must be manually switched on and off by the user when the TEM is "cryo-cycled" (pumped using a turbomolecular pump rather than the usual ion-getter pumps) to remove excess water vapor after a cryo-EM data collection session.

## 10.3 Alignment procedures

To provide phase contrast, the electron beam and LPP must be aligned such that the unscattered electron wave intersects an antinode of the LPP's standing laser wave at the focus of the cavity mode (see chapter 8). This alignment has three degrees of freedom. First, the unscattered wave must be focused in the phase plate plane such that there is no "plane offset" (see section 8.8). Second, the unscattered wave must intersect the optical axis of the cavity mode ("transverse" alignment). Third, it must intersect an antinode of the cavity mode standing wave ("longitudinal" alignment), assuming that the LPP is operated with its polarization axis perpendicular to the electron beam axis (see chapter 9).

If the plane offset alignment condition is not met, then Ronchigrams of the LPP's phase shift profile may become visible in the image, and phase contrast enhancement will be lost (see section 8.8). If the transverse and/or longitudinal alignment is incorrect, the phase shift experienced by the unscattered wave will be less than $\pi/2$ and the amount of contrast enhancement in the image will be reduced.

### 10.3.1 Plane offset

We roughly align the plane offset to zero by first turning on the LPP, setting the C2 and C3 lenses such that the electron beam is collimated in the sample plane ("parallel illumination"), and then setting the strength of the Lorentz lens to its nominal value such that it focuses the unscattered wave in the phase plate plane (based on the manufacturer's simulations of the electron optics). In this condition the plane offset is near, but never exactly, zero. We then move the electron beam in the phase plate plane until we see a laser Ronchigram on the fluorescent screen or camera. This is done using a coordinated actuation of the image and transfer deflectors such that the beam shifts in the phase plate plane but does not move in the camera plane. When first aligning the system the angle between the axes of the image deflectors and the optical axis of the LPP is unknown, so we either move the beam in a two-dimensional raster pattern, or along each axis of the image deflectors. Since the phase profile of the LPP is appreciably non-zero for roughly several Rayleigh ranges around the focus, the latter approach usually still results in intersecting the cavity mode at a point along its axis where the phase shift is still large enough to generate visible Ronchigrams. Once Ronchigrams are seen, the image deflector current drivers can be calibrated such that one of their control axes is aligned with the LPP's optical axis (and the other perpendicular). This is not strictly necessary, but is convenient when operating the TEM with the LPP.

If Ronchigrams are not seen, it is possible that the plane offset is too close to zero. Changing the Lorentz lens strength by a few percent is generally sufficient to see Ronchigrams. It is also possible that the axis of the cavity mode is not close enough to center of the TEM column to lie within the adjustment range of the image deflectors. In our system, this adjustment range is nearly $\pm 1$ mm, so we have not had this problem before. Still, if this were to case, it is in principle possible to move the LPP cavity (see section 4.9) and try again to find the cavity mode, in a blind search pattern.

Once Ronchigrams are seen, the Lorentz lens is adjusted to bring the LPP "on-plane" (zero plane offset). This can be done by eye, by watching the period of the Ronchigram fringes in the camera plane expand to infinity as the plane offset approaches zero. It can also be done quantitatively, by recording images of the Ronchigrams as a function of Lorentz lens setting, determining the spatial frequency of the Ronchigram, and then fitting the resulting data to find the Lorentz lens setting at which the Ronchigram spatial frequency should be zero. This procedure is illustrated in figure 10.3.

Once this rough alignment using the Lorentz lens is completed, we fine-tune the alignment using the C2 and C3 condenser lenses to slightly adjust the collimation of the beam in the sample plane (and thus the vertical position of the focal point of the unscattered wave). We don't use the Lorentz lens for fine-tuning the alignment because it lies downstream of the sample, and changing its strength therefore affects the position of the image on the camera, and the defocus of the image. We first set the strength of the C1 lens such that the plane offset does not change when the C2 and C3 lenses are "zoomed" to change the beam diameter in the sample plane. This is done by adjusting the zoom parameter back and forth ("wobbling") and setting the C1 lens to minimize the concomitant change in Ronchigram magnification. We then similarly adjust the condenser deflectors to minimize any lateral motion of the Ronchigram when zooming the condenser. These alignments help keep the electron beam aligned to the LPP when the sample plane beam diameter is regularly changed during automated data collection (see section 10.4).

Finally, we adjust the condenser stigmation such that the focus of the unscattered wave in the phase plate plane is stigmatic. Astigmatism in the unscattered wave manifests as an apparent rotation of the Ronchigram fringes as the plane offset is moved through zero. This happens because astigmatism of the unscattered wave causes there to be different plane offsets for the principal axes of the unscattered wave, and the magnification of the Ronchigram image is inversely proportional to the plane offset. We adjust the two axes of the condenser stigmator while wobbling the C2/C3 lens plane offset across zero until there is no apparent rotation of the Ronchigram fringes. This process is underdetermined because it only ensures that one of the principal axes of the astigmatic beam is aligned with the LPP's optical axis (not that the beam is stigmatic), but we have not yet found this to cause problems.

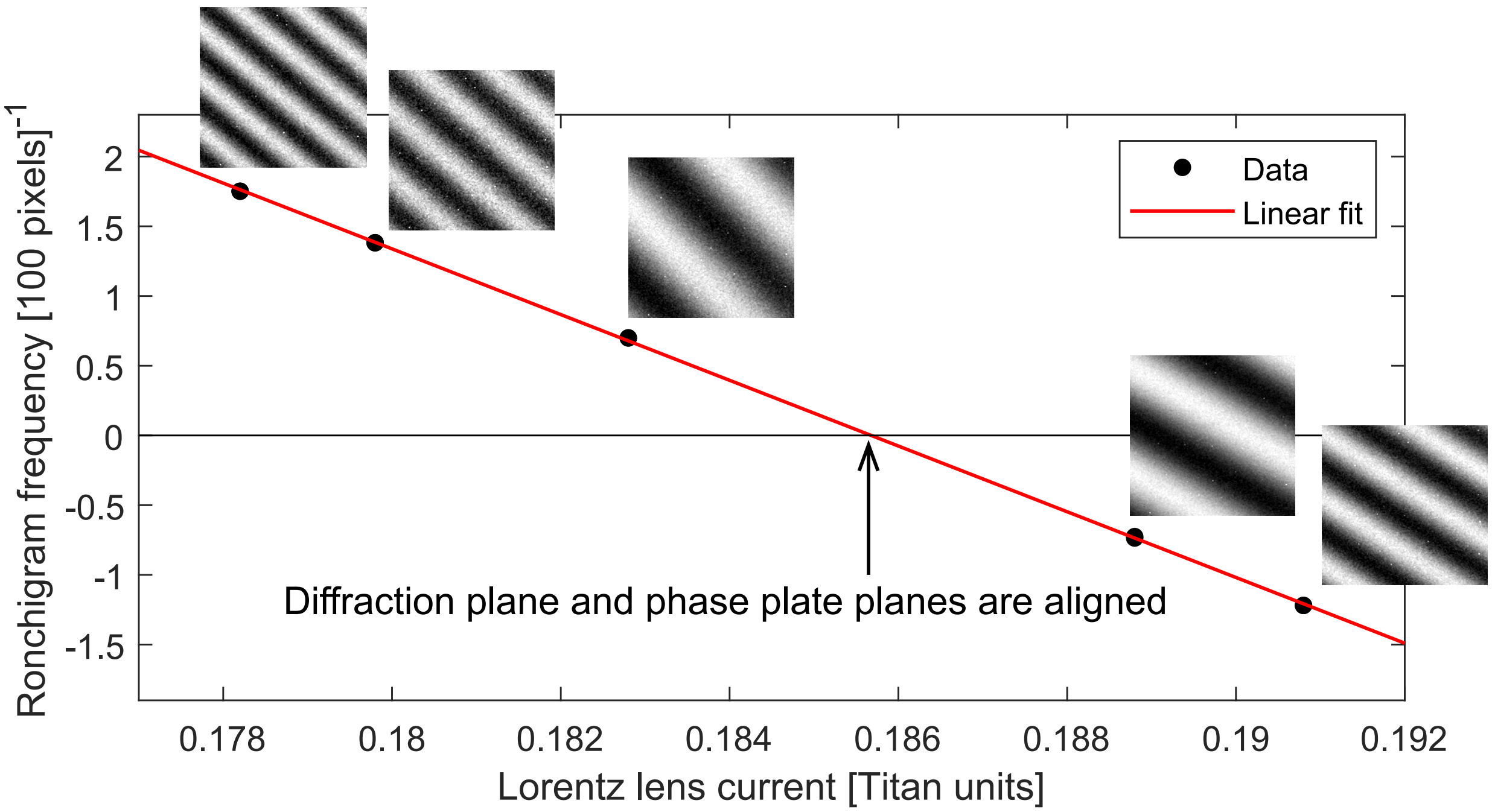

Figure 10.3: Plane offset alignment of the electron beam to the laser phase plate. The Ronchigram's magnification varies with Lorentz lens current as the magnified diffraction plane is moved above/below the laser phase plate. At infinite Ronchigram magnification, the magnified electron diffraction pattern and the laser phase plate are in the same plane. The plot shows the Ronchigram frequency as a function of the Lorentz lens current (in manufacturer-specified units), along with a linear fit to determine the Lorentz lens setting for the on-plane condition. Reproduced from [23].

### 10.3.2 Transverse

The electron beam is aligned to the cavity mode along the axis transverse to its optical axis by first making a small non-zero plane offset with the C2/C3 lenses to generate a laser Ronchigram with just a few fringes visible on the electron camera. Ronchigrams are recorded as a function of beam position in the phase plate plane along the transverse axis, and their amplitude (relative to the mean image intensity) is measured. The electron beam position with the highest amplitude indicates the location of the optical axis of the cavity mode. An example alignment dataset is shown in figure 10.4. A Gaussian fit to the data is also shown for reference. Strictly speaking, the Ronchigram amplitude as a function of transverse beam position is not exactly Gaussian (a more accurate model is given by equation (9.31)) but it is close enough to provide an accurate estimate of the transverse position deflector setting which results in the maximum Ronchigram amplitude. Note that the accuracy of this procedure relies on the condenser deflector alignment described earlier in this section so that the position of the electron beam in the phase plate plane does not change appreciably (on the scale of the cavity mode focal waist $w_0$) when the plane offset is set back to zero after the alignment is complete. This alignment can also be performed by estimating the Ronchigram depth by eye, though naturally the accuracy is lower.

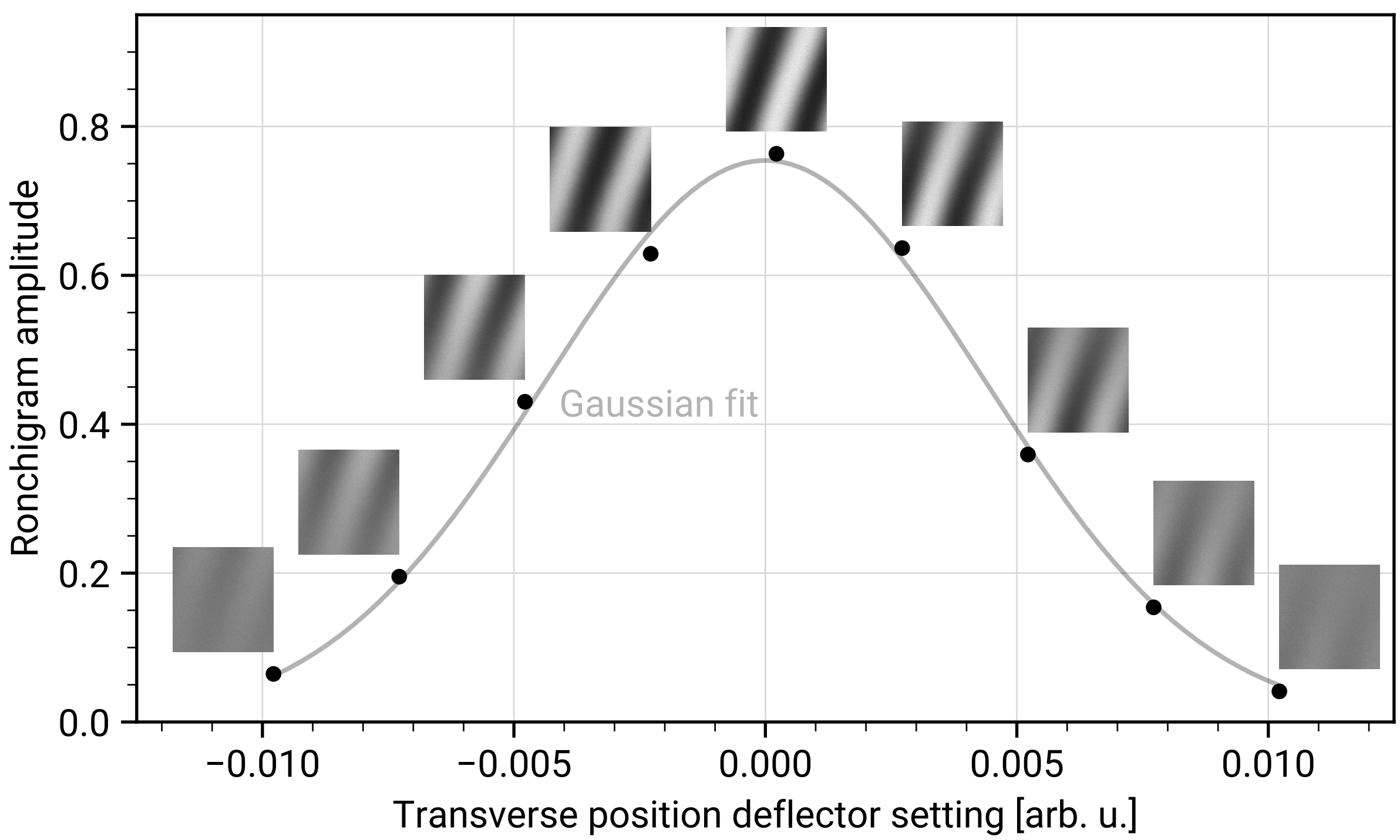

Figure 10.4: Ronchigram amplitude as a function of the setting of the image deflector along the axis transverse to the cavity mode axis. Data is shown with black dots; a Gaussian fit to to the data is shown in gray. Insets show the Ronchigram images collected for each data point. The grayscale of the images ranges from 0 to 2, where 1 is the mean pixel value.

### 10.3.3 Longitudinal

The "longitudinal" alignment of the electron beam along the cavity mode axis is the most critical of the three alignment degrees of freedom, since the phase profile of the LPP varies most rapidly along that axis. On a larger scale, the electron beam must be aligned to well within one Rayleigh range of the cavity mode focus (see section 2.2) so that the unscattered wave phase shift is as close as possible to the maximum possible value for a given circulating power and cavity mode numerical aperture, and that the width of the phase profile near the unscattered beam is as small as possible to reduce the cut-on frequency of the phase plate. This can be achieved by using the C2 and C3 lenses to generate a relatively large amount of plane offset such that a laser Ronchigram can be seen on the electron camera with a $\sim 100\,\mu\text{m}$ field of view (similar to that shown in figure 8.9a). The Ronchigram can be analyzed to extract the local contrast of the fringes, which is larger closer to the focal waist. The electron beam can then be moved in the phase plate plane until this region is at the center of the field of view in the camera image, and the plane offset successively reduced (to "zoom in" on the Ronchigram) while checking that the field of view is still centered on the highest contrast portion of the Ronchigram. This procedure is accurate enough to align to well within one Rayleigh range of the focal waist.

Within a range of several dozen wavelengths about the focal waist the maximum phase shift does not change by much, and so the LPP will work just as well when aligned to any of the standing wave antinodes within this range. Therefore, the fine adjustment of the longitudinal beam alignment is simply a matter of moving the electron beam along the cavity mode axis (with zero plane offset) within a range of one phase profile period ($\pm\lambda_L/4$) until the phase shift is maximized. The phase shift maximum is inferred by setting a small amount of underfocus $\sim 1\,\mu\text{m}$ and observing the Thon rings in the FFT of the image (assuming a sample is present). The minima of the Thon rings are closest to the center of the FFT when the phase shift is maximized (at $\pi/2$, see chapter 8). This can be used to align the longitudinal position deflector setting sufficiently accurately by eye. The procedure can also be automated by fitting the Thon rings (we use CTFFIND4 [116]) to determine the phase shift as a function of longitudinal position deflector setting, and then fitting a cosine model to the resulting data to find the optimal deflector setting. This is illustrated in figure 10.5 using Thon rings from images taken on the thick carbon portion of a Quantifoil sample grid.

The correct longitudinal alignment can also in principle be set by centering a light or dark (depending on the sign of the plane offset) Ronchigram fringe in the field of view, but this relies on the position of the unscattered wave in the phase plate plane moving much less than one Ronchigram fringe period when the plane offset is changed. We have found that is not typically the case, and so to use this method the Ronchigram fringe position corresponding to the aligned, on-plane unscattered wave position must first be calibrated. This approach is discussed further in section 10.4.

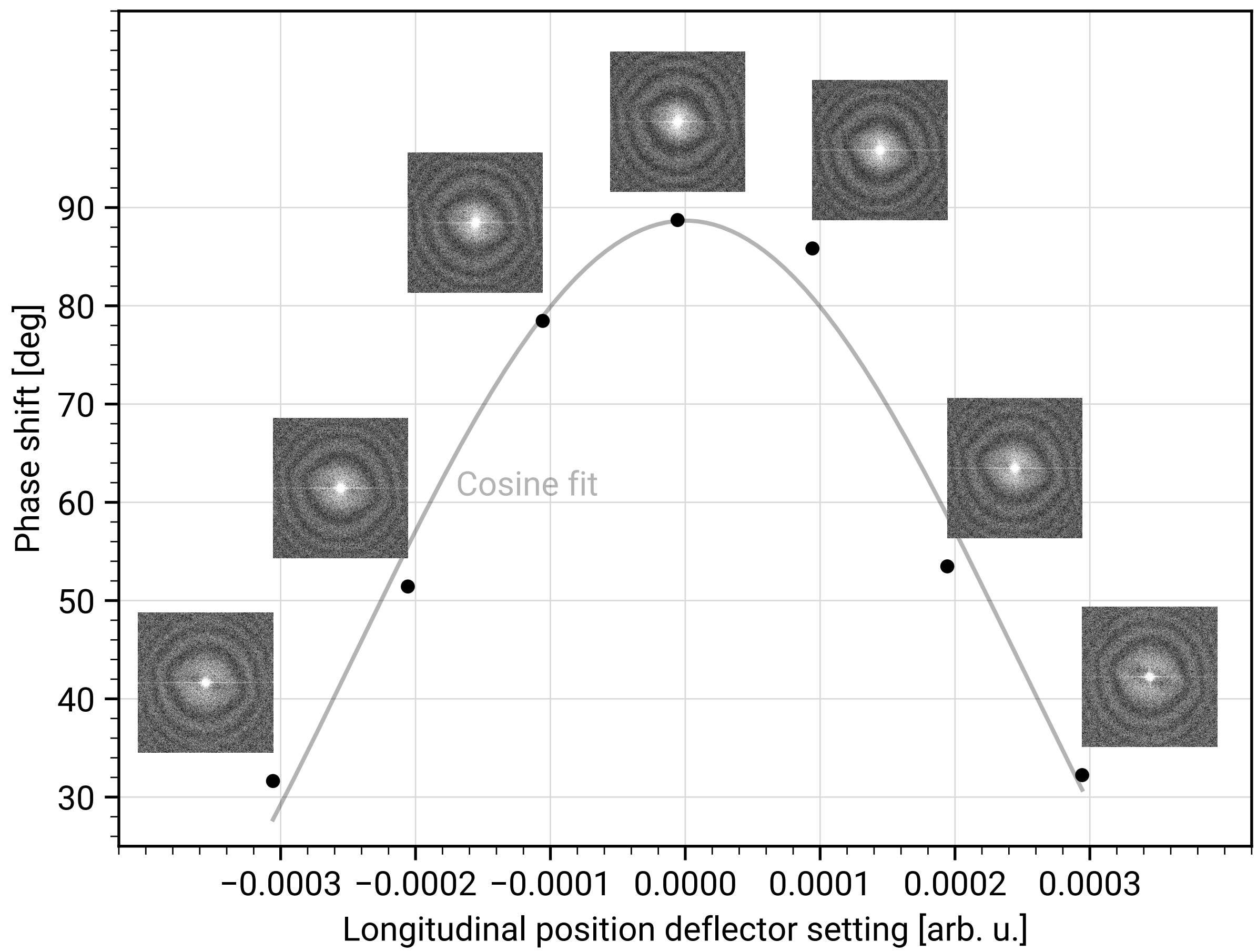

Figure 10.5: Laser phase plate phase shift as a function of the longitudinal position deflector setting. The phase shift is estimated from the Thon rings in the FFT of each image (insets) using CTFFIND4. A fit to a cosine function is shown in gray.

Note that if the LPP is operated with its polarization at the relativistic reversal angle (see chapter 9), the fine longitudinal alignment is unnecessary. However, in this case no laser Ronchigrams will be visible (unless the plane offset is so large as to provide contrast for the very low spatial frequencies remaining in the phase profile) and so the previously described methods for plane offset, transverse, and rough longitudinal alignments will no longer work. Rather than come up with new alignment techniques, it may instead be easier to simply perform these alignments with a horizontal polarization (as we usually do) and then rotate the polarization axis to the relativistic reversal angle before starting to use the LPP for phase contrast imaging.

### 10.3.4 Phase plate tilt

For the standing wave structure in the phase profile to be present, the electron beam must intersect the cavity mode at nearly right angles (tilt angle $\Theta = 0$ in equation (9.25)). Since we use this structure to increase the maximum phase shift in the profile, reduce the cut-on frequency of the phase plate, and provide laser Ronchigrams as an alignment aid, we must first align this tilt degree of freedom before using the LPP for phase contrast imaging.

The LPP is designed such that the cavity mode axis is nominally perpendicular to the electron beam axis, so even before this alignment step it is typically possible to generate (perhaps weak) Ronchigrams. The Ronchigrams are fit to the model described in chapter 9 and the corresponding phase profile modulation depth extracted. This modulation depth is then divided by the circulating power measured at the time the Ronchigram was recorded (since tilting the cavity mode substantially changes the input coupling efficiency and therefore the circulating power). This normalized value is then maximized as a function of the cavity tilt angle, which is measured in terms of the cavity output beam position on the position-sensitive detector and adjusted via the cavity alignment piezos (see section 6.2.5). An example dataset is shown in figure 10.6. The ratio of phase modulation depth to circulating power appears lower than nominal ($1.2\,^{\circ}\mathrm{kW}^{-1}$) because the Ronchigrams are fit (incorrectly) assuming the plane offset is such that the Ronchigram contrast is maximal (see section 8.8).

This procedure requires that that the cavity can sustain a relatively high circulating power ($\sim 10\,\mathrm{kW}$, sufficient to generate easily visible Ronchigrams) over a range of mode positions on the mirror surface. This further increases the requirements on the defect-free area of the mirror surface (see section 7.2). Additionally, it means that the mirrors must be sufficiently defect-free to sustain the nominal LPP mode parameters at the particular position on the mirror surface which results in the optimal cavity tilt angle for operating the LPP in the TEM.

Once the optimal tilt angle is found, the cavity input laser beam is realigned to maximize the input coupling efficiency. The tilt angle is maintained by the cavity mode position feedback system described in section 6.2.5, and does not have to be adjusted again. Changes in the electron beam angle in the phase plate plane which occur during normal operation of the TEM are small enough to not noticeably change this alignment.

### 10.3.5 Hysteresis

The magnetic lenses used in the TEM are hysteretic due to ferromagnetism in the materials they use to enhance and shape their magnetic fields. As such, changing a lens setting from $A$ to $B$ to $A$ only results in the same starting lens strength if the lens was set to $B$ prior to starting the cycle from $A$. That is, for the lens to exhibit consistent behavior, it must first be established in a "hysteresis loop" containing the same sequence of values it is to be used at. This is extremely important for working with the LPP (and for phase plates in general [11]), as it means that all alignments of the electron beam to the cavity mode should be done after all lenses upstream of the LPP have been established in their hysteresis loops. This may not be necessary for some applications which do not require the electron beam parameters (lens settings) to change during imaging. However, cryo-EM applications typically require the sample plane beam diameter and flux to be changed in order to image large areas of the sample at low flux to identify areas of interest, and then image those areas of interest at higher flux. As such, we always make sure to perform the electron beam position alignments described in sections 10.3.1, 10.3.2, and 10.3.3 after establishing the lenses in their hysteresis loops for the particular data collection routine that will be used. Alignments are then performed with lens settings that exist within this loop, and these settings are only accessed by moving the lenses through the appropriate sequence.

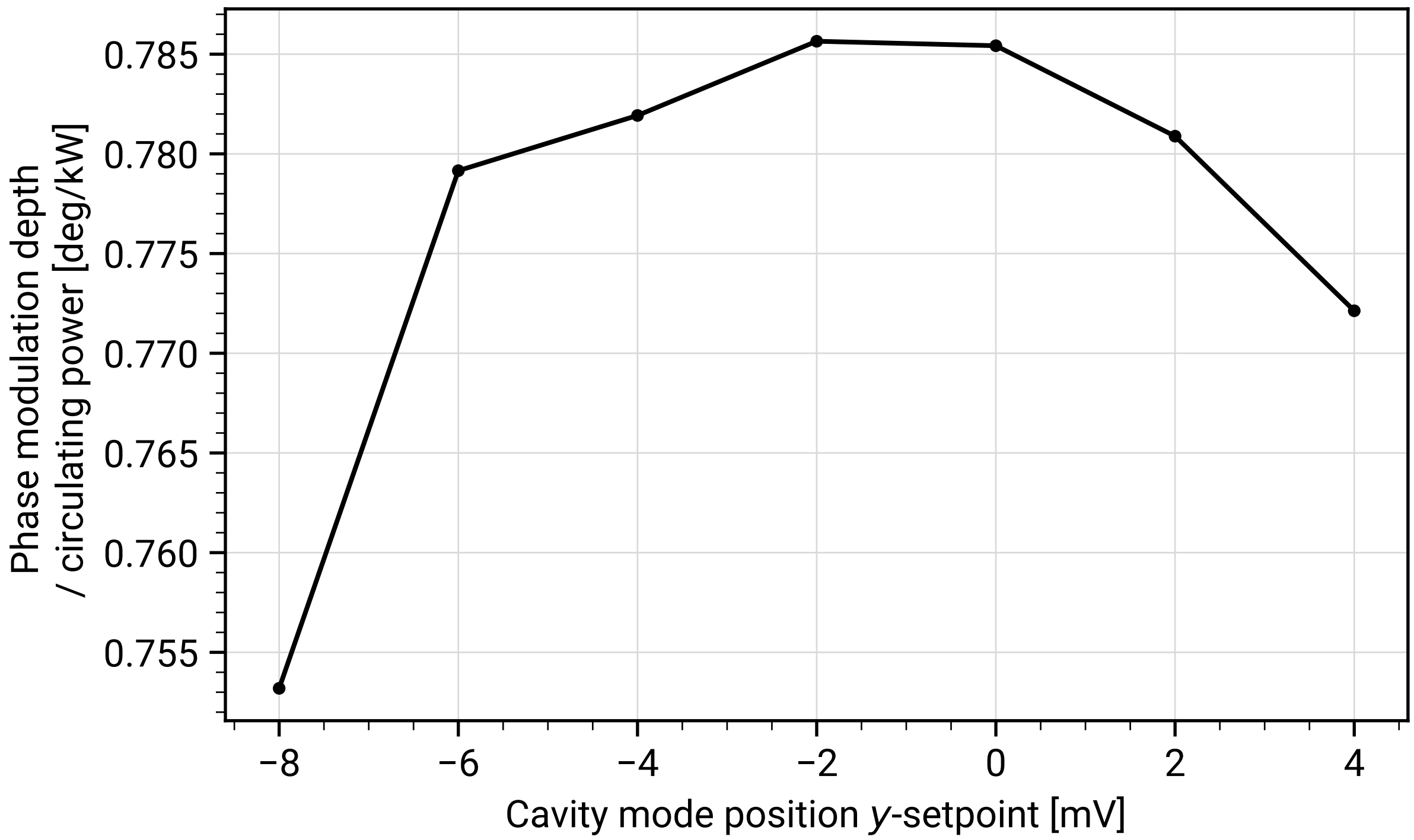

Figure 10.6: Ronchigram phase modulation depth divided by cavity circulating power as a function of the cavity tilt angle $\Theta$, as measured through the setpoint of the cavity mode position feedback system's vertical ($y$-) axis setpoint.

Fortunately, we have not seen hysteresis effects from relatively small changes in deflector settings, which means that their settings can consistently changed independent of their prior values.

The magnetization may also take several seconds to equilibrate when the lens current setting is changed. This must be taken into account when aligning the TEM or collecting data. When using the LPP, we typically wait for $\sim 5\,\mathrm{s}$ after changing lens settings to record any images.

### 10.3.6 Gun blanker

The magnetization equilibration time also affects the gun blanker, which is a magnetic deflector used to divert the beam away from the sample until an image exposure is started. When the beam is "unblanked", the gun blanker current is set to zero and the magnetization of the materials in and around the coil falls. Nearly all of the magnetic field change occurs in $< 25\,\mathrm{ms}$ (compared to the typical exposure time of several seconds), but the remaining field takes several seconds to fully equilibrate, and the equilibrated field strength is slightly different after each unblanking. When a phase plate is not being used, this does not cause any issues, as the residual motion and position reproducibility of the illuminating beam is small enough to not noticeably change any of the imaging properties of the TEM. However, these effects can be large enough to misalign the unscattered wave with the phase plate.

We have largely solved this issue on our system by setting the C1 lens strength such that the phase plate plane is conjugate to the plane of the gun blanker. Equivalently, since the condenser is set to provide parallel illumination in the sample plane, this means that the gun lens must be set to provide a beam focus in the gun blanker plane. This configuration is depicted for the illuminating beam ray shown in figure 10.1. This means that the residual motion of the beam in the phase plate plane is mostly a tilt (rather than a shift) of the beam, which does not noticeably change the profile of the phase shift imparted to the electron beam by the LPP.

However, this solution does restrict the accessible range of electron fluxes at the sample plane, since the fixed gun lens setting means that the beam diameter at the C1 aperture cannot be adjusted. The solution is also likely incompatible with using the monochromator, since the monochromator is designed to provide optimal filtering of the electron beam energy spread when the gun lens focuses the beam on the C1 aperture (rather than the gun blanker plane).

### 10.3.7 Coma-free

To achieve its ultimate resolution, the TEM must also be aligned to minimize image aberrations. For cryo-EM applications, this typically consists of (twofold) stigmating the image (we use the transfer stigmator for this purpose) and aligning the tilt of the beam in the sample plane such that it is aligned with the optical axis of the objective lens, which removes coma ($C_{2,\pm1}$, see table 8.1). These alignments are usually done on a daily basis. On our system, removing the axial coma is slightly more complicated, since the spherical aberration of the system (which generates axial coma when the electron beam is misaligned with the optical axis of the lens) comes from both the objective lens and transfer lens, in roughly equal parts.

This means that to eliminate axial coma, the electron beam should be aligned to the optical axes of each of these lenses individually. We do this by first tilting the beam in the sample plane only using a coordinated actuation of the beam and image deflectors such that the magnitude of image defocus is minimized. The mean defocus (average of defocus along the principal axes of astigmatism) is measured from the Thon rings generated by the sample; the minimum absolute value of defocus corresponds to the coma-free axis of the objective lens. We then repeat the procedure for the transfer lens by tilting the beam in the phase plate plane only using a coordinated actuation of the image and transfer deflectors. For unknown reasons, this procedure does not completely eliminate the axial coma, and so we follow this step by changing the tilt angle in the sample plane (and the phase plate plane) using just the beam deflectors such that the astigmatism induced by the beam tilt is symmetric about the optimal beam tilt axis. This is the standard "Zemlin tableau" coma-free alignment procedure commonly used in cryo-EM [117].

Threefold astigmatism ($C_{2,\pm 3}$, see table 8.1) is typically eliminated by the manufacturer using the threefold objective stigmator. On our system we found that with the system aligned to the LPP there was substantial threefold astigmatism present. We were able to correct for this astigmatism using the objective (threefold) stigmator by removing threefold asymmetry in Zemlin tableaus [118]. The remaining aberrations in our TEM have been characterized via a three-dimensional reconstruction of RuBisCO as shown in figure 12.6—they are negligibly small.

## 10.4 Automated data collection

In cryo-electron microscopy (cryo-EM) it is usually necessary to collect many images of a sample to conclusively determine its structure. For example, in the case of single-particle analysis (SPA) hundreds or thousands of images are recorded, typically with hundreds of individual proteins in each image. Cryo-electron tomography (cryo-ET) requires $\sim 100$ images per tomogram, and it may be necessary to acquire $\sim 100$ tomograms in order to capture a particular feature of interest. In both cases, images must be recorded in particular regions of interest within the sample, and the image defocus must be set within a desired range (typically $1-5\,\mu$m when not using a phase plate). This must be accomplished without exposing the target area to a damaging electron dose ($> 1$ electron $\text{Å}^{-2}$), so alignment images are collected either with a low electron dose or in adjacent areas of the sample such that they do not expose the target area at all.

Automated data collection software routines have been developed for both SPA and cryo-ET to reduce the human workload of data collection and increase the rate at which data can be acquired ("throughput"). State-of-the-art automated data collection procedures can now collect over a hundred images per hour for SPA, or several tomograms per hour for cryo-ET [119], [120].

Automated data collection using the LPP has the additional requirement that the electron beam and laser phase plate be aligned for each image. We have developed an automated data collection routine for SPA which (for each designated target area) moves the target area into the center of the field of view, autofocuses, re-aligns the electron beam unscattered wave to the nearest LPP standing wave antinode (along the longitudinal axis), and records and image.

Before beginning automated data collection, we identify target areas on the sample. The sample consists of an array of regularly-spaced, vitreous ice-filled holes in a $\sim 10\,\text{nm}$ thick carbon film (QUANTIFOIL holey carbon). The target areas are the centers of these holes, where individual proteins are suspended in a thin ($< 100\,\text{nm}$) layer of vitreous ice. The sample stage positions roughly corresponding to the hole centers are calculated by taking a low-dose wide field-of-view image of the sample, identifying holes in the image (areas of higher intensity), and using a calibration of image coordinate to stage position. This is performed using the standard tools in the TEM data acquisition software SerialEM [121], [122]. We use SerialEM's Python scripting interface to execute our automated data collection routine.

The TEM lenses are then established in their hysteresis loops by moving the lenses through the sequence of settings used in the collection routine. An overview of the routine is shown in figure 10.7. The transverse, plane offset, and longitudinal LPP alignments are performed while imaging a sacrificial area of the sample ($\sim 2\,\mu\text{m}$ diameter). At this point the electron beam is optimally aligned to the phase plate, and a phase contrast image is being formed (step 9 of the routine, as seen in figure 10.7). An iteration of the collection routine is then carried out with the purpose of recording a Ronchigram image in step 5 (steps 1, 4, 6, and 9 are skipped since they do not affect the hysteresis state of the lenses). This Ronchigram image serves as a reference for subsequent alignments: it represents the position of the Ronchigram fringes when the electron beam has been properly aligned to an LPP antinode.

The collection routine is then executed as follows:

1. The sample stage is moved to the nominal position of the next target area. The condenser is set to provide a several micrometer beam diameter in the sample plane with an electron flux of $\sim 0.1\,\text{electron}\,\text{Å}^{-2}\,\text{s}^{-1}$. We have found that using a different discrete "spot size" settings (which sets the C1 and C2 lens strengths to change the flux in the sample plane) at different points in the routine disrupts the alignment of the electron beam to the LPP, even when executed within a consistent hysteresis loop—we do not know the reason. Therefore, we maintain a consistent spot size setting throughout the routine. Achieving a low enough flux for this step requires that we use a $50\,\mu\text{m}$ C3 aperture and spread the beam on the aperture. The C3 aperture thus defines the beam diameter in the sample plane during this step. For subsequent steps the beam entirely passes through the C3 aperture. The projector is set to a magnification which provides a several micrometer-wide field of view, and the transfer lens is set to give a large $\sim 25\,\mu\text{m}$ of defocus. This allows the holes to be easily seen in even a low-dose image ($\sim 0.03\,\text{electron}\,\text{Å}^{-2}$). An example image is shown in the inset in step 1 of figure 10.7. The circular feature is the hole. The thin stripe which extends from the upper-left to lower-right of the image is a Ronchigram of the LPP, as changing the condenser settings substantially changes the plane offset. If the center of the hole is not close enough to the center of the field of view, the stage is moved appropriately and another image taken. This process repeats until the hole is aligned in the center of the field of view (typically only one or two iterations are required). The routine is then paused for 10 s to allow the stage position to stabilize.

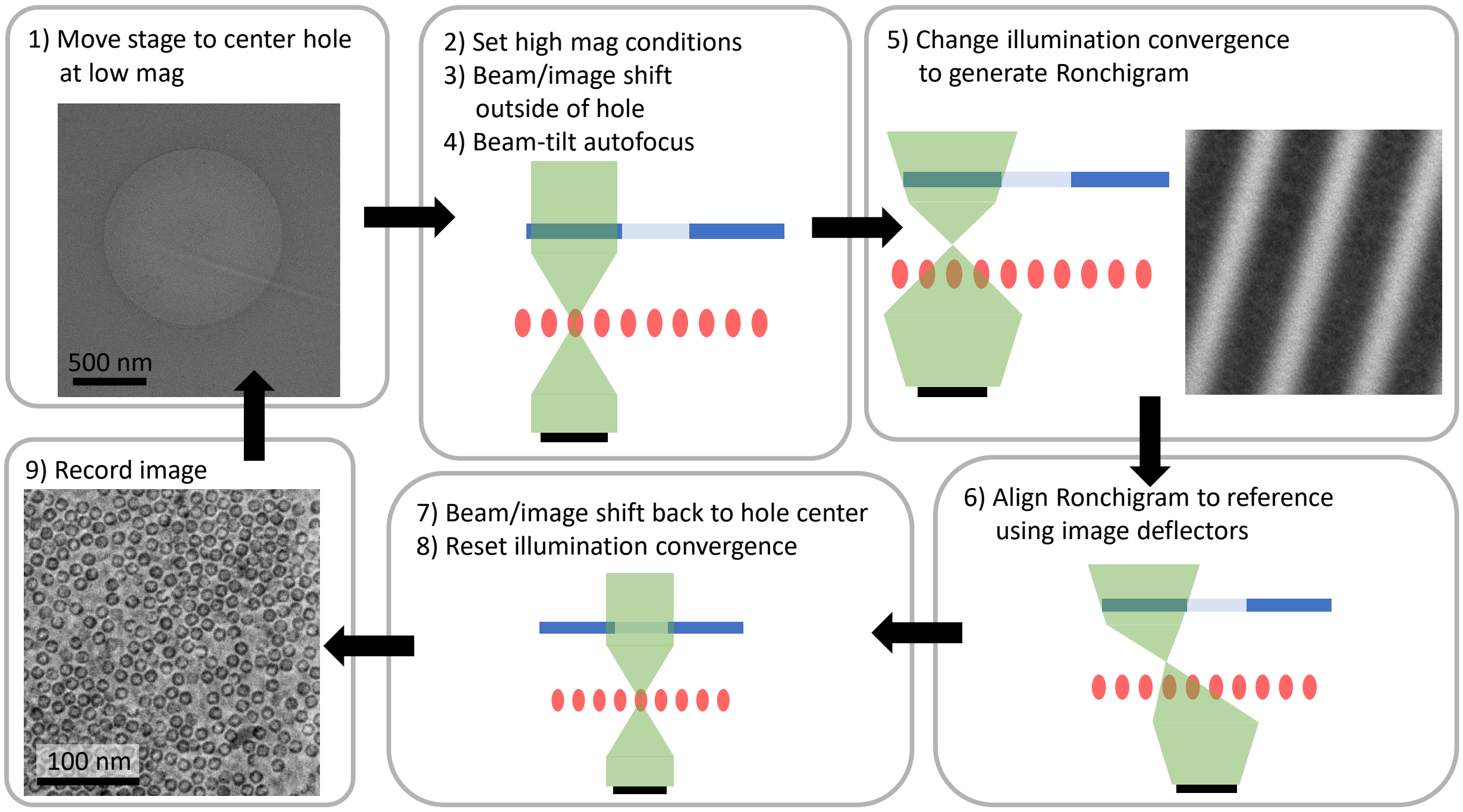

Figure 10.7: Automated data collection routine for single-particle analysis using the laser phase plate. In the depictions of the transmission electron microscope alignment, the electron beam is shown in green, the sample is shown in blue (holes in light blue), the laser phase plate is shown in red, and the electron camera is shown in black.

2. The condenser and projector are set to their nominal imaging settings with a few micrometer sample plane beam diameter, an electron flux of $\sim$ 10 electron $\text{Å}^{-2}\,\text{s}^{-1}$, and an field of view of several hundred nanometers.

3. The beam and image deflectors are used to shift the beam so it is just adjacent to the target area, without moving the beam on the camera. Note that the beam is not actually hitting the sample at this point because the gun blanker is active (sample not exposed) except when an image is being recorded. The beam is also deflected roughly 20 µm away from the LPP in the transverse direction using the image and transfer deflectors such that the beam remains stationary on the camera. This is done so that the phase contrast generated by the LPP does not interfere with the next step.

4. The defocus is measured by tilting the beam in the sample plane using the beam deflectors and measuring the resulting shift in positions of the image of the sample [123]. The defocus is reset as necessary to a user-specified value (or random value within a user-specified range) using the transfer lens. We found that keeping the electron beam roughly aligned to the LPP during this step would generate different amounts of contrast in the images taken at different beam tilt angles (due to the concomitant motion of the beam in the phase plate plane) which reduced the accuracy in the measurement of the change in image position.

5. The beam is returned to its nominal transverse position in the phase plate plane. The condenser is used to change the illumination convergence angle in the sample plane such that a non-zero plane offset is generated and a laser Ronchigram is formed (see inset in step 5 of figure 10.7). The transfer lens strength is set to that used when the reference Ronchigram image was taken. This is done because changing the transfer lens strength in general shifts the image position on the camera, so to make accurate position comparisons between images the same lens strength must be used. In principle, the electron beam could be aligned through the transfer lens such that this shift is minimized—this would subsequently require the LPP cavity to be aligned to the position of the electron beam. The lenses are then allowed to equilibrate for 10 s.

6. The position of the fringes in this Ronchigram are compared to that of the reference Ronchigram. If they differ by more than 1/20th of a period (26.6 nm in the phase plate plane), the electron beam is shifted by the corresponding amount along the longitudinal axis of the LPP in the phase plate plane using the image and transfer deflectors such that the beam remains stationary on the camera. The amount and direction of motion is calibrated by measuring the change in Ronchigram fringe position as a function of longitudinal position deflector setting. Another Ronchigram image is recorded, and the process repeats until the Ronchigram is sufficiently well aligned to the reference. Often, no realignment is required (see section 12.1), and if it is, typically only one or two realignment iterations are required.

7. The beam and image deflectors are used to shift the beam position back to the target area without moving the beam on the camera. The transfer lens strength is set back to its autofocused value (end of step 4).

8. The condenser settings are returned to nominal (those of step 2) with parallel illumination in the sample plane. The lenses are then allowed to equilibrate for 10 s.

9. An image of the target area is recorded. The inset in step 9 in figure 10.7 shows an example of an LPP phase contrast image using an apoferritin SPA sample.

The collection routine continues cyclically until commanded to stop or images have been collected in all target areas. Note that there is no communication between the TEM and LPP control software during the routine.

The imaging performance of this routine is discussed in detail in section 12.1. The throughput is roughly 30 images per hour. This is less than state-of-the-art because we move the stage for every target rather than shift the electron beam in the sample plane to record images from multiple targets with a fixed stage position (referred to as "beam-image shift" acquisition). We found that beam-image shift, even when well-calibrated, inevitably shifts the beam in the phase plate plane by a large enough distance to noticeably change the unscattered wave phase shift. We believe that this shift is reproducible, and therefore should be possible to measure and compensate for. However, throughput has not yet been a serious limitation for our research and so we have not yet fully explored this approach.

This routine will likely need to be modified to be compatible with holey gold (rather than carbon) grids [124], [125], because laser Ronchigrams are only barely visible through such a high contrast material. On possibility is to use an adjacent hole (rather than adjacent area of the grid) for autofocusing and aligning to the LPP. This would sacrifice half of the holes in the sample for alignment purposes, but in practice this likely only somewhat reduces throughput.

This routine can likely be adapted for use in cryo-ET, and in particular we are working to determine if the Ronchigram alignment method will work in that context.

# Chapter 11

# Thermal magnetic field fluctuations

During development of the LPP, we found that thermal magnetic field fluctuations emanating from the electrically conductive cavity mount can limit the resolution of our TEM. All electrically conductive materials at non-zero temperature contain thermally excited stochastic electrical currents, and these currents generate magnetic fields which extend outside of the material. These fluctuating magnetic fields exert a force on a nearby electron beam, which can blur the image formed by the beam, leading to loss of resolution.

Using a previous prototype design which used an electron beam hole (see section 4.4) with a diameter of 2 mm (compared to 8 mm on the current design), we were able to demonstrate the long-term stability of the LPP during data collection as well as the high contrast of images, and produced a 3D density map of 20S proteasome particles at a resolution of 3.8 Å [24]. We initially believed that this rather modest resolution, given the number of asymmetric units included in the final reconstruction, was limited by the increased value of chromatic aberration associated with our TEM's relay optics (described in section 10.1). However, subsequent efforts to improve the resolution, or to reduce the number of particles needed for a given reconstruction, revealed the presence of a further limitation, in addition to the expected falloff of the temporal coherence envelope.

In this chapter, I present data which identifies thermal magnetic field fluctuations as the cause of this loss of resolution. These magnetic field fluctuations are already known to potentially limit the resolution in aberration-corrected transmission electron microscopy [126], [127], and though they have negligible effects in standard TEMs, their effect is enhanced in our system because of the magnification of the diffraction plane from the relay optics and the fact that the electron beam has to pass through a long (23 mm) and relatively narrow (2 mm diameter) hole in our previous LPP prototype. Increasing the electron beam hole diameter to 8 mm on our current LPP prototype has almost entirely eliminated the loss of resolution.

## 11.1 Theoretical model

Uhlemann and co-authors [126], [127] have demonstrated that magnetic field fluctuations [128]–[132] from electrical currents driven by thermal fluctuations in the electrically conductive parts of the TEM (Johnson-Nyquist noise, or Johnson noise for brevity) causes a loss of resolution. A magnetic field $B_\perp$ applied over a length $L$ in a direction transverse to the electron velocity will deflect the electron beam by an angle $\theta = \chi B_\perp L$ where $\chi \coloneqq \frac{e\lambda_e}{2\pi\hbar}$ with $\lambda_e$ being the electron wavelength. This expression is valid for small angles $\theta \ll 1$. If the magnetic field varies stochastically in both space and time, then the variance (over time) in deflection angle $\langle\theta^2\rangle$ will depend on the magnetic field's spatial correlation length $\xi$ and temporal variance $\langle B_\perp^2\rangle$ such that

$$\left\langle\theta^2\right\rangle = \chi^2\xi\left\langle B_\perp^2\right\rangle L \tag{11.1}$$

While detailed models of the spatiotemporal structure of the Johnson noise requires finite-element numerical calculations, an analytic result exists for the product $\xi\left\langle B_\perp^2\right\rangle$ when an electron beam passes through a tube of length $L$ and diameter $D$ where $L \gg D$, and the tube is mode conducting, non-magnetic material [127]:

$$\xi\left\langle B_\perp^2\right\rangle = C\frac{\mu_0 k_B T}{D^2} \tag{11.2}$$

where $C \approx 0.77$ is a dimensionless coefficient, $\mu_0$ is the vacuum permeability, $k_B$ is Boltzmann's constant, and $T$ is the absolute temperature of the tube. In our TEM, the diameter $D$ of the electron beam liner tube (including LPP, if present) changes as a function of distance $z$ along the electron beam optical axis. To obtain a simple model that allows us to estimate the cumulative effect of Johnson noise along the length of the liner tube without resorting to numerical simulations, we postulate a differential formulation of equation (11.1):

$$d\left\langle\theta^2\right\rangle = \chi^2 C\frac{\mu_0 k_B T}{D\left(z\right)^2}dz \tag{11.3}$$

where the diameter of the tube $D\left(z\right)$ is now a function of position $z$.

The angular deflection applied to the electron beam causes an apparent shift in the position of objects in the sample plane. The amount of shift depends not only on the amount of angular deflection, but also on where in the imaging system that deflection is applied. The shift can be quantified using ray transfer matrix analysis. First, consider a ray which emanates from the specimen ($z = 0$) at a distance and angle relative to the optical axis of $x_0$, and $\theta_0$, respectively. represent this ray by a column vector

$$\begin{pmatrix} x_0 \\ \theta_0 \end{pmatrix} \tag{11.4}$$

Further down the optical axis, the optical system transforms this ray into

$$U\left(z\right)\begin{pmatrix} x_0 \\ \theta_0 \end{pmatrix} \tag{11.5}$$

where $U(z)$ is the ray transfer matrix representing the optical system between $z = 0$ and $z$. At this position, we add an angular deflection $\theta$, which transforms the ray into

$$U(z)\begin{pmatrix} x_0 \\ \theta_0 \end{pmatrix} + \begin{pmatrix} 0 \\ \theta \end{pmatrix} \tag{11.6}$$

In order to determine the corresponding apparent change in object position, we apply the inverse ray transfer matrix to determine the column vector that would have produced the same ray in absence of magnetic field fluctuations:

$$\begin{pmatrix} x'_0 \\ \theta'_0 \end{pmatrix} = U^{-1}(z)\left(U(z)\begin{pmatrix} x_0 \\ \theta_0 \end{pmatrix} + \begin{pmatrix} 0 \\ \theta \end{pmatrix}\right) \tag{11.7}$$

This shows that the apparent change in object position $\delta x \coloneqq x'_0 - x_0$ is

$$\delta x = \begin{pmatrix} 1 & 0 \end{pmatrix} U^{-1}(z)\begin{pmatrix} 0 \\ \theta \end{pmatrix} \tag{11.8}$$

$$= \theta \begin{pmatrix} 1 & 0 \end{pmatrix} U^{-1}(z)\begin{pmatrix} 0 \\ 1 \end{pmatrix} \tag{11.9}$$

To simplify this expression further, note that a ray transfer matrix

$$U = \begin{pmatrix} A & B \\ C & D \end{pmatrix} \tag{11.10}$$

between two planes with equal index of refraction has a determinant of unity, and thus its inverse is simply

$$U^{-1} = \begin{pmatrix} D & -B \\ -C & A \end{pmatrix} \tag{11.11}$$

Therefore,

$$X(z) \coloneqq \begin{pmatrix} 1 & 0 \end{pmatrix} U^{-1}(z)\begin{pmatrix} 0 \\ 1 \end{pmatrix} \tag{11.12}$$

$$= -\begin{pmatrix} 1 & 0 \end{pmatrix} U(z)\begin{pmatrix} 0 \\ 1 \end{pmatrix} \tag{11.13}$$

is the distance of the marginal ray from the optical axis, where the marginal ray is defined to be that which leaves the optical axis in the sample plane at an angle of 1 rad. The marginal ray (when using relay optics) in our Titan is plotted schematically in figure 11.1 along with a simplified representation of the liner tube diameter $D(z)$. We thus obtain the simple result that

$$\delta x = -X(z)\,\theta(z) \tag{11.14}$$

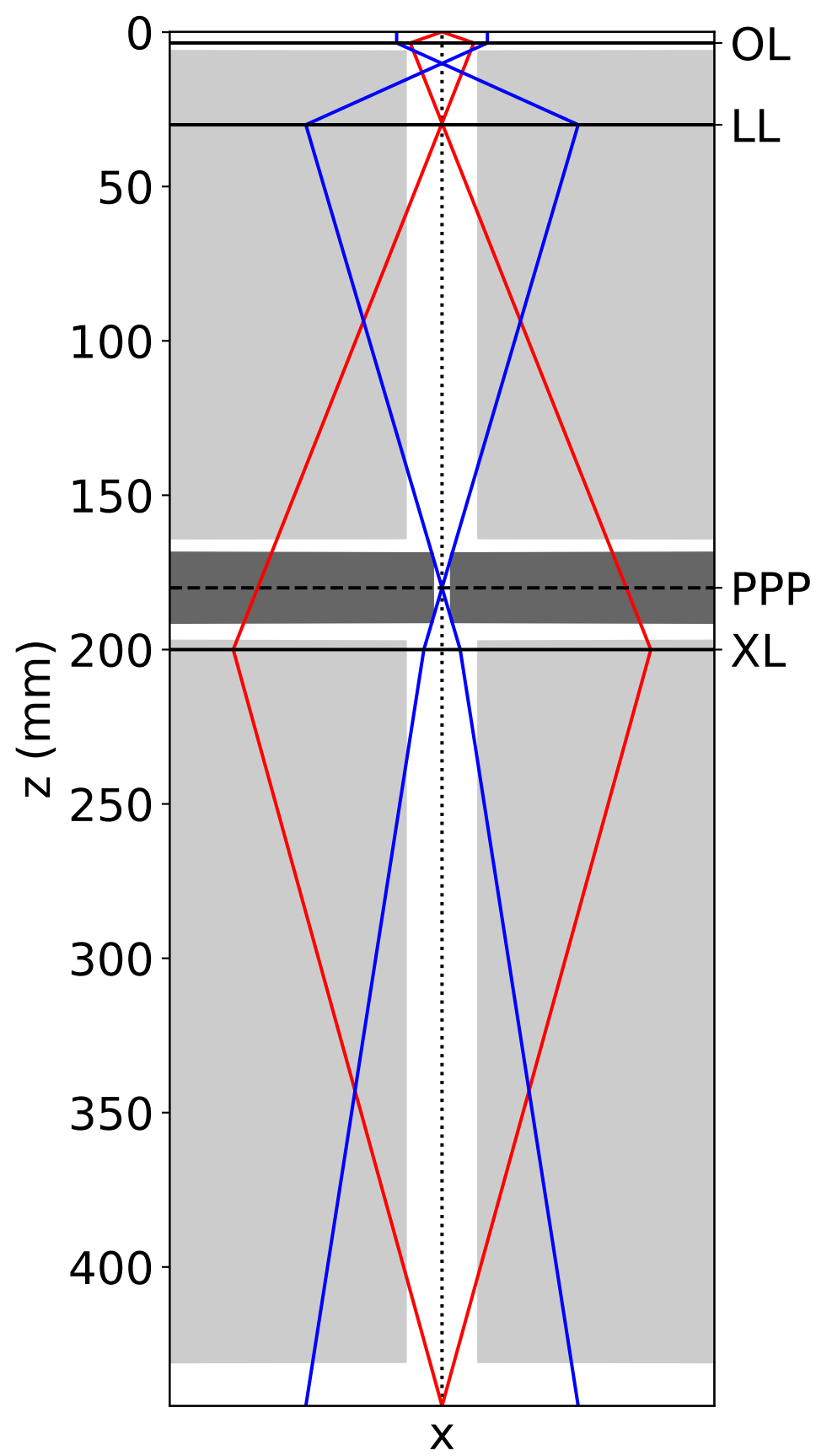

Figure 11.1: Ray diagram of the modified Titan $80-300\,\mathrm{keV}$ TEM used in this work. The marginal ray (red) and paraxial ray (blue) are shown between the sample plane ($z=-0.1\,\mathrm{mm}$, at the top of the figure) and the selected area aperture plane ($z=445\,\mathrm{mm}$, at the bottom of the figure). The marginal ray is drawn so that it leaves the optical axis (vertical dotted line) in the sample plane at an angle of 1 rad. The Lorentz lens (LL) forms a magnified image of the back focal plane of the objective lens (OL) in the phase plate plane (PPP) at $z=180\,\mathrm{mm}$. In that plane, the distance of the marginal ray from the optical axis, $X(z)$, is approximately $20\,\mathrm{mm}$. The transfer lens (XL) forms an image in the selected area aperture plane (bottom of figure). Cross-sections of the electron beam liner tube and LPP dummy ($2\,\mathrm{mm}$ hole diameter) are represented by the light gray and dark gray shaded regions, respectively. The dimensions of the liner tube shown in this figure have been simplified at the request of Thermo Fisher Scientific. Reproduced from [67].

We can then use this equation to sum in quadrature the apparent shifts in object position due to angular deflections caused by Johnson noise throughout the microscope column using equation (11.3), such that the resulting variance in apparent object position due to Johnson noise becomes an integral over the length of the electron beam liner tube:

$$\left\langle \delta x^2 \right\rangle = \int dz\, \chi^2 C \frac{\mu_0 k_\mathrm{B} T}{D\left(z\right)^2} X^2\left(z\right) \tag{11.15}$$

We assume that this variance is generated by a Gaussian distribution $p\left(x,y\right) \propto \exp\left(-\frac{x^2+y^2}{2\left\langle \delta x^2 \right\rangle}\right)$ so that the image $I\left(x,y\right)$—when blurred by Johnson noise—can be written as the convolution

$$I_\mathrm{blurred}\left(x,y\right) = \left[p * I\right]\left(x,y\right) \tag{11.16}$$

Thus, the Fourier transform of the blurred image is

$$\mathcal{F}\left[I_\mathrm{blurred}\right]\left(s_x,s_y\right) = \exp\left(-2\pi^2 \left\langle \delta x^2 \right\rangle \left(s_x^2+s_y^2\right)\right) \mathcal{F}\left[I\right]\left(s_x,s_y\right) \tag{11.17}$$

where $s_x$ and $s_y$ are the Cartesian components of the spatial frequency $\mathbf{s}$. To summarize, the blurring from Johnson noise results in the CTF being multiplied by an additional envelope function (see section 8.7)

$$M_\mathrm{J}\left(\mathbf{s}\right) \coloneqq \exp\left(-2\pi^2 \left\langle \delta x^2 \right\rangle \left|\mathbf{s}\right|^2\right) \tag{11.18}$$

## 11.2 Experimental data

To characterize the effect of different hole diameters on the thermal magnetic field fluctuation induced resolution loss, simplified mechanical dummies were inserted into the port on the microscope's phase plate module. These dummies consisted of a solid 6061 aluminum cylinder (indicated by the dark gray shaded region in figure 11.1) with a diameter identical to that of our actual LPP (23 mm). Each dummy had a single hole through which the electron beam could pass. The axis of that hole was perpendicular to and intersected with the axis of the dummy's cylinder similar to the actual LPP (and such that the length of the hole was equal to the diameter of the holder). Dummies with electron beam hole diameters of 2 mm, 4 mm, and 8 mm were used. The test was also performed with no dummy inserted.

2 nm thick carbon films supported on Quantifoil holey carbon grids (Ted Pella 668-300-Cu) were imaged with a room-temperature specimen holder in order to make quantitative measurements of the microscope's CTF envelope as a function of the diameter of the electron beam hole in the dummies. Images were recorded as movies consisting of 150 frames each, using a Gatan K2 camera, with a pixel size (referred to the specimen) of 0.28 Å with relay optics and 0.32 Å without relay optics. Approximately 50 such images were collected in both modes for each size of hole. The accumulated electron exposure at the sample was $\sim$ 2300 electron $\text{Å}^{-2}$ per movie, while the dose rate at the camera was maintained at $\sim$ $8e\,\mathrm{pix}^{-1}\,\mathrm{s}^{-1}$ in order to ensure minimal coincidence loss [133].

The CTF envelope with relay optics was estimated by computing the ratio of the background-subtracted power spectra of images taken with relay optics to those taken without relay optics. This ratio was used to factor out the unwanted contributions of the camera DQE and the structure factor of the specimen, which are expected to be the same in the two imaging modes. Estimates of the envelope with relay optics were then obtained by multiplying by the theoretical values of the temporal coherence envelope without relay optics. The estimated CTF envelope with relay optics is expected to reflect a fixed contribution due to the increased amount of chromatic aberration and a hole diameter-dependent contribution due to Johnson noise.

The data collection procedure is as follows:

1. 2 nm thick carbon films supported on Quantifoil holey carbon grids (Ted Pella 668-300-Cu) were imaged with a room-temperature specimen holder.

2. The sample-referred pixel size of the images (0.28 Å with relay optics and 0.32 Å without relay optics) was calibrated using a room-temperature nanocrystalline gold sample.

3. Images were taken within the same grid square of the sample both with and without relay optics in order to reduce any systematic bias in the sample structure factor between images taken in each mode.

4. The grid hole locations of the images taken in each mode were made to form an interleaved grid pattern. This was intended to further reduce any systematic bias in the sample structure factor between images taken with and without relay optics.

5. Each grid hole was imaged only once. We found that successive images of the same area resulted in an increase in the sample structure factor (which could be produced by the build-up of hydrocarbon contamination with prolonged exposure).

6. Images were autofocused to approximately 1 µm of defocus using a region of the sample approximately 1 µm away from the target image area. This value of defocus is small enough that the associated spatial coherence CTF envelope is negligible, but large enough to generate a sufficient number of Thon rings for our data analysis procedure.

7. Images were recorded as 150 frame movies using a Gatan K2 direct electron detection camera, with a dose rate at the camera of approximately 8 electron $\mathrm{pix}^{-1}\,\mathrm{s}^{-1}$ in order to ensure minimal coincidence loss.

The data analysis procedure for each dataset (different dummy configurations) is as follows:

1. Movies from both modes were truncated (trailing frames removed) in order that all movies from the dataset had as similar as possible of a total dose per sample area. This was done because we found that higher sample dose images resulted in higher sample structure factors.

2. The truncated movies were motion-corrected into a single image using MotionCor2 [134].

3. The CTF of the resulting motion-corrected images was then fit using CTFFIND4 [116]. In order to optimize the quality of the fit, these fits were performed as a function of the spherical aberration coefficient $C_s$ as well. This was done by running CTFFIND4 for multiple values of $C_s$ and then choosing the value which resulted in the highest "spacing (in Angstroms) up to which CTF rings were fit successfully" output metric from CTFFIND4. The value of $C_s$ chosen this way was constrained to be the same for all subsequent image analysis within a given imaging mode (standard or phase plate).

Then, for each image in the dataset:

4. The magnitude squared of its FFT was calculated and divided by its value at zero frequency (for normalization against any remaining differences in dose between images).

5. From the normalized $|\text{FFT}|^2$, the best CTF fit was used to calculate an equiphase average [135], resulting in a one-dimensional curve as a function of the mean magnitude of spatial frequency around the equiphase averaging contour.

6. The values of this curve at the spatial frequencies of the CTF zero-crossings (as determined by the CTFFIND4 best fit) were used to interpolate a two-dimensional, azimuthally symmetric function representing the frequency-dependent noise spectrum.

7. This noise floor function was then subtracted from the normalized $|\text{FFT}|^2$, and the result was divided by the square of the noise-free, two-dimensional CTFFIND4 best fit function. The result should represent the power spectral density of the sample (square of the structure factor) multiplied by the square of the CTF envelope, since the expectation value of the noise spectrum has been subtracted and the oscillatory part of the CTF has been deconvolved.

8. This two-dimensional function was then equiphase averaged based on the CTFFIND4 best fit. Areas where the magnitude of the oscillatory part of the CTF was less than 0.5 were ignored as they were more sensitive to inaccuracies in the estimate of the noise floor. The resulting one-dimensional curves were linearly interpolated across these regions, and represent the envelope of the noise-subtracted power spectral density in each image.

Then, for each dataset:

9. The set of roughly 50 power spectral density curves from images taken with relay optics were divided by each of the roughly 50 power spectral density curves from images taken without relay optics, and the square root was taken. This resulted in approximately 2500 curves representing the measured ratio of the CTF envelopes, for each possible combination of images taken with and without relay optics in the dataset.

Then, for all datasets:

10. Fits were performed on 10,000 random combinations of curves from each of the four datasets. They were then multiplied by the theoretical envelope without relay optics using the resulting best fit value for the electron beam energy spread $\delta E$ to generate curves representing the measured CTF envelope with relay optics.

11. The mean of these curves along with confidence interval is displayed in figure 11.2a.

12. The mean and confidence interval of the image blur variance fit parameter $\langle \delta x^2 \rangle$ is shown in figure 11.2b.

According to equation (11.18), Johnson noise generates a Gaussian contribution to the falloff of the CTF envelope. We also take into account chromatic aberration by including the usual temporal coherence envelope (see section 8.7.1.1), the exponent of which increases as the fourth power of the spatial frequency. The product of these two envelopes was used to fit the measured CTF envelope data, using the image blur variance $\langle \delta x^2 \rangle$ and electron beam energy spread as fit parameters. The electron beam energy spread fit parameter was forced to be the same for all four measured CTF envelopes (i.e. for the three electron beam hole diameters and the case where no dummy was inserted). To summarize, five parameters were used to fit the four measured CTF envelopes—four representing the different image blur variances, and one representing the identical temporal coherence envelopes.

The results of this measurement are shown in figure 11.2, and confirm that he CTF envelope is sensitive to the diameter of the hole. When a dummy with a 2 mm hole is inserted, for example, the CTF envelope at resolutions close to 3 Å falls almost twofold compared to when no dummy is inserted. Increasing the diameter of the hole to 8 mm restores the CTF envelope to nearly that obtained without a dummy.

Johnson noise in the microscope's liner tube becomes dominant for dummy hole diameters of 8 mm and above. This is visible in the CTF envelope measured with no dummy inserted, as it falls below 0.5 at a frequency of 0.35 $\text{Å}^{-1}$, compared to 0.4 $\text{Å}^{-1}$ for the theoretical temporal coherence envelope. The measured CTF envelopes shown in figure 11.2a are well-fitted by theoretical curves (dashed lines) which includes Johnson noise as well as the increased chromatic aberration of the relay optics.

Figure 11.2b shows the amount of Johnson noise image blur variance $\langle \delta x^2 \rangle$ extracted from the data in figure 11.2a compared to the blur expected from the theoretical model in equation (11.15). The non-zero predicted value that remains when there is no dummy arises from Johnson noise in the liner tube. The model reproduces the observed increase in image blur with decreasing electron beam hole size. However, the measured variance exceeds the prediction by a (roughly constant) 0.045 $\text{Å}^2$. This could be due to an inaccuracy in the simplified model, or additional resolution loss that affects the microscope more strongly when using relay optics than when not (e.g. vibrations of the microscope column, or magnetic field noise from external sources).

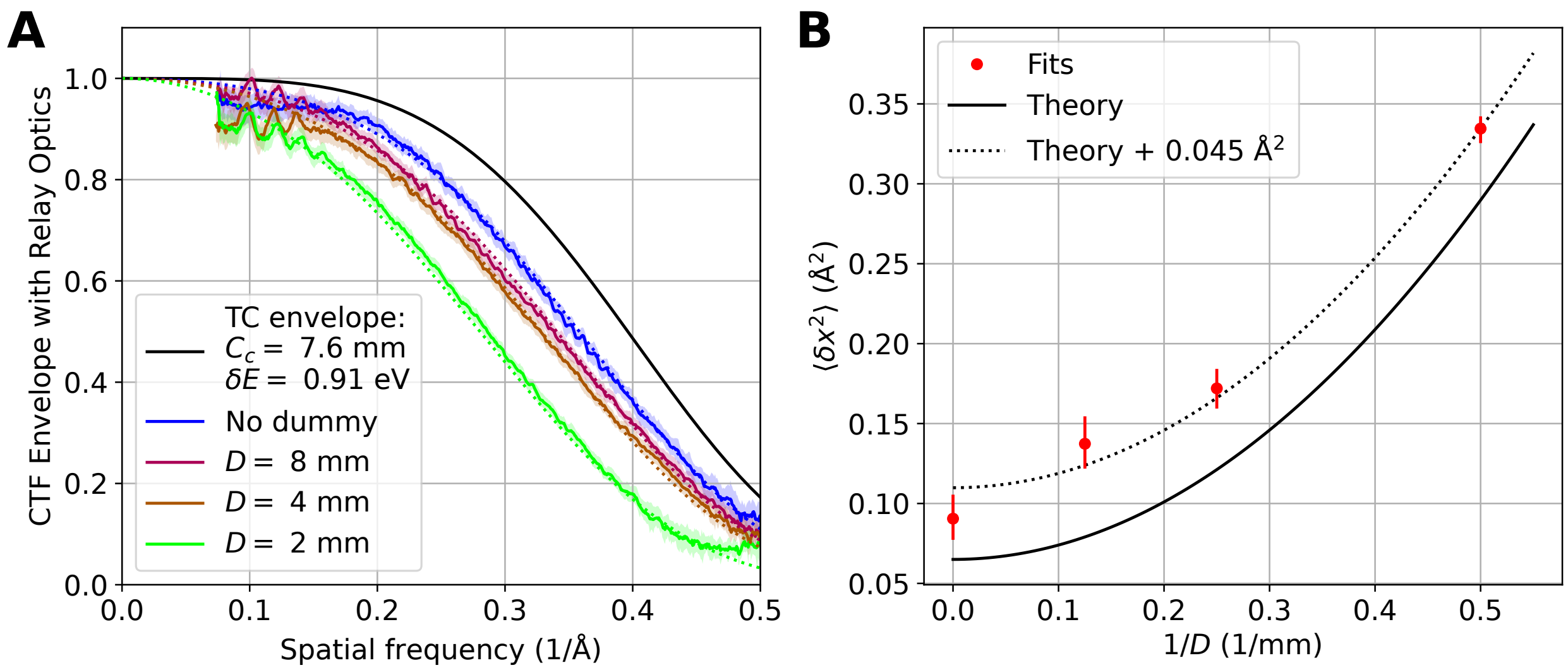

Figure 11.2: **a)** Contrast transfer function (CTF) envelope functions with relay optics measured either without inserting a laser phase plate dummy or when dummies were inserted that had electron beam hole diameters equal to 8 mm, 4 mm, or 2 mm. The colored, solid lines represent the measured CTF envelope, with the associated shaded regions representing a 50% confidence interval. The dashed lines represent a best fit to the data, modeled as the product of the envelopes due to imperfect temporal coherence and Johnson noise. The best fit electron beam energy spread $\delta E$ (full width half maximum) was 0.91 eV (50% confidence interval $[0.88, 0.94]$ eV), somewhat higher than the manufacturer's specification of 0.8 eV for the S-FEG field emission gun used here. The black solid line shows the theoretical temporal coherence (TC) CTF envelope function in the absence of Johnson noise. **b)** The fitted values of the Johnson noise image blur variance (red dots) plotted as a function of the inverse of the dummy's electron beam hole diameter. The error bars show the 50% confidence interval. The solid black line shows the theoretical value of the image blur variance as a function of inverse hole diameter (based on equation (11.15), using manufacturer-supplied values for the marginal ray distance and liner tube diameter). The dotted black line represents a best fit to the data when a constant value is added to the theoretical model as a fit parameter. Reproduced from [67].

## 11.3 Discussion

Thermal fluctuations in the magnetic field adjacent to any electrically conductive component, arising from Johnson-Nyquist noise, are known to be a limiting factor at resolutions beyond $\sim 0.5$ Å [126] but are usually negligible at resolution values achieved in cryo-EM. However, their effect can become appreciable when relay optics are used to magnify the electron diffraction pattern, as is done in our case to accommodate the LPP. In particular, they become a limitation when a long (23 mm), narrow (2 mm diameter) hole is provided for the electron beam to pass though the LPP, as was the case in [24].

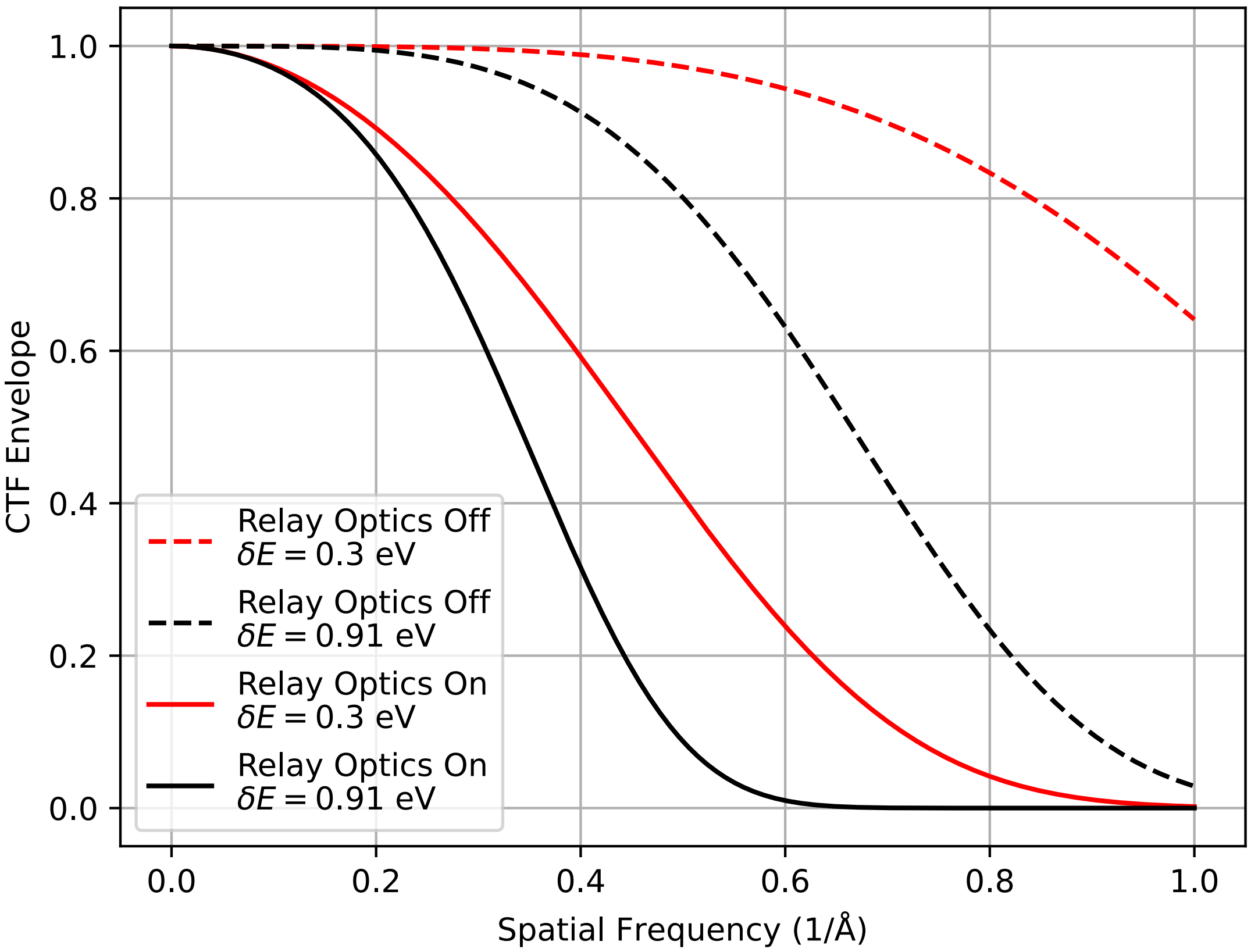

Figure 11.3: Expected improvement in our modified Titan's contrast transfer function (CTF) envelope when using an electron source with a narrower energy spread. The current CTF envelope with relay optics using the measured values of image blur variance $\langle \delta x^2 \rangle$ and electron beam energy spread $\delta E$ for the 8 mm electron beam hole diameter dummy (see figure 11.2) is shown in the solid black line. The solid red line shows this case except that $\delta E$ is reduced to 0.3 eV. For comparison, the dashed black line shows the CTF envelope without relay optics using the measured value of $\delta E$, and the dashed red line shows the CTF envelope without relay optics if $\delta E = 0.3$ eV. Reproduced from [67].

Our theoretical modeling and experimental data indicates that a hole diameter of 8 mm largely resolves this problem, although Johnson noise originating in the liner tube remains a limiting factor. Removing this limitation would require redesigning the TEM liner tube with a larger diameter, especially in regions where the marginal ray deviates most from the optical axis. Alternatively, the TEM could be designed with a relay optics system with a smaller diffraction plane magnification, though this has the side effect of increasing the cut-on frequency of the LPP (see section 8.4).

In addition, the CTF envelope of our TEM is limited by the increased chromatic aberration caused by the relay optics. The CTF envelope resulting from the combination of these two factors is shown in figure 11.3, where it is compared with that obtained when relay optics are not used (i.e. in a standard Thermo Fisher Scientific Titan or Krios).

These two limitations have been addressed in a design for a next-generation phase plate microscope (see chapter 13). Residual Johnson noise has been minimized by increasing the diameter of the beam liner tube and by making it possible to adjust the effective focal length (and thus the marginal ray distance $X(z)$) of the relay optics between $12 - 20\,\mathrm{mm}$. This reduces the influence of Johnson noise proportionally. The resulting increase in cut-on frequency will be negligible when imaging structure sizes smaller than about $50\,\mathrm{nm}$ and can be avoided by reducing the LPP's laser focus waist and/or wavelength. Cooling the liner tube (e.g. with liquid nitrogen to $\sim 77\,\mathrm{K}$) could be another way to substantially reduce the amount of resolution loss.

The effect of chromatic aberration can be improved by reducing the width of the electron energy distribution with a gun monochromator or a cold field emission gun; at a width of 0.3 eV, the temporal coherence envelope with relay optics would become very similar to that achieved without relay optics and without a monochromator. We have found it difficult to use a monochromator on our current TEM (see section 10.1), and plan to instead use a cold field emission gun in future work (see chapter 13)

# Chapter 12

# Phase contrast imaging data

The current LPP prototype which is capable of achieving a $\pi/2$ phase shift for a 300 keV electron beam and is designed to minimize the resolution loss from thermal magnetic field fluctuations (see chapter 11) was installed in our TEM in September 2022. Since then (time of writing is December 2023), we have focused on demonstrating the capabilities of the LPP as applied to cryo-EM single-particle analysis (SPA). We believe that the LPP will also be useful in cryo-electron tomography (cryo-ET). However, we have focused our initial efforts on SPA for several reasons. First, SPA is a well-established technique with more accessible and reproducible sample preparation methods. This means we spent less time (though still plenty!) worrying about the quality and/or consistency of the sample preparation, especially when comparing our results which use the LPP to those that don't. Second, our TEM is not equipped with an imaging energy filter—this substantially reduces the contrast-to-noise ratio in tomographic images of thick ($> 100$ nm) samples, which precludes high-quality results for all but the thinnest samples. Third, our TEM sample stage uses previous-generation "side-entry" sample holders, which have less positional stability during tomographic tilting than modern integrated sample holders. This is not an insurmountable issue but likely requires some amount of additional work to develop an LPP-compatible automated tomographic data collection sequence which can compensate for relatively large stage position drifts and backlash. As is discussed in chapter 13, our group will receive a new TEM in 2024 which has both an imaging energy filter and modern sample stage, among other features—we plan to use this instrument primarily for LPP cryo-ET.

This chapter presents our most recent imaging data which quantifies the performance of the LPP using cryo-EM samples. All data was collected using an electron beam kinetic energy of 300 keV.

## 12.1 Phase shift stability

An important advantage of the LPP is that it can provide a stable $\pi/2$ phase shift which does not change from image to image. We have demonstrated that this is indeed the case using SPA samples and the automated data collection procedure described in section 10.4. Figure 12.1a shows the phase shift of every image taken during a data collection session using an apoferritin SPA sample, plotted as a function of time. 217 images were taken over the course of 6.86 hours, i.e. a throughput of 31.6 images per hour. The phase shift for each image was determined by fitting the Thon rings in the image's FFT using CTFFIND4 [116]. Fitted phase shift values well in excess of 90° are due to fitting errors, usually because of the presence of sample contaminants in the image. The mean phase shift for the entire dataset is 91.8° with a standard deviation of 11.8°. The slow downward drift in the phase shift values over time is because the circulating power feedback system (see section 6.2.2) was not active, and the cavity's input coupling efficiency (and so, circulating power) was slowly dropping. Figure 12.1b shows the change in longitudinal position deflector setting applied between each image to keep the unscattered wave aligned to an antinode of the LPP's standing wave. The deflector setting units have been calibrated to real position units by measuring the change in Ronchigram fringe position as a function of deflector setting, noting that the Ronchigram fringe period is $\lambda_{\mathrm{L}}/2 = 532\,\mathrm{nm}$. This data shows that 37% of the time, it is not necessary (within the tolerance limits discussed in section 10.4) to realign the unscattered wave position between images. The slight positive skew to the data is due to a slow overall longitudinal motion of the LPP relative to the electron beam due to thermal expansion of the cavity support arm. The several realignments of $\sim -532\,\mathrm{nm}$ occur when the system must realign to an adjacent antinode to keep from following a single antinode for more than 532 nm in one direction. This keeps the electron beam from being deflected too far from the optical axis during the course of a data collection session.

Figure 12.2 shows similar data, except that the SPA sample particle was RuBisCO, and the circulating power feedback system was active. 71 images were taken in 1.96 hours (36.2 images per hour). In this case, the mean phase shift was 89.0° with a standard deviation of 4.9°. We believe that this remaining phase shift spread is mostly due to shot noise in the image which limits the accuracy of the Thon ring fitting algorithm—simulated SPA images with shot noise and a fixed phase shift of 90° exhibit a similar spread in fitted phase shift values. If true, this means that further improvements to the phase shift stability of the LPP are unnecessary.

These results represent a significant milestone in the development of TEM phase plates in general: the first realization of a stable phase plate.

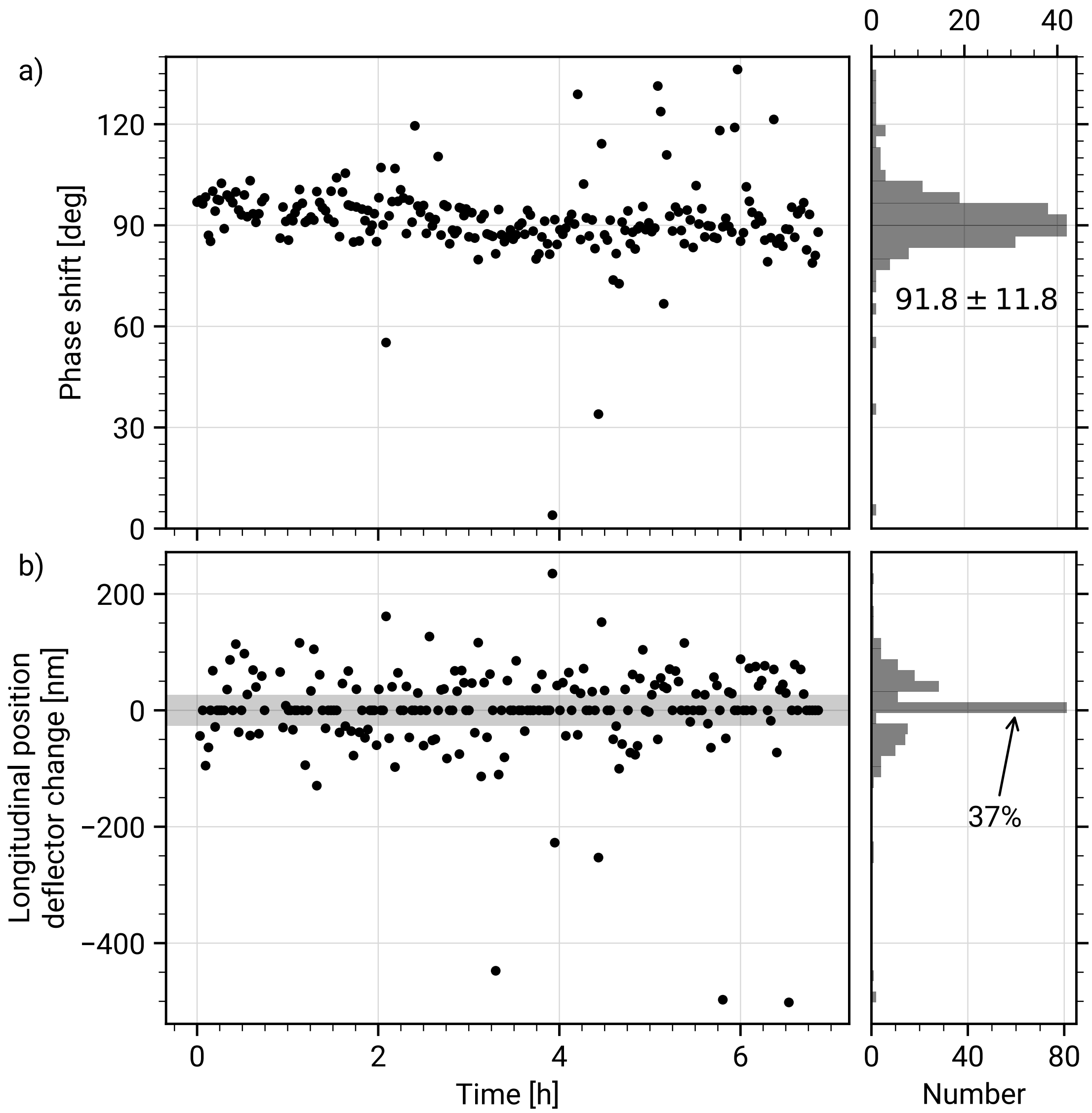

Figure 12.1: **a)** Fitted phase shift values of each image collected in a laser phase plate single-particle analysis dataset of apoferritin, as a function of time. The right panel shows a histogram of the values in the left panel, with a mean phase shift of 91.8° and a standard deviation of 11.8°. **b)** Change between images in the longitudinal position deflector required to maintain alignment of the unscattered wave to an antinode of the laser phase plate's standing wave. The range in which the automated alignment system does not command a realignment is shown as a gray shaded region in the left panel (±26.6 nm). 37% of the time a realignment was not required.

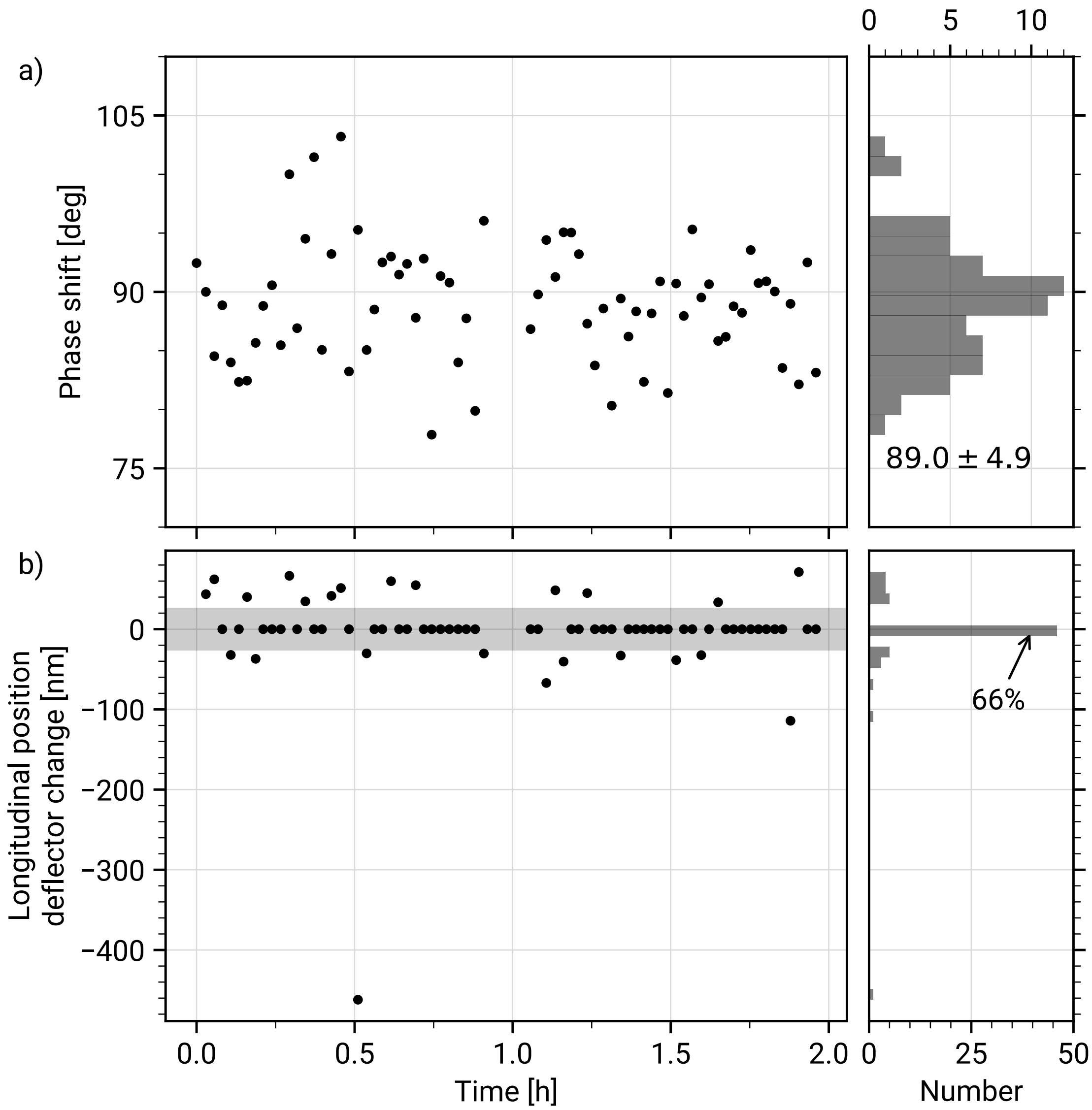

Figure 12.2: **a)** Fitted phase shift values of each image collected in a laser phase plate single-particle analysis dataset of RuBisCO, as a function of time. The right panel shows a histogram of the values in the left panel, with a mean phase shift of 89.0° and a standard deviation of 4.9°. **b)** Change between images in the longitudinal position deflector required to maintain alignment of the unscattered wave to an antinode of the laser phase plate's standing wave. The range in which the automated alignment system does not command a realignment is shown as a gray shaded region in the left panel (±26.6 nm) 66% of the time a realignment was not required.

## 12.2 Laser phase plate single-particle analysis

### 12.2.1 RuBisCO

To evaluate the advantages of using an LPP in SPA, we have performed a reconstruction of the structure of RuBisCO both with and without the LPP. Each dataset ("LPP off" and "LPP on") was taken using the same sample grid on the same day. The "LPP on" dataset was taken first, and no realignment of the TEM was performed between the datasets. We use RuBisCO for this experiment because its high resolution structure is known [136], [137] and our team has prior experience making consistent sample grids using this protein.

A comparison between representative images with the LPP on and off are shown in figure 12.3, along with their FFTs. The "LPP on" image comes from the dataset shown in figure 12.2. The grayscales of both images are the same, and extend from 0.8 to 1.2, where the mean pixel value of the image is normalized to unity. The total electron dose (at the camera) in each image is 43 electron $\text{Å}^{-2}$. The LPP clearly generates substantially more contrast than defocus alone: the "LPP off" image was taken with a defocus of 1.2 µm, while the "LPP on" image was taken with a defocus of 700 nm. The phase shift of the LPP image is 92°, as determined by Thon ring fitting using CTFFIND4. The standing wave features of the LPP are visible in the FFT of the image, where the axis of the laser beam runs between the upper-left and the lower-right of the FFT. This gives the Thon rings a slightly oblate appearance, even in the absence of astigmatism. The image FFTs make it clear that the additional contrast in the LPP image comes from the low spatial frequencies. The concentric rings in the FFT near $\sim 0.015\,\text{Å}^{-1}$ are generated by the tertiary and quaternary structure of the protein, and represent a large component of the contrast enhancement. The "LPP on" image also contains areas with a noticeably brighter (lower-right) and darker (upper-left) background surrounding the particles. We are not yet certain why this is the case, but we believe it may be phase contrast enhancement of inhomogeneity in the vitreous ice thickness or composition in the sample. Bright halos immediately surrounding each particle are a well-known consequence of any phase contrast transfer function [138].

Figure 12.4 shows a random selection of "LPP on" images from the dataset (the same dataset shown in figure 12.2). Similar features are present in these images as to the one shown in figure 12.3. Substantial variability in the sample from image to image can be easily seen with the additional contrast provided by the LPP. Strong "ghost images" (see section 8.8) from a contaminant (dark blob) are evident in the image in the second row, third column. Note that the axis of displacement of the ghost images is aligned with the axis of the laser beam in the FFT of the image (upper-left to lower-right).

Reconstructions were performed using a combination of RELION v4.0 and CryoSPARC v4.4.0 using the "LPP on" and "LPP off" datasets, yielding high-resolution ($\sim 3\,\text{Å}$) structures in both cases. A comparison of the reconstruction results is shown in figure 12.5. Panels a and b show a comparison of the reconstructed density map of a section of the protein including a docked atomic model of the known structure (dark inlaid lines). The Fourier shell correlation (FSC) [139] of each reconstruction is shown in panel c. Both reconstructions used the same number of particles (16 000), and so the resolution of the reconstruction (where the FSC crosses 0.143 [140]) should be comparable between the datasets. Interestingly, the "LPP on" dataset shows a slightly lower resolution than the "LPP off" dataset. This effect is further explored in panel d, which shows a "ResLog" plot of the (inverse-squared) reconstruction resolution as a function of the (natural log) of the number of particles used in the reconstruction. The slope of this curve is commonly used as an indicator of the high-resolution ($< 4$ Å) information content in the images [141]. The "$B$-factor" (two divided by the slope of the curve) is the most commonly quoted metric, where a smaller $B$-factor indicates that the images contain a larger high-resolution signal-to-noise ratio. Here, we see that the $B$-factor of the "LPP off" dataset is smaller than that of "LPP on". This indicates that there is either less high-resolution information content in the LPP images, and/or that the information that is present is not being properly incorporated into the reconstruction by the reconstruction algorithm. Since we have yet to identify a physical mechanism that should reduce the information content in an LPP image (see section 8.7), we favor the latter explanation. Indeed, we have seen preliminary indications that the enhanced low-frequency contrast in the LPP images may bias the "alignment" step of the reconstruction process, in which the relative orientations and centroid positions of individual particle images are determined. Inaccurate particle alignment is known to affect both the $B$-factor and $y$-intercept of the ResLog plot curve [141].

The SPA reconstruction process can also be used to quantify aberrations in the TEM's point-spread function (see section 8.2) [142]. These reconstructed (phase) aberrations are displayed in figure 12.6 as a function of spatial frequency for the "LPP on" (panel a) and "LPP off" (panel b) datasets, as well as for a RuBisCO dataset taken with the TEM's relay optics deactivated (panel c). The contributions of defocus and spherical aberration ($C_{1,0}$ and $C_{3,0}$, respectively, see section 8.3) are not included the aberration plots. The aberrations for all datasets are comparable to those of state-of-the-art cryo-EM reconstructions. The primary aberration in all cases is coma ($C_{2,\pm1}$), with an equivalent beam-tilt angle in each dataset of a) 0.093 mrad, b) 0.0096 mrad, and c) 0.16 mrad. This confirms that it is possible to align our custom TEM to the LPP while also minimizing unwanted phase aberrations. Phase aberrations from the LPP's standing wave are visible near the origin (laser axis runs from upper-left to lower-right) in the "LPP on" data, mostly manifesting around a spatial frequency of 0.015 $\text{Å}^{-1}$ where the tertiary structure of RuBisCO generates a stronger structure factor. Note that the "relay optics off" data extends to a higher spatial frequency before becoming noisy because the lower chromatic aberration coefficient of the TEM in that configuration results in a better reconstruction resolution.

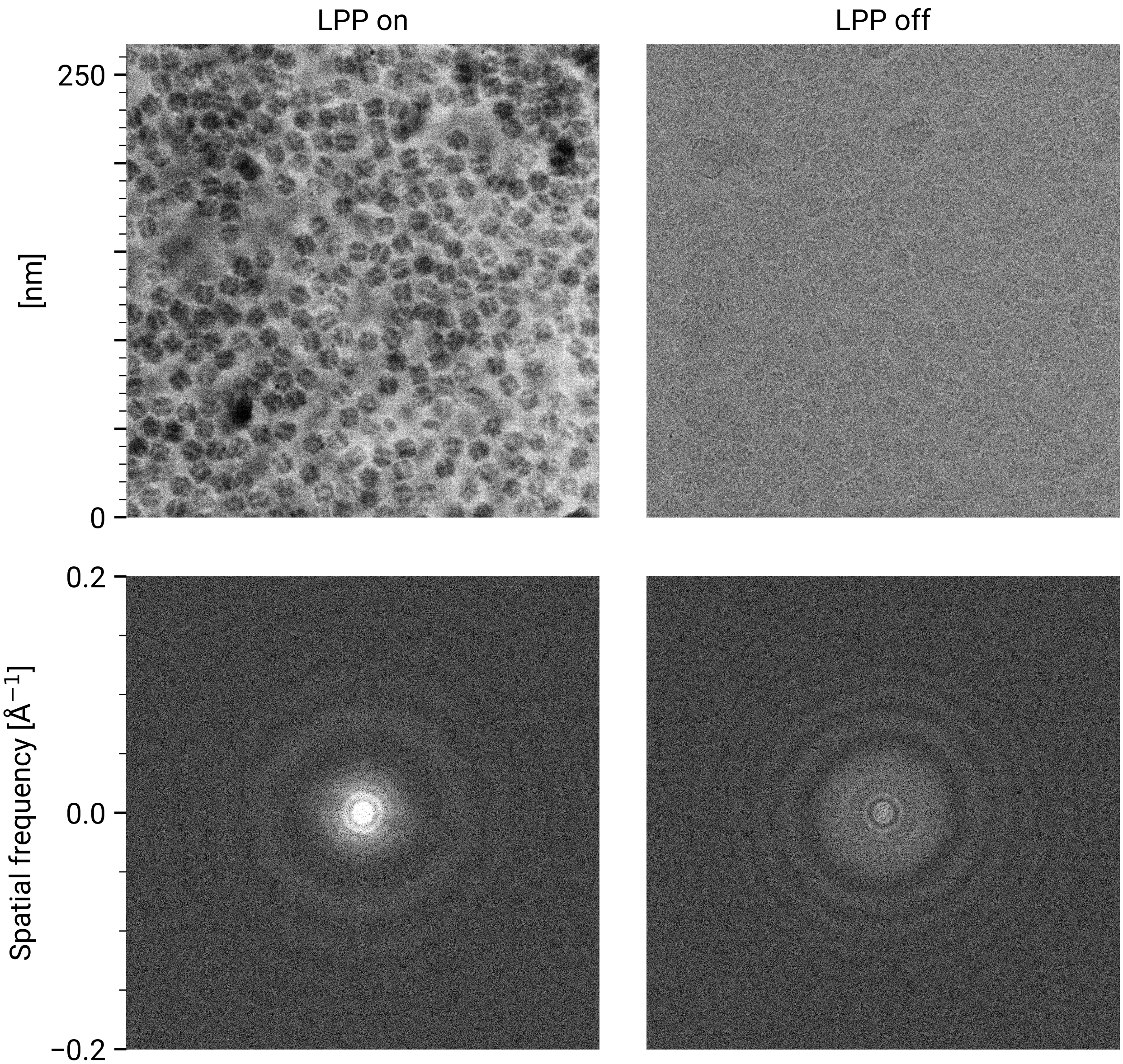

Figure 12.3: Images (top row) and their FFTs (bottom row) of a RuBisCO single-particle analysis sample with the laser phase plate (LPP) on (left column) and off (right column). The defocus of the LPP on image is 700 nm with a phase shift of 92°. The defocus of the "LPP off" image is 1.2 µm. The grayscale of both images extends from 0.8-1.2, where 1 is the mean pixel value of each image.

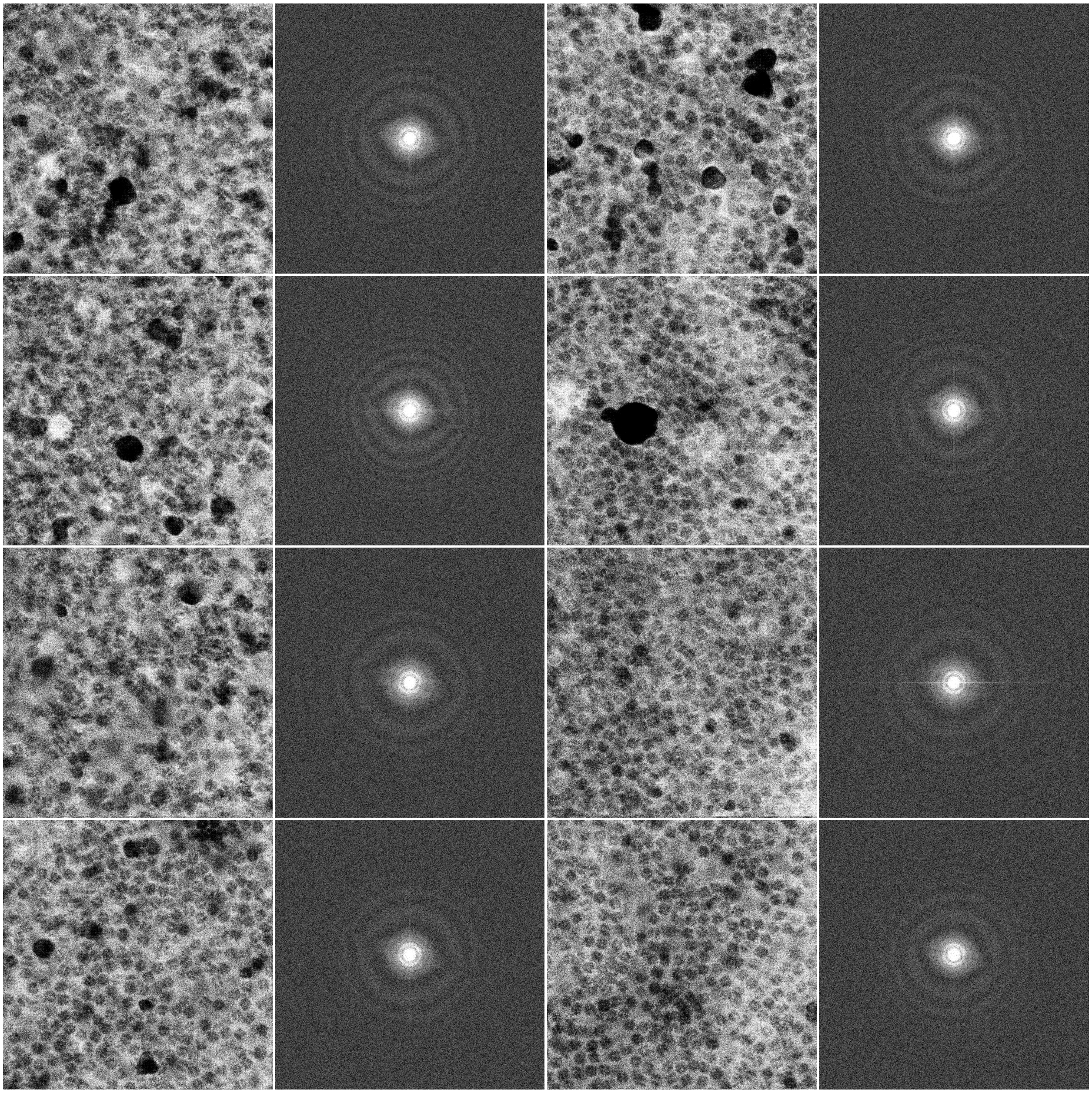

Figure 12.4: Randomly-selected images (left) and corresponding FFTs from the "LPP on" RuBisCO dataset. The width of each image is 267 nm, and the FFTs are shown between $\pm 0.2\,\text{Å}^{-1}$. The image grayscale extends from 0.8-1.2 where 1 is the mean pixel value of each image. The images are downsampled by a factor of 2 to reduce the file size without binning (which otherwise changes the apparent signal-to-noise ratio), and then rendered at 1.1 times that resolution to avoid aliasing. The total electron dose (at the camera) for each image is approximately 43 electron $\text{Å}^{-2}$, and the defocus of all images is $\sim 1\,\mu\text{m}$.

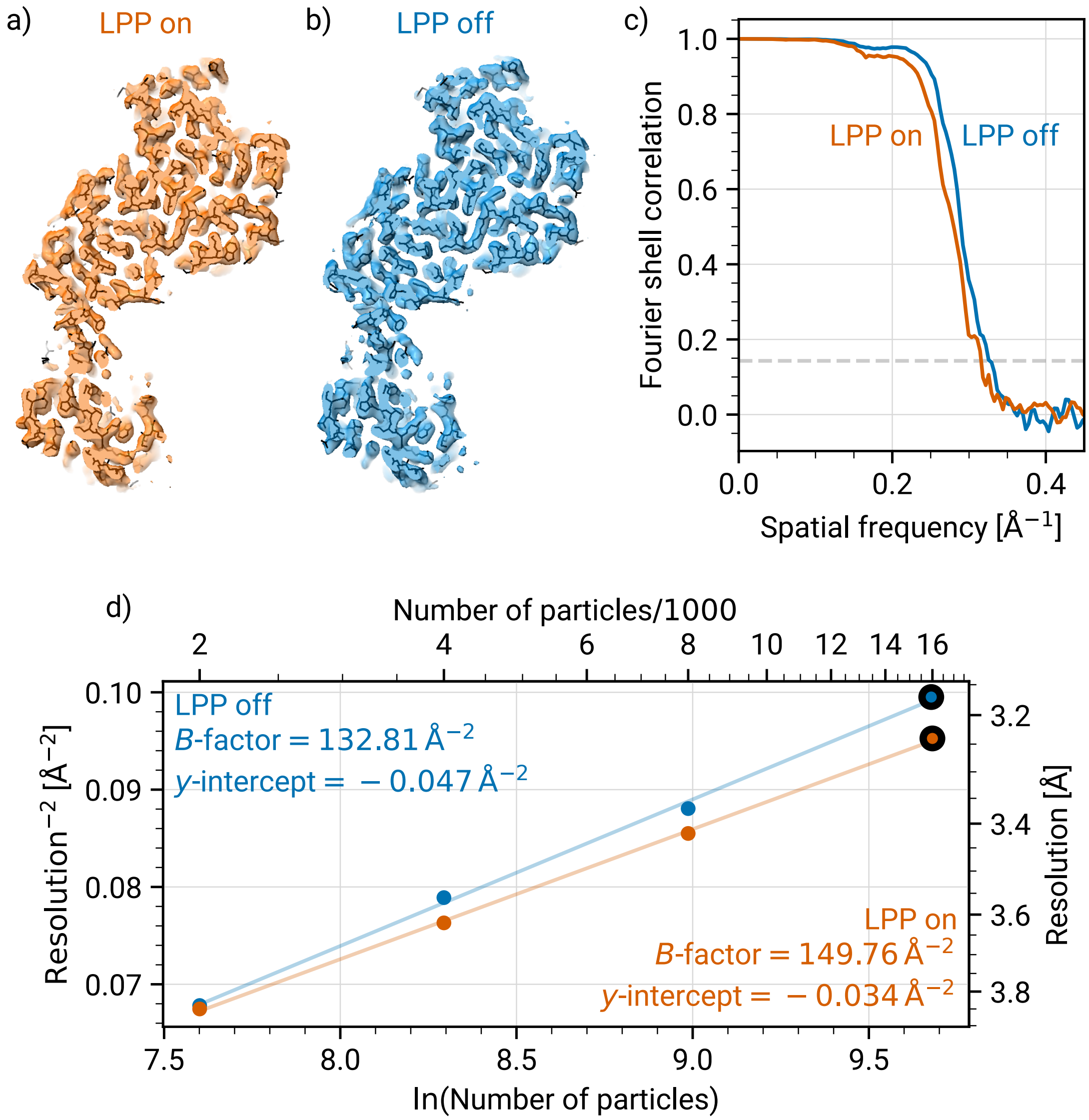

Figure 12.5: Comparison of RuBisCO single-particle analysis reconstructions when using ("LPP on") or not using ("LPP off") the laser phase plate (LPP). **a),b)** Reconstructed density maps of a section of RuBisCO including a docked atomic model of the known structure (dark inlaid lines). **c)** Fourier shell correlation of the two reconstructions. The dashed horizontal line indicates the 0.143 resolution threshold level. **d)** "ResLog" plots of the reconstruction resolution as a function of the number of particles included in the reconstruction. The resolution is defined as the inverse of the spatial frequency at which the Fourier shell correlation first drops below 0.143. The lines are least-squares best-fits, and their corresponding fit parameters are annotated in the upper left and lower right of the plot. The highlighted datapoints are the ones used for the reconstructions in panels a, b, and c.

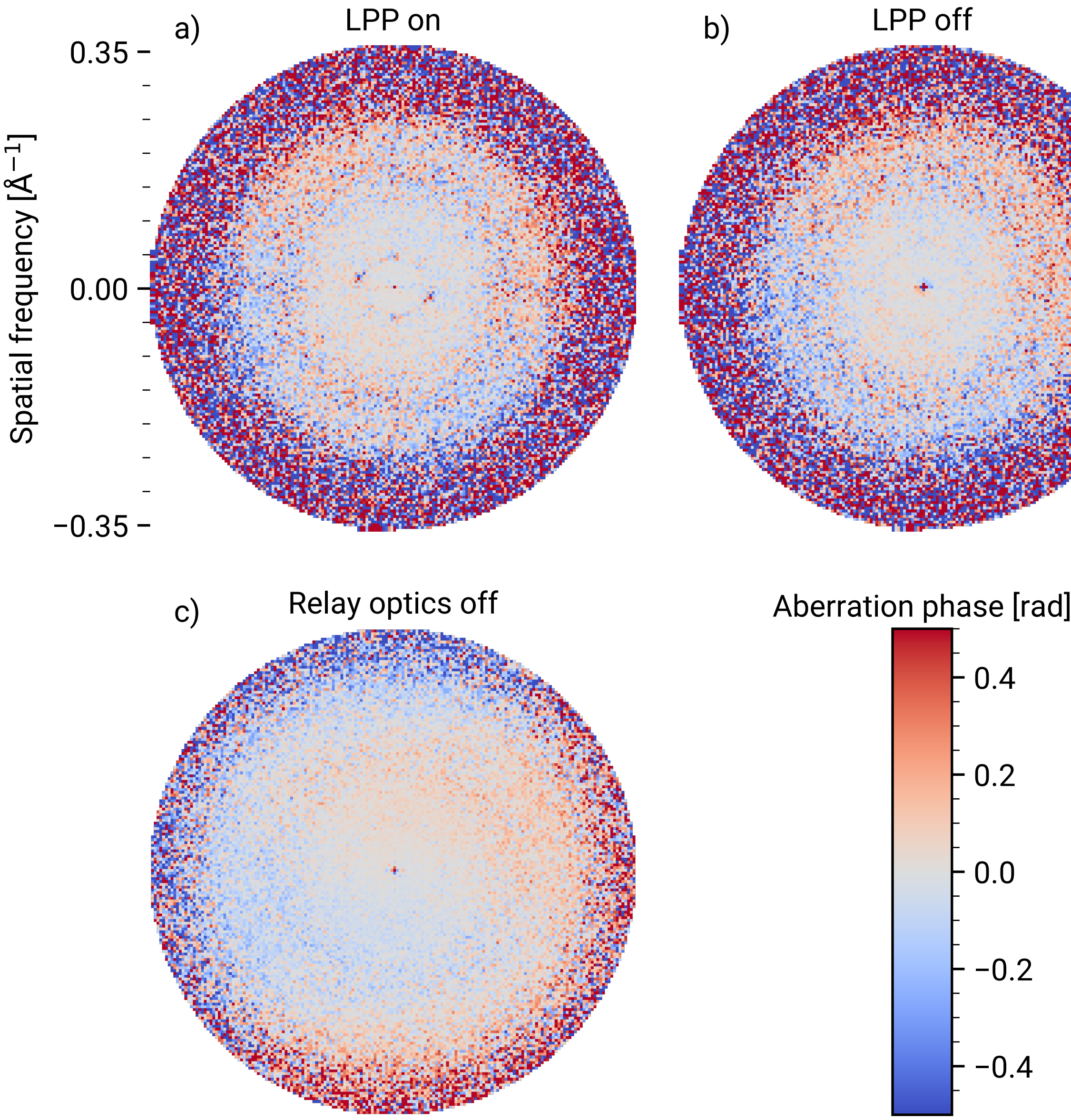

Figure 12.6: Phase aberrations of our transmission electron microscope as a function of spatial frequency, determined as part of the RuBisCO single-particle analysis reconstruction process. Defocus and spherical aberration are not included. Data is shown for reconstructions **a)** using the laser phase plate (LPP), **b)** not using the LPP, and **c)** not using the LPP and with the transmission electron microscope's relay optics turned off.

### 12.2.2 Apoferritin

We have also performed a similar comparison using apoferritin samples. In this case, the "LPP on" and "LPP off" datasets were collected on different days, with different sample grids. Minor realignment of the TEM was performed between the datasets.

A comparison between representative images with the LPP on and off are shown in figure 12.7, along with their FFTs. The "LPP on" image comes from the dataset shown in figure 12.1. The grayscales of both images are the same, and extend from 0.8 to 1.2, where the mean pixel value of the image is normalized to unity. The total electron dose (at the camera) in each image is approximately 30 electron $\text{Å}^{-2}$. The "LPP off" image was taken with a defocus of 755 nm, while the "LPP on" image was taken with a defocus of 638 nm. The phase shift of the LPP image is 96°, as determined by Thon ring fitting using CTFFIND4.

Figure 12.8 shows a random selection of "LPP on" images from the dataset (the same dataset shown in figure 12.1). Like the RuBisCO data, there are sometimes bright and dark background regions. The image in the first row, third column is an example of when the unscattered wave was not properly aligned to a standing wave antinode of the LPP, resulting in a phase shift of only 55°.

Reconstructions were performed in a combination of RELION v4.0 and CryoSPARC v4.4.0 using the "LPP on" and "LPP off" datasets, yielding high-resolution ($\sim 3\,\text{Å}$) structures in both cases. A comparison of the reconstruction results is shown in figure 12.9. Panel a shows the Fourier shell correlations of each reconstruction. They are of comparable resolution, but the "LPP on" reconstruction used more particles—this is illustrated in the ResLog plot in panel b. Like the RuBisCO datasets, the "LPP on" reconstruction has a consistently worse resolution for the same number of particles. In this case, the $B$-factors are nearly the same, but the $y$-intercept of the "LPP on" ResLog data is significantly lower. We do not know if these differences are spurious or are due to differences in the alignment step of the reconstruction process which might arise from the different symmetries of the structures of RuBisCO and apoferritin.

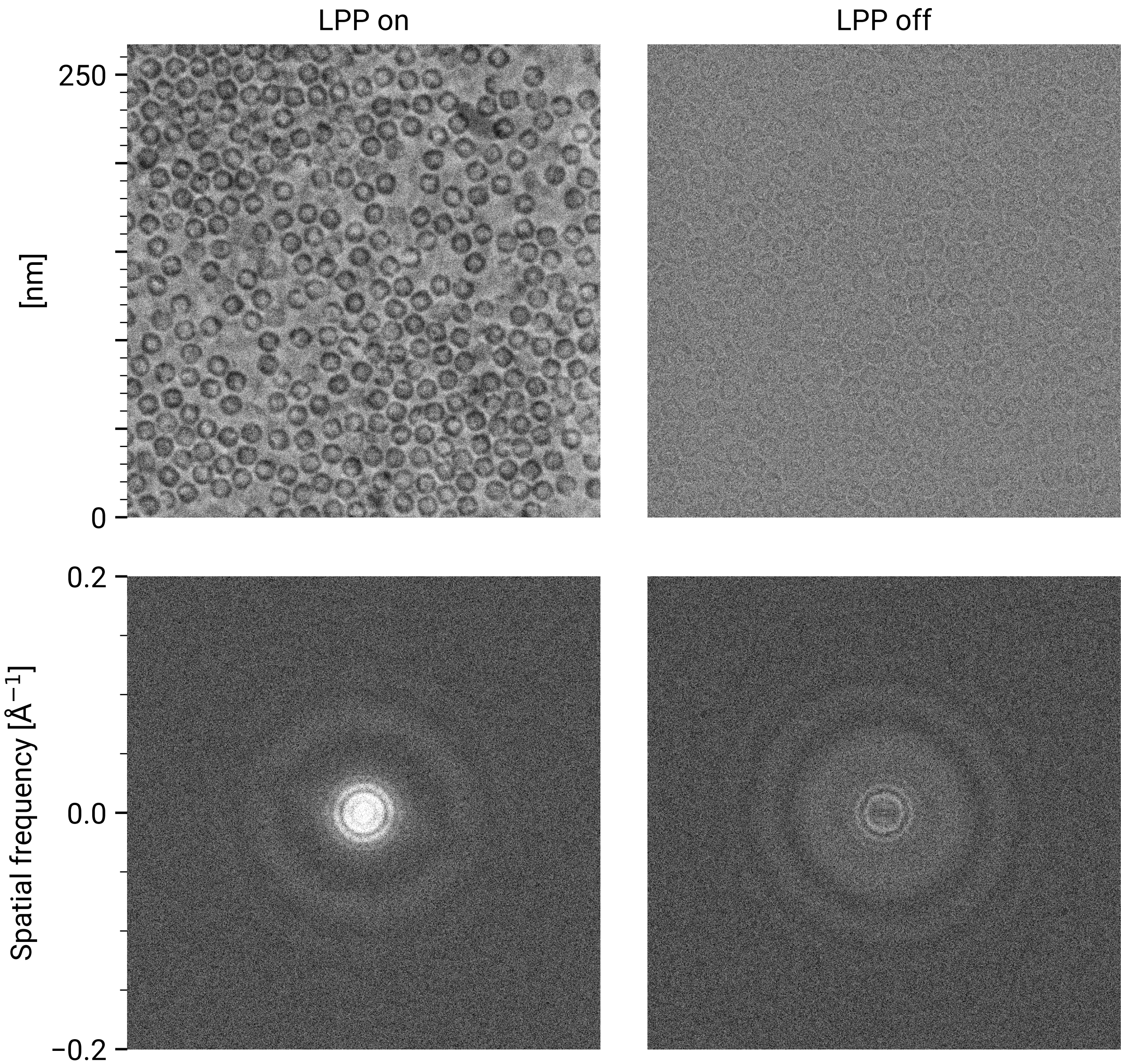

Figure 12.7: Images (top row) and their FFTs (bottom row) of an apoferritin single-particle analysis sample with the laser phase plate (LPP) on (left column) and off (right column). The defocus of the LPP on image is 638 nm with a phase shift of 96°. The defocus of the "LPP off" image is 755 nm. The grayscale of both images extends from 0.8-1.2, where 1 is the mean pixel value of each image.

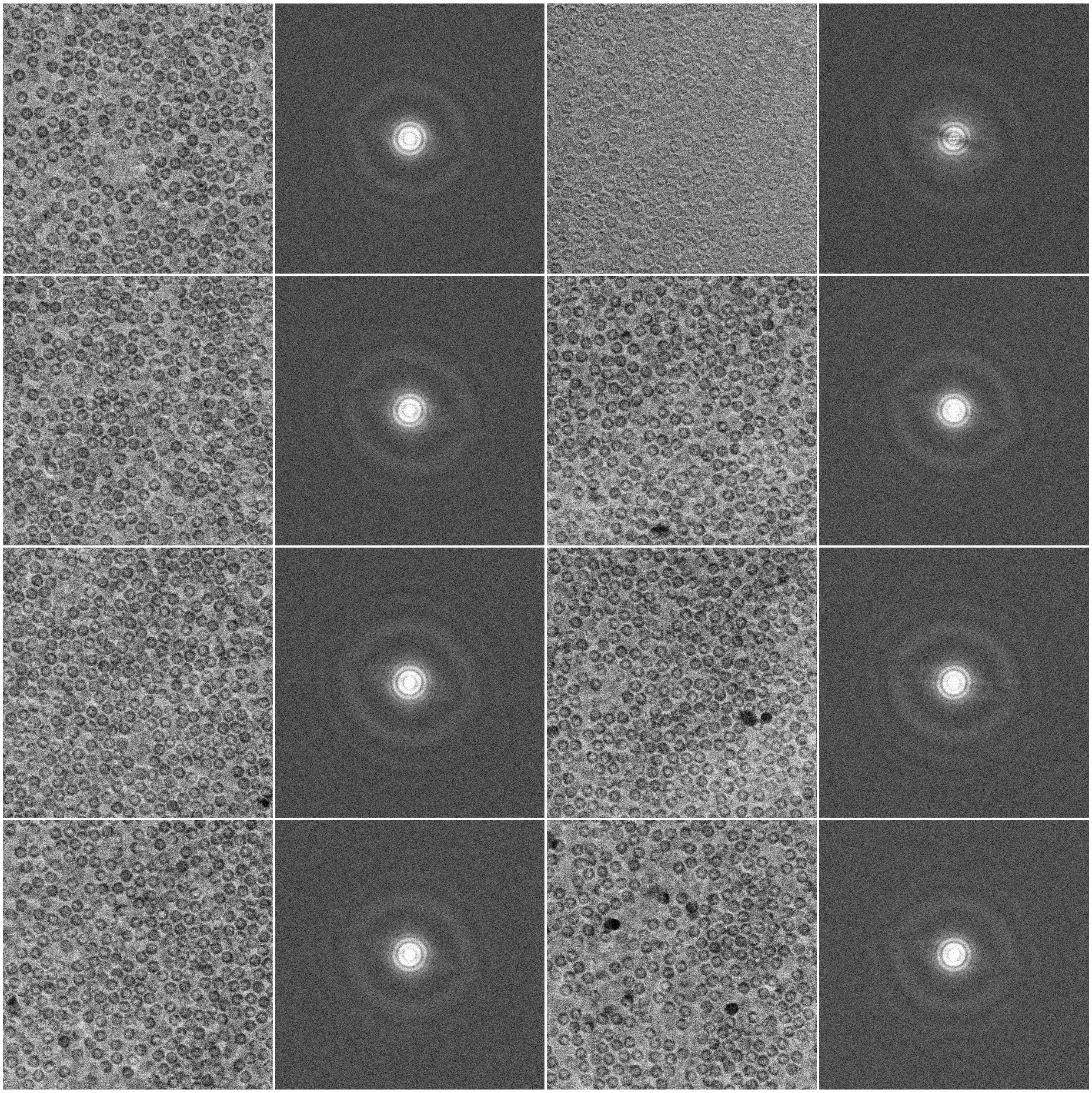

Figure 12.8: Randomly-selected images (left) and corresponding FFTs from the "LPP on" apoferritin dataset. The width of each image is 267 nm, and the FFTs are shown between $\pm 0.2\,\text{Å}^{-1}$. The image grayscale extends from 0.8-1.2 where 1 is the mean pixel value of each image. The images are downsampled by a factor of 2 to reduce the file size without binning (which otherwise changes the apparent signal-to-noise ratio), and then rendered at 1.1 times that resolution to avoid aliasing. The total electron dose (at the camera) for each image is approximately 31 electron $\text{Å}^{-2}$, and the defocus of all images is roughly 650 nm.

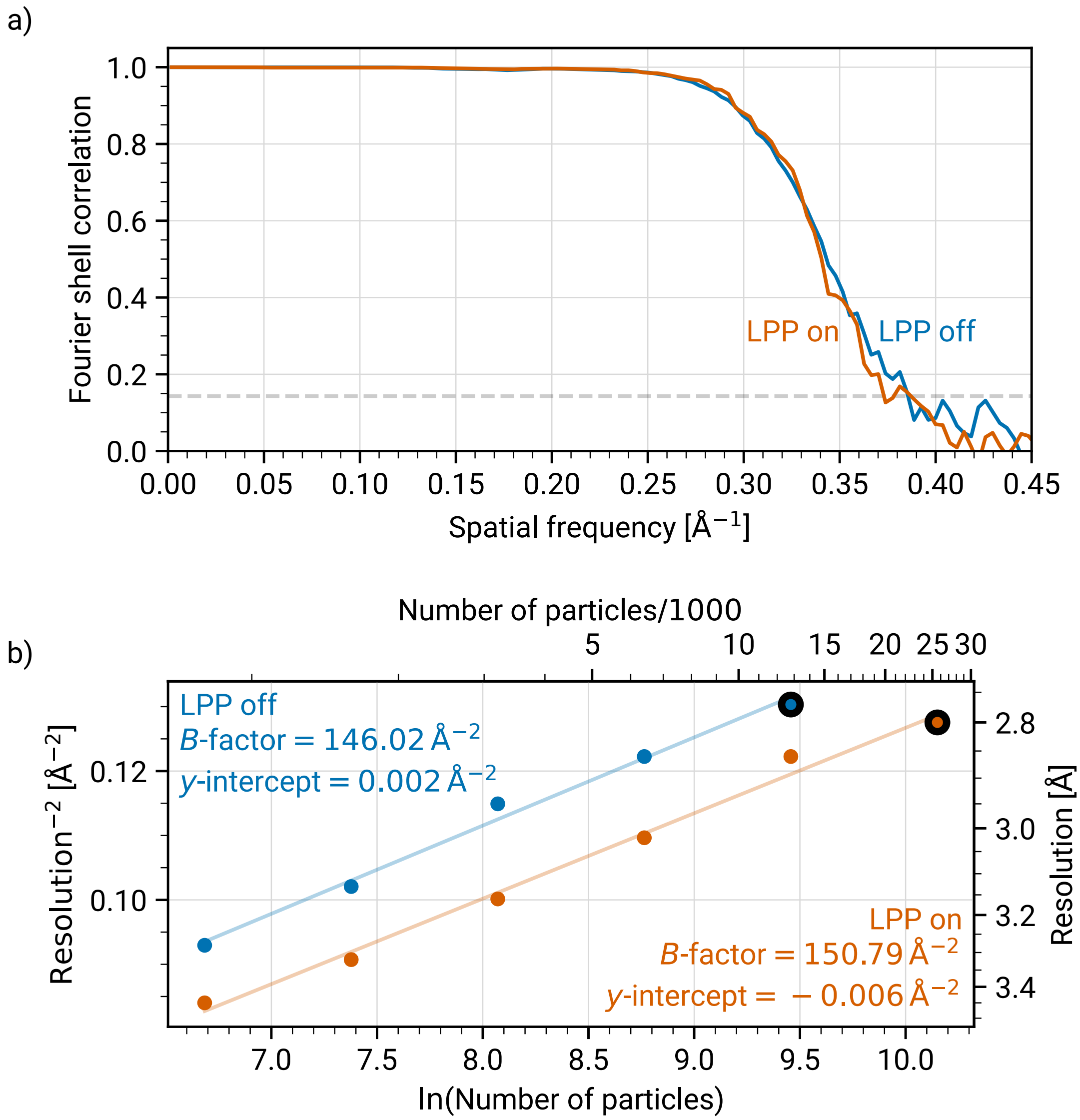

Figure 12.9: Comparison of apoferritin single-particle analysis reconstructions when using ("LPP on") or not using ("LPP off") the laser phase plate (LPP). **a)** Fourier shell correlation of the two reconstructions. The dashed horizontal line indicates the 0.143 resolution threshold level. **b)** "ResLog" plots of the reconstruction resolution as a function of the number of particles included in the reconstruction. The resolution is defined as the inverse of the spatial frequency at which the Fourier shell correlation first drops below 0.143. The lines are least-squares best-fits, and their corresponding fit parameters are annotated in the upper left and lower right of the plot. The highlighted datapoints are the ones used for the reconstructions shown in panel a.

### 12.2.3 Discussion

The fact that the LPP does not improve the resolution of these reconstructions is in hindsight perhaps not surprising. Previous work done using the Volta phase plate (VPP) showed that use of a VPP did not substantially improve, and even degraded, the resolution of SPA reconstructions [11], [12], [143]. Our results may indicate that the phase shift instability and inherent resolution loss [12], [13] of the Volta phase plate is not the only reasons that the VPP does not necessarily lead to better SPA reconstructions. Even though the LPP images show spectacularly higher contrast, the fact that the LPP images were taken at non-zero defocus means the enhancement is only present for low ($< 0.05\,\text{Å}^{-1}$) spatial frequencies which do not directly contribute to the high-resolution reconstruction. Our efforts to use the LPP to increase high-frequency signal in the image by imaging in-focus are discussed in section 12.3. However, the question of why the "LPP on" reconstructions have consistently worse resolution is still open, and answering it may lead to a better understanding of how contrast at different spatial frequencies affects SPA reconstructions.

## 12.3 In-focus phase contrast imaging

In-focus imaging with a phase plate maximizes image contrast by generating a contrast transfer function (CTF) which has nearly unit magnitude for all spatial frequencies. This increases the information content of the image by roughly a factor of 2 relative to a defocused image for spatial frequencies at which the CTF oscillates due to defocus (see section 8.5).

In practice, in-focus imaging in cryo-EM is challenging, since it requires precise autofocusing of the target area without exposing the target area to the electron beam. As is discussed in section 8.5, the defocus must be set to within a range of roughly 10 nm about its optimal value for the CTF to remain above 0.5 for spatial frequencies up to $\sim \left(3\,\text{Å}\right)^{-1}$. Normal out-of-focus autofocusing is typically performed by imaging a "focus" area adjacent to the target area in order to set the defocus. In this scheme, any variation in sample height between the focus and target areas introduces error in the defocus at the target area. In practice, cryo-EM samples may have height variations on the order of $\sim 100$ nm over a lateral distance of $\sim 1$ µm, which is a typical distance between the target and focus areas. Since this variation is too large for consistent in-focus imaging, Danev et al. developed an alternative scheme for in-focus imaging using the VPP in which autofocusing is performed in two focus areas on opposite sides of the target area, and the average focus setting is then used to image the target area [11]. This compensates for linear gradients in the sample height. In addition, some types of sample grid material may produce flatter samples: the primary cause of height variation in SPA cryo-EM samples is presumed to be from crinkling due to differential thermal contraction of the grid material and the holey film when the sample is vitrified [144]. This effect is reduced for grids which use materials with similar coefficients of thermal expansion for the holey foil and support mesh (e.g. gold/gold or amorphous carbon/molybdenum) [124]. Doming of the target area caused by exposure to the electron beam may cause the defocus to change by as much as 25 nm during the course of a typical cryo-EM image exposure [101], [145], which also limits the accuracy of in-focus imaging. Doming can be reduced by using thicker holey foil [124], [146] and reducing the diameter of the foil holes [125].

It can also be difficult to accurately measure the defocus even in the focus areas. The commonly-used beam-tilt method measures the defocus by measuring the shift in image position induced by changing the tilt angle of the electron beam in the sample plane by an amount $\theta$ (equivalently, by shifting the position of the unscattered wave in the diffraction plane). The resulting shift in image position (referred to the object plane) is $-Z\theta + C_\mathrm{s}\theta^3$ where $Z$ is the defocus and $C_\mathrm{s}$ is the coefficient of spherical aberration [11]. Accurately measuring the defocus to within 10 nm therefore requires that the image shift be measured to within 1 Å when a standard beam tilt change of 10 mrad is used. This is comparable to the usual object plane-referred pixel size used when imaging in the focus area, and may be near the lower limit of the shift that can be accurately measured in the presence of shot noise.

It would also be useful to have a way to verify that images recorded in the target area do indeed have the desired value of defocus. Again, this is not straightforward, since the most common method of measuring the image defocus relies on fitting Thon rings—in an in-focus image, Thon rings are only present for spatial frequencies high enough for spherical aberration to dominate ($> \left(3\,\text{Å}\right)^{-1}$). The signal-to-noise ratio at these frequencies in a single cryo-EM image is too low to accurately fit the Thon rings at all, and certainly not with the required accuracy. Instead, we are looking into two-dimensional template matching [147] as an option for determining the defocus of particles within an image. This would require that particles with a known structure be included in the sample. An alternative would be to record the image in the target area, change the focus setting of the TEM to increase the defocus, and then record a second image in the (now radiation damaged) target area. Fitting the Thon rings of the second image and subtracting the known focus setting change could be used to determine the defocus of the first image. This relies on both accurate Thon ring fitting and a reproducible and measurable defocus offset, the latter of which may be difficult due to magnetic lens hysteresis effects (see section 10.3.5).

Furthermore, SPA samples are typically thicker than 10 nm, meaning that even with perfect autofocusing some (perhaps even most) of the proteins will not lie within the in-focus depth of field. We have not yet explored what effects these slightly out-of-focus proteins may have on the reconstruction process.

We have run some initial tests of in-focus imaging methods on an SPA sample (apoferritin), using a method which autofocuses on opposite sides of the target area, similar to [11]. This did result in a successful reconstruction in which we assumed that the defocus of all images was 0. However, the Fourier shell correlation of the reconstruction contained a conspicuous dip around a spatial frequency of $0.16\,\text{Å}^{-1}$ (see figure 12.10). We are not sure why this feature is present, but hypothesize that it is caused by a zero-crossing of the CTF due to the images actually being recorded at a non-zero defocus (but still too small to generate noticeable or fittable Thon rings). To illustrate this hypothesis, the gray line in figure 12.10 shows the squared magnitude of the CTF, assuming a defocus of $-100$ nm (overfocus) and a phase shift of 90°. However, we have not been able to reproduce this result, and we are still working on developing a method for consistently recording in-focus images.

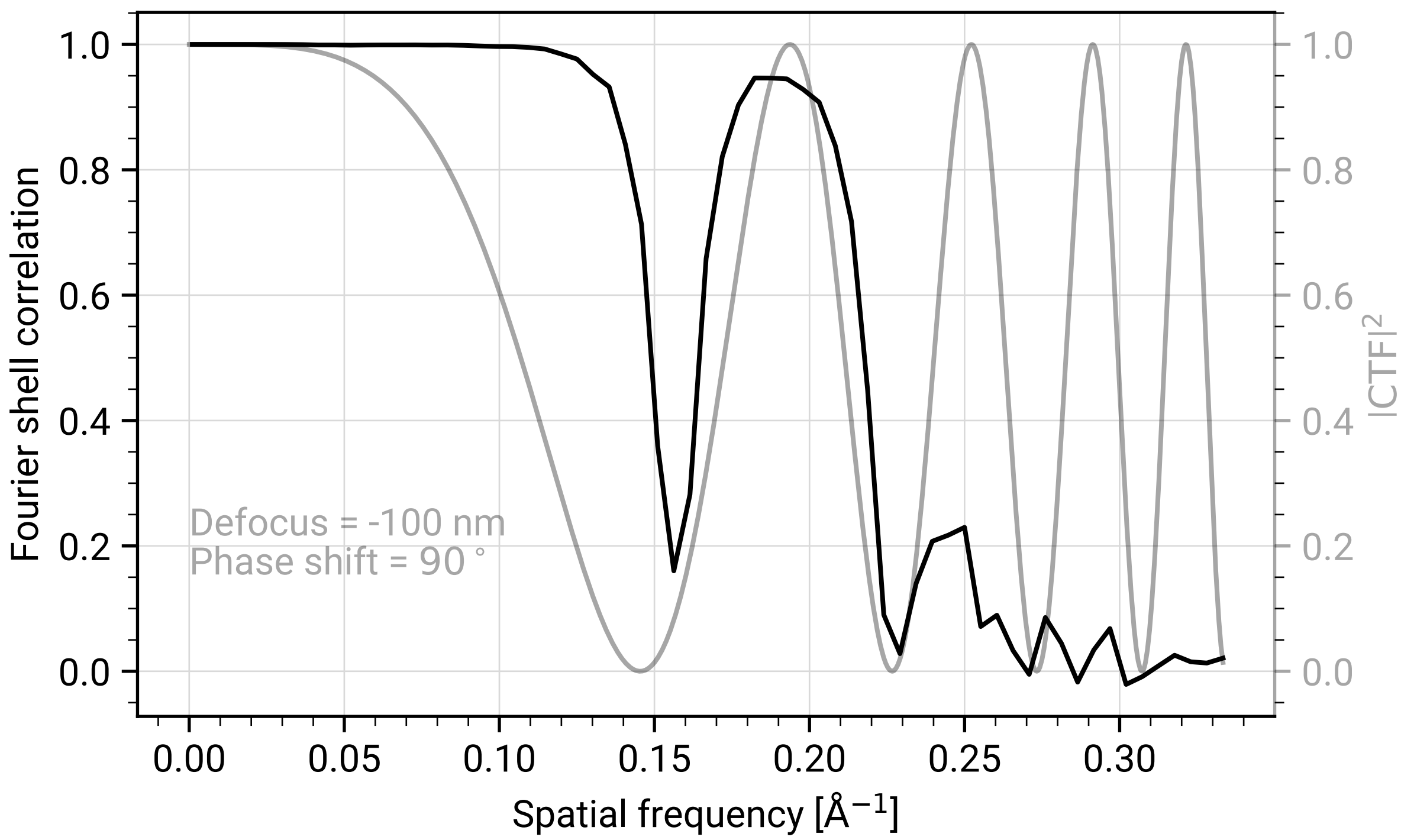

Figure 12.10: Black line: Fourier shell correlation (FSC) of the reconstruction of apoferritin using images taken (nominally) in-focus with the laser phase plate. Gray line: squared magnitude of a theoretical contrast transfer function (CTF) with a defocus of $-100$ nm (overfocus) and a phase shift of 90°. The location of the dips in the FSC roughly correspond to the locations of zeros of the CTF.

## 12.4 Low molecular mass samples

One application where the LPP might be expected to give a substantial advantage is in SPA of low molecular mass particles. In particular, isolated proteins with a mass $< 50$ kDa are difficult to identify at all in defocus SPA images. If particles cannot be identified, then the SPA reconstruction process cannot start to align and average particle images, and a three-dimensional reconstruction cannot be generated. Using the large defocus required to identify the locations of the particles reduces the resolution of the images (see sections 8.7.1.2 and 8.7.2).

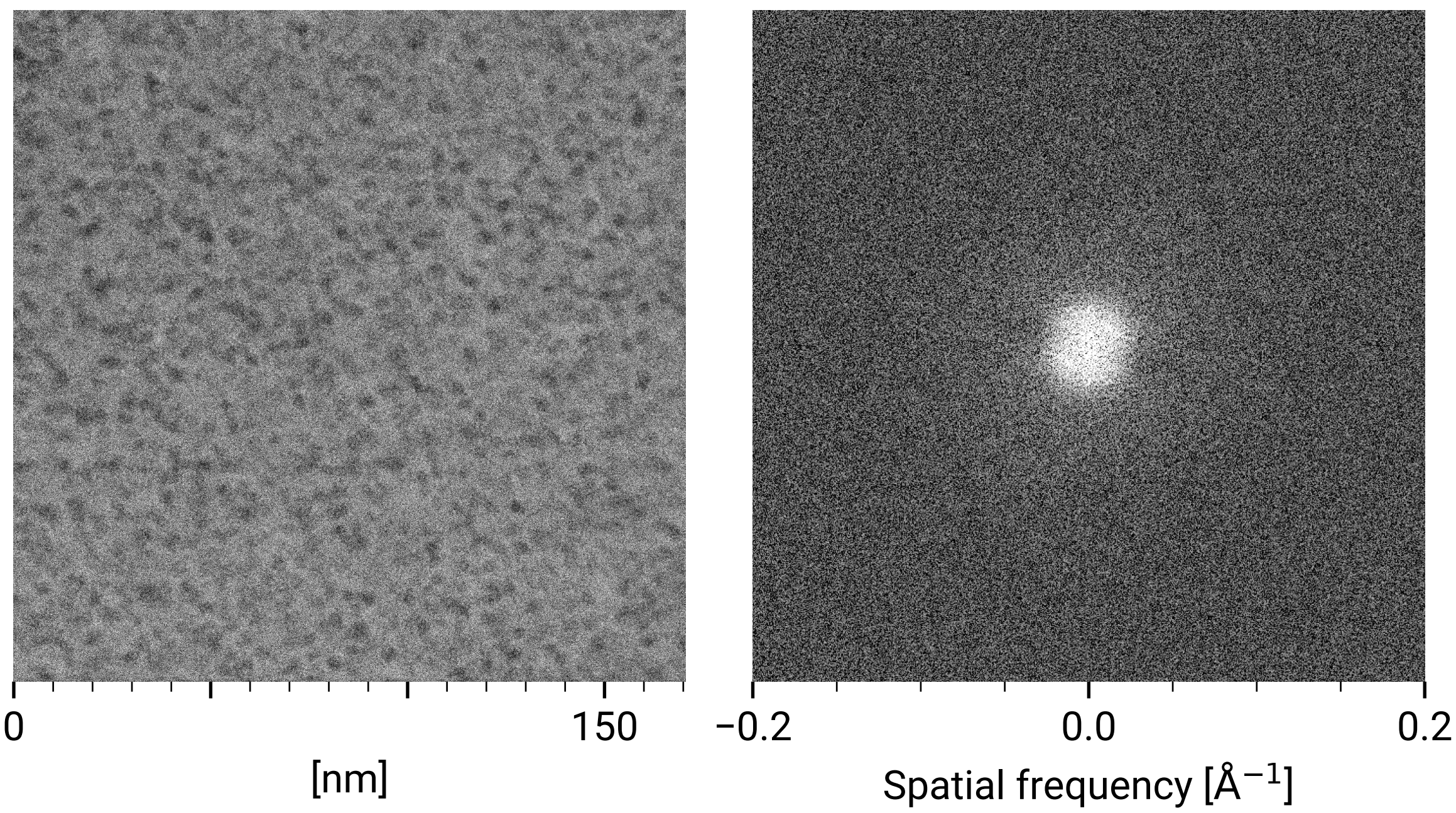

Figure 12.11: Left panel: in-focus laser phase plate image of myoglobin (17 kDa) in a single-particle analysis sample. The image grayscale extends from 0.8-1.2 where 1 is the mean pixel value. Right panel: FFT of the image.

This size class of proteins contains roughly half of the human proteome [38], [39], and in particular includes many membrane proteins, the structure of which are often of interest in pharmaceutical research [34], [38], [40]. While the structure of some smaller proteins may be determined via X-ray crystallography, there is still a large set of small proteins that remain to be reconstructed.

Theoretical estimates of the smallest protein that may be reconstructed using SPA (based on an estimate of the maximum achievable signal-to-noise ratio in perfect phase contrast images) include 14 kDa [148], 17 kDa [149], and 38 kDa [150], with some continuing debate [38], [151]–[153]. So far, the smallest isolated proteins that have been reconstructed to near-atomic ($< 4$ Å) resolution have molecular masses of $\sim 50$ kDa [37].

Following in the footsteps of Khoshouei et al. [154], we have made an initial attempt to reconstruct myoglobin (17 kDa) using the LPP. As shown in figure 12.11, individual particles are readily identifiable in an in-focus LPP image. However, similarly to [154], our reconstructions have failed to yield a structure which resolves any of the secondary structure of the protein. It appears that there is insufficient contrast at the spatial frequencies of the secondary structure ($\sim \left(5\,\text{Å}\right)^{-1}$) to angularly align the particles during the reconstruction process. This issue is present in both nominally in-focus (likely only near-focus, see section 12.3) and defocused LPP datasets. In principle, the even lower-resolution overall shape of the protein can contribute to the accuracy of angular alignments. However, this will only be the case if that structure has no rotational symmetry. Most small proteins are roughly spherical in shape, and so the lowest frequency structural information provides no cues for angular alignment. We are also working to mathematically formalize this hypothesis, and perhaps identify small proteins or other biological targets that can be aligned solely on the basis of their low-resolution structures.

## 12.5 "Superfractionation"

Modern cryo-EM images are generated by recording a movie of the sample during the exposure time. This movie captures the motion of the sample induced by the electron beam—the individual frames of the movie are then registered and summed to form the final image. This process of "motion correction" [134] of a "fractionated" image (i.e. movie) greatly improves the resolution of the final image by removing much of the motion blur resulting from beam-induced motion of the sample. The cumulative amount of beam-induced motion increases with the electron dose at the sample [101], so the electron dose per area per movie frame is the relevant parameter.

However, motion correction is limited by the signal-to-noise ratio of each individual movie frame: if there is not sufficient signal in the frame, the position of objects in the field of view cannot be accurately determined. In this case, higher electron doses per frame must be used (or multiple successive frames summed together without motion correction), which results in more motion blur.

The enhanced contrast provided by the LPP can be used to align movie frames with a lower electron dose per frame, which should reduce resolution loss from beam-induced motion blurring. We call this approach "superfractionation". Even in defocused LPP images, the enhanced low-frequency contrast can be used to register images to a higher precision than the maximum spatial frequency of the contrast enhancement. That is, a diffuse but high contrast object can be registered to a better precision than its width, with the precision limit being set by the contrast-to-noise ratio. This is the core concept of "localization" microscopy [155] and applies here as well.

We expect this advantage to be most relevant during the first few electron $\text{Å}^{-2}$ of dose in the movie, as the beam-induced motion is largest during that time. These frames also contain the highest resolution structural information, which diminishes with increasing electron dose due to radiation damage of the sample.

However, we have found that in SPA of apoferritin, motion correction at a lower dose per frame than typically used in defocus SPA ($\sim 1\,\text{electron}\,\text{Å}^{-2}$) does not noticeably increase the resolution of the final reconstruction. We conducted an experiment by collecting movies with a per-frame dose of $0.25\,\text{electron}\,\text{Å}^{-2}$ and then performing reconstructions using the resulting data by combining adjacent sets of 1, 2, 4, 8, 16, or 32 frames to simulate data taken with a higher per-frame dose. The Fourier shell correlations of the reconstructions are shown in figure 12.12, along with the corresponding resolutions as a function of per-frame dose. The reconstruction resolution is only made worse when the per-frame dose is

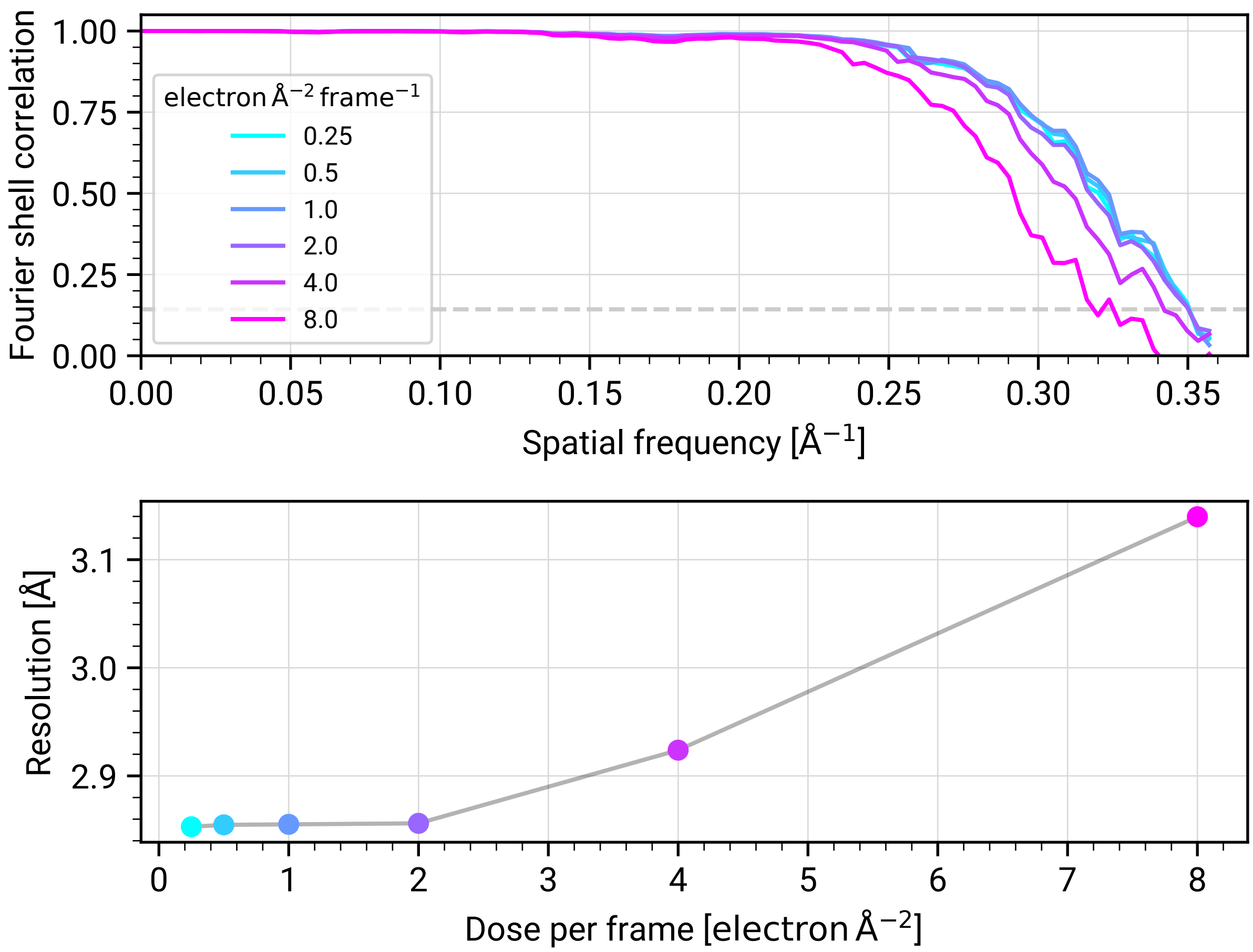

Figure 12.12: Resolution of apoferritin reconstructions using the laser phase plate with different levels of dose fractionation. Top panel: Fourier shell correlation of each reconstruction. Bottom panel: resolution (Fourier shell correlation = 0.143) of each reconstruction as a function of electron dose per frame.

$> 2$ electron Å$^{-2}$. Alignment at this per-frame dose is already usually possible using defocus SPA, and so it is unclear if the LPP provides any benefit in this context. However, it is also possible that the $\sim 3$ Å resolution limit of our TEM is simply not sufficient to see any improvement from superfractionation, and it may in fact help to achieve ultra-high resolution ($< 2$ Å) reconstructions. It could also be the case that superfractionation will benefit SPA of smaller particles, which inherently generate less contrast.

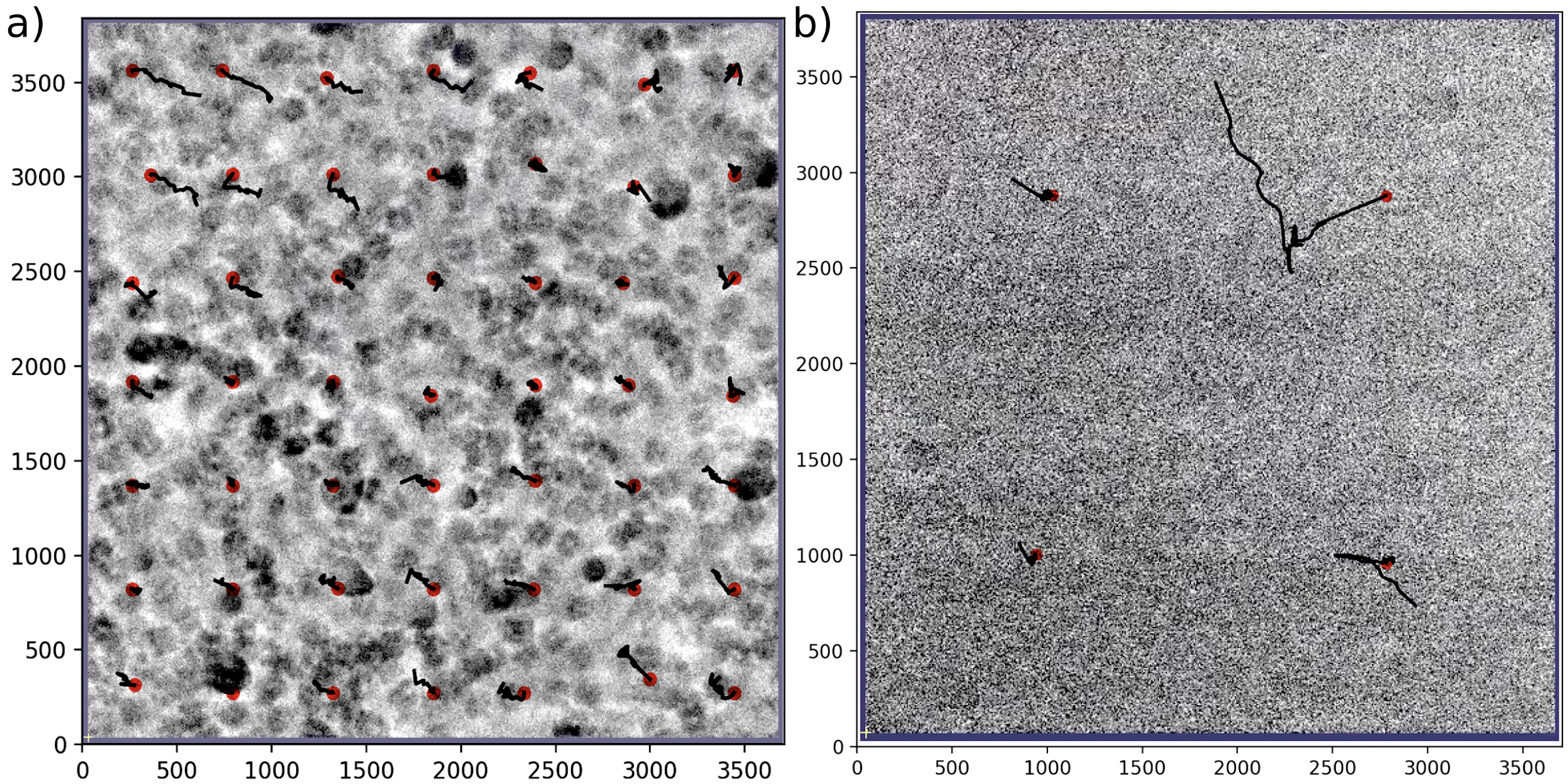

Figure 12.13: Motion correction using MotionCor2 of beam-induced motion on a 30° tilted single-particle analysis sample of RuBisCO. Electron dose per frame is 0.1 electron Å$^{-2}$ with 150 frames per movie (3 s movie exposure, 20 ms frame exposure). Motion trajectories are plotted on top of the center of each patch area, magnified by a factor of 10. Axis labels denote pixel number; the pixel size is 0.72 Å. Both images are defocused by 1.5 µm. **a)** LPP on: successful motion correction using 7 × 7 patches. **b)** LPP off: unsuccessful motion correction using 2 × 2 patches. Courtesy of Shawn Zheng of the Chan Zuckerberg Initiative.

It is also possible to reduce or nearly eliminate beam-induced sample motion by using gold sample grids with small-diameter (∼ 200 nm) holes [125]. This provides a much simpler solution to motion correction for SPA samples than installation of an LPP, though some types of sample may not be compatible with these grids. It is worth noting that such sample grids themselves might not be compatible with the LPP, because strong ghost images formed by the grid material will be projected across the entire width of the hole (depending on the hole width and ghost image separation distance, see section 8.8), corrupting the image of the sample within the hole.

As such, we believe that LPP superfractionation for SPA may only be necessary in niche applications where alternative methods of reducing motion blur from beam-induced motion are not possible.

Cryo-electron tomography, however, greatly suffers from beam-induced motion, both because tilted samples result in more motion in the sample plane and because focused ion beam (FIB)-milled lamellae are subject to more beam-induced deformation than the rel-atively small-diameter and well-supported sample grid holes in SPA [156]. As an initial demonstration of using the LPP for cryo-ET motion correction, we imaged SPA RuBisCO samples with the LPP at a sample tilt angle of 30°. We were successfully able to perform motion correction (using MotionCor2 [134]) on a $7 \times 7$ grid of sub-image patches with a per-frame dose of 0.1 electron $\text{Å}^{-2}$ and 150 frames. In a comparison dataset with the LPP off, motion correction of only $2 \times 2$ patches was unreliable (see figure 12.13). Both datasets used a defocus of 1.5 µm.

# Chapter 13

# Outlook

Now that a fully-functional LPP prototype has been constructed, work on demonstrating its applications is progressing rapidly. Indeed, all of the imaging data presented in chapter 12 were collected in the 12 months prior to the writing of this thesis. Also in progress is, further development work which will result in upgraded designs for the LPP which improve its performance or applicability. This chapter briefly reviews some of these future applications and upgrades.

## 13.1 Single-particle analysis

SPA results using the LPP have so far not indicated an advantage over standard defocus phase contrast. We do not yet understand exactly why this is the case. Perhaps there is no advantage to be had, or maybe we have not yet figured out how to modify the standard SPA data analysis pipeline to take advantage of the additional contrast provided by the LPP. It is also possible that the higher chromatic and spherical aberration of our custom TEM masks the advantage of using the LPP. We will be able to investigate this possibility with the delivery of a new TEM in 2024 (see section 13.2). Additionally, we have not yet figured out a method for consistently collecting in-focus SPA datasets, which if done properly should almost certainly improve the capabilities of SPA by nearly doubling the average information content per image at near-atomic spatial frequencies (as well as by much more at low spatial frequencies). We are also looking into more niche applications which use the LPP's enhanced low-frequency contrast (even when defocused) to determine the tertiary or quaternary structure of biological macromolecules which have a high degree of conformational heterogeneity (i.e. "floppy" molecules).

## 13.2 Cryo-electron tomography

We believe that the LPP has huge potential in cryo-ET, since the enhanced low-frequency contrast can be used to identify the location, orientation, and general structure of macro-molecules within a cellular environment. In pursuit of this application, our team will receive a new custom TEM in 2024. This TEM, based off on a Thermo Fisher Scientific Krios, will include standard modern features like an automatic sample loader, stable sample stage, lower ice contamination rate, and imaging energy filter which will make the instrument more easily usable for cryo-ET as compared to our current TEM. A schematic of the design is shown in figure 13.1. It also incorporates a updated relay optics system to accommodate the LPP which has been designed (based off of the results presented in this work) to minimize the chromatic aberration coefficient and the effects of thermal magnetic field noise. The phase plate module also provides more room inside of the TEM column for the installation of larger LPP designs (see section 13.3). The system will use a cold field emission electron source (rather than a monochromator, which we have struggled to use on our current TEM) to reduce the electron beam energy spread and reduce the influence of chromatic aberration on the resolution. Finally, the TEM will include a spherical aberration coefficient corrector. This will enable true in-focus phase contrast imaging which will provide a near-ideal contrast transfer function (CTF) up to the theoretical resolution limit of nearly 1 Å. We will investigate how this capability can be applied in the context of cryo-ET, where the shallow ($\sim 1\,\mathrm{nm}$) depth of field provided by the near-ideal CTF may allow for three-dimensional structural or position information to be derived from series of image taken at different defocus values near in-focus, or two-dimensional template matching of nominally in-focus images [147], [157].

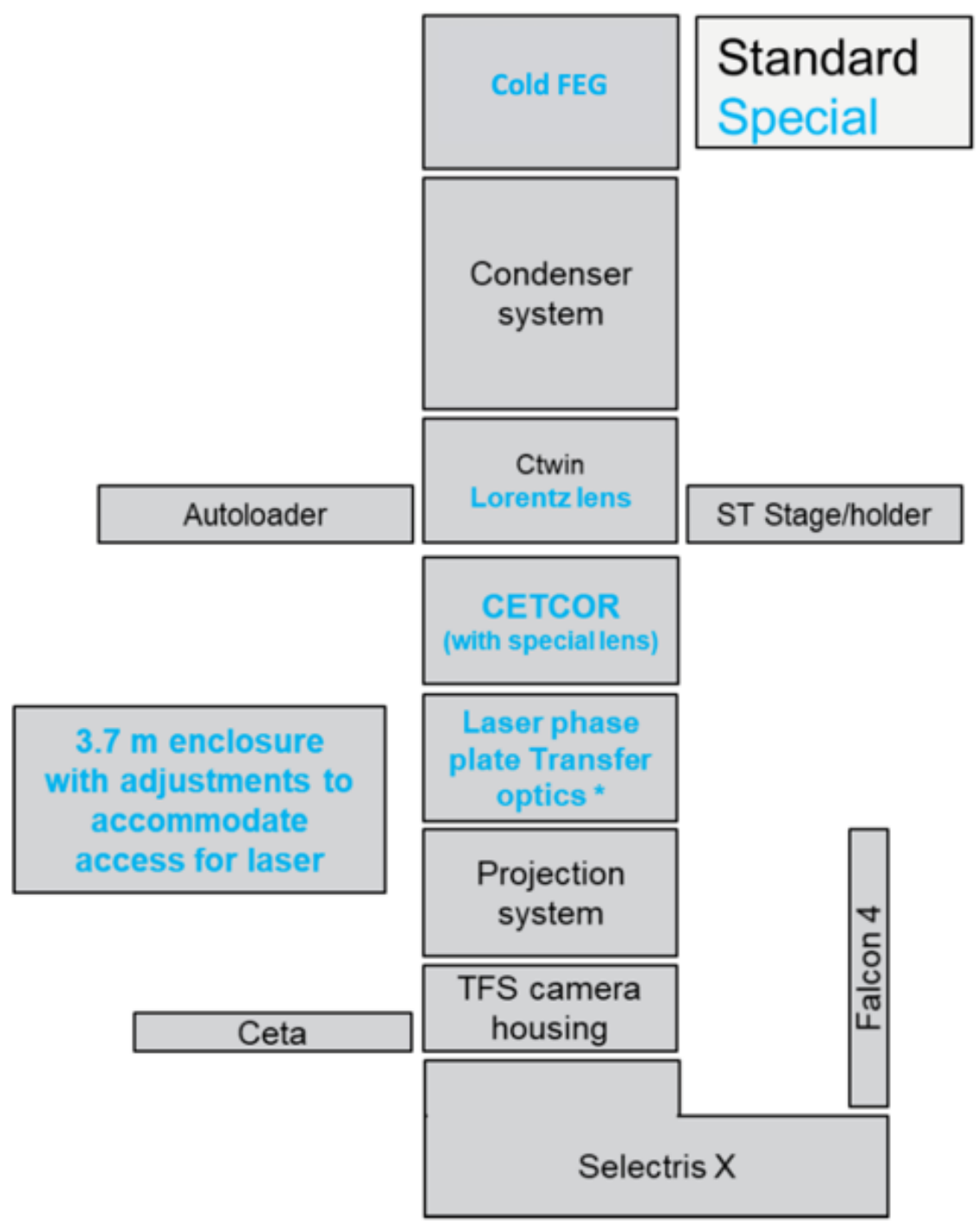

Figure 13.1: Schematic of our new custom laser phase plate-compatible transmission electron microscope, to be installed in 2024. It is based on a Thermo Fisher Scientific Krios, and incorporates a cold field emission electron source, spherical aberration corrector, phase plate module, and imaging energy filter. Courtesy of Bart Buijjse at Thermo Fisher Scientific.

We will need to develop cryo-ET automated data collection procedures compatible with the LPP. We have already conducted preliminary tests using our current TEM which repurpose some of the main alignment steps we have used for automated SPA data collection (see section 10.4). Part of this development work may require designing an LPP-compatible method for using beam-image shift, which would also be useful for increasing throughput in LPP SPA.

## 13.3 Dual laser phase plate

The CTF of the LPP can be made even closer to ideal by using two laser beams which have overlapping foci. If the foci are slighly offset from one another along the electron beam axis, then each beam only need provide a $\pi/4$ phase shift, which increases the average value of the CTF for spatial frequencies which intersect either of the beams outside of the overlap region, further enhancing the contrast for low spatial frequencies as compared to a single LPP. The lower phase shift per beam also has the advantage of substantially reducing the contrast of "ghost" images generated by their standing wave structure, which should be a benefit especially in cryo-ET where ghost images, if strong enough, might complicate the interpretation of the LPP images.

If the foci are in the same plane along the electron beam axis and both laser beams are phase-coherent, then each beam only need provide what would be the equivalent of a $\pi/8$ phase shift, since the amplitudes of the laser waves at the foci interfere constructively. The resulting interference pattern further increase the low spatial frequency contrast. In this configuration, relativistic effects similar to those considered in chapter 9 allow for additional control of the shape of the LPP phase profile. However, such a setup is more technically difficult to implement than the offset dual LPP, where each cavity can operate independently using separate laser systems.

The design and construction of an (offset) dual LPP is currently in progress, and will be tested in the new TEM.

## 13.4 Miniaturized laser phase plate

One disadvantage of the current LPP design is that it requires an additional section in the TEM column which provides space for the LPP cavity and houses the extra relay electron optics required to provide a (magnified) diffraction plane at the location of the LPP. If the LPP cavity could be made substantially smaller to fit within a volume of roughly $1\,\mathrm{mm}\,\times$ $7\,\mathrm{mm} \times 10\,\mathrm{mm}$ then it might be possible to install the system in the objective aperture port of a standard TEM, or at least in a customized version thereof. This would likely reduce the cost of the system and therefore make it more widely accessible.

We are collaborating with Osip Schwartz's group (Weizmann Institute of Science) to design such an LPP. Due to the extremely tight space constraints, it will likely require a radical rethink of the cavity design, perhaps incorporating more than two mirrors and a highly astigmatic cavity mode which will reduces diffraction loss on the necessarily small mirror surfaces and increase the phase shift per circulating power by forming a focus which is extended along the axis of the electron beam.

## 13.5 Inelastic scattering phase contrast

This work has only considered phase contrast imaging using the elastically scattered component of the electron wave function. The inelastically scattered component typically has a large spread in kinetic energies which is tens of electronvolts lower than the elastically scattered component [40]. In a typical TEM with a chromatic aberration coefficient of several millimeters, this severely limits the resolution of the phase contrast images formed by the inelastic component (just as the incident electron beam kinetic energy spread limits the elastic component image resolution as explained in section 8.7.1.1).[1] The blurry image formed by the inelastic component adds to the sharp one formed by the elastic component, which increases the shot noise without increasing the (high resolution) contrast. For this reason, conventional defocus phase contrast imaging often filters out the inelastically scattered component before the detector (especially when imaging thicker samples which generate more inelastic scattering). We did not use an image energy filter in this work because we did not have one installed on our TEM.

The sample phase information carried by the inelastically scattered component can in principle be recovered if the TEM is equipped with a chromatic aberration corrector [159]. This allows both the elastic and inelastic components to be imaged at essentially the same defocus, so that both components form high-resolution images. However, the inelastic component has substantially lower spatial coherence than the elastic component [160]. This limits the resolution of the inelastic component unless the defocus is near zero (see section 8.7.1.2). The useful defocus range for near-atomic resolutions may be as small as $\sim 100\,\mathrm{nm}$ (smaller than the thickness of some samples) [160], meaning that a tradeoff exists with increasing sample thickness: there is more inelastic scattering to recover, but less of it will be usefully in-focus.

Recovering inelastic phase contrast information therefore requires both a chromatic aberration corrector (to provide a consistent defocus for all energy components) and a phase plate (to provide in-focus contrast). The limited spatial coherence of the inelastic component also potentially limits the efficacy of the phase plate, in the same way as is discussed in section

[1]Dickerson and Russo [158] provide an excellent demonstration that the inelastic component is indeed phase modulated by the sample.

8.7.5 in the context of the elastic component. It therefore remains an interesting question as to exactly how much advantage can be derived by using a chromatic aberration corrector with a phase plate, because it is likely substantial [153].

## 13.6 Summary

The LPP is the first realization of a stable phase plate for TEM. It comes 77 years after phase contrast TEM was first proposed [4]. After one year exploring the capabilities of our first fully-functional LPP prototype, there appear to be many potential applications for the technology, as well as many new questions about its ultimate limitations. I hope and expect that the next few years will see a rapid development in the use and applications of the LPP in cryo-electron microscopy, and that the optical technology developed for it will find use in other fields.